%% file: these.tex
\documentclass[11pt]{thloria}

\usepackage{latexsym}
\usepackage{amssymb}
\usepackage[latin1]{inputenc}
\usepackage[all]{xy}

\input{includeThese}

\font\tenmsb=msbm10 scaled\magstep1\font\sevenmsb=msbm7 scaled\magstep1
\font\fivemsb=msbm5 scaled\magstep1\newfam\msbfam
\textfont\msbfam=\tenmsb\scriptfont\msbfam=\sevenmsb
\scriptscriptfont\msbfam=\fivemsb
\newcommand\upun[1]{\uppercase{\underline{\underline{#1}}}}
\FormatHeadingsWith\upun
 
\makeindex

\newcommand\itsfheadings[1]{\textit{\textsf{#1}}}
\FormatHeadingsWith{\itsfheadings}
\newcommand{\comment}[1]{}

\newlength{\largeurtexte}
\begin{document}

\pagestyle{ThesisHeadingsI}

%-------------------------------------------------------------------
%	Premiere page
%-------------------------------------------------------------------
%% Page de titre:  
\ThesisTitle{Calcul de réécriture~: \\
		fondements et applications}
\ThesisDate{25 octobre 2000}  %{\_\_ \_\_\_\_\_\_\_\_\_\_\_\_\_\_ \_\_\_\_}%
\ThesisAuthor{Horatiu Cirstea}  

%% Type de la these (autre solution: \ThesisINPL)  
\ThesisNancyI  

%% Jury: 
\ThesisJuryTitle{Composition du jury\\
} 
\President    = {%
Guy Cousineau, Professeur, 
	Université Denis Diderot - Paris \uppercase\expandafter{\romannumeral7}, France%,\\
}  
\Rapporteurs  = {%
Nachum Dershowitz, Professeur, 
	Université de Tel Aviv, Israel\\
Thérèse Hardin, Professeur, 
	Université Pierre et Marie Curie - Paris \uppercase\expandafter{\romannumeral6}, France\\  
François Lamarche, Directeur de recherche, 
	INRIA, France
}  
\Examinateurs = {%
Alexander Bockmayr, Professeur, 
	Université Henri Poincaré - Nancy 1, France\\%,\\
Claude Kirchner, Directeur de recherche, 
	INRIA, France
} 

%\nthks  %% pas de remerciements automatiques de TheseCRIN

%% Creation de la page de titre:  
\MakeThesisTitlePage  

%-------------------------------------------------------------------
%	DEDICACE
%-------------------------------------------------------------------
%% On peut ne pas encadrer certaines parties.  
%\DontFrameThisInToc  
%\begin{ThesisDedication}   
%À Knuth Bendix
%\end{ThesisDedication}  

%-------------------------------------------------------------------
%	REMERCIEMENTS
%-------------------------------------------------------------------

%\begin{ThesisAcknowledgments} 
\thispagestyle{empty} 
\input remerciement
%\end{ThesisAcknowledgments}  

%-------------------------------------------------------------------
%                TABLE DES MATIERES
%-------------------------------------------------------------------

%\dominitoc[n]
%\renewcommand{\contentsname}{Sommaire}
\tableofcontents

%-------------------------------------------------------------------
%                TABLE DES FIGURES et TABLEAUX
%-------------------------------------------------------------------

%\DontWriteThisInToc
%\listoffigures

%\DontWriteThisInToc
%\listoftables

%-------------------------------------------------------------------
%                CHAPITRES
%-------------------------------------------------------------------

\mainmatter
%-------------------------------------------------------------------
%                CHAPITRES
%-------------------------------------------------------------------

% \input chapitresAll
 \input introduction

 \input chapter_1
 \input chapter_2
 \input chapter_3
 \input chapter_4

 \input chapter_5
 \input chapter_6
 \input chapter_7

 \input conclusion

%%%%%%%%%%%%%%%%%%%%%%%%%%%%%%%%%%%%%%%%%%%%%%%%%%%%%%%%%%%%

\bibliographystyle{myalpha}

%\bibliography{these}
\input{these.bbl}

%-------------------------------------------------------------------
%                INDEX
%-------------------------------------------------------------------

\input{these.ind}

%\printindex
%\BeginIndWith{}
%\printindex[aut]

%-------------------------------------------------------------------
%                RESUME ET ABSTRACT
%-------------------------------------------------------------------

{\small
\begin{ThesisAbstract} 
 
\begin{FrenchAbstract}  
\input abstractFrancais
\end{FrenchAbstract}  

%\newpage

\begin{EnglishAbstract}  
\input abstractAnglais

\end{EnglishAbstract}  

\end{ThesisAbstract} 
}%{\small 

\end{document}

%% file: includeThese.tex
\newcommand{\ra}{\rightarrow}
\newcommand{\ral}{\longrightarrow}

\newcommand{\lal}{\longleftarrow}
\newcommand{\raS}{\rightarrowtail}
\newcommand{\longra}[1]{\stackrel{{#1}}{\longrightarrow}}
\newcommand{\longla}[1]{\stackrel{{#1}}{\longleftarrow}}
\newcommand{\lra}[1]{\stackrel{{#1}}{\longleftrightarrow}}
\newcommand{\lraD}[1]{\longrightarrow_{#1}}
\newcommand{\Longra}[0]{\Longrightarrow}
\newcommand{\Lra}[0]{\Longleftrightarrow}
\newcommand{\Ra}[0]{\Longrightarrow}
\newcommand{\Raro}[0]{\longrightarrow_{\rho}}
\newcommand{\da}[1]{\downarrow_{#1}}
\newcommand{\ov}[1]{\overline{#1}}

\newcommand{\PPos}[0]{{\cal P}os}
\newcommand{\VPos}[0]{{\cal VP}os}
\newcommand{\CPos}[0]{{\cal CP}os}
\newcommand{\FPos}[0]{{\cal FP}os}

\newcommand{\Var}[0]{{\cal V}ar}

\newcommand{\Dom}[0]{{\cal D}om}
\newcommand{\Ran}[0]{{\cal R}an}
\newcommand{\VRan}[0]{{\cal VR}an}

\newcommand{\Mod}[0]{{\cal M}od}

\newcommand{\tpos}[2]{{#1}_{|{#2}}}
\newcommand{\atpos}[2]{{\lceil{#1}\rceil}_{#2}}
\newcommand{\tatpos}[3]{{#1}_{{\lceil{#2}\rceil}_{#3}}}
\newcommand{\tatp}[2]{{#1}{\lceil{#2}\rceil}}
\newcommand{\catpos}[3]{\tatpos{#1}{#2}{#3}}

\newcommand{\subs}[1]{\langle#1\rangle}
\newcommand{\sbs}[0]{\mapsto}
\newcommand{\subS}[1]{\langle\hspace{-2pt}\langle#1\rangle\hspace{-2pt}\rangle}

\newcommand{\eni}[2]{#1=1,\ldots,#2}

\newcommand{\elan}{\textsf{ELAN}}

\newcommand{\rewrite}{\Rightarrow}
\newcommand{\ifwhere}{if-where}
\newcommand{\ifwherechoose}{if-where-choose}
\newcommand{\defgrammar}{::=}
\newcommand{\assign}{:=}

\newcommand{\iif}{\texttt{if}}
\newcommand{\where}{\texttt{where}}

\newcommand{\choice}{\texttt{choose}}
\newcommand{\try}{\texttt{try}}
\newcommand{\dc}{\texttt{dc}}

\newcommand{\dk}{\texttt{dk}}

\newcommand{\first}{\texttt{first}}

\newcommand{\repeate}{\texttt{repeat}}

\newcommand{\id}{\texttt{id}}
\newcommand{\fail}{\texttt{fail}}
\newcommand{\End}{\texttt{end}}

\newcommand{\laCal}[0]{\mbox{$\lambda$-calcul}}

\newcommand{\meqq}[0]{\ll^?}
\newcommand{\meqqT}[0]{\ll^?_T}
\newcommand{\meqqes}[0]{\ll^?_\emptyset}
\newcommand{\ww}[0]{\:\wedge\:}
\newcommand{\fF}[0]{{\mathbf F}}

\newcommand{\ID}[0]{{\mathbb {ID}}}
\newcommand{\Sl}[0]{{\cal S}olution}

\newcommand{\RTT}[0]{{\varrho}(\FF,\XX)}
\newcommand{\RTTE}[0]{{\varrho_{\emptyset}}(\FF,\XX)}
\newcommand{\RTTf}[0]{{\varrho^{1st}}(\FF,\XX)} 	%%%% first %%%%%
\newcommand{\RTL}[0]{{\varrho_{\lambda}}(\FF,\XX)} 	%%%% lambda %%%%%
\newcommand{\RTLp}[0]{{\varrho_{\lambda}^a}(\FF,\XX)} 	%%%% lambda-pur %%%%%
\newcommand{\RTTs}[0]{{\varrho^{\sigma}}(\FF,\XX)} 	%%%% sigma %%%%%
\newcommand{\rtt}[0]{\mbox{$\varrho_{-}(\FF,\XX)$}}
\newcommand{\rttf}[0]{{\varrho^{1st}_{-}}(\FF,\XX)} 	%%%% first %%%%%

\newcommand{\normF}[1]{#1\downarrow}

\newcommand{\paron}[1]{\langle#1\rangle}  %%%% first %%%%%
\newcommand{\paronle}[0]{\langle}
\newcommand{\paronri}[0]{\rangle}
\newcommand{\parod}[1]{\langle\hspace{-2pt}\langle#1\rangle\hspace{-2pt}\rangle}  %%%% dc %%%%%

\newcommand{\conserv}[0]{quasi-régulière}
\newcommand{\conservs}[0]{quasi-régulières}
\newcommand{\linr}[0]{strictement linéaire à droite}
\newcommand{\linrs}[0]{strictement linéaires à droite}
\newcommand{\stable}[0]{stable}
\newcommand{\stables}[0]{stables}

\newcommand{\subf}[0]{{subsume faiblement}}
\newcommand{\subfn}[0]{{subsume pas faiblement}}

\newcommand{\matchD}[0]{\mbox{$\rho$-préfiltrables}}
\newcommand{\matchDs}[0]{\mbox{$\rho$-préfiltrable}}

\newcommand{\safe}[0]{\mbox{$\rho$-safes}}
\newcommand{\safes}[0]{\mbox{$\rho$-safe}}

\newcommand{\ready}[0]{\mbox{$\rho$-calculables}}
\newcommand{\readys}[0]{\mbox{$\rho$-calculable}}

\newcommand{\rhol}[0]{{\rho_\lambda}}
\newcommand{\rholp}[0]{{\rho_\lambda^a}}
\newcommand{\rhoe}[0]{{\rho_{\emptyset}}}
\newcommand{\rhof}[0]{\rho^{1st}} %%%% first %%%%%
\newcommand{\roS}[0]{\rho\sigma}

\newcommand{\sigRo}[0]{\sigma_\rho}

\newcommand{\roCal}[0]{\mbox{$\rho$-calcul}}
\newcommand{\roCalE}[0]{\mbox{$\rho_\emptyset$-calcul}}
\newcommand{\roCalEp}[0]{\mbox{$\rho_\emptyset^{+}$-calcul}}
\newcommand{\roCalT}[0]{\mbox{$\rho_T$-calcul}}
\newcommand{\roCalA}[0]{\mbox{$\rho_A$-calcul}}
\newcommand{\roCalC}[0]{\mbox{$\rho_C$-calcul}}
\newcommand{\roCalAC}[0]{\mbox{$\rho_{AC}$-calcul}}
\newcommand{\roCalfT}[0]{\mbox{$\rho^{1st}_T$-calcul}} %%%% first %%%%%
\newcommand{\roCalf}[0]{\mbox{$\rho^{1st}$-calcul}}
\newcommand{\roCalL}[0]{\mbox{$\rhol$-calcul}} %%%% lambda %%%%%
\newcommand{\roCalLp}[0]{\mbox{$\rholp$-calcul}} %%%% lambda %%%%%
\newcommand{\sigroCal}[0]{\mbox{$\sigma_\rho$-calcul}} %%%% sigma rho %%%%%
\newcommand{\sigCal}[0]{\mbox{$\sigma$-calcul}} %%%% sigma %%%%%
\newcommand{\roSig}[0]{\mbox{$\roS$-calcul}}

\newcommand{\SetN}{S}
\newcommand{\FireN}{F}
\newcommand{\FireNp}{\FireN_\parallel}

\newcommand{\congr}{\longrightarrow_{\SetN}}
\newcommand{\congrxy}{{\SetN}}
\newcommand{\congTR}{\stackrel{*}{\longrightarrow}_{\SetN}}
\newcommand{\congTRxy}{{\SetN}^{*}}
\newcommand{\congE}{\stackrel{*}{\longleftrightarrow}_{\SetN}}
\newcommand{\congExy}{{\SetN}^{*}}

\newcommand{\Cong}[2]{#1 \congr #2}
\newcommand{\CongE}[2]{#1 \congE #2}
\newcommand{\CongTR}[2]{#1 \congTR #2}

\newcommand{\roZ}{\longrightarrow_{\FireN}}
\newcommand{\roZxy}{{\FireN}}
\newcommand{\roTR}{\stackrel{*}{\longrightarrow}_{\FireN}} 
\newcommand{\roZE}{\longrightarrow_{\FireN/\SetN}} 
\newcommand{\roZExy}{{\FireN/\SetN}}
\newcommand{\roZETR}{{\stackrel{*}{\longrightarrow}_{\FireN/\SetN}}}

\newcommand{\RoZ}[2]{#1 \roZ #2}
\newcommand{\RoTR}[2]{#1 \roTR #2}
\newcommand{\RoZE}[2]{#1 \roZE #2}
\newcommand{\RoZETR}[2]{#1 \roZETR #2}

\newcommand{\del}{\longrightarrow_{\FireNp}} 
\newcommand{\delxy}{{\FireNp}} 
\newcommand{\delE}{\longrightarrow_{\FireNp/\SetN}}
\newcommand{\delExy}{{\FireNp/\SetN}}
\newcommand{\delETR}{\stackrel{*}{\longrightarrow}_{\FireNp/\SetN}}

\newcommand{\Del}[2]{#1 \del #2}
\newcommand{\DelE}[2]{#1 \delE #2}

\newcommand{\sigro}{\longrightarrow_{\sigma}}
\newcommand{\sigroxy}{\sigma}
\newcommand{\sigroTR}{\stackrel{*}{\longrightarrow}_{\sigma}}
\newcommand{\sigroTRxy}{{\sigma}^{*}}

\newcommand{\setSig}{\longrightarrow_{\SigSet}}
\newcommand{\setSigxy}{\SigSetxy}
\newcommand{\setSigTR}{\stackrel{*}{\longrightarrow}_{\SigSet}}
\newcommand{\setSigTRxy}{{(\SigSetxy)}^{*}}
\newcommand{\delSig}{\longrightarrow_{\SigDel}}
\newcommand{\delSigxy}{\SigDelxy}
\newcommand{\roZSig}{\longrightarrow_{\SigRoZ}}

\newcommand{\SigSet}{\sigma^*\SetN\sigma^*}
\newcommand{\SigSetxy}{\sigma^*\SetN\sigma^*}
\newcommand{\SigDel}{\sigma^*\FireNp\sigma^*}
\newcommand{\SigDelxy}{\sigma^*\FireNp\sigma^*}
\newcommand{\SigRoZ}{\sigma^*\FireN\sigma^*}

\newcommand{\relR}{\RR}
\newcommand{\relS}{\SS}

\newcommand{\relr}{\longrightarrow_{\relR}}
\newcommand{\relrxy}{{\relR}}

\newcommand{\relrTRxy}{{\relR}^{*}}

\newcommand{\rels}{\longrightarrow_{\relS}}
\newcommand{\relsxy}{{\relS}}

\newcommand{\relsTR}{\stackrel{*}{\longrightarrow}_{\relS}}
\newcommand{\relsTRxy}{{\relS}^{*}}

\newcommand{\rep}{repeat\!*\!}

\newcommand{\NONE}{{\cal NONE}}

\newcommand{\dotctx}[0]{\cdot}
\newcommand{\nilctx}[0]{\emptyset}

\newcommand{\ofenv}[2]{{#1}_{[\hspace{-3pt}[{#2}]\hspace{-3pt}]}}

\newcommand{\rest}[2]{{#1}|_{#2}}
\newcommand{\restpv}[2]{{#1}|_{PV(#2)}}

\newcommand{\DEF}{\begin{definition}}
\newcommand{\FDEF}{\end{definition}}
\newcommand{\PROP}{\begin{proposition}}
\newcommand{\FPROP}{\end{proposition}}
\newcommand{\COR}{\begin{corollary}}
\newcommand{\FCOR}{\end{corollary}}
\newcommand{\EX}{\begin{example}}
\newcommand{\FEX}{\end{example}}
\renewcommand{\TH}{\begin{theorem}}
\newcommand{\FTH}{\end{theorem}}
\newcommand{\LEM}{\begin{lemma}}
\newcommand{\FLEM}{\end{lemma}}
\newcommand{\FACT}{\begin{fact}}
\newcommand{\FFACT}{\end{fact}}
\newcommand{\REM}{\begin{remark}}
\newcommand{\FREM}{\end{remark}}
\newcommand{\THESE}{\begin{these}}
\newcommand{\FTHESE}{\end{these}}

\edef\maxunit{chapter}

\ifx\theorem\undefined
\newtheorem{theorem}{Théorème}[\maxunit]
\fi
\ifx\proposition\undefined
\newtheorem{proposition}{Proposition}[\maxunit]
\fi
\ifx\lemma\undefined
\newtheorem{lemma}{Lemme}[\maxunit]
\fi
\ifx\example\undefined
\newtheorem{example}{Exemple}[\maxunit]
\fi
\ifx\definition\undefined
\newtheorem{definition}{Définition}[\maxunit]
\fi
\ifx\corollary\undefined
\newtheorem{corollary}{Corollaire}[\maxunit]
\fi
\ifx\fact\undefined
\newtheorem{fact}{Remarque}[\maxunit]
\fi
\ifx\remark\undefined
\newtheorem{remark}{Remarque}[\maxunit]
\fi
\ifx\these\undefined
\newtheorem{these}{Thèse}[\maxunit]
\fi

\ifx\proof\undefined
\newcommand{\proof}[1]{\begin{description}
            \item[Preuve~:] #1 \(\Box\) \end{description}}
\fi

\ifx\proofHC\undefined
\newcommand{\proofHC}[1]{\begin{description}
            \item[Preuve en développement~:] #1 \(\Box\) \end{description}}
\fi

\ifx\proofFR\undefined
\newcommand{\proofFR}[1]{\begin{description}
            \item[Preuve~:] #1 \hfill \(\Box\)  \end{description}}
\fi

\newcommand{\Def}[1]{{\em #1}\index{#1}}
\newcommand{\Defi}[2]{{\em #1}\index{#2!#1}}
\newcommand{\Defn}[3]{{\em #1}\index{#2!#3}}

\newcommand{\cond}[3]{{#2} \rightarrow {#3} ~si~ {#1}}

\newcommand{\cnorm}[0]{{c_\rho}}

\newcommand{\fleche}{\mathop{\mapsto\!\!\!\!\!\mapsto}}
\newcommand{\LaFleche}{\mathop{\mapsto\!\!\!\!\!\mapsto}}

\newcommand{\rname}[1]{\mbox{$#1$}}

\newenvironment{truleset}%
{\begin{tabular}{llcl}}{\end{tabular}}%

\newcommand{\tregle}[4]{
           {\small{\rname{#2}}}     & {$#3$}  & $\fleche$       & {$#4$}\\}

\newenvironment{ruleset}%
{\[\begin{array}{llcl}}{\end{array}\]}%

\newcommand{\regle}[3]{
           {\rname{#1}}     & {#2}  & \fleche       & {#3}\\}
\newcommand{\cregle}[4]{
           {\rname{#1}}     & {#2}  & \fleche       & {#3}          \\
                        &       &                       & {\sf si}~ {#4}\\}
\newcommand{\ccregle}[5]{
           {\rname{#1}}     & {#2}  & \fleche       & {#3}          \\
                        &       &                       & {\sf si}~ {#4}\\
                        &       &                       & ~~~ {#5}\\}
\newcommand{\wregle}[4]{
           {\rname{#1}}     & {#2}  & \fleche       & {#3}          \\
                        &       &                       & {\sf avec}~ {#4}\\}

\newcommand{\cwregle}[5]{
           {\rname{#1}}     & {#2}  & \fleche       & {#3}          \\
                        &       &                       & {\sf si}~ {#4}\\
                        &       &                       & {\sf avec}~ {#5}\\}

\newcommand{\hregle}[3]{
           {\rname{#1}}     & {#2}  & \fleche \\
                            & \multicolumn{3}{l}{#3}\\}

\newcommand{\hrregle}[3]{
           {\rname{#1}}     & {#2}  & \fleche & ~ \\
                            & \multicolumn{3}{r}{#3}\\}
\newcommand{\hcrregle}[4]{
           {\rname{#1}}     & {#2}  & \fleche       & {#3}          \\
                        & \multicolumn{3}{r}{{\sf si}~ {#4}}\\}

\newcommand{\hclwlregle}[5]{
           {\rname{#1}}     & {#2}  & \fleche & {#3}\\
                        && \multicolumn{2}{l}{{\sf si}~ {#4}}\\
                        && \multicolumn{2}{l}{{\sf avec}~ {#5}}\\}
\renewcommand{\AA}[0]{{\cal A}}

\newcommand{\EE}[0]{{\cal E}}
\newcommand{\FF}[0]{{\cal F}}
\newcommand{\KK}[0]{{\cal K}}
\newcommand{\LL}[0]{{\cal L}}
\newcommand{\MM}[0]{{\cal M}}
\newcommand{\PP}[0]{{\cal P}}
\newcommand{\RR}[0]{{\cal R}}
\renewcommand{\SS}[0]{{\cal S}}
\newcommand{\TT}[0]{{\cal T}}
\newcommand{\XX}[0]{{\cal X}}
\newcommand{\YY}[0]{{\cal Y}}

\newcommand{\TFX}[0]{\TT_{\FF}(\XX)}
\newcommand{\TF}[0]{\TT_{\FF}}
\newcommand{\TFN}[0]{\TT_{\FF}(\NatP)}

\newcommand{\TR}[0]{{\cal TR}}

\font\bb=msbm10
\def\N{\hbox{\bb N}}

\newcommand{\NatP}[0]{\N^*}

\newcommand{\textBNF}[1]{\textbf{#1}}
\newcommand{\FRAC}[2]{\frac{\displaystyle #1}{\displaystyle #2}}

\def \eqdef {\,\,\mbox{\small $\stackrel{\mbox{\tiny $\triangle$}}{=}$}\,\,}

%% file: remerciement.tex
~\\
\bigskip
\bigskip
\bigskip
\bigskip
\bigskip
\bigskip
\bigskip
\bigskip

{\it 

Je voudrais tout d'abord remercier chaleureusement Claude Kirchner pour m'avoir
aidé et guidé tout au long de la préparation de cette thèse. 
Il a été toujours disponible pour discuter de mes idées, souvent pas claires ou
mal formulées~; ses critiques, commentaires et conseils ont certainement permis à 
cette thèse d'être ce qu'elle est.
Je tiens à lui exprimer ma reconnaissance pour son intérêt  et pour son soutien.

\medskip
Je tiens ensuite à remercier sincèrement ceux qui ont bien voulu prendre part à
ce jury.

\medskip
Thérèse Hardin qui 
a accepté de bien vouloir rapporter cette thèse.
Je la remercie d'avoir lu et disséqué cette thèse jusque
dans les parties les plus techniques. Ses conseils et critiques m'ont permis d'apporter
des améliorations à ce document.

\medskip
Nachum Dershowitz qui 
m'a fait l'honneur de
s'intéresser à ce travail en acceptant, malgré l'obstacle de la langue, 
la lourde tâche de rapporteur.
Ses commentaires et conseils très pertinents permettront encore d'améliorer ce
travail.

\medskip
François Lamarche qui, en tant que rapporteur interne, a accepté d'examiner cette
thèse. Je le remercie pour tout l'intérêt qu'il a manifesté pour ce travail.

\medskip
Alexander Bockmayr pour avoir accepté d'examiner ce document et de participer à
mon jury. 
Je le remercie vivement pour ses questions et commentaires fort judicieux.

\medskip
Guy Cousineau qui m'a fait l'honneur de présider le jury. Par ses questions et
ses remarques, il m'a témoigné un grand intérêt pour les travaux effectués.

\medskip

Je profite de l'occasion qui m'est offerte pour remercier toute l'équipe PROTHEO
pour son accueil.
Je remercie tout particulièrement Hélène Kirchner qui, en tant que directeur de
l'équipe PROTHEO, m'a donné tous les moyens pour poursuivre ma recherche.
Je remercie profondément Christophe Ringeissen pour ses
remarques, critiques et suggestions 
enrichissantes 
qui m'ont aidé pendant la
réalisation de cette thèse. 

\medskip

Je remercie Pierre-Etienne Moreau et Peter Borovanský %\'y 
pour leur aide concernant l'utilisation du langage \elan.

\medskip

Je remercie mes collègues de bureau, Hubert, Laurent et Sorin, pour tous les bons
moments que nous avons passé ensemble. Je leur en suis reconnaissant, à eux ainsi qu'à
Christophe 
et Raulent, pour m'avoir expliqué les finesses de la langue, de la
philosophie et de la cuisine française.

\medskip

Je remercie mon fils Paul pour ses longues nuits pendant la rédaction de cette
thèse. Je remercie Dana et toute ma famille pour leurs encouragements et leur
soutien.

}%{\it

%% file: introduction.tex
%%%%%%%%%%%%%%%%%%%%%%%%%%%%%%%%%%%%%%%%%%%%%%%%%%%%%%%%%%%
% \TLtopbookmark
\chapter*{Introduction}
\label{chap.intro}
%%%%%%%%%%%%%%%%%%%%%%%%%%%%%%%%%%%%%%%%%%%%%%%%%%%%%%%%%%%%

La notion de réécriture est omniprésente en informatique et en logique
mathématique. En effet, le concept de
réécriture apparaît dès
les fondements théoriques jusqu'aux réalisations logicielles.  La
réécriture est utilisée pour définir la sémantique
opérationnelle de langages de programmation~\cite{Kahn-87} aussi bien que pour
décrire la transformation de programmes~\cite{ASF-Industrial-96}. La réécriture
est utilisée pour calculer~\cite{DershowitzIC85}, implicitement ou explicitement
comme dans Mathematica~\cite{Mathematica99} ou OBJ~\cite{obj3-88}, mais
également lorsqu'on décrit par des règles d'inférence une
logique~\cite{GirardLafontTaylor89}, un prouveur de
théorèmes~\cite{JouannaudKirchnerSIAM86} ou un solveur de
contraintes~\cite{JouannaudKirchner-rob91}. La réécriture est naturellement très
importante dans les systèmes où la notion de règle est un objet explicite du
premier ordre, comme les systèmes experts, les langages de programmation basés
sur la logique équationnelle~\cite{Odonnell-77}, les spécifications algébriques
(e.g. OBJ~\cite{obj3-88}) et les systèmes de transition
(e.g. Murphi~\cite{OverviewOfMurphi-1992}).

\section*{La réécriture}
%================================================================

Un exemple d'utilisation de la réécriture très simple mais très souvent
rencontré dans la pratique est le mécanisme ``Rechercher/Remplacer''. Tout
éditeur de texte dispose d'un tel mécanisme permettant le remplacement
(i.e. la réécriture) d'une chaîne de caractères par une autre chaîne de caractères.

Nous pouvons illustrer l'utilisation de ce mécanisme de remplacement sur un
exemple consistant à
transformer automatiquement des programmes écrits en un langage initial ayant une
construction conditionnelle %de la forme 
``{\em si <test> ~alors <instructions> ~fin}'' en des programmes écrits en un
langage cible avec une construction conditionnelle de la forme 
``{\em if (<test>) then \{<instructions>\}}''. Cette transformation peut être
réalisée facilement en remplaçant les mots-clés du langage initial par les
mots-clés du langage cible et ceci est représenté par le système
contenant les trois règles de réécriture~:
    \begin{eqnarray*}
      si & \ra & if ~( \\
      alors & \ra & ~)~ then ~\{ \\
      fin & \ra & ~\} 
    \end{eqnarray*}

La transformation d'un programme en utilisant une telle approche implique trois
applications du mécanisme ``Rechercher/Remplacer'' correspondant à l'application 
des trois règles de réécriture. Mais on dispose habituellement d'un mécanisme de 
remplacement des expressions régulières permettant non seulement le remplacement 
des chaînes de caractères définies explicitement en précisant tous leurs
caractères mais aussi la transformation des chaînes de caractères respectant un
certain motif.

Dans notre exemple de transformation de syntaxe nous voulons remplacer
directement une construction conditionnelle du langage initial par une construction
similaire du langage cible en transformant les mots-clés et la forme
mais sans modifier les conditions de test et les instructions. Les expressions
régulières sons décrites en utilisant des caractères spéciaux et par exemple,
nous pouvons transformer le motif 
``\texttt{si $\backslash(.*\backslash)$ alors $\backslash(.*\backslash)$ fin}''
en ``\texttt{if $(\backslash 1)$ then $\{\backslash 2\}$}''
\footnote{motifs utilisés dans l'outil ``Replace(Regexp)'' d'Emacs correspondant 
à l'exécution de la commande UNIX 
\texttt{sed s/si $\backslash(.*\backslash)$ alors $\backslash(.*\backslash)$ fin/if $(\backslash 1)$ then $\{\backslash 2\}/$}}. 

On précise ainsi que les chaînes de caractères correspondant au test et aux
instructions dans le premier motif peuvent contenir n'importe quel caractère et
que ces chaînes sont utilisées dans le deuxième motif aux positions
correspondantes. Le même comportement est décrit d'une manière plus concise et
plus claire par la règle de réécriture 
      $$si ~C~ alors ~I~ fin  ~~\ra~~  if ~( C )~ then ~\{ I \}$$

L'application de cette règle à un programme consiste à chercher une instruction
du programme qui est obtenue en instanciant les variables $C$ et $I$ du membre gauche
$si ~C~ alors ~I~ fin$ par des expressions appropriées et à
remplacer cette instruction par le membre droit $if ~(C)~ then ~\{I\}$ où les
variables $C$ et $I$ sont instanciées par les expressions obtenues précédemment.
Par exemple, l'application de la règle de réécriture ci-dessus à l'instruction 
$si~ a>b ~alors~ a=a-1 ~fin$ mène à l'instanciation des variables $C$ et $I$ par
$a>b$ et respectivement $a=a-1$ et ainsi l'instruction initiale %$si~ a>b ~alors~ a=a-1 ~fin$
est réécrite en $if~ (a>b) ~then~ \{a=a-1\}$.

Le mécanisme déterminant les instanciations appropriées pour les variables est
appelé {\em filtrage}. Dans le cas des expressions régulières seulement les
motifs linéaires (i.e. les termes contenant une seule fois chacune de leurs
variables) peuvent être utilisés dans le filtrage menant ainsi à un pouvoir
d'expression limité. Dans la réécriture on n'impose pas une telle
restriction et on considère des termes quelconques dans le membre gauche des
règles de réécriture.

L'application d'une règle de réécriture à un terme clos du premier
ordre suppose le filtrage entre le membre gauche de la règle et le terme à
réécrire et ensuite le remplacement des variables du membre droit de la règle
par les termes obtenus par le filtrage. Mais le filtrage peut échouer et dans
ce cas la règle de réécriture ne s'applique pas. D'un autre côté, un système de
règles de réécriture (i.e. un système de réécriture) peut contenir plusieurs
règles de réécriture qui peuvent être appliquées à un même terme.  Les
propriétés des systèmes de réécriture ainsi obtenus, comme la terminaison et la
confluence, ne sont pas toujours vérifiées et par conséquent le comportement de
ces systèmes ne peut pas être garanti.

Considérons par exemple une structure de liste construite en utilisant
l'opérateur $\otimes$ ayant des éléments de la forme $elem(n)$ avec $n$ un
entier. La transformation des listes de telle manière que tout élément $elem(0)$
précédant un autre élément est éliminé peut être réalisée en utilisant la règle
de réécriture suivante~:
	$$elem(0) ~ \otimes ~ x  ~~\ra~~  x$$ 
L'extraction du premier élément d'une telle liste est décrite par l'opérateur
$extract$ avec un comportement défini par la règle de réécriture~:
	$$extract(elem(x)~ \otimes ~ L)  ~~\ra~~  elem(x)$$ 
Dans le système de réécriture contenant les deux règles de réécriture présentées 
ci-dessus le terme $extract(elem(0)~\otimes~elem(1)~\otimes~elem(2))$ est évaluée soit
en $extract(elem(1)~\otimes~elem(2))$ et puis en $elem(1)$, soit directement en
$elem(0)$ et donc, le résultat (la forme normale) n'est pas unique.

Pour parcourir l'espace des résultats, une idée consiste à enrichir
les systèmes de règles par des constructions permettant de contrôler
l'application des règles de réécriture. Ceci est réalisé généralement en
ajoutant du contrôle sous forme de conditions, affectations locales, etc., aux
règles de réécriture ainsi qu'en introduisant une notion de stratégie qui est
utilisée pour décrire le processus de normalisation.  On obtient ainsi la notion
de système de calcul~\cite{KirchnerKV-MIT95} intégrant les règles de
réécriture et leur contrôle exprimé sous forme de stratégies.

Le système de calcul pour les listes consiste en l'ensemble des règles de
réécriture présentées précédemment et la stratégie précisant que la règle de
réécriture décrivant l'élimination de $elem(0)$ est appliquée avant celle pour
l'extraction. Dans ce système de calcul le terme
$extract(elem(0)~\otimes~elem(1)~\otimes~elem(2))$ est nécessairement évalué en
$extract(elem(1)~\otimes~elem(2))$ et puis en $elem(1)$. Une autre possibilité
de contrôler les réécritures consiste à ajouter une condition $x \neq 0$ à la
règle de réécriture pour l'extraction.

Puisque la sémantique opérationnelle des stratégies peut être exprimée en
utilisant la réécriture~\cite{BorovanskyThese98}, il est naturel d'analyser la 
possibilité d'exprimer les deux systèmes de règles et de stratégies au même niveau 
d'un calcul.

On peut considérer qu'une règle de réécriture est une stratégie élémentaire et
que les stratégies générales sont construites en composant de telles stratégies
élémentaires. Mais la réécriture est un calcul du premier ordre et son pouvoir
d'expression n'est pas suffisant pour décrire directement la composition de
règles de réécriture. Les mécanismes permettant de telles opérations sont
offerts par le \laCal, un système de réécriture d'ordre supérieur qui a été
introduit pour exprimer simplement la fonctionnalité.

\section*{Le \laCal}
%================================================================

Le \laCal\  introduit par Church~\cite{Church-41} est un langage expressif possédant
une sémantique simple et suffisamment puissante pour exprimer toutes les
fonctions calculables. Une fonction est représentée dans le
\laCal\  en utilisant une abstraction par rapport à ses arguments et
l'application d'une fonction à un terme est réalisée en substituant le terme à
la variable abstraite correspondante.

Dans le \laCal\  de base l'abstraction est réalisée par rapport à une variable et
on n'impose aucune restriction de contexte pour les variables
abstraites. L'application d'une $\lambda$-abstraction $\lambda x.t$ à un terme
$u$ consiste à substituer la variable abstraite $x$ dans le terme $t$ par le
terme $u$, transformation appelée $\beta$-réduction. On doit mentionner que
cette substitution n'est pas un simple remplacement d'une variable par un terme
mais doit tenir compte des éventuels conflits entre les noms des
variables. Cette opération de substitution utilise l'$\alpha$-conversion qui
permet d'éviter la capture des variables. Elle est définie au méta-niveau du
\laCal.

Le \laCal\  avec motifs~\cite{Peyton87} enrichit le \laCal\  avec une
information de contexte permettant des motifs plus élaborés qu'une simple
variable dans l'abstraction. Dans ce cas, l'application d'une abstraction
$\lambda m.t$ où $m$ est un motif (terme du premier ordre) à un terme $u$
nécessite l'utilisation du même mécanisme de filtrage que dans la réécriture.  La
substitution résultant du filtrage peut impliquer plusieurs ou aucune variable
et non une seule comme dans le \laCal\  de base où le filtrage est trivial.

Dans ce calcul nous pouvons définir les abstractions
$\lambda(elem(0)~\otimes~x)~.~x$ et respectivement 
$\lambda(extract(elem(x)~\otimes ~ L))~.~elem(x)$ correspondant aux
règles de réécriture utilisées précédemment pour les listes. 
Par rapport à la réécriture, les termes du \laCal\  contiennent toute
l'information nécessaire pour leur évaluation et le $\lambda$-terme
décrivant l'extraction d'un élément 
$\lambda(extract(elem(x)~\otimes ~ L))~.~elem(x)~(extract(\lambda(elem(0)~\otimes~x)~.~x~(elem(0)~\otimes~elem(1)~\otimes~elem(2)))$
est évalué en
$\lambda(extract(elem(x)~\otimes ~ L))~.~elem(x)~(extract(elem(1)~\otimes~elem(2)))$
et puis en $elem(1)$.

Comme dans la réécriture, le filtrage peut échouer mais contrairement à la
réécriture ceci est représenté explicitement en introduisant une construction
\textit{FAIL} qui est obtenue comme résultat d'une application avec échec.
Par exemple l'application 
$\lambda(elem(0)~\otimes~x)~.~x~(elem(1)~\otimes~elem(0))$ ainsi que l'application
de la règle de réécriture $elem(0)~\otimes~x~\ra~x$ au terme 
$(elem(1)~\otimes~elem(0))$ mène à un échec
mais tandis que dans le premier cas ceci est représenté par un résultat
\textit{FAIL}, dans le deuxième cas il n'y a aucun résultat.

Il existe d'autres recherches reliés à l'enrichissement du \laCal\  avec des
facilités de filtrage et on peut citer par exemple les travaux de Vincent van
Oostrom~\cite{Oostrom} et Loïc Colson~\cite{Colson88}.

\section*{Le non-déterminisme}
%================================================================

Nous pouvons ajouter une règle de réécriture $x~\otimes~elem(0)~\ra~x$ au
système de réécriture pour les listes permettant l'élimination des éléments
$elem(0)$ en fin de liste et ainsi, le terme $(elem(1)~\otimes~elem(0))$ est
réduit en $elem(1)$ par rapport à ce système.

On dit que le choix de la règle de réécriture à appliquer est non-déterministe
et si plusieurs règles peuvent être appliquées à un terme alors plusieurs
résultats, éventuellement différents, peuvent être obtenus. Un tel comportement
a été déjà obtenu dans le cas du système de réécriture pour les listes.

Une autre source de non-déterminisme est l'utilisation d'une théorie de filtrage
équationnelle dans la réécriture modulo classique~\cite{PS81}. En général, le
filtrage dans une telle théorie n'est pas unitaire et des résultats différents
pour le filtrage mènent à des résultats différents pour l'application d'une
règle de réécriture.

Reprenons le système de réécriture pour les listes mais cette fois-ci en
définissant l'opérateur $\otimes$ comme étant associatif-commutatif, ce qui
donne à notre objet une structure de multi-ensemble. Dans ce cas la règle de
réécriture $elem(0)~\otimes~x~\ra~x$ est suffisante pour décrire l'élimination
des éléments $elem(0)$ quelque soit leur position dans le multi-ensemble.  La
règle de réécriture $extract(elem(x)~\otimes~L)~\ra~elem(x)$ décrit maintenant
l'extraction non du premier élément d'une liste mais d'un élément quelconque
d'un multi-ensemble et ainsi le terme $extract(elem(1)~\otimes~elem(2))$ est
évalué soit en $elem(1)$, soit en $elem(2)$.

On s'intéresse souvent au développement des programmes
déterministes et ceci peut être facilement réalisé en réécriture en imposant un
ordre sur la sélection de la règle à appliquer et en utilisant seulement un
filtrage syntaxique. Mais quand ces programmes sont exécutés dans un
environnement réel on obtient souvent des comportements non-déterministes qu'on
veut représenter et analyser.  Il existe de multiples domaines où le
non-déterminisme et la possibilité de revenir en arrière pour effectuer des
réductions alternatives sont essentiels. On peut ainsi mentionner, sans être
exhaustifs, la démonstration automatique, la résolution de contraintes, la
programmation logique, la recherche des solutions optimales ou encore le
model-checking.

Dans toutes ces situations on s'intéresse aux réductions possibles à chaque étape
d'exécution ou autrement dit à tous les résultats intermédiaires de l'exécution.
Mais dans la réécriture la possibilité d'avoir plusieurs ou aucun résultat pour
l'évaluation d'un terme par rapport à un système de réécriture ne peut pas être
exprimée explicitement.

\section*{Le \roCal}
%================================================================

L'objectif de cette thèse est donc de proposer et d'étudier un calcul permettant
la définition au même niveau de représentation des règles et des stratégies ainsi
que leur application et les résultats obtenus.

Ce calcul doit être suffisamment puissant pour décrire le \laCal\  et la
réécriture. Nous considérons aussi les règles de réécriture enrichis avec des
conditions et affectations locales et nous souhaitons exprimer des stratégies
construites en partant de telles règles de réécriture. Nous nous intéressons
particulièrement à des stratégies de normalisation par rapport à un ensemble de
règles de réécriture. De plus, l'application de règles et stratégies peut
échouer ou mener à plusieurs résultats (différents) et nous voulons exprimer
explicitement ces propriétés dans le calcul.

Nous partons donc des constructions du \laCal, l'abstraction et
l'application. Puisque les membres gauches des règles de réécriture sont des
termes plus élaborés qu'une simple variable il est naturel de considérer des
abstractions avec des motifs autres qu'une variable. Afin de mémoriser les
résultats possibles de l'application nous pouvons utiliser une structure de
liste (de résultats). Mais pour mettre en évidence la nature non-déterministe
de l'application, c'est-à-dire la sélection dans un ordre quelconque des règles
de réécriture à appliquer, une structure de multi-ensemble où l'ordre des
éléments n'est pas important semble plus appropriée. En plus, si le nombre de
solutions identiques n'est pas essentiel, une structure d'ensemble peut être
utilisée. Dans une telle approche l'échec ou autrement dit le fait de n'avoir
aucun résultat pour une application est représenté par l'ensemble vide (de
résultats).

Nous introduisons le calcul de réécriture, appelé aussi \roCal.  Dans ce calcul
l'opérateur d'abstraction ainsi que l'opérateur d'application sont des objets du
calcul. Une $\rho$-abstraction est une règle de réécriture dont le membre gauche 
précise les variables abstraites et une information de contexte.
Le résultat de l'évaluation d'une application (d'une $\rho$-abstraction ou d'un
$\rho$-terme plus élaboré) est toujours un ensemble, qui est également un
\mbox{$\rho$-terme}. Le mécanisme permettant d'instancier les variables avec
leur valeur actuelle est le filtrage qui peut être syntaxique, équationnel ou
d'ordre supérieur.

Les propriétés principales que nous voulons obtenir pour le \roCal\  sont la
confluence et la terminaison. Puisque il existe une correspondance forte entre
le \laCal\  et le \roCal\  on peut s'attendre à un résultat de confluence
similaire dans les deux calculs mais nous remarquons immédiatement que dans le
\roCal\  utilisant une théorie de filtrage du premier ordre cette propriété n'est
pas vérifiée. Néanmoins, la confluence peut être retrouvée si une stratégie
d'évaluation est utilisée pour guider les règles d'évaluation du calcul. Nous
pouvons définir des stratégies d'évaluation très simples au prix de
restrictions relativement fortes sur les réductions possibles ou plus
compliquées mais plus flexibles.

En partant de la non-terminaison du \laCal\  le même résultat est obtenu pour le
\roCal. Afin d'obtenir la terminaison nous procédons comme dans le \laCal\  et
nous définissons un système de types permettant d'éliminer les termes avec des
réductions infinies. Nous partons d'une approche similaire à celle utilisée dans 
le \laCal\  typé et encore une fois les ensembles nécessitent un traitement
spécial. En se limitant à des ensembles ayant tous les éléments d'un même type
et avec un bon choix pour les règles de typage, le \roCal\  typé est terminant.

Une fois que nous avons défini les conditions nous permettant d'obtenir la
confluence et la terminaison du \roCal\  nous pouvons envisager de réaliser une
implantation du calcul. Nous considérons que les solutions du problème de
filtrage sont fournis indépendemment.  Comme dans le \laCal, dans
le \roCal\  l'application de substitution n'est pas une partie du calcul mais
est définie au méta-niveau du calcul. 
La description de l'application de substitution est relativement simple mais
le coût d'exécution de cette opération n'est pas constant. En effet, la
complexité de l'application d'une substitution dépend de la forme du terme dans
lequel elle est effectué. Deuxièmement, la correspondance entre la théorie et
l'implantation devient non-triviale et la correction des implantations peut être
compromise.
Nous utilisons donc une approche similaire aux différentes versions de \laCal\
avec substitutions explicites et nous décrivons l'application de substitution
au même niveau que l'application de règles de réécriture.

Nous montrons que le pouvoir d'expression du \roCal\  est suffisant pour exprimer
les réductions du \laCal\  et de la réécriture du premier ordre. Mais nous ne
nous arrêtons pas là et nous utilisons le \roCal\  pour donner une sémantique
opérationnelle aux règles et stratégies du langage \elan.  En définissant le
$\rho$-terme correspondant à un programme \elan\  nous explicitons non seulement
les opérateurs du langage mais aussi le comportement de certaines
constructions. Ceci nous permet de mieux comprendre les exécutions d'un
programme \elan\  et particulièrement le traitement du non-déterminisme.

\section*{Plan du travail}
%================================================================

Après cette introduction, le Chapitre~\ref{chap.lambda_reec} a pour but de
rappeler les concepts utilisés au cours de cette thèse avec notamment les termes
du premier ordre, les substitutions, le \laCal, la réécriture ainsi que le
langage \elan, un cadre logique dont le noyau est la logique de réécriture
étendue avec la notion de stratégies. \\

Le chapitre~\ref{chap.calcul_non_type} présente le \roCalT\  au travers ses
composants et montre des exemples d'utilisation du calcul général ainsi que des
instances possibles du calcul de base. Nous décrivons la formation des
$\rho$-termes et la façon dont les substitutions sont appliquées sur ces
termes. Les règles d'évaluation du \roCal\  sont ensuite présentées en commentant 
nos choix et donnant des exemples de réductions.
Le \roCalT\  est le \roCal\  paramétré par la théorie de filtrage $T$ et même si
dans le cas général, nous considérons un filtrage d'ordre supérieur, dans les
cas pratiques nous utilisons le filtrage syntaxique ou équationnel. 
Nous illustrons le comportement dans certaines instances du calcul général
obtenues en précisant la théorie $T$ par des exemples simples de
$\rho_{T}$-réductions. \\

Le chapitre~\ref{chap.resultats_confluence} est consacré à l'analyse des
propriétés des relations induites par les règles d'évaluation et en particulier
à la confluence du \roCal.  L'utilisation des ensembles de résultats pour
représenter le non-déterminisme mène immédiatement à des réductions
non-convergentes et ainsi, le \roCal\  n'est pas confluent si les règles
d'évaluation ne sont pas guidées par une stratégie d'évaluation. 

Nous nous limitons à l'analyse de la confluence du \roCalE, c'est-à-dire le
\roCal\  utilisant un filtrage syntaxique. Nous définissons une stratégie
confluente générique simple à comprendre mais pas utilisable dans une
implantation du \roCal\  et ensuite nous proposons plusieurs stratégies
opérationnelles définies en imposant des restrictions structurelles sur les
$\rho$-termes à réduire.
En partant d'une stratégie confluente permettant la réduction de l'application
d'une règle de réécriture seulement à un terme clos du premier ordre, nous
présentons d'autres approches où les conditions sur la structure des termes sont
plus compliquées mais moins restrictives.\\

Le chapitre~\ref{chap.recursion} présente comment nous pouvons décrire dans le
\roCal\  des stratégies de réduction et, principalement, des stratégies de
normalisation. Afin d'exprimer des réductions déterministes, nous introduisons
un nouvel opérateur appelé $first$ qui a le rôle de sélectionner parmi ses
arguments le premier terme dont l'application à un $\rho$-terme n'échoue pas.
Nous définissons le \roCalf\  en ajoutant cet opérateur à la syntaxe et en
décrivant son comportement par des règles d'évaluation.

L'application d'une règle de réécriture en tête ou aux arguments d'un terme avec
un certain symbole de tête est décrite explicitement dans le \roCal\  par un
$\rho$-terme approprié. Nous montrons qu'il est possible de décrire
l'application d'une règle de réécriture aux arguments d'un terme indépendamment
du symbole de tête en utilisant seulement les opérateurs du \roCal. Nous
définissons ensuite, en utilisant le \roCalf, des opérateurs décrivant
l'application répétitive d'un terme en tête ou aux positions les plus profondes
d'un autre terme et finalement nous décrivons la représentation de stratégies
\textit{innermost} et \textit{outermost} dans le \roCalf. \\

Le chapitre~\ref{chap.encodage} est donc consacré à l'utilisation des opérateurs 
définis dans le chapitre précédent pour décrire des réductions réalisées en
\laCal\  et en réécriture. 
La représentation des $\lambda$-termes et des réductions sous-jacentes en
\roCal\  est réalisée en définissant des fonctions de transformation entre les
termes des deux formalismes et en montrant que les réductions dans les deux
calculs sont identiques modulo ces transformations. Nous décrivons ensuite les
$\rho$-termes correspondant à des réductions en réécriture (conditionnelle). Ces
termes peuvent être construits soit en utilisant les preuves pour les réductions
correspondantes en réécriture, soit en utilisant seulement les règles de
réécriture appliquées dans les réductions respectives. 

En partant de la représentation de la réécriture conditionnelle nous analysons
la description en \roCal\  des règles et stratégies du langage \elan. Le langage
\elan\  introduit les affectations locales de variables aux résultats de
sous-dérivations ainsi que des opérateurs permettant la construction de
stratégies à partir des règles de réécriture. Nous décrivons les $\rho$-termes
correspondant aux règles et stratégies \elan\  et nous montrons comment on peut
construire le $\rho$-terme correspondant à un module \elan. \\

Le chapitre~\ref{chap.calcul_type} est dédié à l'étude du \roCal\  typé.  Le
\roCal\  non-typé n'est pas terminant et afin d'éliminer les termes avec une
réduction infinie nous imposons des restrictions sur la formation des
$\rho$-termes en introduisant une information de type pour chaque terme.  Nous
utilisons une approche similaire à celle utilisée dans le \laCal\  typé et nous
ajoutons des règles de typage pour les ensembles. Nous nous concentrons sur le
typage du \roCalE\  et nous montrons que la réduction de tout terme bien typé est
finie et préserve le type du terme initial. \\

Le chapitre~\ref{chap.calcul_explicite} a pour objectif d'étendre le \roCal\  en
rendant explicite l'application de substitution. Nous étendons la syntaxe du
\roCal\  en introduisant les définitions des substitutions et un opérateur
d'application de substitution ainsi que les règles d'évaluation décrivant son
comportement. Nous obtenons ainsi le  \roSig. La présentation du \roSig\  est
basée sur une notation de de Bruijn et le sous-système contenant les règles
d'évaluation décrivant l'application de substitution est inspiré des systèmes
similaires utilisés dans le \laCal. Nous montrons que le \roSig\  est confluent
dans les mêmes conditions que le \roCal. \\

En conclusion, nous résumons les résultats obtenus et nous explorons les 
perspectives de recherche émergeant de ce travail.

%% file: chapter_1.tex
%%%%%%%%%%%%%%%%%%%%%%%%%%%%%%%%%%%%%%%%%%%%%%%%%%%%%%%%%%%
% \TLtopbookmark
\chapter{Notions préliminaires}
\label{chap.lambda_reec}
%%%%%%%%%%%%%%%%%%%%%%%%%%%%%%%%%%%%%%%%%%%%%%%%%%%%%%%%%%%%

Nous présentons dans ce chapitre les notions préliminaires utilisées dans ce
document. Nous introduisons notamment les termes du premier ordre, les algèbres
universelles, le \laCal\  ainsi que la réécriture.  Les bases de ces théories
sont présentées au travers de leurs principales définitions et de leurs
propriétés les plus connues.

Les notions de variable et de substitution du \laCal\  et la notion
de règle de réécriture seront utilisées d'une façon similaire dans le cadre du
\roCal.

\section{Définitions de base}
%================================================================

\subsection{Les algèbres de termes} \label{algebresTermes}
%================================================================

\DEF%------------------------------------------------------------------------
Une \Def{signature} $\FF$ est un ensemble de symboles dont chacun est associé à
un entier naturel qui est appelé son arité. Le sous-ensemble de symboles d'arité
$n$ est noté $\FF_n$ et donc $\FF = \bigcup_{i \geq 0} \FF_i$.  L'arité d'un
symbole $f$ est notée $|f|$.
\FDEF%------------------------------------------------------------------------

\DEF%------------------------------------------------------------------------ 
Soit $\FF = \{f_1,\ldots,f_n\}$ une signature. Soit $\XX$ un ensemble de
variables.  La \mbox{$\FF$-algèbre}%\index{algèbre} 
\Defi{libre}{algèbre}
\Defi{homogène}{algèbre} engendrée par $\XX$, notée $\TFX$ est le plus petit
ensemble tel que~:
	\begin{itemize}
	\item $\XX \subset \TFX$,
	\item pour tout symbole $f$ de $\FF$ d'arité $n$ ($f \in \FF_n$) et pour tous
	$t_1,\ldots,t_n \in \TFX$ alors $f(t_1,\ldots,t_n) \in \TFX$.
	\end{itemize}

Pour désigner $\TFX$ nous parlerons le plus souvent de l'algèbre de
termes \index{terme} engendrée par la signature $\FF$. $\FF$ est appelé la
signature de l'algèbre. Les éléments de $\TFX$ sont appelés termes (du premier
ordre\index{terme!du premier ordre}). Un terme peut être vu comme un arbre fini
étiqueté (cf. Définition~\ref{positionTermeArbre}).
\FDEF%------------------------------------------------------------------------

Ce genre de définitions, dites par clôture, où un ensemble est défini par un
ensemble de base (ici les variables) et des règles de ``construction'' de nouveaux
éléments (ici les symboles de la signature), permettent de faire des définitions
et des démonstrations dites par récurrence structurelle. La démonstration d'une
propriété quelconque (respectivement une définition) se fait en prouvant la
propriété sur les éléments de base et en prouvant que cette propriété est
conservée par les règles de construction. Le principe de récurrence des entiers
naturels en est l'exemple le plus connu.

\DEF%------------------------------------------------------------------------
L'ensemble $\Var(t)$ des variables d'un terme $t$ est défini inductivement par:
	\begin{itemize}
	\item $\Var(t) = \emptyset$ si $t \in \FF_0$,
	\item $\Var(t) = \{ t \}$ si $t \in \XX$,
	\item $\Var(t) = \bigcup_{i=1}^n \Var(t_i)$ si $t=f(t_1,\dots,t_n)$.
	\end{itemize}

Un terme est \Defi{linéaire}{terme} si chacune de ses variables apparaît une
seule fois dans le terme.
\FDEF%------------------------------------------------------------------------

\DEF%------------------------------------------------------------------------
Une algèbre \Defi{initiale}{algèbre}, noté $\TF$, est une algèbre homogène
engendrée par un ensemble vide de variables. Les termes d'une algèbre initiale,
c'est à dire les termes ne contenant pas de variable, sont appelés les termes
\Defi{clos}{terme}. 
\FDEF%------------------------------------------------------------------------

Nous nous sommes donnés, grâce au langage des algèbres de termes, un moyen de
construire des ensembles de termes. Pour pouvoir décrire les opérations sur ces
termes, nous allons définir l'ensemble de positions d'un terme ainsi que la
notion de sous-terme d'un terme à une position donnée.

\DEF%------------------------------------------------------------------------
\label{positionTermeArbre}
Soit $\N_+$ l'ensemble des entiers strictement positifs, $\N_+^*$ le monoïde
libre engendrée par $\N_+$, $\epsilon$ le mot vide et $.$ l'opération de
concaténation.  Pour tous $p,q \in \N_+^*$, $p$ est un préfixe de $q$, ce que
l'on note $p \leq q$, s'il existe $q' \in \N_+^*$ tel que $q=p.q'$. $p$ est un
préfixe strict de $q$, noté $p<q$, si $p \leq q$ et $p \neq q$. 
Si $p \not\leq q$ et $q \not\leq p$, $p$ et $q$ sont disjoints ou
incomparables, ce qu'on note $p \Join q$.

Un arbre sur $\FF \cup \XX$ est une application $t$ d'une partie non vide
$\PPos(t)$ de $\N_+^*$ dans $\FF \cup \XX$ telle que~:
	\begin{enumerate}
        \item $\PPos(t)$ est clos par préfixe.
        \item Pour tout $p \in \PPos(t)$ et tout $i \in \N_+$, 
		$p.i \in \PPos(t)$ si et seulement si $t(p)=f \in \FF$ et $1\leq i \leq |f|$.
	\end{enumerate}
\Defi{$\PPos(t)$}{position} est appelé ensemble des positions \index{position} de
$t$, et $t$ est fini si $\PPos(t)$ l'est. La taille $|t|$ d'un terme $t$ est
dans ce cas le cardinal de $\PPos(t)$.
\FDEF%------------------------------------------------------------------------

L'ensemble des arbres finis sur $\FF \cup \XX$ peut être muni naturellement
d'une structure de \mbox{$\FF$-algèbre} isomorphe à $\TFX$. On parlera donc
dorénavant indifféremment d'arbre ou de terme.

\DEF%------------------------------------------------------------------------
~
        \begin{itemize}
        \item Pour tout terme $t$ et toute position $p \in \PPos(t)$, $t(p)$ est
        appelé symbole à la position $p$ dans $t$.  $t(\epsilon)$ est également
        appelé symbole de tête de $t$.
        \item On appelle \Def{sous-terme} de $t$ à la position $p \in \PPos(t)$, le
        terme noté $\tpos{t}{p}$, et défini par \\
	$\forall p.q \in \PPos(t), q \in \PPos(\tpos{t}{p}), \tpos{t}{p}(q)=t(p.q)$. 
	$\tpos{t}{p}$ est un sous-terme \Defi{strict}{sous-terme} de $t$ si $p \neq \epsilon$.
        \item Si $t(\epsilon)=f \in {\cal F}$, on notera $t$ sous la forme
        $f(\tpos{t}{1},\ldots,\tpos{t}{n})$ où $n=|f|$.
        \item Une position $p$ de $t$ est variable si $t(p) \in \XX$. L'ensemble
        des positions variables \index{position!variable,~$\VPos(t)$} de $t$ est
        noté $\VPos(t)$ alors que l'ensemble des positions non variables
        \index{position!non variable,~$\FPos(t)$} de $t$ est noté $\FPos(t)$.
        Une position $p$ de $t$ est constante si $t(p) \in \FF_0$. L'ensemble
        des positions constantes \index{position!constante,~$\CPos(t)$} de $t$
        est noté $\CPos(t)$.
        \end{itemize}
\FDEF%------------------------------------------------------------------------

La notation $\tatpos{t}{s}{p}$ est utilisée pour signifier que $t$ contient $s$
comme sous-terme à la position $p$ et la notation $\tatp{t}{p \hookleftarrow s}$
pour faire remarquer que le sous-terme $\tpos{t}{p}$ a été remplacé par $s$ dans
$t$. Nous notons par
$t(\ov{x})$ le terme $t$ tel que $\Var(t)=\{x_1,\ldots,x_n\}$.

\DEF%------------------------------------------------------------------------
La relation \Defi{de sous-terme}{relation}, noté $\unlhd_{sub}$, est définie par
$s \unlhd_{sub} t$ si $s$ est un sous-terme de $t$. La relation de sous-terme
strict, noté $\lhd_{sub}$, est définie par $s \lhd_{sub} t$ si 
$s \unlhd_{sub} t$ et $s \neq t$.
\FDEF%------------------------------------------------------------------------

\DEF%------------------------------------------------------------------------
Etant donné un terme $\tatpos{t}{s}{p}$ et un entier $P$ représentant la
longueur du mot $p$ (la longueur du mot vide $\epsilon$ est $0$). On dit que le
terme $s$ est un sous-terme à la \Def{profondeur} $P$ dans $t$. La profondeur de
$t$ est le maximum des profondeurs des sous-termes de $t$.
\FDEF%------------------------------------------------------------------------

Les sous-termes (à la profondeur $1$) $t_1,\dots,t_n$ d'un terme
$t=f(t_1,\dots,t_n)$ sont appelés les arguments de $t$.

\EX%------------------------------------------------------------------------
\label{exampleAlgebre}

Une représentation algébrique possible des expressions de l'arithmétique est
l'algèbre de termes engendrée par la signature $\FF=\FF_0 \cup \FF_1 \cup \FF_2$
contenant~:
        \begin{itemize}
        \item $\FF_0=\{0\}$~: une constante;
        \item $\FF_1=\{succ,-\}$~: deux symboles unaires,
        \item $\FF_2=\{+,\times\}$~: deux symboles binaires.
        \end{itemize}

Le terme $t = +(\times(x,succ(succ(0))), succ(0))$ est le terme représentant en
notation préfixée l'expression $x \times 2 + 0$.

On a $t(1.2)=succ$, $\tpos{t}{1.2}=succ(succ(0))$ et on peut écrire
$\tatpos{t}{succ(succ(0))}{1.2}$.

L'ensemble des variables de $t$ est $\Var(t)=x$ et les positions variables de $t$
sont décrites par $\VPos(t)=\{1\}$. L'ensemble des positions non variables de
$t$ est $\FPos(t)=\{1.2,1.2.1,1.2.1.1,2,2.1\}$ et les positions constantes de
$t$ sont $\CPos(t)=\{1.2.1.1,2.1\}$.

Le terme $succ(succ(0))$ est un sous-terme de profondeur $2$ de $t$. Les
sous-termes de $t$ de profondeur $2$ sont $succ(succ(0))$ et $0$. La profondeur
de $t$ est $4$.
\FEX%------------------------------------------------------------------------

En général, lorsque nous désirons définir une algèbre de termes particulière,
nous utilisons une notation empruntée aux grammaires. L'algèbre des expressions
arithmétiques présentée dans l'Exemple~\ref{exampleAlgebre} se définit avec cette
notation par~:
$$exp ~~ ::= ~~ x ~~|~~ 0  ~~|~~ succ(exp)  ~~|~~ - exp  ~~|~~ + (exp, exp) ~~|~~  \times(exp, exp)$$
où $x \in \XX$.

L'algèbre initiale engendrée par la signature $\FF$ de
l'Exemple~\ref{exampleAlgebre} est définie par~:
$$exp ~~ ::= ~~ 0  ~~|~~ succ(exp)  ~~|~~ - exp  ~~|~~ + (exp, exp) ~~|~~  \times(exp, exp)$$

Cette notation permet de spécifier dans le même temps les symboles d'une
signature et leur arité. Nous pouvons faire implicitement des conventions de
notation et utiliser une syntaxe mixfix pour certains symboles. L'algèbre
initiale précédente peut être ainsi définie par~:
$$exp ~~ ::= ~~ 0  ~~|~~ succ(exp)  ~~|~~ - exp  ~~|~~ exp + exp ~~|~~  exp \times exp$$

\DEF%------------------------------------------------------------------------
Soient $\KK$ un ensemble de symboles de sorte, $\FF$ une signature et $\XX$ un
ensemble de variables. A chaque symbole $f$ de $\FF$ d'arité $n$ est associé une
suite de $n+1$ symboles de sorte $(k_1,\ldots,k_{n+1})$, et à chaque variable
$x$ de $\XX$ est associé un symbole de sorte. La suite de symboles de sortes est
appelée le \Def{profil} du symbole $f$ et on note:
	$$f : k_1 \times \ldots k_n \raS k_{n+1}$$ 
où $f \in \FF_n$ et $k_i \in \KK$.

Les termes de la $\FF$-algèbre \Defi{hétérogène}{algèbre} libre $\TFX$ engendrée
par $\XX$ et la sorte $k$ d'un terme $t$, noté $t:k$, sont définis simultanément
par~:
        \begin{itemize}
        \item pour toute variable $x \in \XX$ ayant associé un symbole de sorte
        $k$, $x \in \TFX$ et la sorte de $x$ est $k$,
        \item pour tout symbole $f : k_1 \times \ldots k_n \raS k_{n+1}$ et
        termes $t_1:k_1,\ldots,t_n:k_n$, $f(t_1,\ldots,t_n) \in \TFX$
        et la sorte de $f(t_1,\ldots,t_n)$ est $k_{n+1}$.
        \end{itemize}
Le sous-ensemble $\TT_k$ de $\TFX$ défini par l'ensemble des termes de sorte $k$ est
appelé une \Def{sorte}.  Une algèbre hétérogène est aussi désignée sous le nom
d'algèbre de termes multi-sortée.
\FDEF%------------------------------------------------------------------------

On peut remarquer que lorsque nous considérons une algèbre de termes
multi-sortée, nous devons distinguer les variables suivant la sorte à laquelle
elles appartiennent.

Nous reprenons l'Exemple~\ref{exampleAlgebre} sur les expressions arithmétiques
et nous distinguons deux sortes: les entiers naturels et les expressions
proprement dites.

\EX%------------------------------------------------------------------------
\label{exampleAlgebreSorte}

Les entiers naturels et les expressions arithmétique construites en utilisant
des entiers sont représentés par l'algèbre initiale engendrée par la signature
$\FF$ de l'Exemple~\ref{exampleAlgebre} et définie par~:
	$$nat ~~ ::= ~~ 0  ~~|~~ succ(nat)$$
	$$exp ~~ ::= ~~ nat ~~|~~ - exp  ~~|~~ + (exp, exp) ~~|~~  \times(exp, exp)$$
\FEX%------------------------------------------------------------------------

L'algèbre définie dans l'Exemple~\ref{exampleAlgebreSorte} n'est pas la même que
l'algèbre mono-sortée donnée dans l'Exemple~\ref{exampleAlgebre}. En effet,
$succ(0 + 0)$ n'est pas un terme de cette algèbre multi-sortée alors qu'il en
est un de la précédente.

D'autre part, la notation par grammaire introduit implicitement un symbole
unaire de type $N:nat \raS exp$. Ce symbole n'a pas une notation explicite, mais
formellement il doit être défini. Si nous avions voulu être explicite, la
définition de la sorte $exp$ aurait été
	$$exp ~~ ::= ~~ N(nat) ~~|~~ - exp  ~~|~~ + (exp, exp) ~~|~~  \times(exp, exp)$$

\subsection{Substitutions du premier ordre}  \label{subst_premier_ordre}
%================================================================

Dans cette section nous allons définir une opération sur les termes, que nous
appellerons substitution, permettant de les modifier. Effectuer une substitution
consiste à remplacer une variable d'un terme par un autre terme et pour bien
comprendre le mécanisme de remplacement nous allons donner une définition
formelle des substitutions et ensuite une définition équivalente plus
opérationnelle et intuitive.

\DEF%------------------------------------------------------------------------
Une \Def{substitution} est un endomorphisme de l'algèbre $\TFX$ dont la
restriction à $\XX$ est l'identité presque partout, c'est-à-dire sauf sur un
sous-ensemble fini de $\XX$.
\FDEF%------------------------------------------------------------------------

Les substitutions seront notées par des lettres grecques minuscules,
$\sigma, ~\mu, ~\gamma, ~\phi,\dots$ La notation préfixe, i.e. $\sigma t$, est
utilisée pour l'application d'une substitution $\sigma$ à un terme $t$.  Une
substitution bijective est un \Defi{renommage}{substitution}\index{renommage}. Une substitution
$\sigma$ est \Defi{idempotente}{substitution} si $\sigma \circ \sigma =\sigma$.

On appelle \Defi{domaine}{substitution} \index{domaine} d'une substitution $\sigma$ l'ensemble
$\Dom(\sigma)=\{ x \in \XX ~|~ \sigma x \neq x\}$ et
\Defi{codomaine}{substitution} \index{codomaine} d'une substitution $\sigma$ l'ensemble
$\Ran(\sigma) = \{ \sigma x ~|~ x \in \Dom(\sigma) \}$.  L'ensemble des
variables introduites par une substitution $\sigma$ est $\VRan(\sigma) = \cup_{
x \in \Dom(\sigma)} {\Var}(\sigma x)$. L'ensemble de toutes les variables
impliquées dans $\sigma$ est $\Var(\sigma)=\Dom(\sigma) \cup \VRan(\sigma)$.

La restriction de $\sigma$ à un ensemble de variables $X$, notée $\sigma_{|_X}$,
est définie par $\sigma_{|_X} x = \sigma x$ si $x \in X$ et $\sigma_{|_X} x = x$
sinon.

Le préordre de filtrage ou de \textit{subsomption} est défini par $s \leq t$ s'il
existe une substitution $\sigma$ telle que $\sigma s = t$. Dans ce cas on dit
que le terme $s$ \Def{subsume} le terme $t$.

Nous avons présenté une définition formelle des substitutions du premier ordre
et pour renforcer l'intuition derrière cette notion nous donnons aussi une
définition plus opérationnelle.

\DEF%------------------------------------------------------------------------
Le remplacement du terme $u$ dans le terme $t \in \TFX$ à la position $p$,
$\tatp{t}{p \hookleftarrow s}$, est définie inductivement par~:
        \begin{itemize}
        \item $\tatp{t}{\epsilon \hookleftarrow u}=u$,
        \item $\tatp{f(\tpos{t}{1},\ldots,\tpos{t}{n})}{i.p \hookleftarrow u}$
	$=$ $f(\tpos{t}{1},\ldots,\tatp{\tpos{t}{i}}{p \hookleftarrow u},\ldots,\tpos{t}{n})$,
	si $f \in \FF$.
        \end{itemize}
\FDEF%------------------------------------------------------------------------

\DEF%------------------------------------------------------------------------
La substitution de la variable $x$ par le terme $u$ dans le terme $t$, notée
$\subs{x \sbs u}t$ est la composition de chacun des remplacements du terme $u$ à
chacune des positions $p$ telle que $t(p)=x$.
\FDEF%------------------------------------------------------------------------

\DEF%------------------------------------------------------------------------
Une substitution est une fonction accomplissant en simultané plusieurs
substitutions de différentes variables par des termes. L'application d'une
substitution à un terme $t$ sera notée 
$\subs{x_1 \sbs u_1,\ldots,x_n \sbs u_n}t$.
\FDEF%------------------------------------------------------------------------

\FACT%------------------------------------------------------------------------
Les substitutions ne commutent pas entre elles dans le cas général.  La
substitution $\subs{x_1 \sbs u_1,\ldots,x_n \sbs u_n}$ représente le
remplacement simultané des variables $x_1,\ldots,x_n$ par les termes
$t_1,\ldots,t_n$ et pas la composition des substitutions 
$\subs{x_1 \sbs t_1}\ldots\subs{x_n \sbs t_n}$.
\FFACT%------------------------------------------------------------------------

Nous avons $\subs{x \sbs y,y \sbs x}f(x,y)=f(y,x)$ mais 
$\subs{x \sbs y}\subs{y \sbs x}f(x,y)=f(y,y)$ et
$\subs{y \sbs x}\subs{x \sbs y}f(x,y)=f(x,x)$.

\subsection{Théories équationnelles}
%================================================================

Une paire de termes $(l, r)$ est appelé {\em égalité}\index{egalite@égalité}, %\Def{égalité}
\Def{axiome} {\em équationnel} ou {\em égalitaire}, ou 
{\em équation} \index{equation@équation} %\Def{équation} 
suivant le contexte, et notée $(l=r)$.

\DEF%------------------------------------------------------------------------
Etant donné un ensemble de variables $\XX$, une algèbre $\AA$ et une
assignation $\alpha: \XX \ra \AA$, on note $\underline{\alpha}$ l'unique
homomorphisme de $\TFX$ vers l'algèbre $\AA$ étendant $\alpha$ tel que
	$$\forall f \in \FF, \underline{\alpha}(f(t_1,\ldots,t_n))=
	f_{\AA}(\underline{\alpha}(t_1),\ldots,\underline{\alpha}(t_n))$$
\FDEF%------------------------------------------------------------------------

\DEF%------------------------------------------------------------------------
Une $\FF$-algèbre $\AA$ {\em valide} une égalité $s=t$, noté $\AA \models s=t$
ou plus simplement $s =_{\AA} t$ si pour toute assignation $\alpha: \XX \ra \AA$,
$\underline{\alpha}(s) = \underline{\alpha}(t)$. L'algèbre $\AA$
{\em satisfait} une égalité $s = t$ s'il existe une assignation $\alpha$ telle
que $\underline{\alpha}(s) = \underline{\alpha}(t)$.  Une $\FF$-algèbre $\AA$
est un modèle d'un ensemble d'égalités $E$ si elle valide toutes les égalités de
$E$.
\FDEF%------------------------------------------------------------------------

On note $\TT h(\AA)$ l'ensemble des égalités valides dans une $\FF$-algèbre
$\AA$ et $\MM od(E)$ la classe des $\FF$-algèbres qui sont modèles de $E$.

Soit $E$ un ensemble d'égalités de $\TFX$, appelés dans ce contexte,
{\em axiomes}.

\DEF%------------------------------------------------------------------------
Etant donnée une signature $\FF$, une \Def{présentation équationnelle} est un
couple $(\FF,E)$ telle que $E$ est un ensemble d'axiomes de $\TFX$.
\FDEF%------------------------------------------------------------------------ 

Le problème de validité dans $\Mod(E)$ consiste à décider si une égalité $s=t$
est valide dans tout modèle de $E$. Ce problème peut se ramener à des
considérations syntaxiques.

\DEF%------------------------------------------------------------------------
Etant donnée une présentation équationnelle $(\FF,E)$, on appelle \Def{théorie
équationnelle} engendrée par $(\FF,E)$ ou {$E$-égalité} et on note $=_E$ la plus
petite congruence sur $\TFX$ contenant toutes les égalités 
$(\sigma l = \sigma r)$ où $(l=r)$ est un axiome de $E$ et $\sigma$ une
substitution quelconque.
\FDEF%------------------------------------------------------------------------

Le théorème suivant est le fondement de la logique équationnelle. Il
relie le problème sémantique de la validité d'une égalité dans une
classe de modèles au problème syntaxique de la $E$-égalité.

\TH%------------------------------------------------------------------------
(Birkhoff~\cite{BirkhoffPCPS35}, Complétude du raisonnement équationnel pour un
ensemble $E$ d'axiomes équationnels)

$s=t$ est valide dans ${\cal M}od(E)$ si et seulement si $s =_E t$.
\FTH%------------------------------------------------------------------------

La $E$-égalité peut encore être obtenue par le \Def{remplacement} d'égal par
égal décrit ci-après.

\DEF%------------------------------------------------------------------------
Etant donné un ensemble d'axiomes $E$, on note $\lra{}_E$ la relation binaire
symétrique sur $\TFX$ définie par $s \lra{}_E t$ s'il existe un axiome $(l=r)$
de $E$, une position $\omega$ de $s$ et une substitution $\sigma$ tels que 
$s_{|\omega} = \sigma l$ et $t = s[\sigma r]_{\omega}$.
\FDEF%------------------------------------------------------------------------

\FACT%------------------------------------------------------------------------
	$s =_E t \Lra s \lra{*}_E t$.
\FFACT%------------------------------------------------------------------------

Par abus de langage et de notation, on confond souvent la théorie équationnelle
$=_E$, la présentation équationnelle $(\FF,E)$ et l'ensemble des axiomes
équationnels $E$.

L'ensemble des classes de congruence de $E$ dans $\TFX$ peut être muni
naturellement d'une structure d'algèbre, notée $\TFX/=_E$, qui est
l'algèbre libre sur $\XX$ de la classe des $\FF$-algèbres modèles de $E$.

\subsection{Propriétés des relations binaires sur un ensemble} \label{relationsBinaires}
%================================================================

Les termes sont connectés entre eux par des relations ou par des transformations
des uns vers les autres. Nous donnons quelques propriétés abstraites liées aux
relations binaires dont nous aurons besoin par la suite.

\DEF%------------------------------------------------------------------------
Une relation \index{relation} binaire $\ral$ sur un ensemble de termes construits
en utilisant un ensemble d'opérateurs $\Phi$ est \Def{compatible} \index{relation!compatible}
(avec les opérateurs) si pour tous termes $u_i,v_i \in T$, $i=1,\ldots,n$ et tout
opérateur $\phi_n$ d'arité $n$ 
	$$u_i \ral v_i, ~ i=1,\ldots,n  ~~ \Longrightarrow ~~ 
	\phi_n(u_1,\ldots,u_n) \ral \phi_n(v_1,\ldots,v_n)$$
\FDEF%------------------------------------------------------------------------

\DEF%------------------------------------------------------------------------
Etant donnée une relation \index{relation} binaire $\ral$ sur un ensemble $T$~:
  \begin{itemize} 
  \item la relation inverse de $\ral$ est notée $\longleftarrow$,
  \item la \Defi{fermeture symétrique}{relation} \index{fermeture
	symétrique} de $\ral$, notée $\lra{}$, est la plus petite relation
	symétrique contenant $\ral$.
  \item la \Defi{fermeture transitive}{relation} \index{fermeture transitive} de $\ral$, notée
	$\longra{+}$, est la plus petite relation transitive contenant $\ral$.
  \item la fermeture réflexive et transitive de $\ral$ est notée
	$\longra{*}$.
  \item la fermeture réflexive, symétrique et transitive de $\ral$ est
	notée $\lra{*}$.
  \item la \Defi{fermeture compatible}{relation} \index{fermeture compatible} 
	(ou \Defi{fermeture par contexte}{relation}\index{fermeture par contexte}) de
	$\ral$ est la plus petite relation contenant $\ral$ et fermée par
	rapport aux règles de formation de termes de $T$.
  \end{itemize}
La composition des relations $\ral_1$ et $\ral_2$ est notée 
$\ral_1 \circ \ral_2$ ou $\ral_1 \ral_2$.
\FDEF%------------------------------------------------------------------------

Une relation binaire $\sim$ réflexive, symétrique et transitive est une relation
\Defi{d'équivalence}{relation}. Un \Def{ordre} $>$ est une relation
binaire irréflexive, antisymétrique et transitive. Un \Defi{préordre}{ordre}
$\ge$ est une relation binaire réflexive et transitive. 

\DEF%------------------------------------------------------------------------
Un ordre $>$ sur $T$ est \Defi{n\oe thérien}{ordre} s'il n'existe pas de suite
infinie $(t_i)_{i \ge 1}$ d'éléments de $T$ telle que $t_1>t_2>\dots$

Un ordre $>$ sur $T$ est \Defi{total}{ordre} si $\forall s,t \in T$ on a $s>t$
ou $t>s$.
\FDEF%------------------------------------------------------------------------

La construction d'ordres n\oe thériens peut éventuellement se faire par
extension.  L'extension lexicographique permet par exemple de comparer des
uplets. Pour comparer des suites d'objets, on introduit la notion de {\em
multi-ensemble} sur $T$ qui est une application de $T$ vers $\N$. Un ordre sur
$T$ peut être facilement étendu à un ordre sur les multi-ensembles sur $T$.

\DEF%------------------------------------------------------------------------
Pour une relation $\ral$, un élément $t$ de $T$ est
{\em réductible} \index{reductible@réductible}%\Def{réductible} 
par $\ral$ s'il existe $t'$ dans $T$ tel que $t \ral t'$. Dans le cas contraire, il est
irréductible.  On appelle \Def{forme normale} de $t$ tout élément $t'$ irréductible
tel que $t \longra{*} t'$. Lorsque un terme $t$ a une unique forme normale,
celle-ci est notée $t\da{}$.
\FDEF%------------------------------------------------------------------------

La question que l'on se pose est de savoir si $t \lra{*} t'$.  L'idéal serait de
calculer une forme normale de chacun des éléments et de tester si elles sont
égales. Cela n'est possible que si d'une part une forme normale existe pour tout
élément, et si d'autre part elle est unique.  Les formes normales existent dès
que $\ral$ {\em termine}\index{terminaison}, c'est-à-dire qu'il n'existe pas de suite infinie
$(t_i)_{i\geq1}$ d'éléments de $T$ telle que $t_1\ral t_2 \ral \cdots$.  Dans le
cas où une forme normale existe, son unicité est assurée par la propriété de
Church-Rosser ou par la confluence qui est une propriété équivalente.

\DEF%------------------------------------------------------------------------
~
	\begin{enumerate}
	\item	$\ral$   a  la  propriété de Church-Rosser si
		$$\lra{*} ~\subseteq~ \longra{*} \circ \longla{*}$$
	\item	$\ral$ est confluente  si
		$$\longla{*} \circ \longra{*} ~\subseteq~ \longra{*} \circ \longla{*}$$
	\item	$\ral$ est localement confluente \index{confluence!locale} si
		$$ \lal \circ  \ral   ~\subseteq~   \longra{*} \circ \longla{*}$$
	\item	$\ral$ est fortement confluente \index{confluence!forte} si
		$$ \lal \circ  \ral   ~\subseteq~   \ral \circ \lal$$
	\item $\ral$ est convergente \index{convergence} si $\ral$ termine et a la
		propriété de Church-Rosser.
	\end{enumerate}
\FDEF%------------------------------------------------------------------------

Ces différentes définitions se représentent chacune par un diagramme. Dès que ce
sera possible, nous adopterons cet artifice typographique pour exprimer les
propriétés des relations. Une flèche pleine figure une hypothèse et une flèche
en pointillé une conclusion.

\begin{tabular}{llll}
%------------
\xymatrix{
\\
~ \ar@{{<}-{>}}[rr]^-{*} \ar@{.{>}}[dr]_{*} & & ~\ar@{.{>}}[dl]^-{*} \\
& ~ &
}~~~~
&%------------
\xymatrix{
&~ \ar[dl]_{*} \ar[dr]^-{*} & \\
~ \ar@{.{>}}[dr]_{*} & & ~\ar@{.{>}}[dl]^-{*} \\
& ~ &
}~~~~
&%------------
\xymatrix{
&~ \ar[dl]_{} \ar[dr]^-{} & \\
~ \ar@{.{>}}[dr]_{*} & & ~\ar@{.{>}}[dl]^-{*} \\
& ~ &
}~~~~
&%------------
\xymatrix{
&~ \ar[dl]_{} \ar[dr]^-{} & \\
~ \ar@{.{>}}[dr]_{} & & ~\ar@{.{>}}[dl]^-{} \\
& ~ &
}
%------------
\end{tabular}

\begin{tabular}{cccc}%{llll}
%------------
\textit{Church-Rosser}
&%------------
~~~~~~~~ \textit{confluence}
&%------------
~~~~~~~~ \textit{confluence locale}
&%------------
~~~~ \textit{confluence forte}
%------------
\end{tabular}\\

Si une relation est fortement confluente alors elle est confluente.  Si une
relation est confluente alors elle est localement confluente.  Une relation est
confluente si et seulement si elle satisfait la propriété de Church-Rosser.

La confluence est une propriété difficile à tester. En pratique, le test de
confluence se fait localement grâce au théorème suivant:

\TH%------------------------------------------------------------------------
(Newman~\cite{Newman42})
\label{Church-Rosser}

Si $\ral$  termine, alors les propriétés suivantes sont équivalentes~: 
	\begin{enumerate}
	\item	$\ral$ a la propriété de Church-Rosser,
	\item	$\ral$ est confluente,
	\item	$\ral$ est localement confluente,
	\item	$\forall t, ~t' \in T: t \lra{*} t'~ \Lra~ t\da{} = t'\da{}$.
	\end{enumerate}
\FTH%------------------------------------------------------------------------

La normalisation (forte ou faible) est la seconde des deux propriétés
importantes pour une relation. Si nous considérons une relation comme un calcul
sur un ensemble, la normalisation forte assure que ce calcul est toujours
fini; la normalisation faible assure qu'il y a un moyen de terminer tout calcul.

\DEF%------------------------------------------------------------------------
Soit une relation binaire $\ral$ sur un ensemble $T$.
  \begin{itemize} 
  \item On dit que $t \in T$ est une forme normale s'il n'existe pas de 
	$u \in T$ tel que $t \ral u$ et on dit que $v \in T$ a une forme normale
	$t$ s'il existe une forme normale $t$ tel que $v \ral t$.
  \item La relation $\ral$ est \Defi{faiblement normalisable}{relation}
	\index{faiblement normalisant}
	(\Def{weakly normalizing}) si tout terme $t \in T$ a une forme normale.
  \item La relation $\ral$ est \Defi{fortement normalisable}{relation}
	\index{fortement normalisant}
	(\Def{strongly normalizing}) ou \Defi{normalisable}{relation} \index{normalisant} s'il
	n'existe pas de suite infinie $(t_i)_{i\geq1}$ d'éléments de $T$ telle que 
	$t_1\ral t_2 \ral \cdots$.
  \end{itemize}
\FDEF%------------------------------------------------------------------------

Dans la pratique, on est souvent amené à analyser les propriétés d'une relation
obtenue en composant deux (ou plusieurs) relations. Plusieurs méthodes ont été
développées pour démontrer la confluence d'une telle relation en fonction des
propriétés des deux relations.

\LEM%------------------------------------------------------------------------
(Hindley-Rosen~\cite{Rosen-JACM73})
\label{HindleyRosen}

Etant données deux relations confluentes $\relr$ et $\rels$
telles que le diagramme suivant est satisfait~:
\begin{center}$~$
\xymatrix{%@C+10pt{ 
& t
\ar[dl]_{\relsTRxy} \ar[dr]^-{\relrTRxy}
& \\
t'
\ar@{.{>}}[dr]_{\relrTRxy} && 
s
\ar@{.{>}}[dl]^-{\relsTRxy} \\
& 
s'
&
}
\end{center}

Alors la relation $\relr \cup \rels$ est confluente.
\FLEM%------------------------------------------------------------------------

Si le diagramme du Lemme~\ref{HindleyRosen} est satisfait on dit que les
relations  $\relr$ et $\rels$ commutent.

\LEM%------------------------------------------------------------------------
(Yokouchi~\cite{YokouchiHikita90})
\label{Yokouchi}

Etant données deux relations $\relr$ et $\rels$ telles que $\rels$ est
confluente et terminante, $\relr$ est fortement confluente et le diagramme
suivant est satisfait~:
\begin{center}$~$
\xymatrix{%@C+10pt{ 
& t
\ar[dl]_{\relrxy} \ar[dr]^-{\relsxy}
& \\
t'
\ar@{.{>}}[dr]_{\relsTRxy} && 
s
\ar@{.{>}}[dl]^-{\relsTRxy~\relrxy~\relsTRxy} \\
& 
s'
&
}
\end{center}

Alors la relation $\relsTR\relr\relsTR$ est confluente.
\FLEM%------------------------------------------------------------------------

Si le diagramme du Lemme~\ref{Yokouchi} est satisfait on dit que les
relations $\relr$ et $\rels$ sont {\em cohérentes}\index{relation!cohérentes}.%\Def{cohérentes}.

On peut aussi analyser les propriétés de confluence et Church-Rosser modulo une
relation d'équivalence.  On considère $\ral_{R}$ et $\lra{}_E$ deux relations
binaires sur l'ensemble $T$, dont l'une, $\lra{}_E$, est une relation
d'équivalence. On note $\ral_{R/E}$ la relation $\lra{*}_E \circ \ral_{R}\circ
\lra{*}_E$ simulant la relation induite par $\ral_R$ sur les classes
d'équivalence de $\lra{*}_E$.

D'habitude, on simule la relation $\ral_{R/E}$ par une relation $\ral_{S}$ plus
faible satisfaisant $\ral_{R} \subseteq \ral_{S} \subseteq \ral_{R/E}$.  On a
alors une propriété de Church-Rosser modulo $E$ pour $\ral_{S}$, ainsi qu'une
notion de confluence qui n'implique plus la propriété de Church-Rosser. Pour une
présentation détaillée des propriétés des relations définies sur des classes
d'équivalence le lecteur peut se référer à~\cite{Huet80},
\cite{JouannaudKirchnerSIAM86} et~\cite{KirchnerKirchner-RSP-99}.

\section{Les systèmes de réécriture} \label{systemeReec}
%================================================================

L'idée centrale de la
réécriture~\cite{DershowitzJouannaud-90,Klop90,BaaderNipkowREW-98} est d'imposer
une direction dans l'utilisation d'axiomes en définissant les règles de
réécriture.

\DEF%------------------------------------------------------------------------
Une {\em règles de réécriture} \index{regle de réécriture@règle de réécriture}
%\Defi{règle de réécriture}{règle} 
est une paire de termes orientée, noté $l \ra r$, où $l$ est le membre gauche de
la règle et $r$ son membre droit. 

Un \Def{système de réécriture} sur les termes est un ensemble de règles de
réécriture.
\FDEF%------------------------------------------------------------------------

Deux conditions sont imposées habituellement sur la construction des règles de réécriture~: 
\begin{enumerate}
\item le membre gauche d'une règle de réécriture n'est pas une variable
	($\forall x \in \XX, ~ l \neq x$),
\item l'ensemble des variables du membre droit est inclu dans l'ensemble des
	variables du membre gauche ($\Var(r) \subseteq \Var(l)$).
\end{enumerate}

L'ensemble des variables d'une règle $l \ra r$, noté $\Var(l \ra r)$, est
défini par $\Var(l) \cup \Var(r)$ et si la condition précédente est satisfaite
alors  $\Var(l \ra r)=\Var(l)$.

Une règle de réécriture est 
{\em linéaire à gauche} \index{regle de réécriture@règle de réécriture!linéaire à gauche}
%\Defi{linéaire à gauche}{règle} 
si son membre gauche est linéaire. Un système de réécriture est linéaire à gauche si toutes ses
règles le sont.

Une règle de réécriture $l \ra r$ est 
{\em régulière} \index{regle de réécriture@règle de réécriture!régulière}
%\Defi{régulière}{règle} 
si $\Var(l)=\Var(r)$. Un système de réécriture est régulier si toutes ses règles le
sont.

La relation \Defi{de réécriture}{relation} $\ral_R$ associée à un
système de réécriture $R$ est définie par~: $t \ral_R t'$ s'il existe une
position $p$ dans $t$, une règle $l \ra r$ dans $R$ et une substitution $\sigma$
telles que $t_{|p}=\sigma l$ et $t'=t[\sigma r]_p$.  Si on veut préciser la
position, la règle et la substitution, alors on écrira 
$t \ral_{R,p,l \ra r,\sigma} t'$.  

Par application du Théorème~\ref{Church-Rosser}, si la relation de réécriture
$\ral_R$ est convergente, alors pour décider de l'égalité $t \lra{*}_R t'$, il
suffit de calculer les formes normales $t\da{R}$ et $t'\da{R}$ puis de les
comparer.

\DEF%------------------------------------------------------------------------
Un système de réécriture $R$ est convergent (resp.  est confluent, termine) si
la relation de réécriture $\ral_R$ est convergente (resp. est confluente,
termine).
\FDEF%------------------------------------------------------------------------

\subsection{Terminaison des systèmes de réécriture}
%================================================================

La convergence requiert la terminaison\index{terminaison}.  Cette propriété est
indécidable en général même pour un système de réécriture réduit à une seule
règle linéaire à gauche, comme l'a montré M.~Dauchet \cite{Dauchet-RTA3-89}.  On
peut néanmoins prouver la terminaison dans certain cas au moyen d'un ordre sur
les termes.

\DEF%------------------------------------------------------------------------
Un ordre de réécriture sur les termes est un ordre $>$ stable par contexte et
par substitution~: pour tous termes $t,t',u$ et toute substitution $\sigma$, 
	$$t > t' \Longra u[\sigma t]_p > u[\sigma t']_p$$ 
Un ordre \Defi{de réduction}{ordre} est un ordre de réécriture n\oe thérien.
\FDEF%------------------------------------------------------------------------

On assure la terminaison de la réécriture en orientant les règles de manière à
ce que toute règle $l \ra r$ vérifie $l>r$ où $>$ est un ordre de réduction sur
les termes.

\TH%------------------------------------------------------------------------
\cite{Lankford-77-ATP-36}
Le système de réécriture $R$ termine si et seulement si $\ral_R$ est contenu dans
un ordre de réduction.
\FTH%------------------------------------------------------------------------

De nombreux auteurs ont décrit des ordres de réduction sur les termes.  Parmi
les plus connus, citons l'ordre de Knuth-Bendix ou $kbo$~\cite{KnuthBendix70},
les ordres sur les chemins
~\cite{Plaisted78,Dershowitz82,BachmairPlaisted85,JouannaudLescanneReinig82} ou
encore les interprétations
polynômiales~\cite{Lankfords75,Cherifa-Lescanne-SCP87}.

Il est souvent approprié de construire un ordre de réduction par interprétation
(polynômiale) en utilisant un homomorphisme $\tau$ de termes clos vers une
$\FF$-algèbre $\AA$ équipée d'un ordre bien fondé $>$. On note $f_\tau$ l'image
de $f \in \FF$ par $\tau$ et on demande que la contrainte de monotonicité
suivante soit satisfaite~:
	$$\forall a, ~b \in \AA, ~\forall f \in \FF,
	~a>b ~~ implique ~~ f_\tau(\ldots,a,\ldots)>f_\tau(\ldots,b,\ldots)$$
Alors, l'ordre $>_\tau$ défini par
	$$\forall s, ~t \in \TF, ~s >_\tau t ~~  si ~~  \tau(s) > \tau(t)$$
est bien fondé.

Afin de comparer les termes contenant des variables, les variables sont
introduites dans $\AA$ menant à $\AA(\XX)$ et aux variables de $\XX$ on fait
correspondre des variables distinctes dans $\AA(\XX)$. L'ordre $>_\tau$ est
étendu en définissant
	$$\forall s, ~t \in \TFX, ~s >_\tau t ~~  si ~~  \alpha(\tau(s)) > \alpha(\tau(t))$$
pour toute assignation $\alpha$ des valeurs dans $\AA$ aux variables de
$\tau(s)$ et $\tau(t)$. Puisque $>$ est supposé bien fondé, on peut montrer la
terminaison d'un système de réécriture si on trouve $\AA$, $\tau$, $\alpha$
satisfaisant les conditions précédentes.

Dans la pratique on utilise très souvent l'algèbre des entiers naturels avec
l'ordre habituel et des interprétations polynômiales et exponentielles.

\EX%------------------------------------------------------------------------
On considère le système de réécriture suivant

$
\begin{array}{lcl}
\ominus \ominus x & \ra & x \\
\ominus (x \oplus y) & \ra & (\ominus x) \oplus (\ominus y) \\
\ominus (x \otimes y) & \ra & (\ominus x) \otimes (\ominus y) \\
x \otimes (y \oplus z) & \ra & (x \otimes y) \oplus (x \otimes z) \\
(x \oplus y) \otimes z & \ra & (x \otimes y) \oplus (x \otimes z) \\
\end{array}
$\\%$

En utilisant l'interprétation exponentielle ci-dessous dans les entiers
supérieurs à $2$

$
\begin{array}{lcl}
\tau(\ominus x) & \ra & 2^{\tau(x)} \\
\tau(x \oplus y) & \ra & \tau(x)+ \tau(y)+1\\
\tau(x \otimes y) & \ra & \tau(x) \times \tau(y)\\
\tau(c) & \ra & 3 \\
\end{array}
$%$

\noindent
pour toute constante $c \in \FF$, le système a été montré terminant
dans~\cite{Filman78}.

Par exemple, pour la première règle on a $2^{2^n} > n$ pour tout entier $n>2$
assigné à la variable $x$. 
Utiliser une interprétation dans les entiers positifs ne serait pas
suffisant pour montrer les inégalités correspondant aux deux dernières règles et
considérer les entiers supérieurs à $1$ ne serait pas suffisant pour montrer
l'inégalité correspondant à la troisième règle.
\FEX%------------------------------------------------------------------------

Un ordre de réduction total contient la relation de sous-terme strict. Dans le cas contraire,
si $t_{|\omega} > t$ pour un terme $t$ et une position $\omega$, alors il existe
une chaîne infinie décroissante 
$t > t[t]_{\omega} > t[t[t]_{\omega}]_{\omega} > \ldots$.  
On appelle ordre \Defi{de simplification}{ordre}, un ordre de réduction contenant
l'ordre sous-terme.

\TH%------------------------------------------------------------------------
\cite{Dershowitz82}
Soit ${\cal F}$ un ensemble fini de symboles de fonctions.  Un système de
réécriture $R$ termine s'il existe un ordre de simplification $>$ tel que pour
toute règle $l \ra r$ de $R$, $l > r$.
\FTH%------------------------------------------------------------------------

Les ordres de simplifications peuvent être construits à partir d'un ordre sur
les symboles de fonctions $\FF$ appelé \Def{précédence}.  Parmi les ordres de
simplifications on peut citer l'{\em ordre multi-ensemble sur les
chemins}~\cite{Dershowitz82} et l'{\em ordre lexicographique sur les
chemins}~\cite{Kamin-Levy}. Pour plus de détails concernant la terminaison, nous
renvoyons le lecteur à~\cite{Dershowitz-JSC-87}.

\subsection{Les systèmes de réécriture conditionnels}
%================================================================

En ajoutant des conditions sur l'application des règles de réécriture, les
systèmes de réécriture sont naturellement étendus à des systèmes de réécriture
\Defi{conditionnels}{système de réécriture}. Plusieurs définitions des systèmes
de réécriture conditionnels ont été proposées et la correspondance entre ces
systèmes et la relation avec les systèmes équationnels a été analysée
dans~\cite{DershowitzOkada90}. La différence essentielle entre les systèmes
conditionnels est l'interprétation des conditions et nous présentons par la
suite quelques approches possibles.

Un système de réécriture \Defi{conditionnel naturel}{système de réécriture} (natural
conditional rewriting system) a des règles de réécriture de la forme
	$$l \ra r ~~ {\sf si} ~~ s_1 \lra{*} t_1 \wedge \ldots \wedge s_n \lra{*} t_n$$
où $s_i \lra{*} t_i$ sont appelées les conditions de la règle.

La règle $l \ra r$ est appliquée dans le sens de la réécriture
non-conditionnelle s'il existe une preuve pour toute condition 
$s_i \lra{*} t_i$, $i=1 \ldots n$, instanciée par la substitution appropriée, où
les preuves peuvent utilisées un nombre quelconque de réécritures dans les deux
directions. Si $n=0$ on obtient une règle non-conditionnelle.

Puisque l'application de telles règles implique des preuves arbitraires
d'égalité où la réécriture n'apporte pas beaucoup de bénéfices par rapport aux
systèmes équationnelles, on peut utiliser une définition plus restrictive de la
réécriture conditionnelle.

Un système de réécriture \Defi{conditionnel standard}{système de réécriture}
(standard (join) conditional rewriting system) a des règles de réécriture de la
forme
	$$l \ra r ~~ {\sf si} ~~ s_1 \downarrow t_1 \wedge \ldots \wedge s_n \downarrow t_n$$

Dans ce cas, une instance $\sigma l$ du membre gauche de la règle est réécrite
en $\sigma r$ seulement si, pour tout $i=1 \ldots n$, $\sigma s_i$ peut être
réduit (en utilisant zéro ou plusieurs réécritures) au même terme que 
$\sigma t_i$.

La condition d'application pour une règle de réécriture peut être affaiblie
encore plus. Un système de réécriture \Defi{conditionnel normal}{système de réécriture}
(normal conditional rewriting system) a des règles de réécriture de la forme
	$$l \ra r ~~ {\sf si} ~~ s_1 \ral^! t_1 \wedge \ldots \wedge s_n \ral^! t_n$$
où $s_i \ral^! t_i$ indique que $t_i$ est une forme normale de $s_i$.

Un système standard contenant des règles de la forme
   $$l \ra r ~~ {\sf si} ~~ s_1 \downarrow t_1 \wedge \ldots \wedge s_n \downarrow t_n$$
peut être transformé dans un système normal où les règles sont remplacées par
   $$l \ra r ~~ {\sf si} ~~ eq(s_1,t_1) \ral^! true \wedge \ldots \wedge eq(s_n,t_n) \ral^! true$$
et la règle 
   $$eq(x,x) \ra true$$
est ajoutée au système.
Les réductions des termes ne contenant pas les symboles $eq$ et $true$ sont
similaires dans les deux systèmes.

\subsection{Logique de réécriture} 	\label{logiqueReec}
%================================================================

La logique de réécriture est proposée dans~\cite{MeseguerTCS92} comme une
manière d'interpréter les systèmes de réécriture.

Une logique est définie en général par une syntaxe, un système de
déduction, une classe de modèles et une relation de satisfaisabilité.
Dans cette section, nous présentons ces quatre composantes dans le cas
de la logique de réécriture.

\subsubsection{Syntaxe}
%================================================================

La syntaxe nécessaire pour définir une logique est spécifiée par sa
signature qui nous permet de construire des formules.

\DEF%------------------------------------------------------------------------
%[Signature de la logique de réécriture]
  
Soit $\XX$ un ensemble de variables et $\LL$ un ensemble de symboles
appelés étiquettes. La signature \Defi{de la logique de réécriture}{signature}
est un triplet
	$$\Sigma = (\SS, \FF, \EE)$$
où $\SS$ est un ensemble de sortes, $\FF$ est un ensemble de symboles de
fonctions et $\EE$ est un ensemble d'axiomes équationnels dans $\TFX$.
\FDEF%------------------------------------------------------------------------

Les axiomes équationnels dans $\EE$ doivent être interprétées
comme étant des axiomes exprimés sur la signature. Les formules
$sen(\Sigma)$ formées sur la signature $\Sigma$ sont définies comme
des {\em séquents} $Seq(\Sigma)$ de la forme suivante
	$$\pi : [t]_{\EE} \ra [t']_{\EE}$$
où $t, ~t^\prime \in \TFX$ et $\pi \in \TT_{\FF \cup \LL \cup \{ ; \}}$.

$\pi$ est appelé un terme de preuve et l'ensemble de tous ces termes de preuve
$\TT_{\FF \cup \LL \cup \{ ; \}}$ est désigné par $\Pi$.

Le sens informel du séquent $\pi : [t]_{\EE} \ra [t']_{\EE}$ est que
$\pi$ permet de dériver les termes de la classe d'équivalence $[t']_{\EE}$ à
partir des termes de la classe d'équivalence $[t]_{\EE}$ et que le terme de
preuve $\pi$ représente une preuve de cette dérivation.

\subsubsection{Système de déduction}
%================================================================

Afin de construire le système de déduction de la logique de
réécriture, on introduit d'abord la notion de {\em théorie de réécriture}.

\DEF%------------------------------------------------------------------------
%[Théorie de réécriture] 
  
Une \Def{théorie de réécriture} est définie par un quadruplet 
$\TR = (\Sigma, \LL, \XX, \RR)$, où $\Sigma = (\SS, \FF, \EE)$ est une signature
composée des sortes $\SS$, des symboles de fonctions $\FF$ et des équations
$\EE$ dans $\TFX$, $\XX$ est un ensemble infini de variables, $\LL$ est un
ensemble d'étiquettes des règles et $\RR$ est un ensemble de règles de
réécriture étiquetées de la forme
	$$[\ell] ~~ l \ra r$$
où l'étiquette $\ell \in \LL$, les membres gauche et droit $l, r \in \TFX$ tels
que $\Var(r) \subseteq \Var(l)$ et l'arité de l'étiquette $\ell$ est égale au
nombre de variables distinctes dans cette règle.
\FDEF%------------------------------------------------------------------------

L'ensemble d'équations $\EE$ définit une relation de congruence modulo laquelle
la réécriture par les règles de $\RR$ est réalisée. Typiquement, l'ensemble
$\EE$ contient des équations qui ne sont pas orientables, i.e., transformables
en un système de réécriture terminant. Cependant, la terminaison, et aussi la
confluence, peuvent être des propriétés souhaitables pour certains
sous-ensembles de règles de $\RR$.

La relation de déduction $\vdash$ est donc définie comme suit.

\DEF%------------------------------------------------------------------------
%[Système de déduction]
  
Étant donnée une théorie de réécriture étiquetée $\TR$, le séquent 
$\pi: [t]_{\EE} \ra [t']_{\EE}$ se déduit à partir de $\TR$ si $\pi$ est
obtenu en appliquant un nombre fini de fois {\em les règles de déduction de la
logique de réécriture} données dans la figure \ref{regles_de_deduction}. Ceci
est désigné par
	$$\TR ~ \vdash ~ \pi: [t]_{\EE} \ra [t']_{\EE}$$ 
\FDEF%------------------------------------------------------------------------

\begin{figure}[!htp]
\centering
\fbox{\begin{minipage}{\largeurtexte}
$$
\begin{array}{ll}

{\textbf{Réflexivité}}   %\\
                    & {\bf \Rightarrow}\\
                    & [t]_\EE : [t]_\EE \ra [t]_\EE \\
                    & {\bf si} ~~ t \in \TFX\\
\\
{\textbf{Congruence}}     %\\
                    & \pi_1 : [t_1]_\EE \ra [{t^\prime}_1]_\EE, \ldots , \pi_n : [t_n]_\EE \ra [{t^\prime}_n]_\EE\\
                    & {\bf \Rightarrow}\\
                    & f(\pi_1 , \ldots , \pi_n) : [f(t_1, \ldots, t_n)]_\EE \ra [f(t'_1, \ldots, t'_n)]_\EE\\
                    & {\bf si} ~~ f \in \FF_n\\
\\
{\textbf{Remplacement}}   %\\
                    & \pi_1 : [t_1]_\EE \ra [{t^\prime}_1]_\EE, \ldots , \pi_n : [t_n]_\EE \ra [{t^\prime}_n]_\EE\\
                    & {\bf \Rightarrow}\\
                    & \ell(\pi_1 , \ldots , \pi_n) : [l(t_1, \ldots, t_n)]_\EE \ra [r(t'_1, \ldots, t'_n)]_\EE\\
                    & {\bf si} ~~ [\ell(x_1 , \ldots , x_n)] l(x_1 , \ldots , x_n) \ra r(x_1 , \ldots , x_n) \in \RR\\
\\
{\textbf{Transitivité}}   %\\
                    & \pi_1:[t_1]_\EE \ra [t_2]_\EE ,  \pi_2:[t_2]_\EE \ra [t_3]_\EE\\
                    & {\bf \Rightarrow}\\
                    & \pi_1 ; \pi_2: [t_1]_\EE \ra [t_3]_\EE\\

\end{array}
$$
\end{minipage}}
\caption{Règles de déduction de la logique de réécriture}
\label{regles_de_deduction}
\end{figure}

\subsubsection{Modèle}
%================================================================

Le modèle de la logique de réécriture présenté ici est basé sur une
axiomatisation algébrique des séquents de réécriture. En particulier, on
s'intéresse à une {\em sémantique algébrique}.

La sémantique algébrique permet de décrire l'idée intuitive d'un système de
réécriture: les états du système sont des classes d'équivalence de termes modulo
$\EE$ et les transitions sont des réécritures utilisant les règles du système de
réécriture. Ainsi, l'espace des calculs de la théorie de réécriture $\TR$ peut
être choisi comme un modèle de la logique de réécriture. Cet espace des calculs
est déterminé par l'ensemble des termes de preuves $\pi$ calculés dans les
séquents $\pi : [t]_{\EE} \ra [t^\prime]_{\EE}$ modulo une équivalence
de calcul. Cette équivalence est donnée par $\EE$ et un ensemble $\EE_{\Pi}$
d'axiomes équationnels sur les termes de preuves décrits dans la
Figure~\ref{equivalence_de_termes_de_preuves}, où
\begin{itemize}
\item les deux premiers axiomes décrivent les équations habituelles
  d'associativité et d'identité;
\item l'axiome de préservation de composition décrit une équivalence
  entre la composition de plusieurs pas de réécriture dans le contexte
  ``$f$'' et la composition de chaque pas de réécriture dans ce
  contexte;
\item l'axiome de préservation d'identités $\EE$ décrit la
  stabilité par contexte de $\EE$;
\item l'équivalence induite par les cinq premières équations définit
  des termes de preuve équivalents tels que les dérivations
  correspondantes diffèrent uniquement par l'ordre de réduction de
  radicaux;
\item l'axiome de permutation parallèle décrit la réduction {\em
    simultanée} de radicaux compatibles.  Ceci peut être simulé par
  une composition d'exécution séquentielle \cite{Gadducci-96}.
  Intuitivement, la réécriture au sommet par une règle $\ell$ et une
  réécriture en dessous sont des processus indépendants ce qui permet
  ainsi leur exécution dans n'importe quel ordre.
\end{itemize}

\begin{figure}[!htp]
\centering\fbox{\begin{minipage}{\largeurtexte}
$$
\begin{array}{lll}

{\textbf{Associativité}} %\\
& \forall \pi_1 , \pi_2 , \pi_3 \in \Pi \\
& \pi_1 ; (\pi_2 ; \pi_3) = (\pi_1 ; \pi_2) ; \pi_3 \\
\\

{\textbf{Identités}} %\\
& \forall \pi: [t]_{\EE} \ra [t^\prime]_{\EE}, \\
& \pi ; [t^\prime]_{\EE} = \pi, {\rm ~~ et ~~} [t]_{\EE}; \pi = \pi \\
\\

{\textbf{Préservation de composition}} %\\
& \forall f \in \FF_n , n = |f|, \forall \pi_1 , \ldots , \pi_n , ~ \pi'_1 , \ldots , \pi'_n : \\
& f(\pi_1 ; \pi'_1 , \ldots , \pi_n ; \pi'_n) =
f(\pi_1 , \ldots , \pi_n) ; f(\pi'_1 , \ldots , \pi'_n)\\
\\

{\textbf{Préservation d'identités}} %\\
& \forall f \in {\cal }F_n , n = |f|: \\
& f([t_1]_{\EE} , \ldots , [t_n]_{\EE}) = [f(t_1,\ldots,t_n)]_{\EE} \\
\\

{\textbf{Axiomes de \EE}} %\\
& \forall u = v \in \EE , \forall \pi_1, \ldots, \pi_n: \\
& u(\pi_1 , \ldots , \pi_n) = v(\pi_1 , \ldots , \pi_n) \\
\\

{\textbf{Permutation parallèle}} %\\
& \forall [\ell]~l \ra r \in \RR,  
\forall \pi_1:[t_1]_{\EE} \ra [{t^\prime}_1]_{\EE}, \ldots ,
\pi_n: [t_n]_{\EE} \ra [{t^\prime}_n]_{\EE} \\
&   \ell(\pi_1 , \ldots , \pi_n) = \ell([t_1]_{\EE} , \ldots , [t_n]_{\EE}) ; r(\pi_1 , \ldots , \pi_n) {\rm ~et~} \\
&   \ell(\pi_1 , \ldots , \pi_n) = l(\pi_1 , \ldots , \pi_n) ; \ell([{t^\prime}_1]_{\EE} , \ldots ,[{t^\prime}_n]_{\EE})

\end{array}
$$
\end{minipage}
}
\caption{Équivalence des termes de preuves -- $\EE_{\Pi}$}
\label{equivalence_de_termes_de_preuves}
\end{figure}

Le modèle considéré est un ensemble quotient noté
	$$\TT_{\TR} = \{ \pi ~ | ~ \TR \vdash \pi : [t]
	\ra [t^\prime]\} / ({\EE \cup \EE_{\Pi}})\}$$

\subsubsection{Satisfaisabilité}
%================================================================

La relation de satisfaisabilité $\models \subseteq \TT_{\TR} \times Seq(\Sigma)$
doit être compatible avec les morphismes des signatures. Elle est définie
dans~\cite{Meseguer-GenLog89}.

\subsection{Systèmes de calcul} \label{systemeCalcul}
%================================================================

Les systèmes de calcul ont été introduits par Kirchner, Kirchner et Vittek
dans~\cite{KirchnerKV-MIT95}, où ils présentent une version plus
élaborée des idées qu'ils avaient proposées originalement
dans~\cite{KirchnerKirchnerVittek-PPCP93}. Un système de calcul enrichit le
formalisme de la logique de réécriture avec une notion de \Def{stratégie}~: un
\Def{système de calcul} est composé d'une théorie de réécriture et d'un {\em
système de stratégies}. Les stratégies contrôlent l'application des règles de
réécriture en spécifiant des parcours dans l'arbre de toutes les dérivations
possibles et de cette façon décrivent quels sont les n\oe uds considérés comme 
des résultats d'un calcul. Elles sont utilisées, d'une part, pour décrire le
déroulement de preuves qui nous intéressent et, d'autre part, pour restreindre
l'espace de recherche de ces preuves.

La première composante d'un système de calcul est une théorie de
réécriture $\TR$ à partir de laquelle on définit la notion de
calcul.

\DEF%------------------------------------------------------------------------
%[Pas de réécriture simple]
  
Étant donnés une théorie de réécriture $\TR$ et un séquent 
$\pi: [t]_{\EE} \ra [t']_{\EE}$ avec le terme de preuve 
$\pi=t[\ell(\sigma x)]_\omega$, un {\em pas de réécriture simple} est défini par
  $$[t]_{\EE} \Rightarrow_{\ell, \sigma, \omega} [t']_{\EE}$$
\FDEF%------------------------------------------------------------------------

Cette définition correspond exactement à la notion traditionnelle d'un
pas de réécriture à la position $\omega$ en utilisant la règle
étiquetée avec l'étiquette $\ell$ et le match $\sigma$.

En plus, on s'intéresse à une représentation canonique de tous les
calculs qui sont équivalents modulo les axiomes $\EE_{\Pi(R)}$.
Puisque tout séquent peut être décomposé en une composition de
séquents élémentaires (séquentiels)
	$$\forall ~ \pi : [t]_{\EE} \ra [t']_{\EE}$$
soit 
	$$\pi = [t]_{\EE} = [t']_{\EE}$$
soit
	$$\exists n \in \N ~~ tel ~ que ~~ [t]_{\EE} = [t_0]_{\EE}
	\Rightarrow_{\ell_0} [t_1]_{\EE} \Rightarrow_{\ell_1}
	[t_2]_{\EE} \ldots \Rightarrow_{\ell_{n-1}} [t_n]_{\EE} =
	[t']_{\EE}$$
et
	$$\pi =_{A(;)} (\pi_0 ; \pi_1 ; \pi_2 ; \ldots ; \pi_{n-1})$$
où $A(;)$ désigne l'associativité du ``;''.

Pour un terme de preuve $\pi$, $[t_n]_{\EE}$ est appelé le {\em résultat de
l'application de $\pi$ sur $[t_0]_{\EE}$} et il est aussi désigné par 
$[t]_{\EE} \stackrel{\pi}{\Rightarrow} [t']_{\EE}$.  La relation d'équivalence
générée par $(\EE \cup \EE_{\Pi(R)})$ sur les termes de preuve induit une
équivalence sur les calculs~: deux calculs sont équivalents s'ils amènent au
même résultat et que leurs termes de preuve sont équivalents.

En général, on ne s'intéresse pas à tous les calculs, on s'intéresse
seulement à ceux guidés par une stratégie, c'est-à-dire une
description de la séquence de pas de réécriture élémentaires permis
par les calculs.  D'un point de vue formel, une stratégie est un
ensemble de termes de preuve, i.e., un sous-ensemble des termes de
preuve $\Pi$, qui est clos par concaténation.

La relation de transition $\pi : [t]_{\EE} \ra [t']_{\EE}$ peut être
étendue pour les stratégies. 

\DEF%------------------------------------------------------------------------
%[Stratégie]
  
Soient $S \subseteq \Pi$ et $t, t^\prime \in \TF$.  La relation
  $$S : [t]_{\EE} \ra [t']_{\EE}$$
est vraie s'il existe un terme de preuve $\pi \in S$ tel que
  $$\pi : [t]_{\EE} \ra [t']_{\EE}$$
  
Le résultat de l'application d'une \Def{stratégie} $S$ sur un terme $t$, désigné
fonctionnellement par $S(t)$, est défini comme suit
  $$S(t) = \{ [t^\prime]_{\EE} | \exists \pi \in S, 
    ~[t]_\EE \stackrel{\pi}{\Rightarrow} [t']_\EE \}$$
\FDEF%------------------------------------------------------------------------

La relation $S : t \ra t'$ exprime la dérivabilité du terme $t$
en $t^\prime$ suivant une certaine stratégie $S$. À partir de cette
définition, on peut noter que l'application d'une stratégie sur un terme
peut retourner plusieurs résultats.

Une première façon de décrire une stratégie est d'énumérer extensivement le
sous-ensemble des termes de preuve. Cette approche n'est pas
satisfaisante en pratique, donc le problème est de définir un langage permettant
de décrire des sous-ensembles des termes de preuve. La différence entre la
représentation d'une stratégie comme un ensemble de termes de preuve et une
expression dans un formalisme de stratégies reflète la différence entre la vue
sémantique des stratégies et la vue syntaxique des stratégies exprimées sous la
forme d'un programme dans un langage de stratégies. Dans la
Section~\ref{langage_elan}, nous présentons les opérateurs de stratégies utilisés
dans le langage \elan\  qui peut être décrit comme un cadre logique pour le
prototypage de systèmes de calcul.

On peut maintenant définir formellement la notion de système de calcul.

\DEF%------------------------------------------------------------------------
%[Système de calcul]
  
Un \Def{système de calcul} est composé d'une théorie de réécriture 
$\TR=(\Sigma, \LL, \XX, \RR)$ et d'une stratégie $S$.
\FDEF%------------------------------------------------------------------------

\subsection{Langage \elan}      \label{langage_elan}
%=============================================================================== 
%

Le langage \elan\  a été conçu au sein du projet Protheo à Nancy au début des
années quatre-vingt-dix. Sa première version est décrite dans la thèse de
Vittek~\cite{VittekThese} et son implantation est détaillée
dans~\cite{ELAN-Manual-95}. Au cours des années, le langage a évolué et depuis
le début de l'année 2000 la version 3.4 est disponible~\cite{ELAN-Manual-00}.

\elan\  a été conçu comme un cadre logique pour le prototypage de systèmes de
calcul. Du point de vue de la programmation, le langage offre la possibilité de
spécifier des systèmes de calcul composés de théories de réécriture
multi-sortées, chacune décrite par une signature et par un ensemble de règles de
réécriture et de stratégies d'exécution.
\begin{itemize}
\item La signature définit les sortes et les symboles de fonctions
  utilisés dans la description de la théorie. \elan\  permet d'utiliser
  des symboles libres et associatifs-commutatifs, qui peuvent être
  spécifiés en utilisant une notation \textit{mixfix}.
\item L'ensemble de règles de réécriture est composé de règles
  non-nommées et de règles nommées ou étiquetées.
\begin{itemize}
	\item Les règles non-nommées sont utilisées pour la normalisation de
	termes.  Leur application n'est pas contrôlée par l'utilisateur, elles
	sont exécutées avec une stratégie pré-définie dans le langage.  Cette
	stratégie pré-définie est la stratégie de normalisation
	\textit{leftmost-innermost}. Puisque la stratégie de normalisation est
	pré-définie dans \elan, elle n'est pas spécifiée dans la théorie de
	réécriture de l'utilisateur, l'ensemble de règles non-nommées doit être
	confluent et terminant.
	\item L'ensemble de règles nommées, qui n'est pas nécessairement
	confluent et terminant, peut être contrôlé par des \textit{stratégies
	élémentaires}.  Les deux raisons principales pour leur utilisation sont
\begin{itemize}
\item si l'ensemble de règles n'est pas terminant, l'utilisateur a la
  possibilité de restreindre l'ensemble de dérivations à un
  sous-ensemble de dérivations finies, afin d'éviter des dérivations
  infinies;
\item si l'ensemble de règles n'est pas confluent, l'utilisateur a la
  possibilité de spécifier certains sous-ensembles de toutes les
  dérivations possibles et obtenir ainsi un sous-ensemble de tous les
  résultats possibles.
\end{itemize}
\end{itemize}
\item Les stratégies sont utilisées en \elan\  de trois façons
  différentes
\begin{itemize}
\item pour séparer dans un programme la partie calcul de la partie
  contrôle;
\item pour exprimer des dérivations non-déterministes;
\item pour spécifier des procédures de normalisation particulières.
\end{itemize}
\end{itemize}

Parmi plusieurs caractéristiques, le calcul non-déterministe d'\elan\  
le différencie d'autres systèmes basés sur la réécriture. L'avantage
de cette option est que cela permet de travailler avec des systèmes de
réécriture non-confluents.

Le style de programmation en \elan, basé sur le paradigme des systèmes
de calcul, unifie certaines caractéristiques de la programmation
fonctionnelle et logique. La programmation par réécriture est
similaire à l'approche fonctionnelle restreinte au premier ordre.
Cependant, la possibilité de spécifier des sous-ensembles de
dérivations par un langage de stratégies joue le rôle du
non-déterminisme de la programmation logique.

La variété des applications qui ont été implantées en \elan\  illustre
la généralité du paradigme des systèmes de calcul et montre
l'expressivité et la puissance du langage comme un outil de
programmation. Parmi elles, on peut citer
\begin{itemize}
\item une implantation de la procédure de résolution de contraintes
	d'ordre pour la preuve de terminaison basée sur l'ordre général sur les
	chemins~\cite{GenetGnaedig-GPO95};
\item deux implantations de la procédure de complétion de
	Knuth-Bendix~\cite{KirchnerM-RTA95,KirchnerLynchScharff-RTA96};
\item une implantation du prouveur de prédicats {\sf B}~\cite{CirsteaKirchner97};
\item vérification du protocole d'authentification de Needham-Schroeder~\cite{Cirstea-Proto-99};
\item la combinaison d'algorithmes d'unification~\cite{Ringeissen-RTA97};
\item un algorithme d'unification d'ordre supérieur~\cite{Borovansky-95};
\item résolution de CSP~\cite{CastroThese98};
\item CLP~\cite{KirchnerRingeissen-FI98};
\item la réécriture du premier ordre~\cite{Reflection-RwLg1996} et d'ordre
	supérieur.
\end{itemize}

La première version d'\elan\  avait offert un interpréteur et un compilateur
restreint~\cite{Vittek-RTA96}. De nouvelles techniques de compilation de systèmes de calcul ont été
étudiées et maintenant il existe un compilateur du langage permettant d'utiliser
des symboles associatifs-commutatifs~\cite{MoreauK-ASF+SDF97,MoreauK-PLILP+ALP98}.

Dans le reste de cette section, nous présentons brièvement le langage
\elan. Nous illustrons la syntaxe des trois composants d'un système
de calcul~: signatures, règles de réécriture et stratégies.  Une description
formelle et détaillée du langage est donnée dans~\cite{BorovanskyThese98}, une
sémantique du point de vue fonctionnelle est présentée dans~\cite{BKK-Fuji-98,BKKR-IJFCS-2001}
et tous les détails nécessaires pour l'utilisation du langage peuvent être
trouvés dans~\cite{ELAN-Manual-00}.

Afin d'illustrer les composants et l'utilisation du langage \elan\  nous
présentons une partie de la spécification du protocole d'authentification
Needham-Schroeder~\cite{NSPK78}. Le but de ce protocole et d'établir une
authentification mutuelle entre plusieurs agents communiquant dans un réseau
non-sécurisé (c'est-à-dire en présence des intrus). Dans les exemples suivants
nous présentons seulement quelques règles décrivant le protocole et une
stratégie recherchant toutes les attaques possibles; une description plus
détaillée est donnés dans~\cite{Cirstea-Proto-99}.

\subsubsection{Signatures \elan}
%=============================================================================== 
%

\elan\  permet de définir des signatures multi-sortées qui sont spécifiées par un
ensemble de sortes $\SS$.  L'exemple \ref{definition_de_sortes_en_elan} présente
la déclaration de différentes sortes utilisées pour prototyper le protocole
d'authentification Needham-Schroeder.

\EX%------------------------------------------------------------------------ 
(Déclaration de sortes en \elan)
\label{definition_de_sortes_en_elan} 

La déclaration des sortes des termes représentant les agents, l'intrus, le
réseau et leur caractéristiques, peut être faite en \elan\  de la manière
suivante

\vbox{
\begin{verbatim}
   sort
      agent intruder SWC AgentId Nonce message network state;
   end
\end{verbatim}
}
\FEX%------------------------------------------------------------------------

Une fois déclarées les sortes de la signature, on peut définir les
symboles de fonctions indexés qui appartiennent à l'ensemble de
symboles de fonctions $\FF$ de la signature. Chaque symbole,
défini par son profil en notation \textit{mixfix}, peut être décoré par
des attributs sémantiques comme étant un symbole libre ou
associatif-commutatif (\texttt{(AC)}) et il peut aussi être décoré par
des attributs syntaxiques tels que
\begin{itemize}
\item sa priorité syntaxique (e.g. \texttt{pri 10});
\item sa visibilité dans d'autres modules (e.g. \texttt{global/local});
\item son associativité syntaxique par défaut à gauche (\texttt{assocLeft}) ou à
	droite (\texttt{assocRight});
\item le fait d'être synonyme avec un autre symbole (e.g. \texttt{alias}).
\end{itemize}

L'Exemple~\ref{definition_de_symboles_de_fonction_en_elan} montre la
définition de symboles de fonctions avec des attributs syntaxiques et
sémantiques.

\EX%------------------------------------------------------------------------
\label{definition_de_symboles_de_fonction_en_elan}
(Définition de symboles de fonctions en \elan)

À partir de la déclaration de sortes de
l'Exemple~\ref{definition_de_sortes_en_elan}, on peut spécifier que toute
constante de la sorte {\em \texttt{int}} est un {\em \texttt{AgentId}}. On peut
également définir les états ({\em \texttt{SWC}}) possibles d'un agent et la
modalité de construire des nonces. Un message est défini en précisant son
expéditeur, son destinataire et deux nonces cryptés avec la clé publique du
destinataire. L'ensemble de messages représentant le réseau ({\em
\texttt{network}}) est défini en utilisant l'attribut {\em \texttt{AC}} pour
l'opérateur $\&$. Le caractère spécial {\em \texttt{@}} est utilisé pour indiquer la 
position d'un argument.
%\vbox{
\begin{verbatim}
   operators   global
      @       : (int)         AgentId;

      SLEEP  : SWC;
      WAIT   : SWC;
      COMMIT : SWC;

      N(@,@)  :  (AgentId AgentId) Nonce;

      @ + @ + @  :  ( AgentId SWC Nonce ) Agent;
      @-->@ K(@)[@,@]  :  (AgentId AgentId AgentId Nonce Nonce) message;

      @  :  (message) network;
      @ & @  :  (network  network) network (AC);

      @ <> @ <> @ <> @  :  ( Agent Agent intruder network) state;
   end 
\end{verbatim}
%}
L'état général consiste en les états des deux agents participant à la
communication, de l'intrus et du réseau.
\FEX%------------------------------------------------------------------------

\subsubsection{Règles de réécriture}
%=============================================================================== 
%

Il existe deux types de règles souvent introduites dans une théorie de
réécriture: les règles non-conditionnelles et les règles conditionnelles. Le
langage \elan\  introduit en plus la notion d'\textit{affectation locale} pour
des variables (locales) non-instanciées pendant le
filtrage~\cite{ELAN-Manual-00}. Cela nous permet d'appliquer une stratégie sur
un terme autre que celui de tête et aussi de garder la valeur d'une variable
lorsqu'elle est utilisée plusieurs fois dans une règle.

La syntaxe des règles conditionnelles avec des affectations locales
est la suivante
$$
\begin{array}{ll}
[\ell] \quad l \rewrite & r \\
                        & \ifwhere
\end{array}
$$
où
\begin{itemize}
\item $\ell \in \LL$ est l'étiquette de la règle (qui est vide
  dans le cas d'une règle non-nommée);
\item $l$ et $r$ sont des termes de $\TFX$
  représentant les membres gauche et droit de la règle;
\item $\ifwhere ~~ \defgrammar ~~ \{\iif ~~ v ~ | ~ \where ~~ y ~
  \assign (S)u ~ | ~ \where ~~ y ~ \assign ()u \}^*$ où
\begin{itemize}
	\item $\iif ~~ v$ est une condition booléenne;
	\item $\where ~~ y ~ \assign (S)u$ est une affectation de la variable 
	$y \in \XX$ par le résultat de l'application de la stratégie $S$ sur le
	terme $u \in \TFX$;
	\item $\where ~~ y ~ \assign () u$ est une affectation de la variable 
	$y \in \XX$ par le résultat de la normalisation du terme $u$.
\end{itemize}
\end{itemize}

L'application d'une règle de réécriture à un terme clos commence par une étape
de filtrage permettant de calculer la substitution associée au problème de
filtrage considéré. Les évaluations locales et les conditions sont alors
évaluées les unes à la suite des autres (de haut en bas) jusqu'à atteindre la
dernière~; c'est seulement à ce moment là que la règle peut s'appliquer et que
le membre droit est construit. Chaque condition $v$ est mise en forme normale
puis comparée à la valeur de vérité~\texttt{true} pré-définie par le système. En
cas d'égalité, on dit que la condition est satisfaisable et le calcul des
évaluations locales se poursuit.
L'affectation locale $\where ~ y ~ \assign (S)u$ permet de déclencher
l'application d'une stratégie. Dans un premier temps, le terme $u$ est mis en
forme normale en n'utilisant que des règles non nommées, la stratégie $S$ est
ensuite appliquée sur le terme en forme normale.
En cas d'échec d'une condition ou (de la stratégie) d'une évaluation locale, un
mécanisme de \textit{retour arrière} (\textit{backtracking}) est déclenché~: les
évaluations locales précédentes sont réévaluées pour en extraire d'autres
solutions. Si aucune autre solution n'est trouvée, on dit que l'application de
la règle courante échoue et une autre règle est sélectionnée.

\EX%------------------------------------------------------------------------
\label{affectation_locales}
(Règle avec affectation locales)

Les règles du protocole Needham-Schroeder décrivent l'évolution de l'état global
pendant une session.  À partir des sortes et des symboles de fonctions définis
dans l'Exemple~\ref{definition_de_sortes_en_elan} et
l'Exemple~\ref{definition_de_symboles_de_fonction_en_elan}, respectivement, la
règle {\em \texttt{initiate}} ci-dessous décrit l'initialisation de la
communication en envoyant en réseau le message construit à partir des identités
des deux agents participants.

\vbox{
\begin{verbatim}
   rules for term
      x,y : AgentId;  m,n : Nonce;  ls : network;
      I : intruder;  mes : message;
   global 
      [initiate] x+SLEEP+n <> y+SLEEP+m <> I <> ls 
                 =>
                 x+WAIT+N(x,y) <> y+SLEEP+m <> I <> mes & ls
                       where mes :=() genMessage(x,y)
      end
   end
\end{verbatim}
}

La variable {\em \texttt{mes}} représentant le message envoyé en réseau par
l'agent avec l'identité $x$ est instanciée en normalisant le terme {\em
\texttt{genMessage(x,y)}} lequel dans ce cas génère un message 
\linebreak
{\em \texttt{x-->y K(y)[N(x,y),N(x,y)]}}. L'émetteur du message change son état
en {\em \texttt{WAIT}} et attend une réponse à son message.
\FEX%------------------------------------------------------------------------

Pour des raisons de confort et d'efficacité d'exécution, le langage dispose de
certaines extensions pour la construction de règles: les \textit{affectations
généralisées} et la \textit{factorisation de règles}.

%\paragraph{Affectation généralisée}

L'\textit{affectation généralisée} est une construction syntaxique
	$$\where ~~ (sort) ~~ p \assign (S) u$$
où $p$ est un terme non-clos de sorte $sort \in \SS$.
     
Le terme $p$, dit \textit{motif}, est composé de constructeurs et de
variables, où un constructeur est un symbole de fonction qui
n'apparaît pas comme opérateur de tête dans un membre gauche d'une
règle de réécriture.
     
Toutes les variables dans le motif $p$ non encore instanciées, sont
instanciées par le filtrage de ce motif $p$ avec le résultat de
l'application de la stratégie $S$ au terme $u$, ou au cas où une
stratégie $S$ n'est pas spécifiée, le résultat de la normalisation du
terme $u$.
     
\EX%------------------------------------------------------------------------
\label{affectation_generalises}
(Règle avec affectation généralisée)

La réponse d'un message envoyé par l'initiateur d'une communication, comme celui
généré dans la règle de l'Exemple~\ref{affectation_locales}, est construite en
utilisant l'information contenue dans le message initial. Une affectation généralisée
est employée afin d'extraire les caractéristiques du message qui était destiné à
l'agent \texttt{y}, identité qui est spécifiée dans le motif à filtrer.

\vbox{
\begin{verbatim}
   rules for term
      y,z,n1,n2,n3,n4 : AgentId;  m : Nonce;  ls : network;
      S : Agent;  I : intruder;  mes : message;
   global 
     [response] S <> y+SLEEP+m <> I <> mes & ls
                =>
                S <> y+WAIT+N(y,z) <> I <> y-->z K(z)[N(n1,n3),N(y,z)] & ls
                       where (message) z-->y K(y)[N(n1,n3),N(n2,n4)] :=() mes
     end
   end
\end{verbatim}
}
\FEX%------------------------------------------------------------------------

%\paragraph{Factorisation de règles de réécriture}

La construction de factorisation permet de mettre en facteur des
parties communes de plusieurs règles ayant les mêmes membres gauche et
droit. Cela permet, d'une part, de supprimer une ou plusieurs règles
de réécriture, et d'autre part, d'éviter le filtrage et des exécutions
communes dans plusieurs règles.

En général, la syntaxe des règles est la suivante
$$
\begin{array}{lll}
\textrm{règle} & \defgrammar & l \rewrite r \\
               &             & \ifwherechoose \\
\ifwherechoose & \defgrammar & \{\ifwhere | \\
               &             & \choice \\
               &             & \{ \try ~~ \ifwherechoose \}^+ \\
               &             & end \}^{*}
\end{array}
$$

Cette construction de factorisation peut être enchaînée au même niveau
que des conditions $\iif$ et des affectations locales $\where$, mais
elle peut également être imbriquée.

\EX%------------------------------------------------------------------------
\label{choose_try}
(Règle avec factorisation)

Si l'agent initiant la communication de l'Exemple~\ref{affectation_locales} reçoit
le message attendu alors il peut passer dans l'état \texttt{COMMIT} représentant
une session accomplie. Si le message ne contient pas le nonce correct alors une
erreur est obtenue.

\vbox{
\begin{verbatim}
rules for term
   x,v,w,n1,n2,n3,n4 : AgentId;  ls : network;
   R : Agent;  I : intruder;
global 
[ack] x+WAIT+N(x,v) <> R <> I <> w-->x K(x)[N(n1,n3),N(n2,n4)] & ls
      =>
      S
   choose
   try
    if x==n1 and v==n3
    where S:=() x+COMMIT+N(x,v) <> R <> I <> x-->v K(v)[N(n2,n4),N(n2,n4)] & ls
   try
    if x!=n1 or v!=n3
    where S:=() ERROR
   end
end end
\end{verbatim}
}

Un dernier message de confirmation est envoyé dans le cas où le nonce reçu
était correct.
\FEX%------------------------------------------------------------------------

\subsubsection{Stratégies élémentaires d'\elan}
%=============================================================================== 
%

Le langage de \textit{stratégies élémentaires} d'\elan\  permet de contrôler
l'application des règles nommées, de définir des exécutions
non-déterministes et de spécifier des dérivations simultanées. 

Le langage de stratégies élémentaires offre plusieurs opérateurs et nous
présentons juste les plus importants en décrivant
\begin{itemize}
\item la construction pour la concaténation de stratégies: ``;''; 
\item les constructions de choix $\dk$, $\dc$, $\first$;
\item la construction d'itération $\repeate*$;
%\item la construction de normalisation $\normalise$;
\item les constructions pour les stratégies identité
  et échec: $\id$ et $\fail$, respectivement.
\end{itemize}

La syntaxe et la sémantique opérationnelle des stratégies élémentaires,
de façon informelle, sont les suivantes.
\begin{itemize}

\item[]\textbf{Construction de concaténation} ~
\begin{description}
    
\item[;] La concaténation de stratégies $S_1 ~ ; ~ S_2$ correspond à
  l'axiome de transitivité de la logique de réécriture. Pour typer une
  concaténation, il faut que les deux stratégies $S_1$ et $S_2$ soient
  de la même sorte, qui devient également la sorte de cette
  concaténation.
  
\end{description}

\item[]\textbf{Constructions de choix}
\begin{description}

\item[\dk] La stratégie $\dk(S_1 ~ , ~ \dots ~ , ~ S_n)$ donne tous
  les résultats de l'application de toutes les stratégies $S_1, \dots,
  S_n$. Si toutes les stratégies $S_i$ échouent alors la stratégie
  \dk\  échoue.
  
\item[\dc] La stratégie $\dc(S_1 ~ , ~ \dots ~ , ~ S_n)$ donne tous
  les résultats de l'application d'une des stratégies $S_1, \dots,
  S_n$ laquelle est choisie de manière aléatoire. Si toutes les
  stratégies $S_i$ échouent alors la stratégie \dc\  échoue.
  
\item[\first] La stratégie $\first(S_1 ~ , ~ \dots ~ , ~ S_n)$ donne
  tous les résultats de l'application de la première stratégie $S_1,
  \dots, S_n$ qui est applicable (en ordre textuel). Si toutes les
  stratégies $S_i$ échouent alors la stratégie \first\  échoue.
  
\end{description}

\item[]\textbf{Constructions d'itération}
\begin{description}
  
\item[$\repeate*$] La stratégie $\repeate*(S)$ correspond à
  $S^{i} = \overbrace{S ~ ; ~ \dots ~ ; ~ S}^{i}$ si $S^{i+1}$ échoue.
  Si la stratégie $S$ échoue alors la stratégie $\repeate*$
  correspond à l'identité, elle n'échoue jamais.
  
\end{description}

\item[]\textbf{Constructions d'identité et d'échec}
\begin{description}
  
\item[\id] La stratégie $\id$ correspond à l'identité, elle retourne
  le même terme d'entrée et pourtant elle peut toujours être
  appliquée.
  
\item[\fail] La stratégie $\fail$ correspond à un échec, elle échoue
  toujours.

\end{description}

\end{itemize}

En utilisant des stratégies non-déterministes nous pouvons explorer
exhaustivement l'espace de recherche d'un problème donné et trouver des schéma
satisfaisant des propriétés spécifiques.

L'exemple \ref{strategies_en_elan} montre la définition d'une stratégie en
\elan.

\EX%------------------------------------------------------------------------
\label{strategies_en_elan}
(Définition de stratégies en \elan)

La stratégie recherchant des attaques possibles applique d'une manière
répétitive et non-déterministe toutes les règles de réécriture décrivant le
comportement des agents honnêtes et de l'intrus et sélectionne seulement les
résultats représentant une attaque.

\vbox{
\begin{verbatim}
    []attStrat => repeat*(
                            dk( initiate, response, ack, ..., intruder )
                  );
                  attackFound
    end
\end{verbatim}
}

Le résultat de la stratégie {\em \texttt{repeat*(...)}} est l'ensemble de tous
les comportements possibles dans une session du protocole où les messages
peuvent être interceptés ou truqués par un intrus. La stratégie {\em
\texttt{attackFound}} vérifie ensuite si le terme reçu en entrée représente une
attaque et choisit donc parmi l'ensemble précédent de résultats seulement ceux
qui représentent une attaque.
\FEX%------------------------------------------------------------------------

Le langage \elan\  utilise implicitement une stratégie de normalisation
avec l'ensemble de toutes les règles non-nommées. Cette normalisation
se déclenche automatiquement après chaque application d'une règle de
réécriture.  

\section{Le \laCal}
%================================================================

Les systèmes de réécriture que nous avons présentés jusqu'à maintenant sont
appelés {\em systèmes de réécriture du premier ordre} et ils permettent
d'exprimer des calculs sur des expressions contenant des variables. Le pouvoir
d'expression de ces systèmes n'est pas suffisant pour décrire directement les
fonctions sur les fonctions comme, par exemple, la composition de fonctions. Le
\laCal\  est un système de réécriture d'ordre supérieur qui a été introduit pour
exprimer simplement la fonctionnalité.

Nous présentons brièvement les concepts et les propriétés du \laCal. Pour 
une présentation détaillée du calcul dans les cas non-typé et typé 
la référence classique est~\cite{Barendregt84} mais on peut citer aussi 
\cite{Hindley-1986} et~\cite{Krivine-book90}.

\subsection{Le \laCal\  non-typé} \label{laCalNonType}
%================================================================

Une fonction est souvent décrite par un terme du premier ordre qui comporte une
ou plusieurs variables. Par exemple, la fonction d'incrementation est représenté
par $x +1$ et on écrit $incr(x)=x+1$. L'application de cette fonction à l'entier
$2$ est notée $f(2)$.

Il y a donc deux constructions indispensables pour exprimer une fonction. La
première est l'abstraction d'une variable, comme $x$ dans $incr(x)$. La seconde
est l'application d'une fonction à une valeur; $incr(2)$ dans le cas précèdent.
L'ensemble des termes du \laCal\  est engendré à partir les termes du premier
ordre en utilisant les deux nouveaux opérateurs, l'abstraction et l'application.

\DEF%------------------------------------------------------------------------
\index{terme!lambda-terme@$\lambda$-terme}
Soit $\XX$ un ensemble de variables et $\FF$ un ensemble de symboles appelés
{\em cons\-tantes}\index{constante}. L'ensemble des termes du \laCal, noté $\Lambda_{\XX}^{\FF}$,
est le plus petit ensemble satisfaisant~:
	\begin{itemize} 
	\item si $x \in \XX$ est une variable, alors $x \in \Lambda_{\XX}^{\FF}$,
	\item si $f \in \FF$ est une constante, alors $f \in \Lambda_{\XX}^{\FF}$,
	\item si $x \in \XX$ et $t \in \Lambda_{\XX}^{\FF}$, alors 
	$\lambda x.t \in \Lambda_{\XX}^{\FF}$,
	\item si $u \in \Lambda_{\XX}^{\FF}$ et $v \in \Lambda_{\XX}^{\FF}$, alors
	$u~v \in \Lambda_{\XX}^{\FF}$.
	\end{itemize}

Les variables et les constantes sont appelés \Def{atomes}. 
\FDEF%------------------------------------------------------------------------

Si l'ensemble de symboles $\FF$ est vide alors l'ensemble des termes du
\laCal\  est noté $\Lambda_{\XX}$ et le calcul est appelé
\textit{pur}. Sinon le calcul est appelé \textit{appliqué} (cf.~\cite{Hindley-1986}).

Les $\lambda$-termes peuvent être définis en utilisant la notation proposée
précédemment pour la définition d'une algèbre de termes:
	$$t ~~ ::= ~~ x ~|~ f ~|~ \lambda x.t ~|~ t~t$$

Intuitivement, $\lambda x.t$ représente la fonction qui associe la valeur $t$ à
la variable $x$. Un terme de la forme $\lambda x.t$ est appelé une
\Def{abstraction}. Le terme $(u~v)$ représente intuitivement le résultat de
l'application de la fonction $u$ a l'argument $v$. Un terme de la forme $(u~v)$
est appelé l'\Def{application} du terme $u$ au terme $v$.

\DEF%------------------------------------------------------------------------
L'ensemble des variables \Defn{libres}{variable}{libre} d'un $\lambda$-terme
$t$, noté $FV(t)$, est défini inductivement par~:
	\begin{itemize} 
	\item $FV(x)=x$,
	\item $FV(f)=\emptyset$,
	\item $FV(\lambda x.t) = FV(t) - \{x\}$,
	\item $FV(u~v) = FV(u) \cup FV(v)$.
	\end{itemize}
\FDEF%------------------------------------------------------------------------

On dit que l'occurrence d'une variable $x$ dans un terme $t$ est
\Defi{liée}{variable} si cette variable apparaît dans un sous-terme de $t$ de la
forme $\lambda x.u$. Dans le cas contraire l'occurrence de la variable $x$ est
\Defi{libre}{variable}. Si la variable $x$ a au moins une occurrence libre dans
le terme $t$ alors $x$ est appelée une variable libre de $t$. L'ensemble des
variables libres de $t$ est exactement $FV(t)$.

Les substitutions du premier ordre présentées dans la
section~\ref{subst_premier_ordre} ne conviennent pas pour le \laCal\  car, par
exemple, la variable $y$ est libre dans le terme $\lambda x.(x~y)$ alors que
son image ne l'est plus dans le terme 
$\subs{y \sbs x} \lambda x.(x~y)=\lambda x.(x~x)$. On dit que la variable $y$ a
été \Defi{capturée}{variable}.

Nous allons donner une définition de la substitution utilisant un mécanisme de
renommage des variables qui évitera les captures éventuelles.

\DEF%------------------------------------------------------------------------
Soit $t$, $u$, $v$, $s$ des $\lambda$-termes et $x$ une variable.  La substitution de la
variable $x$ par le terme $s$ dans le terme $t$, notée $\subs{x/s}t$ est définie
par récurrence sur la structure de t~:
	\begin{itemize} 
	\item si $t$ est la variable $x$ alors $\subs{x/s}t = s$,
	\item si $t$ est un atome $a$ différent de $x$ alors $\subs{x/s}t = a$,
	\item si $t = (u~v)$ alors $\subs{x/s}t = (\subs{x/s}u~\subs{x/s}v)$,
	\item si $t = \lambda x.u$ alors $\subs{x/s}t = t$,
	\item si $t = \lambda y.u$ alors $\subs{x/s}t = \lambda z.(\subs{x/s}\subs{y/z}u)$\\
		où $z$ est une \textit{nouvelle} variable, i.e. $z \neq x$, 
		$z \not \in FV(s)$ et $z \not \in FV(u)$.
	\end{itemize}
\FDEF%------------------------------------------------------------------------

Les règles de ``propagation'' des substitutions ne sont pas des règles du \laCal.
On les appelle des méta-règles.

Nous avons défini l'ensemble des termes du \laCal\  et par la suite nous
définissons les relations classiques de réduction et d'équivalence sur les
$\lambda$-termes.

Pour éviter la capture des variables, on définit une relation, appelée 
{\em $\alpha$-conversion}\index{alpha-conversion@$\alpha$-conversion}, %\Def{$\alpha$-conversion}
dont le rôle est de remplacer le nom d'une variable liée par un nouveau nom.

\DEF%------------------------------------------------------------------------
Soit $t$ un $\lambda$-terme contenant un sous-terme $\lambda x.u$ et soit $y$
une variable telle que $y \not \in FV(u)$. Le remplacement de $\lambda x.u$ par
$\lambda y.\subs{x/y}u$ est appelé renommage de la variable liée $y$ ou
$\alpha$-conversion.

Deux termes $u,v$ sont dits $\alpha$-équivalents, noté $u\equiv_{\alpha} v$, si
$v$ est obtenu en appliquant une série finie (éventuellement vide) de renommages
de variables liées à $u$.
\FDEF%------------------------------------------------------------------------

Par exemple, les termes $\lambda x.x$ et $\lambda y.y$ sont $\alpha$-équivalents 
mais $\lambda x.\lambda y.(x~y)$ et $\lambda x.\lambda x.(x~x)$ ne sont pas
$\alpha$-équivalents parce que $x$ est libre dans $(x~y)$ et donc nous ne
pouvons pas renommer $y$ en $x$.

En \laCal\  on raisonne toujours modulo $\alpha$-équivalence, c'est-à-dire qu'on
ne distingue pas deux termes $\alpha$-équivalents (i.e. on raisonne sur les
classes d'$\alpha$-équivalence).

\DEF%------------------------------------------------------------------------
Un $\lambda$-terme de la forme $(\lambda x.t)~u$ est appelé un
{\em $\beta$-radical} \index{radical!beta-radical@$\beta$-radical} %\Defi{$\beta$-radical}{radical} 
et le terme $\subs{x/u}t$ son réduit.
\FDEF%------------------------------------------------------------------------

Si un terme $t$ contient un sous-terme $(\lambda x.v)~u$ et le terme $t'$ est le
terme $t$ avec le sous-terme $(\lambda x.v)~u$ remplacé par le terme
$\subs{x/u}v$, on dit que $t$ $\beta$-réduit en $t'$. Formellement la
{\em $\beta$-réduction} \index{beta-réduction@$\beta$-réduction} %\Def{$\beta$-réduction} 
est définie par~:

\DEF%------------------------------------------------------------------------
La relation entre termes $t \ral_{\beta} t'$ ($t$ se $\beta$-réduit sur une
étape en $t'$) est la plus petite relation telle que
	\begin{itemize} 
	\item $(\lambda x.t)u \ral_{\beta} \subs{x/u}t$
	\item si $u \ral_{\beta} v$ alors $(u~t) \ral_{\beta} (v~t)$
	\item si $u \ral_{\beta} v$ alors $(t~u) \ral_{\beta} (t~v)$
	\item si $u \ral_{\beta} v$ alors $(\lambda x. u) \ral_{\beta} (\lambda x. v)$
	\end{itemize}

La relation $t \longra{*}_{\beta} t'$ ($t$ se $\beta$-réduit sur $t'$) est
définie comme la fermeture réflexive-transitive de la relation $\ral_{\beta}$.

La relation $t \equiv_{\beta} t'$ ($t$ et $t'$ sont $\beta$-équivalents) est définie
comme la fermeture réflexive-symétrique-transitive de la relation
$\ral_{\beta}$.
\FDEF%------------------------------------------------------------------------

\DEF%------------------------------------------------------------------------
Le \laCal\  est le calcul défini par l'algèbre de termes $\Lambda_{\XX}^{\FF}$ et
la relation $\longra{*}_{\beta}$.
\FDEF%------------------------------------------------------------------------

\TH%------------------------------------------------------------------------
Le \laCal\  est confluent.
\FTH%------------------------------------------------------------------------

Il existe de nombreuses preuves différentes de ce théorème, on peut entre autres
se référer à~\cite{Barendregt84,Hindley-1986,Krivine-book90}.

Le \laCal\  n'est pas fortement normalisable (ni même faiblement normalisable)~:
le contre exemple classique est $\omega~\omega$ où 
$\omega \equiv_{\alpha} \lambda x.(x~x)$. Ce terme se réduit en une étape de
$\beta$-réduction en lui-même, ce qui fournit une chaîne de réduction infinie.

\subsection{Le \laCal\   simplement typé}
%================================================================

Dans l'algèbre de termes du premier ordre présentée en
Section~\ref{algebresTermes}, les symboles de fonctions d'arité non nulle ne
sont pas des termes~; ils sont juste utilisés pour construire des termes. En
\laCal\  appliqué, les symboles de fonctions sont des termes et permettent la
construction de termes dénués de sens. Par exemple, les termes $(succ~succ)$ et
$(succ~0~0)$ sont des termes de $\Lambda_{\XX}^{succ,0}$ mais ils ne
correspondent pas à des fonctions mathématiques ou à des applications de
fonctions.

Il faut donc restreindre les règles de construction des $\lambda$-termes afin de
ne pas générer de tels termes. La notion d'arité n'est plus suffisante, puisque
elle empêcherait la construction des termes $(succ~succ)$ et $(succ~0~0)$ mais
pas du terme $(succ~\lambda x.x)$. Il faut donc associer à chaque terme une
information indiquant sa \textit{fonctionnalité} et cette information est
appelée un \Def{type}.

\DEF%------------------------------------------------------------------------
Etant donné un ensemble de types de base encore appelés types atomiques.
L'ensemble des types est inductivement défini par
	\begin{itemize} 
	\item les types \Defi{de base}{type} sont des types,
	\item si $A$ et $B$ sont des types, alors $(A \raS B)$ est un type.
	\end{itemize}
\FDEF%------------------------------------------------------------------------

Les types de la forme $(A \raS B)$ sont appelés types \Defn{composés}{type}{composé} est 
représentent l'ensemble de fonctions de $A$ vers $B$.

La flèche $\raS$ associe à droite et donc un type de la forme 
$A_1 \raS A_2 \raS \ldots \raS A_n$ est une abréviation pour 
$A_1 \raS (A_2 \raS (\ldots \raS A_n)\ldots)$.

\DEF%------------------------------------------------------------------------
Une variable \Defi{typée}{variable} est un couple $(x: A)$ où $x$ est une
variable et $A$ un type. Un contexte est une liste de variables typées, telle
que chaque variable apparaisse au plus une fois dans cette liste. On note
$\Gamma[x : A]$ le contexte $\Gamma$ contenant la variable typée $(x: A)$.
\FDEF%------------------------------------------------------------------------

\DEF%------------------------------------------------------------------------
Soit $\Gamma$ un contexte, $t$ un terme et $A$ un type. On dit que le terme $t$
est bien typé et a le type $A$ dans le contexte $\Gamma$, noté $\Gamma \vdash
t:A$ si~:
	\begin{itemize} 
	\item $t=x$ et $\Gamma=\Gamma[x : A]$ ou,
	\item $t=u~v$ et $\Gamma \vdash u:B \raS A$ et $\Gamma \vdash v:B$ ou,
	\item $t=\lambda x:B.u$ et $\Gamma[x:B] \vdash u:C$ et $A=B \raS C$.
	\end{itemize}
\FDEF%------------------------------------------------------------------------

S'il existe un type $A$ tel que $\Gamma \vdash t:A$, le terme $t$ est dit
\Defi{bien typé}{terme} dans $\Gamma$. On dit que le terme $t$ est bien typé
s'il existe un contexte $\Gamma$ tel que $t$ est bien typé dans $\Gamma$.

Considérons les profils $0 : A$ et $succ:A \raS A$ , les termes $(succ~succ)$,
$(succ~0~0)$ et \mbox{$(succ~\lambda x:A.x)$} ne sont pas bien typés quelque soit le
contexte utilisé, mais $\Gamma \vdash (succ~0)$ et $\Gamma[x:A] \vdash (succ~x)$.  Il
n'existe pas de contexte tel que le terme $\omega_A=\lambda x:A.(x~x)$ soit bien
typé.

\PROP%------------------------------------------------------------------------
Il existe un algorithme qui prend en argument un contexte $\Gamma$
et un terme $t$ et qui décide si $t$ est bien typé dans $\Gamma$ et retourne le
type de $t$ dans ce cas.
\FPROP%------------------------------------------------------------------------

\subsection{Le formalisme de de Bruijn}
%================================================================

Dans le \laCal, à chaque application de substitution et donc à chaque étape de
$\beta$-réduction il faut gérer explicitement un système de renommage des
variables. Même si d'un point de vue théorique le renommage des variables est
souvent vu (à tort) comme un détail mineur, une implantation du système tel que
est très inefficace.

Le formalisme de de Bruijn (\cite{deBruijn72}, \cite{deBruijn78}) est basé sur
le remplacement de chaque variable du \laCal\  classique par un entier naturel
représentant le nombre de $\lambda$ qui la sépare du $\lambda$ qui la lie. Ce
nombre est appelé un indice de de Bruijn et le calcul est noté
$\lambda_{DB}$-calcul.

Dans le $\lambda_{DB}$-calcul il n'est plus nécessaire d'étiqueter les $\lambda$
par la variable qu'ils lient puisque cette information est déjà contenue dans
chaque variable.  Les variables libres sont traitées comme si elles étaient
liées par des $\lambda$ extérieurs. Donc il n'existe pas une unique
représentation dans le formalisme de de Bruijn d'un $\lambda$-terme contenant
des variables libres mais cette ambiguïté est levée si on précise un ordre sur
les variables. Par exemple,

$\lambda x.\lambda y.(x~y~z)$ devient $\lambda\lambda(2~1~3)$

$\lambda x.(x~(\lambda y.(x~y)))$ devient $\lambda(1~\lambda(2~1))$

La notation de de Bruijn est difficile à lire mais elle s'avère très efficace
pour le traitement mécanique des substitutions. Par la suite nous introduisons
les termes du $\lambda_{DB}$-calcul et la \mbox{$\beta$-réduction} correspondante.

\DEF%------------------------------------------------------------------------
\index{terme!lambdaDB-terme@$\lambda_{DB}$-terme}

L'ensemble des termes du $\lambda_{DB}$-calcul pur, noté $\Lambda_{DB}$, est
l'algèbre de termes induite par la grammaire suivante~:
	$$t ~~ ::= ~~ n ~|~ \lambda t ~|~ t~t$$
où $n$ est un entier naturel non nul.
\FDEF%------------------------------------------------------------------------

Le $\lambda_{DB}$-terme $\subs{1 / u}t$ représente le terme $t$ où la
\textit{variable} $1$ a été remplacée par le terme $u$.
Un $\lambda_{DB}$-terme de la forme $(\lambda t)~u$ est appelé $\beta$-radical et
le terme $\subs{1 / u}t$ son réduit.

L'utilisation d'indices de de Bruijn à la place de noms de variables nécessite
une nouvelle notion de substitution. Dans la réduction d'un $\beta$-radical
$(\lambda t)~u$, il faut d'une part mettre à jour les indices (variables) libres
de $t$ pour prendre en compte la disparition d'un $\lambda$ et d'autre part,
modifier les indices de de Bruijn de $u$ pour tenir compte du nombre de
$\lambda$ supplémentaires traversés lors de la propagation de la substitution.
Nous ne donnons pas la définition classique de la substitution du
$\lambda_{DB}$-calcul mais une formulation équivalente présentée, par exemple,
dans~\cite{PaganoThesis}. 

\DEF%------------------------------------------------------------------------
\label{subs_explicite}
La substitution $\subs{n / u}t$ est définie par induction sur $t$~:

\[
\begin{array}{|l|l|}
\hline
&\\
%------------
\subs{n / u}(t'~t'') =  (\subs{n / u}t')~(\subs{n / u}t'')
&
\uparrow^n (t'~t'') = (\uparrow^n (t'))~(\uparrow^n (t'')) 
\\
&\\%------------
\subs{n / u}(\lambda t) =  \lambda (\subs{n+1 / \uparrow^0 (u)}t)
&
\uparrow^n (\lambda t) = \lambda (\uparrow^{n+1} (t))
\\
&\\%------------
\subs{n / u}m=\left\{
\begin{array}{ll}
m-1,~~si~m>n \\
u,~~si~m=n\\
m,~~si~m<n \\
\end{array}
\right.
&
\uparrow^n (m)=\left\{
\begin{array}{ll}
m+1,~~si~m \geq n \\
m,~~si~m=n\\
\end{array}
\right.
\\
&\\%------------
\hline
\end{array}
\]
\FDEF%------------------------------------------------------------------------

\DEF%------------------------------------------------------------------------
La $\beta$-réduction est la plus petite relation $\ral$ telle que~:
	\begin{itemize} 
	\item $(\lambda t)u \ral \subs{1 / u}t$
	\item si $u \ral v$ alors $(u~t) \ral (v~t)$
	\item si $u \ral v$ alors $(t~u) \ral (t~v)$
	\item si $u \ral v$ alors $(\lambda u) \ral (\lambda v)$
	\end{itemize}
\FDEF%------------------------------------------------------------------------

\PROP%------------------------------------------------------------------------
%\cite{deBruijn72}%preuve dans \cite{Mau85}
Le $\lambda_{DB}$-calcul et le $\lambda$-calcul sont isomorphes.
\FPROP%------------------------------------------------------------------------
\proof{Une preuve est donnée dans \cite{Mau85}.}

Dans la Définition~\ref{subs_explicite} la fonction $\uparrow^n$ a le même rôle
que l'$\alpha$-conversion du \laCal\  classique et incrémente tous les indices
libres du terme à substituer pour éviter la capture par le $\lambda$ que la
substitution vient de passer.

Comme dans le cas du \laCal\  avec des noms de variables, une substitution est une
opération indivisible que l'on effectue en un seul pas et qui est décrite ici au
méta-niveau du calcul.

\subsection{Le $\lambda\sigma$-calcul}
%================================================================

Le $\lambda\sigma$-calcul, introduit dans~\cite{ACCL90}, %\cite{AbadiCCL91JFP}, 
est un formalisme permettant de rendre explicite l'application de substitution.
Dans le $\lambda\sigma$-calcul la substitution est gérée par un constructeur
explicite de la syntaxe et son application est décrite par les règles du calcul.

\DEF%------------------------------------------------------------------------
L'algèbre de termes du $\lambda\sigma$-calcul est une algèbre à deux sortes, une pour
les termes et une pour les substitutions~:

\begin{tabular}{llll}
& \\
\textBNF{Termes} ~~~~ & 
$t$ &::=& 
$x_t ~|~ 1 ~|~ \lambda t ~|~ t~t ~|~ t\subs{s}$\\
& \\
\textBNF{Substitutions} ~~~~ & 
$s$ &::=& 
$x_s ~|~ id ~|~ \uparrow ~|~ t.s ~|~ s \circ s$\\
%& \\
\end{tabular}\\
\FDEF%------------------------------------------------------------------------

Les substitutions sont appelées aussi des \Def{environnements}. Les substitutions
ne sont plus des couples variable (ou indice) et terme, mais des liste de termes.

La composition de $n$ symboles $\uparrow$, i.e. $\uparrow \circ \ldots
\uparrow$, est notée $\uparrow^n$.  Remarquons que le seul indice de de Bruijn
qui apparaît explicitement dans la syntaxe est $1$ mais on peut coder l'indice
$n$ par le terme $1\subs{\uparrow^{n-1}}$.

Nous distinguons deux types d'application, l'application d'un terme à un autre
terme, notée par juxtaposition, et l'application d'une substitution $s$ à un 
terme $t$, notée $t\subs{s}$. Le symbole $\circ$ représente la composition de deux
substitutions. La substitution $id$, appelée \Def{identité}, est interprétée
comme la suite des entiers de de Bruijn $(n)_{n \geq 1}$. La substitution
$\uparrow$, appelée {\em shift}\index{shift,~$\uparrow$}, est interprétée comme la suite des entiers de
de Bruijn $(n+1)_{n \geq 1}$.

L'ensemble de règles du $\lambda\sigma$-calcul est donné dans la
Figure~\ref{fig_lambda_sigma}.
La règle \rname{Beta} est la règle déclenchant la 
{\em $\beta$-réduction}\index{beta-réduction@$\beta$-réduction}. %\Def{$\beta$-réduction} 
Si on considère le système de réécriture où on omet cette règle on obtient le calcul de
substitution appelé $\sigma$-calcul. 

\begin{figure}[!htp]
\noindent
\framebox{\parbox{\largeurtexte}{

\renewcommand{\fleche}{\Rightarrow}

\begin{ruleset}
%===========================================================================

\regle{Beta}{(\lambda t)~u}{t\subs{u.id}}

\\

\regle{App}{(u~v)\subs{s}}{(u\subs{s})~(v\subs{s})}

\regle{VarCons}{1\subs{u.s}}{u}

\regle{Clos}{u\subs{s}\subs{t}}{u\subs{s \circ t}}

\regle{Abs}{(\lambda t)\subs{s}}{\lambda (t\subs{1.(s \circ \uparrow)})}

\regle{IdL}{id \circ s}{s}

\regle{ShiftCons}{\uparrow \circ (u.s)}{s}

\regle{AssEnv}{(s \circ t) \circ v}{s \circ (t \circ v)}

\regle{MapEnv}{(u.s) \circ t}{u\subs{t}.(s \circ t)}

\regle{Id}{u\subs{id}}{u}

\regle{IdR}{s \circ id}{s}

\regle{VarShift}{1.\uparrow}{id}

\regle{SCons}{1\subs{s}.(\uparrow \circ s)}{s}

%===================================
\end{ruleset}

}}
\caption{\label{fig_lambda_sigma}Les règles du $\lambda\sigma$-calcul}
\end{figure}%

\PROP%------------------------------------------------------------------------
Le $\sigma$-calcul est confluent et fortement normalisable.
\FPROP%------------------------------------------------------------------------

Si on considère un terme du $\lambda\sigma$-calcul sans variable et sans
substitution, on obtient un \mbox{$\lambda_{DB}$-terme}. On peut montrer qu'un terme
sans variable se réduit par $\sigma$ en un terme sans substitution.

\TH%------------------------------------------------------------------------
Le $\lambda\sigma$-calcul est confluent pour les termes ne contenant pas de
variable de substitution (termes semi-clos).
\FTH%------------------------------------------------------------------------

La preuve de la confluence du $\lambda\sigma$-calcul avec des termes clos est
donnée dans~\cite{ACCL90}. La confluence dans le cas où les termes sont
semi-clos a été prouvée dans~\cite{RiosTHESE}.

\subsection{Le $\lambda\sigma_{\Uparrow}$-calcul}	\label{lambdaSigma}
%================================================================

L'un des problèmes du $\lambda\sigma$-calcul est la non-confluence sur la
totalité de l'algèbre de termes. Dans~\cite{CurienHardinLevy-JACM96}, Curien,
Hardin et Lévy ont montré qu'en introduisant un nouvel opérateur unaire appelé
{\em lift} \index{lift,~$\Uparrow$} et noté $\Uparrow$ on obtient un calcul
confluent sur l'ensemble de termes ouverts.

L'opérateur $\Uparrow$ est utilisé pour simplifier la règle \rname{Abs} en
remplaçant $1.(s \circ \uparrow)$ par $\Uparrow(s)$. L'idée sous-jacente est de
faire disparaître une paire critique engendrée par la règle \rname{Abs}. En
complétant le $\lambda\sigma$-calcul pour que les paires critiques introduites
par ce nouveau symbole convergent on obtient le
$\lambda\sigma_{\Uparrow}$-calcul.

Dans le $\lambda\sigma_{\Uparrow}$-calcul les indices de de Bruijn sont des
vrais numéros de de Bruijn, i.e. on ne code pas l'indice $n$ par $\subs{\uparrow^{n-1}}1$.

\DEF%------------------------------------------------------------------------
L'algèbre de termes du $\lambda\sigma_{\Uparrow}$-calcul est l'algèbre à deux
sortes définie par~:
%\samepage%\nopagebreak

\begin{tabular}{llll}
& \\
\textBNF{Termes} ~~~~ & 
$t$ &::=& 
$x_t ~|~ n ~|~ \lambda t ~|~ t~t ~|~ t\subs{s}$\\
& \\
\textBNF{Substitutions} ~~~~ & 
$s$ &::=& 
$x_s ~|~ id ~|~ \uparrow ~|~ t.s ~|~ s \circ s ~|~ \Uparrow(s)$\\
%& \\
\end{tabular}\\
\FDEF%------------------------------------------------------------------------

L'ensemble de règle du $\lambda\sigma_{\Uparrow}$-calcul est donné dans la
Figure~\ref{fig_lambda_lift}.

\begin{figure}[!htp]
\noindent
\framebox{\parbox{\textwidth}{

\renewcommand{\fleche}{\Rightarrow}

%\small{

\begin{tabular}{ll}

\begin{truleset}
%===========================================================================

\tregle{Beta}{Beta}{(\lambda t)u}{t\subs{u.id}}

\tregle{App}{App}{(u v)\subs{s}}{(u\subs{s}) (v\subs{s})}

\tregle{Lambda}{Lambda}{(\lambda t)\subs{s}}{\lambda (t\subs{\Uparrow(s)})}

\tregle{Clos}{Clos}{u\subs{s}\subs{t}}{u\subs{s \circ t}}

\tregle{VarShift1}{VS1}{n\subs{\uparrow}}{(n+1)}

\tregle{VarShift2}{VS2}{n\subs{\uparrow \circ s}}{(n+1)\subs{s}}

\tregle{FVarCons}{FVC}{1\subs{u.s}}{u}

\tregle{FVarLift1}{FVL1}{1\subs{\Uparrow(s)}}{1}

\tregle{FVarLift2}{FVL2}{1\subs{\Uparrow(s) \circ t}}{1\subs{t}}

\tregle{RVarCons}{RVC}{(n+1)\subs{u.s}}{n\subs{s}}

\tregle{RVarLift1}{RVL1}{(n+1)\subs{\Uparrow(s)}}{n\subs{s \circ \uparrow}}

\tregle{RVarLift2}{RVL2}{(n+1)\subs{\Uparrow(s) \circ t}}{n\subs{s \circ (\uparrow \circ t)}}

\end{truleset}
&
\begin{truleset}

\tregle{AssEnv}{Ass}{(s \circ t) \circ v}{s \circ (t \circ v)}

\tregle{MapEnv}{Map}{(u.s) \circ t}{u\subs{t}.(s \circ t)}

\tregle{ShiftCons}{SC}{\uparrow \circ (u.s)}{s}

\tregle{ShiftLift1}{SL1}{\uparrow \circ \Uparrow(s)}{s \circ \uparrow}

\tregle{ShiftLift2}{SL2}{\uparrow \circ \Uparrow(s) \circ t}{s \circ(\uparrow \circ t)}

\tregle{Lift1}{L1}{\Uparrow(s) \circ \Uparrow(t)}{\Uparrow(s \circ t)}

\tregle{Lift2}{L2}{\Uparrow(s) \circ (\Uparrow(t) \circ v)}{\Uparrow(s \circ t) \circ v}

\tregle{LiftEnv}{Lift}{\Uparrow(s) \circ (u.t)}{u.(s \circ t)}

\tregle{IdL}{IdL}{id \circ s}{s}

\tregle{IdR}{IdR}{s \circ id}{s}

\tregle{LiftId}{LId}{\Uparrow(id)}{id}

\tregle{Id}{Id}{u\subs{id}}{u}

%===================================
\end{truleset}
\end{tabular}
%}
}}
\caption{\label{fig_lambda_lift}Les règles du $\lambda\sigma_{\Uparrow}$-calcul}
\end{figure}%

Si on considère le système de réécriture où on omet la règle \rname{Beta} on
obtient le calcul de substitution appelé $\sigma_{\Uparrow}$-calcul.

\PROP%------------------------------------------------------------------------
Le $\sigma_{\Uparrow}$-calcul est confluent et fortement normalisable.
\FPROP%------------------------------------------------------------------------

\TH%------------------------------------------------------------------------
(\cite{CurienHardinLevy-JACM96})

Le $\lambda\sigma_{\Uparrow}$-calcul est confluent.
\FTH%------------------------------------------------------------------------

%\DontWriteThisInToc  
%\subsection*{Conclusion}
%================================================================
%~

~\\
Dans ce chapitre nous avons rappelé plusieurs concepts qui seront utilisés au
cours de cette thèse. Nous avons décrit brièvement différents langages
exprimant des calculs et nous avons présenté leurs propriétés principales.

%% file: chapter_2.tex
%%%%%%%%%%%%%%%%%%%%%%%%%%%%%%%%%%%%%%%%%%%%%%%%%%%%%%%%%%%
% \TLtopbookmark
\chapter{Le \roCal\  non typé} 
\label{chap.calcul_non_type}
%%%%%%%%%%%%%%%%%%%%%%%%%%%%%%%%%%%%%%%%%%%%%%%%%%%%%%%%%%%%

Dans le chapitre précédent nous avons brièvement présenté la réécriture du
premier ordre et le \laCal. Les deux concepts ayant de nombreux points
communs et aussi des propriétés complémentaires extrêmement utiles,
beaucoup de travaux s'intéressent à l'intégration de la réécriture et du
\laCal. Ceci a déjà été réalisé soit en enrichissant la réécriture du premier ordre
avec des caractéristiques d'ordre supérieur, soit en ajoutant au \laCal\  des
caractéristiques algébriques permettant, en particulier, le traitement efficace
de l'égalité. Dans le premier cas, on trouve les travaux sur les
CRS~\cite{KlopOostromRaamsdonk} et aussi d'autres systèmes de réécriture
d'ordre supérieur~\cite{WolframLivre,Nipkow-Prehofer-98}; dans le deuxième cas
on peut citer les travaux sur la combinaison du \laCal\  avec la
réécriture~\cite{Breazu-TannenLICS88,Okada1989SNC,GallierBreazu,JouannaudOkada-97}.

En partant des travaux sur le contrôle de la réécriture
(\cite{VittekThese,KirchnerKV-MIT95,BKK-Fuji-98}), nous avons introduit le
\roCal, un calcul intégrant la réécriture du premier ordre, le \laCal\  et les
calculs non-déterministes. Ce chapitre a pour objectifs de présenter le \roCal\
dans un cadre non-typé et puis de montrer des exemples d'utilisation du calcul
général et des instances possibles du calcul de base.

\section{Présentation générale}
%================================================================

Les objets manipulés dans la réécriture du premier ordre sont les termes du
premier ordre et dans une présentation simpliste nous pouvons dire qu'à chaque
pas de réécriture on applique une {\em règle de réécriture} à une position
quelconque d'un {\em terme (initial)} pour obtenir un autre {\em terme (résultat)}. On
peut remarquer que les termes ne sont pas décrits au même niveau que la
description des règles de réécriture qui les transforment et la façon dont cette
application est effectuée est définie au méta-niveau.

De plus, si on ne considère pas une seule règle de réécriture mais un système de
réécriture, la règle de réécriture à appliquer à chaque pas n'est pas
sélectionnée d'une façon déterministe mais on peut appliquer toute règle de
réécriture satisfaisant les conditions d'application au \mbox{(sous-)}terme
considéré. On n'a donc aucun contrôle ni sur la sélection de la règle de
réécriture à appliquer, ni sur la position où cette règle est appliquée dans le
terme à réduire.

Les systèmes de calcul (voir Section~\ref{systemeCalcul}) enrichissent les
systèmes de réécriture en introduisant la notion de stratégie contrôlant
l'application des règles de réécriture. On a cette fois des règles de réécriture
décrivant la transformation des termes et des stratégies décrivant l'application
de ces règles. On peut donc distinguer trois niveaux de description pour la
définition d'un système de calcul~: les termes, les règles de réécriture et les
stratégies.

La caractéristique principale du \roCal\  consiste à rendre explicites les
ingrédients principaux de la réécriture, en particulier, les notions
d'application de règle et de résultat d'une telle application. Les règles et
l'application des règles (ou des $\rho$-termes plus compliqués) sont des objets
du \roCal\  et les résultats des applications sont représentés par des ensembles
qui sont également des \mbox{$\rho$-termes}.

Ainsi, les objets du \roCal\  sont construits en utilisant une signature, un
ensemble de variables, l'opérateur d'abstraction $\ra$ et l'opérateur
d'application $[~](~)$ et nous considérons des ensembles de tels objets. Cela
donne au \roCal\  la capacité de représenter le non-déterminisme au moyen des
ensembles de résultats.

Dans le \roCal\  nous pouvons représenter explicitement l'application d'une règle
de réécriture (par exemple $a \ra b$) à un terme (par exemple la constante
$a$) par l'objet $[a \ra b](a)$ qui est évalué dans le $\rho$-terme $\{b\}$.
Puisque le résultat de l'évaluation est le singleton $\{b\}$ nous pouvons dire
que l'application de la règle $a \ra b$ sur le terme $a$ est déterministe et ceci
signifie, intuitivement, que le terme $a$ peut être évalué par rapport au
système de réécriture contenant la règle $a \ra b$ en un seul résultat, $b$.

L'introduction d'un opérateur d'application dans la syntaxe du \roCal\  nous
permet non seulement de définir explicitement l'application d'une règle de
réécriture à un terme mais aussi de construire des $\rho$-termes décrivant les
stratégies des systèmes de calcul.  Par exemple, l'application de la règle de
réécriture $a \ra b$ suivie de la règle de réécriture $b \ra c$ au terme $a$ est
représentée par le $\rho$-terme $[b \ra c]([a \ra b](a))$ qui est évalué d'abord
en $[b \ra c](\{b\})$ puis en $\{c\}$.

Naturellement, des variables peuvent être utilisées dans les règles de
réécriture et nous pouvons dire qu'une $\rho$-règle de réécriture construite en
utilisant l'opérateur $\ra$ est une \Def{abstraction} dont le rapport avec la
$\lambda$-abstraction pourra fournir une intuition utile~: une
$\lambda$-expression $\lambda x.t$ est représentée dans le \roCal\  par la
règle $x \ra t$.  En effet, le $\beta$-radical $(\lambda x.t~u)$ correspond 
simplement au 
{\em $\rho$-radical} \index{radical!rho-radical@$\rho$-radical} %\Defi{$\rho$-radical}{radical} 
$[x \ra t](u)$ représentant
l'\Def{application} de la règle $x \ra t$ au terme $u$. Le $\lambda$-terme
$(\lambda x.t~u)$ est $\beta$-réduit en $\subs{x/u} t$ tandis que le
$\rho$-terme correspondant est réduit dans le singleton $\{\subs{x/u} t\}$.

Le membre gauche des règles de réécriture peut évidemment être plus élaboré
qu'une constante ou une variable comme dans $[f(x) \ra x](f(a))$. Dans ce cas, le
mécanisme d'évaluation du calcul réduit l'application en $\{a\}$. En fait, en
évaluant cette expression, la variable $x$ est liée à $a$ par le mécanisme de
filtrage et nous retrouvons le même comportement que dans le cas de la
réécriture.

Comme nous l'avons vu ci-dessus, le résultat d'une application est toujours un
ensemble et à un terme résultat obtenu dans la réécriture ou dans le \laCal\
correspond, dans le \roCal, le singleton contenant ce terme. Mais l'application
d'une règle de réécriture peut échouer comme dans $[a \ra b](c)$ et ceci est
représenté explicitement dans le \roCal\  en fournissant l'ensemble vide comme
résultat.

En plus, nous pouvons déclarer que certains symboles de la signature ne sont pas
purement syntaxiques mais nécessitent un filtrage modulo une théorie (par
exemple équationnelle) différente de la théorie vide et dans ce cas l'ensemble
de résultats peut contenir plus d'un élément. Par exemple, si on suppose que le
symbole $+$ est commutatif alors l'application de la règle $x+y \ra x$ au terme
$a+b$ produit l'ensemble $\{a,b\}$. Puisqu'il y a deux manières différentes
d'appliquer (de filtrer) cette règle modulo la commutativité, dans la réécriture
modulo classique (\cite{PS81}) on obtient comme résultat un des deux termes $a$
ou $b$ tandis que dans le \roCal\  le résultat est l'ensemble $\{a,b\}$
représentant le choix non-déterministe entre les deux termes.

Le symbole d'ensemble permet non seulement la description des résultats
non-déterministes mais aussi la représentation de l'application
(non-déterministe) de plusieurs règles. Par exemple, un pas de réécriture à la
position de tête d'un terme $a$ par rapport à un système de réécriture contenant
les règles $a \ra b$ et $a \ra c$ est représenté dans le \roCal\  par le terme
$[\{a \ra b,a \ra c\}](a)$. Encore une fois, le résultat obtenu dans le \roCal\
est une représentation du comportement dans la réécriture du premier ordre~; dans
la réécriture on applique une des deux règles et on obtient soit $b$ soit $c$,
tandis que dans le \roCal\  le résultat est l'ensemble $\{b,c\}$.

Pour résumer, nous pouvons dire que dans le \roCal\  la flèche est un opérateur
binaire utilisé pour abstraire, le filtrage est le mécanisme de passage de
paramètre, la substitution prend en compte la capture de variables et les
ensembles de résultats sont manipulés explicitement.

\section{Les composants du calcul}
%================================================================

Nous avons présenté d'une façon informelle les termes manipulés dans le \roCal\
et nous avons donné des exemples de réductions de tels termes mais nous n'avons
explicité ni les règles d'évaluation du calcul, ni les mécanismes
sous-jacents comme le filtrage et la substitution.

Afin d'obtenir une définition précise du \roCal\  nous présentons les composants
principaux d'un calcul et nous décrivons par la suite chaque composant dans le
cadre du \roCal\  général.

D'abord, la \Def{syntaxe} d'un calcul décrit formellement la formation des objets
manipulés dans le calcul aussi bien que la formation des substitutions qui sont
utilisées par le mécanisme d'évaluation.  Dans le cas du \roCalT, la construction
des objets est basée sur une signature du premier ordre qui est enrichie par un
constructeur de règles de réécriture, un opérateur d'application de règle et les
ensembles de résultats.

La description de l'application de \Defn{substitutions}{substitution}{} aux
termes est souvent donnée au niveau méta du calcul, excepté pour les calculs
utilisant des substitutions explicites.  Pour le \roCal, nous utilisons la
substitution d'ordre supérieur et non le remplacement du premier ordre,
i.e. l'application utilise l'$\alpha$-conversion pour éviter la capture des
variables. Nous présenterons aussi plus tard une version du \roCal\  avec
substitutions explicites, appelé \roSig.

L'algorithme de \textit{filtrage} est employé pour lier les variables à leurs
valeurs actuelles. Nous appelons \roCalT\  le \roCal\  paramétré par la théorie
$T$ de filtrage et les propriétés du \roCalT\  sont fortement influencées par les
propriétés de cette théorie.  Dans le cas général, nous considérons un filtrage
d'ordre supérieur mais dans des cas pratiques nous utilisons le filtrage
d'ordre supérieur avec motif, ou le filtrage équationnel, ou simplement le filtrage
syntaxique.  Si dans un contexte donné la théorie de filtrage est claire, ce
paramètre est omis et nous parlons simplement de \roCal.

Les {\em règles d'évaluation} \index{regle d'évaluation@règle d'évaluation}
%\Defn{règles d'évaluation}{règle}{d'évaluation} 
décrivent le fonctionnement du
calcul. Ces règles représentent le lien entre les composants précédents. La
simplicité et la clarté de ces règles sont fondamentales pour une utilisation
facile du calcul.

Un dernier composant est la \Defn{stratégie d'évaluation}{stratégie}{d'évaluation}
%\Def{stratégie} d'évaluation 
utilisée pour guider l'application des règles d'évaluation du calcul. Selon la
stratégie employée nous obtenons différentes versions et donc différentes
propriétés pour le calcul.

Nous explicitons maintenant tous ces composants pour le \roCalT\  et nous
commentons nos principaux choix.

\section{La syntaxe}
%================================================================

\DEF%------------------------------------------------------------------------
\label{rhoTermesDeBase} 
Etant donnés un ensemble de variables $\XX$ et un ensemble de symboles
$\FF=\bigcup_{i \geq 0} \FF_i$ tel que pour tout $m$, $\FF_m$ est le sous-ensemble
de symboles d'arité $m$. Nous supposons que chaque symbole a une arité unique,
c'est-à-dire que les $\FF_m$ sont disjoints.

L'ensemble de $\rho$-termes \index{terme!rho-terme@$\rho$-terme} de base, noté $\RTT$, est
le plus petit ensemble tel que~:
\begin{itemize}
\item les variables de $\XX$ sont des $\rho$-termes,
\item si $t_1,\ldots,t_n$ sont des $\rho$-termes et $f \in \FF_n$ alors
	$f(t_1,\ldots,t_n)$ est un $\rho$-terme,
\item si $t_1,\ldots,t_n$ sont des $\rho$-termes alors $\{t_1,\ldots,t_n\}$ est un
	$\rho$-terme,
\item si $t$ et $u$ sont des $\rho$-termes alors $[t](u)$ est un $\rho$-terme,
\item si $t$ et $u$ sont des $\rho$-termes alors $t \ra u$ est un $\rho$-terme.
\end{itemize}

On appelle \Defn{position fonctionnelle}{position}{fonctionnelle} d'un
$\rho$-terme $t$, toute position $p$ du terme tel que $t(p)\in \FF$. Les
sous-termes $t_1,\ldots,t_n$ d'un terme fonctionnel $t=f(t_1,\ldots,t_n)$ sont
appelés les \Def{arguments} de $t$ ou, par abus de langage, les arguments de
$f$. Les symboles de $\FF_0$ sont appelés
{\em constantes}\index{constante}.%\Def{constantes}.
\FDEF%------------------------------------------------------------------------

En notation BNF les $\rho$-termes peuvent être définis par~:

\begin{tabular}{llll}
& \\
\textBNF{$\rho$-termes} ~~~~ & 
$t$ &::=&  $x ~~|~~ f(t,\ldots,t) ~~|~~ \{t,\ldots,t\} ~~|~~ [t](t)  ~~|~~ t \ra t$ \\
& \\
\end{tabular}

Dans la syntaxe précédente, le terme représentant l'ensemble vide ($n=0$) est
$\{\}$. Nous considérons que $\{\}$ et $\emptyset$ dénotent tous les deux
l'ensemble vide. Pour les termes de la forme $\{t_1, \ldots, t_n\}$ nous
supposons, comme d'habitude, que la virgule est associative, commutative et
idempotente. Un terme de la forme $t \ra u$ est appelé 
{\em règles de réécriture} \index{regle de réécriture@règle de réécriture}
%\Defn{règle de réécriture}{règle}{de réécriture} 
ou $\rho$-\Def{abstraction}. Le terme $[t](u)$
représente l'\Def{application} du $\rho$-terme $t$ au $\rho$-terme $u$ et si $t$
est de la forme $l \ra r$ alors on dit qu'il est la règle de réécriture de
l'application et $u$ l'argument de l'application.

Nous adoptons une discipline très générale pour la formation des règles de
réécriture et nous n'imposons, \textit{a priori}, aucune des restrictions
standards utilisées souvent en réécriture. Par exemple, nous pouvons avoir une
variable dans le membre gauche d'une règle de réécriture et nous pouvons avoir
dans le membre droit d'une règle de réécriture des variables qui n'apparaissent
pas dans le membre gauche de la règle de réécriture respective. Nous permettons
également des règles de réécriture contenant elles aussi des règles de
réécriture aussi bien que des applications de règles de réécriture.

L'intuition principale derrière cette syntaxe est que le membre gauche d'une
règle de réécriture précise les variables liées et une information structurelle.
Avoir de nouvelles variables dans le membre droit d'une règle de réécriture
donne la possibilité d'avoir des variables libres dans les règles de réécriture.

Les termes du \laCal\  et de la réécriture peuvent être représentés facilement
par des \mbox{$\rho$-termes}. Par exemple, le $\lambda$-terme $\lambda x. (y ~
x)$ correspond au $\rho$-terme $x \ra [y](x)$ et toute règle de réécriture de la
réécriture du premier ordre est représentée par la même règle dans le \roCal.
Dans le Chapitre~\ref{chap.encodage} nous montrons que non seulement les termes
des deux formalismes sont réprésentables dans le \roCal\  mais aussi les
réductions sous-jacentes.

Nous avons choisi les ensembles comme structure de données pour traiter le
non-déterminisme potentiel. Un ensemble de termes peut être vu comme l'ensemble
des résultats distincts obtenus en appliquant une règle de réécriture à un
terme.

Selon l'utilisation souhaitée du calcul, d'autres choix peuvent être faits pour
représenter le non-déterminisme.  Par exemple, si nous voulons fournir tous les
résultats d'une application, y compris les termes identiques, un multi-ensemble
pourrait être utilisé. Quand l'ordre dans lequel les résultats sont obtenus est
important, des listes pourraient être employées. Puisque dans cette présentation
du calcul nous nous concentrons sur les résultats possibles d'un calcul et pas
sur leur nombre ou leur ordre, des ensembles sont utilisés. Les propriétés de
confluence présentées dans la Section~\ref{stratRocal} sont préservées dans une
approche multi-ensemble. Il est clair que pour l'approche utilisant les listes,
seule une confluence modulo la permutation des listes peut être obtenue.

\EX%------------------------------------------------------------------------
\label{exRo}
Etant donnés $\FF_0 = \{a,b,c\}$, $\FF_1 = \{f,g\}$, $\FF=\FF_0 \cup
\FF_1$ et les variables $x,y$ dans $\XX$, nous présentons quelques $\rho$-termes
de $\RTT$~:
\begin{itemize}
\item le terme $[f(x,y) \ra g(x,y)](f(a,b))$ représente une application
	classique de règle de réécriture.
\item le terme $[x \ra x+y](a)$ contient la variable libre $y$ et nous allons
	voir plus tard pourquoi le résultat de cette application est $\{a+y\}$.
\item le terme $[y \ra [x \ra x+y](b)]([x \ra x](a))$ représente le %correspond au
	$\lambda$-terme $(\lambda y.((\lambda x.x+y) ~ b)) ~ ((\lambda x.x) ~ a)$.
	Dans la règle de réécriture $x \ra x+y$ la variable $y$ est libre
	mais dans la règle de réécriture $y \ra [x \ra x+y](b)$ cette variable
	est liée.
\item le terme $[x \ra [x](x)](x \ra [x](x))$ représente le $\lambda$-terme bien 
	connu $(\omega \omega)$.
\item le terme $[[(x \ra x+1) \ra (1 \ra x)](a \ra a+1)](1)$ est un $\rho$-terme 
	plus compliqué sans correspondant dans le \laCal\  ou dans la réécriture.
\end{itemize}
\FEX%------------------------------------------------------------------------

Pour un terme $u$ avec $p_1,\ldots,p_n$ des positions disjointes dans $u$ et
$t_1,\ldots,t_n$ des termes, nous notons
$u_{\atpos{t_1}{p_1}\ldots\atpos{t_n}{p_n}}$ le terme $u$ avec les termes $t_i$
aux positions $p_i$. La position d'un sous-terme dans un $\rho$-terme ensemble
est obtenue en considérant un des arbres représentant le $\rho$-terme
ensemble.

\section{Les substitutions}    \label{subst_ordre_sup}
%================================================================

Dans tous les calculs utilisant des lieurs, comme par exemple le \laCal, la
substitution du premier ordre n'est pas appropriée.  Afin d'obtenir un calcul de
substitutions correct nous devons définir les notions de variables libres,
$\alpha$-conversion et substitution similaires à celles définies dans la
Section~\ref{laCalNonType} pour le \laCal\  mais adaptées aux $\rho$-termes.

Nous utilisons donc un mécanisme de substitution évitant les captures
éventuelles des variables libres et nous considérons une approche similaire à
celle présentée dans~\cite{dowekhardinkirchner-ic} permettant de faire une distinction claire
entre la \Def{substitution} (qui prend en compte les variables liées) et la
\Def{greffe} (qui effectue un remplacement direct des variables).
La \Def{greffe} est appelée habituellement substitution du premier ordre
tandis que la \Def{substitution} désigne habituellement une substitution
d'ordre supérieur.

\DEF%------------------------------------------------------------------------
\label{freeVars} 
L'ensemble des variables \Defn{libres}{variable}{libre} d'un $\rho$-terme $t$,
noté $FV(t)$, est défini inductivement par~:
\begin{enumerate}
  \item si $t = x$ alors $FV(t) = \{x\}$,
  \item si $t = f(u_1,\ldots,u_n)$ alors $FV(t) = \bigcup_{i=1}^n FV(u_i)$,
  \item si $t = \{u_1, \ldots, u_n\}$ alors $FV(t) = \bigcup_{i=1}^n FV(u_i)$,
  \item si $t = [u](v)$ alors $FV(t) = FV(u) \cup FV(v)$,
  \item si $t = u \ra v$ alors $FV(t) = FV(v) \setminus FV(u)$.
\end{enumerate}
\FDEF%------------------------------------------------------------------------

Similairement au \laCal, on dit que l'occurrence d'une variable $x$ dans un
terme $t$ est \Defi{liée}{variable} si cette variable apparaît dans un
sous-terme de $t$ de la forme $v \ra u$ et $x \in FV(v)$. Dans le cas contraire,
l'occurrence de la variable $x$ est \Defi{libre}{variable}. Si la variable $x$ a
au moins une occurrence libre dans le terme $t$ alors $x$ est appelée une
variable libre de $t$. L'ensemble des variables libres de $t$ est exactement
$FV(t)$.

Dans la Définition~\ref{alphaConv} nous introduisons une notion appropriée de
renommage des variables liées afin d'éviter les captures éventuelles. Ainsi,
nous calculons une variante d'un $\rho$-terme qui est équivalente au terme
initial modulo l'$\alpha$-conversion.

\DEF%------------------------------------------------------------------------
\label{alphaConv} 
Etant donné un ensemble de variables $\YY$, l'application $\alpha_\YY$ (appelée
{\em $\alpha$-conversion} \index{alpha-conversion@$\alpha$-conversion} %\Def{$\alpha$-conversion}) 
renomme les variables liées d'un terme qui se trouvent dans $\YY$. Elle est
définie inductivement par~:
\begin{itemize}
\item $\alpha_\YY(x)$ $=$ $x$,
\item $\alpha_\YY(\{t\})$ $=$ $\{\alpha_\YY(t)\}$,
\item $\alpha_\YY(f(u_1,\ldots,u_n))$ $=$
	$f(\alpha_\YY(u_1),\ldots,\alpha_\YY(u_n))$,
\item $\alpha_\YY([t](u))$ $=$ $[\alpha_\YY(t)](\alpha_\YY(u))$,
\item $\alpha_{\YY}(u \ra v)$ $=$ $\alpha_{\YY}(u) \ra \alpha_{\YY}(v)$,
	si $FV(u) \cap \YY = \emptyset$,
\item $\alpha_{\YY}(u \ra v)$ $=$ $(\subs{x_i \mapsto y_i}_{x_i \in FV(u)}\
	\alpha_{\YY}(u)) \ra (\subs{x_i \mapsto y_i}_{x_i \in FV(u)}\
	\alpha_{\YY}(v))$,\\
	si $x_i \in FV(u) \cap \YY$ et $y_i$ sont des variables ``fraîches'' et
	$\subs{x \mapsto y}$ représente le remplacement de la variable $x$ par la
	variable $y$ dans le terme sur lequel il est appliqué.
\end{itemize}
\FDEF%------------------------------------------------------------------------

En utilisant l'$\alpha$-conversion nous pouvons définir la notion usuelle de
\Def{substitution}~:

\DEF%------------------------------------------------------------------------
\label{defSubstGreffe}
Une \Def{valuation} $\theta$ est une correspondance entre les variables
$x_1,\ldots,x_n$ et les termes $t_1,\ldots,t_n$, i.e. un ensemble fini de
couples $\{(x_1,t_1), \ldots, (x_n,t_n)\}$.

Etant donnée une valuation $\theta$ nous pouvons définir les deux notions de
substitution et greffe associées à $\theta$~:
\begin{itemize}
\item la \Def{substitution} étendant $\theta$ est notée
$\Theta=\subs{x_1/t_1,\ldots,x_n/t_n}$,
\item la \Def{greffe} étendant $\theta$ est noté
$\bar{\Theta}=\subs{x_1 \mapsto t_1,\ldots,x_n \mapsto t_n}$.
\end{itemize}

$\Theta$ et $\bar{\Theta}$ sont définies structurellement par~:

\noindent\begin{tabular}{ll}
%\hline 
& \\
-- $\Theta(x) = u$, si $(x,u) \in \theta$ &
-- $\bar{\Theta}(x) = u$, si $(x,u) \in \theta$
\\
%& \\
-- $\Theta(\{t\}) = \{\Theta (t)\}$ &
-- $\bar{\Theta}(\{t\}) = \{\bar{\Theta} (t)\}$
\\
%& \\
-- $\Theta(f(u_1,\ldots,u_n)) = (f(\Theta(u_1),\ldots,\Theta(u_n)))$ &
-- $\bar{\Theta}(f(u_1,\ldots,u_n)) = (f(\bar{\Theta}(u_1),\ldots,\bar{\Theta}(u_n)))$
\\
%& \\
-- $\Theta([t](u)) = [\Theta(t)](\Theta(u))$ &
-- $\bar{\Theta}([t](u)) = [\bar{\Theta}(t)](\bar{\Theta}(u))$
\\
%& \\
-- $\Theta(u \ra v) = \Theta(u') \ra \Theta(v') $ &
-- $\bar{\Theta}(u \ra v) = \bar{\Theta}(u) \ra \bar{\Theta}(v)$
\\
& \\
%\hline
\end{tabular}\\
où on considère que $z_i$ sont des variables fraîches (i.e.  $\theta z_i =z_i$),
que les variables $z_i$ n'apparaissent pas dans $u$ et $v$ et que pour toute
variable $y \in FV(u) \cup FV(v)$, $z_i \not\in FV(\theta y)$, et $u'$, $v'$ sont définis
par~:

$u'=\subs{y_i \mapsto z_i}_{y_i \in FV(u)}\   \alpha_{FV(u) \cup \Var(\theta)}(u)$,
 
$v'=\subs{y_i \mapsto z_i}_{y_i \in FV(u)}\   \alpha_{FV(u) \cup \Var(\theta)}(v)$.
\FDEF%------------------------------------------------------------------------

Les notions introduites pour les substitutions du premier ordre sont utilisées
aussi pour les substitutions présentées ci-dessus. L'ensemble de variables
$\{x_1,\ldots,x_n\}$ s'appelle le \Defi{domaine}{substitution} \index{domaine} de la
substitution $\Theta$ ou de la greffe $\bar{\Theta}$ et est noté respectivement par
$\Dom(\Theta)$ ou $\Dom(\bar{\Theta})$. Le
\Defi{codomaine}{substitution} \index{codomaine} de la substitution $\Theta$ est l'ensemble
$\Ran(\Theta) = \{t_1,\ldots,t_n\}$. L'ensemble de toutes les variables de
$\Theta$ est $\Var(\Theta)= \cup_{i=1}^n\Var(t_i) \cup \Dom(\Theta)$.

Nous rappelons que $\subs{x_1/u_1,\ldots,x_n/u_n}$ représente la substitution
simultanée des variables $x_1,\ldots,x_n$ par les termes $t_1,\ldots,t_n$ et non
la composition des substitutions $\subs{x_1/t_1} \ldots \subs{x_n/t_n}$.

Il y a rien de neuf dans la définition de la substitution et de la greffe sauf
que l'abstraction fonctionne dans le cas du \roCal\  sur des termes et pas
seulement sur des variables.

La complexité de la manipulation de variables pour l'$\alpha$-conversion peut
être évitée en utilisant des indices de de Bruijn et une notion de substitutions
explicites, dans l'esprit de~\cite{CurienHardinLevy-JACM96}. Nous allons
présenter une telle approche dans le Chapitre~\ref{chap.calcul_explicite}.

\section{Le filtrage} \label{filtrageNonType}
%================================================================

La liaison entre paramètres formels et actuels est basée sur le filtrage qui est
donc un composant fondamental du \roCal. Nous définissons d'abord les problèmes
de filtrage dans un cadre général~:

\DEF%------------------------------------------------------------------------
\label{defMa}
Etant donnée une théorie $T$ sur les $\rho$-termes, une 
$T$-{\em équation de filtrage} \index{equation de@équation de filtrage} %\Def{équation de filtrage} 
est une formule de la forme $t \meqqT t'$, où $t$ et
$t'$ sont des $\rho$-termes.  Une substitution $\sigma$ est une solution de
l'équation $t \meqqT t'$ si $T \models \sigma (t) = t'$.  Un $T$-\Def{système
de filtrage} est une conjonction de $T$-équations de
filtrage. Une substitution est une solution d'un $T$-système de filtrage $P$ si
c'est une solution de toutes les $T$-équations de filtrage.  Nous notons par
$\fF$ un $T$-système de filtrage sans solution. Un $T$-système de filtrage est
appelé \Defi{trivial}{système de filtrage} quand toute substitution est une solution du système.

Nous définissons \Sl($\SS$) pour un $T$-système de filtrage $\SS$ comme étant la
fonction qui retourne l'ensemble de toutes les solutions de $\SS$ quand $\SS$
n'est pas trivial et $\{\ID\}$\index{substitution!$\ID$}, où $\ID$ est la
substitution identité, quand $\SS$ est trivial.
\FDEF%------------------------------------------------------------------------

On appelle $T$-\Def{filtre} de $t$ à $t'$ une substitution $\sigma$ qui est une
solution d'un problème de filtrage $t \meqqT t'$. Si une telle substitution existe, on dit
que le terme $t$ \Def{subsume} le terme $t'$.

Remarquons que si l'algorithme de filtrage échoue (i.e. retourne $\fF$) alors la
fonction {\em \Sl} \index{Solution@\Sl} %\Def{\Sl} 
retourne l'ensemble vide.

Puisqu'en général on peut considérer des théories arbitraires sur les
$\rho$-termes, le $T$-filtrage est en général indécidable, même lorsqu'on se
restreint à des théories équationnelles du premier
ordre~\cite{JouannaudKirchner-rob91}. Afin de surmonter ce problème
d'indécidabilité, on peut penser à utiliser des contraintes, comme dans la
résolution d'ordre supérieur avec contraintes~\cite{Huet-IJCAI73} ou dans la
déduction avec contraintes~\cite{KirchnerKirchnerRusinowitch-RIA90}.

Nous sommes principalement intéressés par les cas décidables.  Parmi les
théories décidables, on peut mentionner le filtrage d'ordre supérieur avec motif
qui est décidable et unitaire comme conséquence de la décidabilité de
l'unification avec motif~\cite{MillerLProlog-89-91,DowekHKP-JICSLP96}, le
filtrage linéaire d'ordre supérieur~\cite{deGroote2000}, le
filtrage d'ordre supérieur qui est décidable jusqu'au quatrième
ordre~\cite{PadovaniThese-96,DowekLICS92,HuetLang-78} (le problème de décision
du cas général étant encore ouvert), beaucoup de théories équationnelles du
premier ordre comprenant l'associativité, la commutativité, la distributivité et
la majorité de leur combinaisons~\cite{Nipkow-RTA3-89,RingIC96}.

Par exemple, quand la théorie $T$ est vide, la substitution résultant du
filtrage syntaxique entre les termes $t$ et $t'$, quand elle existe, est unique
et peut être calculée par un algorithme récursif simple donné, par exemple, par
G.~Huet~\cite{HuetThese76}.  Cette substitution peut également être calculée par
l'ensemble de règles \rname{SyntacticMatching} où on suppose que le symbole
$\ww$ est associatif et commutatif.

\begin{figure}[!htp]%----------------------------------------------------------
%\noindent
\framebox{\parbox{\largeurtexte}{

\renewcommand{\fleche}{\LaFleche}
\begin{ruleset}

\regle{Decomposition} 
{(f(t_1, \ldots, t_n) \meqqes f(t'_1, \ldots, t'_n)) \ww P} 
{\bigwedge_{i=1\ldots n} t_i \meqqes t'_i \ww P}

\cregle{SymbolClash} 
{(f(t_1, \ldots, t_n) \meqqes g(t'_1, \ldots, t'_m)) \ww P}
{\fF} {f \neq g}

\cregle{MergingClash} 
{(x \meqqes t) \ww (x \meqqes t') \ww P}
{\fF} {t \neq t'} 

\cregle{SymbolVariableClash~} 
{(f(t_1, \ldots, t_n) \meqqes x) \ww P} 
{\fF} {x \in \XX}

\end{ruleset}

}}
\caption{\label{SyntMatch}\rname{SyntacticMatching} - Règles pour le filtrage syntaxique}
\end{figure}%-----------------------------------------------------------------

\PROP%------------------------------------------------------------------------
\cite{KirchnerKirchner-RSP-99}
\label{normform}
La forme normale de tout problème de filtrage $t \meqqes t'$ calculée par les
règles \rname{SyntacticMatching} existe et est unique. Après avoir enlevé de la
forme normale toute équation dupliquée et toute équation triviale de la forme 
$x \meqqes x$, si le système résultant est~:
\begin{enumerate}
\item $\fF$, alors il n'y a pas de filtre de $t$ à $t'$ et $\Sl(t \meqqes t') =
	\Sl(\fF)=\emptyset$,
\item de la forme $\bigwedge_{i\in I} x_i \meqqes t_i$ avec $I \neq \emptyset$,
	alors la substitution $\sigma = \subs{x_i / t_i}_{i \in I}$ est l'unique
	filtre de $t$ à $t'$ et 
	$\Sl(t \meqqes t') = \Sl(\bigwedge_{i\in I} x_i \meqqes t_i) = \{\sigma\}$,
\item vide, alors $t$ et $t'$ sont identiques et $\Sl(t \meqqes t)=\{\ID\}$.
\end{enumerate}
\FPROP%------------------------------------------------------------------------

\EX%------------------------------------------------------------------------
\label{exMatchSynt} 
Si nous considérons le problème de filtrage $(f(x,g(x,y)) \meqqes f(a,g(a,b))$,
nous appliquons d'abord la règle de filtrage \rname{Decomposition} et nous
obtenons le système avec les deux équations $(x \meqqes a)$ et $(g(x,y) \meqqes
g(a,b))$. Quand nous appliquons la même règle de nouveau pour la deuxième
équation nous obtenons $(x \meqqes a)$ et $(y \meqqes b)$. Ainsi, l'équation
initiale est réduite en $(x \meqqes a) \ww (x \meqqes a) \ww (y \meqqes b)$ et
donc $\Sl(f(x,g(x,y)) \meqqes f(a,g(a,b))=\{\subs{x/a,y/b}\}$.

Pour le problème de filtrage $(f(x,x) \meqqes f(a,b))$ nous appliquons, comme
avant, la règle \rname{Decomposition} et nous obtenons le système $(x \meqqes a)
\ww (x \meqqes b)$.  Ce dernier système est réduit par la règle de filtrage
$MergingClash$ en $\fF$ et ainsi, $\Sl(f(x,x) \meqqes f(a,b))=\emptyset$.
\FEX%------------------------------------------------------------------------

Cet algorithme de filtrage syntaxique peut être étendu d'une façon naturelle
quand on suppose qu'un symbole, comme le $+$ par exemple, est commutatif. Dans
ce cas-ci, l'ensemble précédent de règles devrait être complété avec~:
\renewcommand{\fleche}{\LaFleche}
\begin{ruleset}
\hregle{CommDec} 
{(t_1 + t_2) \meqq (t'_1 + t'_2) \ww P} 
{((t_1 \meqq t'_1 \ww t_2 \meqq t'_2) \vee (t_1 \meqq t'_2 \ww t_2 \meqq t'_1)) \ww P}
\end{ruleset}
où la disjonction a les propriétés habituelles.

Il est clair que dans ce dernier cas le nombre de filtres peut être exponentiel par
rapport à la taille des membres gauches des équations de filtrage initiales.

\EX%------------------------------------------------------------------------
\label{exMatchAC}
Quand on filtre modulo la commutativité un terme comme $x+y$, avec $+$ défini
comme étant commutatif, contre le terme $a+b$, la
règle $CommDec$ mène à
$$((x \meqq a \ww y \meqq b) \vee (x \meqq b \ww y \meqq a))$$ et ainsi,
$\Sl(x+y \meqq a+b)=\{\subs{x/a,y/b},\subs{x/b,y/a}\}$.
\FEX%------------------------------------------------------------------------

Le filtrage associatif-commutatif (AC) est souvent utilisé. Il pourrait être
défini en utilisant une approche basée sur des règles comme
dans~\cite{AdiKirchner-UNIF92,KirchnerRingeissen-FI98} ou une approche
sémantique comme dans~\cite{Eker-AC93}.

Comme nous l'avons déjà dit, la théorie $T$ est un paramètre du \roCal\  et ainsi
nous notons, par exemple, par \roCalE\  le \roCal\  avec une théorie de filtrage
vide (filtrage syntaxique), par \roCalC\  le \roCal\  avec une théorie de filtrage
commutative, par \roCalA\  le \roCal\  avec une théorie de filtrage associative,
par \roCalAC\  le \roCal\  avec une théorie de filtrage associative-commutative.

\section{Les règles d'évaluation} \label{regles_evaluation}
%================================================================

Nous supposons donnée une théorie $T$ sur les $\rho$-termes pour laquelle le
filtrage est décidable. 

Les règles d'évaluation du \roCalT\  décrivent principalement l'application d'un
$\rho$-terme sur un autre $\rho$-terme et indiquent le comportement des
différents opérateurs du calcul quand les arguments sont des ensembles.  En
fonction de leurs spécificités elles sont décrites dans les Figure~\ref{MRA}
à~\ref{MRAflat}.

\subsection{L'application d'une règle en tête dans le \roCalT} 
\label{regles_evaluation_application}
%===============================================================================

L'application d'une règle de réécriture {\em à la position de tête d'un terme}
$t$ est accomplie en utilisant les méta-opérations de filtrage et application de
substitution.  D'abord, on filtre le membre gauche de la règle de réécriture
contre le terme $t$ et puis, les substitutions (aucune, une ou plusieurs)
obtenues sont appliquées au membre droit de la règle de réécriture.
 
Comme nous l'avons déjà mentionné précédemment, dans le cas général, le filtrage
n'est pas unitaire et nous devrons traiter ainsi les ensembles (vides ou
infinis) de substitutions. Nous considérons une application des ensembles de
substitutions au niveau méta du calcul représentée par l'opérateur binaire
``$\_\subS{\_}$'' dont le comportement est décrit par la méta-règle~:

\renewcommand{\fleche}{\leadsto}
\begin{ruleset}
%=================================== 
\regle {Propagate}
{r\subS{\{\sigma_1,\ldots,\sigma_n,\ldots\}}}
{\{\sigma_1 r,\ldots,\sigma_n r,\ldots\}}
%===================================
\end{ruleset}
\vspace{-.5cm}

Le résultat de l'application d'un ensemble de substitutions
$\{\sigma_1,\ldots,\sigma_n,\ldots\}$ sur un terme $r$ est l'ensemble de termes
$\sigma_i r$, où $\sigma_i r$ représente le résultat de la (méta-)application de
la substitution $\sigma_i$ sur le terme $r$ comme détaillé dans la
Section~\ref{subst_ordre_sup}. Notez que lorsque $n$ vaut $0$, c'est-à-dire
l'ensemble de substitutions est vide, l'ensemble de termes instanciés est
également vide.

L'application d'une règle de réécriture $l \ra r$ sur un terme $t$ est décrite
par la règle d'évaluation \rname{Fire} présentée dans la Figure~\ref{MRA}.  La
règle \rname{Fire}, comme toutes les règles d'évaluation du calcul, peut être
appliquée à n'importe quelle position d'un $\rho$-terme.

\begin{figure}[!htp]%------------------------------------------------------------
\noindent
\framebox{\parbox{\largeurtexte}{

\renewcommand{\fleche}{\Longrightarrow}
\begin{ruleset}
%===================================
  \regle 
  {Fire} 
  {[l \ra r](t)} 
  {r \subS{\Sl(l \meqqT t)}} 
%===================================
\end{ruleset}

}}
\caption{\label{MRA}La règle d'évaluation \rname{Fire} du \roCalT}
\end{figure}%---------------------------------------------------------------

L'idée centrale est que l'application d'une règle de réécriture $l \ra r$ à
la position de tête d'un terme $t$, écrite $[l\ra r](t)$, consiste à remplacer
le terme $r$ par $r\subS{\Re}$ où $\Re$ est l'ensemble de substitutions
obtenues en $T$-filtrant $l$ contre $t$ (c'est-à-dire $\Sl(l \meqqT t)$). 
Par conséquent, quand le filtrage échoue menant à un ensemble vide de
substitutions, le résultat de
l'application de la règle \rname{Propagate} et ainsi de la règle \rname{Fire}
est l'ensemble vide.

Une autre façon de décrire l'application d'une règle de réécriture consiste à
expliciter l'application d'un ensemble de substitutions dans la règle
\rname{Fire} qui devient~:

\renewcommand{\fleche}{\Longrightarrow}
\begin{ruleset}
%===================================
  \wregle 
  {Fire'} 
  {[l \ra r](t)} 
  {\{\sigma_1 r,\ldots,\sigma_n r,\ldots\}}
  {\sigma_i \in \Sl(l \meqqT t)}
%===================================
\end{ruleset}

Notons que, comme dans \laCal, une application peut toujours être évaluée, mais
contrairement au \laCal, l'ensemble de résultats peut être vide.  Plus
généralement, quand on filtre modulo une théorie $T$, l'ensemble de filtres
correspondants peut être vide, un singleton (comme dans la théorie vide), un
ensemble fini (comme dans le cas d'une théorie associative-commutative) ou
infini~(\cite{FagesHuet83}). %\cite{Plotkin72} 
Nous avons ainsi choisi de représenter les résultats de l'application d'une règle
de réécriture à un terme par un ensemble.  Un ensemble vide signifie que la
règle de réécriture $l \ra r$ ne s'applique pas sur le terme $t$ dans le sens
d'un échec pour le filtrage entre les termes $l$ et $t$.

Quand nous évaluons, par exemple, l'application $[a \ra b](a)$, le membre gauche
$a$ de la règle de réécriture est filtré contre l'argument $a$ de l'application
menant à un ensemble contenant la substitution identité. L'application de cette
substitution ne modifie pas le membre droit $b$ de la règle et donc,
l'application est évaluée en l'ensemble $\{b\}$.

Le système de filtrage à résoudre peut être non-trivial comme pour la
$\rho$-application
\linebreak
$[f(x) \ra x](f(a))$ où la solution du filtrage entre le membre gauche de la
règle de réécriture et le terme $f(a)$ est la substitution $\subs{x/a}$. Le
membre droit $x$ de la règle est instancié par cette substitution à $a$ et donc,
le résultat final de l'évaluation est le $\rho$-terme $\{a\}$.

Si le filtrage entre le membre gauche de la règle de réécriture et l'argument de
l'application échoue comme, par exemple, pour les applications $[a \ra b](c)$ et
$[f(x) \ra x](g(a))$, alors nous obtenons un ensemble vide de substitutions et
le résultat de l'évaluation de ces applications est l'ensemble vide.

Mais un filtrage équationnel peut générer plusieurs substitutions comme, par
exemple, le problème de filtrage $x \cup y \meqq_{AC} a \cup b$ ayant les
solutions $\subs{x/a,y/b}$ et $\subs{x/b,y/a}$ si le symbole $\cup$ est
considéré associatif-commutatif. Quand de tels opérateurs sont utilisés dans le
membre gauche des règles de réécriture, l'application peut être évaluée à un
ensemble ayant plusieurs éléments. Considérons, par exemple, l'application 
$[x \cup y \ra x](a \cup b)$ représentant la sélection non-déterministe d'un des 
éléments du couple $a \cup b$. Puisque 
$\Sl(x \cup y \meqq_{AC} a \cup b)=\{\subs{x/a,y/b},\subs{x/b,y/a}\}$,
la méta-règle \rname{Propagate} évalue $x\subS{\{\subs{x/a,y/b},\subs{x/b,y/a}\}}$
en $\{a,b\}$ et donc le résultat de l'évaluation de l'application est $\{a,b\}$.

\subsection{Les règles \rname{Congruence}  du \roCalT} 
\label{regles_evaluation_congruence}
%===============================================================================

Afin de pousser l'application de règles de réécriture plus profondément dans les
termes, nous introduisons les deux règles d'évaluation \rname{Congruence}
présentées dans la Figure~\ref{MRAcong}. Ces règles traitent l'application d'un
terme de la forme $f(u_1,\ldots,u_n)$ (où $f \in \FF_n$) à un autre terme d'une
forme semblable.  Quand les deux termes de l'application $[u](v)$ ont le même
symbole de tête, les arguments du terme $u$ sont appliqués sur ceux du terme
$v$.  Si les symboles de tête ne sont pas identiques, un ensemble vide est
obtenu.

\begin{figure}[!htp]%------------------------------------------------------------
\noindent
\framebox{\parbox{\largeurtexte}{

\renewcommand{\fleche}{\Longrightarrow}
\begin{ruleset}
%===================================
  \regle
  {Congruence}   
  {[f(u_1,\ldots,u_n)](f(v_1,\ldots,v_n))} 
  {\{f([u_1](v_1),\ldots,[u_n](v_n))\}}
\\
%===================================
  \regle
  {Congruence\_fail}   
  {[f(u_1,\ldots,u_n)](g(v_1,\ldots,v_m))} 
  {\emptyset}
%===================================
\end{ruleset}

}}
\caption{\label{MRAcong}Les règles d'évaluation \rname{Congruence} du \roCalT}
\end{figure}%------------------------------------------------------------------

Nous pouvons simuler\label{redondantesCongruence} les réductions utilisant les règles d'évaluation
\rname{Congruence} pour les applications d'un terme avec un symbole de tête
fonctionnel à un terme de la même forme en transformant le terme initial et
utilisant la règle d'évaluation \rname{Fire}. En effet, l'application d'un terme
$f(u_1,\ldots,u_n)$ à un autre terme $t$ (i.e., le terme
$[f(u_1,\ldots,u_n)](t)$) est évalué, en utilisant les règles d'évaluation
\rname{Congruence} et \rname{Congruence\_fail}, au même terme que l'application
du $\rho$-terme $f(x_1,\ldots,x_n) \ra f([u_1](x_1),\ldots,[u_n](x_n))$ au terme
$t$ (formellement, le $\rho$-terme 
$[f(x_1,\ldots,x_n) \ra f([u_1](x_1),\ldots,[u_n](x_n))](t)$) en utilisant la
règle \rname{Fire}. Bien que nous puissions exprimer les mêmes calculs en
utilisant seulement la règle d'évaluation \rname{Fire} et en transformant les
termes de départ, nous préférons garder les règles d'évaluation
\rname{Congruence} dans le calcul pour une représentation plus concise des
$\rho$-termes.

La représentation des résultats des applications par des ensembles a des
conséquences importantes en ce qui concerne le comportement du calcul. Nous
pouvons mentionner que lorsqu'on évalue un $\rho$-terme, le nombre de symboles
d'ensemble dans le terme résultat représente le nombre d'applications réduites
dans la dérivation du terme initial. 

L'application de la règle de réécriture $a \ra b$ à l'argument du terme $f(a)$
peut être décrite par le $\rho$-terme $[f(x) \ra f([a \ra b](x))](f(a))$ qui est
évalué en utilisant la règle d'évaluation \rname{Fire} en $\{f([a \ra b](a))\}$
et puis en $\{f(\{b\})\}$. Mais nous pouvons également donner une représentation
plus concise et utiliser le $\rho$-terme $[f(a \ra b)](f(a))$ qui est évalué en
appliquant la règle d'évaluation \rname{Congruence} en $\{f([a \ra b](a))\}$ et
puis, conformément à la règle d'évaluation \rname{Fire}, en $\{f(\{b\})\}$.  On
peut remarquer que les deux symboles d'ensemble dans le terme résultat
correspondent aux deux applications de la règle d'évaluation \rname{Fire} dans
le premier cas et aux applications des règles d'évaluation \rname{Fire} et
\rname{Congruence} dans le deuxième cas.

\subsection{Le traitement des ensembles dans le \roCalT}
%===============================================================================

Les réductions correspondant aux termes contenant des ensembles sont définis par
les règles d'évaluation présentées dans la Figure~\ref{MRAset}. Ces règles
décrivent la propagation des ensembles par rapport aux constructeurs de
$\rho$-termes~: les règles \rname{Distrib} et \rname{Batch} pour l'application,
\rname{Switch_L} et \rname{Switch_R} pour l'abstraction et \rname{OpOnSet}
pour les fonctions.

\begin{figure}[!htp]%---------------------------------------------------
\noindent \framebox{\parbox{\textwidth}{

\renewcommand{\fleche}{\Longrightarrow}
\begin{ruleset}
%=================================== 
\regle {Distrib}
	{[\{u_1,\ldots,u_n\}](v)}
	{\{[u_1](v),\ldots,[u_n](v)\}}
\\
%=================================== 
\regle {Batch}
	{[v](\{u_1,\ldots,u_n\})}
	{\{[v](u_1),\ldots,[v](u_n)\}}
\\
%=================================== 
\regle {Switch_L}
	{\{u_1,\ldots,u_n\} \ra v}
	{\{u_1 \ra v,\ldots,u_n \ra v\}}
\\
%=================================== 
\regle {Switch_R} 
	{u \ra \{v_1,\ldots,v_n\}}
	{\{u \ra v_1,\ldots,u \ra v_n\}}
\\
%=================================== 
\regle {OpOnSet}
	{f(v_1,\ldots,\{u_1,\ldots,u_m\},\ldots,v_n)}
	{\{f(v_1,\ldots,u_1,\ldots,v_n),\ldots,f(v_1,\ldots,u_m,\ldots,v_n)\}}
%===================================
\end{ruleset}

}}
\caption{\label{MRAset}Les règles d'évaluation \rname{Set} du \roCalT}
\end{figure}%---------------------------------------------------------

Le nombre de symboles d'ensemble n'est pas modifié au cours de l'évaluation par
les règles \rname{Distrib}, \rname{Batch}, \rname{Switch_L}, \rname{Switch_R} et
\rname{OpOnSet}. De cette manière, le nombre de symboles d'ensemble dans un
terme (ne contenant pas d'ensemble vide) compte toujours le nombre de règles
\rname{Fire} et \rname{Congruence} qui ont été appliquées afin d'obtenir le
terme respectif.

Un résultat de la forme $\{\}$ (noté habituellement $\emptyset$) représente
l'échec de l'application (d'une règle de réécriture) et les échecs sont
strictement propagés dans les $\rho$-termes en utilisant l'ensemble de
règles d'évaluation \rname{Set}.

La relation de réécriture induite par les règles d'évaluation \rname{Fire},
\rname{Congruence} et les règles \rname{Set} est plus fine (c'est-à-dire
contient plus d'éléments) que la relation standard (sans ensemble) et elle est
non-confluente. Une raison de la non-confluence est l'absence d'une règle
d'évaluation similaire aux règles \rname{Set} pour la propagation des ensemble
par rapport aux ensembles.

L'application de l'ensemble de règles de réécriture $\{a \ra b,a \ra c\}$ au
terme $a$ (i.e. le $\rho$-terme $[\{a \ra b,a \ra c\}](a)$) est réduite en
utilisant la règle d'évaluation \rname{Distrib} en l'ensemble contenant les
applications de chaque règle au terme $a$, c'est-à-dire le $\rho$-terme 
$\{[a \ra b](a),[a \ra c](a)\}$. 
En plus, nous pouvons factoriser un ensemble de règles de réécriture ayant le
même membre gauche et utiliser le $\rho$-terme $a \ra \{b,c\}$ qui est réduit en
appliquant la règle d'évaluation \rname{Switch_R} en $\{a \ra b,a \ra c\}$.
Ainsi, nous pouvons dire que le $\rho$-terme $[a \ra \{b,c\}](a)$ décrit le
choix non-déterministe entre l'application de la règle $a \ra b$ au terme $a$ et 
l'application de la règle $a \ra c$ au même terme $a$ et cette application est
réduite en l'ensemble contenant le résultat de deux applications, c'est-à-dire 
$\{\{b\},\{c\}\}$.

Considérons maintenant le $\rho$-terme $[f(a \ra b)](f(a))$ qui est réduit,
comme montré précédemment, en $\{f(\{b\})\}$ et puis, en utilisant la règle
\rname{OpOnSet} en $\{\{f(b)\}\}$. Les deux symboles d'ensemble correspondent
aux deux applications des règles d'évaluation \rname{Fire} et \rname{Congruence}
sont donc préservés par l'application de la règle \rname{OpOnSet}.

Le résultat $\emptyset$ obtenu pour l'évaluation d'une application comme, par
exemple, $[a \ra b](c)$ est propagé dans les règles de réécriture, les
applications et les termes du premier ordre ayant cette application en
 argument. Le $\rho$-terme $f([a \ra b](c))$ est réduit en utilisant la règle
d'évaluation \rname{Fire} en $f(\emptyset)$ et ensuite, en appliquant la règle
d'évaluation \rname{OpOnSet} en $\emptyset$.

Notons que la propagation d'un ensemble vide dans un terme peut mener à des
résultats non-convergents par rapport aux règles d'évaluation présentées jusqu'à
maintenant. Un $\rho$-terme $g([a \ra b](c),[a \ra b](a))$ est réduit en
$g(\emptyset,\{b\})$ et ensuite, en appliquant la règle d'évaluation
\rname{OpOnSet} nous obtenons soit $\emptyset$, soit $\{g(\emptyset,b)\}$ en
fonction de l'argument considéré. Ce dernier terme est réduit en $\{\{\}\}$, avec 
les deux symboles d'ensemble correspondant aux applications des règles
d'évaluation \rname{Fire}, mais il ne peut pas être réduit en $\emptyset$, 
terme dans lequel l'information sur le nombre de pas d'évaluation est perdue.

\subsection{Aplatir les ensembles dans le \roCalT}
%===============================================================================

La règle d'évaluation \rname{Flat} présentée dans la Figure~\ref{MRAflat} décrit
la propagation des ensembles par rapport aux $\rho$-termes de type ensemble et
l'élimination des symboles d'ensemble redondants.

\begin{figure}[!htp]%---------------------------------------------------------
\noindent \framebox{\parbox{\largeurtexte}{

\renewcommand{\fleche}{\Longrightarrow}
\begin{ruleset}
%=================================== 
\regle {Flat}
	{\{u_1,\ldots,\{v_1,\ldots,v_n\},\ldots,u_m\}}
	{\{u_1,\ldots,v_1,\ldots,v_n,\ldots,u_m\}}
%===================================
\end{ruleset}

}}
\caption{\label{MRAflat}La règle d'évaluation \rname{Flat} du \roCalT}
\end{figure}%------------------------------------------------------------

Puisque chaque élément d'un ensemble représente un des résultats non-déterministes
d'une réduction, nous pouvons dire qu'un échec dans la réduction d'un élément
représente l'échec d'une des réductions possibles et ne doit pas mener à l'échec
de toutes les réductions. Ceci correspond exactement au fonctionnement du
\roCalT\  où un échec (i.e. l'ensemble vide) obtenu dans la réduction d'un élément
d'un $\rho$-terme ensemble \textit{n'est pas} strictement propagé dans le sens
où l'ensemble initial n'est pas nécessairement réduit en l'ensemble vide.

La règle d'évaluation \rname{Flat} est utilisée dans le \roCal\  pour aplatir les
ensembles et on doit noter que, puisque les symboles d'ensemble imbriqués sont
éliminés par cette règle d'évaluation, le nombre d'étapes de réduction n'est
plus indiqué par le nombre de symboles d'ensemble.

La règle d'évaluation qui correspond à la propagation des ensembles pour les
symboles d'ensemble et qui préserveraient correctement le nombre de symboles
d'ensemble est la règle d'évaluation \rname{Flat\_one}~:

\renewcommand{\fleche}{\Longrightarrow}
\begin{ruleset}
%=================================== 
\cregle {Flat\_one}
	{\{ u_1,\ldots,\{v_1,\ldots,v_n\},\ldots,u_m \}}
	{\{ \{u_1,\ldots,v_1,\ldots,v_n,\ldots,u_m\} \}}
	{m \in \textbf{N}^*}
%===================================
\end{ruleset}
\vspace{-.5cm}

L'inconvénient de cette solution est la différence faite entre des termes
identiques, mais obtenus après un nombre différent de pas de réduction. Par
exemple, en utilisant cette approche, le terme $\{\{a\}\}$ ne se réduit pas en
$\{a\}$, ce qui est souhaitable dans un calcul où nous souhaitons que le
résultat ne contienne pas d'information sur la dérivation qui l'a produit.

Par conséquent, dans une approche utilisant la règle d'évaluation
\rname{Flat\_one} nous devons aussi introduire la règle d'évaluation
\rname{Elim} qui élimine les symboles d'ensemble redondants~:

\renewcommand{\fleche}{\Longrightarrow}
\begin{ruleset}
%===================================
\regle
	{Elim}
	{\{\{u_1,\ldots,u_m\}\}}
	{\{u_1,\ldots,u_m\}}
%===================================
\end{ruleset}
\vspace{-.5cm}

En combinant les règles d'évaluation \rname{Flat\_one} et \rname{Elim} nous
obtenons exactement la règle d'évaluation \rname{Flat} présentée ci-dessus.

Dans la section précédente nous avons vu que le $\rho$-terme 
$[a \ra \{b,c\}](a)$ est réduit en $\{\{b\},\{c\}\}$ et maintenant, nous pouvons
utiliser la règle d'évaluation \rname{Flat} pour réduire ce dernier terme en
$\{b,c\}$. De la même manière, le $\rho$-terme $\{\{f(b)\}\}$ est réduit en 
$\{f(b)\}$.

En éliminant les symboles d'ensemble redondants nous pouvons résoudre le
problème de non-convergence pour la réduction du terme $g([a \ra b](c),[a \ra
b](a))$ en $\{\}$ ou $\{\{\}\}$. Puisque le terme $\{\{\}\}$ est réduit en
utilisant la règle d'évaluation \rname{Flat} en $\{\}$ alors le terme initial
ne peut être réduit qu'en $\{\}$.

Le terme $[\{a \ra b, b \ra a\}](a)$ correspondant à l'application d'un ensemble
de règles de réécriture est réduite en le terme $\{[a \ra b](a),[b \ra a](a)\}$
et ensuite, en appliquant la règle d'évaluation \rname{Fire}, nous obtenons le
terme $\{\{b\},\emptyset\}$. Cet ensemble est réduit en utilisant la règle
d'évaluation \rname{Flat} en $\{b\}$ et en analysant ce terme résultat nous ne
pouvons ni déduire le nombre des règles d'évaluation \rname{Fire} et
\rname{Congruence} ni détecter un échec au cours de l'évaluation.

Ce comportement du calcul pourrait être résumé en disant que la propagation de
l'échec décrite par les règles d'évaluation \rname{Set} est stricte sur tous les
opérateurs sauf l'ensemble.  En fait, nous verrons plus tard que la règle
\rname{Fire} peut induire des propagations non-strictes dans quelques cas
particuliers (voir l'Exemple~\ref{ns_failure} à la page~\pageref{ns_failure}).

La décision d'utiliser des ensembles pour représenter les résultats des
réductions a une autre conséquence importante concernant le traitement des
ensembles par rapport au filtrage.

En effet, les ensembles sont juste utilisés pour enregistrer des résultats et
nous ne souhaitons pas qu'ils fassent partie de la théorie de filtrage.  Nous
supposons de ce fait que le filtrage utilisé dans la règle d'évaluation
\rname{Fire} \textit{n'est pas} réalisé modulo une théorie équationnelle
axiomatisant les ensembles.  Ceci exige dans certains cas l'utilisation d'une
stratégie qui pousse les symboles d'ensemble des termes autant que possible vers
les positions les plus petites (c'est-à-dire distribuer les ensembles vers
l'extérieur).

\section{La stratégie d'évaluation}
%================================================================

La définition générale d'une stratégie est donnée dans~\cite{KirchnerKV-MIT95}
mais nous spécialisons cette définition dans le cas du \roCal.

\DEF%--------------------------------------------------------------------------
Une \Defn{stratégie d'évaluation}{stratégie}{d'évaluation} dans le \roCal\   est
un sous-ensemble de toutes les dérivations possibles.
\FDEF%--------------------------------------------------------------------------

La stratégie $\SS$ guidant l'application des règles d'évaluation du \roCalT\  
peut être cruciale pour obtenir les bonnes propriétés pour le calcul. Dans une
première étape, la propriété principale analysée est la confluence du calcul et
nous verrons que si la règle \rname{Fire} est appliquée sans aucune restriction
et à n'importe quelle position d'un $\rho$-terme alors le calcul n'est pas
confluent.

Par exemple, la stratégie $\NONE$ représente l'ensemble de toutes les
dérivations, c'est-à-dire qu'elle n'est pas restrictive. La stratégie vide est
la stratégie qui ne permet aucune réduction. Nous désirons définir la stratégie
la moins restrictive permettant d'obtenir certaines propriétés et en particulier
la confluence du calcul.

Les raisons de la non-confluence du calcul général sont expliquées dans la
Section~\ref{nonconfRocal} et une solution est proposée pour obtenir un calcul
confluent. La stratégie garantissant la confluence du calcul peut être donnée
explicitement ou exprimée comme une condition sur l'application de la règle
d'évaluation \rname{Fire}.

\section{La définition du \roCal}
%================================================================

En utilisant les notions présentées dans les sections précédentes nous donnons la
définition du \roCalT\  général.

\DEF%------------------------------------------------------------------------
\label{roTdef} 

Etant donnés un ensemble de symboles de fonctions $\FF$, un ensemble de variables
$\XX$ et une théorie $T$ sur les termes de $\RTT$ avec un problème de filtrage
décidable, nous appelons \roCalT\  (ou calcul de réécriture) un calcul défini
par~:
\begin{itemize}
\item un sous-ensemble non vide $\rtt$ de l'ensemble de termes $\RTT$,
\item l'application (d'ordre supérieur) de substitution aux termes comme définie 
	dans la Section~\ref{subst_ordre_sup},
\item une théorie $T$,
\item l'ensemble de règles d'évaluation $\EE$: \rname{Fire}, \rname{Congruence},
	\rname{Congruence\_fail}, \rname{Distrib}, \rname{Batch},
	\rname{Switch_L}, \rname{Switch_R}, \rname{OpOnSet}, \rname{Flat}
	(Figure~\ref{MRAgen}),
\item une stratégie d'évaluation $\SS$ qui guide l'application des règles
	d'évaluation.
\end{itemize}

Nous utilisons la notation $(\rtt,T,\SS)$ pour rendre explicite les
composants principaux du calcul de réécriture considéré. Lorsque ces composant
sont clairs suivant le contexte, la notation simplifiée $\rho_T$ est utilisée.
\FDEF%------------------------------------------------------------------------

\begin{figure}[!htp]%------------------------------------------------------------
\noindent
\framebox{\parbox{\largeurtexte}{

\renewcommand{\fleche}{\Longrightarrow}
\begin{ruleset}
%===================================
  \regle 
  {Fire} 
  {[l \ra r](t)} 
  {r \subS{\Sl(l \meqqT t)}} 
\\
%===================================
  \regle
  {Congruence}   
  {[f(u_1,\ldots,u_n)](f(v_1,\ldots,v_n))} 
  {\{f([u_1](v_1),\ldots,[u_n](v_n))\}}
\\
%===================================
  \regle
  {Congruence\_fail}   
  {[f(u_1,\ldots,u_n)](g(v_1,\ldots,v_m))} 
  {\emptyset}
\\
%===================================
\regle {Distrib}
	{[\{u_1,\ldots,u_n\}](v)}
	{\{[u_1](v),\ldots,[u_n](v)\}}
\\
%=================================== 
\regle {Batch}
	{[v](\{u_1,\ldots,u_n\})}
	{\{[v](u_1),\ldots,[v](u_n)\}}
\\
%=================================== 
\regle {Switch_L}
	{\{u_1,\ldots,u_n\} \ra v}
	{\{u_1 \ra v,\ldots,u_n \ra v\}}
\\
%=================================== 
\regle {Switch_R} 
	{u \ra \{v_1,\ldots,v_n\}}
	{\{u \ra v_1,\ldots,u \ra v_n\}}
\\
%=================================== 
\hregle {OpOnSet}
	{f(v_1,\ldots,\{u_1,\ldots,u_m\},\ldots,v_n)}
	{\{f(v_1,\ldots,u_1,\ldots,v_n),\ldots,f(v_1,\ldots,u_m,\ldots,v_n)\}}
\\
%===================================
\regle {Flat}
	{\{u_1,\ldots,\{v_1,\ldots,v_n\},\ldots,u_m\}}
	{\{u_1,\ldots,v_1,\ldots,v_n,\ldots,u_m\}}
%===================================
\end{ruleset}

}}
\caption{\label{MRAgen}Les règles d'évaluation du \roCalT}
\end{figure}%---------------------------------------------------------------

\DEF%------------------------------------------------------------------------
	Un $\rho$-terme $t$ tel qu'il existe une règle d'évaluation de
	l'ensemble $\EE$ applicable à la position de tête de $t$ est appelé un
	{\em $\rho$-radical}\index{radical!rho-radical@$\rho$-radical} %\Defi{$\rho$-radical}{radical} 
	ou \Def{radical}.
\FDEF%------------------------------------------------------------------------

Quand les paramètres du calcul général sont remplacés par certaines valeurs
spécifiques, différentes instances du calcul sont obtenues. Nous pouvons préciser
un ou plusieurs des paramètres et obtenir des calculs (partiellement) instanciés
comme, par exemple, $\rhoe=(\RTTE,\emptyset,\SS)$, où $\RTTE$ est un
sous-ensemble bien défini de $\RTT$, le filtrage employé est syntaxique 
et $\SS$ dénote une stratégie pas encore précisée.

\section{Relation de réécriture versus calcul de réécriture}
%================================================================

Les termes du \roCal\  contiennent toute l'information nécessaire pour leur
évaluation. C'est également le cas pour le \laCal\  mais ceci est tout à fait
différent de la manière habituelle dont les \textit{relations de réécriture}
sont définies.

Comme vu plus haut, la relation de réécriture \index{relation!de réécriture}
engendrée par un système de réécriture $\RR=\{l_1 \ra r_1,\ldots,l_n \ra r_n\}$
est la plus petite relation transitive stable par contexte et substitution et
contenant $(l_1,r_1),\ldots,(l_n,r_n)$.

\EX%------------------------------------------------------------------------
\label{exReductionReec} 
Si $\RR = \{a \ra f(b),b \ra c\}$ alors la relation de réécriture correspondante
contient $(a,f(b))$, $(b,c)$, $(a,f(c))$ et on peut dire que la dérivation 
$a \ra f(b) \ra f(c)$ est engendrée par $\RR$.

On dit que la relation de réécriture associée à un système de réécriture
contenant la règle de réécriture $a \ra a$ ne termine pas puisque la dérivation
$a \ra a \ra a \ra \ldots$ est générée dans un tel système.
\FEX%------------------------------------------------------------------------ 

Dans le \roCal\  la situation est différente puisqu'un $\rho$-terme est évalué
seulement en fonction de l'information de réécriture explicitement contenue dans
le terme. L'ensemble des règles de réécriture utilisées dans la réduction n'est
pas décrit au méta-niveau du calcul mais toutes les règles sont présentes dans
le $\rho$-terme à réduire.

\EX%------------------------------------------------------------------------ 
\label{exReductionRo} 
Dans le \roCal\  une dérivation similaire à celle générée par la règle de
réécriture $a \ra a$ de l'Exemple~\ref{exReductionReec} doit être construite
explicitement en décrivant, par exemple, toutes les application de la règle $a \ra a$. 

Le $\rho$-terme $[a \ra a](a)$ décrit un pas de réécriture et il est réduit en
utilisant la règle d'évaluation \rname{Fire} en
$\{a\}$. Le $\rho$-terme $[a \ra a]([a \ra a](a))$ décrit deux applications de
la règle $a \ra a$ étant réduit d'abord en $[a \ra a](\{a\})$ et ultérieurement
en $\{a\}$.
\FEX%------------------------------------------------------------------------ 

On doit noter que l'utilisation des règles de réécriture avec une variable comme
membre gauche dans les $\rho$-termes ne conduit pas systématiquement à des
réductions non-terminantes.

\EX%------------------------------------------------------------------------ 
La forme normale d'un $\rho$-terme $[x \ra x](a)$ est l'ensemble $\{a\}$ tandis
que la relation engendrée par tout système de réécriture contenant la règle 
$x \ra x$ est non-terminante.
\FEX%------------------------------------------------------------------------ 

Dans l'Exemple~\ref{exReductionRo} on peut remarquer que pour chaque application
d'une règle dans une réduction de la réécriture classique le $\rho$-terme
correspondant doit contenir une application de la règle respective à la position
appropriée.

Une autre approche consiste à définir des $\rho$-opérateurs d'itération
permettant la description des stratégies de normalisation par rapport à un
ensemble de règles de réécriture. La manière dont ces opérateurs sont construits
sera décrite ultérieurement mais pour donner une intuition sur leur utilisation
nous pouvons dire que le $\rho$-terme $[im(\{a \ra a\})](a)$ décrit la réduction
innermost du terme $a$ par rapport à la règle $a \ra a$ et la réduction de ce
terme dans le \roCal\  est très similaire à celle du terme $a$ par rapport au
système de réécriture contenant la règle $a \ra a$.

Nous allons présenter dans la Section~\ref{encodRew} une description plus
détaillée de la correspondance entre la réécriture du premier ordre et le
\roCal\  mais nous pouvons noter immédiatement une similarité entre les termes de
preuve de la logique de réécriture et les $\rho$-termes. 

\EX%------------------------------------------------------------------------ 
Considérons une théorie de réécriture basée sur la règle de réécriture étiquetée
$[a2b] ~~ a \ra b$. Dans cette théorie nous pouvons déduire le séquent
$f(a2b):f(a) \ra f(b)$, où $f$ est un symbole de la signature. 

En partant du terme de preuve $f(a2b)$ pour la réduction $f(a) \ra f(b)$ nous
construisons le \mbox{$\rho$-terme} $[f(a \ra b)](f(a))$ avec la réduction %\\
%$~~~~~~~~$ 
$[f(a \ra b)](f(a)) \longra{}_{\rho} f([a \ra b](a)) \longra{}_{\rho}  f(\{b\})
\longra{}_{\rho} \{f(b)\}$.

Si nous identifions l'étiquette d'une règle avec la règle nous pouvons dire que
l'application du terme de preuve  $f(a2b)$ au terme initial $f(a)$ (i.e. le 
$\rho$-terme $[f(a \ra b)](f(a))$) est réduite en l'ensemble contenant le terme
final $f(b)$ (i.e. $\{f(b)\}$).
\FEX%------------------------------------------------------------------------ 

Nous avons donc une interprétation immédiate des termes de preuve de la logique
de réécriture par des $\rho$-termes mais il existe des réductions des
$\rho$-termes sans terme de preuve correspondant.

Il n'existe pas de correspondance directe entre la réduction des $\rho$-termes
contenant des ensembles et les termes de preuve mais nous pouvons considérer,
par exemple, qu'à la réduction du terme $[\{a \ra b, a\ra c\}](a)$ en $\{b,c\}$
correspondent les deux termes de preuve pour l'application des règles de
réécriture $a \ra b$ et $a \ra c$ au terme $a$. 

Si nous considérons des $\rho$-règles de réécriture ne contenant pas seulement des
termes du premier ordre dans le membre droit nous obtenons immédiatement des
$\rho$-réductions sans aucune correspondance dans les termes de preuve.  Par
exemple, pour la $\rho$-réduction de l'application $[x \ra [y \ra x \cup y](b)](a)$ en 
$\{a \cup b\}$ il n'existe pas de terme de preuve associé.

\section{Instances du calcul de réécriture}	\label{utilisation_rho}
%================================================================

Cette section a pour objectif d'illustrer les concepts que nous avons présentés
auparavant en donnant quelques exemples de $\rho$-termes et de
$\rho$-réductions.

Nous allons commencer par la partie fonctionnelle du calcul et nous donnons les
$\rho$-termes représentant certains $\lambda$-termes. Par exemple, la
$\lambda$-abstraction $\lambda x.(y~x)$, où $y$ est une variable, est
représentée par la $\rho$-règle $x \ra [y](x)$.  L'application de ce
$\lambda$-terme à une constante $a$, notée en \laCal\  $\lambda x.(y~x)~a$, est
représentée dans le \roCal\  par l'application $[x \ra [y](x)](a)$. Cette
application est réduite dans le \laCal\  en $(y~a)$ tandis que dans le
\roCal\  le résultat de la réduction du terme correspondant est le singleton
$\{[y](a)\}$. Nous pouvons être plus précis et dire que les
$\lambda$-termes et réductions ci-dessus sont représentés dans le \roCalE\  où le
filtrage est syntaxique.  Lorsqu'une notation fonctionnelle $f(x)$ est choisie
pour $(f~x)$, le $\lambda$-terme $\lambda x.f(x)$ est représenté par le
$\rho$-terme $x \ra f(x)$. On note que pour les $\rho$-termes de cette
forme (c'est-à-dire des règles de réécriture qui ont une variable comme membre
gauche) le filtrage syntaxique effectué dans le \roCalE\  est trivial, il
n'échoue jamais et fournit toujours un seul résultat.

Il n'y a pas de difficulté à représenter des $\lambda$-termes plus élaborés dans
le \roCalE~: 

\EX%------------------------------------------------------------------------
Considérons le terme $\lambda x.f(x)~(\lambda y.y~a)$ avec la
$\beta$-dérivation suivante~:
\[
\lambda x.f(x)~(\lambda y.y~a) 
\lraD{\beta} \lambda x.f(x)~a
\lraD{\beta} f(a)
\]

La dérivation correspondante dans le \roCalE\  pour le $\rho$-terme correspondant
est la suivante~:
%$~~~~~~~~$
$$[x \ra f(x)]([y \ra y](a)) \lraD{Fire} [x \ra f(x)](\{a\}) \lraD{Batch}$$%\\
%$~~~~~~~~$ 
$$\{[x \ra f(x)](a)\} \lraD{Fire} \{\{f(a)\}\} \lraD{Flat} \{f(a)\}$$%\\
\FEX%------------------------------------------------------------------------

Naturellement, plusieurs stratégies de réduction peuvent être utilisées dans le
\laCal\  et être reproduites en conséquence dans le \roCalE. En effet, nous
verrons dans la Section~\ref{encodageLambda} que le \roCalE\  contient
strictement le \laCal.

Le \laCal\  avec motifs présenté dans~\cite{Peyton87} enrichit le \laCal\  avec
des ``$\lambda$-abstractions filtrantes'' permettant des motifs plus élaborés
qu'une simple variable dans l'abstraction. De la même manière que dans le
\roCal, le mécanisme de filtrage est utilisé pour instancier les variables liées 
d'une abstraction et ce calcul peut être représenté dans le \roCal\  de la même
façon que le \laCal.

\EX%------------------------------------------------------------------------
Considérons, par exemple, le $\lambda$-terme $\lambda(PAIR~x~y).x$ qui
sélectionne le premier élément d'une paire et l'application $\lambda(PAIR~x~y).x
~ (PAIR~a~b)$. En filtrant le motif $(PAIR~x~y)$ contre le terme $(PAIR~a~b)$ la
variable $x$ est liée à $a$ et donc l'application est réduite en $a$.

La représentation dans le \roCal\  du premier $\lambda$-terme est la règle
$Pair(x,y) \ra x$, où $Pair$ est le symbole de fonction qui correspond au
symbole $PAIR$, et l'application correspondante $[Pair(x,y) \ra x](Pair(a,b))$
est réduite dans le \roCal\  en $\{\subs{x/a, y/b}x\}$, c'est-à-dire en $\{a\}$.
\FEX%------------------------------------------------------------------------

Si nous introduisons de l'information contextuelle dans les membres gauches des
règles de réécriture nous obtenons les règles de réécriture classiques comme,
par exemple, $f(a) \ra f(b)$ ou $f(x) \ra g(x)$.  Quand nous appliquons une
telle règle de réécriture le filtrage peut échouer et par conséquent,
l'application peut échouer.  Comme nous l'avons déjà précisé dans les sections
précédentes, l'échec d'une règle de réécriture n'est pas une méta-propriété dans
le \roCal\  mais elle est représentée par un ensemble vide (de résultats).

\EX%------------------------------------------------------------------------
Dans la réécriture de termes classique on dit que l'application de la règle de
réécriture $f(a) \ra f(b)$ ne s'applique pas au terme $f(c)$ tandis que dans le
\roCal\  le terme $[f(a) \ra f(b)](f(c))$ est réduit en $\emptyset$.
\FEX%------------------------------------------------------------------------

Puisque la construction des règles de réécriture est faite sans aucune
restriction, une règle de réécriture peut avoir une variable comme membre gauche
comme, par exemple, $x \ra x+1$ et une règle peut introduire de nouvelles
variables comme dans $f(x) \ra g(x,y)$. L'utilisation des variables libres dans
les $\rho$-règles, permet par exemple la construction dynamique de règles
de réécriture classiques.

\EX%------------------------------------------------------------------------
Le $\rho$-terme $[f(x) \ra g(x,y)](f(a))$ représentant l'application de la règle
de réécriture $f(x) \ra g(x,y)$ au terme $f(a)$ est évalué en $\{g(a,y)\}$ en
gardant donc la variable $y$ libre.

En partant de la $\rho$-règle de réécriture précédente nous pouvons construire
une application $[y \ra (f(x) \ra g(x,y))](a)$ qui est évaluée en 
$\{f(x) \ra g(x,a)\}$. Dans ce cas-ci la variable $y$ est libre dans la règle de
réécriture $f(x) \ra g(x,y)$, mais elle est liée dans la règle 
$y \ra (f(x) \ra g(x,y))$.
\FEX%------------------------------------------------------------------------

On peut aussi remarquer que les variables libres dans le membre droit d'une
$\rho$-règle de réécriture nous permettent la ``paramétrisation'' des règles
de réécriture par des ``stratégies'', comme dans le terme 
$y \ra [f(x) \ra [y](x)](f(a))$ où le terme appliqué à $x$ n'est pas connu
dans la règle $f(x) \ra [y](x)$. La réduction de l'application 
$[y \ra [f(x) \ra [y](x)](f(a))](a \ra b)$ revient à la réduction de
l'application $[f(x) \ra [a \ra b](x)](f(a))$ en $\{b\}$.

Quand le filtrage est effectué modulo une théorie équationnelle nous obtenons
des comportements intéressants nous permettant de décrire d'une manière concise
des opérations complexes. Nous rappelons que le \roCalC, le \roCalA\  et le
\roCalAC\  représentent le \roCalT\  avec une théorie de filtrage commutative,
associative et associative-commutative respectivement.

L'utilisation d'une théorie de filtrage associative nous permet, par exemple,
d'exprimer le fait qu'une expression accepte plusieurs parenthésages. 

\EX%------------------------------------------------------------------------
Prenons l'opérateur binaire $\circ$ qui représente la concaténation de deux
listes avec des éléments d'un certain type $Elem$. Nous considérons que tout
objet de type $Elem$ représente une liste contenant seulement cet objet. 

Si nous définissons l'opérateur $\circ$ comme étant
associatif, la règle de réécriture décrivant la décomposition d'une
liste peut être définie dans le \roCalA\  par~: 
$l \circ l' \ra l.$ Quand nous appliquons cette règle à la liste 
$a \circ b \circ c \circ d$ nous obtenons comme résultat le $\rho$-terme 
$\{a,a \circ b,a \circ b \circ c\}$.  Si l'opérateur $\circ$ n'avait pas été
défini comme étant associatif, nous aurions obtenu comme résultat de
l'application précédente un des singletons $\{a\}$ ou $\{a \circ b\}$ ou 
$\{a \circ (b \circ c)\}$ ou encore $\{(a \circ b) \circ c\}$, en fonction du
parenthésage du terme $a \circ b \circ c \circ d$.
\FEX%------------------------------------------------------------------------

Dans l'exemple précédent nous avons considéré tous les parenthésages possibles
d'un terme construit en utilisant un opérateur associatif mais l'ordre des
arguments était le même dans tous les cas. En utilisant une théorie de filtrage
commutative nous pouvons, par exemple, représenter le fait que l'ordre des
arguments d'un terme n'est pas significatif.

\EX%------------------------------------------------------------------------
\label{exCalculComm}
Considérons un opérateur commutatif $\oplus$ et la règle de réécriture 
$x \oplus y \ra x$ qui sélectionne un des éléments d'un t-uplet 
$x \oplus y$. Dans le \roCalC\  l'application $[x \oplus y \ra x](a \oplus b)$
est réduite en l'ensemble $\{a,b\}$ correspondant au choix non-déterministe
entre les deux résultats.

Dans la réécriture modulo classique, le résultat de l'application peut être soit
$a$ soit $b$.
\FEX%------------------------------------------------------------------------

Nous pouvons également utiliser une théorie associative-commutative comme, par
exemple, dans le cas d'un opérateur décrivant la formation des multi-ensembles.

\EX%------------------------------------------------------------------------
Nous considérons de nouveau l'opérateur $\circ$ mais cette fois-ci nous le
définissons comme étant associatif-commutatif.  Nous introduisons la règle de
réécriture $x \circ x \circ L \ra L$ qui élimine les doublons des listes de
type $Elem$. Puisque le filtrage est fait modulo l'associativité-commutativité,
l'application de cette règle à un ensemble élimine les doublons quelles que soient
leurs positions dans la représentation de l'ensemble.

Par exemple, dans le \roCalAC, l'application 
$[x \circ x \circ L \ra L](a \circ b \circ c \circ a \circ d)$ est réduite en
$\{b \circ c \circ d\}$~: la recherche des deux éléments égaux est faite
grâce au filtrage modulo l'associativité-commutativité de l'opérateur $\circ$.
\FEX%------------------------------------------------------------------------

Une autre facilité est obtenue grâce à l'utilisation des ensembles %comme moyen
pour décrire le non-détermi\-nisme. Ceci nous permet d'exprimer facilement
l'application non-déterministe d'un ensemble de règles de réécriture à un terme.

\EX%------------------------------------------------------------------------
Considérons, par exemple, l'opérateur $\otimes$ en tant qu'opérateur
syntaxique. Si nous voulons le même comportement que pour l'opérateur commutatif
$\oplus$ de l'Exemple~\ref{exCalculComm} et sélectionner chaque élément du terme
$x \otimes y$, deux règles de réécriture devraient être appliquées d'une façon
non-déterministe comme dans la réduction suivante~:%\\
%$~~~~$
$$[\{x \otimes y \ra x, x \otimes y \ra y\}](a \otimes b) \lraD{Distrib}
\{[x \otimes y \ra x](a \otimes b),[x \otimes y \ra y](a \otimes b)\} \lraD{Fire}$$
%$~~~~$
$$\{\{a\},\{b\}\} \lraD{Flat} \{a,b\}$$

Nous pouvons exprimer le même comportement en utilisant une règle de réécriture
avec un membre droit non-déterministe et dans ce cas nous obtenons deux
réductions possibles~:
$$[x \otimes y \ra \{x,y\}](a \otimes b) \lraD{Switch_R}
[\{x \otimes y \ra x, x \otimes y \ra y\}](a \otimes b) \lraD{} \{a,b\}$$
ou
$$[x \otimes y \ra \{x,y\}](a \otimes b) \lraD{Fire}
\{\{a,b\}\} \lraD{Flat} \{a,b\}$$
\FEX%------------------------------------------------------------------------

Le non-déterminisme représenté par un ensemble de résultats peut apparaître non
seulement au niveau des règles de réécriture d'une application mais aussi bien
au niveau des arguments de l'application. L'argument d'une application peut
être un $\rho$-terme quelconque et en particulier un terme se réduisant en un
ensemble. Cet ensemble représente le choix non-déterministe parmi ses éléments
et l'application d'une règle de réécriture à l'ensemble est réduite en
l'ensemble d'applications de la règle respective à chaque élément.

\EX%------------------------------------------------------------------------
Pour le terme $[a \ra b](\{a,b\})$ nous obtenons la réduction 
$$[a \ra b](\{a,b\}) \lraD{Batch} \{[a \ra b](a),[a \ra b](b)\} \lraD{Fire} 
\{\{b\},\emptyset\} \lraD{Flat} \{b\}$$

On doit remarquer que dans la réduction du terme $\rho$-terme $[a \ra b](\{a,b\})$
seuls les résultats des applications ne menant pas à un échec figurent dans le
terme final. Nous pouvons ainsi obtenir un résultat déterministe (ensemble avec
un seul élément) même si le terme de départ contient des ensembles ayant plus
d'un élément ou des termes se réduisant d'une façon non-déterministe (en un
ensemble ayant plus d'un élément).
\FEX%------------------------------------------------------------------------

Nous avons insisté dans cette section sur l'application des règles de réécriture
mais des \mbox{$\rho$-termes} plus élaborés peuvent être utilisés dans les
applications comme décrit, par exemple, par les règles d'évaluation
\rname{Congruence}. Dans la Section~\ref{encodRew} nous détaillons l'encodage de
la réécriture (conditionnelle) de termes et dans la Section~\ref{encodageELAN}
nous montrons l'utilisation des $\rho$-termes pour représenter les règles et
stratégies du langage \elan.

%\DontWriteThisInToc  
\subsection*{Conclusion}
%================================================================
%~

Dans ce chapitre nous avons introduit le \roCal\  en définissant ses
composants~: la syntaxe, l'application de substitution, le filtrage, les règles
d'évaluation et la stratégie d'évaluation. Nous avons discuté nos principaux
choix et nous avons présenté des exemples d'utilisation du calcul.

Dans le \roCal\  l'opérateur d'abstraction est la flèche, l'application est une
opération explicite représentée en utilisant l'opérateur $[~](~)$ et le
non-déterminisme est exprimé en utilisant les ensembles de $\rho$-termes.  Nous
employons la substitution d'ordre supérieur qui utilise l'$\alpha$-conversion
afin d'éviter la capture des variables.  La théorie de filtrage est un paramètre
du \roCal\  et même si dans le cas général nous considérons un filtrage d'ordre
supérieur, nous sommes principalement intéressés par les cas décidables. Les
règles d'évaluation décrivent l'application des $\rho$-termes ainsi que le comportement
des ensembles par rapport aux autres opérateurs du \roCal. La stratégie
d'évaluation guide l'application des règles d'évaluation et elle est utilisée
principalement afin d'obtenir certaines propriétés pour le \roCal\  et en
particulier la confluence du calcul.

Nous avons introduit plusieurs instances du \roCal\  obtenues en remplaçant les
paramètres du calcul par certaines valeurs spécifiques. Nous avons présenté la
translation de certains \mbox{$\lambda$-termes} en des $\rho$-termes et la manière dont nous
pouvons décrire en \roCal\  des réductions de la réécriture de termes
classique. Nous avons aussi montré le pouvoir d'expression des instances du
\roCal\  utilisant une théorie de filtrage équationnelle.
Ces travaux ont été présentés aux conférences
ASIAN'98~\cite{CirsteaKirchner-Asian98} et
FroCoS'98~\cite{CirsteaKirchner-LivreFroCoS99}.

%% file: chapter_3.tex
%%%%%%%%%%%%%%%%%%%%%%%%%%%%%%%%%%%%%%%%%%%%%%%%%%%%%%%%%%%
% \TLtopbookmark
\chapter{Sur la confluence du \roCal}
\label{chap.resultats_confluence}
%%%%%%%%%%%%%%%%%%%%%%%%%%%%%%%%%%%%%%%%%%%%%%%%%%%%%%%%%%%%

Dans le Chapitre~\ref{chap.calcul_non_type} nous avons introduit le \roCalT\
général et nous avons mentionné que le calcul n'est pas confluent si les règles
d'évaluation sont guidées par la stratégie $\NONE$ qui n'impose aucune
restriction. Ce résultat négatif est obtenu même si la théorie $T$ de filtrage
est vide.

Nous analysons dans ce chapitre les raisons conduisant à la non-confluence du
\roCal\  et nous proposons plusieurs stratégies confluentes. En partant des
exemples simples de réductions non-confluentes nous déduisons les conditions à
imposer pour l'application des règles d'évaluation afin d'empêcher de telles
réductions non-convergentes. Nous voulons obtenir d'un côté une stratégie simple
à exprimer et aussi facile à implanter, et d'un autre côté une stratégie qui
n'impose aucune restriction sur les réductions de certaines instances
spécifiques du \roCal, comme par exemple la représentation du \laCal\  dans le
\roCal.

Dans un premier temps, nous introduisons une stratégie confluente générique en
décrivant les réductions interdites. Nous définissons ensuite plusieurs
contraintes sur la structure des termes permettant d'éliminer les réductions
non-confluentes et donc, d'établir la confluence du calcul. Les conditions
imposées sur les termes sont relativement simples mais assez restrictives et
dans la dernière section de ce chapitre nous présentons une autre approche où
les conditions sur la structure des termes sont plus compliquées mais moins
restrictives.

\section{Le \roCalE} \label{roCalEmpty}
%================================================================

Pour une instance spécifique du \roCalT, il y a un rapport fort entre les termes
présents dans le membre gauche d'une règle de réécriture et la théorie
$T$. Intuitivement, la théorie $T$ doit être assez puissante pour permettre
suffisamment d'applications de règles de réécriture.

Par exemple, si nous utilisons le filtrage syntaxique, le $\rho$-terme
$[[a \ra c](a) \ra a]([b \ra c](b))$ est réduit par la règle \rname{Fire}
appliquée à la position de tête en l'ensemble vide.  Si nous commençons par
évaluer les deux applications $[a \ra c](a)$ et $[b \ra c](b)$ en $\{c\}$ et si
les ensembles sont ensuite distribués en utilisant les règles d'évaluation du
\roCal, alors nous obtenons comme résultat de l'application initiale le terme
$\{[c \ra a](c)\}$ qui est ensuite réduit en $\{a\}$.

Afin d'obtenir la confluence du calcul nous pouvons demander que la règle
d'évaluation \rname{Fire} soit appliquée sur un terme $[l \ra r](t)$ seulement
si les termes $l$ et $t$ sont des termes du premier ordre, i.e. 
$l,t \in \TFX$. De cette manière nous garantissons que le résultat $\emptyset$ est
la conséquence d'une incompatibilité (du premier ordre) entre les termes $l$ et
$t$ et pas d'une évaluation incomplète d'un des deux termes.

Il semble plus intéressant d'utiliser le filtrage d'ordre supérieur au lieu du
filtrage syntaxique quand les membres gauches des règles contiennent des
abstractions et des applications.  Si nous utilisons une théorie $T$ d'ordre
supérieur, alors le terme $[[a \ra c](a) \ra a]([b \ra c](b))$ peut être réduit
directement par la règle \rname{Fire} dans le terme $\{a\}$.  Mais la complexité du
calcul et donc de l'analyse de la confluence sont fortement influencées par la
complexité de cette théorie de filtrage.

Puisque le \roCalT\  est très général, nous nous limitons dans l'analyse de la
confluence au \roCalE\  que nous définissons comme le \roCal\  où la théorie de
filtrage est limitée au filtrage syntaxique du premier ordre. Nous obtenons la
définition du \roCalE\  comme une instance de la Définition~\ref{roTdef}~:

\DEF%------------------------------------------------------------------------
\label{roEdef} 

Etant donné un ensemble de symboles de fonctions $\FF$, un ensemble de variables
$\XX$, nous appelons \roCalE\  un calcul défini par~:
\begin{itemize}
\item un sous-ensemble $\RTTE$ de l'ensemble de termes $\RTT$ tel que toutes les règles de 
	réécriture soient de la forme $l \ra r$ avec $l \in \TFX$,
\item l'application (d'ordre supérieur) de substitution aux termes,
\item la théorie $\emptyset$ (filtrage syntaxique),
\item les règles d'évaluation  \rname{Fire}, \rname{Congruence},
	\rname{Congruence\_fail}, \rname{Distrib}, \rname{Batch},
	\rname{Switch_L}, \rname{Switch_R}, \rname{OpOnSet}, \rname{Flat},
\item une stratégie d'évaluation $\SS$ qui guide l'application des règles
	d'évaluation.
\end{itemize}

\FDEF%------------------------------------------------------------------------

Nous obtenons donc le calcul $\rhoe=(\RTTE,\emptyset,\SS)$ où la
stratégie $\SS$ n'est pas encore précisée.

Puisque toutes les règles de réécriture du \roCalE\  ont comme membre gauche un
terme du premier ordre, nous pouvons remarquer immédiatement que la règle
d'évaluation \rname{Switch_L} ne sera jamais utilisée. Pour permettre l'extension
facile du \roCalE\  à un calcul avec un ensemble de termes étendu, nous
considérerons la règle d'évaluation \rname{Switch_L} dans les preuves et nous
préciserons explicitement les points où la restriction aux termes de $\RTTE$ est
nécessaire.

Puisque le filtrage syntaxique du premier ordre est unitaire, la méta-règle
\rname{Propagate} présentée dans la Section~\ref{regles_evaluation_application}
donne toujours comme résultat un singleton $\{\sigma r\}$ ou l'ensemble
vide. Par conséquent, la règle d'évaluation $Fire$ peut être remplacée par la
règle~:

\renewcommand{\fleche}{\Longrightarrow}
\begin{ruleset}
%===================================
  \wregle 
  {Fire_{\emptyset}} 
  {[l \ra r](t)} 
  {\{\sigma r\}}
  {\sigma \in \Sl(l \meqqes t) } 
%===================================
\end{ruleset}

Si le filtrage $l \meqqes t$ échoue alors, il n'existe pas de substitution
$\sigma$ et le résultat de l'application de la règle \rname{Fire_{\emptyset}}
est l'ensemble vide. Nous pouvons simplifier encore plus cette règle en
définissant explicitement les cas d'échec et de succès par les deux règles
suivantes~:

\renewcommand{\fleche}{\Longrightarrow}
\begin{ruleset}
%===================================
  \regle {Fire'} 
	{[l \ra r](\sigma l)} 
	{\{\sigma r\}}
%===================================
  \cregle {Fire''} 
	{[l \ra r](t)} 
	{\emptyset}
	{il~n'existe~aucune~substitution~\sigma~telle~que~\sigma l=t}
%===================================
\end{ruleset}

Le cas des théories équationnelles finitaires décidables est plus technique mais
est conceptuellement similaire au cas de la théorie vide. Le cas des théories
avec des problèmes de filtrage infinitaires ou indécidables pourrait être
traitée sans difficultés majeures initiales en utilisant des $\rho$-termes
contraints dans l'esprit de~\cite{KirchnerKirchnerRusinowitch-RIA90}. 

Dans les sections suivantes nous analysons les problèmes liés à la confluence du
\roCalE\  en présentant d'abord des exemples de réductions non-confluentes dans
le cas où les règles d'évaluation sont guidées par la stratégie $\NONE$ et nous
proposons ensuite des stratégies permettant d'obtenir la confluence.

\section{La non-confluence du calcul de réécriture de base}	\label{nonconfRocal}
%================================================================

Il est facile de remarquer que le \roCalE\  n'est pas confluent et nous
fournissons dans cette section des exemples typiques de dérivations
non-confluentes.

Une première raison de la non-confluence est le conflit entre l'utilisation du
filtrage syntaxique et la représentation des résultats des réductions par des
ensembles. Ceci mène, d'une part, à des échecs de filtrage indésirables dus aux
termes qui ne sont pas complètement évalués ou pas encore instanciés. D'autre
part, nous pouvons avoir soit des ensembles ayant plus d'un élément qui peuvent
mener à des résultats indésirables dans un contexte non linéaire, soit des
ensembles vides qui ne sont pas strictement propagés.

Un premier exemple de non-confluence est obtenu si nous réduisons un
(sous-)terme de la forme $[l \ra r](t)$ en filtrant $l$ et $t$ et le filtrage
échoue~:

\EX%------------------------------------------------------------------------
\label{SymbolVariableClashA}

[Variable potentiellement instanciée]
\begin{center}$~$
\xymatrix{ 
&
[x \ra [a \ra b](x)](a)
\ar[dl]_{Fire}^{(externe)} \ar[dr]^{Fire}_{(interne)}
& \\
\{[a \ra b](a)\}
\ar[d]_{Fire} &&
[x \ra \emptyset](a)
\ar[d]^{Switch_R}
\\
\{\{b\}\}
\ar[d]_{Flat} &&
[\emptyset](a)
\ar[d]^{Distrib}
\\
\{b\}
&&
\emptyset
}
\end{center}

Dans l'évaluation du radical $[a \ra b](x)$ le filtrage $a \meqqes x$ échoue,
menant au résultat $\emptyset$. Par contre, dans le radical extérieur, le filtrage
$x \meqqes a$ réussi et le résultat de l'application permet de filtrer
maintenant $a \meqqes a$ qui réussit et a comme résultat la substitution
identité.
\FEX%------------------------------------------------------------------------

Dans l'Exemple~\ref{SymbolVariableClashA} on peut remarquer qu'un terme peut
être réduit en un ensemble vide à cause d'un échec de filtrage impliquant ses
variables liées. Un résultat différent de l'ensemble vide peut être obtenu si
les réductions des sous-termes contenant les variables respectives sont
effectuées après l'instanciation de ces variables.

Une situation similaire est obtenue lorsque la règle d'évaluation \rname{Fire}
engendre un résultat $\emptyset$ par suite d'un échec de filtrage et
l'application d'une autre règle d'évaluation avant la règle \rname{Fire} mène à
un résultat non vide. Dans l'Exemple~\ref{SymbolVariableClashC} nous illustrons
ce comportement en commençant l'évaluation soit par la règle d'évaluation
\rname{Fire} soit par la règle d'évaluation \rname{Batch}.

\EX%------------------------------------------------------------------------
\label{SymbolVariableClashC}

[Terme non réduit]
\begin{center}$~$
\xymatrix{ 
&
[f(x) \ra x](\{f(a)\})
\ar[dl]_{Batch} \ar[dr]^{Fire} 
& 
\\
\{[f(x) \ra x](f(a))\}
\ar[d]_{Fire}
&&
\emptyset
\\
\{\{a\}\}
\ar[d]_{Flat}
&&
\\
\{a\}
&&
}
\end{center}

Le filtrage $f(x) \meqqes \{f(a)\}$ échoue mais le filtrage $f(a) \meqqes f(a)$
réussit et ainsi, nous obtenons deux réductions non-convergentes.
\FEX%------------------------------------------------------------------------

Dans les exemples précédents les réductions non-confluentes d'un terme sont
provoquées par le filtrage qui peut soit échouer, si la première étape de la
réduction est l'application de la règle \rname{Fire}, soit réussir, si une
autre réduction est effectuée avant l'application de la règle
\rname{Fire}. Une situation similaire survient si un échec de filtrage est
obtenu du à un sous-terme non réduit de l'argument de l'application et le
filtrage réussit quand le sous-terme est réduit. Dans ce cas, des exemples de
non-confluence peuvent être facilement trouvés~:

\EX%------------------------------------------------------------------------
\label{SymbolVariableClashB}

[Sous-terme non réduit]
\begin{center}$~$
\xymatrix{ 
&
[a \ra b]([a \ra a](a))
\ar[dl]_{Fire}^{(interne)} \ar[dr]^{Fire}_{(externe)}
& \\
[a \ra b](\{a\})
\ar[d]_{Batch} 
\ar[dr]^{Fire}&&
\emptyset
\\
\{[a \ra b](a)\} 
\ar[d]_{Fire} &
\emptyset&
\\
\{\{b\}\}
\ar[d]_{Flat}
&&
\\
\{b\}
&&
}
\end{center}

Les filtrages $a \meqqes [a \ra a](a)$ et $a \meqqes \{a\}$ échouent mais le
filtrage $a \meqqes a$ réussit et ainsi, nous obtenons une réduction avec un
résultat qui ne peut pas être réduit en $\emptyset$.
\FEX%------------------------------------------------------------------------

Afin d'éviter ce genre de situation nous ne devrions pas permettre la réduction
d'une application de la forme $[l \ra r](t)$ lorsque l'échec du filtrage entre
les termes $l$ et $t$ est causé par les règles de filtrage $SymbolVariableClash$
(Exemple~\ref{SymbolVariableClashA}) ou $SymbolClash$
(Exemple~\ref{SymbolVariableClashC}, \ref{SymbolVariableClashB}) mais
il existe dans $t$ des variables qui ne sont pas instanciées ou il existe
certains sous-termes (stricts ou non) de $t$ qui soient des ensembles ou qui
ne soient pas suffisamment réduits.

Les règles de filtrage $SymbolVariableClash$ et $SymbolClash$ ne seraient jamais
appliquées dans les conditions précédentes si l'ensemble de positions
fonctionnelles du terme $l$ était un sous-ensemble de l'ensemble de positions
fonctionnelles du terme $t$.  Ce n'est pas le cas dans
l'Exemple~\ref{SymbolVariableClashA} où, dans le terme $[a \ra b](x)$, la
position de $a$ est une position fonctionnelle et la position correspondante
dans l'argument de l'application est la position variable de $x$. Dans
l'Exemple~\ref{SymbolVariableClashC} et dans
l'Exemple~\ref{SymbolVariableClashB} une position fonctionnelle dans le membre
gauche de la règle de réécriture correspond à un ensemble et à une application
respectivement dans le terme à filtrer et la condition n'est pas satisfaite.

Par conséquent, nous pourrions considérer que la règle d'évaluation \rname{Fire}
est appliquée seulement quand la condition sur les positions fonctionnelles est
satisfaite. Malheureusement, une telle condition ne suffira pas pour éviter un
échec indésirable du filtrage à cause de l'application de la règle de filtrage
$MergingClash$ quand le terme $l$ n'est pas linéaire, comme montré dans
l'exemple suivant~:

\EX%------------------------------------------------------------------------
\label{MergingClash}

[Non-linéarité à gauche]
\begin{center}$~$
\xymatrix{
&
[f(x,x) \ra x](f(a,[a \ra a](a)))
\ar[dl]_{Fire}^{(interne)}  \ar[dr]^{Fire}_{(externe)}  & 
\\
[f(x,x) \ra x](f(a,\{a\}))
\ar[d]_{OpOnSet}
\ar[dr]_{Fire} &&
\emptyset
\\
[f(x,x) \ra x](\{f(a,a)\})
\ar[d]_{Batch} &
\emptyset &
\\
\{[f(x,x) \ra x](f(a,a))\}
\ar[d]_{Fire} &&
\\
\{\{a\}\}
\ar[d]_{Flat} &&
\\
\{a\} &&
}
\end{center}

L'ensemble de positions fonctionnelles du terme $f(x,x)$ est un sous-ensemble de
l'ensemble de positions fonctionnelles du terme $f(a,[a \ra a](a))$ mais le
filtrage $f(x,x) \meqqes f(a,[a \ra a](a))$ échoue et le terme est réduit en
$\emptyset$. Similairement, le filtrage $f(x,x) \meqqes f(a,\{a\})$ échoue
menant à $\emptyset$.
\FEX%------------------------------------------------------------------------

Dans l'Exemple~\ref{MergingClash} la présence des ensembles dans le terme $t$
peut conduire à un échec de filtrage qui est évité si les ensembles sont
distribués avant l'application de la règle d'évaluation \rname{Fire}. De la même
manière, des variables non-instanciées ou des termes insuffisamment réduits
peuvent engendrer des échecs de filtrage qui peuvent être évités si les variables
sont instanciées et les termes réduits. Ainsi, une condition suffisante pour
assurer le comportement désiré imposerait que l'argument de l'application soit
un terme du premier ordre clos.

Un autre cas pathologique survient quand le terme $t$ de l'application 
$[l \ra r](t)$ contient un ensemble vide. Plus précisément, l'application de la
règle d'évaluation \rname{Fire} peut mener à la non-propagation de l'échec et
ainsi, à la non-confluence du calcul, comme dans l'Exemple~\ref{ns_failure}.

\EX%------------------------------------------------------------------------
\label{ns_failure}

[Propagation non stricte de l'échec]
\begin{center}$~$
\xymatrix{ 
&
[x \ra b](\emptyset)
\ar[dl]_{Fire} \ar[dr]^{Batch}
&
\\
\{b\} 
&&  
\emptyset
}
\end{center}
\FEX%------------------------------------------------------------------------

Un comportement similaire à celui présenté dans l'Exemple~\ref{ns_failure}
est obtenu si l'argument de l'application peut être réduit en un ensemble vide,
comme dans le terme $[x \ra b]([a \ra b](b))$ qui peut être réduit en $\{b\}$ ou
$\emptyset$. De la même façon, si l'argument de l'application peut être
instancié en un ensemble vide nous obtenons deux résultats non-convergents, comme
pour le terme $[y \ra [x \ra b](y)](\emptyset)$ qui mène soit à un ensemble vide
soit à $\{b\}$. 

Une première approche pour résoudre ce problème consiste à ne pas appliquer la
règle d'évaluation \rname{Fire} si l'argument de l'application de la règle de
réécriture est l'ensemble vide ou peut être réduit ou instancié en l'ensemble
vide. On peut remarquer que la condition d'application de la règle
d'évaluation \rname{Fire} est nécessaire seulement dans le cas des règles dites
non-régulières, c'est-à-dire telles que les variables libres du
membre gauche de la règle de réécriture ne soient pas libres dans le membre droit
de la règle. Par exemple, le terme $[x \ra x](\emptyset)$ sera toujours réduit en
$\emptyset$ indépendemment de la règle d'évaluation utilisée.

L'application de règles de réécriture non-régulières à des ensembles vides peut
donc mener à des réductions non-convergentes. Des échecs de filtrage indésirables
et donc des réductions non-confluentes peuvent être obtenus si des ensembles
avec un (ou plusieurs) élément(s) sont présents dans l'argument d'une
application. Les ensembles ayant plus d'un élément peuvent mener également à des
réductions non-convergentes s'ils apparaissent dans un contexte non-linéaire.

En effet, en appliquant une règle de réécriture non-linéaire à droite sur un
terme qui contient des ensembles ayant plus d'un élément, nous obtenons des
réductions non-convergentes comme dans l'exemple suivant~:

\EX%------------------------------------------------------------------------
\label{RLandFailure}

[Non-linéarité à droite]
\begin{center}$~$
\xymatrix{ 
[x \ra f(x,x)](\{a,b\})
\ar[d]_{Fire} \ar[dr]^{Batch}
&
\\
\{f(\{a,b\},\{a,b\})\}
\ar[d]_{OpOnSet}
&
\{[x \ra f(x,x)](a), [x \ra f(x,x)](b)\}
\ar[d]^{Fire}
\\
\{\{f(a,\{a,b\}),f(b,\{a,b\})\}\} 
\ar[d]_{OpOnSet}
&
\{\{f(a,a)\},\{f(b,b)\}\}
\ar[d]^{Flat}
\\
\{\{\{f(a,a),f(a,b)\},\{f(b,a),f(b,b)\}\}\} 
\ar[d]_{Flat}
&
\{f(a,a),f(b,b)\}
\\
\{f(a,a),f(a,b),f(b,a),f(b,b)\}
&
}
\end{center}
\FEX%------------------------------------------------------------------------

On peut remarquer qu'un comportement similaire à celui présenté dans
l'Exemple~\ref{RLandFailure} est obtenu si l'argument de l'application de la
règle de réécriture peut être réduit ou instancié en un ensemble ayant plus d'un
élément, comme pour les termes $[x \ra f(x,x)]([a \ra \{a,b\}](a))$ ou 
$[y \ra [x \ra f(x,x)](y)](\{a,b\})$.

La solution immédiate pour éviter des réductions non-convergentes comme celles
présentées dans l'Exemple~\ref{RLandFailure} consiste à ne pas appliquer la
règle d'évaluation \rname{Fire} si l'argument de l'application est un ensemble
ayant plus d'un élément ou si l'argument peut être réduit ou instancié en un
ensemble ayant plus d'un élément. Cette restriction peut être allégée en imposant
la condition sur l'argument de l'application seulement quand la règle de
réécriture de l'application est non-linéaire à droite.

Pour résumer les problèmes présentés dans les exemples de cette section, la
non-confluence est due à l'application de la règle d'évaluation \rname{Fire} trop tôt
dans une réduction et les situations typiques que nous voulons éviter sont:
\begin{itemize} 
\item l'application prématurée de la règle d'évaluation \rname{Fire} pour une
	application contenant des variables non-instanciées
	(Exemple~\ref{SymbolVariableClashA}),
\item l'application prématurée de la règle d'évaluation \rname{Fire} pour une
	application contenant des termes non-réduits
	(Exemple~\ref{SymbolVariableClashC}, Exemple~\ref{SymbolVariableClashB}),
\item l'application prématurée de la règle d'évaluation \rname{Fire} pour une
	application contenant une règle de réécriture non-linéaire à gauche
	(Exemple~\ref{MergingClash}),
\item l'application de la règle d'évaluation \rname{Fire} pour l'application d'une
	règle de réécriture (non-linéaire à droite) à un terme contenant un
	ensemble ayant plus d'un élément (Exemple~\ref{RLandFailure}),
\item l'application de la règle d'évaluation \rname{Fire} pour l'application d'une
	règle de réécriture (ne conservant pas les variables du membre gauche
	dans le membre droit) à un terme contenant l'ensemble vide
	(Exemple~\ref{ns_failure}).
\end{itemize}

\section{Stratégies confluentes pour le \roCalE}	\label{stratRocal}
%================================================================

Comme nous venons de le voir, le calcul n'est pas confluent si aucune stratégie
n'est employée pour guider l'application des règles d'évaluation.  Mais la
confluence peut être obtenue avec une stratégie appropriée d'évaluation.  En
particulier, cette stratégie devrait imposer que la propagation de l'échec dans
les termes soit stricte et que les variables non-linéaires des termes ne soient
pas instanciées par des ensembles ayant plus d'un élément.

\subsection{Premières stratégies} \label{strat_confluente_strict}
%================================================================

Nous avons vu dans la section précédente que la possibilité d'avoir des
ensembles vides ou ayant plusieurs éléments mène immédiatement à des réductions
non-confluentes impliquant les règles d'évaluation \rname{Fire} et
\rname{Congruence}.  Une première approche envisagée consiste à réduire un
$\rho$-terme en appliquant d'abord toutes les règles manipulant des ensembles
(\rname{Distrib}, \rname{Batch}, \rname{Switch_L}, \rname{Switch_R},
\rname{OpOnSet}, \rname{Flat}) et seulement quand aucune de ces règles ne
s'applique plus, appliquer une des règles \rname{Fire}, \rname{Congruence},
\rname{Congruence\_fail} à des termes ne contenant pas d'ensemble.  

Mais une
application peut être réduite, en utilisant la règle \rname{Fire}, en un ensemble
vide ou ayant plusieurs éléments générant ainsi des éventuelles réductions
non-confluentes et donc, cette stratégie n'est pas suffisante pour assurer la
confluence du calcul. Un autre inconvénient de cette approche est que pour
aucune instance du \roCal\  la stratégie proposée ne se réduit pas à la stratégie
triviale $\NONE$.

Puisque les ensembles (vide ou ayant plus d'un élément) sont la cause principale
de la non-confluence du calcul, une stratégie naturelle consiste à réduire
l'application d'une règle de réécriture en un terme en respectant les étapes
suivantes~: instancier et réduire d'abord l'argument de l'application,
faire monter les symboles d'ensemble à l'extérieur de l'application en les
distribuant dans les termes et seulement lorsqu'aucune des réductions précédentes
n'est possible, appliquer la règle d'évaluation \rname{Fire}. Nous pouvons
exprimer facilement cette stratégie en imposant une condition simple pour
l'application de la règle d'évaluation \rname{Fire}.

\DEF%----------------------------------------------------------
\label{confStratOpStrict} 

Nous appelons \textit{ConfStratStrict} la stratégie qui consiste à appliquer la
règle d'évaluation \rname{Fire} à un radical $[l \ra r](t)$ seulement si le terme
$t$ est un terme clos du premier ordre.
\FDEF%----------------------------------------------------------

La stratégie \textit{ConfStratStrict} est très restrictive et nous
voudrions définir une stratégie plus générale qui devient triviale (c'est-à-dire
n'impose aucune restriction) pour des restrictions du \roCalE\  général à des
calculs plus simples comme le \laCal.

Nous proposons maintenant une stratégie qui émerge des contre-exemples présentés
dans la Section~\ref{nonconfRocal} et qui permet l'application de la règle
d'évaluation \rname{Fire} seulement si un éventuel échec dans le filtrage est
préservé par les $\rho$-réductions et l'argument de l'application ne
peut pas être réduit en un ensemble vide ou ayant plus d'un élément.

\DEF%----------------------------------------------------------
\label{confStratGen} 

Nous appelons \textit{ConfStratGen} la stratégie qui consiste à appliquer la règle
d'évaluation \rname{Fire} à un radical $[l \ra r](t)$ seulement si~:
\begin{itemize}
\item[]
\begin{itemize}
\item $t \in \TF$ est un terme clos du premier ordre
\end{itemize} 
\item[] ou
\item[]
\begin{itemize}
\item le terme $t$ est tel que si le filtrage $l \meqqes t$ échoue alors pour
	tout terme $t'$ obtenu en instanciant ou $\rho$-réduisant $t$ le filtrage 
	$l \meqqes t'$ échoue et
\item le terme $t$ ne peut pas être $\rho$-réduit en un ensemble vide ou ayant
	plus d'un élément.
\end{itemize}
\end{itemize}
\FDEF%----------------------------------------------------------

Si nous considérons une instance du \roCalE\  telle que tous les ensembles soient
des singletons et toutes les applications soient de la forme $[x \ra u](v)$ alors,
toutes les conditions de la Définition~\ref{confStratGen} sont toujours
satisfaites et ainsi, nous pouvons dire que dans ce cas la stratégie
\textit{ConfStratGen} est équivalente à la stratégie $\NONE$, c'est-à-dire
n'impose aucune restriction sur les réductions. On peut remarquer que les termes
de la représentation du \laCal\  pur dans le \roCal\  satisfont la condition
précédente et donc, la stratégie \textit{ConfStratGen} n'impose aucune
restriction sur les réductions de cette instance du \roCal.

Les conditions imposées dans la Définition~\ref{confStratGen} dans le cas où le
terme $t$ n'est pas un terme clos du premier ordre ne sont clairement pas
utilisables dans une implantation du \roCal\  et donc nous devons définir des
stratégies opérationnelles garantissant la confluence du calcul. Dans les
sections suivantes nous allons définir des restrictions structurelles sur les
$\rho$-termes et des stratégies confluentes définies en utilisant ces notions.

\subsection{Termes \matchD, \safe\  et \ready}	\label{notionsPreliminaires}
%================================================================

Dans les exemples présentés dans la Section~\ref{nonconfRocal} nous avons vu que
l'application de la règle d'évaluation \rname{Fire} à un terme peut mener à des
résultats non-confluents si le terme respectif n'est pas suffisamment
réduit. Afin de résoudre ce problème, dans la Définition~\ref{confStratGen} nous
avons imposé une condition générale de \textit{``filtrage cohérent''} pour les
termes $l$ et $t$ mais nous voulons définir des conditions qui peuvent être
implantées facilement et efficacement.

Nous introduisons maintenant une définition plus opérationnelle et plus
restrictive garantissant la \textit{``cohérence''} du filtrage en imposant des
conditions structurelles sur les termes $l$ et $t$ pouvant apparaître dans un
problème de filtrage $l \meqqes t$. Afin d'assurer la préservation d'un éventuel 
échec de filtrage par les $\rho$-réductions, l'échec doit être généré seulement
par des symboles du premier ordre différents dans les positions correspondantes
des deux termes. Cette propriété est toujours vérifiée si les deux termes sont
des termes du premier ordre mais une condition supplémentaire doit être imposée
si le terme $t$ contient des symboles du \roCal.

\DEF%----------------------------------------------------------
\label{subsumeFaible} 
Un $\rho$-terme $l$ {\em \subf} \index{subsume faiblement@\subf}%\Def{\subf} 
un $\rho$-terme $t$ si 
	$$\forall p \in \FPos(l) \cap \PPos(t) \Rightarrow t(p) \in \FF$$
\FDEF%----------------------------------------------------------

Ainsi, un $\rho$-terme $l$ {\em \subf} un $\rho$-terme $t$ si pour toute
position fonctionnelle du terme $l$, soit cette position n'est pas une position
du terme $t$, soit elle est une position fonctionnelle du terme $t$.

\REM%----------------------------------------------------------
\label{positionsFonctionnelles}
Si $l \in \TFX$ \subf\  $t$, alors pour toute position non-fonctionnelle (i.e. la
position d'une variable, d'une application, d'une abstraction ou d'un ensemble)
dans $t$ la position correspondante dans $l$, si elle existe, est une position
variable. Ainsi, si la position de tête de $t$ n'est pas une position
fonctionnelle alors $l$ est une variable.
\FREM%----------------------------------------------------------

On peut aussi remarquer que si un terme $l$ du premier ordre subsume un terme
$t$ alors $l$ \subf\  $t$.

\EX%----------------------------------------------------------
Le terme $f(a,y,c)$ \subf\  le terme $g(b,[x \ra x](c))$ et le terme $f(a)$
\subf\  le terme $g(b,[x \ra x](c))$. Le terme $f(a,y)$ \subf\  le terme 
$g(b,[x \ra x](c))$ mais le terme $f(a)$ ne \subfn\  le terme 
$g([x \ra x](c))$.
\FEX%----------------------------------------------------------

\DEF%------------------------------------------------------------------------
\label{matchDef}
Le couple de $\rho$-termes $(l,t)$ est {\em \matchDs} si $l \in \TFX$ est un terme du
premier ordre et~:
\begin{itemize}
\item le terme $t \in \TF$ est un terme clos du premier ordre ou,
\item le terme $l \in \TFX$ est linéaire et le terme $l$ \subf\  le terme $t$.
\end{itemize}
Par abus de langage nous disons que les $\rho$-termes $l$ et $t$ sont
{\em \matchD}\index{rho-prefiltrable@\matchD}\index{terme!rho-prefiltrable@\matchD}. 
%\Def{\matchD}. 
\FDEF%------------------------------------------------------------------------

Il est clair que pour tout terme $l$ de la forme $f(l_1,\ldots,l_n)$ et tout
terme $t$ de la forme $t=g(t_1,\ldots,t_m)$ tels que $l,t$ soient \matchD, les
termes $l_i,t_i$ sont \matchD\  pour tout $i=1,\ldots,min(m,n)$. 

\REM%----------------------------------------------------------
\label{echecMatch}
Si les termes $l$ et $t$ sont \matchD, conformément à la
Remarque~\ref{positionsFonctionnelles} le filtrage $l \meqqes t$ ne peut pas
échouer à cause de l'application des règles de filtrage $SymbolClash$ ou
$SymbolVariableClash$ (Figure~\ref{SyntMatch}) pour un symbole non-fonctionnel
de $t$. Puisque $l$ est linéaire le filtrage $l \meqqes t$ ne peut pas échouer à
cause de l'application de la règle de filtrage $MergingClash$. Ainsi, le
filtrage $l \meqqes t$ peut échouer seulement à cause de l'application de la
règle de filtrage $SymbolClash$ et donc, du à des symboles fonctionnels
différents à la même position des termes $l$ et $t$.
\FREM%----------------------------------------------------------

\LEM%------------------------------------------------------------------------
Etant donnés les termes \matchD\  $l,t$. Si le filtrage $l \meqqes t$ échoue
alors pour tout terme $t'$ obtenu en $\rho$-réduisant ou instanciant le terme
$t$, le filtrage $l \meqqes t'$ échoue.
\FLEM%------------------------------------------------------------------------

\proof{
Si le terme $t \in \TF$ est un terme clos du premier ordre alors $t'=t$ et le
lemme est clairement vrai. Pour le cas où $t \not \in \TF$ nous procédons par
induction sur la structure du \mbox{$\rho$-terme} $t$.  Dans ce cas,
conformément à la Remarque~\ref{echecMatch}, le filtrage $l \meqqes t$ peut
échouer seulement à cause de deux symboles de fonctions différents à la même
position des deux termes.

Si $t=f(t_1,\ldots,t_m)$ et $l=g(l_1,\ldots,l_n)$ avec $f \neq g$ alors le terme 
$t'$ ne peut être que de la même forme que $t$ ou un ensemble. Dans les deux cas 
le filtrage $l \meqqes t'$ échoue.
Si $t=f(t_1,\ldots,t_m)$ et $l=f(l_1,\ldots,l_m)$ alors $l_i,t_i$ sont \matchD\
et $t=f(t_1',\ldots,t_m')$ avec $t_i'$ obtenus en $\rho$-réduisant ou
instanciant les termes $t_i$. Puisque $l \meqqes t$ échoue, un des filtrages
$l_i \meqqes t_i$ échoue et par induction le filtrage $l_i \meqqes t_i'$
correspondant échoue et donc $l \meqqes t'$ échoue.
}

Un premier pas pour obtenir un calcul confluent est la réduction du terme
$[l \ra r](t)$ en filtrant $l$ et $t$ seulement si $l$ et $t$ sont
\matchD. Nous pouvons remarquer dans l'Exemple~\ref{RLandFailure} et dans
l'Exemple~\ref{ns_failure} que cette condition ne suffit pas pour établir la
confluence du calcul et nous devons ajouter certaines conditions (structurelles)
afin d'éviter les réductions non-confluentes.

Comme nous l'avons déjà remarqué dans la Section~\ref{nonconfRocal}, les
contraintes sur la réduction de l'application des règles de réécriture doivent
garantir qu'une règle de réécriture (non-régulière) n'est pas appliquée à (un
terme réductible en) un ensemble vide et qu'une règle de réécriture (non-linéaire
à droite) n'est pas appliquée à (un terme réductible en) un ensemble ayant plus
d'un élément. Nous pouvons être encore plus restrictifs et ne pas permettre la
réduction d'une application $[l \ra r](t)$ en utilisant la règle d'évaluation
\rname{Fire} si le terme $t$ est réductible en un ensemble vide ou à un ensemble
ayant plus d'un élément.

\DEF%------------------------------------------------------------------------
\label{safeDef}
Nous disons que le $\rho$-terme $t$ est {\em \safes}
\index{rho-safe@\safes}\index{terme!rho-safe@\safes} %\Def{\safes} 
si les conditions suivantes sont satisfaites~:
\begin{itemize}
\item le terme $t$ ne contient aucun ensemble ayant plus d'un élément et
	aucun ensemble vide et,
\item le terme $t$ ne contient pas de sous-terme de la forme $[u](v)$ où $u$
	n'est pas une abstraction 
	et,
\item pour tout sous-terme $[u \ra w](v)$ de $t$, $u$ subsume $v$.
\end{itemize}

Si tous les termes du codomaine d'une substitution $\sigma$ sont \safe, nous
disons que $\sigma$ est \safes.
\FDEF%------------------------------------------------------------------------

Intuitivement, la première condition permet seulement des ensembles avec un
élément dans le terme $t$ et les deux dernières interdisent la
présence dans $t$ des termes pouvant se réduire en $\emptyset$.

\LEM%------------------------------------------------------------------------
Etant donné le terme \safes\  $t$. Alors, le terme $t$ ne peut pas être
$\rho$-réduit en un ensemble vide ou ayant plus d'un élément.
\FLEM%------------------------------------------------------------------------

\proof{
Le terme $t$ ne contient aucun ensemble ayant plus d'un élément et aucun
ensemble vide et donc, $t$ peut être $\rho$-réduit en tel ensemble seulement en
utilisant les règles d'évaluation \rname{Fire} ou \rname{Congruence\_fail}.
Mais toute application de $t$ est de la forme $[u \ra w](v)$ ou $u$ subsume $v$
et donc la règle \rname{Fire} ne peut pas mener à un échec et la règle
\rname{Congruence\_fail} ne peut pas être appliquée. Puisque le filtrage
est syntaxique et donc unitaire dans le \roCalE, la règle \rname{Fire} ne peut
pas mener à un ensemble ayant plus d'un élément.  
En plus, la forme des applications est préservée par la $\rho$-réduction et
donc, la propriété est vérifiée.
}

\DEF%------------------------------------------------------------------------
\label{readyDef}
Le couple de $\rho$-termes $(l,t)$ est {\em \readys} si les conditions
suivantes sont satisfaites~:
\begin{itemize}
\item les termes $l,t$ sont \matchD\  et
\item le terme $t$ est \safes.
\end{itemize}
Par abus de langage nous disons que les $\rho$-termes $l$ et $t$ sont
{\em \ready}\index{rho-calculables@\ready}\index{terme!rho-calculables@\ready}.
%\Def{\ready}.
\FDEF%------------------------------------------------------------------------

Ainsi, la première condition assure la préservation de l'échec de filtrage par
rapport aux \mbox{$\rho$-réductions} tandis que la deuxième condition permet seulement
des termes qui ne peuvent pas être $\rho$-réduits en un ensemble vide ou ayant plus
d'un élément.

Suivant la définition des termes \ready\  nous déduisons immédiatement des
propriétés pour les substitutions obtenues comme résultat du filtrage impliquant
de tels termes.

\PROP%------------------------------------------------------------------------
Etant donnés les termes \ready\  $l,t$, si $\sigma l = t$, alors $\sigma$ est \safes.
\FPROP%------------------------------------------------------------------------

\PROP%------------------------------------------------------------------------
Etant donnés des termes $l,t$ \ready\  et une substitution $\sigma$ \safes,
les termes $l,\sigma t$ sont \ready.
\FPROP%------------------------------------------------------------------------

\subsection{Une stratégie opérationnelle} \label{strat_confluente}
%================================================================

Les notions de termes \matchD\  et \safe\  nous permettent de faire une
distinction claire entre les problèmes menant à la non-confluence du \roCalE\
sans stratégie~: d'une part, le conflit entre l'utilisation du filtrage
syntaxique et les termes d'ordre supérieur qui peuvent intervenir dans les
problèmes de filtrage et d'autre part, le traitement du non-déterminisme
représenté par des ensembles vides ou ayant plus d'un élément.

Nous introduisons une stratégie appelée \textit{ConfStrat} qui consiste à
appliquer la règle d'évaluation \rname{Fire} à un terme de la forme 
$[l \ra r](t)$ seulement quand les termes $l,t$ sont \ready.  Cette stratégie
peut être vue comme une version opérationnelle de la stratégie
\textit{ConfStratGen} introduite dans la Section~\ref{strat_confluente_strict}
et afin de mettre en évidence les similitudes avec cette stratégie et avec les
stratégies définies plus tard dans la Section~\ref{stratOpConf} nous explicitons les
notions de termes \ready, \matchD\  et \safe~:

\DEF%----------------------------------------------------------
\label{confStratOp} 

Nous appelons \textit{ConfStrat} la stratégie qui consiste à appliquer la règle
d'évaluation \rname{Fire} à un radical $[l \ra r](t)$ seulement si~:
\begin{itemize}
\item[]
\begin{itemize}
\item $t \in \TF$ est un terme clos du premier ordre
\end{itemize} 
\item[] ou
\item[]
\begin{itemize}
\item le terme $l \in \TFX$ est linéaire et le terme $l$ \subf\  le terme $t$ et,
\item le terme $t$ ne contient aucun ensemble ayant plus d'un élément et
	aucun ensemble vide et,
\item pour tout sous-terme $[u \ra w](v)$ de $t$, $u$ subsume $v$ et,
\item le terme $t$ ne contient pas de sous-terme de la forme $[u](v)$ où $u$
	n'est pas une abstraction.
\end{itemize}
 
\end{itemize}
\FDEF%----------------------------------------------------------

On doit remarquer que les conditions imposées par la stratégie
\textit{ConfStrat} sont décidables même dans le cas où le terme $t$ n'est pas un
terme clos du premier ordre. On peut clairement décider si un terme est de la
forme $[u](v)$ ou $[u \ra w](v)$ ainsi que le nombre d'éléments d'un
ensemble fini. La condition que $l$ \subf\  $t$ est simplement une condition sur les
symboles aux mêmes positions des deux termes et puisque le filtrage est
syntaxique alors la condition de subsomption est aussi décidable. Par
conséquent, toutes les conditions utilisées dans la stratégie
\textit{ConfStrat} sont décidables.

L'interdiction d'avoir des sous-termes de $t$ de la forme $[u](v)$ si $u$ n'est
pas une règle de réécriture est imposée afin d'empêcher des réductions en un
ensemble vide en utilisant la règle d'évaluation \rname{Congruence\_fail}. Si on
considère une version du \roCalE\  sans les règles d'évaluation
\rname{Congruence} alors cette dernière condition n'est plus nécessaire dans la
stratégie \textit{ConfStrat}.  Dans ce cas tous les termes de la représentation
du \laCal\  %(pur ou appliqué) 
dans le \roCal\  satisfont les conditions de la
stratégie \textit{ConfStrat} et dans ce cas cette stratégie est équivalente à la
stratégie $\NONE$.

\TH%----------------------------------------------------------
Si la stratégie d'évaluation \textit{ConfStrat} est utilisée alors, le \roCalE\
est confluent.
\FTH%----------------------------------------------------------

\proof{
Nous donnons dans la Section~\ref{confluence_modulo} la preuve de la
confluence pour le \roCalE\  avec la règle d'évaluation \rname{Fire} transformée
dans une règle conditionnelle. Les conditions de cette règle sont exactement les
conditions de la Définition~\ref{confStratOp} et ainsi, les réductions dans le
calcul utilisant la règle \rname{Fire} conditionnelle et dans le \roCalE\  avec
les règles d'évaluation guidées par la stratégie \textit{ConfStrat} sont
identiques.
}

Dans le cas des calculs intégrant des réductions modulo une théorie équationelle
(par exemple associativité et commutativité), comme exemplifié dans la
Section~\ref{utilisation_rho}, la preuve de la confluence est plus compliquée et
dépend fortement des propriétés (décidabilité, ensemble fini de solutions, etc.)
de la théorie de filtrage utilisée.

\subsection{Les relations induites par les règles d'évaluation} \label{relationsInduites}
%================================================================

Chaque fois qu'un $\rho$-terme est réduit en utilisant les règles d'évaluation
\rname{Fire}, \rname{Congruence} et \rname{Congruence\_fail} du \roCalE, un
ensemble est produit.  Ces règles d'évaluation sont celles qui décrivent
l'application d'une règle de réécriture à la position de tête ou plus
profondément dans un terme.  L'ensemble obtenu en utilisant une de ces trois
règles d'évaluation peut déclencher l'application des autres règles d'évaluation
du calcul.  Les règles d'évaluation traitant la propagation des ensembles
calculent \textit{``une forme normale d'ensemble''} pour les $\rho$-termes en poussant vers
l'extérieur les symboles d'ensemble et en aplatissant les ensembles.
Par exemple, l'application d'un ensemble à un
$\rho$-terme est évaluée en l'ensemble d'applications de chacun des éléments de
l'ensemble au $\rho$-terme respectif.

Par conséquent, nous pouvons considérer que l'ensemble des règles d'évaluation
du \roCalT\  est la réunion d'un ensemble de règles de \textit{déduction}
(\rname{Fire}, \rname{Congruence}, \rname{Congruence\_fail}) et d'un ensemble de
règles de \textit{calcul} (\rname{Distrib}, \rname{Batch}, \rname{Switch_L},
\rname{Switch_R}, \rname{OpOnSet}, \rname{Flat}) et que l'évaluation se comporte
comme dans la déduction modulo~\cite{DHK-ENAR-98}.  Cette approche nous permet
de considérer les règles de calcul comme permettant de décrire une congruence
modulo sur laquelle les règles de déduction sont appliquées.

Alternativement, nous pouvons considérer la relation habituelle induite par les
règles d'évaluation du calcul.

Dans cette section nous définissons une relation induite par la règle
\rname{Fire} et les règles \rname{Congruence}, appelée \rname{FireCong}, et une
deuxième relation induite par les règles \rname{Distrib}, \rname{Batch},
\rname{Switch_L}, \rname{Switch_R}, \rname{OpOnSet}, \rname{Flat} appelée
\rname{Set}. Nous analysons les propriétés des deux relations et des relations
obtenues en les composant.

A partir des règles d'évaluation du \roCalE\  définissant la propagation des
ensembles sur les $\rho$-opérateurs et de la règle d'évaluation \rname{Flat} qui
aplatit les ensembles et élimine les symboles d'ensemble (redondants) nous
définissons la relation \rname{Set}.

\DEF%------------------------------------------------------------------------
\label{transCongRel}
Nous considérons la \textit{relation} sur $\RTTE$ appelée \rname{Set}
induite par les règles d'évaluation \rname{Distrib}, \rname{Batch},
\rname{Switch_L}, \rname{Switch_R}, \rname{OpOnSet} et \rname{Flat}.

Les relations suivantes sont induites par la relation \rname{Set}~: 
\begin{itemize}
\item[] $\congr$ est la fermeture compatible de la relation \rname{Set},
\item[] $\congTR$ est la fermeture réflexive, transitive de $\congr$
	(la réduction engendrée par \rname{Set}),
\item[] $\congE$ est la relation d'équivalence engendrée par $\congTR$.
\end{itemize}

\FDEF%------------------------------------------------------------------------

L'application de la règle d'évaluation \rname{Fire} est guidée par une stratégie qui tient
compte des conditions présentées dans la section précédente et qui peut
être exprimée explicitement en transformant la règle \rname{Fire} dans une
règle conditionnelle~:

\renewcommand{\fleche}{\Longrightarrow}
\begin{ruleset}
%===================================
  \cwregle 
  {Fire_c} 
  {[l \ra r](t)} 
  {\{\sigma r\}}
  {l,t~sont~\ready}
  {\sigma \in \Sl(l \meqqes t)}
%===================================
\end{ruleset}

\DEF%------------------------------------------------------------------------
\label{transFire}
Nous considérons la \textit{relation} appelée \rname{FireCong} sur $\RTTE$
induite par la règle d'évaluation \rname{Fire_c} et les règles d'évaluation
\rname{Congruence} et \rname{Congruence\_fail}.

Nous considérons les relations suivantes induites par les relations \rname{FireCong}
et \rname{Set} (Définition~\ref{transCongRel})~:
\begin{itemize}
\item[] $\roZ$ est la fermeture compatible de la relation \rname{FireCong},

\item[] $\roTR$ est la fermeture réflexive, transitive de $\roZ$,

\item[] $\roZE$ est la relation $\roZ$ modulo la relation $\congE$
	définie de façon standard (\cite{AhoSethiUllman72})~: étant donnés deux
	$\rho$-termes $u,v$ nous avons $\RoZE{u}{v}$ ssi il existe deux
	$\rho$-termes $u',v'$ tels que $\CongE{u}{u'}$, $\RoZ{u'}{v'}$ et
	$\CongE{v}{v'}$,

\item[] $\roZETR$ est la fermeture réflexive, transitive de $\roZE$.
\end{itemize}
 
\FDEF%------------------------------------------------------------------------

La relation $\roZ$ et toutes les relations induites par cette relation sont
définies sur l'ensemble de termes $\RTTE$ ne contenant que des règles de
réécriture ayant un terme $l \in \TFX$ du premier ordre comme membre
gauche. Pour permettre l'extension facile du \roCalE\  à un calcul avec un
ensemble de termes étendu, nous considérons, dans les preuves, n'importe quelle
forme de règle de réécriture et nous indiquons les situations où la restriction
à des termes de $\RTTE$ est nécessaire.

\subsection{Les propriétés des relations \rname{Set}}
%================================================================

Dans cette section nous montrerons que la relation $\congr$ est confluente et
terminante. Nous donnons d'abord une preuve de la terminaison et ensuite nous
obtenons la confluence comme une conséquence de la confluence locale.

\LEM%--------------------------------------------------------------------------
\label{CongRterm} 
La relation $\congr$ termine.
\FLEM%--------------------------------------------------------------------------

\proof{

Nous utilisons les interprétations polynômiales suivantes~:\\

$~~~~~~~~$
$P(\{u_1,\ldots,u_n\})$ $=$ $\Sigma_{i=1}^n P(u_i) + 2$~~~
($P(\emptyset)$ $=$ $2$)

$~~~~~~~~$
$P(f(u_1,\ldots,u_n))$ $=$ $\Pi_{i=1}^n P(u_i)$

$~~~~~~~~$
$P(u \ra v)$ $=$ $P([u](v))$ $=$ $P(u) \times P(v)$\\

Nous utilisons l'ordre standard sur les naturels. Puisque l'addition et la
multiplication sont croissantes sur les naturels, la condition de monotonicité
$a > b$ implique $P(a) > P(b)$ est clairement satisfaite. Nous montrons que pour
tous termes $t,t'$ tels que $\Cong{t}{t'}$, l'image de $t$ est strictement
supérieure à celle de $t'$ pour tout remplacement des interprétations des
variables de $t,t'$ avec des naturels supérieurs à $2$
(i.e. $P(u),P(u_i),P(v),P(v_i) > 2$).

Les inégalités correspondant aux règles \rname{Distrib}, \rname{Batch},
\rname{Switch_L}, \rname{Switch_R}, \rname{OpOnSet} et \rname{Flat} sont
présentées ci-dessous~:

\begin{itemize}
\item les inégalités pour les règles \rname{Distrib} et \rname{Batch} sont similaires
	et nous présentons uniquement les interprétations pour la règle
	\rname{Batch}~:

$~~~~~~~~$
$P([u](\{v_1,\ldots,v_m\}))$ $=$

$~~~~~~~~$
$(P(v_1) + \ldots + P(v_m) + 2) \times P(u)$ $=$

$~~~~~~~~$
$((P(v_1) + \ldots + P(v_m)) \times P(u) + 2 \times P(u)$ $>$

$~~~~~~~~~~~~~$
$((P(v_1) + \ldots + P(v_m)) \times P(u) + 2$

$~~~~~~~~~~~~~$
$=$ $P(v_1) \times P(u) + \ldots + P(v_m) \times P(u) + 2$

$~~~~~~~~~~~~~$
$=$ $P(\{[u](v_1),\ldots, [u](v_m)\})$

\item les inégalités pour les règles \rname{Switch_L} et \rname{Switch_R} sont similaires
	et nous présentons seulement les interprétations pour la règle
	\rname{Switch_R}~:

$~~~~~~~~$
$P(u \ra \{v_1,\ldots,v_m\})$ $=$

$~~~~~~~~$
$(P(v_1) + \ldots + P(v_m) + 2) \times P(u)$ $=$

$~~~~~~~~$
$(P(v_1) + \ldots + P(v_m)) \times P(u) + 2 \times P(u)$ $>$

$~~~~~~~~~~~~~$
$(P(v_1) + \ldots + P(v_m)) \times P(u) + 2$

$~~~~~~~~~~~~~$
$=$ $P(v_1) \times P(u) + \ldots + P(v_m) \times P(u) + 2$

$~~~~~~~~~~~~~$
$=$ $P(\{u \ra v_1,\ldots, u \ra v_m\})$

\item pour la règle \rname{OpOnSet} nous obtenons~:

$~~~~~~~~$
$P(f(u_1,\ldots,\{v_1,\ldots,v_n\},\ldots,u_m)) =$

$~~~~~~~~$
$P(u_1) \times \ldots \times (\Sigma_{i=1}^n P(v_i) + 2) \times \ldots \times P(u_m)$ $=$

$~~~~~~~~$
$P(u_1) \times \ldots \times \Sigma_{i=1}^n P(v_i) \times \ldots \times P(u_m) + 2 \times P(u_1) \times \ldots \times P(u_m)$ $>$

$~~~~~~~~~~~~~$
$P(u_1) \times \ldots \times \Sigma_{i=1}^n P(v_i) \times \ldots \times P(u_m) + 2$

$~~~~~~~~~~~~~$
$=$ $\Sigma_{i=1}^n (P(u_1) \times \ldots \times P(v_i) \times \ldots \times P(u_m)) + 2$

$~~~~~~~~~~~~~$
$=$ $P(\{f(u_1,\ldots,v_1,\ldots,u_m),\ldots,f(u_1,\ldots,v_n,\ldots,u_m)\})$

\item pour la règle \rname{Flat} nous avons~:

$~~~~~~~~$
$P(\{u_1,\ldots,\{v_1,\ldots,v_n\},\ldots,u_m\})$ $=$

$~~~~~~~~$
$P(u_1) + \ldots + (\Sigma_{i=1}^n P(v_i) + 2) + \ldots + P(u_m) + 2$ $=$

$~~~~~~~~$
$P(u_1) + \ldots + \Sigma_{i=1}^n P(v_i) + \ldots + P(u_m) +4$ $>$

$~~~~~~~~~~~~~$
$P(u_1) + \ldots + \Sigma_{i=1}^n P(v_i) + \ldots + P(u_m) +2$

$~~~~~~~~~~~~~$
$=$ $P(\{u_1,\ldots,v_1,\ldots,v_n,\ldots,u_m\})$
\end{itemize}
}

\LEM%--------------------------------------------------------------------------
\label{CongRconf}
La relation $\congr$ est localement confluente.
\FLEM%--------------------------------------------------------------------------

\proof{

Nous analysons les paires critiques engendrées par les règles d'évaluation.

La règle \rname{Flat} a une paire critique avec elle-même~:
\begin{center}$~$
\xymatrix{
\{ \{u_1,\ldots,u_n\},\{v_1,\ldots,v_m\}, \ldots, t \}
\ar[d]_{Flat} \ar[dr]^-{Flat} & \\
\{u_1,\ldots,u_n,\{v_1,\ldots,v_m\}, \ldots, t \}
\ar@{.{>}}[dr]_{Flat} &
\{ \{u_1,\ldots,u_n\},v_1,\ldots,v_m, \ldots, t \}
\ar@{.{>}}[d]^-{Flat} \\
& \{ u_1,\ldots,u_n,v_1,\ldots,v_m, \ldots, t \} }
\end{center}

Nous procédons de la même façon pour la règle \rname{OpOnSet}~:
\begin{center}$~$
\xymatrix{ 
f(\{u_1,\ldots,u_n\},\ldots,\{v_1,\ldots,v_m\})
\ar[d]_{OpOnSet} \ar[dr]^-{OpOnSet} & \\
*\txt{$\{f(u_1,\ldots,\{v_1,\ldots,v_m\}),\ldots,$\\
$f(u_n,\ldots,\{v_1,\ldots,v_m\})\}$}
\ar@{.{>}}[d]_{OpOnSet} &
*\txt{$\{f(\{u_1,\ldots,u_n\},\ldots,v_1),\ldots,$\\
$f(\{u_1,\ldots,u_n\},\ldots,v_m)\}$}
\ar@{.{>}}[d]^-{OpOnSet} \\
*\txt{$\{\{f(u_1,\ldots,v_1),\ldots,f(u_1,\ldots,v_m)\},\ldots,$\\
$\{f(u_n,\ldots,v_1),\ldots,f(u_n,\ldots,v_m)\}\}$}
\ar@{.{>}}[dr]_{Flat^*} &
*\txt{$\{\{f(u_1,\ldots,v_1),\ldots,f(u_n,\ldots,v_1)\},\ldots,$\\
$\{f(u_1,\ldots,v_m),\ldots,f(u_n,\ldots,v_m)\}\}$}
\ar@{.{>}}[d]^-{Flat^*} \\
&
*\txt{$\{f(u_1,\ldots,v_1),\ldots,f(u_1,\ldots,v_m),\ldots,$\\
$f(u_n,\ldots,v_1),\ldots,f(u_n,\ldots,v_m)\}$}
}
\end{center}

Les règles \rname{OpOnSet} et \rname{Flat} mènent à une paire critique
convergente~:
\begin{center}$~$
\xymatrix{ 
f(\{\{u_1,\ldots,u_n\},\ldots,v_m \})
\ar[d]_{Flat} \ar[dr]^-{OpOnSet} & \\
f(\{u_1,\ldots,u_n,\ldots,v_m \})
\ar@{.{>}}[d]_{OpOnSet} &
\{f(\{u_1,\ldots,u_n\}),\ldots,f(v_m) \}
\ar@{.{>}}[d]^-{OpOnSet} \\
\{f(u_1),\ldots,f(u_n),\ldots,f(v_m) \}
&
\{\{f(u_1),\ldots,f(u_n)\},\ldots,f(v_m) \}
\ar@{.{>}}[l]^-{Flat} \\
}
\end{center}

Les diagrammes pour les paires critiques de \rname{Switch_L} et \rname{Switch_R} d'une
part et \rname{Flat} et \rname{Switch_R} d'autre part sont présentés ci-dessous. Nous
pouvons montrer de la même manière la convergence de la paire critique due
à \rname{Flat} et \rname{Switch_L}.  
\begin{center}$~$
\xymatrix{
\{u_1,\ldots,u_n\} \ra \{v_1, \ldots, v_m\}
\ar[d]_{Switch_L} \ar[dr]^-{Switch_R} & \\
*\txt{$\{u_1 \ra \{v_1, \ldots, v_m\},\ldots,$\\
$u_n \ra \{v_1, \ldots, v_m\}\}$}
\ar@{.{>}}[d]_{Switch_R} &
*\txt{$\{\{u_1, \ldots, u_n\} \ra v_1,\ldots,$\\
$\{u_1, \ldots, u_n\} \ra v_m\}$}
\ar@{.{>}}[d]^-{Switch_L} \\
*\txt{$\{\{u_1 \ra v_1, \ldots,u_1 \ra v_m\},\ldots,$\\
$\{u_n \ra v_1, \ldots,u_n \ra v_m\}\}$}
\ar@{.{>}}[dr]_{Flat^*} &
*\txt{$\{\{u_1 \ra v_1, \ldots, u_n \ra v_1\},\ldots,$\\
$\{u_1 \ra v_m, \ldots,u_n \ra v_m\}\}$}
\ar@{.{>}}[d]^-{Flat^*} \\
& \{u_1 \ra v_1, \ldots, u_n \ra v_1,\ldots,u_n \ra v_m\} 
}
\end{center}

\begin{center}$~$
\xymatrix{ 
u \ra \{\{v_1,\ldots,v_n\},\ldots,w_m \}
\ar[d]_{Flat} \ar[dr]^-{Switch_R} & \\
u \ra \{v_1,\ldots,v_n,\ldots,w_m \}
\ar@{.{>}}[d]_{Switch_R} &
\{u \ra \{v_1,\ldots,v_n\},\ldots,u \ra w_m\}
\ar@{.{>}}[d]^-{Switch_R} \\
\{u \ra v_1,\ldots,u \ra v_n,\ldots,u \ra w_m\}
&
\{\{u \ra v_1,\ldots,u \ra v_n\},\ldots,u \ra w_m\}
\ar@{.{>}}[l]^-{Flat} \\
}
\end{center}

Les paires critiques pour les règles \rname{Distrib}, \rname{Batch} et
\rname{Flat} sont traitées comme dans les cas précédents~:
\begin{center}$~$
\xymatrix{
[\{u_1,\ldots,u_n\}](\{v_1, \ldots, v_m\})
\ar[d]_{Distrib} \ar[dr]^-{Batch} & \\
*\txt{$\{[u_1](\{v_1, \ldots, v_m\}),\ldots,$\\
$[u_n](\{v_1, \ldots, v_m\})\}$}
\ar@{.{>}}[d]_{Batch} &
*\txt{$\{[\{u_1, \ldots, u_n\}](v_1),\ldots,$\\
$[\{u_1, \ldots, u_n\}](v_m)\}$}
\ar@{.{>}}[d]^-{Distrib} \\
*\txt{$\{\{[u_1](v_1), \ldots, [u_1](v_m)\},\ldots,$\\
$\{[u_n](v_1), \ldots,[u_n](v_m)\}\}$}
\ar@{.{>}}[dr]_{Flat^*} &
*\txt{$\{\{[u_1](v_1), \ldots, [u_n](v_1)\},\ldots,$\\
$\{[u_1](v_m), \ldots,[u_n](v_m)\}\}$}
\ar@{.{>}}[d]^-{Flat^*} \\
& \{[u_1](v_1), \ldots, [u_n](v_1),\ldots,[u_n](v_m)\} 
}
\end{center}

\begin{center}$~$
\xymatrix{ 
[u](\{\{v_1,\ldots,v_n\},\ldots,w_m \})
\ar[d]_{Flat} \ar[dr]^-{Batch} & \\
[u](\{v_1,\ldots,v_n,\ldots,w_m \})
\ar@{.{>}}[d]_{Batch} &
\{[u](\{v_1,\ldots,v_n\}),\ldots,[u](w_m)\}
\ar@{.{>}}[d]^-{Batch} \\
\{[u](v_1),\ldots,[u](v_n),\ldots,[u](w_m)\}
&
\{\{[u](v_1),\ldots,[u](v_n)\},\ldots,[u](w_m)\}
\ar@{.{>}}[l]^-{Flat}
}
\end{center}

}

\LEM%--------------------------------------------------------------------------
\label{CongRconfluent}
La relation $\congr$ est confluente.
\FLEM%--------------------------------------------------------------------------

\proof{ 
Puisque $\congr$ est terminante (Lemme~\ref{CongRterm}) et localement confluente 
(Lemme~\ref{CongRconf}) alors, $\congr$ est confluente.
}

\COR%--------------------------------------------------------------------------
\label{CongTRconf}
La relation $\congTR$ est confluente.
\FCOR%--------------------------------------------------------------------------

\LEM%--------------------------------------------------------------------------
\label{CongContext} 
Les relations $\congr$, $\congTR$, $\congE$ sont compatibles.
\FLEM%--------------------------------------------------------------------------

\proof{
Par la construction de la relation $\congr$.  Pour les deux autres relations la
preuve est faite par induction sur la génération de ces relations.
%Bar/p52
}

\subsection{Les propriétés des relations \rname{FireCong}}
%================================================================

La relation $\roZ$ est trivialement non-terminante comme montré par le contre
exemple classique $[\omega_{\rho}](\omega_{\rho})$ où $\omega_{\rho} =x \ra [x](x)$. 
Ce terme se réduit en une étape en lui-même et on obtient ainsi une
chaîne de réduction infinie. 

Dans cette section nous montrerons que la relation $\roZ$ est confluente et pour 
cela nous nous inspirons de la preuve de confluence du \laCal\  donnée par
exemple dans~\cite{Barendregt84}. 

Nous montrerons que la relation $\roZ$ est confluente,
propriété qui est obtenu immédiatement à partir de la confluence forte de la
relation $\roTR$. Pour prouver la confluence forte de $\roTR$ nous exhibons une
relation $\delta$ qui est fortement confluente et dont la fermeture %réflexive et
transitive est la relation $\roTR$. Ayant trouvé une telle relation, nous
pouvons prouver facilement, en utilisant le lemme suivant, que la relation
$\roTR$ est fortement confluente et ainsi, que $\roZ$ est confluente.

\LEM%--------------------------------------------------------------------------
(\cite{Barendregt84})
\label{diamondTransitive}
Etant données une relation $\longra{}$ et sa fermeture transitive $\longra{*}$.
Si $\longra{}$ est fortement confluente alors $\longra{*}$ est fortement confluente.
\FLEM%--------------------------------------------------------------------------

\proof{
En utilisant un diagramme simple suggéré par~:
\begin{center}$~$
\xymatrix@R-10pt@C-10pt{
&&~ \ar[dl] \ar[dr] && \\
& ~ \ar[dl]\ar@{.{>}}[dr] && ~  \ar[dr]\ar@{.{>}}[dl]& \\
~ \ar@{.{>}}[dr] && \ar@{.{>}}[dl] ~ \ar@{.{>}}[dr] &&  ~  \ar@{.{>}}[dl]\\
& ~ \ar@{.{>}}[dr] && ~  \ar@{.{>}}[dl]& \\
&& ~ &&
}
\end{center}
}

Nous pourrions choisir comme relation $\delta$ la fermeture réflexive de
$\roZ$. Malheureusement, dans l'Exemple~\ref{nonCRdelta} nous pouvons voir que
cette relation n'est pas fortement confluente.

\EX%--------------------------------------------------------------------------
\label{nonCRdelta}

Nous considérons le terme $t = [x \ra [x](x)](r)$ avec $\RoZ{r}{r'}$. Alors,
nous obtenons
$\RoZ{[x \ra [x](x)](r)}{\{[r](r)\}}$ et
$\RoZ{[x \ra [x](x)](r)}{[x \ra [x](x)](r')}$ 
mais il n'existe aucun terme $t'$ tel que
$\RoZ{\{[r](r)\}}{t'}$ et
$\RoZ{[x \ra [x](x)](r')}{t'}$.
\FEX%--------------------------------------------------------------------------

Afin d'obtenir la relation $\delta$ appropriée nous adoptons une approche
similaire à la réduction parallèle introduite par \textit{Tait \& Martin-Löf }.

%\subsubsection{Relation $\del$}
%===============================================================================
%

On doit noter que dans l'Exemple~\ref{nonCRdelta} nous n'obtenons pas la
confluence forte parce que la réduction est effectuée pour seulement un
sous-terme à la fois. Nous recherchons donc une relation effectuant le maximum de
réductions en une étape. La relation qui émerge naturellement est la version
parallèle de la relation $\roZ$ et elle est décrite dans la
Définition~\ref{deltaRel}.

\DEF%--------------------------------------------------------------------------
\label{deltaRel}
La relation $\del$ est définie par les règles suivantes~:

\begin{enumerate}

\item
$\Del{t}{t}$,

\item 
$\Del{u_i}{u_i'}$, $i=1 \ldots n$ $~~~$
$\Rightarrow$%\\
$~~~$ $\Del{\{u_1,\ldots,u_n\}}{\{u_1',\ldots,u_n'\}}$,

\item
$\Del{u_i}{u_i'}$, $i=1 \ldots n$ $~~~$
$\Rightarrow$%\\
$~~~$ $\Del{f(u_1,\ldots,u_n)}{f(u_1',\ldots,u_n')}$

\item 
$\Del{u}{u'}$, $\Del{v}{v'}$ $~~~$
$\Rightarrow$%\\
$~~~$ $\Del{u \ra v}{u' \ra v'}$

\item 
$\Del{u}{u'}$, $\Del{v}{v'}$ $~~~$ 
$\Rightarrow$%\\
$~~~$ 	$\Del{[u](v)}{[u'](v')}$,

\item 
$\Del{l}{l'}$, $\Del{t}{t'}$, $\Del{r}{r'}$ %\\
$\Rightarrow$

   \begin{tabular}{lr}
   $\Del{[l \ra r](t)}{\{\sigma r'\}}$ &   
   ~~~~~~~~~~~~~~~~~~~~~~~~~~~~~~~~~~~~ (\rname{Fire_c})\\
   ~~~~~~	si $l,t$ sont \ready  \\
   ~~~~~~	avec $\sigma \in \Sl(l' \meqqes t')$
   \end{tabular} %\\

\item 
$\Del{u_i}{u_i'}$, $\Del{v_i}{v_i'}$, $i = 1 \ldots n$
$\Rightarrow$

   \begin{tabular}{ll}
   $\Del{[f(u_1,\ldots,u_n)](f(v_1,\ldots,v_n))}{\{f([u_1'](v_1'),\ldots,[u_n'](v_n'))\}}$, &
   ~~~~ (\rname{Congruence})
   \end{tabular} %\\

\item 
$\Del{u_i}{u_i'}$, $\Del{v_i}{v_i'}$, $i = 1 \ldots n$
$\Rightarrow$

   \begin{tabular}{ll}
   $\Del{[f(u_1,\ldots,u_n)](g(v_1,\ldots,v_m))}{\emptyset}$ &	
   ~~~~~~~~~~~~~~~~~~~~~~~~~~~~~~~~~~ (\rname{Congruence\_fail}) \\
   ~~~~~~	si $f \neq g$
   \end{tabular} %\\

\end{enumerate}
D'une façon similaire à la relation $\roZ$, la relation $\del$ modulo la
relation $\congE$ est notée $\delE$ et la fermeture réflexive et transitive de
$\delE$ est notée $\delETR$.

\FDEF%--------------------------------------------------------------------------

Pour le terme $t = [x \ra [x](x)](r)$ présenté dans l'Exemple~\ref{nonCRdelta}
nous avons toujours les réductions
$\Del{[x \ra [x](x)](r)}{\{[r](r)\}}$ et
$\Del{[x \ra [x](x)](r)}{[x \ra [x](x)](r')}$ 
mais les deux termes se réduisent dans un seul pas en le même terme puisque
$\Del{\{[r](r)\}}{\{[r'](r')\}}$ et
$\Del{[x \ra [x](x)](r')}{\{[r'](r')\}}$.

Dans le reste de cette section nous nous concentrons sur la preuve de la
confluence forte de la relation $\del$. Une fois que nous aurons prouvé cette
propriété, nous prouverons que $\roTR$ est la fermeture transitive de $\del$ et
comme corollaire, nous obtiendrons la confluence forte de la relation $\roTR$.

Nous devons préciser que, selon nos restrictions sur les réductions $\del$, tout
membre gauche d'une règle de réécriture $l \ra r$ doit être un terme du premier
ordre ($l \in \TFX$) afin d'appliquer la règle d'évaluation
\rname{Fire_c} pour réduire un terme de la forme $[l \ra r](t)$. Par conséquent, pour
le terme $l'$ tels que $\Del{l}{l'}$ ou $\RoZ{l}{l'}$ nous avons $l'=l$.

Le premier lemme décrit la préservation de la propriété de termes \ready\  par
les relations $\roZ$ et $\del$. Ce lemme est utilisé pour prouver que la
relation $\roTR$ est la fermeture transitive de la relation $\del$ ainsi
que pour montrer la confluence forte de la relation $\del$.

\LEM%------------------------------------------------------------------------
\label{readyLemmaRo}
\comment{dans Lemme~\ref{TransClosure}}

Etant donnés les $\rho$-termes $l,t,l',t'$ tels que $\RoZ{l}{l'}$ et
$\RoZ{t}{t'}$. Alors,
\begin{itemize}
	\item si $l,t$ sont \matchD\  alors $l',t'$ sont \matchD,
	\item si $t$ est \safes\  alors $t'$ est \safes,
	\item si $l,t$ sont \ready\  alors $l',t'$ sont \ready.
\end{itemize}
\FLEM%------------------------------------------------------------------------

\proof{

Par définition du \roCalE\  $l \in \TFX$ et par conséquent $l'=l$. Si le terme 
$t \in \TF$ alors $t=t'$ et le lemme est clairement vrai. Nous considérons par
la suite que $t \not\in \TF$.

Puisque $l,t$ sont \matchD\  alors le terme $l$ \subf\  le terme $t$ et donc toute
position fonctionnelle du terme $l$ est une position fonctionnelle du terme $t$
ou n'est pas une position du terme $t$.  Par définition de la
relation $\roZ$ on peut noter que le symbole de tête d'un terme $u$ n'est pas le
même que celui du terme $u'$ tel que $\RoZ{u}{u'}$ seulement si $u$ est une
application (de la forme $[~](~)$) et dans ce cas $u'$ est un ensemble.  Par
conséquent, l'ensemble des positions fonctionnelles du terme $t$ est le même que
l'ensemble des positions fonctionnelles du terme $t'$, et donc, le terme $l$
\subf\  le terme $t'$. Ainsi, les termes $l',t'$ sont \matchD.

Nous montrons maintenant que $t'$ est \safes\  si $t$ est \safes\  et donc, nous
analysons les types des applications et des ensembles pouvant apparaître dans
le terme $t'$.

Puisque $t$ est \safes, tout sous-terme $[u \ra w](v)$ de $t$ est tel que $u$
subsume $v$. En utilisant la définition de la relation $\roZ$ de la même manière
que précédemment, nous obtenons que $u$ subsume $v'$ pour tout terme $v'$ tel
que $\RoZ{v}{v'}$. Ainsi, tout terme $t'$ contenant le sous-terme $[u \ra w](v')$
et donc tel que $\RoZ{t}{t'}$ est \safes.
Tout sous-terme de $t'$ de la forme $[u](v)$ où $u$ n'est pas une règle de
réécriture peut être engendrer par une réduction $\roZ$ seulement si le terme $t$
contient un sous-terme de cette forme mais ceci n'est pas possible puisque $t$
est \safes.

Aucune règle de réécriture dans $t$ ne peut mener à un ensemble vide parce que
le membre gauche de toute règle de réécriture subsume son argument et donc le
filtrage n'échoue pas. Puisque il n'existe pas de sous-terme de la forme
$[u](v)$ dans $t$, la règle d'évaluation \rname{Congruence\_fail} ne peut pas
engendrer un ensemble vide. La seule règle d'évaluation qui peut introduire des
ensembles ayant plus d'un élément est \rname{Fire_c} mais puisque le filtrage
est syntaxique et donc unitaire dans le \roCalE, l'application d'une règle de
réécriture ne peut pas générer un ensemble ayant plus d'un élément.  Ainsi, un
ensemble ayant plus d'un élément ou un ensemble vide peuvent être engendrés dans
le terme $t'$ seulement si le terme $t$ contient de tels ensembles mais ceci
n'est pas possible puisque $t$ est \safes.

Nous avons donc montré que toutes les conditions de la Définition~\ref{safeDef}
sont satisfaites par le terme $t'$ si elles sont satisfaites pour le
terme $t$ et donc, $t'$ est \safes.

Ainsi, les propriétés de termes \matchD\  et \safe\  sont préservées par la
relation $\roZ$ et donc, la propriété de termes \ready\  est préservée
par la relation $\roZ$.
}%________________________________________________________________________________

\LEM%------------------------------------------------------------------------
\label{readyLemmaDel}
\comment{Lemme~\ref{BasicCase}}

Etant donnés les $\rho$-termes $l,t,l',t'$ tels que $\Del{l}{l'}$ et
$\Del{t}{t'}$. Alors,
\begin{itemize}
	\item si $l,t$ sont \matchD\  alors $l',t'$ sont \matchD,
	\item si $t$ est \safes\  alors $t'$ est \safes,
	\item si $l,t$ sont \ready\  alors $l',t'$ sont \ready.
\end{itemize}
\FLEM%------------------------------------------------------------------------

\proof{
Similaire au Lemme~\ref{readyLemmaRo}.
}

Nous analysons maintenant la correspondance entre les solutions des problèmes de
filtrage $(l \meqqes t)$ et $(l \meqqes t')$ où $\Del{t}{t'}$. Plus précisément,
nous voulons montrer que les termes du codomaine de la substitution obtenue pour
le premier problème de filtrage se réduisent dans les termes du codomaine de la
substitution obtenue pour le deuxième problème de filtrage et qu'un échec dans
le premier cas implique un échec dans le deuxième cas.

\LEM%--------------------------------------------------------------------------
\label{matchDeltaOne}
\comment{dans Lemme~\ref{substRed}}

Etant donnés les $\rho$-termes $l,t$ et $t'$ tels que $l,t$ soient \matchD\   et
$\Del{t}{t'}$.
\begin{itemize}
\item[a.]  Si $\subs{x_1/u_1,\ldots,x_n/u_n}$ est le résultat de 
	$(l \meqqes t)$ et $\subs{x_1/v_1,\ldots,x_n/v_n}$ est le résultat de
	$(l \meqqes t')$ (où $n$ est le nombre de variables de $l$), alors
	$\Del{u_i}{v_i}$.
\item[b.]  $\Sl(l \meqqes t)=\emptyset$ ssi $\Sl(l \meqqes t')=\emptyset$.
\end{itemize}
\FLEM%--------------------------------------------------------------------------

\proof{

Si $t \in \TF$ alors, $t'=t$ et le lemme est trivialement vrai. Pour le cas où
$t \not \in \TF$ nous procédons par induction sur la structure du $\rho$-terme
$t$.

\textit{Le cas de base}~: $t=x$, avec $x \in \XX$.
\begin{itemize}

\item[a.] Puisque $t=t'$ ce cas est trivial.

\item[b.]  Puisque $l,t$ sont \matchD\  alors $l$ \subf\  $t$ et conformément à la
Remarque~\ref{positionsFonctionnelles}, $l$ est une variable et donc, 
$(l \meqqes t)=(l \meqqes t')$ ne peut pas échouer.

\end{itemize}

\textit{Induction}~: 
Les cas que nous devons analyser sont $t=\{t_1,\ldots,t_q\}$,
$t=f(t_1,\ldots,t_q)$, $t=t_1 \ra t_2$ et $t=[t_1](t_2)$. Si la position de tête
de $t$ n'est pas une position fonctionnelle, c'est-à-dire si
$t=\{t_1,\ldots,t_q\}$, $t=t_1 \ra t_2$ ou $t=[t_1](t_2)$, alors $l=x$ 
(cf. Remarque~\ref{positionsFonctionnelles}) et dans
ces cas le lemme est clairement vrai. Le cas où $l=x$ et $t=f(t_1,\ldots,t_q)$
est trivial.

Si $t=f(t_1,\ldots,t_q)$ et $l$ n'est pas une variable nous analysons les deux points~:
\begin{itemize}

\item[a.] Si $(l \meqqes t)$ n'échoue pas alors
le terme $l$ doit être de la forme $l=f(l_1,\ldots,l_q)$ avec $l_i,t_i$ \matchD\
pour tout $i=1,\ldots,q$ et $\subs{x_1/u_1,\ldots,x_n/u_n}$ est la solution de 
$(l \meqqes t)$ est donc, la solution du système 
$(\bigwedge_{i=1,\ldots,q} l_i \meqqes t_i)$.
Par hypothèse d'induction, si $\Del{t_i}{t_i'}$ et la solution de 
$(l_i \meqqes t_i)$ est la substitution 
$\subs{x_{1i}/u_{1i}, \ldots, x_{mi}/u_{mi}}$, 
alors la solution de $(l_i \meqqes t_i')$ est 
$\subs{x_{1i}/u_{1i}', \ldots,x_{mi}/u_{mi}'}$ avec 
$\Del{u_{ki}}{u_{ki}'}$, $k=1 \ldots m$.
Puisque $t'=f(t_1',\ldots,t_q')$ avec $\Del{t_i}{t_i'}$, la
première règle de filtrage appliquée pour filtrer
$l'$ et $t'$ est $Decomposition$~:
$(l \meqqes t')=(\bigwedge_{i=1,\ldots,q} l_i \meqqes t_i')$.

Du à la linéarité de $l$ les variables $x_{ki}$, $i=1 \ldots q$, $k=1 \ldots m$
sont différentes et donc la règle de filtrage $MergingClash$ ne peut pas être
appliquée. Ainsi, la propriété est vérifiée.

\item[b.] Conformément à la Remarque~\ref{echecMatch}, l'échec peut être obtenu
seulement en appliquant la règle de filtrage $SymbolClash$ à la position de tête
ou à des positions plus profondes.

Nous pouvons donc avoir $l=g(l_1,\ldots,l_q)$ et $t'=f(t_1',\ldots,t_q')$ avec
$\Del{t_i}{t_i'}$ et $f \neq g$ et ainsi, $(l \meqqes t)$ et $(l \meqqes t')$
échouent.  Si $l=f(l_1,\ldots,l_q)$ alors l'échec est obtenu à une position plus
profonde. Puisque par induction, le problème $(l_i \meqqes t_i)$ mène à un échec
de filtrage ssi le problème $(l_i \meqqes t_i')$ mène à un échec alors, 
$(l \meqqes t)$ échoue ssi $(l \meqqes t')$ échoue.

\end{itemize}
}%________________________________________________________________________________

Nous analysons maintenant la correspondance entre l'application d'une
substitution à un terme $r$ et l'application de la même substitution ou d'une
substitution en correspondance forte avec cette substitution, à un terme obtenu
en réduisant le terme $r$.

\LEM%--------------------------------------------------------------------------
\label{substRed}
\comment{dans Lemme~\ref{BasicCase}}

Etant donnés les $\rho$-termes $l,t,r$ et $t',r'$ tels que $l,t$ soient \ready\
et $\Del{t}{t'}$, $\Del{r}{r'}$. 
Si les problèmes de filtrage $(l \meqqes t)$ et $(l \meqqes t')$ ont comme solutions
les substitutions $\sigma$ et $\sigma'$ respectivement alors, 
$\Del{\sigma r}{\sigma' r'}$.
\FLEM%--------------------------------------------------------------------------

\proof{
Si nous considérons la substitution $\sigma=\subs{x_1/s_1,\ldots,x_m/s_m}$ alors,
par le Lemme~\ref{matchDeltaOne}, $\sigma'=\subs{x_1/s_1',\ldots,x_m/s_m'}$, avec
$\Del{s_i}{s_i'}$, $i=1 \ldots m$.
Nous procédons par induction sur la structure du terme $r$ en considérant toutes
les réductions possibles $\Del{r}{r'}$. Les cas à analyser correspondent aux
règles de la Définition~\ref{deltaRel}~:
\begin{enumerate}

\item
$r = x$ et $r' = x$

\item 
$r = \{u_1,\ldots,u_n\}$ et $r' = \{u_1',\ldots,u_n'\}$ avec $\Del{u_i}{u_i'}$

\item
$r = f(u_1,\ldots,u_n)$ et $r' = f(u_1',\ldots,u_n')$ avec $\Del{u_i}{u_i'}$

\item 
$r = u \ra v$ et $r' = {u' \ra v'}$ avec $\Del{u}{u'}$, $\Del{v}{v'}$

\item 
$r = [u](v)$ et $r' = [u'](v')$ avec $\Del{u}{u'}$, $\Del{v}{v'}$

\item 
$r = [u \ra v](w)$ avec $u,w$ \ready\  et
$r' = \{\mu v'\}$, avec
$\mu \in \Sl(u' \meqqes w')$ et 
$\Del{u}{u'}$, $\Del{v}{v'}$, $\Del{w}{w'}$.

\item 
$r = [f(u_1,\ldots,u_n)](f(v_1,\ldots,v_n))$ et  
$r' = \{f([u_1'](v_1'),\ldots,[u_n'](v_n'))\}$ avec
$\Del{u_i}{u_i'}$, $\Del{v_i}{v_i'}$ pour tout $i = 1 \ldots n$.

\item 
$r = [f(u_1,\ldots,u_n)](g(v_1,\ldots,v_m)$ et  
$r' = \emptyset$.

\end{enumerate}

Pour le cas de base, $r=x$, nous devons prouver que $\Del{\sigma x}{\sigma' x}$
et cette réduction suit immédiatement par le Lemme~\ref{matchDeltaOne}.

Le seul cas où l'application de l'induction est plus élaborée est le cas $6$
décrivant l'application d'une règle de réécriture. Pour ce cas nous devons
prouver que
\begin{equation}
\label{firstEq}
\Del{\sigma ([u \ra v](w))}{\sigma' \{\mu v'\}}
\end{equation}
avec $\mu \in \Sl(u' \meqqes w')$.

Puisque $u$ est un terme du premier ordre, $u'=u$. Par $\alpha$-conversion, nous
supposons que $u$ ne contient aucune variable de $\sigma$ et aucune variable de
$\sigma'$. Avec cette supposition nous avons 
$\sigma ([u \ra v](w))=[u \ra\sigma v](\sigma w)$.
Puisque $\sigma$ représente la solution d'un problème de filtrage entre deux
termes \ready\  et $u,w$ sont \ready, nous obtenons que $u, \sigma w$ sont
\ready.

Si $\Sl(u \meqqes \sigma w)=\emptyset$ alors, en
utilisant le Lemme~\ref{matchDeltaOne}, nous obtenons la réduction 
\mbox{$\Del{[u \ra \sigma v](\sigma w)}{\emptyset}$} et nous devons prouver que
$\sigma' r'=\emptyset$ et donc, que $r'=\{\mu v'\}=\emptyset$.  Puisque les
termes $u, \sigma w$ sont \ready\  alors ils sont \matchD\  et conformément à la
Remarque~\ref{echecMatch} le filtrage $u \meqqes \sigma w$ peut échouer
seulement à cause des symboles fonctionnels différents à la même position des
termes $u$ et $\sigma w$ et donc, des termes $u$ et $w$. En plus, tout terme de
la forme $f(\ldots)$ peut être réduit en utilisant la relation $\del$ seulement
en un terme de la même forme et donc, le filtrage $u \meqqes w'$ échoue et nous
obtenons $\{\mu v'\}=\emptyset$.

Si le filtrage $u \meqqes \sigma w$ n'échoue pas, nous pouvons appliquer
l'induction aux sous-termes %$\sigma v$ et $\sigma w$ 
$v$ et $w$  et nous avons
$\Del{\sigma v}{\sigma' v'}$ et $\Del{\sigma w}{\sigma' w'}$.
Ainsi, nous obtenons la réduction~:
\begin{equation}
\label{secondEq}
\Del{\sigma ([u \ra v](w))=[u \ra \sigma v](\sigma w)}{\{\mu' (\sigma' v')\}}
\end{equation}
avec  $\mu' \in \Sl(u \meqqes \sigma' w')$.

En utilisant les réductions (\ref{firstEq}) et (\ref{secondEq}), l'égalité
qui nous permettrait de conclure la preuve est: 
	$$\{\mu' (\sigma' v')\} = \sigma' \{\mu v'\}$$
ou
	$$\mu' (\sigma' v') = \sigma' (\mu v').$$

Nous supposons que
$\sigma' = \subs{x_1/t_1, \ldots, x_n/t_n}$
et
$\mu = \subs{y_1/s_1, \ldots, y_m/s_m}$.
Il est clair que
$\mu' = \subs{y_1/\sigma' s_1, \ldots, y_m/\sigma' s_m}$.
Puisque $u$ ne contient aucune variable de $\sigma'$ nous déduisons que $y_j$
n'est pas une variable de $t_i$ et $x_i \neq y_j$ pour tous 
$j=1 \ldots m$, $i=1 \ldots n$.
Ainsi, nous avons
$\mu' (\sigma' v')$ $=$
$\subs{y_1/\sigma' s_1, \ldots, y_m/\sigma' s_m} (\subs{x_1/t_1, \ldots, x_n/t_n} v')$
et puisque $y_j$ n'est pas une variable de $t_i$,
$$\mu' (\sigma' v') =
\subs{y_1/\sigma' s_1, \ldots, y_m/\sigma' s_m,x_1/t_1, \ldots, x_n/t_n} v'.$$

Pour le deuxième terme nous prenons
$\sigma' (\mu v')$ $=$
$\subs{x_1/t_1, \ldots, x_n/t_n} (\subs{y_1/s_1, \ldots, y_m/s_m} v)$
et puisque $x_i \neq y_j$ et $y_j$ n'est pas une variable de $t_i$,\\
$$\sigma' (\mu v') =
\subs{x_1/t_1, \ldots, x_n/t_n, y_1/\subs{x_1/t_1, \ldots, x_n/t_n}s_1,
\ldots, y_m/\subs{x_1/t_1, \ldots, x_n/t_n}s_m} v =$$
$$\subs{x_1/t_1, \ldots, x_n/t_n, y_1/\sigma' s_1, \ldots, y_m/\sigma' s_m} v$$

L'égalité $\mu' (\sigma' v') = \sigma' (\mu v')$ est valide et ainsi, le lemme
est prouvé.
}%________________________________________________________________________________

Si la condition de termes \ready\  de la règle d'évaluation \rname{Fire_c} est
changée en une condition de termes \matchD, nous pouvons montrer facilement que
$\Del{\sigma r}{\sigma' r'}$ si les termes $l,t$ sont seulement \matchD. Pour
prouver cette réduction il faut juste remarquer que si $\sigma$ représente la
solution d'un problème de filtrage entre deux termes $u,w$ \matchD, nous
obtenons que $u, \sigma w$ sont \matchD.

En gardant la condition de termes \ready\  pour la règle d'évaluation
\rname{Fire_c} nous ne pouvons pas démontrer la relation 
$\Del{\sigma r}{\sigma' r'}$ si nous demandons que les termes $l,t$ soient
simplement \matchD\  au lieu de \ready.

La condition que les termes sont non seulement \matchD\  mais \ready\  est
nécessaire essentiellement pour éviter que des termes (non-\safe) $\emptyset$ apparaissent
dans le codomaine de la substitution $\sigma$ sans être propagés strictement.
Par exemple, si $l=y$ et $t=\emptyset$ alors $l,t$ sont \matchD\  mais pas
\ready\  ($t$ n'est pas \safes) et $\sigma=\subs{y/\emptyset}$. Si $r=[x \ra b](y)$ et
$r'=\{b\}$ alors $\sigma r=[x \ra b](\emptyset)$, $\sigma r'=\{b\}$ et il
n'existe pas de réduction $\Del{[x \ra b](\emptyset)}{\{b\}}$ puisque
$x,\emptyset$ ne sont pas \ready\  ($\emptyset$ n'est pas \safes) .

\LEM($\del$ est fortement confluente)%------------------------------------------
\label{BasicCase}
\comment{dans Lemme~\ref{DiamDelE} et Lemme~\ref{ConfCongDEL}}

Etant donnés les termes $t_0,t_1,t_2$ tels que $\Del{t_0}{t_1}$ et
$\Del{t_0}{t_2}$. Alors, il existe un terme $t_3$ tel que $\Del{t_1}{t_3}$ et
$\Del{t_2}{t_3}$~:
\begin{center}$~$
\xymatrix{
&
t_0
\ar[dl]_{\delxy} \ar[dr]^-{\delxy}
& \\
t_1
\ar@{.{>}}[dr]_{\delxy} && 
t_2
\ar@{.{>}}[dl]^-{\delxy} \\
&t_3&
}
\end{center}

\FLEM%--------------------------------------------------------------------------

\proof{
Nous montrons le lemme par induction sur la structure du terme $t_0$.
\begin{enumerate}

\item $t_0 = x$

Par la définition de la relation $\del$, $t_0 = t_1 = t_2$ et nous pouvons
choisir $t_3=t_0$.

\item $t_0 = \{u_1,\ldots,u_n\}$

Nous avons $t_1 = \{u_1',\ldots,u_n'\}$, $t_2 = \{u_1'',\ldots,u_n''\}$ avec
$\Del{u_i}{u_i'}$ et $\Del{u_i}{u_i''}$, $i=1 \ldots n$.  Par induction, il
existe les termes $u_i'''$ tels que $\Del{u_i'}{u_i'''}$,
$\Del{u_i''}{u_i'''}$, $i = 1 \ldots n$. Ainsi, nous pouvons choisir 
$t_3 = \{u_1''',\ldots,u_n'''\}$.

Les diagrammes correspondants sont présentés ci-dessous~:
\begin{center}$~$
\xymatrix{ 
\{u_1,\ldots,u_n\}
\ar[d]_{\delxy} \ar[dr]^-{\delxy}
& \\
\{u_1',\ldots,u_n'\}
\ar@{.{>}}[dr]_{\delxy} &
\{u_1'',\ldots,u_n''\}
\ar@{.{>}}[d]^-{\delxy} \\
& 
\{u_1''',\ldots,u_n'''\}
}
$~~~~$ puisque $~~~~$ 
\xymatrix{ 
&
u_i
\ar[dl]_{\delxy} \ar[dr]^-{\delxy}
& \\
u_i'
\ar@{.{>}}[dr]_{\delxy} &&
u_i''
\ar@{.{>}}[dl]^-{\delxy} \\
&u_i'''&
}
\end{center}

\item $t_0 = f(u_1,\ldots,u_n)$

Nous avons $t_1 = f(u_1',\ldots,u_n')$ et $t_2 = f(u_1'',\ldots,u_n'')$ avec
$\Del{u_i}{u_i'}$ et $\Del{u_i}{u_i''}$, $i = 1 \ldots n$.  Par induction, il
existe les termes $u_i'''$ tels que $\Del{u_i'}{u_i'''}$,
$\Del{u_i''}{u_i'''}$, $i = 1 \ldots n$.  Ainsi, nous pouvons choisir 
$t_3 = \{u_1''',\ldots,u_n'''\}$.

\item $t_0 = u_0 \ra v_0$

Par la définition de la relation $\del$, $t_1 = u_1 \ra v_1$ et $t_2 = u_2 \ra v_2$ 
avec $\Del{u_0}{u_1}$, $\Del{u_0}{u_2}$, $\Del{v_0}{v_1}$,
$\Del{v_0}{v_2}$. Par induction, il existe les termes $u_3, v_3$ tels que
$\Del{u_1}{u_3}$, $\Del{u_2}{u_3}$ et $\Del{v_1}{v_3}$, $\Del{v_2}{v_3}$.  En
fait, du aux conditions sur le membre gauche d'une règle de réécriture,
$u_0=u_1=u_2=u_3$ et nous obtenons $t_3=u_0 \ra v_3$.

\item $t_0 = [u_0](v_0)$

Nous avons plusieurs possibilités pour choisir les termes $t_1,t_2$ selon les
propriétés des termes $u_0,v_0$ décrites dans la définition de la relation
$\del$~:
\begin{enumerate}

	\item $t_1 = [u_1](v_1)$ et $t_2 = [u_2](v_2)$ avec $\Del{u_0}{u_1}$,
	$\Del{u_0}{u_2}$, $\Del{v_0}{v_1}$, $\Del{v_0}{v_2}$. Par induction, il
	existe les termes $u_3, v_3$ tels que $\Del{u_1}{u_3}$,
	$\Del{u_2}{u_3}$ et $\Del{v_1}{v_3}$, $\Del{v_2}{v_3}$. Ainsi, nous
	obtenons $t_3=[u_3](v_3)$.

	\item $t_0 = [l_0 \ra r_0](p_0)$ avec $l_0,p_0$ \ready. 

	Si nous avons $\Del{r_0}{r_1}$, $\Del{p_0}{p_1}$, $\Del{r_0}{r_2}$,
	$\Del{p_0}{p_2}$ alors nous obtenons $t_1 = \{\sigma_1 r_1\}$, 
	$t_2 = \{\sigma_2 r_2\}$, avec $\sigma_1 \in \Sl(l_0 \meqqes p_1)$ et
	$\sigma_2 \in \Sl(l_0 \meqqes p_2)$.

	Par induction, il existe les termes $r_3, p_3$ tels que
	$\Del{r_1}{r_3}$, $\Del{r_2}{r_3}$ et $\Del{p_1}{p_3}$,
	$\Del{p_2}{p_3}$.
	Si le filtrage $(l_0 \meqqes p_1)$ échoue alors, conformément au
	Lemme~\ref{matchDeltaOne}, $\Sl(l_0 \meqqes p_0)=\emptyset$ et 
	$\Sl(l_0 \meqqes p_2)=\emptyset$. Nous obtenons ainsi 
	$\{\sigma_1 r_1\}=\{\sigma_2 r_2\}=\emptyset$ et le lemme est clairement
	vrai.
	Si le filtrage n'échoue pas, puisque $l_0,p_1$ et $l_0,p_2$ sont \ready\
	par le Lemme~\ref{readyLemmaDel} alors, nous pouvons utiliser le
	Lemme~\ref{substRed} et choisir $t_3=\{\sigma_3 r_3\}$, avec
	$\{\sigma_3\} = \Sl(l_0 \meqqes p_3)$~:
\begin{center}$~$
\xymatrix{ 
[l_0 \ra r_0](p_0)
\ar[d]_{\delxy} \ar[dr]^-{\delxy}
& \\
\{\sigma_1 r_1\}
\ar@{.{>}}[dr]_{\delxy} &
\{\sigma_2 r_2\}
\ar@{.{>}}[d]^-{\delxy} \\
& 
\{\sigma_3 r_3\}
}
\end{center}

	Si nous avons $\Del{(l_0 \ra r_0)}{(l_0 \ra r_1)}$, $\Del{p_0}{p_1}$ et
	$\Del{r_0}{r_2}$, $\Del{p_0}{p_2}$ alors nous obtenons $t_1 =
	[u_1](p_1)$ et $t_2 = \{\sigma_2 r_2\}$, avec 
	$\sigma_2 \in \Sl(l_0 \meqqes p_2)$. Par conséquent, nous devons avoir
	$u_1=l_0 \ra r_1$, avec $\Del{r_0}{r_1}$.

	En raisonnant de la même façon que dans le cas précédent nous obtenons, 
	soit $\{\sigma_2 r_2\}=\emptyset$ et $\Del{[l_0 \ra r_1](p_1)}{\emptyset}$, 
	soit~:
\begin{center}$~$
\xymatrix{ 
[l_0 \ra r_0](p_0)
\ar[d]_{\delxy} \ar[dr]^-{\delxy}
& \\
[l_0 \ra r_1](p_1)
\ar@{.{>}}[dr]_{\delxy} &
\{\sigma_2 r_2\}
\ar@{.{>}}[d]^-{\delxy} \\
& 
\{\sigma_3 r_3\}
}
\end{center}

	\item $t_0 = [f(u_1,\ldots,u_n)](f(v_1,\ldots,v_n))$

	Nous considérons $t_1=\{f([u_1'](v_1'),\ldots,[u_n'](v_n'))\}$,
	$t_2=\{f([u_1''](v_1''),\ldots,[u_n''](v_n''))\}$ avec
	$\Del{u_i}{u_i'}$, $\Del{v_i}{v_i'}$ et $\Del{u_i}{u_i''}$,
	$\Del{v_i}{v_i''}$, $i = 1 \ldots n$. Par induction, il existe les
	termes $u_i''',v_i'''$ tels que $\Del{u_i'}{u_i'''}$,
	$\Del{u_i''}{u_i'''}$ et $\Del{v_i'}{v_i'''}$, $\Del{v_i''}{v_i'''}$,
	$i=1 \ldots n$. Par conséquent, nous pouvons choisir
	$t_3=\{f([u_1'''](v_1'''),\ldots,[u_n'''](v_n'''))\}$.
\begin{center}$~$
\xymatrix{ 
[f(u_1,\ldots,u_n)](f(v_1,\ldots,v_n))
\ar[d]_{\delxy} \ar[dr]^-{\delxy}
& \\
\{f([u_1'](v_1'),\ldots,[u_n'](v_n'))\}
\ar@{.{>}}[dr]_{\delxy} &
\{f([u_1''](v_1''),\ldots,[u_n''](v_n''))\}
\ar@{.{>}}[d]^-{\delxy} \\
& 
\{f([u_1'''](v_1'''),\ldots,[u_n'''](v_n'''))\}
}
\end{center}

	Maintenant, nous considérons les termes
	$t_1=[f(u_1',\ldots,u_n')](f(v_1',\ldots,v_n'))$ et
	$t_2=\{f([u_1''](v_1''),\ldots,[u_n''](v_n''))\}$ tels que nous avons
	$\Del{f(u_1,\ldots,u_n)}{f(u_1',\ldots,u_n')}$,
	$\Del{f(v_1,\ldots,v_n)}{f(v_1',\ldots,v_n')}$ et $\Del{u_i}{u_i''}$,
	$\Del{v_i}{v_i''}$, $i = 1 \ldots n$. Par conséquent, nous devons avoir
	$\Del{u_i}{u_i'}$ et $\Del{v_i}{v_i'}$, $i = 1 \ldots n$. Par induction,
	il existe les termes $u_i''',v_i'''$ tels que $\Del{u_i'}{u_i'''}$,
	$\Del{u_i''}{u_i'''}$ et $\Del{v_i'}{v_i'''}$, $\Del{v_i''}{v_i'''}$,
	$i=1\ldots n$. Par conséquent, nous pouvons choisir
	$t_3=\{f([u_1'''](v_1'''),\ldots,[u_n'''](v_n'''))\}$.
\begin{center}$~$
\xymatrix{ 
[f(u_1,\ldots,u_n)](f(v_1,\ldots,v_n))
\ar[d]_{\delxy} \ar[dr]^-{\delxy}
& \\
[f(u_1',\ldots,u_n')](f(v_1',\ldots,v_n'))
\ar@{.{>}}[dr]_{\delxy} &
\{f([u_1''](v_1''),\ldots,[u_n''](v_n''))\}
\ar@{.{>}}[d]^-{\delxy} \\
& 
\{f([u_1'''](v_1'''),\ldots,[u_n'''](v_n'''))\}
}
\end{center}

\item $t_0 = [f(u_1,\ldots,u_n)](g(v_1,\ldots,v_m))$

	Si nous avons $t_1=\emptyset$,
	$t_2=[f(u_1',\ldots,u_n')](g(v_1',\ldots,v_m'))$ avec $\Del{u_i}{u_i'}$,
	$i = 1 \ldots n$, $\Del{v_j}{v_j'}$, $j = 1 \ldots m$ alors nous pouvons
	choisir $t_3=\emptyset$.
\begin{center}$~$
\xymatrix{ 
[f(u_1,\ldots,u_n)](g(v_1,\ldots,v_m)
\ar[d]_{\delxy} \ar[dr]^-{\delxy}
& \\
\emptyset
\ar@{.{>}}[dr]_{\delxy} &
[f(u_1',\ldots,u_n')](g(v_1',\ldots,v_m'))
\ar@{.{>}}[d]^-{\delxy} \\
& 
\emptyset
}
\end{center}
\end{enumerate}

\end{enumerate}
}%________________________________________________________________________________

Nous avons montré que le Lemme~\ref{substRed} est vrai même si les termes d'une
application doivent être seulement \matchD\  afin de réduire l'application en
utilisant la règle d'évaluation \rname{Fire_c} et dans ce cas la condition du
lemme peut être transformée dans une condition de termes \matchD.  Nous pouvons
remarquer que la même restriction de termes \matchD\  a été suffisante pour
montrer le Lemme~\ref{matchDeltaOne}.

Nous pouvons donc montrer facilement que la relation $\del$ est fortement
confluente même si la condition de la règle d'évaluation \rname{Fire_c} impose
que les termes soient seulement \matchD. Cette condition évite les échecs de
filtrage indésirables et elle est suffisante pour empêcher les réductions
$\del$ non-convergentes. Les autres restrictions imposées pour les termes
\ready\  seront nécessaires pour obtenir la cohérence entre les relations $\del$
et $\congr$.

Nous montrons maintenant que la relation $\roTR$ est la fermeture transitive de
la relation $\del$ et nous obtenons comme conséquence immédiate la confluence de
la relation $\roZ$.

\LEM%--------------------------------------------------------------------------
\label{TransClosureRO}
\comment{used in Theorem~\ref{ConfRO} and in Lemme~\ref{TransClosure}}

La relation $\roTR$ est la fermeture transitive de la relation $\del$.
\FLEM%--------------------------------------------------------------------------

\proof{ 

Nous prouvons les inclusions suivantes~:
$$\roZ~\subseteq~\del~\subseteq~\roTR$$ 
et dans ce cas, puisque $\roTR$ est la fermeture transitive de la relation
$\roZ$, $\roTR$ est la fermeture transitive de la relation $\del$.

Nous devons donc prouver les deux inclusions~:\\
$$\roZ~\subseteq~\del ~~ et ~~ \del~\subseteq~\roTR$$

Il est clair que $\roZ~\subseteq~\del$.

Pour prouver que $\del~\subseteq~\roTR$ le seul cas non-trivial est celui
correspondant à la règle d'évaluation \rname{Fire_c}~:
\begin{itemize}
\item[] si $\RoTR{u}{u'}$, $\RoTR{v}{v'}$ et $\RoTR{w}{w'}$ avec les termes $u,w$
	\ready\  alors \mbox{$\RoTR{[u \ra w](v)}{\{\sigma w'\}}$}, avec
	$\sigma \in \Sl(u' \meqqes v')$.
\end{itemize}

Il est clair que $\RoTR{[u \ra w](v)}{[u' \ra w'](v')}$ et puisque, par le
Lemme~\ref{readyLemmaRo}, les termes $u',w'$ sont \ready\  alors, en utilisant la
définition de la relation $\roZ$, nous déduisons que
$\RoZ{[u' \ra w'](v')}{\{\sigma w'\}}$, avec $\sigma \in \Sl(u' \meqqes v')$.
Ainsi, nous obtenons par transitivité $\RoTR{[u \ra w](v)}{\{\sigma w'\}}$ et
donc $\del~\subseteq~\roTR$.
}%________________________________________________________________________________

\TH($\roTR$ est fortement confluente, $\roZ$ est confluente)%------------------
\label{ConfRO}

Si $\RoTR{t}{u}$ et $\RoTR{t}{v}$ alors il existe un terme $w$ tel que
$\RoTR{u}{w}$ et $\RoTR{v}{w}$.

\FTH%--------------------------------------------------------------------------

\proof{

Conformément au Lemme~\ref{TransClosureRO}, la relation $\roTR$ est la fermeture
transitive de la relation $\del$. Le Lemme~\ref{BasicCase} montre la confluence
forte de la relation $\del$. Ainsi, par le Lemme~\ref{diamondTransitive}, la
relation $\roTR$ est fortement confluente et par conséquent, la relation $\roZ$
est confluente.
}

Le même résultat peut être montré pour une relation $\del$ utilisant une règle
d'évaluation \rname{Fire_c} avec la condition d'application imposant des termes
\matchD. La confluence et la terminaison de la relation $\congr$ ont été
obtenues sans imposer aucune restriction sur les termes. Malheureusement, afin
d'obtenir la confluence de la relation construite à partir des deux relations,
la condition de termes \matchD\  n'est pas suffisante et nous devons imposer que
les termes soient \ready\  dans la règle d'évaluation \rname{Fire_c}.

\subsection{La confluence} \label{confluence_modulo}
%================================================================

En début de la Section~\ref{relationsInduites} nous avons présenté deux possibilités
pour décrire les réductions dans le \roCal. La première approche consiste à
considérer deux sous-ensembles de règles d'évaluation~: un premier ensemble
contenant les règles de \textit{déduction} qui décrivent la réduction d'une
application et un deuxième ensemble contenant les règles de \textit{calcul} qui
décrivent le comportement des ensembles par rapport aux autres symboles du
\roCal. Nous considérons donc, la relation induite par l'ensemble de règles de
\textit{déduction} modulo la relation de congruence induite par l'ensemble de
règles de \textit{calcul} et cette relation correspond à la relation $\roZE$
présentée dans la  Définition~\ref{transFire}.

Une autre approche consiste à ne pas faire la distinction entre la
\textit{déduction} et le \textit{calcul} et considérer la relation induite par
l'ensemble de règles d'évaluation du \roCal. Cette relation correspond à la
relation $\congTR\roZ\congTR$ avec $\roZ$ et $\congTR$ introduit dans la
Définition~\ref{transFire} et la Définition~\ref{transCongRel} respectivement.

Cette dernière approche a l'avantage de la simplicité mais l'approche précédente
permet une flexibilité supérieure du calcul. Ainsi, si nous voulons remplacer
les ensembles par des listes pour représenter le non-déterminisme nous devons
juste remplacer la relation de \textit{calcul} avec une relation appropriée qui
doit être confluente et terminante afin d'obtenir la confluence du calcul. Nous
montrons par la suite que la confluence du calcul est obtenue pour les deux
approches.

\subsubsection{Confluence de la relation $\roZE$}
%================================================================

Dans cette section nous nous concentrons sur la preuve de la confluence forte de
la relation $\delE$. Une fois prouvée cette propriété, nous
montrons que $\roZETR$ est la fermeture transitive de la relation $\delE$ et
comme corollaire, nous obtenons la confluence forte de la relation $\roZETR$.

La confluence de la relation $\delE$ est montrée en utilisant les propriétés des 
relations $\del$ et $\congr$ prouvées dans les sections précédentes et en
démontrant que les deux relations sont cohérentes. 

Nous commençons par regarder la correspondance entre les solutions des problèmes
de filtrage $(l \meqqes t)$ et $(l \meqqes t')$ où $\CongTR{t}{t'}$. Nous
voulons obtenir un résultat similaire à celui présenté dans le
Lemme~\ref{matchDeltaOne} pour le cas où $\Del{t}{t'}$ et plus précisément, nous
voulons montrer qu'un échec pour le premier problème de filtrage implique un
échec pour le deuxième problème de filtrage et que pour toute substitution
obtenue comme solution pour le premier problème de filtrage, les termes du
codomaine de la substitution sont réduits dans les termes du codomaine de la
substitution obtenue pour le deuxième problème de filtrage.

Ce résultat ne peut pas être obtenu dans le cas de la relation $\congTR$ mais la
propriété est montrée si une condition supplémentaire est imposée sur le terme
$t'$.

\LEM%--------------------------------------------------------------------------
\label{matchCongOne}
\comment{dans Lemme~\ref{substEquivBasic}}

Etant donnés deux $\rho$-termes $l,t$ et $t'$ tels que $l,t$ soient \matchD,
$\CongTR{t}{t'}$ et $l,t'$ soient \matchD.
\begin{itemize}
\item[a.]  Si $\subs{x_1/u_1,\ldots,x_n/u_n}$ est le résultat de 
	$(l \meqqes t)$ et $\subs{x_1/v_1,\ldots,x_n/v_n}$ est le résultat de
	$(l \meqqes t')$ (avec $n$ le nombre de variables de $l$), alors
	$\CongTR{u_i}{v_i}$.
\item[b.]  $\Sl(l \meqqes t)=\emptyset$ ssi $\Sl(l \meqqes t')=\emptyset$.
\end{itemize}
\FLEM%--------------------------------------------------------------------------
%%%% CEX: f(x) < f({a}) and f(x) < {f(a)} %%%%%%%%%%
%%%%      [f(x) -> x](f({a}))                   %%%%
%%%%%%%%%%%%%%%%%%%%%%%%%%%%%%%%%%%%%%%%%%%%%%%%%%%%

\proof{

Si $t \in \TF$ alors, $t'=t$ et le lemme est trivialement vrai. Pour le cas où
$t \not \in \TF$ nous procédons par induction sur la structure du $\rho$-terme
$t$.

\textit{Le cas de base}~: $t=x$, avec $x \in \XX$.

Puisque $l,t$ sont \matchD\  alors $l$ \subf\  $t$ et conformément à la
Remarque~\ref{positionsFonctionnelles}, $l$ est une variable.
\begin{itemize}

\item[a.] Puisque $t=t'$ ce cas est trivial.

\item[b.] Puisque $l$ est une variable, $(l \meqqes t)=(l \meqqes t')$ ne peut pas échouer.

\end{itemize}

\textit{Induction}~: 
Les cas que nous devons analyser sont $t=\{t_1,\ldots,t_q\}$,
$t=f(t_1,\ldots,t_q)$, $t=t_1 \ra t_2$ et $t=[t_1](t_2)$. Si
$t=\{t_1,\ldots,t_q\}$, $t=t_1 \ra t_2$ ou $t=[t_1](t_2)$ alors la position de
tête de $l$ est une position variable
(cf. Remarque~\ref{positionsFonctionnelles}) et si nous considérons que $l=x$,
le lemme est clairement vrai. Le cas où $l=x$ et $t=f(t_1,\ldots,t_q)$ est
trivial.

Si $t=f(t_1,\ldots,t_q)$ et $l$ n'est pas une variable nous analysons les deux points~:
\begin{itemize}

\item[a.] Si $(l \meqqes t)$ n'échoue pas alors
le terme $l$ doit être de la forme $l=f(l_1,\ldots,l_q)$ et la solution de
$(l \meqqes t)=(\bigwedge_{i=1,\ldots,q} l_i \meqqes t_i)$ est
$\subs{x_1/u_1,\ldots,x_n/u_n}$.

Puisque $l,t'$ sont \matchD\  et $\CongTR{t}{t'}$ alors, conformément à la
Définition~\ref{transCongRel} nous devons avoir $t'=f(t_1',\ldots,t_q')$ avec
$\CongTR{t_i}{t_i'}$ et $l_i,t_i'$ \matchD, pour $i=1 \ldots q$.

Par hypothèse d'induction, si $\CongTR{t_i}{t_i'}$ et la solution de
$(l_i \meqqes t_i)$ est la substitution 
$\subs{x_{1i}/u_{1i}, \ldots, x_{mi}/u_{mi}}$, alors la solution de
$(l_i \meqqes t_i')$ est $\subs{x_{1i}/u_{1i}', \ldots, x_{mi}/u_{mi}'}$ 
avec $\CongTR{u_{ki}}{u_{ki}'}$, $k=1 \ldots m$.

La première règle de filtrage appliquée pour filtrer $l'$ et $t'$ est
$Decomposition$ et nous obtenons 
$(l \meqqes t') = (\bigwedge_{i=1,\ldots,q} l_i \meqqes t_i')$.
Du à la linéarité de $l$ les variables $x_{ki}$, $i=1 \ldots q$, $k=1 \ldots m$
sont différentes et donc la règle de filtrage $MergingClash$ ne peut pas être
appliquée et la propriété est vérifiée.

\item[b.] Conformément à la Remarque~\ref{echecMatch}, l'échec peut être obtenu
seulement en appliquant la règle de filtrage $SymbolClash$ à la position de tête
ou à des positions plus profondes.

Nous pouvons donc avoir $l=g(l_1,\ldots,l_q)$ et $t'=f(t_1',\ldots,t_q')$ avec
$\CongTR{t_i}{t_i'}$ et $f \neq g$ et ainsi, $(l \meqqes t)$ et $(l \meqqes t')$
échouent.  Si $l=f(l_1,\ldots,l_q)$ alors l'échec est obtenu à une position plus
profonde. Puisque par induction, le problème $(l_i \meqqes t_i)$ mène à un échec
de filtrage ssi le problème $(l_i \meqqes t_i')$ mène à un échec alors, 
$(l \meqqes t)$ échoue ssi $(l \meqqes t')$ échoue.

\end{itemize}
}%_____________________________________________________________________________

La preuve de ce lemme est très similaire à celle du
Lemme~\ref{matchDeltaOne}. Puisque la propriété de termes \matchD\  n'est pas
préservée par la relation $\congr$, comme il était le cas pour la relation $\roZ$
(Lemme~\ref{readyLemmaRo}), nous avons ajouté la condition que $l,t'$ soient
\matchD.  Cette condition est cruciale pour le premier point du lemme et un
contre-exemple est obtenu immédiatement si la condition n'est pas satisfaite~:
si $l=f(x)$ et $t=f(\{a\})$ alors $\Sl(l \meqqes t) \neq \emptyset$, tandis que
$t'=\{f(a)\}$ et $\Sl(l \meqqes t') = \Sl(f(x) \meqqes \{f(a)\}) = \emptyset$.

Nous analysons par la suite la relation entre les termes obtenus en appliquant
deux substitutions avec les codomaines liées, comme précédemment, par la relation
$\congTR$, à un même terme. 

\LEM%--------------------------------------------------------------------------
\label{substEquivBasic}
\comment{ dans Lemme~\ref{BasicCoherence}}

Etant donnés les $\rho$-termes $l,t,r$ et $t'$ tels que $l,t$ soient \matchD,
$\CongTR{t}{t'}$ et $l,t'$ soient \matchD.
Si les problèmes de filtrage $(l \meqqes t)$ et $(l \meqqes t')$ ont comme solutions
les substitutions $\sigma$ et $\sigma'$ respectivement alors, 
$\CongTR{\sigma r}{\sigma' r}$.
\FLEM%--------------------------------------------------------------------------

\proof{
Nous procédons par induction sur la structure du terme $r$.

Le cas de base, $r=x$ suit immédiatement par le Lemme~\ref{matchCongOne}.

Tous les autres cas sont traités facilement en utilisant l'hypothèse
d'induction. Par exemple, si $r=\{u_1,\ldots,u_m\}$ nous avons par induction
$\CongTR{\sigma u_i}{\sigma' u_i}$, $i=1 \ldots m$, et puisque la relation
$\congTR$ est fermée par contexte, 
$\CongTR{\sigma \{u_1,\ldots,u_m\}=\{\sigma u_1,\ldots,\sigma u_m\}}{\{\sigma' u_1,\ldots,\sigma' u_m\}=\sigma' \{u_1,\ldots,u_m\}}$.
}%_____________________________________________________________________________

Nous montrons maintenant que la relation $\congTR$ est stable par rapport à
l'application d'une substitution résultant d'un problème de filtrage.

\LEM%--------------------------------------------------------------------------
\label{substEquivInd}
\comment{dans Lemme~\ref{BasicCoherence}}

Etant donnés les $\rho$-termes $l,t,r$ et $r'$ tels que $\CongTR{r}{r'}$.  Si la
substitution $\sigma$ est la solution du problème de filtrage $(l \meqqes t)$,
alors $\CongTR{\sigma r}{\sigma r'}$.
\FLEM%--------------------------------------------------------------------------

\proof{
Nous procédons par induction sur le nombre $n$ de pas dans la réduction 
$r=r_1 \congr \ldots \congr r_n=r'$. Il est suffisant de montrer que si
$\Cong{r_n}{r_{n+1}}$ alors $\CongTR{\sigma r_n}{\sigma r_{n+1}}$.

Nous montrons que $\Cong{t}{t'}$ implique $\CongTR{\sigma t}{\sigma t'}$ en
utilisant une induction sur la structure du terme $t$.

Le cas de base, $t=x$ est évident puisque $t'=x$ et donc, 
$\CongTR{\sigma x}{\sigma x}$.

Nous considérons toute les formes d'un $\rho$-terme et toute les réductions
$\congr$ possibles.
\begin{enumerate}

\item $t=\{u_1,\ldots,u_m\}$
\begin{enumerate}

\item\label{sei11}
Si $t=\{u_1,\ldots,u_m\}$ et $t'=\{u_1',\ldots,u_m\}$, avec
$\Cong{u_1}{u_1'}$,
nous obtenons, par induction,
$\CongTR{\sigma u_1}{\sigma u_1'}$ et donc,
$\sigma t$ $=$
$\{\sigma u_1,\ldots,\sigma u_m\}$ $\congTR$
$\{\sigma u_1',\ldots,\sigma u_m\}$ $=$
$\sigma t'$.
Dans le cas où nous avons à la place de $u_1$ un autre sous-terme $u_k$ tel que
$\Cong{u_k}{u_k'}$ la preuve est similaire.

\item\label{sei12}

Si $t=\{u_1,\ldots,\{v_1,\ldots,v_n\},\ldots,u_m\}$ et
$t'=\{u_1,\ldots,v_1,\ldots,v_n,\ldots,u_m\}$ alors,
$\sigma t$ $=$
$\{\sigma u_1,\ldots,\{\sigma v_1,\ldots,\sigma v_n\},\ldots,\sigma u_m\}$
$\congTR$
$\{\sigma u_1,\ldots,\sigma v_1,\ldots,\sigma v_n,\ldots,\sigma u_m\}$ $=$
$\sigma t'$.

\end{enumerate}

\item $t=f(u_1,\ldots,u_m)$
\begin{enumerate}

\item\label{sei21}

Si $t=f(u_1,\ldots,u_m)$ et $t'=f(u_1',\ldots,u_m)$, avec
$\Cong{u_1}{u_1'}$,
nous obtenons, par induction,
$\CongTR{\sigma u_1}{\sigma u_1'}$ et donc,
$\sigma t$ $=$
$f(\sigma u_1,\ldots,\sigma u_m)$ $\congTR$
$f(\sigma u_1',\ldots,\sigma u_m)$ $=$
$\sigma t'$.
Dans le cas où nous avons à la place de $u_1$ un autre sous-terme $u_k$ tel que
$\Cong{u_k}{u_k'}$ la preuve est similaire.

\item\label{sei22}

Si nous considérons le terme $t=f(u_1,\ldots,\{v_1,\ldots,v_n\},\ldots,u_m)$ et
le terme
$t'=\{f(u_1,\ldots,v_1,\ldots,u_m),\ldots,f(u_1,\ldots,v_n,\ldots,u_m)\}$
alors nous obtenons la réduction
$\sigma t$ $=$
$f(\sigma u_1,\ldots,\{\sigma v_1,\ldots,\sigma v_n\},\ldots,\sigma u_m)$ 
$\congTR$\\
$\{f(\sigma u_1,\ldots,\sigma v_1,\ldots,\sigma u_m),\ldots,
f(\sigma u_1,\ldots,\sigma v_n,\ldots,\sigma u_m)\}$ $=$
$\sigma t'$

\end{enumerate}

\item $t=u \ra v$

Nous utilisons les mêmes arguments que dans les cas précédents.
\begin{enumerate}

\item\label{sei31}
Si $t=u \ra v$ et $t'=u' \ra v$ la preuve est similaire au point \ref{sei21}.

\item\label{sei32}
Si $t=u \ra v$ et $t'=u \ra v'$ la preuve est similaire au point \ref{sei21}.

\item\label{sei33}
Le terme $t=\{u_1,\ldots,u_m\} \ra v$ n'est pas un terme de $\RTTE$. Le lemme ne
serait pas valide si la relation $\congr$ était définie sur l'ensemble de termes
$\RTT$.

\item\label{sei34}
Si $t=u \ra \{v_1,\ldots,v_m\}$ et 
$t'=\{u \ra v_1,\ldots, u \ra v_m\}$ la preuve est similaire au point
\ref{sei22}.

\end{enumerate}

\item $t=[u](v)$
\begin{enumerate}

\item\label{sei41}
Si $t=[u](v)$ et $t'=[u'](v)$
la preuve est similaire au point \ref{sei21}.

\item\label{sei42}
Si $t=[u](v)$ et $t'=[u](v')$
la preuve est similaire au point \ref{sei21}.

\item\label{sei43}
Si $t=[\{u_1,\ldots,u_m\}](v)$ et $t'=\{[u_1](v),\ldots,[u_m](v)\}$
la preuve est similaire au point \ref{sei22}.

\item\label{sei44}
Si $t=[u](\{v_1,\ldots,v_m\})$ et $t'=\{[u](v_1),\ldots,[u](v_m)\}$
la preuve est similaire au point \ref{sei22}.

\end{enumerate}

\end{enumerate}
}%_____________________________________________________________________________

Dans le Lemme~\ref{substEquivInd}, l'appartenance du terme $r$ à l'ensemble de
termes $\RTTE$ est essentielle et le lemme n'est pas valide dans le cas où $r$
est un terme quelconque de $\RTT$. En particulier, la présence d'une règle de
réécriture avec un membre gauche qui n'est pas un terme du premier ordre peut
conduire à des réductions non-confluentes comme montré dans
l'Exemple~\ref{notFirstOrder}.

\EX%------------------------------------------------------------------------
\label{notFirstOrder}
Nous considérons $r= \{x,y\} \ra x$ et donc, $r'=\{x \ra x,y \ra x\}$. Pour une
substitution $\sigma=\subs{x/a}$ nous obtenons $\sigma r=\{x,y\} \ra x$ et
$\sigma r'=\{x \ra x,y \ra a\}$ et il est évident qu'il n'existe pas de
réduction $\CongTR{\sigma r}{\sigma r'}$.
\FEX%------------------------------------------------------------------------

\LEM%--------------------------------------------------------------------------
\label{substEquivRed}

Etant donnés les $\rho$-termes $l,t,r$ et $t',r'$ tels que $l,t$ soient \matchD\
$\CongTR{t}{t'}$, $\CongTR{r}{r'}$ et $l,t'$ soient \matchD. 
Si les problèmes de filtrage $(l \meqqes t)$ et $(l \meqqes t')$ ont comme solutions
les substitutions $\sigma$ et $\sigma'$ respectivement alors, 
$\CongTR{\sigma r}{\sigma' r'}$.
\FLEM%--------------------------------------------------------------------------

\proof{
Le résultat est obtenu immédiatement par transitivité en utilisant le
Lemme~\ref{substEquivBasic} et le Lemme~\ref{substEquivInd}.
}%_____________________________________________________________________________

Ce dernier lemme correspond au Lemme~\ref{substRed} où la même propriété est
obtenue pour la relation $\del$ mais en imposant une condition plus forte sur
les termes $l$ et $t$. Intuitivement, une variable $x$ du terme $r$ peut
disparaître dans le terme $r'$, où $\Del{r}{r'}$, mais ce n'est pas le cas si
$\CongTR{r}{r'}$. Par conséquent, le terme $\emptyset$ d'une substitution
$\sigma=\subs{x/\emptyset}$ peut ne pas apparaître dans le terme $\sigma r'$ dans
le cas de la relation $\del$ mais il est toujours propagé strictement dans le
cas de la relation $\congr$. Ainsi, nous imposons une condition de termes
\matchD\  afin d'obtenir les propriétés ci-dessus pour la relation $\congr$ mais
dans le cas de la relation $\del$ nous demandons en plus l'absence des ensembles 
vides dans $t$ en utilisant une condition de termes \ready.

Nous analysons par la suite les dérivations de certains $\rho$-termes contenant
des ensembles.

\LEM%--------------------------------------------------------------------------
\label{setEquivBasic} 
\comment{ dans Lemme~\ref{funEquivBasic} et Lemme~\ref{BasicCoherence}}

Etant donnés les $\rho$-termes $l,\{t\},r$ tels que $l,\{t\}$ soient \ready.
Si les problèmes de filtrage $(l \meqqes \{t\})$ et $(l \meqqes t)$ ont comme solutions
les substitutions $\sigma$ et $\sigma'$ respectivement alors, 
$\CongTR{\sigma r}{\{\sigma' r\}}$.
\FLEM%--------------------------------------------------------------------------
%%%% CEX: [x -> b](0) %%%%%%%%%%%%
%%%%      b << x < 0 >> != 0  %%%%
%%%%%%%%%%%%%%%%%%%%%%%%%%%%%%%%%%

%%%% CEX: [x -> f(x,x)]({a,b}) %%%%%%%%%%%%%%%%%%%%%%%%%%%%%%%%%%%%%%%%%%%%%%%%%
%%%%    f(x,x) << x < {a,b} >> != {f(x,x) << x < a >>, f(x,x) << x < b >>}  %%%%
%%%%%%%%%%%%%%%%%%%%%%%%%%%%%%%%%%%%%%%%%%%%%%%%%%%%%%%%%%%%%%%%%%%%%%%%%%%%%%%%

\proof{

Puisque les termes $l,\{t\}$ sont \ready, ils sont \matchD\  et par la
Remarque~\ref{positionsFonctionnelles}, $l$ est une variable $x$. Ainsi, la solution
de $(x \meqqes \{t\})$ est $\subs{x/\{t\}}=\sigma$ et la solution
de $(x \meqqes t)$ est $\subs{x/t}=\sigma'$.

Nous procédons par induction sur la structure du $\rho$-terme $r$.

\textit{Le cas de base}~: $r=x$, avec $x \in \XX$.

Nous avons $\sigma x = \subs{x/\{t\}} x = \{t\}$, 
$\{\sigma' x\} = \{ \subs{x/t} x \} = \{t\}$ et $\CongTR{\{t\}}{\{t\}}$.

\textit{Induction}~: 
\begin{enumerate}

\item $r=\{r_1,\ldots,r_m\}$

Nous avons 
$\sigma r=\sigma \{r_1,\ldots,r_m\}=\{\sigma r_1,\ldots,\sigma r_m\}$ et
$\{\sigma' r\}=\{\sigma' \{r_1,\ldots,r_m\}\}=\{\{\sigma' r_1,\ldots,\sigma'r_m\}\}$.
Par induction, $\CongTR{\sigma r_i}{\{\sigma' r_i\}}$ et donc, 
$\CongTR{\{\sigma r_1,\ldots,\sigma r_m\}}{\{\{\sigma' r_1\},\ldots,\{\sigma'r_m\}\}}$.
En plus, nous avons 
$\CongTR{\{\{\sigma' r_1\},\ldots,\{\sigma'r_m\}\}}{\{\{\sigma' r_1,\ldots,\sigma'r_m\}\}}$
et ainsi,
$\CongTR{\{\sigma r_1,\ldots,\sigma r_m\}}{\{\{\sigma' r_1,\ldots,\sigma'r_m\}\}}$.

\item $r=f(r_1,\ldots,r_m)$

Nous procédons de la même façon que dans le premier cas.

\item $r=u \ra v$

Nous procédons de la même façon que dans le premier cas.

\item $r=[u](v)$

Nous procédons de la même façon que dans le premier cas.

\end{enumerate}
}%__________________________________________________________________________________

Nous devons remarquer que la condition que les termes $l,\{t\}$ sont \ready\  est
essentielle pour garantir que $\{t\}$ n'est pas un ensemble vide et donc,
assurer l'existence de la substitution $\sigma'$. Si le nombre d'éléments de
l'ensemble $\{t\}$ n'est pas restreint alors le lemme est reformulé en~:

\textit{
Etant donnés les $\rho$-termes $x,\{t_1,\ldots,t_n\},r$ et les substitutions
$\sigma,\sigma_i$, $i=1\ldots n$, telles que 
$\{\sigma\} = \Sl(x \meqqes \{t_1,\ldots,t_n\})$ et 
$\{\sigma_i\} = \Sl(x \meqqes t_i)$. Alors, 
$\CongTR{\sigma r}{\{\sigma_1 r,\ldots,\sigma_n r\}}$.  
}

Si $n=0$ alors, il faut prouver que $\CongTR{\subs{x/\emptyset}
r}{\emptyset}$. Si $r$ est un terme clos du premier ordre alors
$\subs{x/\emptyset} r=r$ et puisque dans ce cas il n'existe pas de réduction
$\CongTR{r}{\emptyset}$ il est clair que le lemme n'est pas valide.

La possibilité d'avoir des ensembles ayant plus d'un élément mène immédiatement à
des contre-exemples. Si nous avons $r=f(x,x)$, $l=x$, $t=\{a,b\}$ avec $a,b \in
\TF$ alors $\sigma=\subs{x/\{a,b\}}$ et $\sigma_1=\subs{x/a}$,
$\sigma_2=\subs{x/b}$. Ainsi, nous devons trouver un terme $u$ tel que
$\CongTR{\{f(\{a,b\},\{a,b\})\}}{u}$ et $\CongTR{\{f(a,a),f(b,b)\}}{u}$ mais il
est évident que un tel terme n'existe pas.

\LEM%--------------------------------------------------------------------------
\label{funEquivBasic}
\comment{ dans Lemme~\ref{BasicCoherence}}

Etant donnés les $\rho$-termes $l,f(u_1,\ldots,\{t\},\ldots,u_m),r$ tels que
$l,f(u_1,\ldots,\{t\},\ldots,u_m)$ soient \ready. Si les substitutions $\sigma$
et $\sigma'$ sont les solutions des problèmes de filtrage 
\mbox{$(l \meqqes f(u_1,\ldots,\{t\},\ldots,u_m))$} et 
\mbox{$(l \meqqes f(u_1,\ldots,t,\ldots,u_m))$} respectivement alors,
$\CongTR{\sigma r}{\{\sigma' r\}}$.
\FLEM%--------------------------------------------------------------------------

\proof{

Puisque les termes $l,f(u_1,\ldots,\{t\},\ldots,u_m)$ sont \matchD\  alors, soit
$l$ est une variable, soit $l$ est de la forme $f(l_1,\ldots,l_m)$.

Si $l$ est une variable alors la preuve est très similaire à celle du
Lemme~\ref{setEquivBasic}.

Si $l=f(l_1,\ldots,l_m)$, puisque $l,f(u_1,\ldots,\{t\},\ldots,u_m)$ sont
\matchD\  alors $l_i,u_i$ ($i=1 \ldots m$) sont \matchD. Par conséquent, si
$\{t\}$ est le $k$-ième argument alors $l_k=x$ et la preuve est similaire à
celle du Lemme~\ref{setEquivBasic}.
}%_________________________________________________________________________________

\LEM%--------------------------------------------------------------------------
\label{BasicCoherence}
\comment{ dans Lemme~\ref{DiamDelE} et Lemme~\ref{GeneralCoherence} et Lemme~\ref{ConfCongDEL}}

Etant donnés les $\rho$-termes $t,t'$ et $s$ tels que $\Cong{t}{t'}$ et
$\Del{t}{s}$. Alors, il existe un terme $s'$ tel que $t' \congTR \del \congTR s'$
et $\CongTR{s}{s'}$~:
\begin{center}$~$
\xymatrix{%@C+10pt{ 
& t
\ar[dl]_{\congrxy} \ar[dr]^-{\delxy}
& \\
t'
\ar@{.{>}}[dr]_{\congTRxy~\delxy~\congTRxy} && 
s
\ar@{.{>}}[dl]^-{\congTRxy} \\
& 
s'
&
}
\end{center}

\FLEM%--------------------------------------------------------------------------

\proof{

Nous pouvons reformuler le lemme en décrivant toutes les étapes intermédiaires~:
\textit{
Etant donnés les $\rho$-termes $t,t'$ et $s$ tels que $\Cong{t}{t'}$ et
$\Del{t}{s}$. Alors, il existe les termes $s',s''$ et $t''$ tel que
$\CongTR{t'}{t''}$, $\Del{t''}{s''}$, $\CongTR{s''}{s'}$ et $\CongTR{s}{s'}$.}
\begin{center}$~$
\xymatrix{ 
&& t
\ar[dl]_{\congrxy} \ar[dr]^-{\delxy}
& \\
& t'
\ar@{.{>}}[dl]_{\congTRxy} && 
s
\ar@{.{>}}[dl]^-{\congTRxy} \\
t''
\ar@{.{>}}[dr]_{\delxy} &&
s'
& \\
& s''
\ar@{.{>}}[ur]_{\congTRxy}
&&
}
\end{center}

Nous procédons par induction sur la structure du terme $t$ et nous analysons
toutes les réductions $\congr$ et $\del$ possibles. 

\textit{Le cas de base}~: 
Si $t$ est une variable les réductions sont triviales.

\textit{Induction}~: 
Les cas plus élaborés sont obtenus quand la règle d'évaluation \rname{Fire_c}
peut être appliquée au terme $t$ à la position de tête afin d'obtenir le terme
$s$. Nous présentons par la suite tous les cas possibles et nous insistons sur
les plus compliqués.

\begin{enumerate}

\item $t$ est un ensemble.%<<<<<<<<<<<<<<<<<<<<<<<<<<<<<<<<<<<<<<<<<<<<<<<<<

Nous avons deux possibilités pour réduire $t$ en $t'$~:
\begin{enumerate}
\item\label{c11}
$t = \{u_1,\ldots,u_k,\ldots,u_n\}$ et
$t' = \{u_1,\ldots,v_k,\ldots,u_n\}$ avec $\Cong{u_k}{v_k}$.

Si $\Del{u_i}{u_i^1}$, $i=1 \ldots n$, alors, nous avons~:
\begin{center}$~$
\xymatrix{  
\{u_1,\ldots,u_k,\ldots,u_n\}
\ar[d]_{\congrxy}
\ar[r]^-{\delxy}
&
\{u_1^1,\ldots,u_k^1,\ldots,u_n^1\}
\ar@{.{>}}[d]^-{\congTRxy}
\\
\{u_1,\ldots,v_k,\ldots,u_n\}
\ar@{.{>}}[d]_{\congTRxy}
&
\{u_1^1,\ldots,u_k^3,\ldots,u_n^1\}
\\
\{u_1,\ldots,w_k,\ldots,u_n\}
\ar@{.{>}}[r]^-{\delxy}
&
\{u_1^1,\ldots,u_k^2,\ldots,u_n^1\}
\ar@{.{>}}[u]_{\congTRxy}
}
$~~~$ puisque $~~~$% par induction
\xymatrix{  
u_k
\ar[d]_{\congrxy}
\ar[r]^-{\delxy}
&
u_k^1
\ar@{.{>}}[d]^-{\congTRxy}
& \\
v_k
\ar@{.{>}}[d]_{\congTRxy}
&
u_k^3
\\
w_k
\ar@{.{>}}[r]^-{\delxy}
&
u_k^2
\ar@{.{>}}[u]_{\congTRxy}
}
\end{center}%\\
est obtenu par induction.

\item\label{c12}
$t=\{u_1,\ldots,\{v_1,\ldots,v_n\},\ldots,u_m\}$ et 
$t'=\{u_1,\ldots,v_1,\ldots,v_n,\ldots,u_m\}$.

Si $\Del{u_i}{u_i^1}$, $\Del{v_j}{v_j^1}$, $i=1 \ldots m$, $j=1 \ldots n$,
alors, nous avons~:
\begin{center}$~$
\xymatrix{
\{u_1,\ldots,\{v_1,\ldots,v_n\},\ldots,u_m\} 
\ar[d]_{\congrxy}
\ar[r]^-{\delxy}
&
\{u_1^1,\ldots,\{v_1^1,\ldots,v_n^1\},\ldots,u_m^1\}
\ar@{.{>}}[d]^-{\congrxy}
\\
\{u_1,\ldots,v_1,\ldots,v_n,\ldots,u_m\}
\ar@{.{>}}[r]^-{\delxy}
&
\{u_1^1,\ldots,v_1^1,\ldots,v_n^1,\ldots,u_m^1\}
}
\end{center}

\end{enumerate}

\item $t$ est de la forme $f(u_1,\ldots,u_n)$%<<<<<<<<<<<<<<<<<<<<<<
\begin{enumerate}
\item\label{c21}
$t=f(u_1,\ldots,u_k,\ldots,u_m)$ et
$t'=f(u_1,\ldots,v_k,\ldots,u_m)$ avec $\Cong{u_k}{v_k}$

Similaire au cas~\ref{c11}.

\item\label{c22}
$t=f(u_1,\ldots,\{v_1,\ldots,v_n\},\ldots,u_m)$,\\
$t'=\{f(u_1,\ldots,v_1,\ldots,u_m),\ldots,f(u_1,\ldots,v_n,\ldots,u_m)\}$.

Similaire au cas~\ref{c12}.

\end{enumerate}

\item $t$ est de la forme $u \ra v$%<<<<<<<<<<<<<<<<<<<<<<
\begin{enumerate}
\item \label{c31}
$t=u \ra v$, $t'=u' \ra v$ avec $\Cong{u}{u'}$.

Similaire au cas~\ref{c11}.

\item \label{c32}
$t=u \ra v$, $t'=u \ra v'$ avec $\Cong{v}{v'}$.

Similaire au cas~\ref{c11}.

\item \label{c33}
$t=\{l_1,\ldots,l_n\} \ra r$, $t'=\{l_1 \ra r,\ldots,l_n \ra r\}$.

Similaire au cas~\ref{c12}.

\item \label{c34}
$t=l \ra \{r_1,\ldots,r_n\}$, $t'=\{l \ra r_1,\ldots,l \ra r_n\}$.

Similaire au cas~\ref{c12}.

\end{enumerate}

\item $t$ est de la forme $[v](u)$ et les règles d'évaluation \rname{Fire_c} et
\rname{Congruence} ne sont pas appliquées à la position de tête %<<<<<<<<<<<<<<<<<<<<<<
\begin{enumerate}

\item\label{c41}
$t = [v](u)$, $t'=[v'](u)$ avec $\Cong{v}{v'}$.

Similaire au cas~\ref{c11}.

\item\label{c42}
$t = [v](u)$, $t'=[v](u')$ avec $\Cong{v}{v'}$.

Similaire au cas~\ref{c11}.

\item\label{c43}
$t=[\{r_1,\ldots,r_n\}](u)$, $t'=\{[r_1](p),\ldots,[r_n](p)\}$.

Similaire au cas~\ref{c12}.

\item\label{c44}
$t=[r](\{u_1,\ldots,u_n\})$, $t'=\{[r](u_1),\ldots,[r](u_n)\}$.

Similaire au cas~\ref{c12}.

\end{enumerate}

\item $t$ est de la forme $[l \ra r](u)$ avec $l,u$ \ready.%<<<<<<<<<<<<<<<<<<<<<<
\begin{enumerate}

\item\label{c51}
$t = [l \ra r](u)$, $t'=[l' \ra r](u)$ avec $\Cong{l}{l'}$,

Si $l,u$ sont \ready\  alors $l=l'$ et le lemme est trivialement vrai. Si $l,u$
ne sont pas \ready\  alors nous ne pouvons pas appliquer la règle d'évaluation
\rname{Fire_c} à la position de tête et donc, la preuve est similaire à celle du
cas~\ref{c11}.

\item\label{c52}
$t = [l \ra r](u)$, $t'=[l \ra r'](u)$ avec $\Cong{r}{r'}$,

Nous considérons les termes $r^1,u^1$ tels que $\Del{r}{r^1}$, $\Del{u}{u^1}$ et
la substitution $\sigma$ telle que $\sigma \in \Sl(l \meqqes u^1)$. Alors, nous avons~:
\begin{center}$~$
\xymatrix{  
[l \ra r](u)
\ar[d]_{\congrxy}
\ar[r]^-{\delxy}
&
\{\sigma r^1\}
\ar@{.{>}}[d]^-{\congTRxy}
\\
[l \ra r'](u)
\ar@{.{>}}[d]_{\congTRxy}
&
\{\sigma r^3\}
\\
[l \ra r''](u)
\ar@{.{>}}[r]^-{\delxy}
&
\{\sigma r^2\}
\ar@{.{>}}[u]_{\congTRxy}
}
$~~~$ puisque par induction $~~~$
\xymatrix{  
r
\ar[d]_{\congrxy}
\ar[r]^-{\delxy}
&
r^1
\ar@{.{>}}[d]^-{\congTRxy}
& \\
r'
\ar@{.{>}}[d]_{\congTRxy}
&
r^3
\\
r''
\ar@{.{>}}[r]^-{\delxy}
&
r^2
\ar@{.{>}}[u]_{\congTRxy}
}
\end{center}
et nous appliquons le Lemme~\ref{substEquivInd}. 
Le cas où $\Sl(l \meqqes u^1)=\emptyset$ est trivial.

\item\label{c53}
$t = [l \ra r](u)$, $t'=[l \ra r](u')$ avec $\Cong{u}{u'}$,

Nous considérons d'abord que $l,u$ sont \ready\  et $l,u'$ sont \ready\  et nous
obtenons par induction~:
\begin{center}$~$
\xymatrix{  
u
\ar[d]_{\congrxy}
\ar[r]^-{\delxy}
&
u^1
\ar@{.{>}}[d]^-{\congTRxy}
& \\
u'
\ar@{.{>}}[d]_{\congTRxy}
&
u^3
\\
u''
\ar@{.{>}}[r]^-{\delxy}
&
u^2
\ar@{.{>}}[u]_{\congTRxy}
}
\end{center}

Nous voulons avoir $l,u''$ \ready\  et le seul cas où la propriété n'est pas vraie
est obtenu pour les termes de la forme $u=f(\ldots,v,\ldots)$, $u'=f(\ldots,\{v'\},\ldots)$
et $u''=\{f(\ldots,v',\ldots)\}$ mais nous avons
\begin{center}$~$
\xymatrix{  
v
\ar[d]_{\congrxy}
\ar[r]^-{\delxy}
&
w
\ar@{.{>}}[d]^-{\congTRxy}
\\
\{v'\}
\ar@{.{>}}[d]_{\congTRxy}
&
w'
\\
\{v''\}
\ar@{.{>}}[r]^-{\delxy}
&
w''
\ar@{.{>}}[u]_{\congTRxy}
}
$~~~$ et donc $~~~$
\xymatrix{  
f(\ldots,v,\ldots)
\ar[d]_{\congrxy}
\ar[r]^-{\delxy}
&
s
\ar@{.{>}}[d]^-{\congTRxy}
& \\
f(\ldots,\{v'\},\ldots)
\ar@{.{>}}[d]_{\congTRxy}
&
s'
\\
f(\ldots,\{v''\},\ldots)
\ar@{.{>}}[r]^-{\delxy}
&
s''
\ar@{.{>}}[u]_{\congTRxy}
}
\end{center}

Par conséquent, nous pouvons trouver un terme $u''$ satisfaisant le diagramme
précédent et tel que si $l,u'$ sont \ready\  alors $l,u''$ sont \ready. Nous
considérons $\Del{r}{r^1}$ et en utilisant le Lemme~\ref{substEquivBasic}
nous obtenons~:
\begin{center}$~$
\xymatrix{
[l \ra r](u)
\ar[d]_{\congrxy}
\ar[r]^-{\delxy}
&
\{\sigma_1 r^1\}
\ar@{.{>}}[d]^-{\congTRxy}
\\
[l \ra r](u')
\ar@{.{>}}[d]_{\congTRxy}
&
\{\sigma_3 r^1\}
\\
[l \ra r](u'')
\ar@{.{>}}[r]^-{\delxy}
&
\{\sigma_2 r^1\}
\ar@{.{>}}[u]_{\congTRxy}
}
\end{center}

avec  $\sigma_1,\sigma_2,\sigma_3$ telles que 
$\sigma_1 \in \Sl(l \meqqes u^1)$, $\sigma_2 \in \Sl(l \meqqes u^2)$ et
$\sigma_3 \in \Sl(l \meqqes u^3)$.
Si le filtrage $(l \meqqes u^1)$ échoue alors, en utilisant le
Lemme~\ref{matchCongOne} nous obtenons que le filtrage 
$(l \meqqes u^2)$ échoue et le lemme est clairement vrai.

Nous traitons maintenant le cas où $l,u$ sont \ready\  mais $l,u'$ ne sont pas
\ready. Si $t=[l \ra r](f(\ldots,\{u\},\ldots))$ et 
$t'=[l \ra r](\{f(\ldots,u,\ldots)\})$ alors, $l$ doit être une variable ou de
la forme $l=f(l_1,\ldots,l_n)$ avec $l_i,u_i$ ($i=1 \ldots n$) \ready. Dans le
dernier cas, la position dans le terme $l$ correspondant à la position du terme
$\{u\}$ dans le terme $f(\ldots,\{u\},\ldots)$ est une position variable. Ainsi,
dans les deux cas les termes $l,f(\ldots,u,\ldots)$, $i=1 \ldots n$, sont
\ready.

Nous considérons $\Del{r}{r^1}$, $\Del{u}{u^1}$ et les substitutions
$\sigma,\sigma'$ telles que nous avons $\sigma \in \Sl(l \meqqes f(\ldots,\{u^1\},\ldots))$
et $\sigma' \in \Sl(l \meqqes f(\ldots,u^1,\ldots))$.
Puisque $l,f(\ldots,\{u\},\ldots)$ sont \ready, par le
Lemme~\ref{readyLemmaDel} $l,f(\ldots,\{u^1\},\ldots)$ sont \ready\  et
nous obtenons, en utilisant le Lemme~\ref{funEquivBasic},
\begin{center}$~$
\xymatrix{  
[l \ra r](f(\ldots,\{u\},\ldots))
\ar[d]_{\congrxy}
\ar[r]^-{\delxy}
&
\{\sigma r^1\}
\ar@{.{>}}[dd]^-{\congTRxy}
\\
[l \ra r](\{f(\ldots,u,\ldots)\})
\ar@{.{>}}[d]_{\congrxy}
&
\\
\{[l \ra r](f(\ldots,u,\ldots))\}
\ar@{.{>}}[r]^-{\delxy}
&
\{\{\sigma' r^1\}\}
}
\end{center}

Conformément à la Remarque~\ref{echecMatch}, le filtrage $(l \meqqes
f(\ldots,\{u^1\},\ldots))$ peut échouer seulement à cause des symboles
fonctionnels différents à la même position des termes $l$ et
$f(\ldots,\{u^1\},\ldots)$ et dans ce cas, le filtrage 
$(l \meqqes f(\ldots,u^1,\ldots))$ échoue aussi et le lemme est trivialement
vrai.

\item\label{c54}
$t=[l \ra \{r_1,\ldots,r_n\}](u)$, $t'=[\{l \ra r_1,\ldots,l \ra r_n\}](u)$.

Si nous considérons $\Del{r_i}{r_i^1}$, $i=1 \ldots n$, $\Del{u}{u^1}$ et la
substitution $\sigma$ telle que $\sigma \in \Sl(l \meqqes u^1)$, puisque $l,u$
sont \ready, nous obtenons le diagramme~:
\begin{center}$~$
\xymatrix{  
[l \ra \{r_1,\ldots,r_n\}](u)
\ar[d]_{\congrxy}
\ar[r]^-{\delxy}
&
\{\sigma \{r_1^1,\ldots,r_n^1\}\}
\ar@{.{>}}[d]^-{\congTRxy}
\\
[\{l \ra r_1,\ldots,l \ra r_n\}](u)
\ar@{.{>}}[d]_{\congrxy}
&
\{\sigma r_1^1,\ldots, \sigma r_n^1\}
\\
\{[l \ra r_1](u),\ldots,[l \ra r_n](u)\}
\ar@{.{>}}[r]^-{\delxy}
&
\{\{\sigma r_1^1\},\ldots, \{\sigma r_n^1\}\}
\ar@{.{>}}[u]_{\congTRxy}
}
\end{center}

Le cas où $\Sl(l \meqqes u^1)=\emptyset$ est trivial.

\item\label{c55}
$t=[l \ra r](\{u\})$, $t'=\{[l \ra r](u)\}$.

Nous considérons $\Del{u}{u^1}$, $\Del{r}{r^1}$ et les substitutions
$\sigma,\sigma'$ telles que $\sigma \in \Sl(l \meqqes \{u^1\})$ et
$\sigma' \in \Sl(l \meqqes u^1)$.  Puisque $l,\{u\}$ sont \ready, par le
Lemme~\ref{readyLemmaDel} $l,\{u^1\}$ sont \ready\  et par la
Remarque~\ref{positionsFonctionnelles}, $l$ est une variable et donc,
les filtrages $l \meqqes \{u^1\}$ et $l \meqqes u^1$ n'échouent pas.
En utilisant le Lemme~\ref{setEquivBasic}, nous obtenons le diagramme~:
\begin{center}$~$
\xymatrix{  
[l \ra r](\{u\})
\ar[dd]_{\congrxy}
\ar[r]^-{\delxy}
&
\{\sigma r^1\}
\ar@{.{>}}[dd]^-{\congTRxy}
\\
&
\\
\{[l \ra r](u)\}
\ar@{.{>}}[r]^-{\delxy}
&
\{\{\sigma' r^1\}\}
}
\end{center}

\end{enumerate}

\item $t$ est de la forme $[f(u_1,\ldots,u_n)](f(v_1,\ldots,v_n))$ et nous
appliquons la règle d'évaluation \rname{Congruence} à la position de tête.
\begin{enumerate}
\item \label{c61}
$t = [f(u_1,\ldots,u_n)](f(v_1,\ldots,v_n))$, 
$t' = [f(u_1',\ldots,u_n)](f(v_1,\ldots,v_n))$ 
et $\Cong{u_1}{u_1'}$.

Si $\Del{u_i}{u_i^1}$, $\Del{v_i}{v_i^1}$, $i=1 \ldots n$,
alors, nous avons~:
\begin{center}$~$
\xymatrix{  
[f(u_1,\ldots,u_n)](f(v_1,\ldots,v_n))
\ar[d]_{\congrxy}
\ar[r]^-{\delxy}
&
\{f([u_1^1](v_1^1),\ldots,[u_n^1](v_n^1))\}
\ar@{.{>}}[d]^-{\congTRxy}
\\
[f(u_1',\ldots,u_n)](f(v_1,\ldots,v_n))
\ar@{.{>}}[d]_{\congTRxy}
&
\{f([u_1^3](v_1^1),\ldots,[u_n^1](v_n^1))\}
\\
[f(u_1'',\ldots,u_n)](f(v_1,\ldots,v_n))
\ar@{.{>}}[r]^-{\delxy}
&
\{f([u_1^2](v_1^1),\ldots,[u_n^1](v_n^1))\}
\ar@{.{>}}[u]_{\congTRxy}
}
\end{center}

puisque par induction le diagramme est vrai pour le terme $u_1$ et ses réduits.

\item \label{c62}
$t = [f(u_1,\ldots,u_n)](f(v_1,\ldots,v_n))$, 
$t' = [f(u_1,\ldots,u_n)](f(v_1',\ldots,v_n))$
et $\Cong{v_1}{v_1'}$.

Similaire au cas~\ref{c61}.

\item \label{c63}
$t = [f(s_1,\ldots,\{u_1,\ldots,u_n\},\ldots,s_m)](f(v_1,\ldots,v_m))$,\\
$t' = [\{f(s_1,\ldots,u_1,\ldots,s_m),\ldots,f(s_1,\ldots,u_n,\ldots,s_m)\}](f(v_1,\ldots,v_m))$.

Pour simplicité, nous supposons que 
$t = [f(\{u_1,\ldots,u_n\})](f(v))$ et donc,
$t' = [\{f(u_1),\ldots,f(u_n)\}](f(v))$.
Le cas général est traité exactement de la même façon.

Si nous considérons $\Del{v^1}{v^1}$, $\Del{u_i^1}{u_i^1}$,
$i=1 \ldots n$, nous obtenons~:
\begin{center}$~$
\xymatrix{
[f(\{u_1,\ldots,u_n\})](f(v))
\ar[d]_{\congrxy}
\ar[r]^-{\delxy}
&
\{f([\{u_1^1,\ldots,u_n^1\}](v^1))\}
\ar@{.{>}}[d]^-{\congrxy}
\\
[\{f(u_1),\ldots,f(u_n)\}](f(v))
\ar@{.{>}}[ddd]_{\congrxy}
&
\{f(\{[u_1^1](v^1),\ldots,[u_n^1](v^1)\})\}
\ar@{.{>}}[d]^-{\congrxy}
\\
&
\{\{f([u_1^1](v^1)),\ldots,f([u_n^1](v^1))\}\}
\ar@{.{>}}[d]^-{\congrxy}
\\
&
\{f([u_1^1](v^1)),\ldots,f([u_n^1](v^1))\}
\\
\{[f(u_1)](f(v)),\ldots,[f(u_n)](f(v))\}
\ar@{.{>}}[r]^-{\delxy}
&
\{\{f([u_1^1](v^1))\},\ldots,\{f([u_n^1](v^1))\}\}
\ar@{.{>}}[u]_{\congTRxy}
}
\end{center}

\item \label{c64}
$t = [f(u_1,\ldots,u_m)](f(s_1,\ldots,\{v_1,\ldots,v_n\},\ldots,s_m))$,\\
$t' = [f(u_1,\ldots,u_m)](\{f(s_1,\ldots,v_1,\ldots,s_m),\ldots,f(s_1,\ldots,v_n,\ldots,s_m)\})$.

Similaire au cas~\ref{c63}.

\end{enumerate}

\item $t$ est de la forme $[f(u_1,\ldots,u_n)](g(v_1,\ldots,v_m))$ et nous
appliquons la règle d'évaluation \rname{Congruence\_fail} à la position de tête.
\begin{enumerate}
\item \label{c71}
$t = [f(u_1,\ldots,u_n)](g(v_1,\ldots,v_m))$, 
$t' = [f(u_1',\ldots,u_n)](g(v_1,\ldots,v_m))$,
$\Cong{u_1}{u_1'}$.

Nous obtenons immédiatement~:
\begin{center}$~$
\xymatrix{  
[f(u_1,\ldots,u_n)](g(v_1,\ldots,v_m))
\ar[d]_{\congrxy}
\ar[rr]^-{\delxy}
&&
\emptyset
\\
[f(u_1',\ldots,u_n)](g(v_1,\ldots,v_m))
\ar@{.{>}}[urr]_{\delxy}
&&
}
\end{center}

\item \label{c72}
$t = [f(u_1,\ldots,u_n)](g(v_1,\ldots,v_m))$, 
$t' = [f(u_1,\ldots,u_n)](g(v_1',\ldots,v_m))$,
$\Cong{v_1}{v_1'}$.

Similaire au cas~\ref{c71}.

\item \label{c73}
$t = [f(\ldots,\{u_1,\ldots,u_n\},\ldots)](g(v_1,\ldots,v_m))$,\\
$t' = [\{f(\ldots,u_1,\ldots),\ldots,f(\ldots,u_n,\ldots)\}](g(v_1,\ldots,v_m))$.

Nous obtenons immédiatement~:
\begin{center}$~$
\xymatrix{  
[f(\ldots,\{u_1,\ldots,u_n\},\ldots)](g(v_1,\ldots,v_m))
\ar[d]_{\congrxy}
\ar[r]^-{\delxy}
&
\emptyset
\\
[\{f(\ldots,u_1,\ldots),\ldots,f(\ldots,u_n,\ldots)\}](g(v_1,\ldots,v_m))
\ar@{.{>}}[d]_{\congrxy}
&
\\
\{[f(\ldots,u_1,\ldots)](g(v_1,\ldots,v_m)),\ldots,[f(\ldots,u_n,\ldots)](g(v_1,\ldots,v_m))\}
\ar@{.{>}}[r]^-{\delxy}
&
\{\emptyset,\ldots,\emptyset\}
\ar@{.{>}}[uu]_{\congTRxy}
}
\end{center}

\item \label{c74}
$t = [f(u_1,\ldots,u_n)](g(\ldots,\{v_1,\ldots,v_m\},\ldots))$,\\
$t' = [f(u_1,\ldots,u_n)](\{g(\ldots,v_1,\ldots),\ldots,g(\ldots,v_m,\ldots)\})$. 

Similaire au cas~\ref{c73}.

\end{enumerate}

\end{enumerate}
}

En utilisant le lemme précédent et la terminaison de la relation $\congr$
(Lemme~\ref{CongRterm}) nous obtenons le résultat intermédiaire ci-dessous qui
nous permet de prouver la confluence forte de la relation $\delE$.

\LEM%--------------------------------------------------------------------------
\label{GeneralCoherence}
\comment{ dans Lemme~\ref{DiamDelE}}

Etant donnés les termes $t,s$ et $\normF{t}$, avec $\normF{t}$ représentant la
forme normale du terme $t$ par rapport à la relation $\congTR$. Si $\Del{t}{s}$
alors, il existe un terme $s'$ tel que $\Del{\normF{t}}{s'}$ et $\CongE{s}{s'}$~:
\begin{center}$~$
\xymatrix{ 
& t
\ar[dl]_{\congTRxy} \ar[dr]^-{\delxy}
& \\
\normF{t}
\ar@{.{>}}[dr]_{\delxy} && 
s
\ar@{{<}.{>}}[dl]^-{\congExy} \\
& s'
&
}
\end{center}

\FLEM%--------------------------------------------------------------------------

\proof{
Puisque la relation $\congr$ est terminante, nous pouvons utiliser une induction
sur le nombre d'étapes dans la réduction $\CongTR{t}{t'}$ et obtenir, en utilisant le
Lemme~\ref{BasicCoherence}~:
\begin{center}$~$
\xymatrix{%@C+10pt{ 
& t
\ar[dl]_{\congTRxy} \ar[dr]^-{\delxy}
& \\
t'
\ar@{.{>}}[dr]_{\congTRxy~\delxy~\congTRxy} && 
s
\ar@{.{>}}[dl]^-{\congTRxy} \\
& 
s'
&
}
\end{center}

Si $t' = \normF{t}$ alors, $t'$ ne peut pas être réduit en utilisant la relation 
$\congTR$ et donc, nous obtenons le diagramme du lemme.
}

\LEM($\delE$ est fortement confluente)%----------------------------------
\label{DiamDelE}

Etant donnés les termes $t,t_1,t_2$ tels que $\DelE{t}{t_1}$ et
$\DelE{t}{t_2}$. Alors, il existe un terme $w$ tel que $\DelE{u}{w}$ et
$\DelE{v}{w}$~:
\begin{center}$~$
\xymatrix{ 
& t
\ar[dl]_{\delExy} \ar[dr]^-{\delExy}
& \\
t_1
\ar@{.{>}}[dr]_{\delExy} && 
t_2
\ar@{.{>}}[dl]^-{\delExy} \\
& w &
}
\end{center}

\FLEM%--------------------------------------------------------------------------

\proof{

Puisque la relation $\congr$ a la propriété de Church-Rosser, si nous
considérons deux termes $u,v$ tels que $\CongE{u}{v}$ alors, le termes $u,v$ ont
la même forme normale et nous notons cette forme normale par $t$.
Nous utilisons le Lemme~\ref{BasicCase} et Lemme~\ref{GeneralCoherence}  et nous 
obtenons~:
\begin{center}$~$
\xymatrix{%@C+10pt@R+10pt{ 
&& u
\ar@{{<}{-}{>}}[rr]^-{\congExy}
\ar[dll]_{\delxy}
\ar@{.{>}}[dr]^-{\congTRxy}
&& v
\ar[drr]^-{\delxy}
\ar@{.{>}}[dl]_{\congTRxy}
&&
\\
u'
\ar@{{<}.{>}}[drr]_{\congExy}
&&& t
\ar@{.{>}}[dl]_{\delxy}
\ar@{.{>}}[dr]^-{\delxy}
&&& v'
\ar@{{<}.{>}}[dll]^-{\congExy}
\\
& 
& u''
\ar@{.{>}}[dr]_{\delxy}
&& v''
\ar@{.{>}}[dl]^-{\delxy}
& 
& \\
&&&
w
&&&
}
\end{center}

En utilisant ce dernier diagramme et la définition de la relation $\delE$ nous
obtenons~:
\begin{center}$~$
\xymatrix@C+20pt@R+30pt{ 
&& u 
\ar[dl]_{\delxy}
& t 
\ar@{{<}{-}{>}}[l]_{\congExy}
\ar@{{<}{-}{>}}[r]^-{\congExy}
\ar@{={>}}[dl]^-{\delExy}_{def~~~~}
\ar@{={>}}[dr]_{\delExy}^-{~~~~def}
& v 
\ar[dr]^-{\delxy}
&&
\\
u''
\ar@{{<}.{>}}[r]^-{\congExy}
\ar@{.{>}}[drrr]_{\delxy}^-{~~~~def}
& u'
\ar@{{<}{-}{>}}[r]^-{\congExy}
& t_1
\ar@{=={>}}[dr]^-{\delExy}
&& t_2
\ar@{=={>}}[dl]_{\delExy}
& v' 
\ar@{{<}{-}{>}}[l]_{\congExy}
& v''
\ar@{{<}.{>}}[l]_{\congExy}
\ar@{.{>}}[dlll]^-{\delxy}_{def~~~~}
\\
&&& 
w
&&&
}
\end{center}

}%____________________________________________________________________________________

Ayant prouvé la confluence forte de la relation $\delE$, nous devons montrer que
la relation $\roZETR$ est la fermeture transitive de la relation $\delE$ pour
obtenir la confluence de la relation $\roZE$. Nous avons déjà démontré que
la relation $\roTR$ est la fermeture transitive de la relation $\del$ et ce
résultat s'étend naturellement quand nous considérons les deux relations modulo
la relation $\congTR$.

\LEM%--------------------------------------------------------------------------
\label{TransClosure}
\comment{dans Théorème~\ref{ConfROZE} et dans Théorème~\ref{ConfCongRO}}

La relation $\roZETR$ est la fermeture transitive de la relation $\delE$.
\FLEM%--------------------------------------------------------------------------

\proof{ 

Conformément au Lemme~\ref{TransClosureRO} nous avons
$\roZ\   \subseteq\   \del\   \subseteq\   \roTR$.
Nous devons prouver que $\roZE\  \subseteq\  \delE$ et $\delE\  \subseteq\  \roZETR$.

Pour l'inclusion $\roZE\  \subseteq\  \delE$ il suffit de voir que pour tous termes 
$u',v'$ tels que $\RoZ{u'}{v'}$ nous avons $\Del{u'}{v'}$ et donc,
si
%\begin{center}$~$
\xymatrix@C+5pt{ 
u
\ar@{{<}.{>}}[d]_{\congExy}
\ar[r]^{\roZExy}
&
v
\ar@{{<}.{>}}[d]^-{\congExy}
\\
u'
\ar[r]^{\roZxy}
&
v'
}
%$~~~$ 
alors 
%$~~~$
\xymatrix@C+5pt{ 
u
\ar@{{<}.{>}}[d]_{\congExy}
\ar[r]^{\delExy}
&
v
\ar@{{<}.{>}}[d]^-{\congExy}
\\
u'
\ar[r]^{\delxy}
&
v'
}
%\end{center}

Pour $\delE\  \subseteq\  \roZETR$ nous devons prouver que si 
%\begin{center}$~$
\xymatrix@C+5pt{ 
u
\ar[r]^{\delExy}
\ar@{{<}.{>}}[d]_{\congExy}
&
v
\ar@{{<}.{>}}[d]^-{\congExy}
\\
u'
\ar[r]^{\delxy}
&
v'
}
%\end{center}
alors\\
%\begin{center}$~$
\xymatrix@C+5pt{ 
u
\ar@{{<}.{>}}[d]_{\congExy}
\ar[r]^{\roZExy}
&
u_2
\ar@{{<}.{>}}[d]^-{\congExy}
\\
t_1
\ar[r]^{\roZxy}
&
t_2
}
\xymatrix@C+5pt{ 
u_2
\ar@{{<}.{>}}[d]_{\congExy}
\ar[r]^{\roZExy}
&
u_3
\ar@{{<}.{>}}[d]^-{\congExy}
\\
t_2'
\ar[r]^{\roZxy}
&
t_3
}
$\ldots$
\xymatrix@C+5pt{ 
u_{n-1}
\ar@{{<}.{>}}[d]_{\congExy}
\ar[r]^{\roZExy}
&
v
\ar@{{<}.{>}}[d]^-{\congExy}
\\
t_{n-1}'
\ar[r]^{\roZxy}
&
t_n
}
%\end{center}

Puisque $\del\   \subseteq\  \roTR$, nous pouvons choisir $t_i=t_i'$, $i=2 \ldots
n-1$ pour prouver l'inclusion.
}%________________________________________________________________________________

\TH($\roZETR$ est fortement confluente, $\roZE$ est confluente)%------------------
\label{ConfROZE}

Etant donnés les $\rho$-termes $t,u,v$ tels que $\RoZETR{t}{u}$ et
$\RoZETR{t}{v}$. Alors, il existe un terme $w$ tel que $\RoZETR{u}{w}$ et
$\RoZETR{v}{w}$.
\FTH%--------------------------------------------------------------------------

\proof{

Conformément au Lemme~\ref{TransClosure}, la relation $\roZETR$ est la fermeture
transitive de la relation $\delE$. Le Lemme~\ref{DiamDelE} montre la confluence
forte de la relation $\delE$. Ainsi, par le Lemme~\ref{diamondTransitive}, la
relation $\roZETR$ est fortement confluente et par conséquent, la relation
$\roZE$ est confluente.
}%________________________________________________________________________________

\subsubsection{Confluence de la relation $\congTR\roZ\congTR$}
%================================================================

Jusqu'à maintenant nous avons considéré que les règles d'évaluation du \roCal\
sont soit des règles de \textit{déduction} soit des règles de \textit{calcul} et
nous avons analysé la relation induite par les règles de \textit{déduction}
modulo la congruence générée par les règles de \textit{calcul}.

Une deuxième approche consiste à considérer la relation habituelle induite par
les règles d'évaluation du calcul correspondant à la relation
$\congTR\roZ\congTR$. Dans cette section nous nous concentrons sur la confluence
de cette relation. La preuve est réalisée en utilisant le Lemme de
Yokouchi~\cite{YokouchiHikita90} (voir Section~\ref{relationsBinaires}) qui a
déjà été utilisé dans~\cite{CurienHardinLevy-JACM96} pour montrer la confluence
du $\lambda\sigma_{\Uparrow}$-calcul.

\LEM%--------------------------------------------------------------------------
\label{ConfCongDEL}
La relation $\congTR\del\congTR$ est fortement confluente.
\FLEM%--------------------------------------------------------------------------

\proof{

Les hypothèses pour le Lemme de Yokouchi ont été prouvées dans le
Lemme~\ref{CongRconfluent} et le Lemme~\ref{CongRterm} (confluence et
terminaison de la relation $\congr$), le Lemme~\ref{BasicCase} (confluence forte
de la relation $\del$) et le Lemme~\ref{BasicCoherence} (le diagramme de
Yokouchi).

}%________________________________________________________________________________

\TH%--------------------------------------------------------------------------
\label{ConfCongRO}
La relation $\congTR\roZ\congTR$ est confluente.
\FTH%--------------------------------------------------------------------------

\proof{

Conformément au Lemme~\ref{TransClosure} nous avons~:
$$\roZ\   \subseteq\   \del\   \subseteq\   \roTR$$
et donc
$$\congTR\roZ\congTR\   \subseteq\   \congTR\del\congTR\   \subseteq\   (\congTR\roZ\congTR)^*$$
où $(\congTR\roZ\congTR)^*$ représente la fermeture transitive de
$\congTR\roZ\congTR$.

Puisque dans le Lemme~\ref{ConfCongDEL} nous montrons la confluence forte de la
relation $\congTR\del\congTR$ alors, la relation $\congTR\roZ\congTR$ est
confluente.
}%________________________________________________________________________________

\section{Conditions alternatives pour obtenir la confluence} \label{stratOpConf}
%================================================================

Dans la Section~\ref{strat_confluente} nous avons proposé une stratégie
\textit{ConfStrat} pour guider les règles d'évaluation du \roCalE. Nous avons
montré que toutes les réductions obtenues en utilisant cette stratégie sont
confluentes et que la stratégie devient triviale, c'est-à-dire n'impose aucune
restriction, pour certaines instances spécifiques du calcul (e.g. le
\laCal). Afin d'obtenir la confluence nous avons imposé des conditions
relativement restrictives sur les termes du \roCalE\  et sur l'application des
règles d'évaluation et ces restrictions peuvent être relaxées au prix de la
simplicité de la stratégie. Les conditions que nous voulons affaiblir concernent
d'un coté, la forme des règles de réécriture et d'un autre coté, le nombre
d'éléments dans l'argument d'un application.

\subsection{Règles de réécriture \conservs\  et \linrs}  \label{rewRuleConservLin}
%================================================================

Premièrement, l'absence des ensembles ayant plus d'un élément est nécessaire pour
garantir un bon comportement pour les règles de réécriture non-linéaires à
droite.  Par linéarité à droite d'une règle de réécriture nous désignerons ici
des membres droits linéaires par rapport aux variables du membre gauche. Par
exemple, $x \ra f(z,z)$ est linéaire à droite, mais $x \ra f(x,x)$ n'est pas
linéaire à droite.  De plus, la linéarité du membre droit peut être imposée
seulement aux opérateurs différents des symboles d'ensemble ($\{\}$) et ainsi,
la règle de réécriture $x \ra \{f(x),f(x)\}$ peut être considérée linéaire à
droite.  Intuitivement, nous n'avons pas besoin d'imposer la linéarité à droite
pour les ensembles parce que, en raison de la règle d'évaluation \rname{Flat},
ils ne mènent pas à des réductions non-convergentes comme dans
l'Exemple~\ref{RLandFailure}.

\DEF%------------------------------------------------------------------------
\label{linRR}
On dit que la $\rho$-règle de réécriture $l \ra r$ est 
{\em \linr} \index{regle de réécriture@règle de réécriture!strictement@\linr}
%\Defi{\linr}{règle de réécriture}\index{règle de réécriture} 
si tout sous-terme de $r$ n'étant pas un
ensemble est linéaire par rapport aux variables libres de $l$ et toute règle de
réécriture de $r$ est
\linr.
\FDEF%------------------------------------------------------------------------

L'application d'une règle de réécriture qui n'est pas \linr\  à un ensemble avec
plus d'un élément mène à des réductions non-convergentes, comme montré dans
l'Exemple~\ref{RLandFailure} mais ce n'est pas le cas si la règle de réécriture
appliquée est \linr~:

\EX%------------------------------------------------------------------------
\label{RLandSuc}

[\textit{Règle de réécriture \linr\  appliquée à un ensemble ayant plus d'un élément}]
\begin{center}$~$
\xymatrix{ 
[x \ra \{x,f(x)\}](\{a,b\})
\ar[d]_{Fire} \ar[dr]^{Batch}
&
\\
\{\{a,b\},f(\{a,b\})\}
\ar[d]_{OpOnSet}
&
\{[x \ra \{x,f(x)\}](a), [x \ra \{x,f(x)\}](b)\}
\ar[d]^{Fire}
\\
\{\{a,b\},\{f(a),f(b)\}\}
\ar[dr]_{Flat}
&
\{\{\{a,f(a)\}\},\{\{b,f(b)\}\}\}
\ar[d]^{Flat}
\\
&
\{a,b,f(a),f(b)\}
}
\end{center}
\FEX%------------------------------------------------------------------------

D'un autre côté, afin de garantir la propagation stricte de l'échec, nous avons
demandé que la règle d'évaluation \rname{Fire_c} soit appliquée seulement si
l'argument de l'application n'est pas un ensemble vide et il ne peut pas mener à
un ensemble vide.  Dans l'Exemple~\ref{ns_failure} nous pouvons remarquer que
les variables libres du membre gauche de la règle de réécriture ne sont pas
préservées dans le membre droit de la règle. Si la règle de réécriture $l \ra r$
de l'application conserve les variables du membre gauche dans le membre droit
(e.g. $x \ra x$) alors, l'application d'une substitution de la forme
$\subs{x/\emptyset}$, avec $x$ une variable de $l$, au terme $r$ mène à un terme
contenant $\emptyset$ et donc, qui est réduit éventuellement en $\emptyset$.

Nous définissons par la suite plus formellement les règles de réécriture
conservant les variables (libres) et nous présentons une nouvelle stratégie
définie en utilisant cette propriété. D'abord, nous introduisons une notion
similaire à celle de variable libre mais en considérant cette fois-ci le
comportement des opérateurs du \roCal\  et en particulier la nature
non-déterministe des ensembles.

\DEF%------------------------------------------------------------------------
\label{varPresentes}
L'ensemble des variables \Defn{présentes}{variable}{présente} d'un $\rho$-terme $t$,
noté $PV(t)$, est défini inductivement par~:
\begin{enumerate}
  \item si $t = x$ alors $PV(t) = \{x\}$,
  \item si $t = \{u_1, \ldots, u_n\}$ alors $PV(t) = \bigcap_{i=1,\ldots,n} PV(u_i)$
	($PV(\emptyset)=\XX$),
  \item si $t = f(u_1,\ldots,u_n)$ alors $PV(t) = \bigcup_{i=1,\ldots,n} PV(u_i)$
	($PV(c)=\emptyset ~ si ~ c \in \TF$),
  \item si $t = [u](v)$ alors $PV(t) = PV(u) \cup PV(v)$,
  \item si $t = u \ra v$ alors $PV(t) = PV(v) \setminus FV(u)$.
\end{enumerate}
\FDEF%------------------------------------------------------------------------

L'ensemble de \textit{variables libres} d'un $\rho$-terme ensemble est l'union
des ensembles de variables libres de chaque $\rho$-terme de l'ensemble, tandis
que l'ensemble de \textit{variables présentes} d'un $\rho$-terme ensemble est
l'intersection des ensembles de \textit{variables présentes} de chaque
$\rho$-terme de l'ensemble. Nous pouvons dire qu'une variable est
\textit{présente} dans un ensemble seulement si elle est présente dans tous les
éléments de l'ensemble. Par exemple, $PV(\{x,y,x\})=\emptyset$ et
$PV(\{x,f(x,y)\})=\{x\}$.

Nous voulons définir les règles de réécriture telles que les variables libres du
membre gauche de la règle soit inclus dans l'ensemble de variables libres du
membre droit de la règle mais en tenant compte du comportement de tous les
opérateurs du \roCal.

\DEF%------------------------------------------------------------------------
\label{conservRR}
On dit que la $\rho$-règle de réécriture $l \ra r$ est 
{\em \conserv} \index{regle de réécriture@règle de réécriture!quasi@\conserv}
%\Defi{\conserv}{règle de réécriture}\index{règle de réécriture} 
si $FV(l) \subseteq PV(r)$ et toute règle
de réécriture de $r$ est \conserv.
\FDEF%------------------------------------------------------------------------

Intuitivement, à toute variable libre du membre gauche d'une règle de réécriture
\conserv\   correspond, d'une façon déterministe, une variable libre dans le
membre droit de la règle. Pour tout $\rho$-terme ensemble dans $r$, la
correspondance entre les variables libres des termes $l$ et $r$ doit être vérifiée
pour chaque élément de l'ensemble.

%\EX%------------------------------------------------------------------------
Par exemple, la règle de réécriture $x \ra f(x,y)$ est \conserv\  tandis que la
règle de réécriture $x \ra \{x,y\}$ n'est pas \conserv.  La règle de réécriture
$\{f(x),g(x)\} \ra x$ est \conserv\  tandis que la règle de réécriture
$\{f(x),g(y)\} \ra x$ n'est pas \conserv. Si la définition des règles de
réécriture \conservs\  avait demandé la condition $PV(l) \subseteq PV(r)$, alors
la règle de réécriture $\{f(x),g(y)\} \ra x$ serait devenu
\conserv\  aussi. Ceci n'est pas souhaitable puisque la règle de
réécriture $\{f(x),g(y)\} \ra x$ est réduite en $\{f(x) \ra x,g(y) \ra x\}$ et
seulement la première des deux règles est \conserv. Notez que ces dernières
règles de réécriture contiennent des ensembles dans leur membre gauche et donc,
ne sont pas des termes de $\RTTE$.
Remarquons que la règle de réécriture $x \ra \emptyset$ est \conserv\  aussi bien
que la règle $\emptyset \ra x$.
%\FEX%------------------------------------------------------------------------

Puisque les variables du membre gauche d'une règle de réécriture \conserv\  se
retrouvent dans le membre droit, en appliquant une telle règle à un terme
$\emptyset$ nous garantissons qu'au moins une variable du membre droit de la
règle est instanciée à $\emptyset$ et donc, la propagation stricte de
l'échec. Du point de vue des variables présentes d'une règle de réécriture,
l'ensemble vide peut être vu comme un terme du premier ordre contenant toutes
les variables utilisées afin de construire les $\rho$-termes et donc, le
comportement approprié est assuré pour toute règle de réécriture \conserv\  avec
un $\emptyset$ dans le membre droit.

Dans l'Exemple~\ref{ns_failure} nous avons montré que l'application d'une règle
de réécriture qui n'est pas \conserv\  à un ensemble vide peut mener à des
résultats non-convergents. Si la règle de réécriture est \conserv, l'échec est
distribué strictement est nous obtenons des réductions confluentes comme dans
l'Exemple~\ref{conserv failure}.

\EX%------------------------------------------------------------------------
\label{conserv failure}

[\textit{Règle de réécriture \conserv\  appliquée à un ensemble vide}]
\begin{center}$~$
\xymatrix{ 
& [x \ra \{x,f(x,a)\}](\emptyset)
\ar[dl] \ar[d] \ar[dr] &
\\
[\{x \ra x,x \ra f(x,a)\}](\emptyset)
\ar[d] &
\{\emptyset,f(\emptyset,a)\}   
\ar[d] &
\emptyset
\\
\{[x \ra x](\emptyset),[x \ra f(x,a)](\emptyset)\}
\ar[r]
& \{\emptyset,\emptyset\}
\ar[ur]  &
}
\end{center}
\FEX%------------------------------------------------------------------------

En utilisant les notions de règle \conserv\  et de règle linéaire à droite nous
introduisons une nouvelle stratégie plus générale que la stratégie
\textit{ConfStrat} (\ref{strat_confluente}).

\DEF%----------------------------------------------------------
\label{confStratOper} 

Nous appelons \textit{ConfStratLin} la stratégie qui consiste à appliquer la
règle d'évaluation \rname{Fire} à un radical $[l \ra r](t)$ seulement si 
$t \in \TF$ est un terme clos du premier ordre ou~:
\begin{itemize}
\item le terme $l$ est linéaire
\item[] et
\item le terme $l$ \subf\  le terme $t$
\item[] et
	\begin{itemize}
	\item la règle $l \ra r$ est \conserv
	\item[] ou
	\item le terme $t$ ne contient aucun ensemble vide et,
	\item le terme $t$ ne contient pas de sous-terme de la forme $[u](v)$ où $u$
	n'est pas une abstraction et,
	\item pour tout sous-terme $[u \ra w](v)$ de $t$, $u$ subsume $v$,
	\end{itemize}
\item[] et
	\begin{itemize}
	\item la règle $l \ra r$ est \linr
	\item[] ou
	\item le terme $t$ ne contient aucun ensemble ayant plus d'un élément.
	\end{itemize}
\end{itemize}

\FDEF%----------------------------------------------------------

Par rapport à la stratégie \textit{ConfStrat} nous permettons l'application de
la règle d'évaluation \rname{Fire} à une application avec un argument
$\emptyset$ ou qui peut être réduit en l'ensemble vide si la règle de réécriture
de l'application est \conserv. En plus, si la règle de réécriture est linéaire à
droite nous permettons des arguments contenant un ensemble ayant plus d'un
élément.
Puisque on peut clairement décider si une règle est \conserv, toutes les
conditions utilisées dans la stratégie \textit{ConfStratLin} sont décidables.

Pour prouver la confluence dans le cas de la stratégie \textit{ConfStratLin}
nous procédons de la même façon que pour la stratégie \textit{ConfStrat} et nous
considérons que la relation $\del$est modifiée afin de représenter la nouvelle
stratégie.  Nous montrons d'abord une propriété de préservation pour les notions
de règle \conserv\  et règle \linr\  par rapport à la relation $\del$ et que les
lemmes utilisés dans la Section~\ref{confluence_modulo} restent valides dans le
cas de la stratégie \textit{ConfStratLin}.

\LEM%------------------------------------------------------------------------
\label{preservConserv}
Etant donné une règle de réécriture \conserv\  $u \ra v$ et le terme $v'$ tel que
$\Del{v}{v'}$ ou $\Cong{v}{v'}$. Alors, la règle de réécriture $u \ra v'$ est
\conserv.
\FLEM%------------------------------------------------------------------------

\proof{

Afin d'obtenir ce résultat il suffit de prouver que pour tous $\rho$-termes
$v,v'$ tels que $\Del{v}{v'}$ ou $\Cong{v}{v'}$ nous avons $PV(v) \subseteq PV(v')$.
Nous donnons juste une idée de la preuve qui utilise une induction sur la
structure du terme $v$ et considère toutes les réductions possibles.

Considérons d'abord une règle de réécriture $[l \ra r](t)$ satisfaisant les
bonnes conditions d'application. Dans le cas où le filtrage $l \meqqes t$
échoue, nous obtenons immédiatement un terme $\emptyset$ et donc 
$PV([l \ra r](t)) \subseteq \XX$.  Sinon, nous considérons que la substitution
$\sigma$ est la solution du problème de filtrage $l \meqqes t$ et nous avons
l'union des ensembles de variables présentes de tous les termes de son codomaine
égale à $PV(t)$. Puisque $FV(l) \subseteq PV(r)$ et nous avons $\Dom(\sigma)=FV(l)$ 
alors $PV(\sigma r)=(PV(r) \setminus FV(l)) \cup PV(\Ran(\sigma))=(PV(r)
\setminus FV(l)) \cup PV(t)=PV([l \ra r](t))$.

Pour les réductions $\Cong{v}{v'}$ le résultat est obtenu immédiatement par
induction en remarquant que un $\emptyset$ dans $v$ mène à $v'=\emptyset$ si $v$ 
n'est pas un ensemble.
}

Nous pouvons montrer facilement que la linéarité des membres gauches des règles
de réécriture est préservée par les réductions respectant la stratégie
\textit{ConfStratLin}. 

\LEM%------------------------------------------------------------------------
\label{preservLin}
Etant donnée une règle de réécriture \linr\  $u \ra v$ et le terme $v'$ tel que
$\Del{v}{v'}$ ou $\Cong{v}{v'}$. Alors, la règle de réécriture $u \ra v'$ est
\linr.
\FLEM%------------------------------------------------------------------------

\proof{
La preuve est réalisée par induction sur la structure du terme $v$ en
considérant toutes les réductions possibles.

Pour toute réduction $\Cong{v}{v'}$ la linéarité de $v'$ par rapport à $u$ est
clairement préservée. Par exemple, pour un terme de la forme $v=f(\{v_1,v_2\})$, 
$v_1$ et $v_2$ sont linéaires par rapport aux variables de $u$ et donc, $f(v_1)$ 
et $f(v_2)$ sont aussi linéaires. Par conséquent, la règle de réécriture 
$u \ra \{f(v_1),f(v_2)\}$ est \linr.

Nous devrions vérifier si le terme $v'$ peut devenir non-linéaire
par rapport à $u$ pour une réduction $\Del{v}{v'}$. Ceci serait possible seulement
en appliquant une règle de réécriture non-linéaire à droite dans un sous-terme
de $v$ mais il n'existe pas de telle règle dans $v$.
}

En utilisant les propriétés précédentes nous pouvons adapter les preuves de
confluence de la Section~\ref{confluence_modulo} et considérant les nouveaux cas
introduits par une stratégie \textit{ConfStratLin}, c'est-à-dire l'application
d'une règle de réécriture \conserv\  à un terme contenant éventuellement
l'ensemble vide et l'application d'une règle de réécriture \linr\  à un terme
contenant éventuellement des ensembles ayant plus d'un élément.

\TH%----------------------------------------------------------
Si une stratégie d'évaluation \textit{ConfStratLin} est utilisée, alors le
\roCalE\  est confluent.
\FTH%----------------------------------------------------------

\proof{

La preuve du Lemme~\ref{setEquivBasic} est facilement adaptée en considérant des
substitutions de la forme $\sigma=\subs{x/\{t_1,\ldots,t_n\}}$ appliquées à un
terme $r$ tel que $x \ra r$ soit \linr\  et des substitutions de la forme
$\sigma=\subs{x/\emptyset}$ appliquées à un terme $r$ tel que $x \ra r$ soit
\conserv.

Dans le Lemme~\ref{BasicCoherence} nous n'avons pas considéré des applications
de règle de réécriture \conserv\  à un argument $\emptyset$ et des applications de règle
de réécriture \linr\  à un terme ayant plus d'un élément.
Le Lemme~\ref{preservConserv} et le Lemme~\ref{preservLin} nous permettent
d'étendre le Lemme~\ref{setEquivBasic} dans le cas de la stratégie
\textit{ConfStratLin} et donc d'une règle d'évaluation \rname{Fire_c} avec les
conditions modifiées pour décrire cette stratégie.

Si nous considérons le terme $r$ telle que la règle de réécriture $x \ra r$ soit
\conserv\  et $\Del{r}{r^1}$ alors, nous obtenons~:
\begin{center}$~$
\xymatrix{  
[x \ra r](\emptyset)
\ar[d]_{\congrxy}
\ar[r]^-{\delxy}
&
\{\subs{x/\emptyset} r^1\}
\ar@{.{>}}[dl]^-{\congTRxy}
\\
\emptyset
&
}
$~~~$ et $~~~$
\xymatrix{  
[x \ra r](\emptyset)
\ar[d]_{\congrxy}
\ar[r]^-{\delxy}
&
\{\subs{x/\emptyset} r^1\}
\ar@{.{>}}[d]^-{\congTRxy}
\\
[x \ra r'](\emptyset)
\ar@{.{>}}[r]_{\congTRxy}
&
\emptyset
}
\end{center}

Pour l'application d'une règle de réécriture \linr\  à un terme ayant plus d'un
élément nous obtenons~:
\begin{center}$~$
\xymatrix{  
[x \ra r](\{p_1,\ldots,p_n\})
\ar[d]_{\congrxy}
\ar[r]^-{\delxy}
&
\{\subs{x/\{p_1,\ldots,p_n\}} r^1\}
\ar@{.{>}}[d]^-{\congTRxy}
\\
\{[x \ra r](p_1),\ldots,[x \ra r](p_n)\}
\ar@{.{>}}[d]_{\congTRxy}
&
u
\\
\{\{\subs{x/p_1} r\},\ldots,\{\subs{x/p_n} r\}\}
\ar@{.{>}}[r]^-{\delxy}
&
\{\{\subs{x/p_1} r^1\},\ldots,\{\subs{x/p_n} r^1\}\}
\ar@{.{>}}[u]_{\congTRxy}
}
\end{center}

Nous pouvons ainsi montrer que la relation $\del$ induite par la nouvelle règle
\rname{Fire_c} utilisant les conditions de la stratégie \textit{ConfStratLin}
et la relation $\congr$ sont cohérentes. Puisque les autres lemmes utilisés pour
prouver la confluence des relation induites par les règles d'évaluation du
calcul n'utilisent pas les restrictions sur le nombre d'éléments des ensembles,
nous déduisons la confluence des réductions guidées par la stratégie
\mbox{\textit{ConfStratLin}}.
}

\subsection{Règles de réécriture \stables} \label{rewRuleStable}
%================================================================

Nous avons montré la confluence du calcul $(\RTTE,\emptyset,\SS)$ (i.e. le
\roCalE) où la stratégie $\SS$ est une des stratégies \textit{ConfStrat},
\textit{ConfStratLin} et nous présentons brièvement par la suite les problèmes
liés à la confluence du calcul $(\RTT,\emptyset,\SS)$ que nous appelons le
\roCalEp.

\DEF%------------------------------------------------------------------------
\label{roPdef} 

Etant donné un ensemble de symboles de fonctions $\FF$, un ensemble de variables
$\XX$, nous appelons \roCalEp\  un calcul défini par~:
\begin{itemize}
\item l'ensemble de termes $\RTT$,
\item l'application (d'ordre supérieur) de substitution aux termes,
\item la théorie $\emptyset$ (filtrage syntaxique),
\item l'ensemble de règles d'évaluation $\EE$: \rname{Fire}, \rname{Congruence},
	\rname{Congruence\_fail}, \rname{Distrib}, \rname{Batch},
	\rname{Switch_L}, \rname{Switch_R}, \rname{OpOnSet}, \rname{Flat},
\item une stratégie d'évaluation $\SS$ qui guide l'application des règles
	d'évaluation.
\end{itemize}
\FDEF%------------------------------------------------------------------------

Rappelons que les règles de réécriture de l'ensemble de termes $\RTTE$ ne
peuvent avoir qu'un terme du premier ordre dans le membre gauche. Cette
restriction a du être imposée à cause de l'instabilité de la notion de variable
libre par rapport aux règles d'évaluation du calcul qui mène immédiatement à
l'instabilité des réductions par rapport à la substitution. 
Par exemple, l'ensemble des variables libres d'un terme $\{x,y\} \ra x$ est
$FV(\{x,y\} \ra x)=\emptyset$ mais pour le terme résultant de l'application de
la règle d'évaluation \rname{Switch_L} sur le terme initial nous obtenons
$FV(\{x \ra x,y \ra x\})=$ $FV(x \ra x) \cup FV(y \ra x) =$ $\{x\}$. Si le
membre gauche des règle de réécriture contient de tels termes alors, nous
pouvons obtenir facilement des réductions non-confluentes, comme dans
l'Exemple~\ref{notFirstOrderRule}.

\EX%------------------------------------------------------------------------
\label{notFirstOrderRule}
[Règle avec un ensemble dans le membre gauche]
\samepage
\begin{center}$~$
\xymatrix{ 
&
[x \ra [\{x,y\} \ra x](b)](a)
\ar[dl]_{Fire} \ar[dr]^{Switch_L}
& \\
\{[\{x,y\} \ra x](b)\}
\ar[d]_{Switch_L} &&
[x \ra [\{x \ra x,y \ra x\}](b)](a)
\ar[d]^{Distrib}
\\
\{[\{x \ra x,y \ra x\}](b)\}
\ar[d]_{Distrib} &&
[x \ra \{[x \ra x](b),[y \ra x](b)\}](a)
\ar[d]^{Fire}
\\
\{\{[x \ra x](b),[y \ra x](b)\}\}
\ar[d]_{Fire} &&
[x \ra \{\{b\},\{x\}\}](a)
\ar[d]^{Fire}
\\
\{\{\{b\},\{x\}\}\}
\ar[d]_{Flat} &&
\{\{\{b\},\{a\}\}\}
\ar[d]^{Flat}
\\
\{b,x\}
&&
\{b,a\}
}
\end{center}
\FEX%------------------------------------------------------------------------

Une première méthode utilisée pour obtenir la confluence du \roCalEp\  consiste à
définir une stratégie qui combine les conditions sur les termes de $\RTTE$ et
les restrictions imposées par les stratégies déjà définies.

\DEF%----------------------------------------------------------
\label{confStratPlus} 

Nous appelons \textit{ConfStratPlus} la stratégie qui consiste à appliquer la
règle d'évaluation \rname{Fire} à un radical $[l \ra r](t)$ seulement si les
conditions de la stratégie \textit{ConfStratLin} sont satisfaites et tous les
membres gauches des règles de réécriture de $r$ sont des termes du premier
ordre.
\FDEF%----------------------------------------------------------

Il est clair que si une stratégie d'évaluation \textit{ConfStratPlus} est
utilisée, alors le \roCalEp\  est confluent.

La restriction sur la forme des règles de réécriture est relativement
restrictive et nous voulons alléger cette restriction en gardant la confluence
du calcul. Afin d'obtenir la confluence nous devons imposer une condition
garantissant la stabilité des réductions par rapport à la substitution,
c'est-à-dire un résultat similaire au Lemme~\ref{substEquivInd} pour les termes
de $\RTT$.

En utilisant la notion de variable présente nous définissons les règles de
réécriture conservant l'ensemble de variables libres et ainsi, ne menant pas à
des réductions non-convergentes comme dans l'Exemple~\ref{notFirstOrderRule}.

\DEF%------------------------------------------------------------------------
\label{stableRR}
On dit que la $\rho$-règle de réécriture $l \ra r$ est 
{\em \stable} \index{regle de réécriture@règle de réécriture!stable@\stable}
%\Defi{\stable}{règle de réécriture}\index{règle de réécriture} 
si $FV(r) \cap FV(l) = FV(r) \cap PV(l)$
et toute règle de réécriture de $r$ est \stable.
\FDEF%------------------------------------------------------------------------

Intuitivement, le membre gauche d'une règle de réécriture \stable\  peut lier
seulement ses variables présentes dans le membre droit de la règle. Ainsi, toute
règle de réécriture avec le membre gauche un terme du premier ordre est \stable.

\EX%------------------------------------------------------------------------
Les règles de réécriture $x \ra f(x,y)$ et $\{f(x,y),g(x)\} \ra x$ sont \stables\
et la première règle est \conserv\  tandis que la deuxième règle n'est pas
\conserv. La règle de réécriture $x \ra \{x,y\}$ n'est pas \conserv\  mais elle
est \stable.
\FEX%------------------------------------------------------------------------

Nous définissons une nouvelle stratégie en utilisant la notion de règle de
réécriture \stable\  pour remplacer la restriction imposée sur la forme des
règles de réécriture dans la définition de la stratégie \textit{ConfStratPlus}.

\DEF%----------------------------------------------------------

Nous appelons \textit{ConfStratStable} la stratégie qui consiste à appliquer la
règle d'évaluation \rname{Fire} à un radical $[l \ra r](t)$ seulement si les
conditions pour la stratégie \textit{ConfStratLin} sont satisfaites et toutes les
règles de réécriture dans $r$ sont \stables.
\FDEF%----------------------------------------------------------

La condition qu'une règle de réécriture soit \stable\  est clairement décidable
et donc, toutes les conditions utilisées dans la stratégie
\textit{ConfStratStable} sont décidables.

Pour prouver la confluence du \roCalEp\  avec les règles d'évaluation guidées par
une stratégie \textit{ConfStratStable} il suffit de remarquer que parmi les lemmes
prouvés dans la Section~\ref{stratRocal} seulement le Lemme~\ref{substEquivInd}
utilise la restriction imposant des règles de réécriture ayant un terme du
premier ordre dans le membre gauche. Nous démontrons un lemme similaire pour la
stratégie \textit{ConfStratStable} et nous déduisons immédiatement la confluence du
calcul dans ce cas.

\LEM%------------------------------------------------------------------------
\label{preservStable}

Etant donnés le terme $t$ ne contenant que des règles de réécriture \stables\  et une
substitution $\sigma$. Si $\Cong{t}{t'}$ alors, $\CongTR{\sigma t}{\sigma t'}$.
\FLEM%------------------------------------------------------------------------

\proof{

Cette propriété a été prouvée dans le Lemme~\ref{substEquivInd} pour les termes
appartenant à $\RTTE$ et donc, seulement les règles de réécriture de la forme
$t=\{u_1,\ldots,u_n\} \ra v$ avec $t'=\{u_1 \ra v,\ldots,u_n \ra v\}$ n'ont pas
été considérées.

Nous avons $FV(\{u_1,\ldots,u_n\} \ra v)=FV(v) \setminus (\cup_i FV(u_i))$,
$FV(\{u_1 \ra v,\ldots,u_n \ra v\})=\cup_i (FV(v) \setminus FV(u_i))$ et
$FV(v) \cap FV(\{u_1,\ldots,u_n\}) = FV(v) \cap PV(\{u_1,\ldots,u_n\})$.
Ainsi, toute variable de $FV(\{u_1,\ldots,u_n\})$ éliminée de $FV(v)$
doit être une variable de $PV(\{u_1,\ldots,u_n\})$ et donc, elle doit être
présente dans tous les termes $u_i$. Nous obtenons donc, \\
$FV(v) \setminus (\cup_i FV(u_i))=$
$FV(v) \setminus FV(u_1)=\ldots=FV(v) \setminus FV(u_n)=$
$\cup_i (FV(v) \setminus FV(u_i))$.

Nous pouvons donc  considérer une substitution $\sigma$ telle que
$\Dom(\sigma) \cap PV(\{u_1,\ldots,u_n\})=\emptyset$ et ainsi, nous obtenons
$\sigma t=\{u_1,\ldots,u_n\} \ra \sigma v$. Par conséquent, nous pouvons utiliser une
induction sur la structure des termes comme dans le Lemme~\ref{substEquivInd}.
}

\TH%----------------------------------------------------------
Si une stratégie d'évaluation \textit{ConfStratStable} est utilisée, alors le
\roCalEp\  est confluent.
\FTH%----------------------------------------------------------

Nous pouvons ainsi dire que le calcul
$(\RTT,\emptyset,\textit{ConfStratStable})$ est confluent.

\subsection{Stratégies dans le \roCalT}
%================================================================

Dans les stratégies que nous avons proposées jusqu'à maintenant notre but
principal était de permettre le maximum de réductions en gardant des conditions
décidables relativement simples à comprendre et à implanter.

La théorie de filtrage vide du \roCalE\  nous a permis d'imposer des conditions
d'application pour la règle d'évaluation \rname{Fire} relativement simple à
décider mais dès que la théorie $T$ de filtrage est plus élaborée les notions
utilisées doivent être modifiées en conséquence.

Si nous considérons une théorie $T$ non-unitaire par exemple, l'absence des
ensembles ayant plus d'un élément dans un terme ne garantit pas l'absence des
telles ensembles dans dans son réduit. Par exemple, dans une théorie de filtrage
avec un opérateur commutatif $\oplus$, le terme $[x \oplus y \ra x](a \oplus b)$ 
ne contient pas d'ensemble ayant plus d'un élément mais son réduit $\{a,b\}$ ne
satisfait plus cette condition. Ainsi, la condition sur la réduction d'une
application $[l \ra r](t)$ en utilisant la règle d'évaluation \rname{Fire} doit
imposer non seulement que $t$ ne contient pas d'ensemble ayant plus d'un élément
mais aussi qu'il ne contient pas de radical de la forme $[u](v)$. En plus, les
notions de subsomption doivent être adaptées afin de tenir compte des
caractéristiques de la théorie $T$.

Des conditions similaires à celles utilisées dans les stratégies pour le
\roCalE\  sont difficile à définir dans le cas général du
\roCalT\  mais nous pouvons imposer des conditions plus restrictives sur
l'application de la règle d'évaluation \rname{Fire} pour une application 
$[l \ra r](t)$ impliquant les conditions utilisées précédemment.

\DEF%----------------------------------------------------------
\label{confStratFirstOrder} 

Nous appelons \textit{ConfStratFirstOrder} la stratégie qui consiste à appliquer
la règle d'évaluation \rname{Fire} à un radical $[l \ra r](t)$ seulement si $l,r$
sont des termes du premier ordre et $t$ est un terme clos du premier ordre.
\FDEF%----------------------------------------------------------

La stratégie \textit{ConfStratFirstOrder} est assez restrictive mais, puisque
les conditions imposées sont purement structurelles, elle peut être implantée
d'une manière très efficace et ses propriétés, avec en particulier la
confluence, sont plus facile à analyser.

\TH%----------------------------------------------------------
Si une stratégie d'évaluation \textit{ConfStratFirstOrder} est utilisée, alors le
\roCalT\  est confluent.
\FTH%----------------------------------------------------------

\proof{
En partant de la règle d'évaluation \rname{Fire} qui est appliquée pour un
$\rho$-terme $[l \ra r](t)$ si les conditions de la stratégie
\textit{ConfStratFirstOrder} sont imposées nous considérons la relation
$\del$ induite par cette règle d'évaluation. 

La preuve de la cohérence de cette relation $\del$ avec la relation $\congr$
(Définition~\ref{transCongRel}) peut être facilement adaptée pour le \roCalT\
avec les règles d'évaluation guidées par la stratégie
\textit{ConfStratFirstOrder}.
Nous obtenons immédiatement la confluence forte de la relation $\del$ et en
utilisant une approche similaire que pour le \roCalE\  nous obtenons la
confluence du \roCalT.  
}

Nous pouvons ainsi dire que le calcul $(\RTT,T,\textit{ConfStratFirstOrder})$ est confluent.

Nous pouvons aussi voir les stratégies définies précédemment pour le \roCalE\
comme des raffinements de la stratégie \textit{ConfStratFirstOrder} utilisables
dans un \roCal\  avec une théorie de filtrage vide. De la même façon, des
raffinements de la stratégie \textit{ConfStratFirstOrder} peuvent être réalisées
pour des théories équationnelles mais ceci devient plus difficile pour des
théories de filtrage d'ordre supérieur.

%\DontWriteThisInToc  
\subsection*{Conclusion}
%================================================================
%~

Nous avons étudié dans ce chapitre la confluence du \roCal. En analysant des
exemples de réductions non-convergentes dans le \roCalE, nous avons remarqué que
les raisons de la non-confluence de ce calcul sont le pouvoir limité du filtrage
syntaxique et la manipulation des ensembles.

Afin de résoudre ce problème nous avons proposé des stratégies d'évaluation
garantissant la confluence. Une stratégie simple consiste à réduire
l'application d'une règle de réécriture seulement si le sujet de l'application
est un terme clos du premier ordre, mais cette stratégie est relativement
restrictive.
Nous avons donc proposé une stratégie moins restrictive définie en imposant des
conditions sur les $\rho$-réductions possibles des termes impliqués dans une
application. Cette stratégie devient triviale (c'est-à-dire n'impose aucune
restriction) pour certaines instances du calcul, mais les conditions imposées ne
sont pas utilisables dans une implantation du \roCal.  Par conséquent, nous
avons défini des conditions décidables sur les $\rho$-termes et des stratégies
confluentes définies en utilisant ces restrictions. Nous avons présenté ces
résultats dans~\cite{CirsteaKirchner-LivreFroCoS99}
et~\cite{CirsteaKirchnerINRIA99}.

%% file: chapter_4.tex
%%%%%%%%%%%%%%%%%%%%%%%%%%%%%%%%%%%%%%%%%%%%%%%%%%%%%%%%%%%
% \TLtopbookmark
\chapter{Récursivité dans le \roCal\ -
le \roCalfT}
\label{chap.recursion}
%%%%%%%%%%%%%%%%%%%%%%%%%%%%%%%%%%%%%%%%%%%%%%%%%%%%%%%%%%%%

Le pouvoir d'expression du \roCal\  nous permet de représenter des
\mbox{$\lambda$-termes} et leur dérivation ainsi que l'application d'une règle
de réécriture de la réécriture de termes classique. Les $\lambda$-termes
contiennent toute l'information nécessaire pour leur réduction tandis que dans
la réécriture cette information est implicite et différentes stratégies peuvent
être utilisées pour guider l'application des règles de réécriture.  En effet,
les travaux de Huet et Levy~\cite{HuetLevy79,HuetLevyCL} et
d'O'Donnell~\cite{HoffmannOD-83} ont inspiré l'étude de stratégies efficaces de
réduction.

La description explicite des stratégies de réduction nécessite l'utilisation
d'un opérateur testant l'échec de l'application d'une règle de réécriture.
Nous introduisons donc un opérateur $first$ décrivant l'application du premier
argument ne menant pas à un échec.
Nous appelons \roCalf\  le calcul obtenu en ajoutant la description de cet
opérateur au \roCal. Nous présentons une définition des stratégies de réduction
\textit{innermost} et \textit{outermost} basée sur des opérateurs définis dans le \roCalf.

\section{Présentation de la problématique}
%================================================================

L'application d'une règle de réécriture de la réécriture standard est décrite
explicitement dans le \roCal\  par un $\rho$-terme approprié. Par exemple,
l'application de la règle de réécriture $a \ra b$ au terme $a$ est représentée
par le \mbox{$\rho$-terme} $[a \ra b](a)$. La réduction de cette application en
le terme $b$ dans la réécriture correspond à la réduction du $\rho$-terme \
$[a \ra b](a)$ en le $\rho$-terme $\{b\}$ dans le \roCal. Mais une réduction en
réécriture peut impliquer l'application de plusieurs règles et puisque nous
voulons représenter dans le \roCal\  toute réduction de la réécriture et pas
seulement les réductions nécessitant un seul pas de réécriture alors nous
voulons répondre à la question~:

~\\
\textit{Etant donnée une théorie de réécriture $\RR$ existe-il un $\rho$-terme
$\xi_{\RR}$ tel que pour tout terme $u$, si $u$ se réduit en le terme $v$ dans la
théorie de réécriture $\RR$ alors $[\xi_{\RR}](u)$ se $\rho$-réduit en un
ensemble contenant le $\rho$-terme $v$~?}
~\\

Nous allons voir dans la Section~\ref{encodRewCond} que pour toute réduction dans une
théorie de réécriture, il existe une réduction correspondante dans le \roCal~: si
le terme $u$ se réduit en le terme $v$ dans une théorie de réécriture $\RR$ nous
pouvons construire un $\rho$-terme $\xi_{\RR}(u)$ qui se réduit en le terme $\{v\}$
dans le \roCal. La méthode utilisée permet de construire le terme $\xi_{\RR}(u)$
à partir des étapes de réduction de $u$ en $v$ dans la théorie $\RR$.

Nous voulons aller plus loin et donner une méthode pour construire le terme
$\xi_{\RR}(u)$ sans connaître la dérivation de $u$ en $v$ dans la théorie
$\RR$. Ce problème s'avère plus difficile parce que les propriétés du système de
réécriture, comme la terminaison et la confluence, ne sont pas connues \textit{a
priori}.

Ceci signifie que nous souhaitons décrire dans le \roCal\  des stratégies
de réduction et, principalement, des stratégies de normalisation. Les
$\rho$-termes représentant des stratégies de normalisation nous permettrons
d'obtenir, entre autres, un codage naturel de la réécriture conditionnelle.
La question que nous nous sommes posée précédemment peut
être ainsi reformulée en~:

~\\
\textit{Etant donnée une théorie de réécriture $\RR$ existe-il un $\rho$-terme
$\xi_{\RR}$ tel que pour tout terme $u$, si $u$ se normalise en le terme $v$ dans la
théorie de réécriture $\RR$ alors $[\xi_{\RR}](u)$ se $\rho$-réduit en un
ensemble contenant le $\rho$-terme $v$~?}
~\\

La définition de stratégies (de normalisation) est faite en général au
méta-niveau et nous voulons montrer que le \roCal\  est assez puissant pour nous
permettre la représentation de telles dérivations au niveau objet du calcul.
Dans la Section~\ref{encodageLambda} nous montrerons que le \roCalE\  contient le
\laCal\  et donc, toute fonction calculable, comme la normalisation par exemple,
est exprimable dans le formalisme.

Ce que nous apportons ici, grâce à la puissance du filtrage et à l'utilisation
du non-détermi\-nisme, est la facilité d'exprimer en utilisant les $\rho$-termes,
des stratégies (de normalisation) guidant l'application d'une ou plusieurs
règles de réécriture. Nous pouvons ainsi exprimer les stratégies dans un
formalisme uniforme combinant les mécanismes standard de réécriture et les
techniques d'ordre supérieur.

Pour calculer la forme normale d'un terme $u$ par rapport à un système de
réécriture $\RR$, les règles de réécriture de $\RR$ sont appliquées
\textit{répétitivement} à une position \textit{quelconque} de $u$ jusqu'à ce
qu'aucune règle de $\RR$ ne soit plus
\textit{applicable}.  Par conséquent, les ingrédients nécessaires pour définir
une telle stratégie sont~:
\begin{itemize}
	\item un opérateur d'itération qui applique \textit{répétitivement} des
	règles de réécriture,
	\item un opérateur de parcours de termes qui applique une règle de
	réécriture à une position \textit{quelconque} d'un terme,
	\item un opérateur testant si un ensemble de règles de réécriture est
	\textit{applicable} sur un terme.
\end{itemize}

Il existe évidement plusieurs stratégies de réduction obtenues en précisant la
position d'application des règles de réécriture à chaque pas de réduction. Par
exemple, si les règles de réécriture sont appliquées d'abord à la position la
plus profonde alors une stratégie \textit{innermost} est obtenue tandis que si
les règles de réécriture sont appliquées en tête alors une stratégie
\textit{outermost} est obtenue.

Nous décrivons par la suite une possibilité de définir dans le \roCal\  les
opérateurs utilisés pour la définition de stratégies. Nous commençons par
quelques opérateurs auxiliaires et ensuite nous présentons les $\rho$-opérateurs
qui correspondent aux fonctionnalités énumérées ci-dessus.

\section{Opérateurs auxiliaires}   \label{opAux}
%================================================================

Nous définissons d'abord trois opérateurs auxiliaires qui seront utilisés dans
les sections suivantes. Ces opérateurs sont simplement utilisés pour
définir et nommer des $\rho$-termes plus complexes et ont été introduits pour donner des
définitions plus compactes et plus intuitives pour les opérateurs de récursion.

Le premier de ces trois opérateurs est appelé l'\Def{identité} et noté
\Defi{$id$}{identité}. L'application de cet opérateur sur un 
$\rho$-terme quelconque $t$ est réduite en le singleton contenant ce terme,
c'est-à-dire $[id](t) \lraD{\rho} \{t\}$. Le $\rho$-terme $id$ n'est rien
d'autre que la règle de réécriture $x \ra x$~:
	$$id \eqdef x \ra x.$$

De la même manière nous pouvons définir la stratégie $fail$ qui échoue
toujours~:
	$$fail \eqdef x \ra \emptyset.$$

Nous introduisons aussi l'opérateur
binaire ``$;$'' qui représente l'application séquentielle de deux
$\rho$-termes.  Un $\rho$-terme de la forme $[u;v](t)$ représente
l'application du terme $v$ sur le résultat de l'application du terme $u$ sur le
terme $t$. Par conséquent, nous définissons l'opérateur ``$;$'' par~:
	$$u;v \eqdef x \ra [v]([u](x)).$$

Dans les sections suivantes nous employons généralement la forme abrégée de ces
opérateurs et pas leur forme étendue mais nous montrons éventuellement les
réductions correspondantes.

\section{L'opérateur $first$ et le \roCalfT}  \label{opFirst}
%================================================================

Nous analysons maintenant la possibilité de représenter dans le \roCal\  le test
d'applicabilité d'une règle sur un terme, c'est-à-dire tester si le résultat
n'est pas $\emptyset$. Ce test sera utilisé principalement pour appliquer une
règle de réécriture seulement si elle ne mène pas au résultat $\emptyset$ et
nous définissons un opérateur qui cherche les termes qui ne mènent pas à un
échec.

\subsubsection{Description de l'opérateur $first$}
%================================================================

Nous introduisons un nouvel opérateur $n$-aire appelé $first$
qui a le rôle de sélectionner parmi ses arguments le premier terme dont
l'application à un $\rho$-terme donné n'est pas réduite en $\emptyset$.

Nous voulons donc que l'application d'un $\rho$-terme $first(s_1,\ldots,s_n)$ à
un terme $t$ retourne le résultat de la première application sans échec d'un de
ses arguments au terme $t$. Si la réduction de tout terme $[s_i](t)$,
$\eni{i}{k-1}$, mène à $\emptyset$ et que le terme $[s_k](t)$ ne se
réduit pas en $\emptyset$, alors $[first(s_1,\ldots,s_n)](t)$ est réduit en le même
terme que le terme $[s_k](t)$.

Ainsi, lorsque la réduction de tout terme $[s_i](t)$, $\eni{i}{k-1}$, mène à
$\emptyset$ et que la réduction de $[s_k](t)$ ne termine pas alors la réduction
du terme $[first(s_1,\ldots, s_n)](t)$ ne termine pas.

Nous pouvons résumer la description de l'opérateur $first$ donné ci-dessus par
les deux règles d'évaluation \rname{First'} et \rname{First''}~:
\renewcommand{\fleche}{\Longrightarrow}
\begin{ruleset}
%===================================
  \ccregle 
  {First'}
  {[first(s_1,\ldots,s_n)](t)}
  {\{u_k\da{}\}} 
  {[s_i](t) \longra{*}_{\rho} \emptyset,~\eni{i}{k-1}} 
  {[s_k](t) \longra{*}_{\rho} u_k\da{} \neq \emptyset,~u_k\da{}~clos, ~sans~radical} 
%===================================
  \cregle 
  {First''}
  {[first(s_1,\ldots,s_n)](t)}
  {\emptyset} 
  {[s_i](t) \longra{*}_{\rho} \emptyset,~\eni{i}{n}}
%===================================
\end{ruleset}

Dans les deux règles précédentes la condition qu'un terme soit réduit en
$\emptyset$ ou en un terme clos en forme normale ($u_k\da{}$) est testée au
méta-niveau du calcul. En procédant de cette manière, des opérations intrinsèques
au niveau objet doivent être exécutées au méta-niveau et donc, les réductions au
niveau objet ne contiennent plus toute l'information sous-jacente.

Par exemple, en utilisant la règle \rname{First'} le $\rho$-terme 
$[first(a \ra b,b \ra c)](b)$ est réduit dans un seul pas en $\{c\}$~:
	$$[first(a \ra b,b \ra c)](b) \longra{}_{First'} \{c\}$$ 
et donc les réductions $[a \ra b](b) \ra \emptyset$ et $[b \ra c](b) \ra \{c\}$
ne sont plus visibles au niveau objet.

\subsubsection{Le \roCalfT}
%================================================================

Pour rendre explicite le fonctionnement de l'opérateur $first$ nous introduisons
un nouvel opérateur $n$-aire ``$\paron{}$'' qui, intuitivement, sélectionne les termes non
vides. Contrairement à un terme de la forme $\{t_1,\ldots,t_n\}$ où la virgule
est supposée respecter les axiomes des ensembles (associativité, commutativité,
idempotence), nous ne faisons aucune supposition sur un terme
$\paron{t_1,\ldots,t_n}$ où l'ordre des arguments est essentiel dans
l'évaluation.

\DEF%------------------------------------------------------------------------
L'ensemble des $\rhof$-termes étend l'ensemble $\RTT$ de $\rho$-termes de base
(Définition~\ref{rhoTermesDeBase}) comme étant le plus petit ensemble tel que~:
\begin{itemize}
  \item les éléments de $\TFX$ sont des $\rhof$-termes,
  \item si $t_1,\ldots,t_n$ sont des $\rhof$-termes et $f \in \FF_n$ alors
	$f(t_1,\ldots,t_n)$ est un $\rhof$-terme,
  \item si $t_1,\ldots,t_n$ sont des $\rhof$-termes alors $\{t_1,\ldots,t_n\}$ est un
	$\rhof$-terme,
  \item si $t$ et $u$ sont des $\rhof$-termes alors $[t](u)$ est un $\rhof$-terme,
  \item si $t$ et $u$ sont des $\rhof$-termes alors $t \ra u$ est un $\rhof$-terme.
  \item si $t_1,\ldots,t_n$ sont des $\rhof$-termes alors
	$first(t_1,\ldots,t_n)$ est un $\rhof$-terme,
  \item si $t_1,\ldots,t_n$ sont des $\rhof$-termes alors
	$\paron{t_1,\ldots,t_n}$ est un $\rhof$-terme.
\end{itemize}
Cet ensemble est noté $\RTTf$.
\FDEF%------------------------------------------------------------------------

Les règles d'évaluation décrivant l'opérateur $first$ et l'opérateur auxiliaire
$\paron{}$ sont présentées dans la Figure~\ref{MRAfirst}.
La description des deux opérateurs utilise une condition qui teste si un terme
est (évalué en) un ensemble vide et nous ne savons pas actuellement exprimer ces
opérateurs dans le \roCal\  de base.

\begin{figure}[!htp]%------------------------------------------------------------------------
\noindent \framebox{\parbox{\largeurtexte}{

\renewcommand{\fleche}{\Longrightarrow}
\begin{ruleset}
%===================================
  \regle 
  {First}
  {[first(s_1,\ldots,s_n)](t)}
  {\paron{[s_1](t),\ldots,[s_n](t)}} \\
%===================================
  \regle 
  {First\_fail} 
  {\paron{\emptyset,t_1,\ldots,t_n}} 
  {\paron{t_1,\ldots,t_n}}
  %{n > 0}
%===================================
  \ccregle 
  {First\_success} 
  {\paron{t,t_1,\ldots,t_n}}
  {\{t\}}
  {t~ne~contient~pas~de~radical~et~de~variable}
  {libre~et~n'est~pas~\emptyset}
%===================================
  \regle 
  {First\_single} 
  {\paron{}}
  {\{\}}
%===================================
\end{ruleset}
}}
\caption{\label{MRAfirst}Les règles d'évaluation de l'opérateur $first$}
\end{figure}%------------------------------------------------------------------------

L'opérateur $first$ lance l'évaluation de l'application de ses arguments sur le
terme donné et l'opérateur $\paron{}$ teste explicitement l'échec de ces
applications.  Les conditions testées par les règles d'évaluation présentées dans
la Figure~\ref{MRAfirst} sont des conditions structurelles sur les termes et
n'impliquent aucune réduction au méta-niveau.

Avec cette dernière approche toutes les applications des règles d'évaluation
sont effectuées au niveau objet du calcul et la réduction du terme 
$[first(a \ra b,b \ra c)](b)$ présentée précédemment devient~:
\begin{tabbing}
\indent \indent  \= $\lraD{First\_success}$ ~~~ \=  \kill
\> \> $[first(a \ra b,b \ra c)](b)$
\\
\> $\lraD{First}$ 
\> $\paron{[a \ra b](b),[b \ra c](b)}$
\\
\> $\longra{*}_{Fire}$ 
\> $\paron{\emptyset,\{c\}}$
\\
\> $\lraD{First\_fail}$ 
\> $\paron{\{c\}}$
\\
\> $\lraD{First\_success}$ 
\> $\{\{c\}\}$
\\
\> $\lraD{Flat}$ 
\> $\{c\}$

\end{tabbing}

Nous étendons maintenant la Définition~\ref{roTdef} du \roCalT\  en considérant
les nouveaux opérateurs et les règles d'évaluation correspondantes.

\DEF%------------------------------------------------------------------------
Etant donnés un ensemble de symboles de fonctions $\FF$, un ensemble de variables
$\XX$, une théorie $T$ sur les termes du $\RTTf$ avec un problème de filtrage
décidable, nous appelons \roCalfT\  un calcul défini par~:
\begin{itemize}
\item un sous-ensemble non vide $\rttf$ de l'ensemble de termes $\RTTf$,
\item l'application (d'ordre supérieur) de substitution aux termes,
\item une théorie $T$,
\item l'ensemble de règles d'évaluation noté $\EE_{\rhof}$~: \rname{Fire}, \rname{Congruence},
	\rname{Congruence\_fail}, \rname{Distrib}, ~\rname{Batch},
	~\rname{Switch_L}, ~\rname{Switch_R}, ~\rname{OpOnSet}, ~\rname{Flat},
	~\rname{First}, ~\rname{First\_fail}, ~\rname{First\_success} et
	\rname{First\_single},
\item une stratégie d'évaluation $\SS$ qui guide l'application des règles
	d'évaluation.
\end{itemize}
\FDEF%------------------------------------------------------------------------

Les exemples suivants présentent la réduction de certains $\rhof$-termes
contenant les opérateurs du calcul étendu.

\EX%------------------------------------------------------------------------
L'application non-déterministe d'une des règles $a \ra b$, $a \ra c$, $a \ra d$
au terme $a$ est représentée dans le \roCal\  par l'application $[\{a \ra b,a \ra
c,a \ra d\}](a)$.  Ce dernier $\rho$-terme est réduit en l'ensemble $\{b,c,d\}$ qui
représente un choix non-déterministe parmi les trois termes. Si nous voulons
appliquer les règles $a \ra b$, $a \ra c$, $a \ra d$ d'une façon déterministe et 
dans l'ordre précisé, nous utilisons le $\rho$-terme $[first(a \ra b,a \ra c,a
\ra d)](a)$ avec par exemple la réduction~:

\begin{tabbing}
\indent \indent \= $\lraD{First\_success}$ ~~~ \=  \kill
\> \> $[first(a \ra b,a \ra c,a \ra d)](a)$
\\
\> $\lraD{First}$ 
\> $\paron{[a \ra b](a),[a \ra c](a),[a \ra d](a)}$
\\
\> $\lraD{Fire}$ 
\> $\paron{\{b\},[a \ra c](a),[a \ra d](a)}$
\\
\> $\lraD{First\_success}$ 
\> $\{\{b\}\}$
\\
\> $\lraD{Flat}$ 
\> $\{b\}$

\end{tabbing}

Nous pouvons remarquer que même si toutes les règles de réécriture utilisées
s'appliquent au terme $a$ avec succès (pas de $\emptyset$), le résultat final
est donné par la première règle de réécriture essayée.
\FEX%------------------------------------------------------------------------

\EX%------------------------------------------------------------------------
Nous considérons maintenant le cas où certaines règles données en argument
à $first$ mènent à un échec~:

\begin{tabbing}
\indent \indent \= $\lraD{First\_success}$ ~~~ \=  \kill
\> \> $[first(a \ra b,b \ra c,a \ra d)](b)$
\\
\> $\lraD{First}$ 
\> $\paron{[a \ra b](b),[b \ra c](b),[a \ra d](b)}$
\\
\> $\lraD{Fire}$ 
\> $\paron{\emptyset,[b \ra c](b),[a \ra d](b)}$
\\
\> $\lraD{First\_fail}$
\> $\paron{[b \ra c](b),[a \ra d](b)}$
\\
\> $\lraD{Fire}$ 
\> $\paron{\{c\},[a \ra d](b)}$
\\
\> $\lraD{First\_success}$ 
\> $\{\{c\}\}$
\\
\> $\lraD{Flat}$ 
\> $\{c\}$

\end{tabbing}
\FEX%------------------------------------------------------------------------

\EX%------------------------------------------------------------------------
Si aucune des règles données en argument à $first$ n'est appliquée avec succès le
résultat est bien sur l'ensemble vide~:

\begin{tabbing}
\indent \indent\= $\lraD{First\_single}$ ~~~ \=  \kill
\> \> $[first(a \ra b,a \ra c,a \ra d)](b)$
\\
\> $\lraD{First}$ 
\> $\paron{[a \ra b](b),[a \ra c](b),[a \ra d](b)}$
\\
\> $\longra{*}_{Fire}$ 
\> $\paron{\emptyset,\emptyset,\emptyset}$
\\
\> $\longra{*}_{First\_fail}$ 
\> $\paron{}$
\\
\> $\lraD{First\_single}$ 
\> $\emptyset$

\end{tabbing}
\FEX%------------------------------------------------------------------------

En utilisant l'opérateur $first$ nous introduisons d'autres opérateurs permettant
une description plus concise et intuitive des $\rho$-termes.  Par exemple, nous
pouvons définir un terme 
	$$try(s) \eqdef first(s,id)$$ 
tel que l'application de ce terme au terme $t$ est réduite en le résultat de
l'application $[s](t)$ si $[s](t)$ n'est pas réduite en $\emptyset$ et à $\{t\}$
si $[s](t)$ est réduite en $\emptyset$.

\subsubsection{L'opérateur $dc$}
%================================================================

L'ordre des arguments de l'opérateur $first$ dans un terme
$first(t_1,\ldots,t_n)$ est essentiel pour la réduction d'une application
$[t](u)$. Nous pouvons définir facilement deux opérateurs $n$-aire $dc$
(\textit{dont care choose}) et ``$\parod{}$'' avec un comportement similaire à celui
des opérateurs $first$ et $\paron{}$ mais cette fois-ci sans imposer un ordre
sur l'application et sur l'évaluation des arguments respectivement. Pour cela,
dans les termes de la forme $\parod{t_1,\ldots,t_n}$ nous considérons que la
virgule est associative, commutative et idempotente.  Les règles d'évaluation
caractérisant ces opérateurs sont présentées dans la Figure~\ref{MRAdc}.

\begin{figure}[!htp]%------------------------------------------------------------------------
\noindent \framebox{\parbox{\largeurtexte}{

\renewcommand{\fleche}{\Longrightarrow}
\begin{ruleset}
%===================================
  \regle 
  {DC}
  {[dc(s_1,\ldots,s_n)](t)}
  {\parod{[s_1](t),\ldots,[s_n](t)}} \\
%===================================
  \regle 
  {DC\_fail} 
  {\parod{t_1,\ldots,t_{k-1},\emptyset,t_{k+1},\ldots,t_n}} 
  {\parod{t_1,\ldots,t_n}}
  %{n > 0}
%===================================
  \ccregle 
  {DC\_success} 
  {\parod{t_1,\ldots,t,\ldots,t_n}}
  {\{t\}}
  {t~ne~contient~pas~de~radical~et~de}
  {variable~libre~et~il~n'est~pas~\emptyset}
%===================================
  \regle 
  {DC\_single} 
  {\parod{}}
  {\emptyset}
%===================================
\end{ruleset}

}}
\caption{\label{MRAdc}Les règles d'évaluation de l'opérateur $dc$}
\end{figure}%------------------------------------------------------------------------

De la même manière que l'opérateur $first$, l'opérateur $dc$ lance l'évaluation
de l'application de ses arguments sur le terme donné. L'opérateur $\parod{}$
teste l'échec de la réduction de ses arguments mais, contrairement à l'opérateur
$\paron{}$, pas dans un ordre précis. Dès que nous trouvons un terme $t_k$ clos
en forme normale parmi les arguments d'un terme $\parod{t_1,\ldots,t_n}$ nous
pouvons fournir $\{t_k\}$ comme résultat et ignorer les autres arguments. Toute
réduction en $\emptyset$ d'un argument est ignorée.

\EX%------------------------------------------------------------------------
En utilisant les règles d'évaluation précédentes nous obtenons deux réductions
possibles pour le terme $[dc(a \ra b,a \ra c)](a)$ menant soit au terme $\{b\}$,
soit au terme $\{c\}$. \\

\begin{tabular}{lcl}
$~~~~~$
$[dc(a \ra b,a \ra c)](a)$
& ~~~~~~ & 
$~~~~~$
$[dc(a \ra b,a \ra c)](a)$
\\

$\lraD{}$ 
$\parod{[a \ra b](a),[a \ra c](a)}$
& ~~~~~~ & 
$\lraD{}$ 
$\parod{[a \ra b](a),[a \ra c](a)}$
\\

$\lraD{}$ 
$\parod{\{b\},\{c\}}$
& ~~~~~~ & 
$\lraD{}$ 
$\parod{\{b\},\{c\}}$
\\

$\lraD{}$ 
$\{\{b\}\}$
& ~~~~~~ & 
$\lraD{}$ 
$\{\{c\}\}$
\\

$\lraD{}$ 
$\{b\}$
& ~~~~~~ & 
$\lraD{}$ 
$\{c\}$
\end{tabular}
\FEX%------------------------------------------------------------------------

Nous obtenons ainsi pour l'application d'un terme $dc(t_1,\ldots,t_n)$ sur un
autre terme $u$ un comportement non-déterministe dans le sens où plusieurs
réductions peuvent mener à des résultats différents si plusieurs réductions sans
échec sont possibles. Cette approche permettrait la description de l'application
d'une règle de réécriture à une position \textit{quelconque} dans un terme mais
le calcul obtenu serait évidemment non-confluent. L'inconvénient de cette méthode
est le manque de contrôle et donc l'impossibilité de décrire des stratégies
de réduction précises et particulièrement des stratégies de normalisation du type
\textit{leftmost innermost} ou \textit{leftmost outermost}.

Dans le reste de ce chapitre nous utilisons l'opérateur $first$ pour la
descriptions de nouveaux opérateurs, mais cet opérateur peut être remplacé avec
l'opérateur $dc$ si nous désirons permettre des éventuelles réductions
non-convergentes.

\section{Congruence générique}
%================================================================

Nous définissons maintenant les opérateurs qui permettent l'application d'un
$\rho$-terme $r$ à une certaine position d'un autre $\rho$-terme $t$ en
descendant à chaque étape de réduction d'un niveau dans la profondeur du terme
$t$. 

La première étape est la définition de deux opérateurs qui décrivent
l'application d'un terme non au sommet d'un autre terme mais à ses
sous-termes directs.
Il est déjà possible de décrire cette opération dans le \roCal\  en utilisant la
règle d'évaluation \rname{Congruence}. Ainsi, pour tout symbole $f \in \FF$, nous pouvons
représenter l'application d'un terme $r$ aux sous-termes d'un terme
$f(u_1,\ldots,u_n)$ par le $\rho$-terme $[f(r,\ldots,r)](f(u_1,\ldots,u_n))$ qui
est réduit en $\{f([r](u_1),\ldots,[r](u_n))\}$.  Mais nous voulons définir un
opérateur générique qui applique un \mbox{$\rho$-terme} $r$ aux sous-termes
$u_i$, $\eni{i}{n}$ d'un terme de la forme $F(u_1,\ldots,u_n)$ indépendamment
du symbole de tête $F$.

Pour cela, nous introduisons deux opérateurs unaire $\Phi$ et $\Psi$ dont le
comportement est décrit par les règles d'évaluation présentées dans la
Figure~\ref{MRAtrav}. Ces opérateurs permettent la représentation de
l'application d'un $\rho$-terme aux arguments d'un autre $\rho$-terme et ils
sont inspirés des opérateurs similaires du \textit{System S} décrit
dans~\cite{VISSERwrla98}. Nous montrerons que le comportement de ces opérateurs
peut être exprimé en utilisant seulement les opérateurs du \roCalf.

\begin{figure}[!htp]%--------------------------------------------------------------
\noindent \framebox{\parbox{\largeurtexte}{

\renewcommand{\fleche}{\Longrightarrow}
\begin{ruleset}
%===================================
  \regle 
  {Traverse\_seq} 
  {[\Phi(r)](f(u_1,\ldots,u_n))} 
  {\paron{\{f([r](u_1),\ldots,u_n)\},\ldots,\{f(u_1,\ldots,[r](u_n))\}}} 
\\
%===================================
  \regle 
  {Traverse\_par} 
  {[\Psi(r)](f(u_1,\ldots,u_n))} 
  {\{f([r](u_1),\ldots,[r](u_n))\}} 
%===================================
\end{ruleset}

}}
\caption{\label{MRAtrav}Congruence générique}
\end{figure}%------------------------------------------------------------------------

L'évaluation de l'application du $\rho$-terme $\Phi(r)$ au terme
$t=f(u_1,\ldots,u_n)$ revient à l'évaluation de l'application du terme $r$ à un
des sous-termes $u_i$. Plus précisément, $r$ est appliqué sur le premier $u_i$,
$\eni{i}{n}$ tel que l'application $[r](u_i)$ n'est pas réduite en l'ensemble vide. S'il
n'\textit{existe} pas un tel sous-terme $u_i$ et en particulier, si $t$ est une
fonction sans arguments ($t$ est une constante), alors le terme
$[\Phi(r)](t)$ se réduit en l'ensemble vide.

Le comportement dans le cas des constantes devient plus clair
si le membre droit de la règle \rname{Traverse\_seq}
est écrit $\paron{\{f(\ldots,[r](u_i),\ldots)\} ~|~ i=1 \ldots n}$ 
et si nous remarquons qu'une constante est un terme de la forme
$c(u_1,\ldots,u_n)$ avec $n=0$ obtenant ainsi la réduction suivante~:
	$$[\Phi(r)](c) ~\longra{}_{Traverse\_seq}~ 
	\paron{\{\}} ~\longra{}_{First\_fail}~ 
	\paron{} ~\longra{}_{First\_single}~ \emptyset$$ 

Lorsque le $\rho$-terme $\Psi(r)$ est appliqué à un terme $t=f(u_1,\ldots,u_n)$,
le terme $r$ est appliqué à tous les arguments $u_i$, $\eni{i}{n}$ si
\textit{pour tout} $i$, $[r](u_i)$ n'est pas réduite en $\emptyset$. S'il existe un
$u_i$ tel que $[r](u_i)$ est réduit en $\emptyset$, alors le résultat est
l'ensemble vide. Si nous appliquons $\Psi(r)$ sur une constante $c$, puisque il
n'y a aucun sous-terme, nous obtenons la réduction~:
	$$[\Psi(r)](c) ~\longra{}_{Traverse\_par}~\{c\}$$ 

Si nous considérons un \roCal\  avec une signature finie $\FF$ et si nous
notons par $\FF_{0}=\{c_1,\ldots,c_n\}$ l'ensemble de symboles de fonctions
constantes et par $\FF_{+}=\{f_1,\ldots,f_m\}$ l'ensemble de symboles de
fonctions non constantes, les deux opérateurs $\Phi(r)$ et
$\Psi(r)$ peuvent être exprimés par certains $\rho$-termes appropriés.

Si les deux définitions suivantes sont considérées
$$\Phi'(r) \eqdef first(f_1(r,id,\ldots,id),\ldots,f_1(id,\ldots,id,r),\ldots,
                        f_m(r,id,\ldots,id),\ldots,f_m(id,\ldots,id,r))$$
$$\Psi(r) \eqdef \{c_1,\ldots,c_n,f_1(r,\ldots,r),\ldots,f_m(r,\ldots,r)\}$$
avec $c_i \in \FF_{0}$, $\eni{i}{n}$, et $f_j \in \FF_{+}$,
$\eni{j}{m}$, nous obtenons les réductions suivantes~:

\begin{tabbing}
\indent\indent \=  $\longra{*}_{Congruence}$ ~~~ \=  \kill
\> \> $[\Phi'(r)](f_k(u_1,\ldots,u_p))$
\\
\> $\eqdef$ 
\> $[first(f_1(r,id,\ldots,id),\ldots,f_m(id,\ldots,id,r))](f_k(u_1,\ldots,u_p))$
\\
\> $\longra{}_{First}$ 
\> $\paron{[f_1(r,id,\ldots,id)](f_k(u_1,\ldots,u_p)),\ldots,[f_m(id,\ldots,id,r)](f_k(u_1,\ldots,u_p))}$
\\
\> $\longra{*}_{Congruence}$ 
\> $\paron{\emptyset,\ldots,\emptyset,\{f_k([r](u_1),\ldots,u_p)\},\ldots,
         \{f_k(u_1,\ldots,[r](u_p))\},\emptyset,\ldots,\emptyset}$
\\
\> $\longra{*}_{First\_fail}$ 
\> $\paron{\{f_k([r](u_1),\ldots,u_p)\},\ldots,
         \{f_k(u_1,\ldots,[r](u_p))\},\emptyset,\ldots,\emptyset}$

\end{tabbing}
et
\begin{tabbing}
\indent\indent \=  $\longra{*}_{Congruence}$ ~~~ \=  \kill
\> \> $[\Psi(r)](f_k(u_1,\ldots,u_p))$
\\
\> $\eqdef$ 
\> $[\{c_1,\ldots,c_n,f_1(r,\ldots,r),\ldots,f_m(r,\ldots,r)\}](f_k(u_1,\ldots,u_p))$
\\
\> $\longra{}_{Distrib}$ 
\> $\{[c_1](f_k(u_1,\ldots,u_p)),\ldots\ldots,[f_m(r,\ldots,r)](f_k(u_1,\ldots,u_p))\}$
\\
\> $\longra{*}_{Congruence}$ 
\> $\{\emptyset,\ldots,\emptyset,\{f_k([r](u_1),\ldots,[r](u_p))\},\emptyset,\ldots,\emptyset\}$
\\
\> $\longra{*}_{Flat}$ 
\> $\{f_k([r](u_1),\ldots,[r](u_p))\}$

\end{tabbing}

L'opérateur $\Phi'$ ne correspond donc pas exactement à la description de l'opérateur
$\Phi$ donnée dans la Figure~\ref{MRAtrav} mais le même résultat est obtenu
en appliquant les termes $\Phi(r)$ et $\Phi'(r)$ à un terme
$f_k(u_1,\ldots,u_p)$ comme montré ci-dessous.

\LEM%------------------------------------------------------------------------
Les applications des termes $\Phi(r)$ et $\Psi(r)$ à un terme $t$ peuvent être
décrites dans le \roCalfT.
\FLEM%------------------------------------------------------------------------

\proof{
Si nous considérons $t=f_k(u_1,\ldots,u_p)$ et si pour tout $\eni{i}{p}$ nous
avons les réductions $[r](u_i) \longra{*}_{\rho} \emptyset$ alors, selon les
règles d'évaluation décrivant le comportement de $\Phi(r)$, nous obtenons~:

\begin{tabbing}
\indent\indent \=  $\longra{*}_{Congruence}$ ~~~ \=  \kill
\> \> $[\Phi(r)](f_k(u_1,\ldots,u_p))$
\\
\> $\longra{}_{Traverse\_seq}$ 
\> $\paron{\{f_k([r](u_1),\ldots,u_p)\},\ldots,\{f_k(u_1,\ldots,[r](u_p))\}}$
\\
\> $\longra{*}_{\rho}$ 
\> $\paron{\{f_k(\emptyset,\ldots,u_p)\},\ldots,\{f_k(u_1,\ldots,\emptyset)\}}$
\\
\> $\longra{*}_{OpOnSet}$ 
\> $\paron{\{\emptyset\},\ldots,\{\emptyset\}}$
\\
\> $\longra{*}_{Flat}$ 
\> $\paron{\emptyset,\ldots,\emptyset}$
\\
\> $\longra{*}_{First\_fail}$ 
\> $\paron{}$
\\
\> $\longra{}_{First\_single}$ 
\> $\emptyset$

\end{tabbing}

Autrement, s'il existe un $l$ tel que $[r](u_i) \longra{*}_{\rho} \emptyset$,
$\eni{i}{l-1}$ et $[r](u_l) \longra{*}_{\rho} v_l\da{}$, avec $ v_l\da{}$ un
terme clos ne contenant pas de radical, la réduction suivante est obtenue~:

\begin{tabbing}
\indent\indent \=  $\longra{*}_{Congruence}$ ~~~ \=  \kill
\> \> $[\Phi(r)](f_k(u_1,\ldots,u_p))$
\\
\> $\longra{}_{Traverse\_seq}$ 
\> $\paron{\{f_k([r](u_1),\ldots,u_p)\},\ldots,\{f_k(u_1,\ldots,[r](u_p))\}}$
\\
\> $\longra{*}_{\rho}$ 
\> $\paron{\{f_k(\emptyset,\ldots,u_p)\},\ldots,\{f_k(u_1,\ldots,v_l \da{},\ldots,u_p)\},\ldots,\{f_k(u_1,\ldots,\emptyset)\}}$
\\
\> $\longra{*}_{OpOnSet}$ 
\> $\paron{\emptyset,\ldots,\emptyset,\{f_k(u_1,\ldots,v_l \da{},\ldots,u_p)\},\emptyset,\ldots,\emptyset}$
\\
\> $\longra{*}_{First\_fail}$ 
\> $\paron{\{f_k(u_1,\ldots,v_l \da{},\ldots,u_p)\},\emptyset,\ldots,\emptyset}$
%\\
%\> $\longra{*}_{First\_single}$ 
%\> $\{f_k(u_1,\ldots,v_l \da{},\ldots,u_p)\}$

\end{tabbing}

Maintenant, si nous considérons la définition de $\Phi'(r)$ et si pour tout
$\eni{i}{p}$ nous avons $[r](u_i) \longra{*}_{\rho} \emptyset$ alors, nous
obtenons~:

\begin{tabbing}
\indent\indent \=  $\longra{*}_{Congruence}$ ~~~ \=  \kill
\> \> $[\Phi'(r)](f_k(u_1,\ldots,u_p))$
\\
\> $\longra{*}_{\rho}$ 
\> $\paron{\{f_k([r](u_1),\ldots,u_p)\},\ldots,\{f_k(u_1,\ldots,[r](u_p))\},\emptyset,\ldots,\emptyset}$
\\
\> $\longra{*}_{\rho}$ 
\> $\paron{\{f_k(\emptyset,\ldots,u_p)\},\ldots,\{f_k(u_1,\ldots,\emptyset)\},\emptyset,\ldots,\emptyset}$
\\
\> $\longra{*}_{OpOnSet}$ 
\> $\paron{\{\emptyset\},\ldots,\{\emptyset\},\emptyset,\ldots,\emptyset}$
\\
\> $\longra{*}_{Flat}$ 
\> $\paron{\emptyset,\ldots,\emptyset,\ldots,\emptyset}$
\\
\> $\longra{*}_{First\_fail}$ 
\> $\paron{}$
\\
\> $\longra{}_{First\_single}$ 
\> $\emptyset$

\end{tabbing}

Pour le même terme $[\Phi'(r)](f_k(u_1,\ldots,u_p))$, s'il existe un $l$ tel
que $[r](u_i) \longra{*}_{\rho} \emptyset$, $\eni{i}{l-1}$ et $[r](u_l)
\longra{*}_{\rho} v_l\da{}$, avec $ v_l\da{}$ un terme clos ne contenant pas de
radical, la réduction suivante est obtenue~:

\begin{tabbing}
\indent\indent \=  $\longra{*}_{Congruence}$ ~~~ \=  \kill
\> \> $[\Phi'(r)](f_k(u_1,\ldots,u_p))$
\\
\> $\longra{*}_{\rho}$ 
\> $\paron{\{f_k([r](u_1),\ldots,u_p)\},\ldots,\{f_k(u_1,\ldots,[r](u_p))\},\emptyset,\ldots,\emptyset}$
\\
\> $\longra{*}_{\rho}$ 
\> $\paron{\{f_k(\emptyset,\ldots,u_p)\},\ldots,\{f_k(u_1,\ldots,v_l \da{},\ldots,u_p)\},\emptyset,\ldots,\emptyset}$
\\
\> $\longra{*}_{OpOnSet}$ 
\> $\paron{\{\emptyset\},\ldots,\{\emptyset\},\{f_k(u_1,\ldots,v_l \da{},\ldots,u_p)\},\emptyset,\ldots,\emptyset}$
\\
\> $\longra{*}_{Flat}$ 
\> $\paron{\emptyset,\ldots,\emptyset,\{f_k(u_1,\ldots,v_l \da{},\ldots,u_p)\},\emptyset,\ldots,\emptyset}$
\\
\> $\longra{*}_{First\_fail}$ 
\> $\paron{\{f_k(u_1,\ldots,v_l \da{},\ldots,u_p)\},\emptyset,\ldots,\emptyset}$

\end{tabbing}

Nous pouvons remarquer que les résultats des réductions pour l'application d'un
terme $r$ aux arguments d'un terme $f_k(u_1,\ldots,u_p)$ en utilisant les deux
opérateurs, $\Phi$ et $\Phi'$, sont identiques. Si les termes $u_i$, 
$i=1 \ldots p$, sont des termes clos ne contenant pas de radical alors le
résultat final des deux réductions dans le cas sans échec est
$\{f_k(u_1,\ldots,v_l \da{},\ldots,u_p)\}$.

Quand les opérateurs sont appliqués à une constante $c_k \in \FF_{0}$ nous obtenons~:
$$[\Phi'(r)](c_k) \longra{*}_{\rho} \paron{} \longra{}_{\rho} \emptyset,$$
$$[\Psi(r)](c_k) \longra{*}_{\rho} \{c_k\}.$$
}

L'application d'un terme $r$ à tous les sous-termes de profondeur $n$ d'un terme
$t$ est représentée par le $\rho$-terme $[\Psi(\ldots(\Psi(r))\ldots)](t)$ avec
$n$ niveaux d'imbrication de l'opérateur $\Psi$. L'application à seulement un
sous-terme de profondeur $n$ est représentée par le $\rho$-terme
$[\Phi(\ldots(\Phi(r))\ldots)](t)$ avec $n$ niveaux d'imbrication de l'opérateur
$\Phi$.

Pour une présentation plus concise et plus claire, dans les réductions décrites
ultérieurement nous utilisons la description de l'opérateur $\Phi$ donné par les
règles d'évaluation dans la Figure~\ref{MRAtrav}.

\section{Application en profondeur} \label{opPointFixe}
%================================================================
 
La définition des stratégies d'évaluation (normalisation) comme, par exemple, 
\textit{top-down} ou \textit{bottom-up}, est basée sur l'application d'un terme
à la position de tête ou aux positions les plus profondes d'un autre terme.
Pour le moment nous avons la possibilité d'appliquer un $\rho$-terme $r$ soit à
un ou à tous les arguments $u_i$ d'un $\rho$-terme $t=f(u_1,\ldots,u_n)$, soit
aux sous-termes se trouvant à une profondeur de $t$ précisée explicitement.
Mais la profondeur d'un terme n'est pas connue \textit{a priori} et donc, nous
ne pouvons pas appliquer un terme $r$ aux positions les plus profondes d'un
terme $t$. Si nous voulons appliquer le terme $r$ aux sous-termes à la
profondeur maximale d'un terme $t$ (i.e. les sous-termes n'ayant pas de
sous-terme strict) nous devons définir un opérateur récursif qui réitère
l'application des termes $\Phi(r)$ et $\Psi(r)$ et ainsi, pousse l'application
le plus profondément possible dans les termes.

Nous commençons par présenter le $\rho$-terme permettant la description des applications
récursives dans le \roCal.
En s'inspirant des opérateurs de point fixe du \laCal\  nous pouvons
définir un $\rho$-terme qui applique récursivement un $\rho$-terme donné. Nous
utilisons le combinateur de point fixe classique du \laCal\  (\cite{Barendregt84}),
$\Theta_{\lambda}=(A_{\lambda}~A_{\lambda})$ où
	$$A_{\lambda}=\lambda x y.y(x x y)$$
$\Theta_{\lambda}$ est appelé le combinateur de point fixe de Turing (\cite{Turing37jsl}).

Ce terme correspond dans le \roCal\  au $\rho$-terme $\Theta=[A](A)$ avec
	$$A=x \ra (y \ra [y]([[x](x)](y))).$$

Dans le \laCal, pour tout $\lambda$-terme $G$ nous avons la réduction~:
	$$\Theta_{\lambda}~G \longra{*}_{\beta} G(\Theta_{\lambda}~G).$$

Dans le \roCal\  nous avons une réduction similaire~:

\begin{flushright}
$[\Theta](G) \longra{*}_{\rho} \{[G]([\Theta](G))\}$
~~~~~~~~~~~~~~~~~~~~~~~~~~~~~~
$(Point\_Fixe)$
\end{flushright} 
qui est détaillée par~:

\begin{tabbing}
\indent\indent \= $\lraD{Distrib}$ ~~~ \=  \kill
\> \> $[\Theta](G) \eqdef [[A](A)](G) \eqdef [[x \ra (y \ra [y]([[x](x)](y)))](A)](G)$
\\
\> $\lraD{Fire}$
\> $[\{y \ra [y]([[A](A)](y))\}](G)$
\\
\> $\lraD{Distrib}$
\> $\{[y \ra [y]([[A](A)](y))](G)\}$
\\
\> $\lraD{Fire}$
\> $\{\{[G]([[A](A)](G))\}\}$
\\
\> $\lraD{Flat}$
\> $\{[G]([[A](A)](G))\}$
\\
\> $\eqdef$
\> $\{[G]([\Theta](G))\}$

\end{tabbing}

Nous avons obtenu le résultat souhaité mais la dernière application de la règle
\rname{Fire} dans la réduction ci-dessus peut être remplacée par une réduction
dans le sous-terme $[[A](A)](y)$. Nous pouvons réduire ainsi
$[[A](A)](y) \eqdef [[x' \ra (y' \ra [y']([[x'](x')](y')))](A)](y)$ en le terme 
$\{[y]([[A](A)](y))\} \eqdef \{[y]([\Theta](y))\}$. Nous obtenons donc la
réduction~:

\begin{tabbing}
\indent\indent \= $\longra{*}_{\rho}$ ~~~ \=  \kill
\> \> $[\Theta](G)$
\\
\> $\longra{*}_{\rho}$
\> $\{[y \ra [y]([[A](A)](y))](G)\} \eqdef \{[y \ra [y]([\Theta](y))](G)\}$
\\
\> $\longra{*}_{\rho}$
\> $\{[y \ra [y](\{[y]([\Theta](y))\})](G)\}$
\\
\> $\longra{*}_{\rho}$
\> $\{[y \ra [y]([y]([\Theta](y)))](G)\}$
\\
\> $\longra{*}_{\rho}$
\> $\ldots$

\end{tabbing}
qui ne termine jamais si à chaque fois le même radical $[\Theta](y)$ est
sélectionné pour réduction.

Dans une approche opérationnelle nous ne voudrions pas que les nouvelles
constructions conduisent à des réductions qui ne terminent pas. Puisque le
$\rho$-terme $[\Theta](G)$ peut évidemment mener à des réductions infinies, une
stratégie devrait être imposée afin d'obtenir la terminaison et donc le
comportement souhaité. 

Nous devrions ainsi utiliser une stratégie qui applique les règles d'évaluation
à un terme de la forme $[\Theta](G)$ seulement lorsqu'aucune autre réduction
n'est possible. D'un point de vue opérationnel, cette stratégie est assez
difficile à implanter et évidemment pas très efficace dans un calcul où le terme
$\Theta$ est représenté par sa forme étendue et donc, plus difficile à
identifier. Dans le cas où $\Theta$ est considéré comme un $\rho$-terme
indépendant avec le comportement décrit par une règle d'évaluation correspondant
à la réduction $Point\_Fixe$, la stratégie proposée précédemment est facilement
réalisable.

Une stratégie satisfaisant la condition de terminaison et plus facile à
implanter pourrait appliquer les règles d'évaluation d'abord aux positions de
tête des termes et seulement au moment où aucune règle d'évaluation n'est
applicable en tête, réduire les sous-termes à des positions plus
profondes. Il est clair que cette stratégie \textit{outermost} empêche
seulement les réductions infinies dues à l'opérateur $\Theta$ mais elle ne peut
pas assurer la terminaison du calcul non-typé.

Comme nous l'avons dit précédemment, le but principal est la représentation des
stratégies de normalisation par des $\rho$-termes et donc, nous voulons décrire
l'application d'un terme $r$ à toutes les positions d'un autre terme $t$. Par
conséquent, nous devons définir le terme $G$ approprié qui propage l'application
d'un $\rho$-terme dans les sous-termes d'un autre $\rho$-terme.

\subsection{Applications multiples}
%=============================================================================== 
%

Dans un premier temps, nous voulons définir les opérateurs $BottomUp$ et
$TopDown$ décrivant l'application d'un terme $r$ à tous les sous-termes d'un
terme $t$ en commençant avec les positions les plus profondes et respectivement
avec la position de tête.  Nous voudrions donc trouver un terme qui appliquerait
récursivement le terme $r$ à tous les sous-termes de $t$ et ultérieurement au
sommet du terme résultat et un autre terme qui appliquerait le terme $r$ d'abord
au sommet du terme $t$ et ensuite aux sous-termes du terme résultat.
Le terme $r$ doit être appliqué à un sous-terme seulement si cette application
ne mène pas à un échec.

Nous proposons d'abord deux définitions ``naïves'' pour ces opérateurs et nous
commentons les problèmes rencontrés. En analysant les réductions obtenues nous
définissons finalement des opérateurs décrivant exactement le comportement
souhaité.

Une première approche consiste à définir le $\rho$-terme~:
	$$G_{sds}(r) \eqdef f \ra (x \ra [\Psi(f);r](x))$$
qui décrit l'application d'un terme $f$ (qui sera instancié par 
le mécanisme de filtrage) a tous les arguments du terme qui
instancie la variable $x$, suivie de l'application du terme $r$ à la position de
tête du terme résultant de l'application précédente.

Nous définissons l'opérateur $SDS$ (pour $SpreadDownSimple$)~:
	$$SDS(r) \eqdef [\Theta](G_{sds}(r))$$
et nous obtenons la réduction suivante pour l'application de ce terme au terme
$t=f(t_1,\ldots,t_n)$~:

\begin{tabbing}
\indent\indent \= $\longra{*}_{\rho}$ ~~~ \=  \kill
\> \> $[SDS(r)](t) \eqdef [[\Theta](G_{sds}(r))](t)$
\\
\> $\longra{*}_{\rho}$
\> $\{[[G_{sds}(r)]([\Theta](G_{sds}(r)))](t)\}$ 
\\
\> $\eqdef$ 
\> $\{[[G_{sds}(r)](SDS(r))](t)\}$ 
\\
\> $\eqdef$ 
\> $\{[[f \ra (x \ra [\Psi(f);r](x))](SDS(r))](t)\}$
\\
\> $\longra{*}_{\rho}$
\> $\{[\{x \ra [\Psi(SDS(r));r](x)\}](t)\}$
\\
\> $\longra{*}_{\rho}$
\> $\{[\Psi(SDS(r));r](f(t_1,\ldots,t_n))\}$
\\
\> $\longra{*}_{\rho}$
\> $\{[r]([\Psi(SDS(r))](f(t_1,\ldots,t_n)))\}$
\\
\> $\longra{*}_{\rho}$
\> $\{[r](f([SDS(r)](t_1),\ldots,[SDS(r)](t_n)))\}$

\end{tabbing}

Ainsi, le terme $SDS(r)$ est appliqué à tous les arguments du terme
initial et le terme $r$ est appliqué en tête. Si le terme initial
est une constante $c$ alors, le résultat de la réduction est $\{[r](c)\}$ et
nous obtenons donc le comportement souhaité.

L'inconvénient de cette méthode est la non-confluence des dérivations quand le
terme $r$ est un ensemble ayant plus d'un élément. 
Par exemple, dans le cas où $r=\{a,b\}$ le terme $G_{sds}(r)=G_{sds}(\{a,b\})$
peut être réduit en le terme $\{G_{sds}(a),G_{sds}(b)\}$ comme montré par la
réduction~:

\begin{tabbing}
\indent\indent \= $\longra{*}_{Switch_R}$ ~~~ \=  \kill
\> \> $G_{sds}(\{a,b\})$
 $\eqdef$ 
 $f \ra (x \ra [\Psi(f);\{a,b\}](x))$
\\
\> $\eqdef$ 
\> $f \ra (x \ra [z \ra [\{a,b\}]([\Psi(f)](z))](x))$
\\
\> $\longra{*}_{Distrib}$
\> $f \ra (x \ra [z \ra \{[a]([\Psi(f)](z)),[b]([\Psi(f)](z))\}](x))$
\\
\> $\longra{*}_{Switch_R}$
\> $f \ra (x \ra [\{z \ra [a]([\Psi(f)](z)),z \ra [b]([\Psi(f)](z))\}](x))$
\\
\> $\longra{*}_{Distrib}$
\> $f \ra (x \ra \{[z \ra [a]([\Psi(f)](z))](x),[z \ra [b]([\Psi(f)](z))](x)\})$
\\
\> $\longra{*}_{Switch_R}$
\> $f \ra (\{x \ra [z \ra [a]([\Psi(f)](z))](x),x \ra [z \ra [b]([\Psi(f)](z))](x)\})$
\\
\> $\longra{*}_{Distrib}$
\> $\{f \ra (x \ra [z \ra [a]([\Psi(f)](z))](x),f \ra (x \ra [z \ra [b]([\Psi(f)](z))](x)\})$
\\
\> $\eqdef$ 
\> $\{G_{sds}(a),G_{sds}(b)\}$

\end{tabbing}

En utilisant le résultat précédent, nous obtenons la réduction suivante pour le
$\rho$-terme $SDS(\{a,b\})$~:

\begin{tabbing}
\indent\indent \= $\longra{*}_{\rho}$ ~~~ \=  \kill
\> \> $SDS(\{a,b\})$
\\
\> $\eqdef$ 
\> $[\Theta](G_{sds}(\{a,b\}))$
\\
\> $\longra{*}_{\rho}$
\> $[\Theta](\{G_{sds}(a),G_{sds}(b)\}))$ 
\\
\> $\longra{*}_{\rho}$
\> $\{[\Theta](G_{sds}(a)),[\Theta](G_{sds}(b))\})$ 

\end{tabbing}
qui représente l'ensemble des applications récursives des $a$ et $b$ et non
l'application récursive de l'ensemble $\{a,b\}$ qui est obtenue si l'ensemble
$\{a,b\}$ n'est pas distribué dans les termes $G_{sds}(\{a,b\})$ et $SDS(\{a,b\})$.

Nous pouvons donc remarquer facilement que la raison de la non-confluence est la
propagation de symboles d'ensemble dans les termes $G_{sds}(\{a,b\})$ et
$SDS(\{a,b\})$.

Nous avons discuté dans le Chapitre~\ref{chap.resultats_confluence} les
problèmes liés à la confluence du calcul et nous avons proposé des stratégies
pour obtenir cette propriété. Une des conditions imposées afin d'obtenir la
confluence interdit l'application d'une règle de réécriture (non-linéaire à
droite) si l'argument de l'application contient un ensemble ayant plus d'un
élément. Cette restriction n'est évidemment pas satisfaite dans la réduction d'un
terme $[\Theta](G)$ en $\{[G]([\Theta](G))\}$ dans le cas où $G$ contient un
ensemble ayant plus d'un élément et donc dans la réduction du terme $SDS(\{a,b\})$.

La condition de ne pas évaluer en utilisant la règle d'évaluation \rname{Fire}
une application contenant un ensemble ayant plus d'un élément dans son argument à
été imposée afin d'éviter l'évaluation de l'argument à un tel ensemble. Nous
pouvons donc réduire le terme $[\Theta](G)$ en $\{[G]([\Theta](G))\}$ si le
terme $G$ contient un ensemble mais il ne peut pas être réduit en un ensemble
ayant plus d'un élément.

Dans le cas de l'opérateur $SDS$ nous proposons une solution qui
consiste à empêcher l'évaluation du terme $G_{sds}(r)$ en un ensemble même si $r$
ne satisfait pas cette condition. Nous définissons l'opérateur $G_{sd}$~:
	$$G_{sd}(r) \eqdef f \ra (x \ra \paron{[\Psi(f);r](x)})$$
et respectivement $SD$ (pour $SpreadDown$)~:
	$$SD(r) \eqdef [\Theta](G_{sd}(r)).$$

Si $r=\{a,b\}$ alors, le terme $G_{sd}(r)=G_{sd}(\{a,b\})$ n'est pas réduit en le
terme $\{G_{sd}(a),G_{sd}(b)\}$ comme il était le cas pour le terme $G_{sds}(r)$~:

\begin{tabbing}
\indent\indent \= $\longra{*}_{Distrib}$ ~~~ \=  \kill
\> \> $G_{sd}(r)$ $\eqdef$ $f \ra (x \ra \paron{[\Psi(f);\{a,b\}](x)})$ 
\\
\> $\longra{*}_{\rho}$
\> $f \ra (x \ra \paron{[\{a,b\}]([\Psi(f)](x))})$ 
\\
\> $\longra{*}_{Distrib}$
\> $f \ra (x \ra \paron{\{[a]([\Psi(f)](x)),[b]([\Psi(f)](x))\}})$ 

\end{tabbing}

Dans ce dernier terme le premier argument de l'opérateur $\paron{}$ contient la
variable libre $x$ et donc il ne peut pas être réduit en utilisant la règle
d'évaluation \rname{First\_success}.

Puisque ce dernier terme n'est pas un ensemble, la propagation des symboles
d'ensemble n'intervient plus dans le cas de l'opérateur $G_{sd}$ et nous pouvons
réduire le terme $[\Theta](G_{sd}(r))$ en $\{[G_{sd}(r)]([\Theta](G_{sd}(r)))\}$.
Par conséquent, nous obtenons la réduction~:

\begin{tabbing}
\indent\indent \= $\longra{*}_{\rho}$ ~~~ \=  \kill
\> \> $[SD(r)](t) \eqdef [[\Theta](G_{sd}(r))](t)$
\\
\> $\longra{*}_{\rho}$
\> $\{[[G_{sd}(r)]([\Theta](G_{sd}(r)))](t)\}$ 
\\
\> $\eqdef$ 
\> $\{[[G_{sd}(r)](SD(r))](t)\}$
\\
\>  $\eqdef$ 
\>  $\{[[f \ra (x \ra \paron{[\Psi(f);r](x)})](SD(r))](t)\}$
\\
\> $\longra{*}_{\rho}$
\> $\{[\{x \ra \paron{[\Psi(SD(r));r](x)}\}](t)\}$
\\
\> $\longra{*}_{\rho}$
\> $\{\paron{[\Psi(SD(r));r](f(t_1,\ldots,t_n))}\}$
\\
\> $\longra{*}_{\rho}$
\> $\{\paron{[r](f([SD(r)](t_1),\ldots,[SD(r)](t_n)))}\}$

\end{tabbing}

\EX%------------------------------------------------------------------------
\label{exempleSpreadDown}
Si nous utilisons une stratégie qui applique les règles d'évaluation d'abord
en tête alors, la réduction suivante est obtenue~:

\begin{tabbing}
\indent\indent \= $\longra{*}_{Distrib}$  \= ~~~~~   \= \kill
\> \> $[SD(\{a \ra b, id\})](f(a,g(a)))$
\\
\> $\longra{*}_{\rho}$ 
\> $\{\paron{[\{a \ra b, id\}](f([SD(\{a \ra b, id\})](a),[SD(\{a \ra b, id\})](g(a))))}\}$
\\
\> $\longra{}_{Distrib}$ 
\> $\{\paronle\{[a \ra b](f([SD(\{a \ra b, id\})](a),[SD(\{a \ra b, id\})](g(a)))),$
\\
\> \>\> $[id](f([SD(\{a \ra b, id\})](a),[SD(\{a \ra b, id\})](g(a))))\}\paronri\}$
\\
\> $\longra{}_{Fire}$ 
\> $\{\paron{\{\emptyset,[id](f([SD(\{a \ra b, id\})](a),[SD(\{a \ra b, id\})](g(a))))\}}\}$
\\
\> $\longra{}_{Flat}$ 
\> $\{\paron{\{f([SD(\{a \ra b, id\})](a),[SD(\{a \ra b, id\})](g(a)))\}}\}$
\\
\> $\longra{*}_{\rho}$ 
\> $\{\paron{\{f(\{\paron{[\{a \ra b, id\}](a)}\},[SD(\{a \ra b, id\})](g(a)))\}}\}$
\\
\> $\longra{*}_{\rho}$ 
\> $\{\paron{\{f(\{\{b, a\},[SD(\{a \ra b, id\})](g(a)))\}}\}$
\\
\> $\longra{*}_{\rho}$ 
\> $\{\paron{\{f(\{b, a\},\paronle[\{a \ra b, id\}](g([SD(\{a \ra b, id\})](a))))\}}\}$
\\
\> $\longra{*}_{\rho}$ 
\> $\{\paron{\{f(\{b,a\},g(\{b,a\}))\}}\}$
\\
\> $\longra{*}_{\rho}$ 
\> $\{f(b,g(b)),f(a,g(b)),f(b,g(a)),f(a,g(a))\}$

\end{tabbing}
\FEX%------------------------------------------------------------------------

Nous pouvons remarquer que l'application $[SD(r)](t)$ ne garantit pas que les
applications du terme $r$ aux sous-termes les plus profonds de $t$ sont les
premières à être réduites.
Par exemple, puisque nous essayons d'appliquer les règles d'évaluation d'abord
en tête, dans la réduction de l'Exemple~\ref{exempleSpreadDown} nous
obtenons, en appliquant la règle d'évaluation \rname{Fire},
	$$[a \ra b](f([SD(\{a \ra b, id\})](a),
	[SD(\{a \ra b,id\})](g(a)))) \longra{}_{Fire} \emptyset$$
et non
	$$[a \ra b](f([SD(\{a \ra b, id\})](a),	[SD(\{a \ra b,id\})](g(a))))$$ 
	$$\longra{*}_{\rho} [a \ra b](f(\{b, a\},\{g(\{b, a\})\})) \longra{*}_{\rho} \emptyset$$
comme dans une réduction \textit{innermost}.

L'inconvénient de la non-confluence dans le cas de l'opérateur
$SDS$ a été éliminé en utilisant l'opérateur $\paron{}$ dans la
définition de l'opérateur $SD$, mais nous n'avons pas encore obtenu le
comportement souhaité pour ce type d'iterateur. Dans l'évaluation du terme
$[SD(r)](t)$, si une des applications du terme $r$ à un sous-terme de
$t$ est évaluée en $\emptyset$ alors, cet échec est propagé et l'ensemble vide
est obtenu comme résultat de la réduction.

Si nous voulons garder inchangés les sous-termes de $t$ tels que l'application
de $r$ mène à un échec nous pouvons utiliser le terme $id$ soit de la même façon
que dans l'Exemple~\ref{exempleSpreadDown}, soit en définissant l'opérateur
$G_{bu}$~:
	$$G_{bu}(r) \eqdef f \ra (x \ra [first(\Psi(f),id);first(r,id)](x))$$
De la même manière que dans les cas précédents nous obtenons l'opérateur 
$BottomUp$~:
	$$BottomUp(r) \eqdef [\Theta](G_{bu}(r))$$
correspondant à la description présentée en début de cette section.

\LEM%------------------------------------------------------------------------
L'opérateur $BottomUp$ décrivant l'application d'un terme à tous les sous-termes
d'un autre terme d'une manière \textit{bottom-up} est exprimable dans le
\roCalfT.
\FLEM%------------------------------------------------------------------------

\proof{
Il est facile de remarquer que pour l'application de l'opérateur $BottomUp$ à une
constante $c$ nous obtenons~:
	$$[BottomUp(r)](c) \longra{*}_{\rho} \{[first(r,id)](c)\}$$
et donc le résultat de l'application est soit $\emptyset$ si $[r](c)$ mène à un
échec, soit le résultat de la réduction de l'application $[r](c)$ dans le cas
contraire.

Pour un terme $t=f(t_1,\ldots,t_n)$ nous obtenons la réduction suivante~:

\begin{tabbing}
\indent\indent \= $\longra{*}_{\rho}$  \=  \kill
\> \> $[BottomUp(r)](t) \eqdef [[\Theta](G_{bu}(r))](t)$
\\
\> $\longra{*}_{\rho}$ 
\> $\{[[G_{bu}(r)]([\Theta](G_{bu}(r)))](t)\}$ 
\\
\> $\eqdef$ 
\> $\{[[G_{bu}(r)](BottomUp(r))](t)\}$
\\
\>  $\eqdef$ 
\>  $\{[[f \ra (x \ra [first(\Psi(f),id);first(r,id)](x))](BottomUp(r))](t)\}$
\\
\> $\longra{*}_{\rho}$ 
\> $\{[\{x \ra [first(\Psi(BottomUp(r)),id);first(r,id)](x)\}](t)\}$
\\
\> $\longra{*}_{\rho}$ 
\> $\{[first(\Psi(BottomUp(r)),id);first(r,id)](f(t_1,\ldots,t_n))\}$
\\
\> $\longra{*}_{\rho}$ 
\> $\{[first(r,id)]([first(\Psi(BottomUp(r)),id)](f(t_1,\ldots,t_n)))\}$
\\
\> $\longra{*}_{\rho}$ 
\> $\{[first(r,id)](\paron{[\Psi(BottomUp(r))](f(t_1,\ldots,t_n)),[id](f(t_1,\ldots,t_n))})\}$
\\
\> $\longra{*}_{\rho}$ 
\> $\{[first(r,id)](\paron{f([BottomUp(r)](t_1),\ldots,[BottomUp(r)](t_n)),\{f(t_1,\ldots,t_n)\}})\}$

\end{tabbing}

Grâce à l'utilisation de l'opérateur $id$ dans la définition de $G_{bu}(r)$ et
donc de $BottomUp(r)$, l'application de $BottomUp(r)$ à un terme ne mène jamais
à un échec. Ainsi, nous obtenons pour le terme ci-dessus la même réduction que pour le terme
	$$\{[first(r,id)](f([BottomUp(r)](t_1),\ldots,[BottomUp(r)](t_n)))\}$$
représentant l'application de $r$, ou de l'identité si l'application de $r$ mène 
à un échec, à tous les sous-termes et
ensuite au sommet du terme $f(t_1,\ldots,t_n)$.
}

\EX%------------------------------------------------------------------------
\label{exempleBottomUp}
L'application $[BottomUp(a \ra b)](a)$ est réduite en 
$\{[first(a \ra b,id)](a)\}$ et donc, en le terme $\{b\}$.

En considérant l'application de la règle de réécriture $a \ra b$ au terme
$g(a)$ nous obtenons~:

\begin{tabbing}
\indent\indent \= $\longra{*}_{\rho}$  \= ~~~~~   \= \kill
\> \> $[BottomUp(a \ra b)](g(a))$
\\
\> $\longra{*}_{\rho}$ 
\> $\{[first(a \ra b,id)](\paron{g([BottomUp(a \ra b)](a)),\{g(a)\}})\}$
\\
\> $\longra{*}_{\rho}$ 
\> $\{\paron{[a \ra b](\paron{g([BottomUp(a \ra b)](a)),\{g(a)\}}),
             \{\paron{g([BottomUp(a \ra b)](a)),\{g(a)\}}\}}\}$
\\
\> $\longra{*}_{\rho}$ 
\> $\{\paron{[a \ra b](\paron{g(\{b\}),\{g(a)\}}),
             \{\paron{g([BottomUp(a \ra b)](a)),\{g(a)\}}\}}\}$
\\
\> $\longra{*}_{\rho}$ 
\> $\{\paron{\{[a \ra b](g(b))\},
             \{\paron{g([BottomUp(a \ra b)](a)),\{g(a)\}}\}}\}$
\\
\> $\longra{*}_{\rho}$ 
\> $\{\paron{\emptyset,
             \{\paron{g([BottomUp(a \ra b)](a)),\{g(a)\}}\}}\}$
\\
\> $\longra{*}_{\rho}$ 
\> $\{\paron{\{\paron{g(\{b\}),\{g(a)\}}\}}\}$
\\
\> $\longra{*}_{\rho}$ 
\> $\{g(b)\}$

\end{tabbing}

De la même façon nous obtenons~:

\begin{tabbing}
\indent\indent \= $\longra{*}_{\rho}$  \= ~~~~~   \= \kill
\> \> $[BottomUp(a \ra b)](f(a,g(a)))$
\\
\> $\longra{*}_{\rho}$ 
\> $\{f(b,g(b))\}$

\end{tabbing}

Le résultat de cette dernière réduction n'est pas le même que celui obtenu dans
l'Exemple~\ref{exempleSpreadDown} pour le même terme de départ, mais l'ensemble
obtenu dans cet exemple est inclus dans celui obtenu dans
l'Exemple~\ref{exempleSpreadDown}.

\FEX%------------------------------------------------------------------------

Une réduction du type \textit{top-down} peut être définie de la même façon
qu'une réduction \textit{bottom-up}. Nous introduisons un opérateur
$G_{td}(r)$~:
	$$G_{td}(r) \eqdef f \ra (x \ra \paron{[first(r,id);first(\Psi(f),id)](x)})$$
et nous obtenons immédiatement
	$$TopDown(r) \eqdef [\Theta](G_{td}(r)).$$

\LEM%------------------------------------------------------------------------
L'opérateur $TopDown$ décrivant l'application d'un terme à tous les sous-termes
d'un autre terme d'une manière \textit{top-down} est exprimable dans le
\roCalfT.
\FLEM%------------------------------------------------------------------------

\subsection{Applications singulières}
%=============================================================================== 
%

En utilisant les opérateurs $BottomUp$ et $TopDown$ nous pouvons décrire
l'application d'une règle de réécriture $r$ à tous les sous-termes d'un autre
terme $t$ mais non la normalisation des sous-termes et en particulier du terme
$t$ par rapport à la règle $r$. Nous voulons donc appliquer une règle de
réécriture $r$ jusqu'à ce qu'elle ne soit plus applicable et ceci peut être
réalisé en réitérant l'application des termes $BottomUp(r)$ ou
$TopDown(r)$. Mais une telle approche ne permettrait pas la description des
stratégies de normalisation \textit{innermost} ou \textit{outermost} et nous
voulons donc définir des opérateurs plus fins décrivant l'application d'une
règle de réécriture à un seul sous-terme et réitérer l'application de tels
opérateurs.

En utilisant l'opérateur $\Phi$ nous pouvons définir des $\rho$-termes
similaires à ceux introduits dans la section précédente mais qui appliquent un
terme donné à une seule position d'un terme d'une manière \textit{bottom-up} ou
\textit{top-down}. 

Dans le cas \textit{bottom-up} nous introduisons un terme qui essaie
l'application d'un terme $r$ à tous les arguments d'un terme donné et en cas
d'échec applique le terme $r$ en tête~:
	$$H_{bu}(r) \eqdef f \ra (x \ra [first(\Phi(f),r)](x))$$
En utilisant ce terme nous définissons l'opérateur $Once_{bu}$~:
	$$Once_{bu}(r) \eqdef [\Theta](H_{bu}(r)).$$

Comme dans le cas de l'opérateur $SD$, le terme 
$[Once_{bu}(r)](t) \eqdef [[\Theta](H_{bu}(r))](t)$ peut mener à des réductions
infinies si aucune stratégie n'est employée. La même stratégie que dans le cas
précédent est suffisante pour assurer la terminaison et elle consiste à
appliquer les règles d'évaluation d'abord en tête d'un terme de la
forme $[u](v)$ et seulement si aucune règle d'évaluation n'est applicable à
cette position, réduire les sous-termes $v$ et $u$.

\LEM%------------------------------------------------------------------------
L'opérateur $Once_{bu}$ décrivant l'application d'un terme une fois d'une
manière \textit{bottom-up} est exprimable dans le \roCalfT.
\FLEM%------------------------------------------------------------------------

\proof{
Nous pouvons remarquer que l'application d'un terme $Once_{bu}(r)$ à une
constante mène à une réduction~:
	$$[Once_{bu}(r)](c) \longra{*}_{\rho} \{\paron{[r](c)}\}.$$
et donc le résultat de l'application est soit $\emptyset$ si $[r](c)$ mène à un
échec, soit le résultat de la réduction de l'application $[r](c)$ dans le cas
contraire.

La réduction suivante montre le comportement de l'application d'un terme $t=f(t_1,\ldots,t_n)$
$Once_{bu}(r)$ à un terme $t=f(t_1,\ldots,t_n)$~:

\begin{tabbing}
\indent\indent \= $\longra{*}_{\rho}$ ~~~ \= ~~~~~ \= \kill
\> \> $[Once_{bu}(r)](t) \eqdef [[\Theta](H_{bu}(r))](t)$
\\
\> $\longra{*}_{\rho}$ 
\> $\{[[H_{bu}(r)]([\Theta](H_{bu}(r)))](t)\}$
\\
\> $\eqdef$ 
\> $\{[[f \ra (x \ra [first(\Phi(f),r)](x))](Once_{bu}(r))](t)\}$
\\
\> $\longra{*}_{\rho}$ 
\> $\{[\{x \ra [first(\Phi(Once_{bu}(r)),r)](x)\}](t)\}$
\\
\> $\longra{*}_{\rho}$ 
\> $\{[first(\Phi(Once_{bu}(r)),r)](f(t_1,\ldots,t_n))\}$
\\
\> $\longra{*}_{\rho}$ 
\> $\{\paron{[\Phi(Once_{bu}(r))](f(t_1,\ldots,t_n)),[r](f(t_1,\ldots,t_n))}\}$
\\
\> $\longra{*}_{\rho}$ 
\> $\{\paron{\paron{f([Once_{bu}(r)](t_1),\ldots,t_n),\ldots,f(t_1,\ldots,[Once_{bu}(r)](t_n))},[r](f(t_1,\ldots,t_n))}\}$

\end{tabbing}

Si tous les termes $f(t_1,\ldots,[Once_{bu}(r)](t_k),\ldots,t_n)$,
$k=1,\ldots,n$, sont réduits en l'ensemble vide (l'application de $Once_{bu}(r)$
à tout sous-terme $t_i$ est réduite en $\emptyset$) et $[r](f(t_1,\ldots,t_n))$
est aussi réduit en l'ensemble vide, alors le dernier terme de la réduction
ci-dessus est réduit en $\emptyset$. Cette situation correspond à un terme
$t=f(t_1,\ldots,t_n)$ tel que $r$ n'est pas appliqué avec succès (résultat
différent de $\emptyset$) à aucun de ses sous-termes.

Si tous les termes $f(t_1,\ldots,[Once_{bu}(r)](t_k),\ldots,t_n)$,
$k=1,\ldots,n$, sont réduits en l'ensemble vide mais le terme
$[r](f(t_1,\ldots,t_n))$ n'est pas réduit en l'ensemble vide alors, le résultat
de la réduction est le même que pour le terme $\{[r](f(t_1,\ldots,t_n))\}$.  

Si un des termes $t_i$ est réduit en $\emptyset$ alors le résultat final est
évidemment $\emptyset$. Si aucun des termes $t_i$ n'est réduit en $\emptyset$ et
s'il existe un terme $[Once_{bu}(r)](t_k)$ qui n'est pas réduit en l'ensemble
vide alors, nous obtenons le même résultat que pour la réduction de %u terme
$\{f(t_1,\ldots,[Once_{bu}(r)](t_k),\ldots,t_n)\}$ tel que pour tout $i<k$,
$[Once_{bu}(r)](t_i) \longra{*}_{\rho} \emptyset$ et $[Once_{bu}(r)](t_k)$
n'est pas réduit en $\emptyset$.

Par conséquent, si $p$ est la position la plus profonde et la plus à gauche d'un
terme clos du premier ordre $\tatpos{t}{u}{p}$ telle que l'application du terme
$r$ au terme $u$ n'est pas réduite en $\emptyset$ alors, le terme
$[Once_{bu}(r)](t)$ est réduit en le même terme que
$\tatpos{t}{[r](u)}{p}$. S'il n'existe pas de telle position $p$ alors, le
résultat est $\emptyset$.
}

\EX%------------------------------------------------------------------------
L'application $[Once_{bu}(a \ra b)](a)$ est réduite en 
$\{\paron{[(a \ra b)](a)}\}$ et donc, en le terme $\{b\}$.

L'application de la règle de réécriture $a \ra b$ une fois à la position la plus
profonde et la plus à gauche du terme $f(a,g(a))$ est représentée par le terme
$[Once_{bu}(a \ra b)](f(a,g(a)))$ avec la réduction~:
\begin{tabbing}
\indent\indent \= $\longra{*}_{\rho}$ ~~ \= ~~~~~ \= \kill
\> \> $[Once_{bu}(a \ra b)](f(a,g(a)))$
\\
\> $\longra{*}_{\rho}$ 
\> $\{\paron{\paron{f([Once_{bu}(a \ra b)](a),g(a)),f(a,[Once_{bu}(a \ra b)](g(a)))},[a \ra b](f(a,g(a)))}\}$
\\
\> $\longra{*}_{\rho}$ 
\> $\{\paron{\paron{f(\{b\},g(a)),f(a,[Once_{bu}(a \ra b)](g(a)))},[a \ra b](f(a,g(a)))}\}$
\\
\> $\longra{*}_{\rho}$ 
\> $\{\paron{\{f(b,g(a))\},[a \ra b](f(a,g(a)))}\}$
\\
\> $\longra{*}_{\rho}$ 
\> $\{f(b,g(a))\}$

\end{tabbing}
\FEX%------------------------------------------------------------------------

Si nous voulons définir un opérateur décrivant l'application d'un terme à une
seule position d'un autre terme d'une manière \textit{top-down} alors nous
pouvons utiliser le terme~:
	$$H_{td}(r) \eqdef f \ra (x \ra [first(r,\Phi(f))](x))$$
et nous obtenons immédiatement l'opérateur $Once_{td}$~:
	$$Once_{td}(r) \eqdef [\Theta](H_{td}(r)).$$

Dans ce cas, l'application du terme $r$ est essayée d'abord au sommet du terme
$t$ et dans le cas d'un échec le terme $r$ est appliqué plus profondément dans
le terme $t$.

\LEM%------------------------------------------------------------------------
L'opérateur $Once_{td}$ décrivant l'application d'un terme une fois d'une
manière \textit{top-down} est exprimable dans le \roCalfT.
\FLEM%------------------------------------------------------------------------

\section{Répétition et opérateur de normalisation} \label{opNormalisation}
%================================================================
 
Dans les sections précédentes nous avons défini les opérateurs qui décrivent
l'application (avec succès) d'un terme à la position la plus profonde ou la plus
haute d'un autre terme. Dans cette section nous voulons définir un opérateur qui
applique à plusieurs reprises un terme donné $r$ à un $\rho$-terme $t$.

D'abord, nous voulons définir un opérateur décrivant l'application répétitive
d'un $\rho$-terme $r$ au sommet d'un terme $t$, c'est-à-dire un
terme correspondant au $\rho$-terme $[r;r;\ldots;r](t)$. En remplaçant le terme
$r$ par un des termes d'application en profondeur décrits dans la section
précédente (e.g. $Once_{bu}$ ou $Once_{td}$) nous obtenons un $\rho$-terme
décrivant une procédure de normalisation (e.g. \textit{innermost} ou
\textit{outermost}). 

Nous appelons $repeat$ l'opérateur de répétition et son comportement peut être
décrit par la règle d'évaluation suivante~:
\renewcommand{\fleche}{\Longrightarrow}
\begin{ruleset}
%===================================
  \regle 
  {Repeat} 
  {[repeat(r)](t)} 
  {[repeat(r)]([r](t))} 
%===================================
\end{ruleset}
Nous pouvons utiliser l'opérateur de point fixe $\Theta$ pour introduire le terme
	$$I(r) \eqdef f \ra (x \ra [r;f](x))$$
permettant la définition de l'opérateur $repeat$~:
	$$repeat(r) \eqdef [\Theta](I(r)).$$

Cette approche a deux inconvénients évidents. Premièrement, la terminaison de la
réduction n'est pas garantie même dans le cas où la stratégie d'évaluation
utilisée pour les opérateurs précédents est utilisée.  Lorsque la stratégie
d'évaluation employée applique les règles d'évaluation d'abord en tête d'une
application $[u](v)$ et puis au sous-terme gauche $u$, nous n'obtenons pas le
résultat désiré. En utilisant cette stratégie \textit{rightmost outermost} pour
la réduction précédente, nous obtenons~:
	$$[repeat(r)](t) \longra{*}_{\rho} \{[repeat(r)]([r](t))\} 
	\longra{*}_{\rho}  \ldots 
	\longra{*}_{\rho} \{[repeat(r)]( [r]([r](\ldots [r](t) \ldots)) )\} 
	\longra{*}_{\rho}  \ldots$$
et donc, la réduction est infinie.
Une méthode permettant de résoudre ce problème consiste à évaluer d'abord
l'argument $v$ de l'application dans le cas où les règles d'évaluation ne
peuvent pas être appliquées en tête de l'application $[u](v)$. 

Deuxièmement, quand la réduction termine, le résultat est toujours l'ensemble
vide. Si à un certain point dans la réduction, l'application du terme
$r$ au terme $t$ est réduite en l'ensemble vide, alors, $\emptyset$ est
strictement propagé et le terme $[repeat(r)](t)$ est réduit ainsi en l'ensemble
vide.

Afin de résoudre ces problèmes, nous pouvons définir un opérateur appelé $\rep$
avec un comportement défini par les règles d'évaluation présentées dans la
Figure~\ref{MRArep}.

\begin{figure}[!htp]
\noindent \framebox{\parbox{\largeurtexte}{

\renewcommand{\fleche}{\Longrightarrow}
\begin{ruleset}
%===================================
  \cregle 
  {Repeat*'} 
  {[\rep(r)](t)} 
  {[\rep(r)]([r](t))}
  {\mbox{$[r](t)$ n'est pas évalué en $\emptyset$}}
%===================================
  \cregle 
  {Repeat*''} 
  {[\rep(r)](t)} 
  {t}
  {\mbox{$[r](t)$ est évalué en $\emptyset$}}
%===================================
\end{ruleset}

}}
\caption{\label{MRArep}L'opérateur $\rep$}
\end{figure}

Nous devons donc définir un opérateur similaire à $repeat$ qui permet la
mémorisation du résultat précédent et effectue un retour arrière à la réduction
de ce terme si un échec est obtenu pour le terme courant.
Nous modifions le terme $I(r)$ qui devient
	$$J(r) \eqdef f \ra (x \ra [first(r;f,~id)](x))$$
et nous l'utilisons comme d'habitude pour la définition de l'opérateur $\rep(r)$
	$$\rep(r) \eqdef [\Theta](J(r))$$

Il ne faut pas oublier que la réduction d'une application $[u](v)$ s'effectue en
appliquant les règles d'évaluation en tête, ensuite à son argument $v$ et
seulement après au terme $u$.

\LEM%------------------------------------------------------------------------
L'opérateur $\rep$ décrivant l'application répétée d'un terme tant que le
résultat de l'application n'est pas $\emptyset$ est exprimable dans le
\roCalfT.
\FLEM%------------------------------------------------------------------------

\proof{
Nous obtenons la réduction suivante pour l'application d'un terme $\rep(r)$ à un
terme $t$~:

\begin{tabbing}
\indent\indent \= $\longra{*}_{\rho}$ ~~ \=  \kill
\> \> $[\rep(r)](t) \eqdef [[\Theta](J(r))](t)$
\\
\> $\longra{*}_{\rho}$ 
\> $\{[[J(r)]([\Theta](J(r)))](t)\}$
\\
\> $\eqdef$ 
\> $\{[[f \ra (x \ra [first(r;f,~id)](x))](\rep(r))](t)\}$
\\
\> $\longra{*}_{\rho}$ 
\> $\{[\{x \ra [first(r;\rep(r),~id)](x)\}](t)\}$
\\
\> $\longra{*}_{\rho}$ 
\> $\{[first(r;\rep(r),~id)](t)\}$
\\
\> $\longra{*}_{\rho}$ 
\> $\{\paron{[r;\rep(r)](t),[id](t)}\}$
\\
\> $\longra{*}_{\rho}$ 
\> $\{\paron{[\rep(r)]([r](t)),[id](t)}\}$

\end{tabbing}

Si le terme $[r](t)$ est réduit en l'ensemble vide alors, le terme
$[\rep(r)]([r](t))$ est réduit immédiatement en $\emptyset$ et donc, pour la
réduction précédente nous obtenons~:

\begin{tabbing}
\indent\indent \= $\longra{*}_{\rho}$ ~~ \=  \kill
\> \> $\{\paron{[\rep(r)]([r](t)),[id](t)}\}$
\\
\> $\longra{*}_{\rho}$ 
\> $\{\paron{[\rep(r)](\emptyset),[id](t)}\}$
\\
\> $\longra{*}_{\rho}$ 
\> $\{\paron{\emptyset,[id](t)}\}$
\\
\> $\longra{*}_{\rho}$ 
\> $\{\paron{[id](t)}\}$
\\
\> $\longra{*}_{\rho}$ 
\> $\{\paron{\{t\}}\}$

\end{tabbing}
et si $t \longra{*}_{\rho} t'$, avec $t'$ un terme clos ne contenant pas de radical
alors, le résultat final est $\{t'\}$.

Nous considérons maintenant que le terme $[r](t)$ est réduit en un terme $\{t'\}$
différent de $\emptyset$ et irréductible. Dans ce cas nous obtenons la réduction~:

\begin{tabbing}
\indent\indent \= $\longra{*}_{\rho}$ ~~ \=  \kill
\> \> $\{\paron{[\rep(r)]([r](t)),[id](t)}\}$
\\
\> $\longra{*}_{\rho}$ 
\> $\{\paron{[\rep(r)](\{t'\}),[id](t)}\}$
\\
\> $\longra{*}_{\rho}$ 
\> $\{\paron{\{[\rep(r)](t')\},[id](t)}\}$
\\
\> $\longra{*}_{\rho}$ 
\> $\{\paron{\{\paron{[\rep(r)]([r](t')),[id](t')}\},[id](t)}\}$

\end{tabbing}
et le résultat final est soit un terme $\{t''\}$ tel que
$[\rep(r)]([r](t'))$$\longra{*}_{\rho}$$\{t''\}$, soit $\{t'\}$ si
$[r](t') \longra{*}_{\rho} \emptyset$.
}

\EX%------------------------------------------------------------------------
L'application répétitive des deux règles de réécriture $a \ra b$ et $b \ra c$ à
un terme $a$ est représentée par le $\rho$-terme $[\rep(\{a \ra b, b \ra c\})](a)$ 
et la réduction suivante est obtenue~:

\begin{tabbing}
\indent\indent \= $\longra{*}_{\rho}$ ~~ \=  \kill
\> \> $[\rep(\{a \ra b, b \ra c\})](a)$
\\
\> $\longra{*}_{\rho}$ 
\> $\{\paron{[\rep(\{a \ra b, b \ra c\})]([\{a \ra b, b \ra c\}](a)),[id](a)}\}$
\\
\> $\longra{*}_{\rho}$ 
\> $\{\paron{[\rep(\{a \ra b, b \ra c\})](\{b\}),[id](a)}\}$
\\
\> $\longra{*}_{\rho}$ 
\> $\{\paron{
\{\paron{[\rep(\{a \ra b, b \ra c\})]([\{a \ra b, b \ra c\}](b)),[id](b)}\},[id](a)}\}$
\\
\> $\longra{*}_{\rho}$ 
\> $\{\paron{
\{\paron{[\rep(\{a \ra b, b \ra c\})](\{c\}),[id](b)}\}
,[id](a)}\}$
\\
\> $\longra{*}_{\rho}$ 
\> $\{\paron{
\{\paron{
\{\paron{[\rep(\{a \ra b, b \ra c\})]([\{a \ra b, b \ra c\}](c)),[id](c)}\}
,[id](b)}\}
,[id](a)}\}$
\\
\> $\longra{*}_{\rho}$ 
\> $\{\paron{
\{\paron{
\{\paron{[\rep(\{a \ra b, b \ra c\})](\emptyset),\{c\}}\}
,[id](b)}\}
,[id](a)}\}$
\\
\> $\longra{*}_{\rho}$ 
\> $\{\paron{
\{\paron{
\{\paron{\emptyset,\{c\}}\}
,[id](b)}\}
,[id](a)}\}$
\\
\> $\longra{*}_{\rho}$ 
\> $\{\paron{
\{\paron{
\{c\}
,[id](b)}\}
,[id](a)}\}$
\\
\> $\longra{*}_{\rho}$ 
\> $\{\paron{
\{\{c\}\}
,[id](a)}\}$
\\
\> $\longra{*}_{\rho}$ 
\> $\{c\}$

\end{tabbing}
\FEX%------------------------------------------------------------------------

Si le terme $r$ est un ensemble de règles de réécriture non-confluentes alors nous
obtenons un ensemble ayant plusieurs éléments comme résultat de la réduction du
terme $[\rep(r)](t)$.

\EX%------------------------------------------------------------------------
Si nous considérons l'ensemble de règles de réécriture
$\SS=\{a \ra b, a \ra c, b \ra d\}$ alors, nous obtenons facilement les
réductions $[\SS](a) \longra{*}_{\rho} \{b,c\}$,
$[\SS](b) \longra{*}_{\rho} \{d\}$,$[\SS](c) \longra{*}_{\rho} \emptyset$,
$[\SS](d) \longra{*}_{\rho} \emptyset$.
Par conséquent, la réduction suivante est obtenue~:

\begin{tabbing}
\indent\indent \= $\longra{*}_{\rho}$ ~~ \= \kill
\> \> $[\rep(\SS)](a) \eqdef [\rep(\{a \ra b, a \ra c, b \ra d\})](a)$
\\
\> $\longra{*}_{\rho}$ 
\> $\{\paron{[\rep(\SS)]([\SS](a)),[id](a)}\}$
\\
\> $\longra{*}_{\rho}$ 
\> $\{\paron{[\rep(\SS)](\{b,c\}),[id](a)}\}$
\\
\> $\longra{*}_{\rho}$ 
\> $\{\paron{\{[\rep(\SS)](b),[\rep(\SS)](c)\},[id](a)}\}$
\\
\> $\longra{*}_{\rho}$ 
\> $\{\paron{\{\{\paron{[\rep(\SS)]([\SS](b)),[id](b)}\},\{\paron{[\rep(\SS)]([\SS](c)),[id](c)}\}\},[id](a)}\}$
\\
\> $\longra{*}_{\rho}$ 
\> $\{\paron{\{\{\paron{[\rep(\SS)](\{d\}),[id](b)}\},\{\paron{[\rep(\SS)](\emptyset),[id](c)}\}\},[id](a)}\}$
\\
\> $\longra{*}_{\rho}$ 
\> $\{\paron{\{\{\paron{\{\paron{[\rep(\SS)]([\SS](d)),[id](d)}\},[id](b)}\},\{\paron{\emptyset,[id](c)}\}\},[id](a)}\}$
\\
\> $\longra{*}_{\rho}$ 
\> $\{\paron{\{\{\paron{\{\paron{[\rep(\SS)](\emptyset),[id](d)}\},[id](b)}\},\{\paron{[id](c)}\}\},[id](a)}\}$
\\
\> $\longra{*}_{\rho}$ 
\> $\{\paron{\{\{\paron{\{\paron{\emptyset,[id](d)}\},[id](b)}\},\{c\}\},[id](a)}\}$
\\
\> $\longra{*}_{\rho}$ 
\> $\{\paron{\{\{\paron{\{d\},[id](b)}\},\{c\}\},[id](a)}\}$
\\
\> $\longra{*}_{\rho}$ 
\> $\{\paron{\{d,c\},[id](a)}\}$
\\
\> $\longra{*}_{\rho}$ 
\> $\{d,c\}$

\end{tabbing}
\FEX%------------------------------------------------------------------------

En utilisant les opérateurs de répétition et d'application en profondeur
présentés précédemment nous pouvons définir des stratégies spécifiques de
normalisation. Par exemple, la stratégie \textit{innermost} est représentée par
le terme
	$$im(r) \eqdef \rep(Once_{bu}(r))$$
et la stratégie \textit{outermost} par le terme
	$$om(r) \eqdef \rep(Once_{td}(r)).$$

\COR%------------------------------------------------------------------------
\label{normOpLemma}
Les opérateurs $im$ et $om$ décrivant la normalisation de type
\textit{innermost} et \textit{outermost} respectivement sont exprimables
dans le \roCalfT.
\FCOR%------------------------------------------------------------------------

Nous avons maintenant tous les ingrédients nécessaires pour représenter par un
$\rho$-terme la normalisation d'un terme $t$ dans une théorie de réécriture
$\RR$. Le terme $\xi_{\RR}$ décrit au début de ce chapitre peut être défini par
un des $\rho$-termes $im(\RR)$ ou $om(\RR)$ et donc, nous pouvons représenter la
normalisation d'un terme $u$ dans une théorie de réécriture $\RR$ par les
$\rho$-termes
	$$\xi_{\RR}(u) \eqdef [im(\RR)](u)$$
ou
	$$\xi_{\RR}(u) \eqdef [om(\RR)](u).$$

\EX%------------------------------------------------------------------------
Si nous notons l'ensemble de règles de réécriture $\{a \ra b, f(x,g(x)) \ra x\}$ 
par $\RR$, nous représentons par $[im(\RR)](f(a,g(a)))$ la normalisation du
terme $f(a,g(a))$ dans une théorie de réécriture contenant les deux règles et la
réduction suivante est obtenue~:

$~$
%\samepage

\begin{tabbing}
\indent\indent \= $\longra{*}_{\rho}$  \= ~~~~~ \=  \kill
\> \> $[im(\RR)](f(a,g(a)))$
\\
\> $\eqdef$ 
\> $[\rep(Once_{bu}(\RR))](f(a,g(a)))$
\\
\> $\longra{*}_{\rho}$ 
\> $\{\paron{[\rep(Once_{bu}(\RR))]([Once_{bu}(\RR)](f(a,g(a)))),[id](f(a,g(a)))}\}$
\\
\> $\longra{*}_{\rho}$ 
\> $\{\paron{[\rep(Once_{bu}(\RR))](\{f(b,g(a))\}),[id](f(a,g(a)))}\}$
\\
\> $\longra{*}_{\rho}$ 
\> $\{\paron{\{[\rep(Once_{bu}(\RR))](f(b,g(a)))\},[id](f(a,g(a)))}\}$
\\
\> $\longra{*}_{\rho}$ 
\> $\{\paron{\{
\{\paron{[\rep(Once_{bu}(\RR))]([Once_{bu}(\RR)](f(b,g(a)))),$
\\    \>\>\> $[id](f(b,g(a)))}\}\},[id](f(a,g(a)))}\}$
\\
\> $\longra{*}_{\rho}$ 
\> $\{\paron{\{\paron{[\rep(Once_{bu}(\RR))](\{f(b,g(b))\}),$
\\    \>\>\> $[id](f(b,g(a)))}\},[id](f(a,g(a)))}\}$
\\
\> $\longra{*}_{\rho}$ 
\> $\{\paron{\{\paron{\{\paron{\{[\rep(Once_{bu}(\RR))]([Once_{bu}(\RR)](f(b,g(b))))\},$
\\    \>\>\> $[id](f(b,g(b)))}\},[id](f(b,g(a)))}\},[id](f(a,g(a)))}\}$
\\
\> $\longra{*}_{\rho}$ 
\> $\{\paron{\{\paron{\{\paron{\{[\rep(Once_{bu}(\RR))](\{b\})\},$
\\    \>\>\> $[id](f(b,g(b)))}\},[id](f(b,g(a)))}\},[id](f(a,g(a)))}\}$
\\
\> $\longra{*}_{\rho}$ 
\> $\{\paron{\{\paron{\{\paron{\{\paron{[\rep(Once_{bu}(\RR))]([Once_{bu}(\RR)](b)),[id](b)},$
\\    \>\>\> $[id](f(b,g(b)))}\},[id](f(b,g(a)))}\},[id](f(a,g(a)))}\}$
\\
\> $\longra{*}_{\rho}$ 
\> $\{\paron{\{\paron{\{\paron{\{\paron{[\rep(Once_{bu}(\RR))](\emptyset),[id](b)},$
\\    \>\>\> $[id](f(b,g(b)))}\},[id](f(b,g(a)))}\},[id](f(a,g(a)))}\}$
\\
\> $\longra{*}_{\rho}$ 
\> $\{\paron{\{\paron{\{\paron{\{\paron{\emptyset,[id](b)},$
\\    \>\>\> $[id](f(b,g(b)))}\},[id](f(b,g(a)))}\},[id](f(a,g(a)))}\}$
\\
\> $\longra{*}_{\rho}$ 
\> $\{\paron{\{\paron{\{\paron{\{\{\{b\}\}\},[id](f(b,g(b)))}\},[id](f(b,g(a)))}\},[id](f(a,g(a)))}\}$
\\
\> $\longra{*}_{\rho}$ 
\> $\{\paron{\{\paron{\{\{b\}\},[id](f(b,g(a)))}\},[id](f(a,g(a)))}\}$
\\
\> $\longra{*}_{\rho}$ 
\> $\{\paron{\{\{b\}\},[id](f(a,g(a)))}\}$
\\
\> $\longra{*}_{\rho}$ 
\> $\{\paron{\{b\},[id](f(a,g(a)))}\}$
\\
\> $\longra{*}_{\rho}$ 
\> $\{b\}$

\end{tabbing}
\FEX%------------------------------------------------------------------------

Etant donné un terme $u$, si la théorie de réécriture $\RR$ n'est pas confluente
alors le résultat de la réduction du terme $[im(\RR)](u)$ est un ensemble
représentant les résultats possibles de la réduction du terme $u$ dans la théorie
de réécriture $\RR$.

\EX%------------------------------------------------------------------------
Nous considérons l'ensemble $\RR=\{a \ra b, a \ra c, f(x,x) \ra x\}$ de règles
de réécriture non-confluentes. Le terme $[im(\RR)](f(a,a))$ représentant la
normalisation \textit{innermost} du terme $f(a,a)$ par rapport à l'ensemble de
règles de réécriture $\RR$ est réduit en $\{b,f(c,b),f(b,c),c\}$.  Le terme
$[om(\RR)](f(a,a))$ représentant la normalisation \textit{outermost} est réduit
en $\{b,c\}$.
\FEX%------------------------------------------------------------------------

%\DontWriteThisInToc  
\subsection*{Conclusion}
%================================================================
%~

Nous avons défini dans ce chapitre plusieurs opérateurs décrivant
principalement l'application d'un terme à une position donnée d'un autre
terme. Nous pouvons décrire facilement l'application d'un terme aux sous-termes
d'un autre terme en utilisant uniquement les opérateurs du \roCal\  de
base mais si nous voulons appliquer un terme seulement si ceci ne mène pas à un
échec alors nous devons utiliser les opérateurs du \roCalf.

Ces opérateurs nous permettent de tester l'échec d'une réduction et en utilisant
cette facilité nous pouvons appliquer uniquement les termes ne menant pas à un
échec. Afin de descendre dans la profondeur du terme $t$ nous
utilisons un $\rho$-terme correspondant au combinateur de point fixe de Turing
du \laCal. Cet opérateur nous permet aussi de décrire l'application répétitive
d'un $\rho$-terme donné.

En combinant ces opérateurs nous pouvons décrire d'une manière concise la
normalisation d'un terme par rapport à une théorie de réécriture. Nous
n'imposons pas la terminaison ou la confluence de l'ensemble de règles de
réécriture mais dans ces cas nous pouvons obtenir des \mbox{$\rho$-réductions}
non-terminantes ou menant à un ensemble ayant plusieurs éléments. En utilisant
les opérateurs du \roCalf\  nous allons donner une description des règles et
stratégies du langage \elan~\cite{BorovanskyThese98,BKK-Fuji-98}. La même
approche peut être utilisée pour d'autres langages basés sur la réécriture comme
ASF+SDF~\cite{overview-ASF-BOOK}, CafeOBJ~\cite{FutatsugiN-IEEE97},
Maude~\cite{Maude-RwLg1996}, ML~\cite{MilnerML84} ou Stratego~\cite{Vis99} mais
également pour des démonstrateurs de théorèmes utilisant la réécriture comme
par exemple Coq~\cite{CoqMan96} ou HOL~\cite{gormel93}.

%% file: chapter_5.tex
%%%%%%%%%%%%%%%%%%%%%%%%%%%%%%%%%%%%%%%%%%%%%%%%%%%%%%%%%%%
% \TLtopbookmark
\chapter{L'expressivité du \roCal}
\label{chap.encodage}
%%%%%%%%%%%%%%%%%%%%%%%%%%%%%%%%%%%%%%%%%%%%%%%%%%%%%%%%%%%%

La syntaxe du \roCal\  général nous permet la représentation des abstractions du
\laCal\  \cite{Barendregt84} et des règles de réécriture utilisées en logique de réécriture. Dans
cette section nous analysons la correspondance entre la réduction d'un terme du
premier ordre par rapport à un ensemble de règles de réécriture
(\cite{DershowitzJouannaud-90,Klop90,BaaderNipkowREW-98}) et la réduction du
$\rho$-terme correspondant. Nous présentons aussi des fonctions de traduction
entre les $\lambda$-termes et les $\rho$-termes et nous montrons que des
réductions similaires sont obtenues dans les deux cas.

La représentation des $\lambda$-termes et des réductions sous-jacentes est
réalisée en considérant une restriction de la syntaxe et des règles d'évaluation
du \roCal. Les réductions des termes par rapport à un système de réécriture sont
représentées par des $\rho$-termes construits, soit en utilisant les termes de
preuve des réductions dans la réécriture, soit en utilisant seulement les règles
de réécriture et les opérateurs du \roCalfT.

En partant de la représentation des règles de réécriture conditionnelles, nous
utilisons le \roCal\  pour donner une sémantique opérationnelle aux règles et
stratégies du langage \elan\  dont une description est donnée
dans~\cite{BorovanskyThese98} et une sémantique du point de vue fonctionnelle
est présentée dans~\cite{BKK-Fuji-98}. La représentation d'un programme \elan\
par un $\rho$-terme nous permet de mieux comprendre le comportement des
constructions du langage et particulièrement le traitement du non-déterminisme.

\section{Expression du \laCal\  en \roCal} \label{encodageLambda}
%================================================================

Nous analysons d'abord l'encodage du \laCal\  pur~\cite{Hindley-1986} où les
$\lambda$-termes sont construits en utilisant seulement les variables,
l'opérateur d'abstraction et l'opérateur d'application
(cf. Section~\ref{laCalNonType}) et ensuite nous adaptons cet encodage au cas du
\laCal\  appliqué.

\subsection{Expression du \laCal\  pur} \label{encodageLambdaPur}
%================================================================

Nous considérons une restriction de l'ensemble $\RTT$ de $\rho$-termes noté
$\RTL$ et définie inductivement par~:

\begin{tabular}{llll}
& \\
\textBNF{$\rhol$-termes} ~~~~ & 
$t$ &::=&  $x ~~|~~ \{t\} ~~|~~ [t](t) ~~|~~ x \ra t$ \\
& \\
\end{tabular}\\
ou $x \in \XX$.

Par rapport à la syntaxe du \roCal\  général, les règles de réécriture du \roCalL\
sont restreintes à des règles avec une variable en tant que membre gauche. En 
plus, les ensembles sont toujours des singletons.

\DEF%------------------------------------------------------------------------
\label{roLdef} 

Etant donné un ensemble de variables $\XX$, nous appelons \roCalL\  l'instance du
\roCal\  défini par~:
\begin{itemize}
\item l'ensemble de termes $\RTL$,
\item l'application (d'ordre supérieur) de substitution aux termes,
\item la théorie $\emptyset$ (filtrage syntaxique),
\item l'ensemble de règles d'évaluation $\EE_{\lambda}$:
	\rname{Fire},\rname{Congruence}, \rname{Congruence\_fail},
	\rname{Distrib}, \rname{Batch}, \rname{Switch_L}, \rname{Switch_R},
	\rname{OpOnSet}, \rname{Flat},
\item la stratégie d'évaluation $\NONE$ qui n'impose aucune restriction
	sur l'application des règles d'évaluation.
\end{itemize}
\FDEF%------------------------------------------------------------------------

En raison des restrictions de syntaxe que nous venons d'imposer, les règles
d'évaluation du \roCalL\  peuvent être spécialisées en celles décrites dans la
Figure~\ref{MRAlambda}. 

\begin{figure}[!htp]
\noindent
\framebox{\parbox{\largeurtexte}{

\renewcommand{\fleche}{\Longrightarrow}
\begin{ruleset}
%===================================
  \regle {Fire_{\lambda}} 
	{[x \ra r](t)} 
	{\{\subs{x/t} r\} }
%\\
%===================================
\regle {Distrib_{\lambda}}
	{[\{u\}](v)}
	{\{[u](v)\}}
%=================================== 
\regle {Batch_{\lambda}}
	{[v](\{u\})}
	{\{[v](u)\}}
%=================================== 
\regle {Switch_{\lambda}} 
	{x \ra \{v\}}
	{\{x \ra v\}}
%\\
%=================================== 
\regle {Flat_{\lambda}}
	{\{\{v\}\}}
	{\{v\}}
%===================================
\end{ruleset}

}}
\caption{\label{MRAlambda} Les règles d'évaluation du \roCalL}
\end{figure}

La règle d'évaluation \rname{Fire_{\lambda}} initialise, dans le \roCalL, (comme
la règle $\beta$ dans le \laCal) l'application d'une substitution sur un
terme.  
Une conséquence immédiate de la syntaxe restreinte imposée pour les termes de
$\RTL$ est que le filtrage effectué dans la règle d'évaluation \rname{Fire_{\lambda}}
réussit toujours et la solution de l'équation de filtrage qui est nécessairement
de la forme $x \meqqes t$ est simplement $\Sl(x \meqqes t)=\{\subs{x/t}\}$.

Puisque le membre gauche d'une règle de réécriture est toujours une variable, la
règle d'évaluation \rname{Switch_L} n'est pas nécessaire et puisque nous
n'utilisons pas des symboles de fonctions, les règles \rname{Congruence} et
\rname{Congruence\_fail} ainsi que la règle \rname{OpOnSet} ne sont jamais utilisées.

Dans le \roCalL\  nous aurions pu ajouter une règle d'évaluation éliminant tous
les symboles d'ensemble. Mais dès que l'échec, représenté par l'ensemble vide,
et le non-déterminisme, représenté par des ensembles ayant plus d'un élément,
sont introduits, une telle règle d'évaluation ne serait plus significative.

La confluence du \laCal\  est obtenu indépendamment de la stratégie de réduction
et nous voudrions obtenir le même résultat pour sa $\rho$-représentation.  Afin
d'assurer la confluence du \roCalL\  nous devons utiliser la stratégie
d'évaluation \textit{ConfStrat} définie dans la Section~\ref{strat_confluente}
pour le \roCalE. Cette stratégie permet la réduction d'un terme de la forme $[l
\ra r](t)$ seulement si~:
\begin{itemize}
\item le terme $l$ est linéaire,
\item le terme $l$ \subf\  le terme $t$ et,
\item le terme $t$ ne contient aucun ensemble ayant plus d'un élément et
	aucun ensemble vide et,
\item le terme $t$ ne contient pas de sous-terme de la forme $[u](v)$ où $u$
	n'est pas une règle de réécriture et,
\item pour tout sous-terme $[u \ra w](v)$ de $t$, $u$ subsume $v$.
\end{itemize}

Nous voulons déterminer maintenant quelles sont les conditions satisfaites
implicitement par les termes de $\RTL$ et quelles sont les conditions que nous
devons toujours imposer pour garantir la confluence du \roCalL. 

La condition interdisant les termes de la forme $[u](v)$ dans le terme $t$ si
$u$ n'est pas une règle de réécriture est imposée dans le seul but de ne pas
permettre des termes réductibles en $\emptyset$ dans l'argument $t$ de
l'application. Par exemple, dans le \roCalE\  un sous-terme $[a](b)$, avec $a,b$
des constantes, est réduit en $\emptyset$ par la règle d'évaluation
\rname{Congruence\_fail} et si nous supposons une propagation stricte de
l'échec alors le terme $t$ est réduit en $\emptyset$.

Il est évident que dans $\RTL$ il existe des termes de la forme $[u](v)$ où $u$
n'est pas une règle de réécriture, mais les symboles de tête de $u$ et $v$ ne
sont pas fonctionnels et donc, un échec ne peut pas être obtenu à cause d'une
application de la forme $[u](v)$ avec $u,v$ ayant des symboles de tête
fonctionnels différents.

\PROP%----------------------------------------------------------
Le \roCalL\  est confluent.
\FPROP%----------------------------------------------------------

\proof{
Les preuves de confluence de la Section~\ref{stratRocal} peuvent être facilement
adaptées pour montrer que le \roCalL\  est confluent si les règles d'évaluation
sont guidées par une stratégie permettant la réduction d'un terme de la forme
$[l \ra r](t)$ seulement si~:
\begin{itemize}
\item le terme $l$ est linéaire et,
\item le terme $l$ \subf\  le terme $t$ et,
\item le terme $t$ ne contient aucun ensemble ayant plus d'un élément et
	aucun ensemble vide et,
\item pour tout sous-terme $[u \ra w](v)$ de $t$, $u$ subsume $v$.
\end{itemize}

Nous montrons que toutes les conditions imposées dans la stratégie précédente
sont satisfaites par tous les termes du $\RTL$.

Tous les membres gauches des règles de réécriture du \roCalL\  sont des variables
et donc, ils sont linéaires. Puisque une variable subsume et donc \subf\  tout terme,
la deuxième et la dernière condition sont satisfaites aussi pour tout terme de
$\RTL$. La troisième condition est trivialement vraie grâce à la construction des
termes du \roCalL. 

Toutes les conditions imposées dans la stratégie ci-dessus sont donc
implicitement satisfaites dans le \roCalL\  permettant ainsi l'application de la
règle d'évaluation \rname{Fire_{\lambda}} à tout terme de la forme $[u](v)$.
Cette stratégie est donc équivalente à la stratégie $\NONE$ dans le cas du
\roCalL\  et par conséquent, nous pouvons conclure que le \roCalL\  est confluent
quelle que soit la stratégie d'évaluation utilisée.  
}

Maintenant nous pouvons noter que tout $\lambda$-terme peut être représenté par
un $\rho$-terme.  La fonction $\varphi$ décrivant la correspondance entre les
termes représentés dans la syntaxe du \laCal\  et les termes représentés dans la
syntaxe du \roCalL\  est définie par les règles de transformation suivantes~:
$$
\begin{array}{l@{~~\leadsto~~}l}

\varphi(x) &  x, ~\mbox{si $x$ est une variable} \\

\varphi(\lambda x.t) &  x \ra \varphi(t) \\

\varphi(t~u) &  [\varphi(t)](\varphi(u))

\end{array}
$$

Une fonction de traduction similaire peut être employée pour transformer les
termes représentés dans la syntaxe du \roCalL\  en des termes représentés dans la
syntaxe du \laCal~:
$$
\begin{array}{l@{~~\leadsto~~}l}

\delta(x) &  x, ~\mbox{si $x$ est une variable} \\

\delta(\{t\}) &  \delta(t) \\

\delta(x \ra t) &  \lambda x.\delta(t) \\

\delta([t](u)) &  \delta(t)~\delta(u)

\end{array}
$$

Les réductions dans le \laCal\  et dans le \roCalL\  sont équivalentes modulo les
notations pour l'application et l'abstraction et la gestion des ensembles~:

\PROP%------------------------------------------------------------------------
\label{inclLamNE}
Etant donnés deux $\lambda$-termes $t$ et $t'$. Si $t \longra{}_{\beta} t'$
alors $\varphi(t) \longra{*}_{\rhol} \{\varphi(t')\}$.

Etant donnés deux $\rhol$-termes $u$ et $u'$. Si $u \longra{}_{\rhol} u'$
alors $\delta(u) \longra{*}_{\beta} \delta(u')$.
\FPROP%------------------------------------------------------------------------

\proof{

Nous utilisons une induction structurelle sur le terme $t$~:
\begin{itemize}

\item
Si $t$ est une variable $x$, alors $t'=x$ et $\varphi(t)=\varphi(t')=x$.

\item
Si $t=\lambda x.u$ alors $t'=\lambda x.u'$ avec $u \longra{}_{\beta} u'$ et
nous avons $\varphi(t)=x \ra \varphi(u)$.  
Par induction, nous avons $\varphi(u)$ $\longra{*}_{\rhol}$ $\{\varphi(u')\}$, et donc
	$$\varphi(t)=x \ra \varphi(u) ~~ \longra{*}_{\rhol} ~~ 
	x \ra \{\varphi(u')\} ~~ \longra{}_{\rname{Switch_{\lambda}}} ~~ 
	\{x \ra \varphi(u')\}=\{\varphi(t')\}$$

\item
Si $t=(u~v)$ alors nous avons soit $t'=(u'~v)$ avec $u \longra{}_{\beta} u'$,
soit $t'=(u~v')$ avec $v \longra{}_{\beta} v'$, soit $t=\lambda x.u~v$ et
$t'=\subs{x/v}u$.

Dans le premier cas, en appliquant l'induction nous obtenons
	$$\varphi(t)=[\varphi(u)](\varphi(v)) \longra{*}_{\rhol}
	[\{\varphi(u')\}](\varphi(v))
	\longra{}_{\rname{Distrib_{\lambda}}}
	\{[\varphi(u')](\varphi(v))\}=\{\varphi(t')\}.$$
Le deuxième cas est similaire~:
	$$\varphi(t)=[\varphi(u)](\varphi(v)) \longra{*}_{\rhol}
	[\{\varphi(u)\}](\varphi(v'))
	\longra{}_{\rname{Distrib_{\lambda}}}
	\{[\varphi(u)](\varphi(v'))\}=\{\varphi(t')\}.$$
Dans le troisième cas $\varphi(t)=[x \ra \varphi(u)](\varphi(v))$ et 
	$$\varphi(t)=[x \ra \varphi(u)](\varphi(v)) \longra{}_{Fire_{\lambda}} 
	\{\subs{x/\varphi(v)} \varphi(u)\}=\varphi(\subs{x/v} u)=\varphi(t').$$

Puisque l'application de la  substitution est similaire dans le \laCal\  et le
\roCal, nous obtenons, en utilisant la définition de la fonction $\varphi$,
$\varphi(\subs{x/v} u)=\subs{x/\varphi(v)}\varphi(u)$
et donc, la propriété est vérifiée.

\end{itemize}

Puisque dans le \roCalL\  nous n'avons que des singletons et la transformation
$\delta$ seulement enlève les symboles d'ensemble, alors l'application d'une des règles
d'évaluation \rname{Distrib_{\lambda}}, \rname{Batch_{\lambda}},
\rname{Switch_{\lambda}} et \rname{Flat_{\lambda}} correspond à l'identité
dans le \laCal.
\begin{itemize}

\item
Si $t=[\{u\}](v)$ alors nous avons $t \longra{}_{Distrib_{\lambda}} \{[u](v)\}$.
Comme $\delta([\{u\}](v))=\delta(u)~\delta(v)$ et
$\delta(\{[u](v)\})=\delta(u)~\delta(v)$, la propriété est vérifiée.

\item
Si $t=[x \ra u](v)$ alors $t \longra{}_{Fire_{\lambda}} \{\subs{x/v}u\}$. Nous avons
	$$\delta(t)=\lambda x.\delta(u)~\delta(v) \longra{}_{\beta} 
	\subs{x/\delta(v)} \delta(u)\}=\delta(\subs{x/v} u)=\delta(t').$$

\end{itemize}
Les autres cas sont très similaires au premier cas ou à leurs correspondants de
la première partie.
}

\EX%------------------------------------------------------------------------
\label{exLambdaRep}
Nous considérons les trois combinateurs $I=\lambda x.x$, $K=\lambda x y.x$ et
$S=\lambda x y z. xz (yz)$ et leur représentation dans le \roCal:
\begin{itemize}
	\item $I_{\rho}=x \ra x$,
	\item $K_{\rho}=x \ra (y \ra x)$,
	\item $S_{\rho}=x \ra (y \ra (z \ra [[x](z)]([y](z))))$.
\end{itemize}
et nous vérifions qu'à l'égalité $SKK=I$ correspond une $\rhol$-réduction
$[[S_{\rho}](K_{\rho})](K_{\rho}) \longra{*}_{\rhol} \{I_{\rho}\}$~:

\begin{tabbing}
$[[S_{\rho}](K_{\rho})](K_{\rho})$ $=$  \= 
	$[[x \ra (y \ra (z \ra [[x](z)]([y](z))))](x \ra (y \ra x))](x \ra (y \ra x))$
$\lraD{\rhol}$ \\
\> $[\{y \ra (z \ra [[x \ra (y \ra x)](z)]([y](z)))\}](x \ra (y \ra x))$
$\lraD{\rhol}$ \\
\> $\{[y \ra (z \ra [[x \ra (y \ra x)](z)]([y](z)))](x \ra (y \ra x))\}$
$\lraD{\rhol}$ \\
\> $\{[y \ra (z \ra [\{y \ra z\}]([y](z)))](x \ra (y \ra x))\}$
$\lraD{\rhol}$ \\
\> $\{\{[y \ra (z \ra [y \ra z]([y](z)))](x \ra (y \ra x))\}\}$
$\lraD{\rhol}$ \\
\> $\{\{[y \ra (z \ra \{z\})](x \ra (y \ra x))\}\}$
$\lraD{\rhol}$ \\
\> $\{\{\{[y \ra (z \ra z)](x \ra (y \ra x))\}\}\}$
$\lraD{\rhol}$ \\
\> $\{\{\{\{z \ra z\}\}\}\}$
$\lraD{\rhol}$ \\
\> $\{z \ra z\}$ $=$ $\{I_{\rho}\}$
\end{tabbing}

\FEX%------------------------------------------------------------------------

\subsection{Expression du \laCal\  avec symboles de fonctions} \label{encodageLambdaApp}
%================================================================

Dans la section précédente nous avons considéré la syntaxe du \laCal\  pur mais
si nous voulons analyser le cas du \laCal\  appliqué alors, les constantes
doivent être considérées. Nous allons analyser par la suite le cas plus général
où les symboles de fonctions sont non seulement des constantes mais aussi des
symboles de fonctions d'une arité non-nulle. Nous obtenons ainsi un ensemble de
termes $\RTLp$ défini par la syntaxe~:

\begin{tabular}{llll}
& \\
\textBNF{$\rholp$-termes} ~~~~ & 
$t$ &::=&  $x ~~|~~ \{t\} ~~|~~ f(t,\ldots,t) ~~|~~ [t](t) ~~|~~ x \ra t$ \\
& \\
\end{tabular}\\
ou $x \in \XX$ et $f \in \FF$.

Nous considérons cette fois-ci une instance du \roCal, appelé \roCalLp, ne
contenant pas les règles d'évaluation \rname{Congruence} et \rname{Congruence\_fail}.
Ainsi, les règles d'évaluation du \roCalLp\  sont celles de la
Figure~\ref{MRAlambda} et la règle d'évaluation \rname{OpOnSet} spécialisée pour
des singletons~:
\renewcommand{\fleche}{\Longrightarrow}
\begin{ruleset}
%=================================== 
\regle {OpOnSet_{\lambda}}
	{f(v_1,\ldots,\{u\},\ldots,v_n)}
	{\{f(v_1,\ldots,u,\ldots,v_n)\}}
%===================================
\end{ruleset}
\vspace{-.5cm}

Puisque nous avons précisé explicitement que les règles 
\rname{Congruence} et \rname{Congruence\_fail} ne sont pas utilisées dans le
\roCalLp\  alors nous obtenons le même résultat de confluence que dans le cas du
\roCalL.

\PROP%----------------------------------------------------------
Le \roCalLp\  est confluent.
\FPROP%----------------------------------------------------------

Les fonctions de translation sont complétées pour prendre en compte la syntaxe 
du \roCalLp\  et nous ajoutons donc les règles de transformation suivantes~:
$$
\begin{array}{l@{~~\leadsto~~}l}

\varphi(f(u_1,\ldots,u_n)) &  f(\varphi(u_1),\ldots,\varphi(u_n)) \\

\end{array}
$$
et 
$$
\begin{array}{l@{~~\leadsto~~}l}

\delta(f(u_1,\ldots,u_n)) &  f(\delta(u_1),\ldots,\delta(u_n)) \\

\end{array}
$$

Les réductions dans le \laCal\  appliqué et dans le \roCalLp\  sont équivalentes
modulo la syntaxe des deux calculs~:

\PROP%------------------------------------------------------------------------
\label{inclLamNEapp}
Etant donnés deux $\lambda$-termes $t$ et $t'$. Si $t \longra{}_{\beta} t'$
alors $\varphi(t) \longra{*}_{\rholp} \{\varphi(t')\}$.

Etant donnés deux $\rhol$-termes $u$ et $u'$. Si $u \longra{}_{\rholp} u'$
alors $\delta(u) \longra{*}_{\beta} \delta(u')$.
\FPROP%------------------------------------------------------------------------

\proof{

Nous utilisons une induction structurelle sur le terme $t$. Les cas où $t$ est
une variable, une abstraction ou une application sont traités de la même manière 
que pour le \roCalL. Nous devons donc analyser le cas d'un terme fonctionnel~:
\begin{itemize}

\item
Si $t=f(u_1,\ldots,u_n)$ alors
nous avons $t'=f(u_1,\ldots,u_k',\ldots,u_n)$ avec $u_k \longra{}_{\beta} u_k'$
et $\varphi(t)=f(\varphi(u_1),\ldots,\varphi(u_k),\ldots,\varphi(u_n))$. 
Par induction, $\varphi(u_k) \longra{*}_{\rholp} \{\varphi(u_k')\}$, et donc,
nous obtenons la réduction
$\varphi(t)$ 
$\longra{*}_{\rholp}$ $f(\varphi(u_1),\ldots,\{\varphi(u_k')\},\ldots,\varphi(u_n))$ 
$\longra{}_{OpOnSet_{\lambda}}$ 
$\{f(\varphi(u_1),\ldots,\varphi(u_k'),\ldots,\varphi(u_n))\}$.
Comme $\varphi(t')=f(\varphi(u_1),\ldots,\varphi(u_k'),\ldots,\varphi(u_n))$ 
alors la propriété est vérifiée. 

\end{itemize}

Le cas de la transformation inverse est similaire.
}

\section{Expression de la réécriture de termes en \roCal}	 \label{encodRew}
%================================================================

Nous considérons d'abord le cas des règles de réécriture de la forme $(l \ra r)$
avec $l,r \in \TFX$ avec une traduction directe dans des $\rho$-règles de
réécriture. Nous ajoutons ensuite des conditions et nous obtenons des règles de
réécriture conditionnelles de la forme $(\cond{c}{l}{r})$. Dans les deux cas
nous pouvons construire à partir des dérivations d'un terme $t$ dans un système
de réécriture un $\rho$-terme $t_{\rho}$ avec une $\rho$-réduction similaire à
celle de $t$ dans la réécriture. Afin de construire le $\rho$-terme approprié,
sans connaître \textit{a priori} les dérivations dans la théorie de réécriture,
nous utilisons les opérateurs de normalisation du \roCalfT.

\subsection{Expression de la réécriture non-conditionnelle}
%================================================================

Dans cette section nous décrivons la correspondance entre les dérivations d'un
terme $t$ dans une théorie de réécriture $\TT_{\RR} = (\XX,\FF,\EE,\LL,\RR)$
\cite{MeseguerTCS92} et les réductions d'un $\rho$-terme $t_{\rho}$ construit à
partir du terme $u$ et de l'ensemble de règles de réécriture $\RR$. Nous
considérons que la réécriture est effectuée modulo une théorie vide
($\EE=\emptyset$) mais les résultats peuvent être étendus sans difficulté à une
théorie équationnelle.

Les membres gauches des règles de réécriture $l \ra r$ du \roCalE\  sont des
termes du premier ordre ($l \in \TFX$) et donc, les règles de $\RR$ sont
trivialement traduites dans le \roCalE.

Nous voulons montrer que pour toute dérivation dans une théorie de réécriture,
une réduction correspondante peut être trouvée dans le \roCalE. Si nous
considérons qu'un sous-terme $w$ d'un terme $t$ est réduit en $w'$ en appliquant 
la règle de réécriture  $(l \ra r)$ et donc, 
	$$\catpos{t}{w}{p} \longra{}_{\RR} \catpos{t}{w'}{p}$$
alors, nous pouvons construire immédiatement le $\rho$-terme 
$\catpos{t}{[l \ra r](w)}{p}$ avec la réduction~:
	$$\catpos{t}{[l \ra r](w)}{p}
	\longra{}_{\rho}
	\catpos{t}{\{w'\}}{p}
	\longra{*}_{\rho}
	\{\catpos{t}{w'}{p}\}.$$

La méthode ci-dessus de construction du $\rho$-terme avec une $\rho$-réduction
similaire à celle du terme $t$ par rapport à la règle $l \ra r$ est
très facile mais permet seulement de trouver la correspondance pour un pas de
réécriture. Cette représentation est difficilement étendue pour un nombre
quelconque de pas de réduction par rapport à un ensemble de règles de réécriture
et une méthode systématique pour la construction du $\rho$-terme correspondant
est souhaitable.

\PROP%------------------------------------------------------------------------
\label{inclRewNE} 

Etant donnés une théorie de réécriture $\TT_{\RR}$ et deux termes clos du premier
ordre $t, t' \in \TF$ tels que $t \longra{*}_{\RR} t'$. Alors, il existe des
$\rho$-termes $u_1,\ldots,u_n$ construits en utilisant les règles de réécriture
de $\RR$ et les termes intermédiaires utilisés dans la dérivation
$t \longra{*}_{\RR} t'$ tels que nous ayons 
$[u_n](\ldots[u_1](t)\ldots) \longra{*}_{\rhoe} \{t'\}$.
\FPROP%------------------------------------------------------------------------

\proof{
Nous utilisons une induction sur la longueur de la dérivation $t \longra{*}_{\RR} t'$.

\textit{Le cas de base}~:  $t \longra{0}_{\RR} t$ (dérivation en $0$ étapes)

Nous avons immédiatement $[id](t) \longra{0}_{\rhoe} \{t\}$.

\textit{Induction}~: $t \longra{n}_{\RR} t'$ (dérivation en $n$ étapes)

Nous considérons que la règle de réécriture $l \ra r$ est appliquée à la
position $p$ du terme $\catpos{t'}{w}{p}$ obtenu après $n-1$ étapes de réductions~:
	$$t ~~ \longra{n-1}_{\RR} ~~ 
	\catpos{t'}{w}{p} ~~ \longra{}_{l \ra r,p} ~~ \catpos{t'}{\theta r}{p}$$
où $\theta$ est la greffe telle que $\theta l = w$.

Par induction il existe des $\rho$-termes $u_1,\ldots,u_{n-1}$ tels que 
$[u_{n-1}](\ldots[u_1](t)\ldots) \longra{*}_{\rhoe} \{\catpos{t'}{w}{p}\}$.
Nous considérons le $\rho$-terme $u_n=\catpos{t'}{l \ra r}{p}$ et nous
obtenons la réduction~:
	$$
	[u_n](\ldots[u_1](t)\ldots)  ~~ \longra{*}_{\rhoe}  ~~ 
	[\catpos{t'}{l \ra r}{p}](\{\catpos{t'}{w}{p}\})  ~~ \longra{}_{Batch} ~~ 
	\{[\catpos{t'}{l \ra r}{p}](\catpos{t'}{w}{p})\}  ~~ \longra{*}_{Congruence} ~~ 
	$$
	$$
	\{\{\catpos{t'}{[l \ra r](w)}{p}\}\}  ~~ \longra{}_{Fire} ~~ 
	\{\{\catpos{t'}{\{\theta' r\}}{p}\}\}  ~~ \longra{*}_{OpOnSet}  ~~ 
	\{\{\{\catpos{t'}{\theta' r}{p}\}\}\}  ~~ \longra{*}_{Flat} ~~ 
	\{\catpos{t'}{\theta' r}{p}\}
	$$
où la substitution $\theta'$ est telle que $\{\theta'\}=\Sl(l \meqqes w)$.

Puisque $\theta=\theta'$ et que dans ce cas la substitution et la greffe sont
identiques ($r$ ne contient pas de règle de réécriture), nous obtenons 
$\catpos{t'}{\theta' r}{p}=\catpos{t'}{\theta r}{p}$.
}

Jusqu'à maintenant nous avons utilisé la règle d'évaluation \rname{Congruence}
pour obtenir la \mbox{$\rho$-réduction} 
	$$[\catpos{t^n}{l_n \ra r_n}{p_n}](\ldots[\catpos{t^2}{l_2 \ra r_2}{p_2}]([\catpos{t^1}{l_1 \ra r_1}{p_1}](t))\ldots)
	 ~~ \longra{*}_{\rho} ~~ 
	\{t'\}$$
correspondant à une dérivation
	$$t=\catpos{t^1}{w_1}{p_1} ~~ \longra{}_{l_1 \ra r_1,p_1} ~~ 
	\catpos{t^2}{w_2}{p_2} ~~ \longra{}_{l_2 \ra r_2,p_2} ~~ \ldots ~~
	\catpos{t^{n-1}}{w_{n-1}}{p_{n-1}} ~~ 
	\longra{}_{l_n \ra r_n,p_n} ~~ \catpos{t^{n}}{w_{n}}{p_{n}}=t'$$

Dans la Section~\ref{regles_evaluation} nous avons montré que à toute
$\rho$-réduction d'un terme $u$ impliquant l'utilisation de la règle
d'évaluation \rname{Congruence} nous pouvons faire correspondre la réduction
d'un terme $u'$, construit à partir du terme $u$, telle que la règle
d'évaluation \rname{Fire} est utilisée à la place de la règle d'évaluation
\rname{Congruence}. En appliquant cette méthode, la dérivation ci-dessus
s'exprime par~:
	$$[\catpos{t^n}{l_n}{p_n} \ra \catpos{t^n}{r_n}{p_n}]
	(\ldots([\catpos{t^1}{l_1}{p_1} \ra \catpos{t^1}{r_1}{p_1}](t))\ldots)
	 ~~ \longra{*}_{\rho} ~~ 
	\{t'\}$$
mais dans ce cas les réductions sont effectuées en utilisant la règle
d'évaluation \rname{Fire} au lieu de la règle \rname{Congruence}.

Nous pouvons remarquer que les termes $u_i$ de la Proposition~\ref{inclRewNE}
sont similaires aux termes de preuve utilisés en logique de réécriture
(Section~\ref{logiqueReec}). En effet, les \mbox{$\rho$-termes} sont une
généralisation des termes de preuve de la logique de réécriture. Considérons la
fonction suivante de transformation de termes de preuve en des $\rho$-termes~:
$$
\begin{array}{lcl}

\varphi(f(\pi_1 , \ldots , \pi_n)) & ~~\leadsto~~ &
f(\varphi(\pi_1),\ldots,\varphi(\pi_n)) \\

\varphi(\ell(\pi_1 , \ldots , \pi_n)) & ~~\leadsto~~ & 
l(\varphi(\pi_1),\ldots,\varphi(\pi_n));l(x_1,\ldots,x_n) \ra r(x_1,\ldots,x_n)\\
&& ~~~~ si ~ [\ell(x_1,\ldots,x_n)] l(x_1,\ldots,x_n) \ra r(x_1,\ldots,x_n) \in \RR \\

\varphi(\pi_1 ; \pi_2) & ~~\leadsto~~ &  \varphi(\pi_1);\varphi(\pi_2) 

\end{array}
$$ 
où le $\rho$-opérateur ``;'' décrit dans la Section~\ref{opAux} représente
l'application successive de deux \mbox{$\rho$-termes}.
Une simple vérification permet de s'assurer 
que pour un séquent $\pi: t \ra t'$ obtenu dans une
théorie de réécriture nous obtenons une $\rho$-réduction~:
	$$[\varphi(\pi)](t) \longra{*}_{\rho} \{t'\}$$

Ceci nous permet d'obtenir une seconde preuve de la Proposition~\ref{inclRewNE}
en passant par les termes de preuve de la logique de réécriture.
Nous pouvons donc construire de plusieurs manières un $\rho$-terme décrivant la
réduction d'un terme clos du premier ordre par rapport à un ensemble de règles
de réécriture. 

\subsubsection{Recherche de dérivation}
%================================================================ 

Ce que nous obtenons ici est un encodage en \roCal\  d'une dérivation de la
réécriture. Il est souvent plus intéressant de \textit{trouver} une telle
dérivation. Nous nous intéressons donc maintenant à l'élaboration d'une méthode
permettant la construction du $\rho$-terme ayant la même réduction qu'un terme
$t$ par rapport à un ensemble de règles de réécriture mais sans connaître les
pas intermédiaires de la dérivation de $t$.

Dans le Chapitre~\ref{chap.recursion} nous avons introduit le \roCalfT\  et des
opérateurs permettant la définitions des stratégies de normalisation de type
\textit{innermost} et \textit{outermost}\footnote{les deux opérateurs $im$ et
$om$ sont définis dans la Section~\ref{opNormalisation}}. Nous allons utiliser
ces stratégies pour obtenir une représentation concise des réductions par
rapport à un ensemble de règles de réécriture, qui est considéré comme un
paramètre de la stratégie.

\PROP%------------------------------------------------------------------------
\label{inclRewNEim} 

Etant donnés une théorie de réécriture $\TT_{\RR}$ et deux termes clos du premier
ordre $t, t\!\!\da{} \in \TF$ tels que $t$ est normalisé en $t\!\!\da{}$ par
rapport à l'ensemble de règles $\RR$. Alors, $[im(\RR)](t)$ est $\rho$-réduit en
un ensemble contenant le terme $t\!\!\da{}$.
\FPROP%------------------------------------------------------------------------

\proof{
Par induction sur le nombre d'étapes de dérivation du terme $t$.
Pour le cas de base il suffit de remarquer que si
$t=t\!\!\da{}$ alors $[im(\RR)](t) \longra{*}_{\rhoe} \{t\}$. 
Les autres cas sont obtenus en utilisant la construction du $\rho$-terme
$im(\RR)$.
}

On doit noter que pour une dérivation de type \textit{innermost} par rapport à
l'ensemble de règles de réécriture $\RR$, $t \longra{*}_{\RR} t'$, le
$\rho$-terme $[im(\RR)](t)$ est réduit en un $\rho$-terme avec un sous-terme
$[im(\RR)](t')$.

\EX%------------------------------------------------------------------------
\label{condNorm}

Etant donné un ensemble de règles de réécriture  $\RR=\{(x=x) \ra True,b \ra a\}$.
En utilisant les règles de $\RR$, le terme $a=b$ est réduit en $True$ et en
partant de cette réduction, nous pouvons construire des $\rho$-termes comme, par 
exemple,
	$$[(x=x) \ra True](a=[b \ra a](b))$$
ou
	$$[(x=x) \ra True]([a=(b \ra a)](a=b))$$
ou
	$$[(x=x) \ra True]([(a=b) \ra (b=b)](a=b))$$
qui sont réduits en $\{True\}$ dans le \roCalE.
En utilisant l'opérateur de normalisation $im$ nous pouvons
définir le terme
	$$[im(\{(x=x) \ra True,b \ra a\})](a=b)$$
qui est réduit en $\{True\}$ dans le \roCalfT.
\FEX%------------------------------------------------------------------------

Puisque dans ce cas nous pouvons obtenir des ensembles vide dus à un échec de
filtrage ou des ensembles ayant plus d'un élément si l'approche est étendue pour
un filtrage non-unitaire, des stratégies d'évaluation spécifiques, comme celles
présentées dans le Chapitre~\ref{chap.resultats_confluence}, doivent être
utilisées afin d'obtenir la confluence des réductions de tout $\rho$-terme.

\subsection{Expression de la réécriture conditionnelle} \label{encodRewCond}
%================================================================ 

Dans la section précédente nous avons montré que toute dérivation d'un terme par
rapport à un ensemble de règles de réécriture peut être décrite par une
réduction dans le \roCal. Maintenant, nous analysons la possibilité d'étendre ce
résultat dans le cas de la réécriture conditionnelle.

Nous pouvons utiliser la même approche que pour la représentation des
dérivations dans la réécriture non-conditionnelle mais les $\rho$-termes employés
afin de décrire une réduction utilisant des règles conditionnelles deviennent
très compliqués.  Alternativement, une représentation concise de la
normalisation des conditions peut être obtenue en utilisant les opérateurs de
normalisation $im$ et $om$ présentés dans la Section~\ref{opNormalisation} et
déjà utilisés dans la section précédente.

Nous rappelons que les règles de réécriture conditionnelles définies dans la
Section~\ref{systemeReec} sont de la forme $(\cond{c}{l}{r})$ avec les termes du
premier ordre $l,r,c \in \TFX$ tels que $\Var(r) \cup \Var(c) \subseteq \Var(l)$. 
Etant donné un ensemble $\RR$ de règles de réécriture conditionnelles,
l'application d'une règle $(\cond{c}{l}{r})$ à un terme $t$ consiste à trouver
une substitution $\sigma$ telle que $\sigma l=t$, vérifier la validité de
$\sigma c$ et remplacer $t$ par $\sigma r$.

Dans le cas d'un système de réécriture conditionnel
normal~\cite{DershowitzOkada90}, la vérification de la validité de la condition
est un processus de normalisation et la difficulté principale dans la
représentation d'un tel système réside dans le fait que la relation de réduction
est récursivement appliquée afin d'évaluer les conditions d'une règle de
réécriture conditionnelle.

Nous considérons les termes $c$ et $u$ et une greffe $\theta$ tels que $\theta
c$ est normalisé par rapport à un ensemble $\RR$ de règles de réécriture en
$u$. Si $c$ est une condition booléenne et l'ensemble $\RR$ est complètement
défini sur les booléens (\cite{BouhoulaR-JAR95}) alors le terme $u$ doit être
une des constantes $True$ ou $False$. Nous supposons qu'il existe un
$\rho$-terme $\cnorm$ construit à partir de $c$ et tel que $\theta \cnorm$ est
$\rho$-réduit en $\{u\}$.
La règle de réécriture conditionnelle $(\cond{c}{l}{r})$ est alors représentée par le
$\rho$-terme~:
	$$l \ra [\{True \ra r, False \ra \emptyset\}](\cnorm)$$
ou par le $\rho$-terme suivant qui est plus simple, mais peut-être moins suggestif~:
	$$l \ra [True \ra r](\cnorm).$$

L'utilisation de la règle d'évaluation \rname{Fire} pour l'application des deux
termes précédents à un terme $t$ mène aux termes 
$\{[\{True \ra \theta r, False \ra \emptyset\}](\theta \cnorm)\}$ et respectivement
$\{[True \ra \theta r](\theta \cnorm)\}$ où $\theta$ représente la greffe obtenue en
filtrant les termes $l$ et $t$.
Dans le cas où $\theta \cnorm$ est $\rho$-réduit en $\{False\}$, dans le deuxième
terme le filtrage échoue et le résultat de l'application est, comme dans le
premier cas, l'ensemble vide.  Quand $\theta \cnorm$ est réduit en $\{True\}$ le
résultat de la réduction est clairement $\theta r$ dans les deux cas.

En utilisant la représentation ci-dessus, nous étendons la
Proposition~\ref{inclRewNE} et nous montrons que toute dérivation dans une
théorie de réécriture conditionnelle est représentable par un $\rho$-terme
approprié~:

\PROP%------------------------------------------------------------------------
\label{inclRewConNE} 

Etant donnés une théorie de réécriture conditionnelle $\TT_{\RR}$ et deux termes
clos du premier ordre $t, t' \in \TF$ tels que $t \longra{*}_{\RR} t'$. Alors,
il existe des $\rho$-termes $u_1,\ldots,u_n$ construits en utilisant les règles
de réécriture de $\RR$ et les termes intermédiaires utilisés dans la dérivation
$t \longra{*}_{\RR} t'$ tels que nous ayons 
$[u_n](\ldots[u_1](t)\ldots) \longra{*}_{\rhoe} \{t'\}$.
\FPROP%------------------------------------------------------------------------

\proof{
Nous procédons comme pour la Proposition~\ref{inclRewNE} et nous utilisons une
induction sur la longueur de la dérivation $t \longra{*}_{\RR} t'$.

\textit{Le cas de base}~:  si $t \longra{0}_{\RR} t$

Dans ce cas nous avons immédiatement $[id](t) \longra{0}_{\rhoe} \{t\}$.

\textit{Induction}~: $t \longra{n}_{\RR} t'$

Nous considérons une règle de réécriture de la forme $(\cond{c}{l}{r})$
appliquée à la position $p$ du terme $\catpos{t'}{w}{p}$ obtenu après $m$ ($m<n$)
étapes de réductions et ainsi~:
	$$t ~~ \longra{m}_{\RR} ~~ 
	\catpos{t'}{w}{p} ~~ \longra{}_{(\cond{c}{l}{r}),p} ~~ \catpos{t'}{\theta r}{p}$$
où $\theta$ est la greffe telle que $\theta l = w$ et $\theta c \longra{k}_{\RR} True$.

Par induction il existe des $\rho$-termes $t_{\rho}$, $\cnorm$ tels que 
$t_{\rho} \longra{*}_{\rhoe} \{\catpos{t'}{w}{p}\}$ et 
$\theta \cnorm \longra{*}_{\rho} \{True\}$.
Nous considérons le $\rho$-terme $v=\catpos{t'}{l \ra [True \ra r](\cnorm)}{p}$ et nous
obtenons la réduction suivante~:
\begin{tabbing}
$~~~~~~~~$ \= $\longra{*}_{Congruence}$  \=  \kill
$[v](t_{\rho})$
\> $\longra{*}_{\rho}$
\> 
$[\catpos{t'}{l \ra [True \ra r](\cnorm)}{p}](\{\catpos{t'}{w}{p}\})$
\\
\> $\longra{}_{Batch}$
\> 
$\{[\catpos{t'}{l \ra [True \ra r](\cnorm)}{p}](\catpos{t'}{w}{p})\}$
\\
\> $\longra{*}_{Congruence}$
\> 
$\{\{\catpos{t'}{[l \ra [True \ra r](\cnorm)](w)}{p}\}\}$
\\
\> $\longra{}_{Fire} $
\> 
$\{\{\catpos{t'}{\{\theta [True \ra r](\cnorm)\}}{p}\}\}$
$=$
$\{\{\catpos{t'}{\{[True \ra \theta r](\theta \cnorm)\}}{p}\}\}$
\\
\> $\longra{*}_{OpOnSet}$
\> 
$\{\{\{\catpos{t'}{[True \ra \theta r](\theta \cnorm)}{p}\}\}\}$
\\
\> $\longra{*}_{\rho}$
\> 
$\{\{\{\catpos{t'}{[True \ra \theta r](\{True\})}{p}\}\}\}$
\\
\> $\longra{}_{Batch}$
\> 
$\{\{\{\catpos{t'}{\{[True \ra \theta r](True)\}}{p}\}\}\}$
\\
\> $\longra{*}_{OpOnSet}$
\> 
$\{\{\{\{\catpos{t'}{[True \ra \theta r](True)}{p}\}\}\}\}$
\\
\> $\longra{}_{Fire}$
\> 
$\{\{\{\{\catpos{t'}{\{\theta r\}}{p}\}\}\}\}$
\\
\> $\longra{*}_{OpOnSet}$
\> 
$\{\{\{\{\{\catpos{t'}{\theta r}{p}\}\}\}\}\}$
\\
\> $\longra{*}_{Flat}$
\> 
$\{\catpos{t'}{\theta r}{p}\}$
\end{tabbing}
}

\subsubsection{Recherche de dérivation}
%================================================================ 

Nous pouvons donc construire un terme $t_{\rho}$ qui est $\rho$-réduit en
$\{t'\}$ si le terme $t$ est réduit en $t'$ en utilisant un ensemble $\RR$ de
règles de réécriture conditionnelles. Néanmoins, la construction du terme
$t_{\rho}$ dépend fortement non seulement du terme $t$ mais aussi de tous les
termes intermédiaires obtenus dans la dérivation $t \longra{*}_{\RR} t'$ et,
comme pour la réécriture non-conditionnelle, nous voulons définir un
$\rho$-terme permettant de \textit{trouver} une telle dérivation.

Afin de construire le terme $t_{\rho}$ en partant seulement du terme $t$ et des
règles de réécriture de $\RR$, nous utilisons les opérateurs de
normalisation $im$ et $om$. Par exemple, nous pouvons définir~:
	$$\cnorm \eqdef [im(\RR)](c).$$

\EX%------------------------------------------------------------------------
\label{condNormCond}

Supposons que l'ensemble de règles de réécriture décrivant l'ordre sur les
entiers est noté par $\RR_{<}$. Nous considérons la règle de réécriture
$(\cond{x \geq 1}{f(x)}{g(x)})$ qui appliquée au terme $f(2)$ mène à $g(2)$
puisque $x$ est instancié à $2$ et la condition $(2 \geq 1)$ est réduite en
$True$ en utilisant la règle $(2 \geq 1) \ra True$.

Si nous considérons que la condition est normalisée par rapport à l'ensemble de
règles de réécriture $\RR_{<}$ alors nous obtenons la réduction suivante dans le
\roCal~:

\begin{tabbing}
\indent \= $\longra{}_{Batch}$  \= ~~~~~ \=  \kill
\> \> $[f(x) \ra [True \ra g(x)]([im(\RR_{<})](x \geq 1))](f(2))$
\\
\> $\longra{}_{Fire}$
\> $\{[True \ra g(2)]([im(\RR_{<})](2 \geq 1))\}$
\\
\> $\longra{*}_{\rho}$
\> $\{[True \ra g(2)](\{True\})\}$
\\
\> $\longra{}_{Batch}$
\> $\{\{[True \ra g(2)](True)\}\}$
\\
\> $\longra{}_{Fire}$
\> $\{\{\{g(2)\}\}\}$
\\
\> $\longra{*}_{Flat}$
\> $\{g(2)\}$

\end{tabbing}
\FEX%------------------------------------------------------------------------

Les conditions des règles de réécriture peuvent être normalisées par rapport à
un ensemble de règles de réécriture conditionnelles, y compris la règle
courante, et donc la définition des $\rho$-règles de réécriture représentant
cette normalisation est intrinsèquement récursive et ne peut pas être réalisée
en utilisant seulement l'opérateur $im$.

Nous utilisons l'opérateur de point fixe $\Theta$ décrit dans la
Section~\ref{opPointFixe} pour représenter l'application du même ensemble de
règles de réécriture pour la normalisation de toutes les conditions.

Etant donné un ensemble de règles de réécriture $\RR=\RR_{n} \cup \RR_{c}$ où
$\RR_{n}$ et $\RR_{c}$ représentent le sous-ensemble de règles de réécriture
non-conditionnelles et respectivement le sous-ensemble de règles de réécriture
conditionnelles de la forme $(\cond{c}{l}{r})$. Nous définissons le terme~:
	$$R \eqdef f \ra (y \ra 
	[im(\{l_i \ra [True \ra r_i]([f](c_i))~|~i=1 \ldots m\} \cup \RR_{n})](y))$$
où $\RR_{c}=\{l_i \ra r_i ~ si ~ c_i ~|~ i=1 \ldots m\}$,
$\RR_{n}=\{l'_i \ra r'_i ~|~ i=1 \ldots n\}$ et respectivement
	$$IM(R) \eqdef [\Theta](R).$$

Ainsi, pour décrire la normalisation du terme $t$ par rapport aux règles de
réécriture de $\RR$ nous utilisons le $\rho$-terme $[IM(R)](t)$.

Nous obtenons donc un résultat similaire à la Proposition~\ref{inclRewConNE}
mais avec une méthode de construction du $\rho$-terme correspondant basée
seulement sur le terme initial et sur l'ensemble de règles de réécriture.

\PROP%------------------------------------------------------------------------
\label{inclRewConNEoper} 

Etant donnés une théorie de réécriture conditionnelle $\TT_{\RR}$ et deux termes
clos du premier ordre $t, t\!\!\da{} \in \TF$ tels que $t$ est normalisé en
$t\!\!\da{}$ par rapport à l'ensemble de règles $\RR$. Alors, $[IM(\RR)](t)$ est
$\rho$-réduit en un ensemble contenant le terme $t\!\!\da{}$.
\FPROP%------------------------------------------------------------------------

\EX%------------------------------------------------------------------------
\label{condNormComplex}
Nous considérons un ensemble de règles de réécriture $\RR$ contenant la règle de
réécriture $(x=x) \ra True$ et les règles de réécriture conditionnelles
$(\cond{h(x) = b}{f(x)}{g(x)})$ et $(\cond{x = a}{h(x)}{b})$. Le terme $f(a)$
est réduit en $g(a)$ en utilisant les règle de réécriture de $\RR$ et nous
analysons la réduction correspondante dans le \roCalfT.

En utilisant la méthode présentée ci-dessus nous obtenons le terme

\begin{tabbing}
\indent \= $R \eqdef f \ra (y \ra$ \= $[im($  \= ~~~~~ \=  \kill
\> $R \eqdef f \ra (y \ra [im(\{f(x) \ra [True \ra g(x)]([f](h(x)=b)),$
\\
\>\>\> $h(x) \ra [True \ra b]([f](x=a)),$
\\
\>\>\> $(x=x) \ra True$
\\
\>\> $\})](y))$
\end{tabbing}

Nous montrons les étapes principales de la réduction du terme
$[IM(R)](f(a))$. Nous obtenons immédiatement la réduction
	$$[IM(R)](f(a)) \eqdef [[\Theta](R)](f(a)) 
	~~ \longra{*}_{\rho} ~~
	[[R]([\Theta](R))](f(a)) \eqdef [[R](IM(R))](f(a))$$
et le résultat final est le même que celui obtenu pour le terme

\begin{tabbing}
\indent \=  $[im($  \= ~~~~~ \=  \kill
\> $[im(\{f(x) \ra [True \ra g(x)]([IM(R)](h(x)=b)),$
\\
\>\> $h(x) \ra [True \ra b]([IM(R)](x=a)),$
\\
\>\> $(x=x) \ra True$
\\
\> $\})](f(a)))$
\end{tabbing}
et donc pour
$$[f(x) \ra [True \ra g(x)]([IM(R)](h(x)=b))](f(a)) 
~~ \longra{*}_{\rho} ~~
\{[True \ra g(a)]([IM(R)](h(a)=b))\}$$

Pour le terme $[IM(R)](h(a)=b)$ nous procédons comme précédemment et donc, nous
sommes amenés à réduire le terme

\begin{tabbing}
\indent \= $[im($  \= ~~~~~ \=  \kill
\> $[im(\{f(x) \ra [True \ra g(x)]([IM(R)](h(x)=b)),$
\\
\>\> $h(x) \ra [True \ra b]([IM(R)](x=a)),$
\\
\>\> $(x=x) \ra True\}$
\\
\> $)](h(a)=b)$
\end{tabbing}
avec la réduction intermédiaire
$$[h(x) \ra [True \ra b]([IM(R)](x=a))](h(a))
~~ \longra{*}_{\rho} ~~
\{[True \ra b]([IM(R)](a=a))\}$$

Puisque nous obtenons facilement 
$[IM(R)](a=a) \longra{*}_{\rho} \{True\}$
alors le terme précédent est réduit en
$\{[True \ra b](\{True\})\} \longra{*}_{\rho} \{b\}$
et nous avons
	$$[IM(R)](h(a)=b)
	~~ \longra{*}_{\rho} ~~ [im(\ldots)](\{b\}=b) 
	~~ \longra{*}_{\rho} ~~ \{True\}$$

Nous revenons à la réduction du terme initial et nous obtenons
	$$\{[True \ra g(a)]([IM(R)](h(a)=b))\}
	~~ \longra{*}_{\rho} ~~ \{[True \ra g(a)](\{True\})\}
	~~ \longra{*}_{\rho} ~~ \{g(a)\}$$

Nous avons ainsi obtenu le même résultat que dans la réécriture conditionnelle.

\FEX%------------------------------------------------------------------------

En utilisant les méthodes de représentation des dérivations de la réécriture
conditionnelle par des $\rho$-termes appropriés nous allons présenter dans la
section suivante un encodage des règles de réécriture utilisées dans \elan, un
langage basé sur les règles de réécriture conditionnelles avec affectations
locales.

\section{Réécriture d'ordre supérieur}
%================================================================

Dans le Chapitre~\ref{chap.lambda_reec} nous avons présenté brièvement la
réécriture du premier ordre et nous avons vu que son pouvoir d'expression ne
permet pas de décrire facilement la fonctionnalité. Le \laCal\  est le système de
réécriture d'ordre supérieur permettant la représentation de toute fonction
calculable. Mais l'encodage n'est pas toujours trivial et intuitif et les
entiers de Church en sont un exemple. Dans ce codage tout entier naturel $n$ est
codé par un $\lambda$-terme $\lambda f x. f(f\ldots(f x)\ldots)$ avec $n$
occurrences de $f$ et le symbole $+$ est décrit par le $\lambda$-terme $\lambda
n p. (\lambda f x. m f (n f x))$. Si nous utilisons la représentation algébrique
des expressions de l'arithmétique donnée dans la Section~\ref{algebresTermes}
l'entier $n$ est représenté par $succ(\ldots(succ(0))\ldots)$ et le comportement
du symbole $+$ est décrit par les règles de réécriture $0+a \ra a$ et $succ(a)+b
\ra succ(b)+a$.

Il est évident que du point de vue de la simplicité et de la lisibilité la
deuxième approche est plus attractive et on veut donc bénéficier des avantages
des deux méthodes. C'est ce qui motive les systèmes de réécriture d'ordre
supérieur qui combinent le mécanisme d'abstraction du \laCal\  et la
représentation des types abstraits avec la syntaxe des systèmes de réécriture du
premier ordre.  Ce systèmes sont définis soit en étendant le \laCal\  avec des
symboles fonctionnels, soit en ajoutant aux systèmes de réécriture du premier
ordre un abstracteur et un mécanisme similaire à la $\beta$-réduction du \laCal.

\subsection*{Extensions algébriques du \laCal}
%================================================================

L'extension du \laCal\  avec des règles de réécriture du premier ordre a été
initiée par Breazu-Tannen (\cite{Breazu-TannenLICS88}) pour l'étude de la
confluence et ensuite analysée en parallèle par Breazu-Tannen et Gallier
(\cite{GallierBreazu}) et Okada (\cite{Okada-ISSAC89}) pour la normalisation.

Les termes d'un \laCal\  étendu sont les termes du \laCal\  auxquels on ajoute les
termes du premier ordre construits à partir d'une signature. On considère la
relation de réduction définie par le mélange de la $\beta$-réduction et de la
relation de réduction induite par les règles de réécriture. Les signatures des
deux composants sont apparemment disjoints mais en fait ils partagent
l'opérateur d'application qui est explicite en \laCal\  mais implicite en
réécriture.

Par exemple, l'extension du \laCal\  avec le système de réécriture de
l'arithmétique se définit par~\cite{PaganoThesis}~:\\
~\\
$\textBNF{termes} ~~~~~~ 
t ~~ ::= ~~ x ~|~ \lambda x.t ~|~ t~t ~|~ 0 ~|~ succ(t) ~|~ t + t ~|~ t \times t$ \\
~\\
$\textBNF{règles} ~~~~~~ 
0+a \ra a ~~~~~~ succ(a)+b \ra succ(b)+a ~~~~~~ 
0 \times a \ra 0 ~~~~~~ succ(a) \times b \ra b + (a \times b)$ \\
$~~~~~~~~~~~~~~~~~~~~~~~~~~~~~~~~~~~~~~~~~~~~~~~~~~~~~~~ (\lambda x.t)~u \ra \subs{x/u}t$ \\

Dans le \roCal\  la notion d'application devient explicite et les règles de
réécriture deviennent des abstractions. En effet, les abstractions du \roCal\
peuvent être plus élaborées que les règles de réécriture du premier ordre des
extensions algébriques.

Dans les sections précédentes nous avons montré qu'aux réductions dans la
réécriture du premier ordre et dans le \laCal\  correspondent les réductions de
certains $\rho$-termes appropriés. Nous pouvons procéder de la même manière dans
le cas des extensions algébriques du \laCal\  et construire des $\rho$-termes
avec des réductions correspondantes. Puisque l'application est explicite en
\roCal, les termes ainsi construits contiendront les règles du premier ordre
utilisées dans la réduction.

\subsection*{Systèmes de réécriture avec abstracteur}
%================================================================

En partant des travaux réalisés par Aczel~\cite{Aczel78}, Klop a introduit les
CRSs (Combinatory Rewrite Systems) généralisant les systèmes de réécriture du
premier ordre et les systèmes de réécriture avec des variables liées comme le
\laCal. Après le résultat sur la décidabilité de l'unification d'ordre supérieur avec
motifs~\cite{MillerLProlog-89-91}, Nipkov introduit les HRSs (Higher-order
Rewrite Systems)~\cite{Nipkow-LICS91} utilisés principalement pour étudier les
propriétés des systèmes tels que $\lambda$Prolog et Isabelle. D'autres systèmes
ont été proposés et sans être exhaustifs, on peut mentionner les HOTRSs de
Wolfram~\cite{WolframLivre} et les ERSs de Khasidashvili~\cite{Khasidashvili90}.
Pour une comparaison de différents formalismes le lecteur peut se référer à la
thèse de van Raamsdonk (\cite{vanRaamsdonkTH96}) et pour une comparaison des
CRSs et HRSs on peut citer les travaux de van Oostrom et van Raamsdonk~\cite{OostromRamsdonk}.

Les systèmes de réécriture d'ordre supérieur ainsi que le \roCal\  combinent la
réécriture du premier ordre et le \laCal\  mais dans les systèmes de réécriture
d'ordre supérieur ceci est réalisé en introduisant un abstracteur dans les
règles de réécriture tandis que dans le \roCal\  les règles deviennent des
abstractions au niveau objet du calcul.

L'application d'une règle de réécriture implique un mécanisme de filtrage
d'ordre supérieur qui est généralement restreint au filtrage d'ordre supérieur
avec motifs.

Puisque la théorie de filtrage est un paramètre du \roCal\  nous pensons que les
réductions dans un système de réécriture d'ordre supérieur peuvent être
représentées dans le \roCalT\  où $T$ est la théorie de filtrage d'ordre
supérieur avec motifs. La construction des $\rho$-termes décrivant ces
réductions est faite en fonction de la syntaxe de chaque formalisme.

\section{Une sémantique d'\elan\  en \roCal}		\label{encodageELAN}
%=============================================================================== 
%

\elan\  peut être vu comme un cadre logique dont le noyau est la logique de
réécriture étendue avec la notion fondamentale de stratégies. L'enrichissement
des règles de réécriture avec des constructions pour tenir compte des stratégies
rend plus complexe la sémantique opérationnelle du langage~\cite{BKKM-1999}. Dans
cette section nous donnons une sémantique aux règles et stratégies d'\elan\  en
utilisant le \roCal.

\subsubsection{Sémantique d'une règle \elan\  conditionnelle avec affectations locales}
%================================================================

Une règle de la forme $l \ra r$, sans aucune condition et aucune affectation
locale, est représentée directement par la même $\rho$-règle de réécriture et
une règle de réécriture conditionnelle est représentée par un $\rho$-terme
construit en utilisant la méthode proposée dans la
Section~\ref{encodRewCond}. Les règles \elan\  avec des affectations locales mais
sans conditions de la forme 
$$
\begin{array}{ll}
[\ell] \quad l(x) \rewrite & r(x,y) \\
                        & \where ~~ y ~ \assign (S)u
\end{array}
$$
peuvent être représentées facilement par le $\rho$-terme
	$$l(x) \ra r(x,[S_{\rho}](u))$$
ou le $\rho$-terme
	$$l(x) \ra [y \ra r(x,y)]([S_{\rho}](u))$$
avec $S_{\rho}$ le $\rho$-terme correspondant à la stratégie $S$ dans le \roCal.

La première représentation remplace syntaxiquement toute variable du membre
droit de la règle définie dans une affectation locale avec le terme qui
instancie la variable respective.  Dans la deuxième représentation chaque
variable définie dans une affectation locale est liée dans une $\rho$-règle de
réécriture qui est appliquée au terme correspondant.

\EX%------------------------------------------------------------------------
\label{reprLAfst}
La règle \elan

\begin{verbatim}
    [deriveSum] p_1 + p_2  =>  p_1' + p_2'
                                where p_1' := (derive)p_1
                                where p_2' := (derive)p_2       
    end
\end{verbatim}
peut être représentée par un des deux termes suivants

$p_1 + p_2 \ra [derive_{\rho}](p_1) +[derive_{\rho}](p_2),$

$p_1 + p_2 \ra [p_1' \ra [p_2' \ra p_1' + p_2']([derive_{\rho}](p_2))]([derive_{\rho}](p_1))$\\
où $derive_{\rho}$ est le $\rho$-terme correspondant à la stratégie {\em \texttt{derive}}.

\FEX%------------------------------------------------------------------------

On peut noter l'utilité des variables libres dans les règles de
réécriture. La dernière représentation d'une règle \elan\  avec des affectations
locales ne serait pas possible si on ne permettait pas que la variable $p_1'$
soit libre dans la $\rho$-règle $p_2' \ra p_1' + p_2'$. 

Si nous considérons des règles \elan\  plus générales contenant des affectations
locales aussi bien que des conditions sur les variables locales, la combinaison
des méthodes utilisées pour les règles purement conditionnelles et pour les
règles contenant que des affectations locales doit être faite soigneusement. Si
nous avions utilisé une représentation similaire à la première approche de
représentation d'une règle avec affectations locales nous aurions obtenu
certains résultats incorrects comme dans l'Exemple~\ref{exIfWrong}.

\EX%------------------------------------------------------------------------
\label{exIfWrong}

Nous considérons la description d'un automate par un ensemble de règles de
réécriture décrivant chacune la transition d'un état à un autre. L'exécution
potentielle d'une double transition d'un état initial à un état final en passant
par un état intermédiaire non-final, peut être décrite par la règle \elan\  suivante~:
\begin{verbatim}
    [double] x => next(y)    
                       where y := (dk(s1 => s2,s1 => s3)) x 
                       if not final(y)
    end
\end{verbatim}

Le terme {\em \texttt{next(y)}} représente l'état obtenu en effectuant une
transition à partir de $y$ et ce comportement peut être facilement implanté en
\elan\  par un ensemble de règles non-nommées. Nous notons par $\RR_{f}$
l'ensemble de règles de réécriture décrivant les états finaux et nous supposons
que $s2$ est un état final mais $s3$ ne l'est pas.

En utilisant la première approche de représentation d'une règle avec
affectations locales et la méthode de codage pour les règles conditionnelles
présentée dans la Section~\ref{encodRewCond} nous obtenons le $\rho$-terme
correspondant à la règle \elan\  précédente~:
	$$x \ra [True \ra next([\{s1 \ra s2,s1 \ra s3\}](x))]
	([im(\RR_{f})](not~final([\{s1 \ra s2,s1 \ra s3\}](x))))$$ 

Ce terme appliqué au terme $s1$ mène à la réduction suivante~:

\begin{tabbing}
$\longra{*}_{\rho}$ ~~  \= ~~~~~ \=  \kill

$[x \ra [True \ra next([\{s1 \ra s2,s1 \ra s3\}](x))]([im(\RR_{f})](not~final([\{s1 \ra s2,s1 \ra s3\}](x))))](s1)$
\\
$\longra{}_{Fire}$
\> $\{[True \ra next([\{s1 \ra s2,s1 \ra s3\}](s1))]([im(\RR_{f})](not~final([\{s1 \ra s2,s1 \ra s3\}](s1))))\}$
\\
$\longra{*}_{\rho}$
\> $\{[True \ra next(\{s2,s3\})]([im(\RR_{f})](not~final(\{s2,s3\})))\}$
\\
$\longra{*}_{\rho}$
\> $\{[True \ra \{next(s2),next(s3)\}]([im(\RR_{f})](\{not~final(s2),not~final(s3)\}))\}$
\\
$\longra{*}_{\rho}$
\> $\{[True \ra \{next(s2),next(s3)\}](\{False,True\})\}$
\\
$\longra{*}_{\rho}$
\> $\{\{[True \ra \{next(s2),next(s3)\}](False),[True \ra \{next(s2),next(s3)\}](True)\}\}$
\\
$\longra{*}_{\rho}$
\> $\{\emptyset,[True \ra \{next(s2),next(s3)\}](True)\}$
\\
$\longra{*}_{\rho}$
\> $\{\emptyset,\{next(s2),next(s3)\}\}$
\\
$\longra{*}_{\rho}$
\> $\{next(s2),next(s3)\}$
\end{tabbing}
tandis qu'en \elan\  nous obtenons le seul résultat {\em \texttt{next(s3)}} qui
serait représenté par le $\rho$-terme~:\\
$\{next(s3)\}$.

\FEX%------------------------------------------------------------------------

Le problème dans l'Exemple~\ref{exIfWrong} est la double évaluation du terme
$[\{s1 \ra s2,s1 \ra s3\}](s1)$ remplaçant la variable locale $y$~: une fois
dans la condition et une fois dans le membre droit de la règle de réécriture. Si
ce terme est évalué en un ensemble ayant plus d'un élément et un de ses éléments
satisfait la condition de la règle alors, cet ensemble remplace les variables
correspondantes dans le membre droit de la règle tandis que seulement les
éléments satisfaisant la condition devraient être considérés. Par conséquent, nous
avons besoin d'un mécanisme permettant d'évaluer une seule fois chaque terme
instanciant une variable locale.

Nous utilisons une représentation combinant la deuxième approche de
représentation d'une règle avec affectations locales et la méthode utilisée pour
les règles de réécriture conditionnelles. Sans perdre la généralité, nous
pouvons considérer qu'une règle \elan\  conditionnelle avec des affectations
locales a la forme suivante~:

\vbox{
\begin{tabbing}
\indent\indent \= [$label$] ~~~ $l \Ra \tatpos{r}{x}{q}$ \=  \\
\>\> $\where~x:= (s) t$ \\
\>\> $\iif~\tatpos{C}{x}{p}$ \\
\> \End
\end{tabbing}
}

Nous obtenons alors la représentation de la règle précédente par le $\rho$-terme
$$l \ra [x \ra  [\{True \ra \tatpos{r}{x}{q},   
                False \ra \emptyset \}]([im(\RR)](\tatpos{C}{x}{p}))
        ]([s](t))$$
ou même plus simple~:
$$l \ra [x \ra  [True \ra \tatpos{r}{x}{q}]([im(\RR)](\tatpos{C}{x}{p}))
        ]([s](t))$$
où $\RR$ représente l'ensemble de règles de réécriture par rapport auquel les
conditions sont normalisées.

Les affectations généralisées sont traitées exactement de la même manière mais
les motifs des affectations généralisées sont utilisés à la place des variables
locales.

Afin de simplifier la présentation nous avons supposé que les règles de
l'ensemble $\RR$ sont des règles de réécriture de la forme $l \ra r$ et donc
l'opérateur $im$ est suffisant pour définir la normalisation par rapport à un
tel ensemble. Si nous considérons des règles non-nommées conditionnelles alors
l'opérateur $IM$ doit être employé.

La manière dont la transformation est appliquée à une règle de réécriture \elan\
et la réduction correspondante sont illustrées en reprenant
l'Exemple~\ref{exIfWrong} avec la bonne représentation.

\EX%------------------------------------------------------------------------
\label{ELANwhere}

La règle de réécriture \elan\  de l'Exemple~\ref{exIfWrong} est représentée par
le $\rho$-terme
	$$x \ra [y \ra [True \ra next(y)]([im(\RR_{f})](not~final(y)))]
        ([\{s1 \ra s2,s1 \ra s3\}](x))$$
lequel, appliqué au terme $s1$ mène à la réduction suivante~:

\begin{tabbing}
$\longra{*}_{\rho}$ ~~  \= ~~~~~ \=  \kill
$[x \ra [y \ra [True \ra next(y)]([im(\RR_{f})](not~final(y)))]([\{s1 \ra s2,s1 \ra s3\}](x))](s1)$
\\
$\longra{}_{Fire}$
\> $\{[y \ra [True \ra next(y)]([im(\RR_{f})](not~final(y)))]([\{s1 \ra s2,s1 \ra s3\}](s1))\}$
\\
$\longra{*}_{\rho}$
\> $\{[y \ra [True \ra next(y)]([im(\RR_{f})](not~final(y)))](\{s2,s3\})\}$
\\
$\longra{*}_{\rho}$
\> $\{[y \ra [True \ra next(y)]([im(\RR_{f})](not~final(y)))](s2),$
\\
\> $[y \ra [True \ra next(y)]([im(\RR_{f})](not~final(y)))](s3)\}$
\\
$\longra{*}_{Fire}$
\> $\{\{[True \ra next(s2)]([im(\RR_{f})](not~final(s2)))\},$
\\
\> $\{[True \ra next(s3)]([im(\RR_{f})](not~final(s3)))\}\}$
\\
$\longra{*}_{\rho}$
\> $\{[True \ra next(s2)](False),[True \ra next(s3)](True)\}$
\\
$\longra{*}_{\rho}$
\> $\{\emptyset,\{next(s3)\}\}$
\\
$\longra{*}_{\rho}$
\> $\{next(s3)\}$
\end{tabbing}
qui est la représentation du résultat obtenue en \elan.

\FEX%------------------------------------------------------------------------

Le même résultat que dans l'Exemple~\ref{exIfWrong} est obtenu si la règle
d'évaluation \rname{Fire} est appliquée avant la distribution de l'ensemble
$\{s2,s3\}$. Mais les stratégies confluentes présentées dans le
Chapitre~\ref{chap.resultats_confluence} interdisent une telle réduction et
donc, dans ce cas le résultat correct est obtenu.

\subsubsection{Sémantique d'une règle \elan\  factorisée}
%================================================================

La factorisation des règles de réécriture \elan\  est un moyen de définir
plusieurs réductions possibles pour un même terme de départ. Nous pouvons
toujours transformer une règle factorisée en plusieurs règles avec le membre
gauche initial et avec les membres droits correspondant aux choix de la
factorisation. Il est donc naturel de représenter les règles de réécriture
\elan\  factorisées dans le \roCal\  en utilisant les ensembles. 

Nous considérons la règle \elan\  factorisée suivante~:

\vbox{
\begin{tabbing}
\indent\indent \= [$label$] ~~~ $l \Ra \tatpos{r}{x}{q}$ \=  \\
\>\> $\choice$  \= \\
\>\> $\try$ \\
\>\>\> $\iif~C_1$ \\
\>\>\> $\where~x:= (s1) t_1$ \\
\>\> $\try$ \\
\>\>\> $\iif~C_2$ \\
\>\>\> $\where~x:= (s2) t_2$ \\
\>\> \End \\
\> \End
\end{tabbing}
}
et nous obtenons la représentation de la règle par le $\rho$-terme

\begin{tabbing}
\indent \= $l \ra \{$ \= $[True \ra  [x \ra \tatpos{r}{x}{q}]([s1](t_1))]([im(\RR)](C_1)),$\\
\>\> 	  $[True \ra  [x \ra \tatpos{r}{x}{q}]([s2](t_2))]([im(\RR)](C_2))\}$

\end{tabbing}

On peut évidemment avoir des règles \elan\  plus compliquées comportant des
conditions et des affectations locales extérieures à la factorisation mais une
approche similaire est utilisée dans ces cas.

En fait, le mécanisme d'évaluation \elan\  est plus complexe que représenté
jusqu'ici. L'évaluation d'une affectation locale de la forme \texttt{where
v:=(S) t} implique la normalisation du terme $t$ par rapport à l'ensemble de
règles non-nommées $\RR$ avant l'application de la stratégie $S$. En plus, le
membre droit calculé par l'application d'une règle \elan\  est normalisé par rapport à $\RR$
avant d'être retourné.

Par conséquent, la règle \elan\  de l'Exemple~\ref{exIfWrong} devrait être
représentée par le $\rho$-terme 
	$$x \ra [im(\RR_{f})]([y \ra [True \ra next(y)]([im(\RR_{f})](nf(y)))]
        ([\{s1 \ra s2,s1 \ra s3\}]([im(\RR_{f})](x))))$$
où $\RR$ représente l'ensemble de règles de réécriture non-nommées définies dans
le programme \elan\  contenant la règle respective.

\subsubsection{Stratégies générales dans les affectations locales}
%================================================================

Jusqu'à maintenant nous n'avons considéré dans les affectations locales que des
stratégies n'utilisant pas la règle de réécriture respective. La représentation
d'une règle \elan\  avec des appels locaux à des stratégies définies en utilisant 
cette règle doit être paramétrée par la définition des stratégies
respectives. Par exemple, une règle avec affectations locales de la forme 

\begin{tabbing}
\indent \indent\= [$label$] ~~~ $l \Ra r$ \=  \\
\>\> $\where~x:= (s) t$ 
\end{tabbing}
est représentée par le $\rho$-terme
	$$label(f) \eqdef l \ra [x \ra  r]([[f](s)](t))$$
où la variable libre $f$ sera instanciée par l'ensemble de stratégies du
programme contenant la règle étiquetée par $label$.

\subsubsection{Sémantique des stratégies et d'un programme \elan}
%================================================================

Les stratégies élémentaires d'\elan\  présentées dans la
Section~\ref{langage_elan} ont, dans la plupart des cas, une représentation
directe dans le \roCal. Les stratégies identité ($\id$) et échec ($\fail$) et
l'opérateur de concaténation (\texttt{;}) sont représentées dans le \roCal\  par
les opérateurs $id$, $fail$ et $;$ respectivement, définis dans la
Section~\ref{opAux}.  La stratégie $\dk(S_1,\dots,S_n)$ est représentée dans le
\roCal\  par l'ensemble $\{S_1,\dots,S_n\}$, la stratégie $\first(S_1,\dots,S_n)$
par le $\rho$-terme $first(S_1,\dots,S_n)$ défini dans la Section~\ref{opFirst}
et de façon similaire, le $\rho$-opérateur $dc$ est utilisé pour les stratégies
\elan\  $\dc$. La construction d'itération $\repeate*$ est représentées
facilement en utilisant le $\rho$-opérateur $\rep$.

\EX%------------------------------------------------------------------------
La stratégie \texttt{} de l'Exemple~\ref{strategies_en_elan} est représentée
immédiatement par le $\rho$-terme 
	$$attStrat_{\rho} \ra repeat\!*(\{initiate_{\rho}, \ldots ,intruder_{\rho}\}); 
			attackFound_{\rho}$$
où nous supposons que $initiate_{\rho}$, $intruder_{\rho}$, $attackFound_{\rho}$
sont les représentations des stratégies \elan\  correspondantes dans le \roCal.
\FEX%------------------------------------------------------------------------

Pour la définition des stratégies définies pas l'utilisateur dans un programme
\elan\  nous utilisons une approche basée sur l'opérateur de point fixe et
similaire à celle utilisée dans le cas des règles conditionnelles dans la
Section~\ref{encodRewCond}. Si nous considérons un programme \elan\  contenant
les stratégies $S_1,\ldots,S_n$ et les règles nommées avec les étiquettes alors
le $\rho$-terme représentant le programme est~:
	$$P \eqdef [\Theta](S)$$
où 
	$$S \eqdef f \ra (y \ra [\{S_i \ra Body_i~|~i=1 \ldots n\}](y))$$
et $Body_i$ représentent les membres droits des stratégies avec chaque stratégie
$S_i$ remplacée par $[f](S_i)$, chaque étiquette de règle remplacée par la
$\rho$-représentation de la règle et chaque opérateur de stratégie \elan\
remplacé par son correspondant dans le \roCal.

Pour résumer, nous présentons la transformation d'un programme \elan\  en un
$\rho$-terme~:

\DEF%------------------------------------------------------------------------
Nous considérons un programme \elan\  sans importations.

\begin{enumerate}

\item %Signature \\
La signature du \roCal\  est obtenue en utilisant les symboles définis dans les
parties \linebreak {\em \texttt{operators}} et {\em \texttt{stratop}} du programme \elan.

\item %Règles non-nommées \\
En partant des règles non-nommées de la forme~:
{\em 
\begin{tabbing}
$~~$$~~$ \= [] ~~~ $l_i(\ov{x}) \Ra r_i(\ov{x},\ov{y})$ \=  \\
\>\> $\where~(sort)~u_i(\ov{y}):= ()t_i(\ov{x})$ \\
\>\> $\iif~c_i(\ov{x},\ov{y})$ \\
\> \End
\end{tabbing}
}
nous construisons le terme~:
\begin{tabbing}
$~~$$~~$ \= 
$R_{nn} \eqdef$
$f \ra$ \= $(z \ra $ \= $[im($ \= $\{ l_i(\ov{x}) \ra$ \= $[ u_i(\ov{y}) \ra$ \= \\
\>	\>           \>        \>\>\> $[True \ra r_i(\ov{x},\ov{y})]([f](c_i(\ov{x},\ov{y})))$ \\
\>	\>           \>        \>                      \> $]( t_i(\ov{x}) )$ \\
\>	\>           \>        \> $~|~i=1 \ldots n \} )$ \\
\>	\>	     \> $]( z )$ \\
\>	\> $)$
\end{tabbing}

La normalisation \textit{innermost} par rapport à l'ensemble de règles
non-nommées est représentée par le terme~:
	$$IM_{nn} \eqdef [\Theta](R_{nn})$$

L'encodage est étendu d'une manière incrémentale à des règles contenant
plusieurs conditions et affectations locales. L'encodage peut être simplifié si
le programme ne contient pas des règles conditionnelles non-nommées; dans ce cas
le terme $IM_{nn}$ dévient~:
$$IM_{nn} \eqdef im(\{ l_i(\ov{x}) \ra [ u_i(\ov{y}) \ra r_i(\ov{x},\ov{y})]( t_i(\ov{x}) )
                      ~|~i=1 \ldots n \} )$$
où les règles avec affectations locales peuvent être simplifiées en des règles
élémentaires.

\item %Règles nommées \\
Pour chaque règle nommée de la forme~:
{\em 
\begin{tabbing}
$~~$$~~$ \= [$label$] ~~~ $l(\ov{x}) \Ra r(\ov{x},\ov{y})$ \=  \\
\>\> $\where~(sort)~u(\ov{y}):= (s)t(\ov{x})$ \\
\>\> $\iif~c(\ov{x},\ov{y})$ \\
\> \End
\end{tabbing}
}
nous construisons le terme~:
\begin{tabbing}
$~~$$~~$ \= 
$label(f) \eqdef$
$f \ra$ \= $(l(\ov{x}) \ra$ \= $[IM_{nn}]($ \= $[ u(\ov{y}) \ra$ \= \\
\>	\>        \>\>\> $[ True \ra r(\ov{x},\ov{y}) ]( [IM_{nn}](c(\ov{x},\ov{y})) )$ \\
\>	\>                  \>              \> $]( [[f](s)]( [IM_{nn}](t(\ov{x})) ) )$ \\
\>	\>                  \> $)$ \\
\>	\> $)$
\end{tabbing}

\item %Règles factorisées
Les règles \elan\  factorisées sont représentées de la même manière; la partie
factorisée dans la règle \elan\  est encodée par un ensemble dans le $\rho$-terme 
correspondant.

\item %Stratégies \\

Pour chaque stratégie de la forme~:
{\em 
\begin{tabbing}
$~~$$~~$ \= [] ~~~ $S \Ra Body$ \\
\> \End
\end{tabbing}
}
nous construisons le terme~:
\begin{tabbing}
$~~$$~~$ \= $S \ra BodyRho(f)$
\end{tabbing}
où $BodyRho$ représente le membre droit $Body$ de la stratégie avec chaque
symbole de stratégie $S_i$ remplacée par $[f](S_i)$, chaque étiquette de règle
$label$ remplacée par la $\rho$-représentation $label(f)$ de la règle et chaque
opérateur de stratégie \elan\  remplacé par son correspondant dans le \roCal.

\end{enumerate}

\noindent
Le programme \elan\  définissant les stratégies $S_1,\ldots,S_n$ est représenté
par le  $\rho$-terme~:
	$$P \eqdef [\Theta](S)$$
où
	$$S \eqdef f \ra (z \ra [\{S_i \ra BodyRho_i(f) ~|~i=1 \ldots n\}](z))$$
et $BodyRho_i(f)$ représente l'encodage de la  stratégie $S_i$.

\FDEF%------------------------------------------------------------------------

L'application d'une stratégie $\SS$ d'un programme \elan\  $\PP$ à un terme $t$
est représentée par le $\rho$-terme $[[P](s)](t)$ où $P$ est le $\rho$-terme
représentant le programme $\PP$ et $s$ est le nom de la stratégie $\SS$. Si
l'exécution du programme $\PP$ pour évaluer le terme $t$ par rapport à la
stratégie $\SS$ mène aux résultats $u_1,\ldots,u_n$ alors le $\rho$-terme
$[[P](s)](t)$ est réduit en l'ensemble de termes $\{u_1,\ldots,u_n\}$.

Dans l'Exemple~\ref{moduleELAN} nous présentons un module
\elan\  et les $\rho$-interprétations de toutes les règles et
stratégies de réécriture et donc, du programme \elan.

\EX%------------------------------------------------------------------------
\label{moduleELAN}
Le module {\em \texttt{automate}} décrit un automate avec les états {\em
\texttt{s1,s2,s3,s4,s5}} et avec des transitions non-déterministes décrites par
un ensemble de règles nommées contenant les règles étiquetées avec {\em
\texttt{r12,r13,r25,r32,r34,r41}}. L'opérateur {\em \texttt{next}} définit
l'état suivant d'une manière déterministe et son comportement est décrit par un
ensemble de règles non-nommées. Les états peuvent être ``finaux'' ({\em
\texttt{final}}) ou ``fermés'' ({\em \texttt{closed}}).
Les transitions doubles avec un état intermédiaire non-final et non-fermé sont
décrites par les règles {\em \texttt{double\_f}} et respectivement {\em
\texttt{double\_c}}.

%\begin{figure}[!htp]%----------------------------------------------------------
%\framebox{\parbox{\largeurtexte}{

\begin{verbatim}
module  automate
import global bool;end
sort state ;end
operators global
    s1 : state; s2 : state; s3 : state; s4 : state; s5 : state;
    next(@) : (state)  state;
    final(@) : (state)  bool;
    closed(@) : (state)  bool;
end
stratop  global
    follow   : <state -> state>   bs;
    gen_double   : <state -> state>   bs;
    cond_double   : <state -> state>   bs;
end

rules for bool
global
    [] final(s_1) =>  false         end
    [] final(s_2) =>  true          end
    [] final(s_3) =>  false         end
    [] final(s_4) =>  false         end
    [] final(s_5) =>  true          end

    [] closed(s_1) =>  false        end
    [] closed(s_2) =>  false        end
    [] closed(s_3) =>  true         end
    [] closed(s_4) =>  true         end
    [] closed(s_5) =>  true         end
end

rules for state
    x,y : state;
global
    [] next(s1) =>  s3            end
    [] next(s2) =>  s5            end
    [] next(s3) =>  s2            end
    [] next(s4) =>  s1            end
    [] next(s5) =>  s5            end

    [r12]  s1 =>  s2            end
    [r13]  s1 =>  s3            end
    [r25]  s2 =>  s5            end
    [r32]  s3 =>  s2            end
    [r34]  s3 =>  s4            end
    [r41]  s4 =>  s1            end

    [double_f]  x => next(y)    
                       where y := (follow) x 
                       if not final(y)                end
    [double_c]  x => next(y)    
                       where y := (follow) x 
                       if not closed(y)               end
end

strategies for state
implicit
    []follow         =>   dk(r12,r13,r25,r32,r34,r41)             end 
    []gen_double     =>   follow;follow                           end 
    []cond_double    =>   dk(double_f,double_c)                   end 
end end
\end{verbatim}

%}}
%\caption{\label{automateELAN}Le module \elan\  \texttt{automate}}
%\end{figure}%-----------------------------------------------------------------

Nous notons par $B$ l'ensemble de règles non-nommées définies dans les modules
importés {\em \texttt{bool}} et décrivant les opérations sur les booléens.

L'ensemble de règles non-nommées du module {\em \texttt{automate}} est représenté par le
$\rho$-terme 

\begin{tabbing}
\indent\indent \= $R ~~\eqdef~~$ \= 
$\{next(s1) \ra  s3, \ldots, next(s5) \ra  s5,$ \\
\>\> $final(s1) \ra  false, \ldots, final(s5) \ra  true,$ \\
\>\> $closed(s1) \ra  false, \ldots, closed(s5) \ra  true\}$
\end{tabbing}
et nous notons $RC = R \cup B$.

Les règles étiquetées avec {\em \texttt{double\_f}} et {\em \texttt{double\_c}} sont
représentées par les $\rho$-règles

\begin{tabbing}
\indent\indent  \= $double\_f(f) ~~\eqdef~~ x \ra [im(RC)]($ 
        	\= $[y \ra [True \ra next(y)]([im(RC)](not~final(y)))]$ \\
\>\> $([[f](follow)]([im(RC)](x))))$
\end{tabbing}
et respectivement

\begin{tabbing}
\indent\indent \= $double\_c(f) ~~\eqdef~~ x \ra [im(RC)]($ 
        \= $[y \ra [True \ra next(y)]([im(RC)](not~closed(y)))]$ \\
\>\> $([[f](follow)]([im(RC)](x))))$
\end{tabbing}

Les stratégies du module {\em \texttt{automate}} sont représentées par les
$\rho$-termes

\begin{tabbing}
\indent\indent \= cond\_double(f) \= ~~\eqdef~~ \= \kill 
\> $follow$ \> $~~\eqdef~~$ \> $follow \ra \{s1 \ra  s2, s1 \ra  s3, s2 \ra  s5, 
                             s3 \ra  s2, s3 \ra  s4, s4 \ra  s1\}$ \\
\> $gen\_double(f)$  \> $~~\eqdef~~$ \> $gen\_double \ra [f](follow);[f](follow)$ \\
\> $cond\_double(f)$ \> $~~\eqdef~~$ \> $cond\_double \ra \{double\_f(f),double\_c(f)\}$
\end{tabbing}
et nous obtenons le terme représentant le programme \elan\  {\em \texttt{automate}}~:

\begin{tabbing}
\indent\indent \= $automate ~~\eqdef~~ [\Theta](S)$
\end{tabbing}
où 

\begin{tabbing}
\indent\indent \= $S ~~\eqdef~~ f \ra (y \ra [\{follow,gen\_double(f),cond\_double(f)\}](y))$
\end{tabbing}

L'exécution du programme {\em \texttt{automate}} pour évaluer le terme {\em \texttt{s1}}
avec la stratégie {\em \texttt{cond\_double}} correspond à la réduction du terme
	$$[[automate](cond\_double)](s1)$$
En \elan, nous obtenons pour une telle exécution les résultats {\em \texttt{2}} et
{\em \texttt{5}} et la réduction du $\rho$-terme correspondant mène à l'ensemble
$\{2,5\}$.

\FEX%------------------------------------------------------------------------

Dans l'Exemple~\ref{moduleELAN} nous avons présenté un module \elan\
relativement simple mais, en suivant la même méthodologie, des règles et
stratégies plus compliquées peuvent être traitées.

%\DontWriteThisInToc  
\subsection*{Conclusion}
%================================================================
%~

Dans ce chappitre nous avons définit une traduction des $\lambda$-termes en des
$\rho$-termes et nous avons montré que les réductions des termes correspondants
dans les deux calculs sont équivalentes. Nous avons aussi présenté deux méthodes
de construction d'un $\rho$-terme avec une réduction similaire à la réduction
d'un terme par rapport à un système de réécriture. La première approche est
basée sur les termes de preuves de la logique de réécriture tandis que la
deuxième utilise les opérateurs du \roCalf\  afin d'obtenir un codage plus concis.

Nous avons utilisé cet encodage comme point de départ pour représenter les
règles et stratégies du langage \elan\  en \roCal. Nous avons représenter les
constructions d'\elan\  par des $\rho$-termes et nous avons ainsi obtenu une
sémantique complète du langage.

%% file: chapter_6.tex
%%%%%%%%%%%%%%%%%%%%%%%%%%%%%%%%%%%%%%%%%%%%%%%%%%%%%%%%%%%
% \TLtopbookmark
\chapter{Le \roCal\  typé}
\label{chap.calcul_type}
%%%%%%%%%%%%%%%%%%%%%%%%%%%%%%%%%%%%%%%%%%%%%%%%%%%%%%%%%%%%

Nous avons présenté jusqu'à maintenant le \roCal\  dans un cadre non-typé et nous
avons analysé ses propriétés et notamment la propriété de confluence.
Une deuxième propriété que l'on souhaite avoir pour le calcul est la terminaison,
caractéristique qui nous permettrait de conclure immédiatement de l'unicité des
formes normales.

En partant de la non-terminaison du \laCal\  et de la relation forte entre ce
calcul et le \roCal,  nous pouvons trouver immédiatement des réductions
non-terminantes dans le \roCal. Afin d'obtenir un calcul sans réduction infinie,
nous procédons comme d'habitude et nous imposons des restrictions sur la
formation des $\rho$-termes en introduisant une information de type pour chaque
terme.

La définition d'un système de types dans le cas du \roCal\  général s'avère une
tâche plus difficile que dans le cas du \laCal\  en raison de l'utilisation des
ensembles pour représenter le non-déterminisme et spécialement de la possibilité
d'avoir des ensembles dans le membre gauche des règles de réécriture. Nous
allons donc nous concentrer d'abord sur le typage du \roCalE\  et proposer une
approche similaire à celle utilisée dans le \laCal\  typé. Nous obtenons ainsi un
système de types qui nous permet de prouver que la réduction de tout terme
\textit{bien typé} est finie et préserve le type du terme initial.

Les mêmes propriétés peuvent être obtenues pour le \roCalEp\  (voir
Section~\ref{rewRuleStable}) permettant des règles de réécriture avec un
ensemble dans le membre gauche. Le système de types nécessite dans ce cas
l'utilisation de la notion de \textit{variable présente} introduite dans la
Section~\ref{rewRuleConservLin} et nous présentons brièvement à la fin de ce
chapitre une description du \roCalEp\  typé.

Les notations et les définitions classiques de ce chapitre sont inspirées de
celles utilisées pour le \laCal\  typé dans~\cite{Hindley-1997}
et~\cite{Hindley-1986}.

\section{La syntaxe du \roCalE\  typé} \label{syntaxeType}
%=============================================================================== 
%

Considérons un ensemble $\KK$ de \textit{types atomiques} \index{type!atomique}
$K_1, K_2,\ldots$ et l'ensemble des \textit{types} \index{type} inductivement
défini par~:  
\begin{itemize}
  \item tout \textit{type atomique} est un \textit{type},
  \item si $A$ et $B$ sont des \textit{types} alors $A \raS B$ est un \textit{type}.
\end{itemize}

La flèche des définitions de type associe à droite. Donc, un type de la forme
$A_1 \raS A_2 \raS \ldots \raS A_n$ est une abréviation pour 
$A_1 \raS (A_2 \raS (\ldots \raS A_n)\ldots)$.

Les \textit{types atomiques} sont utilisés généralement pour désigner un ensemble
particulier comme, par exemple, les naturels ou les booléens. Un \textit{type
composé} \index{type!composé} de la forme $A \raS B$ est utilisé pour désigner
l'ensemble des $\rho$-termes qui peuvent être appliqués à des $\rho$-termes de
type $A$ donnant comme résultat des $\rho$-termes de type $B$. En effet, nous
appelons également \Def{stratégie} un $\rho$-terme de type composé.

\DEF%------------------------------------------------------------------------------
Pour un type $A$, une \textit{variable typée} \index{variable!typée} est notée
$x:A$ et on dit que la variable $x$ a le type $A$. Un \Def{contexte} est un
ensemble de variables typées.  Les affectations de type $x:A$ d'un contexte
s'appellent également des définitions de type de variable.
\FDEF%------------------------------------------------------------------------------

\DEF%------------------------------------------------------------------------------
\label{syntaxeRoType}
Etant donnés un ensemble de variables $\XX$ et un ensemble de symboles
$\FF=\bigcup_{i \geq 0} \FF_i$, si nous notons par $K$ tout type atomique,
alors la syntaxe du \roCalE\  simplement typé est définie récursivement par la
grammaire suivante~:

\begin{tabular}{llll}
& \\
\textBNF{Types} ~~~~ & 
$T$ &::=&  $K ~~|~~ T \raS T $\\
& \\
\textBNF{Contextes} ~~~~ & 
$E$ &::=&  $x:T ~~|~~ E \dotctx \ldots \dotctx E $\\
& \\
\textBNF{Termes} ~~~~ & 
$t$ &::=& $x ~~|~~ f(t,\ldots,t) ~~|~~ \{t,\ldots,t\} ~~|~~ \ofenv{u}{E} \ra t ~~|~~[t](t)$\\
& \\
\end{tabular}\\
où $x \in \XX$, $u \in \TFX$ et $f \in \FF$.

Les contextes $E_1 \dotctx \ldots \dotctx E_n$ avec $n=0$ sont représentés par
$\emptyset$. Un contexte $E$ restreint à l'ensemble des variables d'un terme $t$
est noté $\rest{E}{t}$. Le contexte $E$ d'une règle de réécriture $\ofenv{l}{E}
\ra r$ est appelé le contexte \Defi{local}{contexte} de la règle.
\FDEF%------------------------------------------------------------------------------

Si $A_1,\ldots,A_n,A$ sont des types, nous notons 
$\FF_{A_1 \times \ldots \times A_n \raS A}$ l'ensemble des symboles de fonctions
prenant $n$ arguments de type $A_i$ et donnant comme résultat un terme de type
$A$.  Quand un symbole de fonction $f$ appartient à un ensemble 
$\FF_{A_1 \times \ldots \times A_n \raS A}$ nous disons que le symbole $f$ a le
\Def{profil} $A_1 \times \ldots \times A_n \raS A$.  Nous surchargeons les
symboles de fonctions et nous considérons pour tout symbole
$f \in \FF$ que si $f \in \FF_{A_1 \times \ldots \times A_n \raS A}$ et 
$f \in \FF_{B_1 \times \ldots \times B_n \raS B}$ alors nous avons également les
appartenances 
$f \in \FF_{(A_1 \raS B_1) \times \ldots \times (A_n \raS B_n) \raS (A \raS B)}$
et 
$f \in \FF_{(B_1 \raS A_1) \times \ldots \times (B_n \raS A_n) \raS (B \raS
A)}$.  
En plus, si nous considérons qu'un symbole $f$ a un des profils %avons le profil 
$(A_1 \raS B_1) \times \ldots \times (A_n \raS B_n) \raS (A \raS B)$
ou %le profil 
$(B_1 \raS A_1) \times \ldots \times (B_n \raS A_n) \raS (B \raS A)$
alors $f \in \FF_{A_1 \times \ldots \times A_n \raS A}$ et 
$f \in \FF_{B_1 \times \ldots \times B_n \raS B}$.  Afin d'éliminer les
ambiguïtés éventuelles, chaque symbole de fonction peut être annoté avec son
profil mais, lorsqu'il peut être déduit du contexte, cette annotation est
omise.  La raison de cette surcharge deviendra claire en analysant la règle
d'évaluation \rname{Congruence} présentée dans la Figure~\ref{MRAtype} où les
symboles de fonctions ayant le même nom doivent avoir différents profils afin
d'obtenir des termes bien typés ayant le même type dans les deux membres de la
règle.

\DEF%------------------------------------------------------------------------------
\label{freeVarsType} 
L'ensemble des variables \Defn{libres}{variable}{libre} d'un terme $t$ du
\roCalE\  typé, noté $FV(t)$, est défini inductivement par~:
\begin{enumerate}
  \item si $t = x$ alors $FV(t) = \{x\}$,
  \item si $t = \{u_1, \ldots, u_n\}$ alors $FV(t) = \bigcup_{i=1}^n FV(u_i)$,
  \item si $t = f(u_1,\ldots,u_n)$ alors $FV(t) = \bigcup_{i=1}^n FV(u_i)$,
  \item si $t = [u](v)$ alors $FV(t) = FV(u) \cup FV(v)$,
  \item if $t = \ofenv{u}{E} \ra v$ alors $FV(t) = FV(v) \setminus FV(u)$.
\end{enumerate}
\FDEF%------------------------------------------------------------------------------

Le contexte local d'une règle de réécriture n'est en effet pas significatif pour
la définition de l'ensemble des variables libres d'un terme bien typé et nous
obtenons donc une description très similaire à la Définition~\ref{freeVars}
utilisée dans le cas non-typé. Les variables \Defn{liées}{variable}{liée} sont
définies exactement de la même façon que dans le cas non-typé.

\DEF%------------------------------------------------------------------------------
Nous disons qu'un contexte est \Defi{consistant}{contexte} s'il ne contient pas
deux définitions de type différentes pour la même variable.
\FDEF%------------------------------------------------------------------------------

\section{Les règles de typage du \roCalE} \label{reglesTypage}
%=============================================================================== 
%

En s'inspirant des règles de typage utilisées dans le \laCal, nous définissons les
règles de typage pour les $\rho$-abstractions (les règles de réécriture) et les
applications et nous ajoutons des règles de typage pour les ensembles. Nous
commentons ensuite notre choix pour la règle permettant de typer les règles de
réécriture et nous présentons d'autres approches possibles.

\subsection{Présentation des règles de typage}
%=============================================================================== 
%

Les règles de typage du \roCalE\  sont présentées dans la Figure~\ref{typesRoVdash}
où tous les contextes sont supposés être consistants et nous notons par
$ST_{\rho}$ le système de types ainsi obtenu.

\begin{figure}[!htp]
\framebox{\parbox{\largeurtexte}{
$
\begin{array}{lll}
Var & E \vdash x:A
%\\
& ~~~si~x:A \subseteq E \\
&  \\% 

Rule
& \FRAC{E \vdash l:A ~~~~ \rest{E}{l} \dotctx F \vdash r:B}
  %----------------------------------------------------------------------
  {F \vdash (\ofenv{l}{\rest{E}{l}} \ra r) : A \raS B}
\\
%& ~~~si~E \dotctx F~est~consistant \\
&  \\% 

App 
& \FRAC{E \vdash u : A \raS B ~~~~ E \vdash v : A}
  %-----------------------------------------------------------------
  {E \vdash [u](v) : B} \\

&  \\% 

Op 
& \FRAC{E \vdash t_1:A_1 \ldots E \vdash t_n:A_n}
  %--------------------------------------------------------------
  {E \vdash f(t_1,\ldots,t_n) : A}  
%\\
& ~~~si~f \in \FF_{A_1 \times \ldots \times A_n \raS A} \\
&  \\% 

Set 
& \FRAC{E \vdash t_1:A \ldots E \vdash t_n:A}
  %--------------------------------------------------------------------------
  {E \vdash \{t_1, \ldots, t_n\}:A} \\
&  \\% 

Empty ~~~~ & E \vdash \emptyset:A 
%\\
& ~~~pour~tout~type~A \\
&  \\% 
\end{array}
$
}}
\caption{\label{typesRoVdash}Les règles de typage du \roCalE}
\end{figure}%

\DEF%------------------------------------------------------------------------------
\label{typeableDef}
Etant donnés une formule $E \vdash t:A$ déduite en utilisant les règles de typage
du \roCalE\  et un contexte $E'$ tel que $E \subseteq E'$, on dit que le terme
$t$ est \Defi{typable}{terme} (ou \Defi{bien typé}{terme}) de type $A$ dans
le contexte $E'$. On dit que le terme $t$ est \Defi{typable}{terme} dans le
contexte $E'$ s'il existe un type $A$ tel que $t$ a le type $A$ dans le contexte
$E'$. Un terme $t$ est \Defi{typable}{terme} s'il existe un contexte dans lequel
il est typable.
\FDEF%------------------------------------------------------------------------------

\REM%------------------------------------------------------------------------------
\label{subCTX}
Etant donnés deux contextes $E,E'$ tels que $E' \subseteq E$. Si $E' \vdash t:A$
alors $E \vdash t:A$.
\FREM%------------------------------------------------------------------------------

En analysant la règle de typage \rname{Op} on peut noter que le type d'un terme
avec un symbole de tête fonctionnel dépend du profil du symbole et des types de
ses arguments. On doit préciser que la règle de typage \rname{Op} est valable
aussi pour les constantes et donc, si $a \in \FF_A$ alors $\nilctx \vdash a:A$.

La règle de typage \rname{Set} indique qu'un ensemble de termes est bien typé si
tous ses éléments ont le même type et la règle de typage \rname{Empty} précise
que l'ensemble vide peut avoir un type quelconque.

\subsection{Discussion sur le typage d'une règle de réécriture}
%=============================================================================== 
%

L'ensemble de variables liées d'une $\lambda$-abstraction $\lambda x.t$ contient
seulement la variable $x$. Il suffit donc d'éliminer la variable typée $x:A$ du
contexte permettant de typer le terme $t$ afin d'obtenir le contexte utilisé
pour typer l'abstraction $\lambda x:A.t$. 
Le membre gauche d'une $\rho$-règle de réécriture peut être plus élaboré qu'une
simple variable et donc, nous ne pouvons pas utiliser directement l'approche du
\laCal\  typé.

Quand nous déterminons le type d'une règle de réécriture $l \ra r$ en utilisant
la règle de typage \rname{Rule}, nous considérons le fait que les variables
libres du membre droit de la règle de réécriture sont liées par les variables
libres de son membre gauche.  

Les variables libres du membre droit d'une règle de réécriture sont liées dans
la règle de réécriture par les variables libres du membre gauche de la règle
ayant le même nom.  En raison de ce rapport fort entre les variables de même nom des
deux membres d'une règle de réécriture $l \ra r$, le même contexte
($\rest{E}{l}$) doit préciser les types de ces variables. De plus, puisque dans
le \roCalE\  le membre gauche d'une règle de réécriture $l \ra t$ est toujours un
terme du premier ordre, les variables du terme $l$ sont exactement les variables
libres de $l$ et donc, les variables liées de la règle de réécriture. Ainsi, le
contexte nous permettant de typer une règle de réécriture $l \ra t$ ne doit pas
forcement inclure ces variables typées (i.e. $\rest{E}{l}$) mais doit indiquer
précisément les types des variables libres de $r$ qui ne sont pas libres dans
$l$ (i.e. $F$).

On doit noter que l'ensemble des variables libres du membre gauche d'une règle de
réécriture $l \ra r$ n'est pas nécessairement identique à l'ensemble des
variables typées du contexte nous permettant de typer le terme $l$.  Ce contexte
peut éventuellement contenir des variables n'appartenant pas à $l$ mais libres
dans $r$ et qui doivent donc appartenir au contexte permettant de typer la règle
de réécriture $l \ra r$. Dans la règle de typage \rname{Rule}, la restriction du
contexte $E$ à l'ensemble des variables de $l$ évite l'élimination des variables
qui ne sont pas libres dans le terme $l$ et donc, pas liées dans la règle $l \ra r$, du
contexte utilisé pour typer cette règle de réécriture.

Supposons, par exemple, que dans la règle de typage \rname{Rule} le
contexte $\rest{E}{l}$ soit remplacé par $E$ obtenant ainsi la règle de typage
suivante~: 
$$
\begin{array}{lll}
Rule'
& \FRAC{E \vdash l:A ~~~~ E \dotctx F \vdash r:B}
  %----------------------------------------------------------------------
  {F \vdash (\ofenv{l}{E} \ra r) : A \raS B}
\end{array}
$$%\\

En utilisant une telle règle de typage nous pourrions inférer
$\nilctx \vdash \ofenv{x}{x:A \dotctx y:A} \ra y:A \raS A$
et si  $a \in \FF_A$ nous obtenons
$\nilctx \vdash [\ofenv{x}{x:A \dotctx y:A} \ra y](a) : A$.
Cette dernière application se réduit, comme dans le
\roCal\  non-typé, en $\{y\}$ et puisque nous ne pouvons pas inférer
$\nilctx \vdash \{y\} : A$, le type n'est pas préservé par la réduction.

Ce problème peut être résolu en gardant toutes les variables typées du contexte
permettant de typer le membre droit d'une règle de réécriture dans le contexte
utilisé pour typer la règle. Ce comportement est obtenu en utilisant
une règle de typage \rname{Rule''} qui n'élimine pas les variables typées du
membre gauche du contexte de la règle de réécriture~: 
$$
\begin{array}{lll}
Rule''
& \FRAC{E \vdash l:A ~~~~ F \vdash r:B}
  %----------------------------------------------------------------------
  {F \vdash (\ofenv{l}{E} \ra r) : A \raS B}
\end{array}
$$%\\

Les types des variables du membre gauche d'une règle de
réécriture doivent être utilisées pour typer le membre droit de la règle et
donc, dans le cas de la règle de typage \rname{Rule''} nous devons imposer une
condition explicite de consistance pour le contexte $E \dotctx F$. Si cette
condition n'est pas satisfaite nous pouvons obtenir des variables liées dans le
membre droit d'une règle de réécriture qui n'ont pas le même type que les
variables correspondantes (i.e. avec le même nom) du membre gauche de la règle
de réécriture respective.

\EX%------------------------------------------------------------------------
Si nous considérons $x:A \vdash x:A$ et $x:B \vdash x:B$ alors, en
utilisant la règle de typage \rname{Rule''}, nous pouvons inférer 
$x:B \vdash \ofenv{x}{x:A} \ra x : A \ra B$.
Il est clair que nous ne voulons pas que le terme
$\ofenv{x}{x:A} \ra x$ soit de type $A \ra B$ mais de type $A \ra A$ quel que
soit le contexte utilisé.
\FEX%------------------------------------------------------------------------

D'autre part, la consistance du contexte $E \dotctx F$ est une condition trop
restrictive qui ne nous permettrait pas de typer tous les $\rho$-termes
qui sont bien typés selon les règles de typage présentées dans la
Figure~\ref{typesRoVdash}.

\EX%------------------------------------------------------------------------
\label{exRuleSec}
Considérons un symbole de fonction $f \in \FF_{B \raS A}$.  En utilisant la
règle de typage \rname{Rule''} nous  obtenons $x:A \vdash \ofenv{x}{x:A} \ra x : A \ra A$
mais, bien que nous ayons $x:B \vdash f(x):A$,  le terme 
$[\ofenv{x}{x:A} \ra x](f(x))$ ne peut pas être typé dans le contexte $x:B$ car
le contexte $x:A \dotctx x:B$ est inconsistant. Néanmoins, le terme 
$[\ofenv{y}{y:A} \ra y](f(x))$ est bien typé dans le contexte $x:B$.
\FEX%------------------------------------------------------------------------

En généralisant l'Exemple~\ref{exRuleSec} nous pouvons dire que tout terme typé
en utilisant les règles de typage dans la Figure~\ref{typesRoVdash} peut être
typé, modulo l'$\alpha$-conversion, en utilisant une approche basée sur la règle
de typage \rname{Rule''}.

En plus, puisque selon la règle de typage \rname{Rule''}, le contexte d'une
règle de réécriture contient les variables typées de son membre gauche, nous ne
devons pas enregistrer ces variables et donc, nous n'avons plus besoin
d'utiliser un contexte local pour les règles de réécriture.

Bien que la règle de typage \rname{Rule} initiale soit légèrement plus difficile
à manipuler, nous avons donc préféré cette approche qui n'impose aucune restriction
sur les termes qui peuvent être bien typés.

\section{Substitutions typées}
%=============================================================================== 
%

Nous devons définir maintenant les \Defn{substitutions
typées}{substitution}{typée} et la façon dont elles s'appliquent à un
$\rho$-terme typé.

Si $A_1,\ldots,A_n$ sont des types, une substitution typée a la
forme \mbox{$\sigma = \subs{x_1:A_1/t_1, \ldots,x_n:A_n/t_n}$} avec $x_i \in \XX$
et $t_i$ des $\rho$-termes typés, $i=1,\ldots,n$.  Le
domaine de la substitution $\sigma$ est défini comme d'habitude mais en
utilisant une notation empruntée aux contextes~:
$Dom(\sigma)=$
\linebreak
$x_1:A_1 \dotctx \ldots \dotctx x_n:A_n$.

\DEF%------------------------------------------------------------------------------
Nous disons qu'une substitution typée $\sigma=\subs{x_1:A_1/t_1,\ldots,x_n:A_n/t_n}$
est \Defi{bien typée}{substitution} dans un contexte $E$, et nous le
notons par $E \vdash \sigma$, si pour toutes les variables typées
\mbox{$x_i:A_i \in Dom(\sigma)$} nous avons $E \vdash t_i:A_i$.
\FDEF%------------------------------------------------------------------------------

L'application d'une substitution bien typée à un terme typé est définie
similairement au cas non-typé (voir Définition~\ref{defSubstGreffe}) mais en
imposant la condition que le domaine de la substitution et du contexte du terme
doivent être consistants. Dans ce cas, le terme résultant de l'application d'une
substitution à un terme a le même type que ce dernier terme~:

\LEM%********************************************************************
\label{subsTypes}
\comment{used in Lemma~\ref{SCsubs}, Lemma~\ref{subjRed}}
Etant donnés un terme $t$, une substitution typée $\sigma$ et le contexte $E$
tels que $E \vdash \sigma$ et $Dom(\sigma) \dotctx E \vdash t:B$, avec
$Dom(\sigma) \dotctx E$ consistant, alors $E \vdash \sigma t:B$.
\FLEM%*******************************************************************

\proof{

Nous nous limitons à seulement une variable dans le domaine de la substitution,
c'est-à-dire $\sigma = \subs{x:A/u}$.  Nous avons $E \vdash \sigma$ et par conséquent 
$E \vdash u:A$. Nous devons prouver que si $x:A \dotctx E \vdash t:B$ alors
$x:A \dotctx E \vdash \subs{x:A/u}t:B$.  Le cas général où nous avons plus d'une 
variable typée dans le domaine de la substitution est traité de la même
manière.

Nous procédons par induction sur la structure du terme $t$.

\textit{Le cas de base}~: $t=y$, avec $y \in \XX$.

Si $y \neq x$ alors $\subs{x:A/u}y = y$ et ainsi, $E \vdash y:B$.

Si $y = x$, puisque $x:A \dotctx E$ est consistant et $x:B \subseteq E$
alors $B=A$. Nous avons $\subs{x:A/u}y = {\subs{x:A/u}x} = u$ et $E \vdash u:A$ ou
d'une manière équivalente $E \vdash u:B$.

\textit{Induction}~: $t$ n'est pas une variable.
\begin{itemize}

\item  $t = f(t_1,\ldots,t_n)$  avec
$f \in \FF_{B_1 \times \ldots \times B_n \raS B}$.

Puisque $x:A \dotctx E \vdash t:B$ alors, par la règle de typage \rname{Op} nous avons
$x:A \dotctx E \vdash t_i:B_i$, $i=1,\ldots,n$.
Par induction $E \vdash \subs{x:A/u}t_i:B_i$, $i=1,\ldots,n$ et
en utilisant la règle de typage \rname{Op} nous obtenons\\$~~~~~$ 
$E \vdash f(\subs{x:A/u}t_1,\ldots,\subs{x:A/u}t_n):B$.

\item $t = \{t_1,\ldots,t_n\}$.

Puisque $x:A \dotctx E \vdash t:B$ alors, par la règle de typage \rname{Set} nous avons
$x:A \dotctx E \vdash t_i:B$, $i=1,\ldots,n$.
Par induction $E \vdash \subs{x:A/u}t_i:B$, $i=1,\ldots,n$ et
en utilisant la règle de typage \rname{Set} nous obtenons \\
$~~~~~$ 
$E \vdash \{\subs{x:A/u}t_1,\ldots,\subs{x:A/u}t_n\}:B$.

\item $t = [u](v)$.

Puisque $x:A \dotctx E \vdash t:B$ alors, par la règle de typage \rname{App},
$x:A \dotctx E \vdash u:C \raS B$ et $x:A \dotctx E \vdash v:C$.
Par induction, nous avons
$E \vdash \subs{x:A/u}u:C \raS B$ et
$E \vdash \subs{x:A/u}v:C$ et en utilisant la règle de typage \rname{App} nous obtenons\\
$~~~~~$ 
$E \vdash [\subs{x:A/u}u](\subs{x:A/u}v):B$.

\item $t = \ofenv{l}{\rest{F}{l}} \ra r$.

Puisque $x:A \dotctx E \vdash t:B$ alors, par la règle de typage \rname{Rule}, 
$F \vdash l:C$ et $E \vdash r:D$ avec $B=C \raS D$.

Nous supposons $x \not \in FV(l)$ et $FV(l) \cap FV(u) = \emptyset$ (sinon une
$\alpha$-conversion est effectuée) et dans ce cas
$\subs{x:A/u}(\ofenv{l}{\rest{F}{l}} \ra r) = \ofenv{l}{\rest{F}{l}} \ra \subs{x:A/u}r$.

Par induction
$E \vdash \subs{x:A/u}r:D$ et par la règle de typage \rname{Rule} nous obtenons 
$E \vdash \ofenv{l}{\rest{F}{l}} \ra \subs{x:A/u}r:C \raS D$
ou d'une manière équivalente\\
$~~~~~$ 
$E \vdash \ofenv{l}{\rest{F}{l}} \ra \subs{x:A/u}r:B$.

\end{itemize}
}

\section{Filtrage typé}
%=============================================================================== 
%

Nous modifions la Définition~\ref{defMa} en introduisant les contextes contenant
les variables typées des termes présents dans une équation de filtrage~:

\DEF%------------------------------------------------------------------------
\label{defMaType}
Etant donnée une théorie $T$ sur les $\rho$-termes, une 
$T$-{\em équation de filtrage} \index{equation de@équation de filtrage}
%\Def{équation de filtrage} 
est une formule de la forme 
$E \vdash t \meqqT E' \vdash t'$, où $t$ et $t'$ sont des $\rho$-termes
bien typés dans les contextes $E$ et respectivement $E'$.  Une substitution $\sigma$ 
bien typée dans le contexte $E'$ est une solution de l'équation 
$E \vdash t \meqqT E' \vdash t'$ si $T \models \sigma (t) = t'$.  Un
$T$-\Def{système de filtrage} est une conjonction de
$T$-équations de filtrage. Une substitution est une solution d'un $T$-système de
filtrage $P$ si c'est une solution de toutes les $T$-équations de filtrage.
Nous notons par $\fF$ un $T$-système de filtrage sans solution. Un $T$-système
de filtrage est appelé \Defi{trivial}{système de filtrage} quand toute
substitution bien typée est une solution du système.

Nous définissons \Sl($\SS$) pour un $T$-système de filtrage $\SS$ comme étant la
fonction qui retourne l'ensemble de toutes les solutions de $\SS$ quand $\SS$
n'est pas trivial et $\{\ID\}$\index{substitution!$\ID$}, où $\ID$ est la
substitution identité, quand $\SS$ est trivial.
\FDEF%------------------------------------------------------------------------

Selon la définition précédente, si la substitution $\sigma$ est une solution de
l'équation de filtrage $E \vdash t \meqqes E' \vdash t'$ et $t \in \TFX$
alors $Dom(\sigma) \subseteq E$.

Par exemple, quand la théorie $T$ est vide, la substitution résultant du
filtrage syntaxique entre les termes $t$ et $t'$, peut toujours être calculée en
utilisant l'ensemble de règles \rname{SyntacticMatching} présenté dans la
Section~\ref{filtrageNonType} du Chapitre~\ref{chap.calcul_non_type} que nous
reprenons ci-dessous~:

\begin{figure}[!htp]%----------------------------------------------------------
%\noindent
\framebox{\parbox{\largeurtexte}{

\renewcommand{\fleche}{\LaFleche}
\begin{ruleset}

\regle{Decomposition} 
{(f(t_1, \ldots, t_n) \meqqes f(t'_1, \ldots, t'_n)) \ww P} 
{\bigwedge_{i=1\ldots n} t_i \meqqes t'_i \ww P}

\cregle{SymbolClash} 
{(f(t_1, \ldots, t_n) \meqqes g(t'_1, \ldots, t'_m)) \ww P}
{\fF} {f \neq g}

\cregle{MergingClash} 
{(x \meqqes t) \ww (x \meqqes t') \ww P}
{\fF} {t \neq t'} 

\cregle{SymbolVariableClash~} 
{(f(t_1, \ldots, t_n) \meqqes x) \ww P} 
{\fF} {x \in \XX}

\end{ruleset}

}}
\caption{\label{SyntMatchBis}\rname{SyntacticMatching} - Règles pour le filtrage syntaxique}
\end{figure}%-----------------------------------------------------------------

Le résultat de la Proposition~\ref{normform} peut être étendu pour une équation
de filtrage avec les termes bien typés~:

\PROP%------------------------------------------------------------------------
\label{normformType}
La forme normale de tout problème de filtrage $t \meqqes t'$ calculée par les
règles \rname{SyntacticMatching} existe et est unique. Après avoir enlevé de la
forme normale toute équation dupliquée, si le système résultant est~:
\begin{enumerate}
\item $\fF$, alors il n'y a pas de filtre de $t$ à $t'$ et
	$\Sl(E \vdash t \meqqes E' \vdash t')=\emptyset$,
\item de la forme $\bigwedge_{i\in I} x_i \meqqes t_i'$ avec $I \neq \emptyset$,
	alors la substitution $\sigma = \subs{x_i / t_i'}_{i \in I}$ est l'unique
	filtre de $t$ à $t'$.
	Si $E \vdash x_i:A_i$ et $E' \vdash t_i':A_i$ alors
	$\Sl(E \vdash t \meqqes E' \vdash t')=\{\subs{x_i:A_i/t_i'}_{i \in I}\}$. Si pour un
	certain $i \in I$ nous avons $E \vdash x_i:A_i$ et $E' \vdash t_i':B_i$
	avec $A_i \neq B_i$ alors nous obtenons 
	$\Sl(E \vdash t \meqqes E' \vdash t')=\emptyset$,
\item vide, alors $t$ et $t'$ sont identiques et 
	$\Sl(E \vdash t \meqqes E' \vdash t')=\{\ID\}$.
\end{enumerate}
\FPROP%------------------------------------------------------------------------

On doit noter que nous n'enlevons pas les équations triviales $x \meqqes x$ de
la forme normale d'un système de filtrage mais nous vérifions l'égalité des
types de la variable $x$ des deux côtés dans les contextes correspondants. Une
équation de filtrage triviale mais avec les variables typées différemment
mène à un échec de filtrage comme par exemple dans l'équation de filtrage
$\Sl(x:A \vdash x \meqqes x:B \vdash x)=\emptyset$.

Puisque dans l'ensemble de règles de filtrage
\rname{SyntacticMatching} la consistance entre les types des membres des
équations de filtrage n'est pas vérifiée, les équations triviales ne sont pas
enlevées et nous pouvons ainsi obtenir une solution % de la forme
$\Sl(E \vdash t \meqqes E' \vdash t')=\{\subs{x:A/x,\ldots}\}$.

Nous pouvons, naturellement, intégrer les contraintes de type dans les règles de
filtrage en utilisant l'ensemble des règles dans la
Figure~\ref{SyntMatchContext}.

\begin{figure}[!htp]%----------------------------------------------------------
%\noindent
\framebox{\parbox{\largeurtexte}{

\renewcommand{\fleche}{\LaFleche}
\begin{ruleset}

\hrregle{Decomposition} 
{(E \vdash f(t_1, \ldots, t_n) \meqqes E' \vdash f(t'_1, \ldots, t'_n)) \ww P} 
{\bigwedge_{i=1\ldots n} E \vdash t_i \meqqes E' \vdash t'_i \ww P}

\hcrregle{SymbolClash} 
{(E \vdash f(t_1, \ldots, t_n) \meqqes E' \vdash g(t'_1, \ldots, t'_m)) \ww P}
{\fF} {f \neq g}

\hcrregle{MergingClash} 
{(E \vdash x \meqqes F \vdash t) \ww (E \vdash x \meqqes F' \vdash t') \ww P}
{\fF} {t \neq t'} 

\hcrregle{SymbolVariableClash~} 
{(E \vdash f(t_1, \ldots, t_n) \meqqes E' \vdash x) \ww P} 
{\fF} {x \in \XX}

\hcrregle{VariableClash} 
{(E \vdash x \meqqes E' \vdash t) \ww P} 
{\fF} {E \vdash x:A ~et~  E' \vdash t:B ~et~ A \neq B}

\end{ruleset}

}}
\caption{\label{SyntMatchContext}\rname{SyntacticMatchingType} - 
Règles pour le filtrage syntaxique typé}
\end{figure}%-----------------------------------------------------------------

La règle de typage \rname{VariableClash} vérifie juste que les types des deux
membres d'une équation de filtrage avec une variable à gauche sont identiques et 
la Proposition~\ref{normformType} peut être reformulée en~:

\PROP%------------------------------------------------------------------------
\label{normformContext}
La forme normale de tout problème de filtrage $E \vdash t \meqqes E' \vdash t'$
calculée par les règles \rname{SyntacticMatchingType} existe et est
unique. Après avoir enlevé de la forme normale toute équation dupliquée et toute
équation triviale de la forme $E \vdash x \meqqes E' \vdash x$, si le système
résultant est~:
\begin{enumerate}
\item $\fF$, alors il n'y a pas de filtre de $t$ à $t'$ et
	$\Sl(E \vdash t \meqqes E' \vdash t')=\emptyset$,
\item de la forme $\bigwedge_{i\in I} E \vdash x_i \meqqes E' \vdash t_i'$ avec
	$I \neq \emptyset$ et $E \vdash x_i:A_i$, alors la substitution 
	$\sigma = \subs{x_i:A_i / t_i'}_{i \in I}$ est l'unique filtre des 
	termes $t$ à $t'$ bien typés dans les contextes $E$ et respectivement $E'$ et 
	$\Sl(E \vdash t \meqqes E' \vdash t')=\{\subs{x_i:A_i/t_i'}_{i \in I}\}$.
\item vide, alors $t$ et $t'$ sont identiques et 
	$\Sl(E \vdash t \meqqes E' \vdash t')=\{\ID\}$.
\end{enumerate}
\FPROP%------------------------------------------------------------------------

Si l'ensemble de règles de filtrage \rname{SyntacticMatchingType} est utilisé
alors la forme normale d'un problème de filtrage ne peut pas contenir
d'équation de la forme $E \vdash x \meqqes E' \vdash x$ avec la variable $x$
typée différemment dans les contextes $E$ et $E'$ et ainsi, nous pouvons enlever
les équations de cette forme afin d'obtenir la substitution résultat.

L'avantage de maintenir séparées les contraintes de type et les règles de
filtrage comme dans la méthode utilisant les règles \rname{SyntacticMatching}
nous permet d'utiliser le même ensemble de règles de filtrage dans les approches
typées et non-typées.

Quand les contextes $E$, $E'$ des termes $t$, $t'$ sont clairs, nous les omettons et
nous abrégeons la fonction $\Sl(E \vdash t \meqqes E' \vdash t')$ par 
$\Sl(t \meqqes t')$.

\section{Les règles d'évaluation du \roCalE\  typé} \label{infRules}
%=============================================================================== 
%

Les règles d'évaluation du \roCalE\  non-typé présentées dans le
Chapitre~\ref{chap.resultats_confluence} sont enrichies avec l'information de
type et nous obtenons l'ensemble de règles d'évaluation dans la
Figure~\ref{MRAtype}. Les règles qui sont modifiées sont celles qui décrivent
l'évaluation des règles de réécriture~: \rname{Fire} et \rname{Switch_R}.

Les règles d'évaluation du \roCal\  typé général sont obtenues de la même manière à
partir des règles d'évaluation du \roCal\  non-typé mais la règle
\rname{Switch_L} décrivant le comportement des règles de réécriture avec un
ensemble dans le membre gauche nécessite une analyse plus approfondie et nous
allons proposer dans la Section~\ref{typeGeneral} une approche possible et un
système de types approprié.

\begin{figure}[!htp]
\noindent
\framebox{\parbox{\largeurtexte}{

\renewcommand{\fleche}{\Longrightarrow}
\begin{ruleset}
%===================================
  \hclwlregle 
  {Fire} 
  {[\ofenv{l}{E} \ra r](t)} 
  {\{\sigma r\}}
  {le~contexte~est~F} 
  {\sigma \in \Sl(E \vdash l \meqqes F \vdash t)~~}
\\
%===================================
  \regle
  {Congruence}   
  {[f(u_1,\ldots,u_n)](f(v_1,\ldots,v_n))} 
  {\{f([u_1](v_1),\ldots,[u_n](v_n))\}}
\\
%===================================
  \regle
  {Congruence\_fail}   
  {[f(u_1,\ldots,u_n)](g(v_1,\ldots,v_m))} 
  {\emptyset}
\\
%=================================== 
\regle {Distrib}
	{[\{u_1,\ldots,u_n\}](v)}
	{\{[u_1](v),\ldots,[u_n](v)\}}
\\
%=================================== 
\regle {Batch}
	{[v](\{u_1,\ldots,u_n\})}
	{\{[v](u_1),\ldots,[v](u_n)\}}
\\
%=================================== 
\regle {Switch_R} 
	{\ofenv{u}{E} \ra \{v_1,\ldots,v_n\}}
	{\{\ofenv{u}{E} \ra v_n,\ldots,\ofenv{u}{E} \ra v_n\}}
\\
%=================================== 
\hregle {OpOnSet}
	{f(v_1,\ldots,\{u_1,\ldots,u_m\},\ldots,v_n)}
	{\{f(v_1,\ldots,u_1,\ldots,v_n),\ldots,f(v_1,\ldots,u_m,\ldots,v_n)\}}
\\
%===================================
\regle {Flat}
	{\{u_1,\ldots,\{v_1,\ldots,v_n\},\ldots,u_m\}}
	{\{u_1,\ldots,v_1,\ldots,v_n,\ldots,u_m\}}
%===================================
\end{ruleset}

}}
\caption{\label{MRAtype}Les règles d'évaluation du \roCalE\  typé}
\end{figure}

Comme nous l'avons déjà mentionné, l'interprétation des symboles de fonctions est
surchargée et un symbole de fonction peut avoir plusieurs profils.  Si dans le
membre gauche de la règle d'évaluation \rname{Congruence} le premier $f$ a le
profil 
${(A_1 \raS B_1) \times \ldots \times (A_n \raS B_n) \raS (A \raS B)}$
et le deuxième $f$ le profil ${A_1 \times \ldots \times A_n \raS A}$
alors le symbole $f$ du membre droit de la règle d'évaluation a le profil 
${B_1 \times \ldots \times B_n \raS B}$.  Les arguments d'un terme
construit en utilisant le premier $f$ sont des règles de réécriture (ou d'autres
termes de type composé) qui sont appliquées aux arguments d'un terme construit
en utilisant le deuxième $f$ et la surcharge du symbole $f$ est évidemment
nécessaire afin de typer correctement ces applications.

Les réductions utilisant les règles d'évaluation \rname{Congruence} peuvent être
simulées par des réductions utilisant la règle d'évaluation \rname{Fire} et nous
avons montré dans le Chapitre~\ref{chap.calcul_non_type}
(page~\pageref{redondantesCongruence}) que le terme $[f(u_1,\ldots,u_n)](t)$ est
évalué, en utilisant les règles \rname{Congruence} et \rname{Congruence\_fail},
au même terme que le terme 
$[f(x_1,\ldots,x_n) \ra f([u_1](x_1),\ldots,[u_n](x_n))](t)$ en utilisant la
règle \rname{Fire}. Par conséquent, les règles \rname{Congruence} représentent
une forme d'expansion $\eta$ du \roCal\  qui serait définie par~:
\renewcommand{\fleche}{\Longrightarrow}
\begin{ruleset}
%===================================
  \regle 
  {Eta} 
  {f(t_1,\ldots,t_n)} 
  {f(x_1,\ldots,x_n) \ra f([t_1](x_1),\ldots,[t_n](x_n))} 
%===================================
\end{ruleset}
qui, appliquée dans le cas particulier d'une constante $a$, mène à
	$$a \Longrightarrow x \ra [a](x).$$

Pour des raisons de simplicité nous avons omis le contexte local de la règle de
réécriture du membre droit de la règle \rname{Eta} mais il est clair que si le
symbole $f$ du membre gauche de la règle \rname{Eta} a le profil 
$(A_1 \raS B_1) \times \ldots \times (A_n \raS B_n) \raS (A \raS B)$
alors ce contexte est $x_1:A_1 \dotctx \ldots \dotctx x_n:A_n$.

Il y a principalement deux propriétés que nous voulons prouver pour le \roCalE\
typé.  D'abord, nous montrons que les réductions du \roCalE\  préservent le type,
propriété habituellement appelée \textit{préservation du type} (\textit{subject
reduction}).  Deuxièmement, nous prouvons que dans le \roCalE\  typé il n'y a pas
de réduction infinie.  

Par rapport aux preuves de terminaison pour des calculs
similaires, comme le \laCal, nous devons prouver que la manipulation du
non-déterminisme représenté par des ensembles de termes est faite correctement.
Dans le \roCal\  nous avons des singletons correspondant à un résultat
déterministe, comme dans le \laCal, mais nous traitons aussi explicitement les
échecs possibles représentés par $\emptyset$ et les résultats non-déterministes
représentés par des ensembles ayant plus d'un élément.

Comme nous l'avons vu, sans typer le \roCal\  ne termine pas~:

\EX%------------------------------------------------------------------------
Le terme $\omega_{\rho} \omega_{\rho}=[x \ra [x](x)](x \ra [x](x))$ est évalué
comme décrit ci-dessous \\

$[x \ra [x](x)](x \ra [x](x))$\\
$\lraD{Fire}$ $~~$
$\{ \subs{x/(x \ra [x](x))} [x](x)\} \equiv \{[x \ra [x](x)](x \ra [x](x))\}$\\
$\lraD{Fire}$ $~~$
$\{\{[x \ra [x](x)](x \ra [x](x))\}\}$\\
$\lraD{Flat}$ $~~$
$\{[x \ra [x](x)](x \ra [x](x))\}$\\
$\lraD{}$  $~~$ $~~$
\ldots

\FEX%------------------------------------------------------------------------

Dans le \roCalE\  typé  le $\rho$-terme correspondant
$[\ofenv{x}{x:A} \ra [x](x)](\ofenv{x}{x:A'} \ra [x](x))$
n'est pas bien typé, indépendemment du type de la variable $x$ des membres
gauches des règles de réécriture.  Ceci découle de la règle de
typage \rname{App} qui nécessite d'une part un type $B \raS C$ pour le premier
$x$ de l'application et d'autre part un type $B$ pour le deuxième $x$ afin
d'obtenir le type du terme $[x](x)$.
Mais le même contexte (consistant) est utilisé pour typer les deux variables et
donc nous ne pouvons pas avoir deux définitions de type différentes pour la même
variable.

\section{La préservation du type}
%=============================================================================== 
%

Nous montrons maintenant que le système de types que nous avons proposé dans la
Section~\ref{reglesTypage} est cohérent par rapport aux règles
d'évaluation du \roCalE\  dans le sens où le type d'un terme est le même que le
type de tout terme obtenu en le réduisant.

\TH%----------------------------------------------------------
\label{subjRed}
\comment{uses Lemma~\ref{subsTypes}}
Pour tous $\rho$-termes $a$ et $a'$, si $a \Raro a'$ et $E \vdash a:A$, alors 
$E \vdash a':A$.
\FTH%----------------------------------------------------------

\proof{

Nous examinons les règles d'évaluation du \roCalE\  typé une par une et nous
prouvons que le membre gauche et le membre droit de chaque règle ont le même
type dans un contexte donné.  Pour chaque règle d'évaluation de la forme 
$lhs \Longrightarrow rhs$ nous montrons donc que si $E \vdash lhs:A$ alors 
$E \vdash rhs:A$.

\begin{itemize}

\item la règle \rname{Fire}
\renewcommand{\fleche}{\Longrightarrow}
\begin{ruleset}
%===================================
  \cwregle 
  {Fire} 
  {[\ofenv{l}{E'} \ra r](t)} 
  {\{\sigma r\}}
  {le~contexte~est~E} 
  {\sigma \in \Sl(E' \vdash l \meqqes E \vdash t)~~}
\\
%===================================
\end{ruleset}
\vspace{-.5cm}

Considérons $A$ tel que $E \vdash [\ofenv{l}{\rest{F}{l}} \ra r](t):A$. 
En utilisant la règle de typage \rname{App} nous inférons $E \vdash t:B$
et $E \vdash (\ofenv{l}{\rest{F}{l}} \ra r):B \raS A$.
Par la règle de typage \rname{Rule} nous avons $\rest{F}{l} \vdash l:B$ et
$\rest{F}{l} \dotctx E \vdash r:A$ avec $\rest{F}{l} \dotctx E$ consistant.

Si $\sigma$ est la solution de l'équation de filtrage 
$(\rest{F}{l} \vdash l \meqqes E \vdash t)$ alors,
conformément à la définition du filtrage typé, $E \vdash \sigma$ et 
$Dom(\sigma)=\rest{F}{l}$. Puisque $\rest{F}{l} \dotctx E$ est consistant alors 
$Dom(\sigma) \dotctx E$ est consistant et en utilisant le Lemme~\ref{subsTypes}
nous obtenons $E \vdash \sigma r:A$. Par la règle de typage  \rname{Set} nous avons\\
$~~~~$ $E \vdash \{\sigma r\}:A$.

Si le filtrage $(\rest{F}{l} \vdash l \meqqes E \vdash t)$ échoue alors le
membre droit $\emptyset$ satisfait la propriété par la règle de typage
\rname{Empty}.
\\

\item la règle \rname{Congruence}
\begin{ruleset}
%===================================
  \regle
  {Congruence}   
  {[f(u_1,\ldots,u_n)](f(v_1,\ldots,v_n))} 
  {\{f([u_1](v_1),\ldots,[u_n](v_n))\}}
%===================================
\end{ruleset}
\vspace{-.5cm}

Considérons $A$ tel que $E \vdash [f(u_1,\ldots,u_n)](f(v_1,\ldots,v_n)):A$.
En utilisant la règle de typage \rname{App} nous inférons
$E \vdash f(u_1,\ldots,u_n):B \raS A$ et
$E \vdash f(v_1,\ldots,v_n):B$.
Nous supposons les profils suivants pour le symbole de fonction  $f$:%\\
%$~~~~$ 
${A_1 \times \ldots \times A_n \raS A}$, %\\
%$~~~~$ 
${B_1 \times \ldots \times B_n \raS B}$ et %\\
%$~~~~$ 
${(A_1 \raS B_1) \times \ldots \times (A_n \raS B_n) \raS (A \raS B)}$.
Par la règle de typage \rname{Op} appliquée deux fois nous avons~: 
$E \vdash u_i : B_i \raS A_i$, $E \vdash v_i:B_i$.  
Nous appliquons $n$ fois la règle de typage \rname{App} et nous obtenons
$E \vdash [u_i](v_i) : A_i$, $i=1,\ldots,n$
et par la règle de typage \rname{Op} nous avons
$E \vdash f([u_1](v_1),\ldots,[u_n](v_n)):A$.
Finalement, nous
appliquons la règle de typage \rname{Set} et nous obtenons~: \\
$~~~~$ $E \vdash \{f([u_1](v_1),\ldots,[u_n](v_n))\}:A$.
\\

\item  la règle \rname{Congruence\_fail}
\begin{ruleset}
%===================================
  \regle
  {Congruence\_fail}   
  {[f(u_1,\ldots,u_n)](g(v_1,\ldots,v_m))} 
  {\emptyset}
%===================================
\end{ruleset}
\vspace{-.5cm}

L'application de la règle de typage \rname{Empty} prouve immédiatement la propriété.
\\

\item la règle \rname{Distrib} 
\begin{ruleset}
%===================================
\regle {Distrib}
	{[\{u_1,\ldots,u_n\}](v)}
	{\{[u_1](v),\ldots,[u_n](v)\}}
%===================================
\end{ruleset}
\vspace{-.5cm}

Considérons $A$ tel que $E \vdash [\{u_1,\ldots,u_n\}](t):A$. 
En utilisant la règle de typage \rname{App} nous inférons
$E \vdash \{u_1,\ldots,u_n\}:B \raS A$ et
$E \vdash t:B$.
Par la règle de typage \rname{Set} nous avons 
$E \vdash u_i:B \raS A$, $i=1,\ldots,n$ et 
en appliquant la règle de typage \rname{App} avec $E \vdash t:B$ nous obtenons
$E \vdash [u_i](t):A$, $i=1,\ldots,n$. 
Finalement, la règle de typage \rname{Set} mène à~: \\
$~~~~$ $E \vdash \{[u_1](t),\ldots,[u_n](t)\}:A$.
\\

\item  la règle \rname{Batch}
\begin{ruleset}
%===================================
\regle {Batch}
	{[v](\{u_1,\ldots,u_n\})}
	{\{[v](u_1),\ldots,[v](u_n)\}}
%===================================
\end{ruleset}
\vspace{-.5cm}

Considérons $A$ tel que $E \vdash [u](\{t_1,\ldots,t_n\}):A$. 
En utilisant la règle de typage \rname{App} nous inférons
$E \vdash u:B \raS A$,
$E \vdash \{t_1,\ldots,t_n\}:B$.
Par la règle de typage \rname{Set} nous avons 
$E \vdash t_i:B$, $i=1,\ldots,n$ et 
en appliquant la règle de typage \rname{App} $n$ fois nous obtenons
$E \vdash [u](t_i):A$, $i=1,\ldots,n$. 
Finalement, la règle de typage \rname{Set} mène à~: \\
$~~~~$ $E \vdash \{[u_1](t),\ldots,[u_n](t)\}:A$.
\\

\item la règle \rname{Switch_R} 
\begin{ruleset}
%===================================
\regle {Switch_R} 
	{\ofenv{u}{\rest{F}{u}} \ra \{v_1,\ldots,v_n\}}
	{\{\ofenv{u}{\rest{F}{u}} \ra v_n,\ldots,\ofenv{u}{\rest{F}{u}} \ra v_n\}}
%===================================
\end{ruleset}
\vspace{-.5cm}

Considérons $A$ tel que $E \vdash \ofenv{u}{\rest{F}{u}} \ra \{v_1,\ldots,v_n\}:A$. 
En utilisant la règle de typage \rname{Rule} nous inférons $A=B \raS C$ et
$F \vdash u:B$,
$\rest{F}{u} \dotctx E \vdash \{v_1,\ldots,v_n\}:C$.
Par la règle de typage \rname{Set} nous avons 
$\rest{F}{u} \dotctx E \vdash v_i:C$, $i=1,\ldots,n$ et 
par la règle de typage \rname{Rule} appliquée $n$ fois nous obtenons
$E \vdash \ofenv{u}{\rest{F}{u}} \ra v_i:B \raS C$, $i=1,\ldots,n$.
Finalement, la règle de typage \rname{Set} mène à~: \\
$~~~~$ $E \vdash \{\ofenv{u}{\rest{F}{u}} \ra v_1,\ldots,\ofenv{u}{\rest{F}{u}} \ra v_n\}:A$.
\\

\item  la règle \rname{OpOnSet} 
\begin{ruleset}
%===================================
\hregle {OpOnSet}
	{f(v_1,\ldots,\{u_1,\ldots,u_m\},\ldots,v_n)}
	{\{f(v_1,\ldots,u_1,\ldots,v_n),\ldots,f(v_1,\ldots,u_m,\ldots,v_n)\}}
%===================================
\end{ruleset}
\vspace{-.5cm}

Considérons $A$ tel que $f \in \FF_{A_1 \times \ldots \times A_n \raS A}$ et
$E \vdash f(v_1,\ldots,\{u_1,\ldots,u_m\},\ldots,v_n):A$. 
En utilisant la règle de typage \rname{Op} nous inférons
$E \vdash v_i:A_i$, $i=1,\ldots,n$, $i \neq k$, et
$E \vdash \{u_1,\ldots,u_m\}:A_k$.
Par la règle de typage \rname{Set} nous avons 
$E \vdash u_j:A_k$, $j=1,\ldots,m$ et 
par la règle de typage \rname{Op} appliquée pour $j=1,\ldots,m$ nous obtenons
$E \vdash f(v_1,\ldots,u_j,\ldots,v_n):A$. 
Finalement, la règle de typage \rname{Set} mène à~: \\
$~~~~$ $E \vdash \{f(v_1,\ldots,u_1,\ldots,v_n),\ldots,f(v_1,\ldots,u_m,\ldots,v_n)\}:A$.
\\

\item  la règle \rname{Flat}
\begin{ruleset}
%===================================
\regle {Flat}
	{\{u_1,\ldots,\{v_1,\ldots,v_n\},\ldots,u_m\}}
	{\{u_1,\ldots,v_1,\ldots,v_n,\ldots,u_m\}}
%===================================
\end{ruleset}
\vspace{-.5cm}

Considérons $A$ tel que $E \vdash \{u_1,\ldots,\{v_1,\ldots,v_n\},\ldots,u_m\}:A$. 
En utilisant la règle de typage \rname{Set} nous inférons
$E \vdash u_j:A$, $j=1,\ldots,m$ et
$E \vdash \{v_1,\ldots,v_n\}:A$.
Par la règle de typage \rname{Set} nous avons 
$E \vdash u_j:A$, $j=1,\ldots,m$ et
$E \vdash \{v_1,\ldots,v_n\}:A$.
La même règle \rname{Set} mène à
$E \vdash v_i:A$, $i=1,\ldots,n$ et  
finalement, par la règle de typage \rname{Set} nous avons~: \\
$~~~~$ $E \vdash \{u_1,\ldots,v_1,\ldots,v_n,\ldots,u_m\}:A$. 

\end{itemize}

}%--------------------------------------------------------------------------------------

Comme nous l'avons précisé dans la Section~\ref{infRules}, les réductions utilisant
les règles d'évaluation \rname{Congruence} peuvent être simulées par des
réductions utilisant la règle d'évaluation \rname{Fire} et nous pouvons
obtenir une équivalence similaire au niveau des types.  Considérons un symbole $f$
tel que
$f \in \FF_{(A_1 \raS B_1) \times \ldots \times (A_n \raS B_n) \raS (A \raS B)}$
et donc
%$~~~~~~$ 
$f \in \FF_{A_1 \times \ldots \times A_n \raS A}$
et
%$~~~~~~$ 
$f \in \FF_{B_1 \times \ldots \times B_n \raS B}$.
%$~~~~~~$ 

Alors nous pouvons typer le terme $f(t_1,\ldots,t_n)$ dans le contexte $E$~:
$$
\begin{array}{lll}
\FRAC{E \vdash t_1 : A_1 \raS B_1  ~ \ldots ~  E \vdash t_n : A_n \raS B_n}
     %--------------------------------------------------------------------------------------
     {E  \vdash f(t_1,\ldots,t_n) : A \raS B}
& \rname{Op} \\
\end{array}
$$

Si nous considérons le terme équivalent
$\ofenv{f(x_1,\ldots,x_n)}{E'} \ra f([t_1](x_1),\ldots,[t_n](x_n))$
avec le contexte
$E' = x_1:A_1 \dotctx \ldots \dotctx x_n:A_n$,
tel que $E'$, $E$ sont disjoints alors nous obtenons

$
   \FRAC{E' \dotctx E \vdash t_i : A_i \raS B_i ~~~ E' \dotctx E \vdash x_i : A_i}
        {E' \dotctx E \vdash [t_i](x_i):B_i}
   ~~~ App
$
$~~~~~~$
$
   \FRAC{E' \vdash x_1:A_1 ~ \ldots ~ E' \vdash x_n:A_n}
        {E' \vdash f(x_1,\ldots,x_n):A}
   ~~~ Op
$\\

$
 \FRAC{E' \dotctx E \vdash [t_i](x_i):B_i}
      {E' \dotctx E \vdash f([t_1](x_1),\ldots,[t_n](x_n)) : B}
   ~~~ Op
$\\

\noindent
et par la règle de typage \rname{Rule}
$$
\FRAC{
E' \vdash f(x_1,\ldots,x_n):A  ~~~~~
E' \dotctx E \vdash f([t_1](x_1),\ldots,[t_n](x_n)) : B}
%--------------------------
{E \vdash \ofenv{f(x_1,\ldots,x_n)}{E'} \ra f([t_1](x_1),\ldots,[t_n](x_n)) 
   : A \raS B
}
$$

En utilisant les déductions précédentes nous pouvons conclure que si
$t_i : A_i \raS B_i, ~ i=1,\ldots,n$ alors
$$E \vdash \ofenv{f(x_1,\ldots,x_n)}{x_1:A_1 \dotctx \ldots \dotctx x_n:A_n} 
\ra f([t_1](x_1),\ldots,[t_n](x_n)) : A \raS B
$$

Nous avons ainsi induit le même type pour un terme $f(t_1,\ldots,t_n)$
et pour le terme étendu correspondant 
$\ofenv{f(x_1,\ldots,x_n)}{x_1:A_1 \dotctx \ldots \dotctx x_n:A_n} \ra f([t_1](x_1),\ldots,[t_n](x_n))$
dans le même contexte $E$. On doit noter que dans le deuxième terme nous devons
définir explicitement dans le contexte local de la règle de réécriture les types
des variables liées de la règle.

Ceci prouve que notre choix pour le système de types est approprié puisque il est
consistant avec l'expansion implicite du \roCal.

\section{La normalisation forte du \roCalE}
%=============================================================================== 
%

Nous nous concentrons maintenant sur la preuve de la normalisation forte du
\roCalE\  typé.  Cette propriété garantit l'existence d'une forme normale pour
tout $\rho$-terme bien typé et donc, assure que les réductions dans le \roCalE\
sont finies. Quand le \roCalE\  est confluent (cf. Section~\ref{stratRocal}) nous
pourrons conclure en l'unicité du résultat pour la réduction de tout $\rho$-terme
bien typé.

\DEF%------------------------------------------------------------------------------
\label{SNterm}
Un $\rho$-terme $t$ typé ou non-typé est \Def{fortement normalisant}
(\Def{strongly normalizing} ou \Def{SN}) par rapport à une relation de réduction
si toute réduction commençant par $t$ est finie. Un $\rho$-terme est
\Def{faiblement normalisant} si au moins une réduction issue de $t$ est finie.
\FDEF%------------------------------------------------------------------------------

Il n'est pas surprenant qu'en raison du rapport fort entre le \roCal\  et le
\laCal, notre preuve de la normalisation forte du \roCalE\  soit inspirée de la
preuve de la normalisation forte du \laCal.

Il y a plusieurs approches pour prouver la normalisation forte du \laCal. Une
méthode s'appelle \textit{internalization} et a été employée la première fois par
Gandy~\cite{Gandy80}. Une autre est celle appelée habituellement la méthode des
\textit{candidats de réductibilité} et basée sur les notions introduites par
Tait~\cite{Tait67}. Cette dernière approche a été généralisée
dans~\cite{Girard-72} et~\cite{JouannaudOkada-97}.

Nous utilisons par la suite les notations, les définitions et l'approche
de~\cite{Hindley-1986} qui est une variation de la méthode de Tait. Par rapport
à cette méthode, dans notre approche nous devons manipuler correctement les
termes du premier ordre et les termes ensemble.

Lorsque le contexte $E$ dans lequel nous typons les termes est clair, nous
l'omettons et dans ce cas-ci nous abrégeons $E \vdash t:A$ par $t:A$.

\DEF%------------------------------------------------------------------------------
\label{SCterm}
Nous définissons la \Def{calculabilité forte} (\Def{strong computability} ou
\Def{SC}) d'un terme $t$ par induction sur le nombre de flèches de type ``$\raS$''
dans le type de $t$~:
\begin{itemize}
\item[a.] un terme de type atomique est SC s'il est SN,
\item[b.] un terme $t$ tel que $E \vdash t:A \raS B$ est SC si, pour tout
	terme SC $u$ tel que $E \vdash u:A$, le terme $[t](u)$ avec
	$E \vdash [t](u):B$ est SC.
\end{itemize} 

La définition est étendue pour les substitutions typées et nous disons qu'une
substitution de la forme $\subs{x_1:A_1/u_1,\ldots,x_n:A_n/u_n}$ est SC si tous
les termes $u_i$ sont SC.
\FDEF%------------------------------------------------------------------------------

\REM%----------------------------------------------------------
\label{remarksTypes}
En partant en particulier des définitions données ci-dessus nous pouvons remarquer que~:
\begin{enumerate}
\item \label{remarksTypes:first} 
Tout type $A$ est de la forme $A_1 \raS \ldots \raS A_n \raS K$, avec $K$ un
type atomique.

\item \label{remarksTypes:second} 
Etant donné un type $A = A_1 \raS \ldots \raS A_n \raS K$; $(t:A)$ est SC ssi pour
tous termes typés SC, $t_1:A_1, \ldots, t_n:A_n$, le terme
$[\ldots[[t](t_1)](t_2)\ldots](t_n):K$ est SC (par la
Définition~\ref{SCterm}(b)). $[\ldots[[t](t_1)](t_2)\ldots](t_n):K$ est SC ssi
il est SN (par la Définition~\ref{SCterm}(a)).

\item \label{remarksTypes:third} 
Si $(t:A)$ est SC (SN), alors tout terme qui diffère de $t$ seulement au
niveau des noms des variables liées est SC (SN).

\item \label{remarksTypes:fourth} 
Si $t:A \raS B$ est SC et $u:A$ est SC, alors $[t](u):B$ est SC
(par la Définition~\ref{SCterm}(b)).

\item \label{remarksTypes:fifth} 
Si $t:A$ est SN, alors tout sous-terme de $t:A$ est SN, puisque toute
réduction infinie d'un sous-terme de $t$ mène à une réduction infinie de $t$.
\end{enumerate} 

\FREM%----------------------------------------------------------

En utilisant ces remarques nous prouvons d'abord quelques lemmes préliminaires
et ensuite nous concluons par la proposition énonçant la normalisation forte du
\roCalE\  typé.

La preuve de normalisation est faite en deux étapes. D'abord nous montrons que
tout terme typable qui est SC est SN. Ensuite nous prouvons que tous les termes
typables sont SC et nous concluons que tous les termes typables sont SN.

\LEM%------------------------------------------------------------------------------
\label{SCisSN} (tout terme SC est SN)
\comment{used in Lemma~\ref{SCpresSet}, Lemma~\ref{SCpresFun},
Lemma~\ref{SCsubs}, Lemma~\ref{allSC}, Proposition~\ref{strongNormTh}; uses remark~(4,5)}

Pour tout type $A$ nous avons les propriétés suivantes~:
\begin{itemize}
	\item[a.] Etant donnés un atome $t$ et les termes $u_1,\ldots,u_n$. Si
	$u_1,\ldots,u_n$ sont SN alors le terme $[\ldots[[t](u_1)](u_2)\ldots](u_n):A$
	est SC.
	\item[b.] Tout terme SC de type $A$ est SN.
\end{itemize} 
\FLEM

\proof{

Nous procédons par induction sur le nombre de flèches de type ``$\raS$'' dans le type $A$~:

\textit{Le cas de base}~: $A$ est un type atomique.
\begin{itemize}

\item[a.] Puisque $t$ est un atome et donc une variable,  toute réduction de
$[\ldots[[t](u_1)](u_2)\ldots](u_n)$ est une réduction d'un $u_i$. Puisque
$u_1,\ldots,u_n$ sont SN, alors $[\ldots[[t](u_1)](u_2)\ldots](u_n)$ est
SN. Ainsi, par la Définition~\ref{SCterm}~(a), le terme
$[\ldots[[t](u_1)](u_2)\ldots](u_n)$ est SC.

\item[b.] Par la Définition~\ref{SCterm}~(a).

\end{itemize} 

\textit{Induction}~: $A = B \raS C$
\begin{itemize}

\item[a.] Nous considérons un terme SC $v$ de type $B$. 
Par l'hypothèse d'induction (b), $(v:B)$ est SN. Par l'hypothèse d'induction
(a), le terme $[[\ldots[[t](u_1)](u_2)\ldots](u_n)](v):C$ est SC.  
Ainsi, par la Définition~\ref{SCterm}~(b),
$[\ldots[[t](u_1)](u_2)\ldots](u_n):A$ est SC.

\item[b.] Nous considérons un terme SC $u$ de type $A$ et une variable $x$ de
type $B$ qui n'apparaît pas (libre ou liée) dans $(u:A)$. Par l'hypothèse
d'induction (a) et en considérant $n=0$, $x$ est SC. Selon la
Remarque~\ref{remarksTypes}(\ref{remarksTypes:fourth}), $[u](x):C$ est SC et par
l'hypothèse d'induction (b), $[u](x):C$ est SN.  Conformément à la
Remarque~\ref{remarksTypes}(\ref{remarksTypes:fifth}), $(u:a)$ est SN.

\end{itemize} 

}%------------------------------------------------------------------------------

Nous prouvons maintenant que tous les termes typables sont SC.  Pour ceci nous
avons besoin des lemmes montrant la stabilité des propriétés SN et SC pour les
termes du premier ordre et pour les termes ensemble.  Puisque dans le \laCal\
nous n'avons pas de tels termes, ce genre de propriété n'est pas nécessaire pour
la preuve de normalisation du \laCal.

\LEM%------------------------------------------------------------------------------
\label{SNpresSet}
\comment{used in Lemma~\ref{SCpresSet}; uses remark~(5)}
Etant donné un ensemble de termes SN $t_i$, $i=1,\ldots,n$, le terme
$t=\{t_1,\ldots,t_n\}$ est SN.
\FLEM

\proof{

Pour simplifier nous considérons que $i=1$ et donc $t=\{t_1\}$. Nous notons par
$\{u\}^k$ le terme $\{\ldots\{u\}\ldots\}$ avec $k$ symboles d'ensemble
imbriqués.

Nous utilisons une induction sur le nombre maximum d'étapes de réduction
du terme $t_1$.

\textit{Le cas de base}~: Le termes $t_1$ est complètement normalisé. Si
$t_1=\{u_1\}^0$ alors $t=\{u_1\}$ est en forme normale. Si $t_1=\{u_1\}^1$ alors
la seule réduction possible de $t$ est $\{t\}=\{\{u_1\}\} \lraD{Flat} \{u_1\}$
et puisque $\{u_1\}$ est en forme normale alors $\{t\}$ est SN.

\textit{Induction}~: $t_1$ se normalise en maximum $n>0$ étapes en $\overline{t_1}$.

Nous considérons que le terme $t_1$ est de la forme $\{u_1\}^k$ avec $u_1$ SN.
Nous avons deux réductions possibles~:
	$$\{t_1\}=\{\{u_1\}^k\}	\lraD{\rho} \{\{u_1'\}^k\}$$
avec $u_1 \Raro u_1'$ et $\{u_1'\}^k$ SN et normalisant en $(n-1)$ étapes ou $k
\geq 1$ et
	$$\{t_1\}=\{\{u_1\}^k\}	\lraD{Flat} \{\{u_1\}^{k-1}\}.$$
Si $k \geq 2$ alors $\{u_1\}^{k-1}$ est SN et normalisant en $(n-1)$ étapes.
Si $k=1$ alors la seule réduction possible est 
	$$\{t_1\}=\{\{u_1\}\} \lraD{Flat} \{u_1\}\lraD{\rho} \{u_1'\}$$
avec $u_1 \Raro u_1'$ et $u_1'$ SN et normalisant en $(n-1)$ étapes.

Par induction, $\{\{u_1'\}^k\}$ et $\{\{u_1\}^{k-1}\}$ sont SN et par
conséquent, $\{\{u_1\}^k\}$ est SN.
}%------------------------------------------------------------------------------

\LEM%------------------------------------------------------------------------------
\label{SCpresSet}
\comment{used in Lemma~\ref{SCsubs}, Lemma~\ref{allSC}; uses
Lemma~\ref{SNpresSet}, Lemma~\ref{SCisSN} remark~(2,4)}
Etant donnés un contexte $E$ et un ensemble de termes
$u_i$ tels que $E \vdash u_i:A$, $i=1,\ldots,n$. Si les
termes $u_i$ sont SC alors le terme $\{u_1,\ldots,u_n\}$ est SC.
\FLEM

\proof{

Tous les termes utilisés dans la preuve sont typés dans le contexte $E$ qui est
omis quand les types des termes sont donnés.

Si le type $A$ est atomique alors, par la Définition~\ref{SCterm},
$u_i:A$ sont SN et ainsi $\{u_1,\ldots,u_n\}:A$ est SN par le
Lemme~\ref{SNpresSet}. Puisque $\{u_1,\ldots,u_n\}:A$ est de
type atomique et SN alors, par la Définition~\ref{SCterm}, il est SC.

Si le type $A$ est de la forme $A_1 \raS \ldots A_m \raS K$ avec $K$ atomique
alors par la Remarque~\ref{remarksTypes}(\ref{remarksTypes:second}) nous devons
montrer que pour tous les termes SC $t_j$ tels que $E \vdash t_j:A_j$ nous avons
le terme $[\ldots[[\{u_1,\ldots,u_n\}](t_1)](t_2)\ldots](t_m):K$ qui est SN.

Puisque $t_j:A_j$ sont SC alors, par la
Remarque~\ref{remarksTypes}(\ref{remarksTypes:fourth}), les termes
$[\ldots[[u_i](t_1)](t_2)\ldots](t_m):K$ sont SC et puisqu'ils sont de type
atomique alors ils sont SN. Par conséquent, nous déduisons par le
Lemme~\ref{SNpresSet} que le terme
$\{[\ldots[[u_1](t_1)](t_2)\ldots](t_m),\ldots,[\ldots[[u_n](t_1)](t_2)\ldots](t_m)\}$
est SN.

Puisque $u_i$, $t_j$ sont SC alors, par le Lemme~\ref{SCisSN}, $u_i$, $t_j$ sont
SN pour $i=1,\ldots,n$, $j=1,\ldots,m$.  Par conséquent, si le terme
$[\ldots[[\{u_1,\ldots,u_n\}](t_1)](t_2)\ldots](t_m)$ n'était pas SN alors une
réduction infinie n'est pas due seulement à ses sous-termes et aurait la
forme~:\\
$[\ldots[[\{u_1,\ldots,u_n\}](t_1)](t_2)\ldots](t_m)$ \\
$~~~~$ $\longra{*}_{\rho}~~~~~~~~$
$[\ldots[[\{u_1',\ldots,u_n'\}](t_1')](t_2')\ldots](t_m')$ \\
$~~~~$ $\longra{}_{Distrib}~~$
$\{[\ldots[[u_1'](t_1')](t_2')\ldots](t_m'),\ldots,[\ldots[[u_n'](t_1')](t_2')\ldots](t_m')\}$ \\
$~~~~$ $\longra{*}_{\rho}~~~~~~~~$ $\ldots$

Puisque nous avons la réduction \\
$\{[\ldots[[u_1](t_1)](t_2)\ldots](t_m),\ldots,[\ldots[[u_n](t_1)](t_2)\ldots](t_m)\}$\\
$~~~~$ $\longra{*}_{\rho}~~$
$\{[\ldots[[u_1'](t_1')](t_2')\ldots](t_m'),\ldots,[\ldots[[u_n'](t_1')](t_2')\ldots](t_m')\}$\\
alors le terme
$\{[\ldots[[u_1](t_1)](t_2)\ldots](t_m),\ldots,[\ldots[[u_n](t_1)](t_2)\ldots](t_m)\}$
n'est pas SN, ce qui contredit le résultat obtenu ci-dessus.

Par conséquent, $[\ldots[[\{u_1,\ldots,u_n\}](t_1)](t_2)\ldots](t_m)$ est SN et
selon la Remarque~\ref{remarksTypes}(\ref{remarksTypes:second}), nous obtenons
que $\{u_1,\ldots,u_n\}$ est SC.
}%------------------------------------------------------------------------------

\LEM%------------------------------------------------------------------------------
\label{SNpresFun}
\comment{used in Lemma~\ref{SCpresFun}; uses remark~(5)}
Etant donné un ensemble de termes SN $t_i$, $i=1,\ldots,n$,
le terme $t=f(t_1,\ldots,t_n)$ est SN.
\FLEM

\proof{
La preuve est très similaire à celle du Lemme~\ref{SNpresSet}.
}%------------------------------------------------------------------------------

\LEM%------------------------------------------------------------------------------
\label{SCpresFun}
\comment{used in Lemma~\ref{allSC}; uses Lemma~\ref{SCisSN}, uses
Lemma~\ref{SNpresFun}, remark~(2,4)}
Etant donnés un contexte $E$, les termes $u_i$ tels que $E \vdash u_i:A_i$ pour
$i=1,\ldots,n$ et un symbole de fonction 
$f \in \FF_{A_1 \times \ldots \times A_n \raS A}$, alors le terme
$f(u_1,\ldots,u_n)$ est SC ssi les termes $u_i$ sont SC.
\FLEM

\proof{
Tous les termes utilisés dans la preuve sont typés dans le contexte $E$ qui est
omis par la suite.%

Les symboles $f \in \FF$ sont symboles du premier ordre et ne peuvent pas mener
à des termes de la forme $f(u_1,\ldots,u_n)$ avec un type \textit{plus grand}
que le type de leurs arguments.  Ici, \textit{plus grand} est utilisé dans le
sens d'un nombre \textit{plus grand} de symboles ``$\raS$'' dans le type.

Nous procédons par induction sur le nombre de flèches de type ``$\raS$'' dans les types 
$A_i$~:

\textit{Le cas de base}~: $A_i$, $i=1,\ldots,n$ et $A$ sont des types atomiques.

Si les types $A_i$ sont atomiques et les $u_i$ sont SC alors les $u_i$ sont SN
par le Lemme~\ref{SCisSN} et ainsi $f(u_1,\ldots,u_n):A$ est SN par le
Lemme~\ref{SNpresFun}. Puisque $f(u_1,\ldots,u_n):A$ est de type atomique et SN
alors il est SC.

Si $f(u_1,\ldots,u_n):A$ est SC alors, par le Lemme~\ref{SCisSN}, il est SN et
ainsi tous ses sous-termes $u_i$ sont SN.  Puisque $A_i$ sont des types
atomiques alors $u_i$ sont SC.

\textit{Induction}~: $A_i$ de la forme $B_i \raS C_i$,
$i=1,\ldots,n$ et $A$ de la forme $B \raS C$.

Puisque $f \in \FF_{(B_1 \raS C_1) \times \ldots \times (B_n \raS C_n) \raS (B \raS C)}$
nous pouvons considérer %
que nous avons également
$f \in \FF_{B_1 \times \ldots \times B_n \raS B}$ et
$f \in \FF_{C_1 \times \ldots \times C_n \raS C}$.
\begin{itemize}

\item[($\Leftarrow$)] si $u_i:A_i$, $i=1,\ldots,n$ sont SC alors $f(u_1,\ldots,u_n):A$ est SC

Afin de prouver que $f(u_1,\ldots,u_n)$ est SC, conformément à la
Remarque~\ref{remarksTypes}(\ref{remarksTypes:second}), nous devons montrer que
pour tout terme SC $t:B$ nous avons $[f(u_1,\ldots,u_n)](t):C$ qui est SN.

Pour tous les termes $t$ tels que $[f(u_1,\ldots,u_n)](t)$ est réduit dans une
étape à $\emptyset$ la propriété est évidente.  Le seul terme $t$ qui ne mène
pas à un tel résultat doit être de la forme $f(t_1,\ldots,t_n)$.

Puisque $t:B$ est SC alors, par hypothèse d'induction, $t_i:B_i$ sont SC. Ainsi,
par la Remarque~\ref{remarksTypes}(\ref{remarksTypes:fourth}), $[u_i](t_i):C_i$
sont SC et donc SN. Par conséquent, par le Lemme~\ref{SNpresFun},
$f([u_1](t_1),\ldots,[u_n](t_n))$ est SN. De plus, par le Lemme~\ref{SCisSN},
les termes $u_i$ et $t_i$, $i=1,\ldots,n$ sont SN.

Supposons que $[f(u_1,\ldots,u_n)](t):C$ n'est pas SN.  Puisque les termes
$u_i$, $t_i$, $i=1,\ldots,n$ sont SN alors, une réduction infinie aurait la
forme~:

$[f(u_1,\ldots,u_n)](f(t_1,\ldots,t_n))$ \\
$~~~~$ $\longra{*}_{\rho}~~~~~~~~~~~~~$
$[f(u_1',\ldots,u_n')](f(t_1',\ldots,t_n'))$ \\
$~~~~$ $\longra{}_{Congruence}~~$
$f([u_1'](t_1'),\ldots,[u_n'](t_n'))$ \\
$~~~~$ $\longra{*}_{\rho}~~~~~~~~~~~~~$ $\ldots$

Puisque nous avons la réduction\\
$~~~~$ $f([u_1](t_1),\ldots,[u_n](t_n)) \Raro f([u_1'](t_1'),\ldots,[u_n'](t_n'))$ \\
alors $f([u_1](t_1),\ldots,[u_n](t_n))$ n'est pas SN et nous obtenons donc une
contradiction.

Par conséquent, $[f(u_1,\ldots,u_n)](t)$ est SN et selon la
Remarque~\ref{remarksTypes}(\ref{remarksTypes:second}), $f(u_1,\ldots,u_n)$ est SC.

\item[($\Rightarrow$)] si $f(u_1,\ldots,u_n):A$ est SC alors $u_i:A_i$, $i=1,\ldots,n$ sont SC

Afin de prouver que $u_i$, $i=1,\ldots,n$ sont SC, par la
Remarque~\ref{remarksTypes}(\ref{remarksTypes:second}), nous devons montrer que
pour tous les termes SC $t_i$ nous avons $[u_i](t_i):C_i$ qui est SN.

Par hypothèse d'induction, $f(t_1,\ldots,t_n):B$ est SC si $t_i:B_i$ sont
SC. Nous avons par la Remarque~\ref{remarksTypes}(\ref{remarksTypes:fourth}) que
$[f(u_1,\ldots,u_n)](f(t_1,\ldots,t_n)):C$ est SC et donc SN.

Si nous supposons qu'un des $[u_i](t_i)$ n'est pas SN alors nous obtenons que le 
terme  $f([u_1](t_1),\ldots,[u_n](t_n))$ n'est pas SN.
Puisque nous avons la réduction

$[f(u_1,\ldots,u_n)](f(t_1,\ldots,t_n))$ \\
$~~~~$ $\longra{}_{Congruence}~~$ $f([u_1](t_1),\ldots,[u_n](t_n))$ \\
alors 
$[f(u_1,\ldots,u_n)](f(t_1,\ldots,t_n))$ 
n'est pas SN et nous obtenons donc une contradiction.

Par conséquent, $[u_i](t_i)$, $i=1,\ldots,n$ sont SN et d'après la
Remarque~\ref{remarksTypes}(\ref{remarksTypes:second}), $u_i$, $i=1,\ldots,n$
sont SC.

\end{itemize}
}%------------------------------------------------------------------------------

\LEM%------------------------------------------------------------------------------
\label{SCsubs} 
\comment{used in Lemma~\ref{allSC}; uses Lemma~\ref{subsTypes},
Lemma~\ref{SCpresSet}, Lemma~\ref{SCisSN}, Lemma~\ref{substRed}, remark (2,5)}
Etant donné les contextes $E$ et $F$ et les termes
$l,r$ et $t$ tels que  $\rest{F}{l} \dotctx E \vdash r:B$, 
$F \vdash l:A$ et $E \vdash t:A$, nous considérons la substitution typée
$\sigma$ telle que $\{\sigma\}=\Sl(l \meqqes t)$.

Si les termes $\sigma r$ et $l,t$ sont SC, alors le terme
$[\ofenv{l}{\rest{F}{l}} \ra r](t)$ est SC.
\FLEM

\proof{
Conformément à l'algorithme de filtrage, nous avons $E \vdash \sigma$.
Puisque $Dom(\sigma)=\rest{F}{l}$ alors,
par le Lemme~\ref{subsTypes}, $E \vdash \sigma r:B$ et puisque $\sigma r$ est SC alors,
par le Lemme~\ref{SCpresSet}, $\{\sigma r\}:B$ est SC.

Nous considérons $B = B_1 \raS \ldots B_l \raS K$, avec $K$
atomique.  Nous considérons les termes SC $v_1:B_1,\ldots,v_l:B_l$.
Selon la Remarque~\ref{remarksTypes}(\ref{remarksTypes:second}),
\begin{equation}
\label{SCsubs:first1} 
[\ldots[[\sigma r](v_1)](v_2)\ldots](v_l):K, 
\end{equation}
est SN et
\begin{equation}
\label{SCsubs:first2} 
[\ldots[[\{\sigma r\}](v_1)](v_2)\ldots](v_l):K
\end{equation}
est SN.

Nous avons $E \vdash [\ofenv{l}{\rest{F}{l}} \ra r](t):B$ et
nous devons montrer que
\begin{equation}
\label{SCsubs:second} 
[\ldots[[[\ofenv{l}{\rest{F}{l}} \ra r](t)](v_1)](v_2)\ldots](v_l):K
\end{equation}
est SN et par la Remarque~\ref{remarksTypes}(\ref{remarksTypes:second}), nous
pouvons conclure la preuve du lemme.

Puisque (\ref{SCsubs:first1}) est SN, d'après la
Remarque~\ref{remarksTypes}(\ref{remarksTypes:fifth}), tous ses sous-termes sont
SN. Ainsi, les termes $\sigma r$ et $v_1, v_2,\ldots, v_l$ sont SN.  Puisque $l$
et $t$ sont SC alors, par le Lemme~\ref{SCisSN}, $l$ et $t$ sont SN.  Nous avons
immédiatement que $r$ est SN.

Par conséquent, une réduction infinie de (\ref{SCsubs:second}) ne peut pas
consister entièrement en contractions dans $l, r, t, v_1, \ldots, v_l$ et une
telle réduction doit avoir la forme~:

\begin{tabular}{lll}

$[\ldots[[[\ofenv{l}{\rest{F}{l}} \ra r](t)](v_1)](v_2)\ldots](v_l)$ 
      & $\longra{*}_{\rho}$ & $[\ldots[[[\ofenv{l}{\rest{F}{l}} \ra r'](t')](v_1')](v_2')\ldots](v_l')$\\

      & $\longra{}_{Fire}$ & $[\ldots[[\{\sigma' r'\}](v_1')](v_2')\ldots](v_l')$\\

      & $\longra{*}_{\rho}$ & $\ldots$ \\

\end{tabular}\\
où $r \longra{*}_{\rho} r'$, $t \longra{*}_{\rho} t'$, 
$v_k \longra{*}_{\rho} v_k'$ et le filtrage $l \meqqes t'$ n'échoue pas et a la
solution $\sigma'$.

Puisque $l \in \TFX$ et que nous utilisons le filtrage syntaxique alors,
pour tous les termes $l$, $t$ et $t'$ tels que $t \Raro t'$ et 
$\Sl(l \meqqes t)=\{\mu\}$ avec $\mu=\subs{x_1/t_1,\ldots,x_n/t_n}$, si 
$l \meqqes t'$ n'échoue pas, alors nous avons $\Sl(l \meqqes t')=\{\mu'\}$ avec
$\mu'=\subs{x_1/{t_1}',\ldots,x_n/{t_n}'}$ et $t_i \Raro {t_i}'$, $i=1,\ldots,n$.

Selon cette dernière remarque, si $\sigma = \subs{x_1/u_1,\ldots,x_n/u_n}$ alors
$\sigma' = \subs{x_1/{u_1}',\ldots,x_n/{u_n}'}$ avec $u_i \Raro {u_i}'$ et par le
Lemme~\ref{substRed} nous obtenons $\sigma r \Raro \sigma' r'$.

Ainsi, nous pouvons construire une réduction infinie à partir de~(\ref{SCsubs:first2})~:

\begin{tabular}{lll}
$[\ldots[[\{\sigma r\}](v_1)](v_2)\ldots](v_l)$ & $\longra{*}_{\rho}$ &
$[\ldots[[\{\sigma' r'\}](v_1')](v_2')\ldots](v_l')$ \\
      & $\longra{*}_{\rho}$ & $\ldots$ \\
\end{tabular}

Ceci contredit le fait que~(\ref{SCsubs:first2}) est SN et
par conséquent, (\ref{SCsubs:second}) doit être SN.

}%------------------------------------------------------------------------------

\LEM%------------------------------------------------------------------------------
\label{allSC}  (tout terme typable est SC)
\comment{used in Proposition~\ref{strongNormTh}; uses Lemma~\ref{SCisSN},
Lemma~\ref{SCpresFun}, Lemma~\ref{SCpresSet}, Lemma~\ref{allSC},
Lemma~\ref{SCsubs}, remark~(4)}

Pour tout $\rho$-terme typé $t$ tel que $E \vdash t:B$ nous avons~:
\begin{itemize}
\item[a.] $t$ est SC,

\item[b.] Pour toutes les substitutions SC
$\sigma_j$, $j=1,\ldots,m$ telles que $E' \vdash \sigma_j$, le terme
$t^*=\sigma_1 \ldots \sigma_m t$ avec $E' \vdash t^*:B$ est SC,
où $E=Dom(\sigma_1) \dotctx \ldots \dotctx Dom(\sigma_m) \dotctx E'$.

\end{itemize} 

\FLEM

\proof{

Le point (a) est un cas spécial de (b) quand $\sigma_i$, $i=1,\ldots,n$ est la
substitution identité. 

Nous prouvons (b) par induction structurelle sur $t$.
\begin{enumerate}

\item
%\item[{\it Case 1:\  \  }]
$t$ est une variable $x_i^j \in Dom(\sigma_j)$,
et $\sigma_j=\subs{x_1^j/u_1^j,\ldots,x_n^j/u_n^j}$.

Dans ce cas $t^*$ est $u_i^j$ et donc, est SC.

\item
%\item[{\it Case 2:\  \  }]
$t$ est une variable distincte de $x \in \bigcup Dom(\sigma_j)$.

Dans ce cas $t^*$ est $t$ et donc, il est SC par le Lemme~\ref{SCisSN} avec
$n=0$.

\item
%\item[{\it Case 3:\  \  }]
$t = f(t_1, \ldots, t_k)$.

Dans ce cas $t^* = f(t_1^*, \ldots, t_k^*)$. Par hypothèse d'induction,
$t_1^*, \ldots, t_k^*$ sont SC, et ainsi, d'après le Lemme~\ref{SCpresFun}, $t^*$
est SC.

\item
%\item[{\it Case 4:\  \  }]
$t = \{t_1, \ldots, t_k\}$.

Dans ce cas $t^* = \{t_1^*, \ldots, t_k^*\}$. Par hypothèse d'induction,
$t_1^*, \ldots, t_k^*$ sont SC, et ainsi, d'après le Lemme~\ref{SCpresSet}, $t^*$
est SC.

\item
%\item[{\it Case 5:\  \  }]
$t = [t_1](t_2)$.

Dans ce cas $t^* = [t_1^*](t_2^*)$. Par hypothèse d'induction, $t_1^*, t_2^*$
sont SC, et ainsi, conformément à la
Remarque~\ref{remarksTypes}(\ref{remarksTypes:fourth}), $t^*$ est SC.

\item
%\item[{\it Case 6:\  \  }]
$t:B = \ofenv{t_1}{\rest{F}{t_1}} \ra t_2$.

Nous avons $F \vdash t_1:A$ et $\rest{F}{t_1} \dotctx E \vdash t_2:C$ et 
$B=A \raS C$.

Dans ce cas $t^* = t_1^* \ra t_2^*$ si nous négligeons les changements de
noms des variables liées. Conformément à la définition de l'application de
substitution, $t_1^*=t_1$.

Nous devons montrer que pour tous les termes SC $u$ tels que $E' \vdash u:A$, le
terme $[t^*](u)$ est SC ou d'une manière équivalente que $[t_1 \ra t_2^*](u)$
est SC.

Si le filtrage $(t_1 \meqqes u)$ échoue alors le résultat est
$\emptyset$ et la propriété est évidemment vraie.

On doit noter que si $l \in \TFX$ et si le filtrage syntaxique est
considéré alors, par le Lemme~\ref{SCpresFun}, pour tous termes SC $l$ et $t$
et la substitution $\{\mu\}=\Sl(l \meqqes t)$ nous obtenons que $\mu$ est SC.

Nous considérons $\{\mu\} = \Sl(t_1 \meqqes u)$ et conformément à l'algorithme
de filtrage, $E' \vdash \mu$ et puisque $t_1 \in \TFX$ et $u$ sont SC alors 
$\mu$ est SC. 
Par le Lemme~\ref{subsTypes}, $\rest{F}{t_1} \dotctx E' \vdash t_2^*:C$ et
puisque $Dom(\mu)=\rest{F}{t_1}$ alors $E' \vdash \mu t_2^*:C$.  Par l'hypothèse
d'induction (b) 
en utilisant la substitution $\mu \sigma_1 \ldots \sigma_m$
nous obtenons que $\mu t_2^*$
est SC et ainsi, par le Lemme~\ref{SCsubs}, $[t^*](u)$ est SC.

\end{enumerate} 

}%------------------------------------------------------------------------------

\TH%------------------------------------------------------------------------------
\label{strongNormTh} 
\comment{uses Lemma~\ref{allSC},Lemma~\ref{SCisSN}}
Le \roCalE\  est fortement normalisable.
\FTH

\proof{
Le résultat est obtenu immédiatement par les Lemmes~\ref{allSC} et~\ref{SCisSN}.
}

\section{Le typage des $\rho$-termes non-restreints} \label{typeGeneral}
%================================================================

Nous avons proposé dans les sections précédentes un système de types pour les
termes de $\RTTE$ et nous allons étendre ce système pour le cas général de
termes de $\RTT$. Nous considérons le calcul $(\RTT,\emptyset,\SS)$, appelé
aussi \roCalEp, généralisant le \roCalE\  avec une stratégie $\SS$ arbitraire.

\subsubsection{La syntaxe du \roCalEp\  typé}
%================================================================

Nous obtenons donc une nouvelle syntaxe pour le \roCalEp\  typé généralisant
celle du \roCalE\  typé présentée dans la Section~\ref{syntaxeType}~:

\DEF%------------------------------------------------------------------------------
\label{syntaxeRoTypeGen}
Etant donnés un ensemble de variables $\XX$ et un ensemble de symboles
$\FF=\bigcup_{i \geq 0} \FF_i$. Si nous notons $K$ tout type atomique, alors la
syntaxe du calcul $(\RTT,\emptyset,\SS)$ simplement typé est définie
récursivement par la grammaire suivante~:

\begin{tabular}{llll}
& \\
\textBNF{Types} ~~~~ & 
$T$ &::=&  $K ~~|~~ T \raS T $\\
& \\
\textBNF{Contextes} ~~~~ & 
$E$ &::=&  $x:T ~~|~~ E \dotctx \ldots \dotctx E $\\
& \\
\textBNF{Termes} ~~~~ & 
$t$ &::=& $x ~~|~~ f(t,\ldots,t) ~~|~~ \{t,\ldots,t\} ~~|~~ \ofenv{t}{E} \ra t ~~|~~[t](t)$\\
& \\
\end{tabular}\\
où $x \in \XX$ et $f \in \FF$.
\FDEF%------------------------------------------------------------------------------

La seule différence par rapport à la Définition~\ref{syntaxeRoType} est
l'élimination de la condition que le membre gauche des règles de réécriture
est un terme du premier ordre et nous analysons l'influence de cette nouvelle
syntaxe sur les règles de typage et d'évaluation du calcul.

\subsubsection{Discussion sur les règles du \roCalEp}
%=============================================================================== 
%

Puisque nous considérons tous les termes de $\RTT$ et donc les règles de
réécriture ayant un ensemble dans le membre gauche, nous devons introduire et
analyser le comportement de la règle d'évaluation \rname{Switch_L} qui devient
dans un cadre typé~:
\renewcommand{\fleche}{\Longrightarrow}
\begin{ruleset}
%=================================== 
\regle {Switch_L}
	{\ofenv{\{u_1,\ldots,u_n\}}{E} \ra v}
	{\{\ofenv{u_1}{E} \ra v,\ldots,\ofenv{u_n}{E} \ra v\}}
%=================================== 
\end{ruleset}
\vspace{-.5cm}

Les autres règles d'évaluation sont celles présentées dans la
Figure~\ref{MRAtype} et puisque nous souhaitons avoir un calcul confluent, nous
considérons une des stratégies d'évaluation confluentes proposées dans le
Chapitre~\ref{chap.resultats_confluence}. La règle d'évaluation
\rname{Fire} est donc appliquée pour une application de la forme $[l \ra r](t)$
seulement si $l$ est un terme du premier ordre et ainsi, le même mécanisme de
filtrage que dans le \roCalE\  peut être utilisé.

Nous analysons maintenant la possibilité d'employer le même ensemble de règles
de typage que pour le \roCalE. 

Nous devons mentionner d'abord que dans la règle de typage \rname{Rule} il est
essentiel que seule la restriction $\rest{E}{l}$ du contexte $E$ de $l$ soit
éliminée du contexte $F$ de $r$ afin d'obtenir le contexte pour la règle de
réécriture $l \ra r$. Néanmoins, le contexte local de cette règle de réécriture
peut être $\rest{E}{l}$, comme imposé par la règle de typage \rname{Rule},
mais il peut être un tout autre contexte incluant ce contexte, comme dans la
nouvelle règle de typage \\
$$
Rule_s ~~~
  \FRAC{E \vdash l:A ~~~~ \rest{E}{l} \dotctx F \vdash r:B}
  %----------------------------------------------------------------------
  {F \vdash (\ofenv{l}{E} \ra r) : A \raS B}
$$

Les mêmes résultats sur la préservation du type et sur la normalisation forte du
calcul sont obtenus si cette dernière règle de typage est utilisée à la place de
la règle de typage \rname{Rule}.

Malheureusement, si nous utilisons une approche basée sur la
règle de typage \rname{Rule} il est clair que la règle d'évaluation
\rname{Switch_L} ne préserve pas le type.

Considérons, par exemple, le terme $\ofenv{\{x,y\}}{x:A \dotctx y:A} \ra x$ qui
a le type $A \raS A$ dans le contexte vide conformément à la règle de typage
\rname{Rule}. Ce terme est réduit par la règle d'évaluation \rname{Switch_L} en
$\{\ofenv{x}{x:A \dotctx y:A} \ra x,\ofenv{y}{x:A \dotctx y:A} \ra x\}$ et
aucune des deux règles de réécriture dans l'ensemble ne peut être typée par
la règle \rname{Rule} en raison du contexte $x:A \dotctx y:A$ qui devrait
contenir seulement la variable $x$ et respectivement $y$.

L'utilisation de la règle de typage \rname{Rule_s} nous permet d'éliminer cette
restriction et dans ce cas nous pouvons typer la règle de réécriture
$\ofenv{x}{x:A \dotctx y:A} \ra x$ dans le contexte vide. Mais le membre gauche
de la deuxième règle de réécriture de l'ensemble précédent est la variable $y$
et donc le contexte nous permettant de typer cette règle est obtenu en enlevant
la variable $y$ du contexte utilisé pour typer le membre droit $x$. Ainsi, nous
pouvons typer la règle de réécriture $\ofenv{y}{x:A \dotctx y:A} \ra x$
dans tout contexte contenant $x:A$ mais pas dans le contexte vide et donc même
en utilisant la règle de typage \rname{Rule_s} le type n'est pas préservé.

\subsubsection{Les règles de typage du \roCalEp}
%=============================================================================== 
%

La non-préservation du type dans une approche utilisant les règles présentées
dans la section précédente est due à la non-préservation des variables libres
par la réduction. La solution naturelle est l'utilisation de la notion de
variables présentes introduite dans la Définition~\ref{varPresentes}.

Nous modifions ainsi la règle de typage \rname{Rule} et nous introduisons la
règle~:
$$
Rule_{pv} ~~~
  \FRAC{E \vdash l:A ~~~~ \restpv{E}{l} \dotctx F \vdash r:B}
  %----------------------------------------------------------------------
  {F \vdash (\ofenv{l}{E} \ra r) : A \raS B}
$$

Cette fois-ci nous n'avons pas imposé la restriction $\restpv{E}{l}$ du contexte
$E$ dans la règle de réécriture et ceci nous permet l'utilisation de la règle
d'évaluation \rname{Switch_L} distribuant le même contexte local dans toutes les
règles de réécriture.

Nous considérons ainsi l'ensemble de règles de typage de la
Figure~\ref{typesRoVdash} où la règle \rname{Rule} est remplacée par la règle
\rname{Rule_{pv}} et nous notons par $ST_{\rho}^{+}$ le système de types ainsi
obtenu.  Nous montrons qu'en utilisant la même notion de terme \Defi{bien
typé}{terme} que pour le \roCalE\  (Définition~\ref{typeableDef}), le type est
préservé par toutes les règles d'évaluation du calcul $(\RTT,\emptyset,\SS)$.

\subsubsection{La préservation du type}
%=============================================================================== 
%

Si nous reprenons l'exemple de la règle de réécriture 
$\ofenv{\{x,y\}}{x:A \dotctx y:A} \ra x$ nous pouvons remarquer que ce
$\rho$-terme n'est pas bien typé dans le contexte vide puisque 
$\restpv{x:A \dotctx y:A}{\{x,y\}}=\emptyset$. Ainsi, le contexte $F$ de la
règle de typage \rname{Rule_{pv}} est instancié en $x:A$ qui est donc le contexte
nous permettant d'inférer le type $A \raS A$ pour la règle de réécriture 
$\ofenv{\{x,y\}}{x:A \dotctx y:A} \ra x$. Comme
nous l'avons déjà précisé, ce dernier contexte nous permet aussi d'obtenir le
type $A \raS A$ pour les deux règles de réécriture 
$\ofenv{x}{x:A \dotctx y:A} \ra x$ et $\ofenv{y}{x:A \dotctx y:A} \ra x$ et
donc, le type est préservé en utilisant la règle de typage \rname{Rule_{pv}}.

\TH%----------------------------------------------------------
\label{subjRedGen}
Dans le système de types $ST_{\rho}^{+}$, pour tous $\rho$-termes $a$ et $a'$ de
$\RTT$, si $a \Raro a'$ et $E \vdash a:A$, alors $E \vdash a':A$.
\FTH%----------------------------------------------------------

\proof{ 

Nous procédons de la même manière que dans le cas du \roCalE\  et nous prouvons
que le membre gauche et le membre droit de chaque règle d'évaluation ont le même
type dans un contexte donné.

Pour toutes les règles d'évaluation sauf \rname{Switch_L} le résultat est obtenu 
exactement comme dans le Théorème~\ref{subjRed}.

Pour la règle d'évaluation \rname{Switch_L} nous procédons similairement en
utilisant la définition des variables présentes.

\begin{ruleset}
%===================================
\regle {Switch_L}
	{\ofenv{\{u_1,\ldots,u_n\}}{E} \ra v}
	{\{\ofenv{u_1}{E} \ra v,\ldots,\ofenv{u_n}{E} \ra v\}}
%===================================
\end{ruleset}
\vspace{-.5cm}

Nous notons $u=\{u_1,\ldots,u_n\}$ et nous considérons le type $A$ tel que 
$F \vdash \ofenv{u}{E} \ra v:A$.  
En utilisant la règle de typage \rname{Rule_{pv}} nous inférons $A=B \raS C$ et 
$\restpv{E}{u} \dotctx F \vdash v:C$, $E \vdash u:B$.

Par la règle de typage \rname{Set} nous avons $E \vdash u_i:B$,
$i=1,\ldots,n$ et par la règle de typage \rname{Rule_{pv}} appliquée $n$ fois nous
obtenons $F_i \vdash \ofenv{u_i}{E} \ra v:B \raS C$, $i=1,\ldots,n$,
avec $\restpv{E}{u} \dotctx F = \restpv{E}{u_i} \dotctx F_i$.

Conformément à la Définition~\ref{varPresentes} nous avons 
$PV(\{u_1,\ldots,u_n\}) \subseteq PV(u_i)$ et donc,  nous obtenons
$\restpv{E}{u} \subseteq \restpv{E}{u_i}$ et $F_i \subseteq F$. Conformément à la
Définition~\ref{typeableDef} nous avons
$F \vdash \ofenv{u_i}{E} \ra v:B \raS C$, $i=1,\ldots,n$.

Finalement, la règle de typage \rname{Set} mène à~: \\
$~~~~$ $F \vdash \{\ofenv{u_1}{E} \ra v,\ldots,\ofenv{u_n}{E} \ra v\}:A$.

}%--------------------------------------------------------------------------------------

Nous pouvons imposer la restriction aux variables présentes pour le contexte
local dans la règle de typage \rname{Rule_{pv}} qui deviendrait~:
$$
Rule_{pv}' ~~~
  \FRAC{E \vdash l:A ~~~~ \restpv{E}{l} \dotctx F \vdash r:B}
  %----------------------------------------------------------------------
  {F \vdash (\ofenv{l}{\restpv{E}{l}} \ra r) : A \raS B}
$$

Comme nous l'avons expliqué au début de cette section une telle règle de typage
n'est pas suffisante pour assurer la préservation du type dans le cas de la
règle d'évaluation \rname{Switch_L} mais nous pourrions introduire une nouvelle
règle

\renewcommand{\fleche}{\Longrightarrow}
\begin{ruleset}
%=================================== 
\regle {Switch_L'}
	{\ofenv{\{u_1,\ldots,u_n\}}{E} \ra v}
	{\{\ofenv{u_1}{\restpv{E}{u_1}} \ra v,\ldots,\ofenv{u_n}{\restpv{E}{u_n}} \ra v\}}
%=================================== 
\end{ruleset}

Cette règle d'évaluation est plus complexe que la règle \rname{Switch_L} et donc
plus difficile à implanter mais peut être employée si nous souhaitons avoir une
information plus précise sur les variables utilisées dans le contexte local des
règles de réécriture.

La normalisation forte du calcul $(\RTT,\emptyset,\SS)$ typé avec $\SS$
représentant une des stratégies confluentes est montrée exactement de la même
façon que pour le \roCalE\  et nous obtenons ainsi plusieurs instances du calcul
$(\RTT,\emptyset,\SS)$ garantissant la terminaison et l'unicité des formes normales.

%\DontWriteThisInToc  
\subsection*{Conclusion}
%================================================================
%~

Nous avons proposé un système de types pour le \roCalE\  en précisant la syntaxe, 
les règles de typage et les règles d'évaluation.

Nous avons adapté la syntaxe du \roCal\  non-typé en ajoutant les types et les
contextes ainsi qu'en introduisant les règles de réécriture avec contexte local.
Ce contexte correspond à l'information de type donnée dans une
$\lambda$-abstraction et il est employé pour déterminer le contexte global
permettant de typer la règle de réécriture l'utilisant.
Les règles d'évaluation du calcul non-typé sont modifiées afin de prendre en
compte la nouvelle syntaxe et particulièrement le contexte local des règles de
réécriture.
En se limitant à des ensembles ayant tous les éléments d'un même type et avec un
bon choix pour les règles de typage, le \roCalE\  typé est terminant et préserve
le type.

Ces résultats ont été présentés à WRLA2000~\cite{CirsteaKirchnerWRLA2000}.

%% file: chapter_7.tex
%%%%%%%%%%%%%%%%%%%%%%%%%%%%%%%%%%%%%%%%%%%%%%%%%%%%%%%%%%%
% \TLtopbookmark
\chapter{Le calcul de réécriture avec substitutions explicites - le \roSig}
\label{chap.calcul_explicite}
%%%%%%%%%%%%%%%%%%%%%%%%%%%%%%%%%%%%%%%%%%%%%%%%%%%%%%%%%%%%

Dans le \roCal\  tel que nous l'avons présenté jusqu'ici l'application des
substitutions n'était pas une partie du calcul mais était décrite à son
méta-niveau.  Afin de rendre explicite l'application de substitution, nous utilisons
une approche similaire aux différentes versions de \laCal\  avec substitutions
explicites~\cite{HardinLevy89,ACCL90} et nous définissons une version explicite
du \roCal, appelée le \roSig.

Nous étendons la syntaxe du \roCal\  en introduisant les définitions des
substitutions et un opérateur d'application de substitution.
Nous nous limitons dans cette présentation à une version avec substitutions
explicites du \roCalE\  où les membres gauches des règles de réécriture sont des
termes du premier ordre.  Les règles d'évaluation du \roCalE\  ne sont pas
modifiées mais des nouvelles règles sont introduites pour décrire le
comportement des substitutions. Le calcul de substitution et la preuve de
confluence du \roSig\  sont basés sur les approches utilisées
dans~\cite{CurienHardinLevy-JACM96} et~\cite{PaganoThesis}. 

\section{La syntaxe}
%=============================================================================== 

La présentation du \roSig\  est basée sur une notation de de Bruijn
(\cite{deBruijn72}) où les noms des variables sont remplacés par des entiers
représentant le ``degré'' d'abstraction de la variable. Ainsi,
l'$\alpha$-conversion utilisée dans le \roCal\  pour éviter la capture des
variables n'est plus nécessaire, le renommage étant décrit par des
incrémentations et des décrémentations des entiers représentant les variables.

Nous notons par $\TFN$ l'ensemble de termes construits en utilisant une
signature $\FF$ et l'ensemble d'entiers naturels non nuls $\NatP$ de la même
manière que les termes de $\TFX$ sont construits en utilisant les variables d'un
ensemble $\XX$.

\DEF%------------------------------------------------------------------------
\label{rhoTermesSigma} 
Etant donnés un ensemble de variables $\XX=\XX_t \cup \XX_s$ et un ensemble de symboles
$\FF=\bigcup_{i \geq 0} \FF_i$ tel que pour tout $m$, $\FF_m$ est le
sous-ensemble de symboles d'arité $m$.  L'ensemble de termes et substitutions du
\roSig, noté $\RTTs$, est le plus petit ensemble tel que~:
\begin{itemize}
\item les variables de $\XX_t$ sont des $\rho$-termes (méta-variables de termes),
\item $n$ ($n \in \NatP$) est un terme (indice de de Bruijn),
\item si $t_1,\ldots,t_m$ sont des termes alors $\{t_1,\ldots,t_m\}$ est un
	terme,
\item si $t_1,\ldots,t_m$ sont des termes et $f \in \FF_m$ alors
	$f(t_1,\ldots,t_m)$ est un terme,
\item si $t$ et $u$ sont des termes alors [$t$]($u$) est un terme,
\item si $t \in \TFN$ et $u$ sont des termes alors 
	$t \ra_n u$ est un terme où $n$ est l'indice maximal dans $t$,
\item si $t$ est un terme et $s$ est une substitution alors $t\subs{s}$ est un
	terme (application de substitution),
\end{itemize}
et
\begin{itemize}
\item les variables de $\XX_s$ sont des $\rho$-termes (méta-variables de substitutions),
\item $\ID$ est une substitution (identité),
\item $\uparrow$ est une substitution ($shift$),
\item si $s$ est une substitution  $\Uparrow(s)$ est une substitution ($lift$),
\item si $t$ est un terme et $s$ est une substitution alors ($t . s$) est une
	substitution,
\item si $s_1$ et $s_2$ sont des substitutions alors ($s_1 \circ s_2$) est une
	substitution.
\end{itemize}
\FDEF%------------------------------------------------------------------------

La syntaxe du \roSig\  est donc définie récursivement par la grammaire suivante~:

\begin{tabular}{llll}
& \\
\textBNF{Termes} ~~~~ & 
$t$ &::=& 
$x_t ~~|~~ n ~~|~~ f(t,\ldots,t) ~~|~~ \{t,\ldots,t\} ~~|~~ u \ra_m t ~~|~~ [t](t) ~~|~~ t\subs{s}$\\
& \\
\textBNF{Substitutions} ~~~~ & 
$s$ &::=& 
$x_s ~~|~~ \ID ~~|~~ \uparrow ~~|~~ \Uparrow(s) ~~|~~ t.s ~~|~~ s \circ s$\\
& \\
\end{tabular}\\
où $x_t\in\XX_t$, $x_s\in\XX_s$, $n \in \NatP$, $u \in \TFN$, $m$ est l'indice
maximal dans $u$ et $f \in \FF$.

Nous abrégeons par $\uparrow^n$ la composition de $n$ symboles $\uparrow$
(i.e. $\uparrow \circ \ldots \uparrow$) et par $\Uparrow^n(s)$ l'application $n$
fois de $\Uparrow$ (i.e. $\Uparrow(\ldots(\Uparrow(s)\ldots))$).

La transformation d'un terme avec des noms de variables du \roCal\  dans un terme
du \roSig\  utilisant des indices entiers est similaire à la transformation d'un
$\lambda$-terme en un \mbox{$\lambda_{DB}$-terme}.  A cause de la possibilité
d'avoir plusieurs variables liées dans une règle de réécriture par rapport à une
seule dans une $\lambda$-abstraction, la transformation d'une règle de
réécriture est légèrement plus élaborée que la transformation de l'abstraction
du \laCal.

\DEF%=================================================================
\label{defTr}
Nous considérons une liste de variables  $x_1.\ldots.x_n$ appelé
référentiel. Etant donné un référentiel $\RR$, nous définissons récursivement
la traduction d'un $\rho$-terme $t$, noté $tr(t,\RR)$~:
\begin{enumerate}
	\item $tr(x,\RR)$ = $j$, où $j$ représente la première position de $x$ dans $\RR$,
	\item $tr(f(t_1,\ldots,t_n),\RR)=f(tr(t_1,\RR),\ldots,tr(t_n,\RR))$,
	\item $tr([a](b),\RR)=[tr(a,\RR)](tr(b,\RR))$,
	\item $tr(l \ra r,\RR)$ = $tr(l,Var(l).\RR) \ra_{\|Var(l)\|} tr(r,Var(l).\RR)$,
\end{enumerate}
avec $Var(l)$ représentant la liste des variables libres du terme $l$
(i.e. l'ensemble $\Var(l)$ transformé en liste) et $\|Var(l)\|$ sa longueur.
\FDEF%=================================================================

Il est évident qu'en fonction de l'ordre des variables libres du membre gauche
d'une règle de réécriture on peut obtenir plusieurs représentations de 
la règle de réécriture.

\EX%=================================================================
Si nous considérons le $\rho$-terme $[f(x,y) \ra g(x,y,z)](f(a,b))$, sa
transformation dans le référentiel $z.Nil$ est 
$[f(1,2) \ra_2 g(1,2,3)](f(a,b))$ ou bien $[f(2,1) \ra_2 g(2,1,3)](f(a,b))$.
\FEX%=================================================================

\section{Le filtrage}
%=============================================================================== 
%

Nous considérons un algorithme de filtrage qui retourne un résultat de la
forme $t_1.\ldots.t_n.\ID$ avec $t_1,\ldots,t_n$ des termes de $\RTTs$.

Une approche possible consiste à transformer les termes $l,t \in \RTTs$ d'un
problème de filtrage $(l \meqqes t)$ en $l^{\XX},t^{\XX} \in \RTT$ en
remplaçant les indices $1,\ldots,n$ par les variables $x_1,\ldots,x_n$ et
résoudre le problème de filtrage $(l^{\XX} \meqqes t^{\XX})$. Si l'ensemble de règles
\rname{SyntacticMatching} de la Section~\ref{filtrageNonType} appliqué pour ce
dernier problème de filtrage mène à un résultat $\subs{x_1/t^x_1,\ldots,x_n/t^x_n}$
(où nous considérons $t^x_i=x_i$ pour les variables n'apparaissant pas dans la
solution) alors le problème $(l \meqqes t)$ initial a comme solution la
substitution $t_1.\ldots.t_n.\ID$ où $t_i$ représente le terme $t^x_i$ avec les
variables $x_1,\ldots,x_n$ remplacées par les indices $1,\ldots,n$.
Comme dans le cas non-explicite nous définissons la fonction $\Sl(l \meqqes t)$
qui retourne l'ensemble de toutes les solutions de $(l \meqqes t)$ et donc
l'ensemble vide quand l'algorithme de filtrage échoue.

Nous pouvons adapter l'algorithme de filtrage présenté dans la
Section~\ref{filtrageNonType} afin de manipuler des termes utilisant des entiers
à la place des variables comme montré dans la Figure~\ref{SyntMatchSig}.

\begin{figure}[!htp]%----------------------------------------------------------
%\noindent
\framebox{\parbox{\largeurtexte}{

\renewcommand{\fleche}{\LaFleche}
\begin{ruleset}

\regle{Decomposition} 
{(f(t_1, \ldots, t_n) \meqqes f(t'_1, \ldots, t'_n)) \ww P} 
{\bigwedge_{i=1\ldots n} t_i \meqqes t'_i \ww P}

\cregle{SymbolClash} 
{(f(t_1, \ldots, t_n) \meqqes g(t'_1, \ldots, t'_m)) \ww P}
{\fF} {f \neq g}

\cregle{MergingClash} 
{(m \meqqes t) \ww (m \meqqes t') \ww P}
{\fF} {t \neq t', ~m \in \NatP} 

\cregle{SymbolVariableClash~} 
{(f(t_1, \ldots, t_n) \meqqes m) \ww P} 
{\fF} {m \in \NatP}

\end{ruleset}

}}
\caption{\label{SyntMatchSig}\rname{SyntacticMatchingSigma} - Règles pour le filtrage syntaxique}
\end{figure}%-----------------------------------------------------------------

Le résultat de la Proposition~\ref{normform} peut être étendu pour une équation
de filtrage avec les termes de $\RTTs$~:

\PROP%------------------------------------------------------------------------
\label{normformSigma}
La forme normale de tout problème de filtrage $t \meqqes t'$ calculée par les
règles \rname{SyntacticMatchingSigma} existe et est unique. Après avoir enlevé de la
forme normale toute équation dupliquée, si le système résultant est~:
\begin{enumerate}
\item $\fF$, alors il n'y a pas de filtre de $t$ à $t'$ et
	$\Sl(t \meqqes t')=\emptyset$,
\item de la forme $1 \meqqes t_1 \wedge \ldots \wedge n \meqqes t_n$
	alors la substitution $\sigma=t_1.\ldots.t_n.\ID$ est l'unique
	filtre de $t$ à $t'$ et $\Sl(t \meqqes t')=\{\sigma\}$
\end{enumerate}
\FPROP%------------------------------------------------------------------------

Nous pouvons noter que nous n'enlevons pas les équations triviales $n \meqqes n$
de la forme normale d'un système de filtrage. De plus, 
si le système ne contient aucune équation ($n=0$) alors la substitution obtenue
est $\ID$.

\section{Les règles d'évaluation}	\label{nr} 
%=============================================================================== 
%

Les règles d'évaluation de base du \roSig\  présentées dans la
Figure~\ref{MRAsig} sont celles du \roCalE\  et nous introduisons les règles
d'évaluation dans la Figure~\ref{sigmaro} décrivant l'application de
substitution.
Ce sous-système est appelé le \sigroCal\  ou le \sigCal.

\begin{figure}[!htp]%------------------------------------------------------------
\noindent
\framebox{\parbox{\largeurtexte}{

\renewcommand{\fleche}{\Longrightarrow}
\begin{ruleset}
%===================================
  \wregle
  {Fire}    
  {[l \ra_n r](t)} 
  {\{r\subs{\sigma}\}}
  {\sigma \in \Sl(l \meqqes t)}
\\
%===================================
  \regle
  {Congruence}   
  {[f(u_1,\ldots,u_n)](f(v_1,\ldots,v_n))} 
  {\{f([u_1](v_1),\ldots,[u_n](v_n))\}}
\\
%===================================
  \regle
  {Congruence\_fail}   
  {[f(u_1,\ldots,u_n)](g(v_1,\ldots,v_m))} 
  {\emptyset}
\\
%===================================
\regle {Distrib}
	{[\{u_1,\ldots,u_n\}](v)}
	{\{[u_1](v),\ldots,[u_n](v)\}}
\\
%=================================== 
\regle {Batch}
	{[v](\{u_1,\ldots,u_n\})}
	{\{[v](u_1),\ldots,[v](u_n)\}}
\\
%=================================== 
\regle {Switch_R} 
	{u \ra_n \{v_1,\ldots,v_m\}}
	{\{u \ra_n v_1,\ldots,u \ra_n v_m\}}
\\
%=================================== 
\hregle {OpOnSet}
	{f(v_1,\ldots,\{u_1,\ldots,u_m\},\ldots,v_n)}
	{\{f(v_1,\ldots,u_1,\ldots,v_n),\ldots,f(v_1,\ldots,u_m,\ldots,v_n)\}}
\\
%===================================
\regle {Flat}
	{\{u_1,\ldots,\{v_1,\ldots,v_n\},\ldots,u_m\}}
	{\{u_1,\ldots,v_1,\ldots,v_n,\ldots,u_m\}}
%===================================
\end{ruleset}

}}
\caption{\label{MRAsig}Les règles d'évaluation de base du \roSig}
\end{figure}%---------------------------------------------------------------

La règle \rname{Fire} est celle qui déclenche l'application de la substitution
obtenue en filtrant le membre gauche de la règle de réécriture et l'argument de
l'application au membre droit de la règle de réécriture. Le \sigroCal\  décrit
explicitement l'application de cette substitution. Les autres règles
d'évaluation de base ont exactement la même fonctionnalité que dans le \roCalE.

\begin{figure}[!ht]%------------------------------------------------------------
\noindent
\framebox{\parbox{\largeurtexte}{

\renewcommand{\fleche}{\Rightarrow}

\begin{ruleset}
%===========================================================================  
\regle{lam}{(u \ra_n v)\subs{s}}{u\subs{\Uparrow^n(s)} \ra_n v\subs{\Uparrow^n(s)}}
       
\regle{app}{[u](v)\subs{s}}{[u\subs{s}](v\subs{s})}

\regle{clos}{u\subs{s}\subs{t}}{u\subs{s \circ t}}

\regle{vs1}{n\subs{\uparrow}}{(n+1)}

\regle{vs2}{n\subs{\uparrow \circ s}}{(n+1)\subs{s}}

\regle{fvc}{1\subs{u.s}}{u}

\regle{fvl1}{1\subs{\Uparrow(s)}}{1}

\regle{fvl2}{1\subs{\Uparrow(s) \circ t}}{1\subs{t}}

\regle{rvc}{(n+1)\subs{u.s}}{n\subs{s}}

\regle{rvl1}{(n+1)\subs{\Uparrow(s)}}{n\subs{s \circ \uparrow}}

\regle{rvl2}{(n+1)\subs{\Uparrow(s) \circ t}}{n\subs{s \circ (\uparrow \circ t)}}

\regle{id}{u\subs{\ID}}{u}

%\vspace{1cm}  \\
\\

\regle{set}{\bigcup_{i=1}^n \{u_i\} \subs{s}}{\bigcup_{i=1}^n \{u_i \subs{s}\}}

%\vspace{-1cm} \\
\\

\regle{ass}{(s \circ t) \circ v}{s \circ (t \circ v)}

\regle{map}{(u.s) \circ t}{u\subs{t}.(s \circ t)}

\regle{sc}{\uparrow \circ (u.s)}{s}

\regle{sl1}{\uparrow \circ \Uparrow(s)}{s \circ \uparrow}

\regle{sl2}{\uparrow \circ \Uparrow(s) \circ t}{s \circ(\uparrow \circ t)}

\regle{l1}{\Uparrow(s) \circ \Uparrow(t)}{\Uparrow(s \circ t)}

\regle{l2}{\Uparrow(s) \circ (\Uparrow(t) \circ v)}{\Uparrow(s \circ t) \circ v}

\regle{le}{\Uparrow(s) \circ (u.t)}{u.(s \circ t)}

\regle{il}{\ID \circ s}{s}

\regle{ir}{s \circ \ID}{s}

\regle{li}{\Uparrow(\ID)}{\ID}

%\vspace{-1cm} \\
\\

\regle{op}{f(u_1,\ldots,u_n)\subs{s}}{f(u_1\subs{s},\ldots,u_n\subs{s})}

%\vspace{-1cm}  \\
%===================================
\end{ruleset}

}}
\caption{\label{sigmaro}Les règles d'évaluation du \sigroCal}
\end{figure}%------------------------------------------------------------

Les règles d'évaluation du \sigroCal\  sont similaires à celles du
$\sigma_{\Uparrow}$-calcul de substitution du $\lambda\sigma_{\Uparrow}$-calcul
(voir Section~\ref{lambdaSigma}). Par rapport à ce dernier calcul nous
adaptons les règles \rname{lambda} et \rname{app} pour la syntaxe du \roSig\  et
nous ajoutons les règles \rname{set} et \rname{op}. 

Nous pouvons aussi voir les règles d'évaluation mentionnées précédemment comme
des cas particuliers de la règle \rname{f_\Uparrow} du $\Gamma_\Uparrow$-calcul
(\cite{PaganoCADE15}). Dans ce calcul, à chaque symbole d'arité $n$ est associée une
liste de $n$ entiers appelés l'{\em arité de liaison} du symbole. Intuitivement,
l'arité de liaison d'un opérateur représente le nombre d'indices de de Bruijn
liés dans chacun de ses arguments.  Pour une discussion plus détaillée sur
l'arité de liaison on peut se référer à~\cite{PaganoCADE15}.

Dans le $\lambda\sigma_{\Uparrow}$-calcul, si nous considérons le terme $\lambda t$ nous
pouvons dire que l'opérateur $\lambda$ lie l'indice $1$ dans le terme $t$ et
donc, a l'arité de liaison $(1)$. Le symbole unaire $\lambda$ du
$\lambda\sigma_{\Uparrow}$-calcul est remplacé dans le \roSig\  par le symbole binaire
$\ra_n$ dont l'arité de liaison est $(n,n)$. Tout opérateur fonctionnel a une arité
de liaison $(0,\ldots,0)$.

\DEF%------------------------------------------------------------------------
\label{roSdef} 

Etant donné un ensemble de symboles de fonctions $\FF$, un ensemble de variables
$\XX$, nous appelons \roSig\  un calcul défini par~:
\begin{itemize}
\item l'ensemble de termes $\RTTs$,
\item la théorie $\emptyset$ (filtrage syntaxique),
\item les règles d'évaluation de base de la Figure~\ref{MRAsig},
\item les règles d'évaluation pour l'application de substitution de la
	Figure~\ref{sigmaro},
\item une stratégie d'évaluation $\SS$ qui guide l'application des règles d'évaluation.
\end{itemize}

\FDEF%------------------------------------------------------------------------

\section{Propriétés du \roSig} \label{proprRoSig}
%===============================================================================
%

Dans cette section nous analysons quelques propriétés du \roSig\  et de ses
sous-calculs et nous montrons en particulier que le \roSig\  est confluent dans
les mêmes conditions que le \roCalE.

Toutes les définitions de termes \matchD\  et \ready\  de la
Section~\ref{notionsPreliminaires} peuvent être utilisées dans un contexte
explicite en considérant des indices de de Bruijn à la place des variables. Les
mêmes remarques que dans le cas non-explicite peuvent être faites pour les
termes contenant des indices de de Bruijn et des applications de substitution.

\REM%----------------------------------------------------------
\label{positionsFonctionnellesExp}
Si $l \in \TFN$ \subf\  $t$, alors pour toute position non-fonctionnelle (i.e. la
position d'un indice, d'une application, d'une abstraction, d'un ensemble ou
d'une application de substitution) dans $t$ la position correspondante dans
$l$, si elle existe, est une position variable, c'est-à-dire la position d'un
indice. Ainsi, si la position de tête de $t$ n'est pas une position
fonctionnelle alors $l$ est un indice.
\FREM%----------------------------------------------------------

\REM%----------------------------------------------------------
\label{echecMatchExp}
Si les termes $l$ et $t$ sont \matchD, le filtrage $(l \meqqes t)$ peut échouer
seulement à cause des symboles fonctionnels différents à la même position des
termes $l$ et $t$.
\FREM%----------------------------------------------------------

L'application de la règle d'évaluation \rname{Fire} est guidée par une stratégie
\textit{ConfStrat} décrite dans la Section~\ref{strat_confluente} qui peut
être exprimée explicitement en transformant la règle \rname{Fire} en la règle
conditionnelle \rname{Fire_c} déjà utilisée pour prouver la confluence du
\roCalE~:

\renewcommand{\fleche}{\Longrightarrow}
\begin{ruleset}
%===================================
  \cwregle 
  {Fire_c} 
  {[l \ra_n r](t)} 
  {\{r\subs{\sigma}\}}
  {l,t~sont~\ready}
  {\sigma \in \Sl(l \meqqes t)}
%===================================
\end{ruleset}

Nous considérons que les relations $\roZ$, $\del$ et $\congr$ de même que
leurs fermetures sont définies exactement de la même façon que dans le \roCalE\
(Section~\ref{stratRocal}) mais en utilisant la règle \rname{Fire_c} ci-dessus.

\DEF%------------------------------------------------------------------------
\label{transSigma}
Nous considérons la \textit{relation} sur $\RTTs$ appelée $\sigRo$ induite par
les règles d'évaluation dans la Figure~\ref{sigmaro}.

Les relations suivantes sont induites par la relation $\sigRo$~: 
\begin{itemize}
\item[] $\sigro$ est la fermeture compatible de la relation $\sigRo$,
\item[] $\sigroTR$ est la fermeture réflexive, transitive de $\sigro$.
\end{itemize}
\FDEF%------------------------------------------------------------------------

Nous notons par $\setSig$ la relation $(\sigroTR\congr\sigroTR)$.
Nous notons par $\delSig$ la relation $(\sigroTR\del\sigroTR)$ et
par $\roZSig$ la relation $(\sigroTR\roZ\sigroTR)$.

Nous analysons d'abord les propriétés des relations induites par $\sigRo$ et
ensuite nous nous concentrons sur la confluence du \roSig.

\subsection{Propriétés de la relation $\sigRo$}		\label{propSigma} 
%=============================================================================== 
%

La relation engendrée par les règles décrivant l'application de substitution
est confluente et terminante~:

\LEM%=================================================================
\label{confSigma}	\comment{used in Lemma~\ref{setSigterm}}
$\sigro$ est localement confluente.
\FLEM%=================================================================

\proof{

Dans~\cite{CurienHardinLevy-JACM96} il est montré que le système
$\sigma_\Uparrow$ ($\sigRo$ sans les règles $lam$, $op$ et $set$) est
localement confluent.
Nous prouvons que les paires critiques induites par les nouvelles
règles d'évaluation sont convergentes.  

La règle \rname{set} a des paires critiques seulement avec les règles \rname{clos} et
\rname{id}. La réduction suivante est obtenue pour la paire critique entre
\rname{set} et \rname{clos}~:

\begin{center}$~$
\xymatrix{ 
& \bigcup_{i=1}^n \{u_i\} \subs{s}\subs{t} 
\ar[dl]_{set} \ar[dr]^-{clos} & \\
\bigcup_{i=1}^n \{u_i \subs{s}\} \subs{t}
\ar@{.{>}}[d]_{set} 
& & \bigcup_{i=1}^n \{u_i\} \subs{s \circ t} 
\ar@{.{>}}[ddl]^-{set} \\
 \bigcup_{i=1}^n \{u_i \subs{s} \subs{t}\}
\ar@{.{>}}[dr]_{clos} & & \\
& \bigcup_{i=1}^n \{u_i \subs{s \circ t}\}   & 
}
\end{center}

Pour les règles \rname{set} et \rname{id} et respectivement \rname{op} et
\rname{id} nous obtenons~:
\begin{center}%$~$
\begin{tabular}{cc}
\xymatrix{ 
& \bigcup_{i=1}^n \{u_i\} \subs{\ID} 
\ar[dl]_{set} \ar[dd]^-{id} \\
\bigcup_{i=1}^n \{u_i \subs{\ID}\} 
\ar@{.{>}}[dr]_{id} & \\
& \bigcup_{i=1}^n \{u_i\} 
}
%\end{center}

&

%\begin{center}$~$
\xymatrix{ 
& f(u_1,\ldots,u_n)\subs{\ID} 
\ar[dl]_{op}
\ar[dd]^-{id} &
\\
f(u_1\subs{\ID},\ldots,u_n\subs{\ID}) 
\ar@{.{>}}[dr]_{id}
&\\
& f(u_1,\ldots,u_n) 
}
\end{tabular}
\end{center}

Pour les règles \rname{op} et \rname{clos} nous avons~:
\begin{center}$~$
\xymatrix{ 
& f(u_1,\ldots,u_n)\subs{s}\subs{t} \ar[dl]_{op}
\ar[dr]^-{clos} & 
\\
f(u_1\subs{s},\ldots,u_n\subs{s})\subs{t}
\ar@{.{>}}[d]_{op} & & 
f(u_1,\ldots,u_n)\subs{s \circ t}
\ar@{.{>}}[ddl]^-{op} 
\\
f(u_1\subs{s}\subs{t},\ldots,u_n\subs{s}\subs{t}) 
\ar@{.{>}}[dr]_{clos} & &
\\
& f(u_1\subs{s \circ t},\ldots,u_n\subs{s \circ t}) & 
}
\end{center}

La réduction suivante est obtenue pour la paire critique entre \rname{lam} et
\rname{clos}~:
\begin{center}$~$
\xymatrix{ 
&  (u \ra_n v)\subs{s}\subs{t} 
\ar[dl]_{lam} \ar[dr]^-{clos} & \\
(u\subs{\Uparrow^n(s)} \ra_n v\subs{\Uparrow^n(s)}) \subs{t}
\ar@{.{>}}[d]_{lam} 
& & (u \ra_n v) \subs{s \circ t} 
\ar@{.{>}}[dd]^-{lam} \\
u\subs{\Uparrow^n(s)}\subs{\Uparrow^n(t)} \ra_n v\subs{\Uparrow^n(s)}\subs{\Uparrow^n(t)}
\ar@{.{>}}[d]_{clos} & & \\
u\subs{\Uparrow^n(s) \circ \Uparrow^n(t)} \ra_n v\subs{\Uparrow^n(s) \circ \Uparrow^n(t)}
\ar@{.{>}}[rr]_{l1} & & 
u\subs{\Uparrow^n(s \circ t)} \ra_n v\subs{\Uparrow^n(s \circ t)}   & 
}
\end{center}

Pour les règles \rname{lam} et \rname{id} nous avons~:
\begin{center}$~$
\xymatrix{ 
& (u \ra_n v) \subs{\ID} 
\ar[dl]_{lam} \ar[dd]^-{id} \\
u \subs{\Uparrow^n(\ID)} \ra_n v \subs{\Uparrow^n(\ID)}
\ar@{.{>}}[d]_{li} & \\
u \subs{\ID} \ra_n v \subs{\ID}
\ar@{.{>}}[r]_{id}
& u \ra_n v
}
\end{center}

}

\LEM%*************************************************************************
\label{termSigma}	\comment{ used in Lemma~\ref{setSigterm}}
$\sigro$ est terminante.
\FLEM%************************************************************************

\proof{ 

En partant de la preuve proposée dans~\cite{CurienHardinLevy-JACM96}
et~\cite{PaganoCADE15} pour $\sigma_\Uparrow$ et $\Gamma_\Uparrow$
respectivement nous utilisons un ordre lexicographique sur les deux
interprétations polynômiales
ci-dessous et nous pouvons montrer que~:
\begin{itemize}
\item $P_1$ est décroissant sur toutes les règles et strictement décroissant sur 
	\rname{lam},
\item $P_2$ est strictement décroissant sur toutes les règles sauf \rname{lam}.
\end{itemize}

\hspace{-.7cm}
%\begin{figure}[!ht]%------------------------------------------------------------
%\begin{center}
{\small
\begin{tabular}{|l|l|}

\hline & \\
$P_1(n) = 2^n$ & 
$P_2(n) = 1$ \\
& \\
$P_1([u](v)) = P_1(u) * D_1(v) + P_1(v) * D_1(u)$ & 
$P_2([u](v)) = P_2(u) * D_2(v) + P_2(v) * D_2(u)$ \\
& \\
& 
\multicolumn{1}{r|}
{
$+ D_2(u) * D_2(v)$
} \\
& \\
$P_1(u \ra_n v) = P_1(u) * D_1(v) + P_1(v) * D_1(u)$ & 
$P_2(u \ra_n v) = 2 * (P_2(u) * D_2(v) + P_2(v) * D_2(u))$ \\
& \\
\multicolumn{1}{|r|}
{
$+ 2 * D_2(u) * D_2(v)$
} 
& 
\\
& \\
$P_1(u\subs{s}) = P_1(u) * P_1(s)$ & 
$P_2(u\subs{s}) = P_2(u) * (P_2(s) + 1)$ \\
& \\
$P_1(\ID) = 2$ &
$P_2(\ID) = 1$ \\
& \\
$P_1(\{u_1,\ldots,u_n\}) = P_1(u_1) + \ldots + P_1(u_n) + 2$ & 
$P_2(\{u_1,\ldots,u_n\}) = P_1(u_1) + \ldots + P_1(u_n) + 1$ \\
& \\
$P_1(\uparrow) = 2$ & 
$P_2(\uparrow) = 1$ \\
& \\
$P_1(\Uparrow(s)) = P_1(s)$ & 
$P_2(\Uparrow(s)) = 4 * P_2(s)$ \\
& \\
$P_1(u.s) = P_1(u) + P_1(s)$ & 
$P_2(u.s) = P_1(u) + P_1(s) +1$ \\
& \\
$P_1(s \circ t) = P_1(s) * P_1(t)$ & 
$P_2(s \circ t) = P_2(s) * (P_2(t) + 1)$ \\
& \\
\multicolumn{2}{|l|}
{$P_1(f(u_1,\ldots,u_n))$ $=$ } \\
\multicolumn{2}{|r|}
{$P_1(u_1) * D_1(u_2) * \ldots * D_1(u_n) + \ldots + P_1(u_n) * D_1(u_1) * \ldots * D_1(u_{n-1}) + D_1(u_1) * \ldots * D_1(u_n)$} \\
& \\
\multicolumn{2}{|l|}
{$P_2(f(u_1,\ldots,u_n))$ $=$ } \\
\multicolumn{2}{|r|}
{$P_2(u_1) * D_2(u_2) * \ldots * D_2(u_n) + \ldots + P_2(u_n) * D_2(u_1) * \ldots * D_2(u_{n-1}) + D_2(u_1) * \ldots * D_2(u_n)$} \\
& \\
$P_1(f)$ $=$ $2$ &
$P_2(f)$ $=$ $1$ \\
& \\
%\hline
%& \\
$D_1(\{u_1,\ldots,u_n\})$ $=$ $D_1(u_1) + \ldots + D_1(u_n) + 1$ &
$D_2(\{u_1,\ldots,u_n\})$ $=$ $D_2(u_1) + \ldots + D_2(u_n) + 1$ \\
& \\
$D_1(u)$ $=$ $1$ &
$D_2(u)$ $=$ $1$ \\
& \\
\hline

\end{tabular}
}%end small
%\end{center}
%\caption{\label{LPO}Polynomial interpretations}
%\end{figure}%------------------------------------------------------------

}

Puisque $\sigro$ est confluente et terminante nous pouvons énoncer un lemme
sur les formes normales des termes réduits par cette relation.

\LEM%**********************************************************************
\label{sigmaNorm}
Etant donnée une substitution $s$ en $\sigRo$-forme normale. Alors, $s$ a
nécessairement une des formes suivantes~:
\begin{itemize}
\item[] $s = \ID$,
\item[] $s = t . s'$, avec $t,s'$ en $\sigRo$-forme normale,
\item[] $s = \uparrow^n$,
\item[] $s = \Uparrow(s') \circ \uparrow^n$, avec $s'$ en $\sigRo$-forme normale 
	et $n \geq 0$.
\end{itemize}

Etant donné un terme $t$ en $\sigRo$-forme normale. Alors, $t$ a
nécessairement une des formes suivantes~:
\begin{itemize}
\item[] $t = n$,
\item[] $t = \{t_1,\ldots,t_n\}$, avec $t_1,\ldots,t_n$ en $\sigRo$-forme normale,
\item[] $t = f(t_1,\ldots,t_n)$, avec $t_1,\ldots,t_n$ en $\sigRo$-forme normale,
\item[] $t = [u](v)$, avec $u, v$ en $\sigRo$-forme normale,
\item[] $t = u \ra v$, avec  $u, v$ en $\sigRo$-forme normale.
\end{itemize}
\FLEM%*******************************************************************

\proof{

Nous procédons par induction sur la structure des substitutions.  Le seul cas
qui n'est pas immédiat est $s = s_1 \circ s_2$. Nous avons les
possibilités suivantes~:
\begin{enumerate}
\item $s_1=\ID$ n'est pas possible, autrement $s$ serait un radical pour \rname{il}.

\item $s_1=t.s'$ n'est pas possible, autrement $s$ serait un radical pour \rname{map}.

\item $s_1=\uparrow^n$. Dans ce cas $n=1$, autrement $s$ serait un radical pour \rname{ass}.
	$s_2$ ne peut pas être $\ID$ (autrement $s$ serait un radical pour \rname{ir}),
	$s_2$ ne peut pas être $t.s'$ (autrement $s$ serait un radical pour \rname{sc}),
	$s_2$ ne peut pas être $\Uparrow(s') \circ \uparrow^n$ (autrement $s$
	serait un radical pour \rname{sl2}). Ainsi, $s_2$ doit être $\uparrow^p$ et
	donc, $s=\uparrow^{p+1}$.

\item $s_1=\Uparrow(s') \circ \uparrow^n$, avec $n=0$, c'est-à-dire 
	$s=\Uparrow(s')$. $s_2$ ne peut pas être $\ID$ (autrement $s$ serait un
	radical pour \rname{ir}), $s_2$ ne peut pas être $t.s'$ (autrement $s$ serait
	un radical pour \rname{le}), $s_2$ ne peut pas être 
	$\Uparrow(s') \circ \uparrow^n$ (autrement $s$ serait un radical
	pour \rname{l2}). Ainsi, $s_2$ doit être $\uparrow^p$ et donc,
	$s=\Uparrow(s') \circ \uparrow^p$.
\end{enumerate}

Pour la deuxième partie de la preuve nous procédons similairement, par induction
sur la structure des termes.  Nous avons les possibilités suivantes pour un
terme $t$~:
\begin{enumerate}
\item $t=n\subs{s}$. Pour toute substitution $s$, le terme $t$ contient un radical.

\item $t=\{t_1,\ldots,t_n\}$, $t=f(t_1,\ldots,t_n)$, $t=[u](v)$, $t=u \ra v$ 
	sont facilement prouvés par induction.

\item $t = u\subs{s}$. Ce cas ne peut pas apparaître puisque $u\subs{s}$
	ne doit pas contenir de radical pour $set$, $op$, $app$ ou $lam$.
\end{enumerate}

}

Ainsi, les termes en $\sigRo$-forme normale ne peuvent contenir aucune
substitution et donc un terme $t \subs{s}$ mène à une forme normale ne contenant 
pas de substitution.

\subsection{La confluence du \roSig }\label{propRo} 
%=============================================================================== 
%

Nous devons remarquer que les substitutions que nous manipulons dans le \roSig\
sont obtenues exclusivement par la résolution d'un problème de filtrage déclenché par
l'application de la règle d'évaluation \rname{Fire_c}. Puisque tout problème de
filtrage est de la forme $(l \meqqes t)$ avec $l,t$ \ready\  alors toute
substitution obtenue comme solution d'un tel filtrage ne contient que des termes
satisfaisant les mêmes conditions que $t$.

Nous allons considérer par la suite uniquement des substitutions de cette forme,
c'est-à-dire qui ne contiennent aucun ensemble vide ou ayant plus d'un élément et
qui contiennent que des applications de la forme $[u \ra w](v)$ où $u$ subsume
$v$. 

\LEM%------------------------------------------------------------------------
\label{readyLemmaSigma}
\comment{dans Lemme~\ref{}}

Etant donnés les $\rho$-termes $l,t,t'$ tels que $t \sigro t'$. Alors,
\begin{itemize}
	\item si $l,t$ sont \matchD\  alors $l,t'$ sont \matchD,
	\item si $t$ est \safes\  alors $t'$ est \safes,
	\item si $l,t$ sont \ready\  alors $l,t'$ sont \ready.
\end{itemize}
\FLEM%------------------------------------------------------------------------

\proof{

Si le terme $t \in \TF$ alors $t=t'$ et le lemme est clairement vrai. Nous
considérons par la suite que $t \not\in \TF$.

Puisque $l,t$ sont \matchD\  alors le terme $l$ \subf\  le terme $t$ et donc toute
position fonctionnelle du terme $l$ est une position fonctionnelle du terme $t$
ou n'est pas une position du terme $t$. Ainsi, pour toute position d'une
application de substitution dans $t$, la position correspondante dans $l$, si
elle existe, est la position d'un indice. Les applications de substitution dans
$t$ sont éventuellement réduites par $\sigro$ mais les positions correspondantes
dans $l$ sont des positions variables et donc, $l,t'$ sont \matchD.

En regardant les règles d'évaluation de la Figure~\ref{sigmaro}, nous pouvons
noter qu'un ensemble vide ou ayant plus d'un élément peut être engendré dans le
terme $t'$ seulement si un tel ensemble existe dans le terme $t$. Puisque  $t$
est \safes\  ceci n'est pas possible et donc $t'$ est \safes.

Ainsi, les propriétés de termes \matchD\  et \safe\  sont préservées par la
relation $\sigro$ et donc, la propriété de termes \ready\  est préservée
par la relation $\sigro$.
}%________________________________________________________________________________

La confluence du \roSig\  est obtenue en montrant que les conditions pour le
lemme de Yokouchi sont satisfaites par les relations $\delSig$ et
$\setSig$. Nous montrons d'abord que la relation $\setSig$ est confluente et
terminante, ensuite nous prouvons la confluence forte de la relation $\delSig$
et finalement nous montrons que le diagramme de Yokouchi est satisfait par les
deux relations.

\LEM%--------------------------------------------------------------------------
\label{BasicCoherenceSS}
\comment{ used in Lemma~\ref{setSigterm}, Lemma~\ref{BasicCoherenceGeneral}}

Si $t \sigro t'$ et $\Cong{t}{u}$, alors il existe $u'$ tel que 
$t' \sigroTR u'$ et $u \setSig u'$.
\begin{center}$~$
\xymatrix{%@C+10pt{
& t
\ar[dl]_{\congrxy} \ar[dr]^-{\sigroxy}
& \\
t'
\ar@{.{>}}[dr]_{\sigroTRxy} && 
u
\ar@{.{>}}[dl]^-{\setSigxy} \\
& 
u'
&
}
\end{center}
\FLEM%--------------------------------------------------------------------------

\proof{

Puisque les $\sigRo$-radicaux ne peuvent pas apparaître dans un $Set$-radical nous
devons seulement analyser les cas générés par une $\sigRo$-réduction. Nous montrons
seulement le cas de la règle \rname{Distrib}, toutes les autres preuves sont faites
d'une manière similaire.
\begin{center}$~$
\xymatrix{%@C+10pt{ 
[\{u_1,\ldots,u_n\}](v)\subs{s} 
\ar[d]_{Distrib} \ar[r]^-{app} 
& [\{u_1,\ldots,u_n\}\subs{s}](v\subs{s})
\ar@{.{>}}[d]^-{set}\\
\{[u_1](v),\ldots,[u_1](v)\} \subs{s}
\ar@{.{>}}[d]_{set} 
& [\{u_1\subs{s},\ldots,u_n\subs{s}\}](v\subs{s})
\ar@{.{>}}[d]^-{Distrib}\\
\{[u_1](v)\subs{s},\ldots,[u_1](v)\subs{s}) \}
\ar@{.{>}}[r]_{app} 
& \{[u_1\subs{s}](v\subs{s}),\ldots,[u_1\subs{s}](v\subs{s})\}
} 
\end{center}

Nous utilisons les règles d'inférence \rname{Distrib} (relation $\congr$) et
\rname{app}, \rname{set} (relation $\sigro$).
}

\LEM%--------------------------------------------------------------------------
\label{setSigterm} 	\comment{ used in Theorem~\ref{roSigConf}}
La relation $\setSig$ est confluente et terminante.
\FLEM%--------------------------------------------------------------------------

\proof{

La relation $\congr$ est confluente par le Lemme~\ref{CongRconf}.  La
terminaison de la relation $\congr$ a été montré dans le Lemme~\ref{CongRterm}
en utilisant une méthode basée sur des interprétations polynômiales. Nous
pouvons aussi utiliser les interprétations polynômiales suivantes afin
d'obtenir la même propriété.

{\small
\begin{tabular}{|l|}

\hline \\
$P_1([u](v)) = P_1(u) * D_1(v) + P_1(v) * D_1(u)$ \\
\\
$P_1(u \ra_n v) = P_1(u) * D_1(v) + P_1(v) * D_1(u) + 2 * D_2(u) * D_2(v)$ \\
\\
$P_1(\{u_1,\ldots,u_n\}) = P_1(u_1) + \ldots + P_1(u_n) + 2$ \\
\\
$P_1(f(u_1,\ldots,u_n))$ $=$ 
$P_1(u_1) * D_1(u_2) * \ldots * D_1(u_n) + \ldots + P_1(u_n) * D_1(u_1) * \ldots * D_1(u_{n-1})+$ \\
\multicolumn{1}{|r|}
{$D_1(u_1) * \ldots * D_1(u_n)$} \\
\\
$P_1(f)$ $=$ $2$\\
\\
%\hline
%\\
$D_1(\{u_1,\ldots,u_n\})$ $=$ $D_1(u_1) + \ldots + D_1(u_n) + 1$ \\
\\
$D_1(u)$ $=$ $1$ \\
\\
\hline
\end{tabular}
}\\
\\%ou\\
{\small
\begin{tabular}{|l|l|}

\hline \\
$P_2([u](v)) = P_2(u) * D_2(v) + P_2(v) * D_2(u) + D_2(u) * D_2(v)$ \\
\\
$P_2(u \ra_n v) = 2 * (P_2(u) * D_2(v) + P_2(v) * D_2(u))$ \\
\\
$P_2(\{u_1,\ldots,u_n\}) = P_1(u_1) + \ldots + P_1(u_n) + 1$ \\
\\
$P_2(f(u_1,\ldots,u_n))$ $=$ 
$P_2(u_1) * D_2(u_2) * \ldots * D_2(u_n) + \ldots + P_2(u_n) * D_2(u_1) * \ldots * D_2(u_{n-1}) +$\\
\multicolumn{1}{|r|}
{$D_2(u_1) * \ldots * D_2(u_n)$} \\
\\
$P_1(f)$ $=$ $1$\\
\\
%\hline
%\\
$D_2(\{u_1,\ldots,u_n\})$ $=$ $D_2(u_1) + \ldots + D_2(u_n) + 1$ \\
\\
$D_2(u)$ $=$ $1$ \\
\\
\hline
\end{tabular}
}

La relation $\sigro$ est confluente et terminante par le Lemme~\ref{confSigma}
et le Lemme~\ref{termSigma}.

Afin de prouver la terminaison de la relation $\setSig$ nous utilisons la même
méthode que dans le Lemme~\ref{termSigma}. Puisque les interprétations
polynômiales présentées ci-dessus pour les règles de $\congr$ sont incluses
dans celles du Lemme~\ref{termSigma} pour les règles de $\sigro$, ces dernières
peuvent être utilisées pour montrer la terminaison de la relation globale
$\setSig$.

En partant de la cohérence entre les deux relations montrée dans le
Lemme~\ref{BasicCoherenceSS} et de la terminaison de $\congr$ et $\sigro$ nous
obtenons immédiatement par induction sur le nombre de réductions~:
\begin{center}$~$
\xymatrix{%@C+10pt{
& t
\ar[dl]_{\congrxy} \ar[dr]^-{\sigroTRxy}
& \\
t'
\ar@{.{>}}[dr]_{\sigroTRxy} && 
u
\ar@{.{>}}[dl]^-{\setSigxy} \\
& 
u'
&
}
$~~~~~~$
et
$~~~~~~$
\xymatrix{%@C+10pt{
& t
\ar[dl]_{\congTRxy} \ar[dr]^-{\sigroTRxy}
& \\
t'
\ar@{.{>}}[dr]_{\sigroTRxy} && 
u
\ar@{.{>}}[dl]^-{\setSigTRxy} \\
& 
u'
&
}
\end{center}

En utilisant ce dernier diagramme et la confluence des deux relations
nous obtenons la confluence de $\setSig$ par induction sur le nombre de
réductions $\sigro$~:
\begin{center}$~$
\xymatrix@C+30pt@R+30pt{
\ar[r]^-{\congrxy} 
\ar[d]_{\congrxy} 
&
\ar[r]^-{\sigroTRxy}
\ar[d]_{\congTRxy}
&
\ar[d]^-{\setSigTRxy}
\\
\ar[r]^-{\congTRxy} 
\ar[d]_{\sigroTRxy} 
&
\ar[r]^-{\sigroTRxy}
\ar[d]^-{\sigroTRxy}
&
\ar[d]^-{\sigroTRxy}
\\
\ar[r]_{\setSigTRxy}
&
\ar[r]_{\sigroTRxy}
&
}
\end{center}

\begin{center}$~$
\xymatrix@C+30pt@R+30pt{
\ar[r]^-{\sigroTRxy}
\ar[d]_{\sigroxy}
&
\ar[r]^-{\congrxy} 
\ar[d]_{\sigroTRxy} 
&
\ar[r]^-{\sigroTRxy}
\ar[d]_{\sigroTRxy}
&
\ar[d]^-{\sigroTRxy}
\\
\ar[r]^-{\sigroTRxy} 
\ar[d]_{\sigroTRxy~\congrxy~\sigroTRxy}
&
\ar[r]^-{\sigroTRxy~\congrxy~\sigroTRxy}
\ar@{{}{}}[d]^-{induction}
&
\ar[r]^-{\sigroTRxy}
&
\ar[d]^-{\setSigTRxy}
\\
\ar[rrr]_{\setSigTRxy}
&
&
&
}
\end{center}
}

Nous procédons d'une manière similaire afin de prouver la confluence forte de
la relation $\delSig$. Nous montrons d'abord que la relation $\del$ est
fortement confluente et ensuite nous prouvons qu'un diagramme de cohérence
similaire à celui du Lemme~\ref{BasicCoherenceSS} est obtenu pour les relations
$\del$ et $\sigro$.

\LEM%--------------------------------------------------------------------------
\label{BasicCaseSigma}
\comment{ used in Lemme~\ref{ConfDelSigma} and ?? Theorem~\ref{roSigConf} ??}

La relation $\del$ est fortement confluente~:
\begin{center}$~$
\xymatrix{%@C+10pt{
&
t_0
\ar[dl]_{\delxy} \ar[dr]^-{\delxy}
& \\
t_1
\ar@{.{>}}[dr]_{\delxy} && 
t_2
\ar@{.{>}}[dl]^-{\delxy} \\
&t_3&
}
\end{center}
\FLEM%--------------------------------------------------------------------------

\proof{
Les règles \rname{Fire},\rname{Congruence} et \rname{Congruence\_fail} sont
linéaires à gauche et sans paires critiques (\cite{Huet80}).
}

Nous analysons maintenant la correspondance entre les solutions des problèmes de
filtrage $(l \meqqes t)$ et $(l \meqqes t')$ où $t \sigroTR t'$. Nous obtenons un
résultat similaire à ceux obtenus dans le \roCalE\  pour la relation
$\del$, dans le Lemme~\ref{matchDeltaOne}, ou pour la relation $\congTR$, dans le
Lemme~\ref{matchCongOne}.

\LEM%--------------------------------------------------------------------------
\label{equivMatch} \comment{ used in Lemma~\ref{matchSubs}, Lemma~\ref{BasicCoherenceSigma}}

Etant donnés les $\roS$-termes $l,t$ et $t'$ et une substitution $s$ tels que
$l,t$ sont des termes \matchD\  et $t\subs{s} \sigroTR t'$. Alors, nous avons~:
\begin{itemize}
\item[a.]  Si $u_1.\ldots.u_n.\ID$ est la solution de $(l \meqqes t)$
	et $v_1.\ldots.v_n.\ID$ est la solution de $(l \meqqes t')$ (avec $n$
	l'indice maximal de $l$), alors $u_i\subs{s} \sigroTR v_i$.
\item[b.]  Si $\Sl(l \meqqes t)=\emptyset$, alors $\Sl(l \meqqes t')=\emptyset$.
\end{itemize}
\FLEM%--------------------------------------------------------------------------

\proof{

Si $t \in \TF$ alors, $t'=t$ est le lemme est trivialement vrai. Pour le cas où
$t \not \in \TF$ nous procédons par induction sur la structure du $\roS$-terme
$t$.

\textit{Le cas de base}~: $t=n$, avec $n \in \NatP$.

Puisque $l,t$ sont \matchD\  alors $l$ \subf\  $t$ et conformément à la
Remarque~\ref{positionsFonctionnellesExp}, nous pouvons
considérer $l=1$.
\begin{itemize}

\item[a.] $t=n$, avec $t \in \NatP$

Nous obtenons $(l \meqqes t) = (1 \meqqes n)$ avec la solution $n.\ID$ et 
$(l \meqqes t') = (1 \meqqes t')$ avec la solution $t'.\ID$. Puisque 
$n\subs{s} \sigroTR t'$ alors la propriété est vérifiée.

\item[b.] Puisque $l$ est un indice $(l \meqqes t)$ ne peut pas échouer.

\end{itemize}

\textit{Induction}~:

Conformément à la Remarque~\ref{positionsFonctionnellesExp} le seul
cas où $l \neq 1$ est obtenu pour $t=f(t_1,\ldots,t_n)$. Les cas où $l=1$ sont
immédiats.
\begin{itemize}

\item[a.] Si $(l \meqqes t)$ n'échoue pas alors le terme $l$ doit être de la
forme $l=f(l_1,\ldots,l_n)$.

Si $t=f(t_1,\ldots,t_n)$ alors
$t\subs{s}=f(t_1,\ldots,t_n)\subs{s} \sigro f(t_1\subs{s},\ldots,t_n\subs{s})$ et
$t'=f(t_1',\ldots,t_n')$ avec $t_i\subs{s} \sigroTR t_i'$ et puisque les termes $l,t$ sont
\matchD\  alors $l_i,t_i$ sont \matchD.  L'équation $(l \meqqes t)$ est
transformé en $(\bigwedge_{i=1,\ldots,n} l_i \meqqes t_i)$ avec la solution
$u_1.\ldots.u_n.\ID$.
Par hypothèse d'induction, si la solution de $(l_i \meqqes t_i)$ est la substitution 
$u_1.\ldots.u_m.\ID$ alors la solution de $(l_i \meqqes t_i')$ est la substitution 
$v_1.\ldots.v_m.\ID$ avec $u_k\subs{s} \sigroTR v_k$, $k=1 \ldots m$.
L'équation $(l \meqqes t')$ est transformée en 
$(\bigwedge_{i=1,\ldots,n} l_i \meqqes t_i')$ avec la solution
$v_1.\ldots.v_n.\ID$ et par la linéarité de $l$ les solutions des 
$(l_i \meqqes t_i')$ sont disjoints et donc, la propriété est vérifiée.

\item[b.] Conformément à la Remarque~\ref{echecMatchExp}, l'échec peut être obtenu
seulement en appliquant la règle de filtrage $SymbolClash$ à la position de tête
ou à des positions plus profondes.

Nous pouvons avoir $l=f(l_1,\ldots,l_n)$ et soit $t=f(t_1,\ldots,t_n)$ et
$t'=f(t_1,\ldots,t_n)\subs{s}$, soit $t=g(t_1,\ldots,t_m)$ et
$t'=g(t_1',\ldots,t_m')$ avec $f \neq g$ et $t_i\subs{s} \sigroTR t_i'$.
Dans les deux cas $(l \meqqes t')$ échoue aussi. 
Si $l=f(l_1,\ldots,l_n)$, $t=f(t_1,\ldots,t_n)$ et $t'=f(t_1',\ldots,t_n')$
avec $t_i\subs{s} \sigroTR t_i'$ alors, la règle de filtrage $SymbolClash$
appliquée pour un problème $(l_i \meqqes t_i)$ mène à un échec de filtrage et
par induction $(l_i \meqqes t_i')$ échoue et ainsi, $(l \meqqes t')$ échoue.

\end{itemize}
}

Nous devons noter que le filtrage $(l \meqqes t')$ peut échouer même si le
filtrage $(l \meqqes t)$ n'échoue pas et donc, dans le premier point du
Lemme~\ref{equivMatch} nous supposons que $t'$ est suffisamment $\sigRo$-réduit pour
garantir que le filtrage $(l \meqqes t')$ n'échoue pas. Nous pouvons par exemple 
considérer le terme $t'$ tel que toutes les applications de substitution ont
été propagées par les règles d'évaluation du \sigroCal\  (Figure~\ref{sigmaro})
jusqu'aux positions des indices du terme initial $t$.

\LEM%--------------------------------------------------------------------------
\label{BasicCoherenceSigma}
\comment{ used in Lemma~\ref{BasicCoherenceGeneral}}

Etant donnés les $\roS$-termes $t,t',u,u'$ tels que $t \sigro t'$ et $\Del{t}{u}$,
alors il existe un $\roS$-terme $u'$ tel que $t' \delSig u'$ et $u \sigroTR u'$.
\begin{center}$~$
\xymatrix{%@C+10pt{
& t
\ar[dl]_{\sigroxy} \ar[dr]^-{\delxy}
& \\
t'
\ar@{.{>}}[dr]_{\delSigxy} && 
s
\ar@{.{>}}[dl]^-{\sigroTRxy} \\
& 
s'
&
}
\end{center}

\FLEM%--------------------------------------------------------------------------

\proof{
Puisque les radicaux $\sigro$ ne peuvent pas apparaître dans un radical $\del$ nous
devons seulement analyser les cas générés par une $\sigRo$-réduction. 

Nous devons mentionner d'abord que si nous considérons deux entiers (indices de
de Bruijn) $m,n$ tels que $m \le n$, alors $m\subs{\Uparrow^n(s)}=m$ et ceci est
montré par la dérivation suivante~:

\begin{tabular}{ll}
$m\subs{\Uparrow^n(s)}$ & 
$\stackrel{rvl1}{\longrightarrow} (m-1)\subs{\Uparrow^{n-1}(s) \circ \uparrow}$ \\
& $\ldots$ \\&
$\stackrel{rvl1}{\longrightarrow} 1\subs{\Uparrow^{n-m+1}(s) \circ \uparrow^{m-1}}$ \\
& $\stackrel{fvl2}{\longrightarrow} 1\subs{\uparrow^{m-1}}$ \\
& $\stackrel{vs1}{\longrightarrow} m$
\end{tabular}

Il est facile de prouver par induction que pour tout terme $t$ en $\sigRo$-forme
normale qui contient seulement des indices inférieurs ou égal à $n$ nous avons
$t\subs{\Uparrow^n(s)} \sigroTR t$.

Nous considérons les $\roS$-termes $l,t,r$ et $t'$ et une substitution $s$ tels
que $l,t$ sont \matchD\  et $t\subs{s} \sigroTR t'$ et nous montrons que si
$\{\sigma\}=\Sl(l \meqqes t)$ et $\{\mu\}=\Sl(l \meqqes t')$, alors il existe un
$\roS$-terme $w$ tel que~:
\begin{center}$~$
\xymatrix{%@C+10pt{ 
r \subs{\Uparrow^n(s)}\subs{\mu} 
\ar@{.{>}}[dr]_{\sigRo^*} 
&& 
r \subs{\sigma}\subs{s}
\ar@{.{>}}[dl]^-{\sigRo^*} \\
& w &
}
\end{center}

Si nous considérons $\sigma=\{u_1.\ldots.u_n.\ID\}$ et 
$\mu=\{v_1.\ldots.v_n.\ID\}$, nous avons les réductions suivantes~:

\begin{tabular}{ll}
$r\subs{\Uparrow^n(s)}\subs{\mu}$ 
& 
$\stackrel{}{\longrightarrow}$
$\{r\subs{\Uparrow^n(s)}\subs{v_1.\ldots.v_n.\ID}\}$ \\
&
$\stackrel{clos}{\longrightarrow} \{r\subs{\Uparrow^n(s) \circ
v_1.\ldots.v_n.\ID}\}$ \\
& 
$\stackrel{le}{\longrightarrow}$
$\{r\subs{v_1.\Uparrow^{n-1}(s) \circ v_2.\ldots.v_n.\ID}\}$ \\
& 
$\ldots$ \\
& 
$\stackrel{le}{\longrightarrow}$
$\{r\subs{v_1.\ldots.v_{n-1}.\Uparrow(s) \circ v_n.\ID}\}$ \\
&
$\stackrel{le}{\longrightarrow} \{r\subs{v_1.\ldots.v_n.s}\}$
\end{tabular}

\begin{tabular}{ll}
$r\subs{\sigma}\subs{s}$ 
&
$\stackrel{}{\longrightarrow}$ 
$\{r\subs{u_1.\ldots.u_n.\ID}\}\subs{s}$ \\
&
$\stackrel{set}{\longrightarrow}$
$\{r\subs{u_1.\ldots.u_n.\ID}\subs{s}\}$ \\
&
$\stackrel{clos}{\longrightarrow} \{r\subs{u_1.\ldots.u_n.\ID \circ
  s}\}$ \\
& $\stackrel{map}{\longrightarrow}$
$\{r\subs{u_1\subs{s}.u_2\ldots.u_n.\ID \circ s}\}$ \\
&
$\ldots$ \\
& 
$\stackrel{map}{\longrightarrow}$
$\{r\subs{u_1\subs{s}.\ldots.u_n\subs{s}.s}\}$ 
\end{tabular}\\
et puisque $u_i\subs{s} \sigroTR v_i$ (Lemme~\ref{equivMatch}) nous
obtenons le résultat souhaité.

Par conséquent, si nous considérons les mêmes notations que précédemment et si 
$r \del r'$ alors nous montrons immédiatement qu'il existe un $\roS$-terme $w$
tel que~:
\begin{center}$~$
\xymatrix{%@C+10pt{ 
([l \ra_n r](t))\subs{s} 
\ar[d]_{\delxy} \ar[r]^-{\sigroxy}
& 
[(l \ra_n r)\subs{s}](t\subs{s})
\ar@{.{>}}[d]^-{\sigroxy} 
\\
\{r'\subs{\sigma}\}\subs{s}
\ar@{.{>}}[d]_{\sigroxy}
&
[l\subs{\Uparrow^n(s)} \ra_n r\subs{\Uparrow^n(s)}](t\subs{s})
\ar@{.{>}}[d]^-{\sigroTRxy}
\\
\{r'\subs{\sigma}\subs{s}\}
\ar@{.{>}}[d]_{\sigroTRxy}
& 
[l \ra_n r\subs{\Uparrow^n(s)}](t')
\ar@{.{>}}[d]^-{\delxy}
\\
w
& 
\{r'\subs{\Uparrow^n(s)}\subs{\mu}\}
\ar@{.{>}}[l]^-{\sigroTRxy}
}
\end{center}

Si le filtrage $(l \meqqes t)$ échoue alors le filtrage $(l \meqqes t\subs{s})$
échoue et le lemme est clairement vrai.

Pour les règles \rname{Congruence} et \rname{Congruence\_fail} les diagrammes
sont similaires et nous le montrons seulement pour le premier cas~:
\begin{center}$~$
\xymatrix{%@C+10pt{ 
([f(u_1,\ldots,u_n)](f(v_1,\ldots,v_n)))\subs{s} 
\ar[d]_{\delxy} \ar[r]^-{\sigroxy}
& 
[f(u_1,\ldots,u_n)\subs{s}](f(v_1,\ldots,v_n)\subs{s})
\ar@{.{>}}[d]^-{\sigroTRxy} \\
\{f([u_1](v_1),\ldots,[u_n](v_n))\}\subs{s}
\ar@{.{>}}[dr]_{\sigroTRxy} &
[f(u_1\subs{s},\ldots,u_n\subs{s})](f(v_1\subs{s},\ldots,v_n\subs{s}))
\ar@{.{>}}[d]^-{\delxy} &
\\
&
\{f([u_1\subs{s}](v_1\subs{s}),\ldots,[u_n\subs{s}](v_n\subs{s}))\}
}
\end{center}
}

\LEM%--------------------------------------------------------------------------
\label{ConfDelSigma}
\comment{ used in Theorem~\ref{roSigConf}}

La relation $\delSig$ est fortement confluente~:
\begin{center}$~$
\xymatrix{%@C+10pt{
&
t_0
\ar[dl]_{\delSigxy} \ar[dr]^-{\delSigxy}
& \\
t_1
\ar@{.{>}}[dr]_{\delSigxy} && 
t_2
\ar@{.{>}}[dl]^-{\delSigxy} \\
&t_3&
}
\end{center}
\FLEM%--------------------------------------------------------------------------

\proof{

En utilisant la cohérence de $\del$ avec $\sigro$ (Lemme~\ref{BasicCoherenceSigma})
et la terminaison de $\sigro$ (Lemme~\ref{termSigma}) nous obtenons facilement
par induction sur la réduction $\sigro$~:
\begin{center}$~$
\xymatrix{%@C+10pt{
& t
\ar[dl]_{\sigroTRxy} \ar[dr]^-{\delxy}
& \\
t'
\ar@{.{>}}[dr]_{\sigroTRxy~\delxy~\sigroTRxy} && 
s
\ar@{.{>}}[dl]^-{\sigroTRxy} \\
& 
s'
&
}
\end{center}

En utilisant ce dernier résultat et la confluence des deux relations $\del$ et $\sigro$,
nous obtenons la confluence de $\delSig$ par induction sur le nombre de
réductions $\sigro$~:
\begin{center}$~$
\xymatrix@C+30pt@R+30pt{
\ar[r]^-{\delxy} 
\ar[d]_{\delxy} 
&
\ar[r]^-{\sigroTRxy}
\ar[d]_{\delxy}
&
\ar[d]^-{\sigroTRxy~\delxy~\sigroTRxy}
\\
\ar[r]^-{\delxy} 
\ar[d]_{\sigroTRxy} 
&
\ar[r]^-{\sigroTRxy}
\ar[d]^-{\sigroTRxy}
&
\ar[d]^-{\sigroTRxy}
\\
\ar[r]_{\sigroTRxy~\delxy~\sigroTRxy}
&
\ar[r]_{\sigroTRxy}
&
}
\end{center}

\begin{center}$~$
\xymatrix@C+30pt@R+30pt{
\ar[r]^-{\sigroTRxy}
\ar[d]_{\sigroxy}
&
\ar[r]^-{\delxy} 
\ar[d]_{\sigroTRxy} 
&
\ar[r]^-{\sigroTRxy}
\ar[d]_{\sigroTRxy}
&
\ar[d]^-{\sigroTRxy}
\\
\ar[r]^-{\sigroTRxy} 
\ar[d]_{\sigroTRxy~\delxy~\sigroTRxy} 
&
\ar[r]^-{\sigroTRxy~\delxy~\sigroTRxy}
\ar@{{}{}}[d]^-{induction}
&
\ar[r]^-{\sigroTRxy}
&
\ar[d]^-{\sigroTRxy~\delxy~\sigroTRxy}
\\
\ar[rrr]_{\sigroTRxy~\delxy~\sigroTRxy}
&
&
&
}
\end{center}
}

Nous avons prouvé la confluence de la relation $\setSig$ et la confluence forte
de la relation $\delSig$. Il nous reste donc à montrer la cohérence entre ces
deux relations. Pour cela nous avons besoin d'un résultat de cohérence entre les 
relations $\del$ et $\congr$ modulo la relation $\sigro$ similaire à celui
obtenu dans le Lemme~\ref{BasicCoherence} pour le \roCalE.
La preuve de ce dernier lemme aussi bien que celles des lemmes auxiliaires
utilisés sont facilement adaptées pour le cas de substitutions explicites.

\LEM%--------------------------------------------------------------------------
\label{BasicCoherenceSet}
\comment{ used in Lemma~\ref{BasicCoherenceGeneral}}

Etant donnés les $\roS$-termes $t,t',u,u'$ tels que $\Cong{t}{t'}$ et $\Del{t}{u}$,
alors il existe un $\roS$-terme $u'$ tel que $t' \setSigTR \del \setSigTR u'$
et $u \setSigTR u'$.
\begin{center}$~$
\xymatrix{%@C+10pt{
& t
\ar[dl]_{\congrxy} \ar[dr]^-{\delxy}
& \\
t'
\ar@{.{>}}[dr]_{\setSigTRxy~\delxy~\setSigTRxy} && 
u
\ar@{.{>}}[dl]^-{\setSigTRxy} \\
& 
u'
&
}
\end{center}
\FLEM%--------------------------------------------------------------------------

\proof{
La preuve est très similaire à celle du Lemme~\ref{BasicCoherence} du
\roCal\  mais dans le cas du \roSig\  l'application de substitution est
décrite par les règles d'évaluation du \sigroCal\  et les réductions
correspondantes sont illustrées explicitement dans le diagramme du lemme.
}

\LEM%--------------------------------------------------------------------------
\label{BasicCoherenceGeneral}
\comment{ used in Theorem~\ref{roSigConf}}

Les relations $\delSig$ et $\setSig$ satisfont le diagramme de cohérence suivant~:
\begin{center}$~$
\xymatrix{%@C+10pt{
& t
\ar[dl]_{\setSigxy} \ar[dr]^-{\delSigxy}
& \\
t'
\ar@{.{>}}[dr]_{\setSigTRxy~\delSigxy~\setSigTRxy} && 
s
\ar@{.{>}}[dl]^-{\setSigTRxy} \\
& 
s'
&
}
\end{center}
\FLEM%--------------------------------------------------------------------------

\proof{

En utilisant la cohérence de $\del$ avec $\sigro$ (Lemme~\ref{BasicCoherenceSigma})
et la terminaison de $\sigro$ (Lemme~\ref{termSigma}) nous obtenons facilement
par induction sur la réduction $\sigro$~:
\begin{center}$~$
\xymatrix{%@C+10pt{
& t
\ar[dl]_{\sigroTRxy} \ar[dr]^-{\delxy}
& \\
t'
\ar@{.{>}}[dr]_{\sigroTRxy~\delxy~\sigroTRxy} && 
s
\ar@{.{>}}[dl]^-{\sigroTRxy} \\
& 
s'
&
}
%$~~~~~~$
et par la confluence de $\sigro$ (Lemme~\ref{confSigma}) 
%$~~~~~~$
\xymatrix{%@C+10pt{
& t
\ar[dl]_{\sigroTRxy} \ar[dr]^-{\delSigxy}
& \\
t'
\ar@{.{>}}[dr]_{\delSigxy} && 
s
\ar@{.{>}}[dl]^-{\sigroTRxy} \\
& 
s'
&
}
\end{center}

De la même manière, en utilisant la cohérence de $\congr$ avec $\sigro$
(Lemme~\ref{BasicCoherenceSS}) et la terminaison de $\sigro$ nous obtenons~:
\begin{center}$~$
\xymatrix{%@C+10pt{
& t
\ar[dl]_{\congrxy} \ar[dr]^-{\sigroTRxy}
& \\
t'
\ar@{.{>}}[dr]_{\sigroTRxy} && 
s
\ar@{.{>}}[dl]^-{\sigroTRxy~\congrxy~\sigroTRxy} \\
& 
s'
&
}
\end{center}

Par la confluence de $\sigro$ et la terminaison de $\setSig$
(Lemme~\ref{setSigterm}) nous pouvons étendre le diagramme précédent à
\begin{center}$~$
\xymatrix{%@C+10pt{
& t
\ar[dl]_{\setSigxy} \ar[dr]^-{\sigroTRxy}
& \\
t'
\ar@{.{>}}[dr]_{\sigroTRxy} && 
s
\ar@{.{>}}[dl]^-{\setSigxy} \\
& 
s'
&
}
$~~~~~~$
et respectivement
$~~~~~~$
\xymatrix{%@C+10pt{
& t
\ar[dl]_{\setSigTRxy} \ar[dr]^-{\sigroTRxy}
& \\
t'
\ar@{.{>}}[dr]_{\sigroTRxy} && 
s
\ar@{.{>}}[dl]^-{\setSigTRxy} \\
& 
s'
&
}
\end{center}

En utilisant les diagrammes ci-dessus et le Lemme~\ref{BasicCoherenceSet} nous
obtenons la propriété souhaitée par induction sur le nombre de réductions
$\sigro$~:
\begin{center}$~$
\xymatrix@C+30pt@R+30pt{
\ar[rrr]^-{\delxy} 
\ar[d]_{\congrxy} 
&&&
\ar[r]^-{\sigroTRxy}
\ar[d]_{\setSigTRxy}
&
\ar[d]^-{\setSigTRxy}
\\
\ar[r]^-{\setSigTRxy} 
\ar[d]_{\sigroTRxy} 
&
\ar[r]^-{\delxy} 
\ar[d]^-{\sigroTRxy}
&
\ar[r]^-{\setSigTRxy} 
\ar[d]^-{\sigroTRxy}
&
\ar[r]^-{\sigroTRxy}
\ar[d]^-{\sigroTRxy}
&
\ar[d]^-{\sigroTRxy}
\\
\ar[r]_{\setSigTRxy} 
&
\ar[r]_{\delSigxy} 
&
\ar[r]_{\setSigTRxy} 
&
\ar[r]_{\sigroTRxy}
&
}
\end{center}

\begin{center}$~$
\xymatrix@C+30pt@R+30pt{
\ar[rr]^-{\delSigxy}
\ar[d]_{\sigroxy}
&
&
\ar[d]^-{\sigroTRxy}
\\%----------------------
\ar[rr]^-{\delSigxy}
\ar[d]_{\setSigxy}^-{~~~~~~~~~~induction}
&
&
\ar[d]^-{\setSigTRxy}
\\%----------------------
\ar[rr]_{\setSigTRxy\delSigxy\setSigTRxy}
&
&
}
\end{center}

\begin{center}$~$
\xymatrix@C+30pt@R+30pt{
\ar[r]^-{\sigroxy}
\ar[d]_{\setSigxy}
&
\ar[rr]^-{\delSigxy} 
\ar[d]_{\setSigxy}^-{~~~~~~~~~~induction}
&
&
\ar[d]^-{\setSigTRxy}
\\%----------------------
\ar[r]_{\sigroTRxy}
&
\ar[rr]_{\setSigTRxy\delSigxy\setSigTRxy} 
&
&
}
\end{center}

}

\TH%--------------------------------------------------------------------------
\label{roSigConf}
Le \roSig\  est confluent.
\FTH%--------------------------------------------------------------------------

\proof{

Nous appliquons le Lemme de Yokouchi (Lemme~\ref{Yokouchi}) pour les relations
$\delSig$ et $\setSig$ et nous obtenons ainsi la confluence de la relation
$\setSigTR\delSig\setSigTR$.

Les propriétés suivantes sont vérifiées pour les deux relations~:
\begin{itemize}

\item[] $\setSig$ est confluente et terminante (Lemme~\ref{setSigterm}),

\item[] $\delSig$ est fortement confluente (Lemme~\ref{BasicCaseSigma}),

\item[] $\delSig$ et $\setSig$ vérifient le diagramme de Yokouchi 
(Lemme~\ref{BasicCoherenceGeneral}).

\end{itemize}

Nous procédons de la même manière que dans le Lemme~\ref{TransClosureRO} et nous
obtenons les inclusions $\roZ\subseteq\del\subseteq\roTR$, et donc, la relation
$(\setSigTR\roZSig\setSigTR)^*$ est la fermeture transitive de la relation
$\setSigTR\delSig\setSigTR$.

Ainsi, par le Lemme~\ref{diamondTransitive} nous obtenons la confluence de 
$\setSigTR\roZSig\setSigTR$.
}

\section{Le \roSig\   comme calcul de réécriture d'ordre supérieur}
%=============================================================================== 

Nous avons présenté le \roSig\   comme une version explicite du \roCal\  où
l'application de substitution est définie au niveau objet calcul. Nous avons
adapté le $\sigma_{\Uparrow}$-calcul de substitution utilisé dans le
$\lambda\sigma_{\Uparrow}$-calcul \footnote{présenté brièvement dans dans la
Section~\ref{lambdaSigma}} pour la syntaxe du \roSig\  et nous avons utilisé une
approche similaire à celle proposée dans~\cite{CurienHardinLevy-JACM96} pour
prouver la confluence du calcul ainsi obtenu. Bruno Pagano a introduit dans sa
thèse les \textit{``eXplicit Rewrite Systems''}~\cite{PaganoThesis}, des
systèmes de réécriture d'ordre supérieur utilisant le $\sigma_{\Uparrow}$-calcul
comme mécanisme de substitution et a donné des conditions assurant la confluence
de tels systèmes. La preuve de confluence d'un XRS revient ainsi à tester que
les règles de réécriture vérifient certaines propriétés, parfois simplement
structurelles ou locales. Nous avons donc représenté le \roSig\  comme un XRS et
analysé les conditions de confluence dans ce cas.

\subsection{eXplicit Rewrite Systems - XRSs}
%=============================================================================== 

L'objectif des XRSs est de définir des extensions du
$\lambda\sigma_{\Uparrow}$-calcul permettant d'ajouter de nouveaux lieurs et
donc, d'associer à certains symboles de la signature des règles de propagation
plus générales que celles utilisées dans le
$\lambda\sigma_{\Uparrow}$-calcul. Le calcul de substitution pour une signature
quelconque est ainsi inspiré du $\sigma_{\Uparrow}$-calcul.

\DEF%------------------------------------------------------------------------
Etant donnée une signature $\Gamma$, l'ensemble de termes $\Gamma_{\Uparrow}$
est définie par~:

\begin{tabular}{llll}
& \\
\textBNF{Termes} ~~~~ & 
$t$ &::=& $x_t ~~|~~ n ~~|~~ f(t,\ldots,t) ~~|~~ t\subs{s}$\\
& \\
\textBNF{Substitutions} ~~~~ & 
$s$ &::=& $x_s ~~|~~ \ID ~~|~~ \uparrow ~~|~~ \Uparrow(s) ~~|~~ t.s ~~|~~ s \circ s$\\
& \\
\end{tabular}\\
où $n \in \NatP$ et $f \in \FF$.
\FDEF%------------------------------------------------------------------------

Un symbole $f$ de $\Gamma$ peut être un symbole du premier ordre ou un lieur
(symbole d'ordre supérieur) et ceci est décrit au niveau des règles du
calcul de substitution.
Pour chaque symbole $f$ de $\Gamma$ d'arité $n$, on associe un $n$-uplet
d'entiers $(p_1,\ldots,p_n)$ appelé l'{\em arité de liaison} de $f$ et une règle
$f_{\Uparrow}$~:
	$$f(t_1,\ldots,t_n)\subs{s} ~\longra{{f_{\Uparrow}}}~
	f(t_1\subs{\Uparrow^{p_1}(s)},\ldots,t_n\subs{\Uparrow^{p_n}(s)})$$

Un symbole dont l'arité de liaison est de la forme $(0,\ldots,0)$ est un symbole
du premier ordre. Un symbole dont l'arité de liaison contient un $p_i$ non-nul
est un symbole d'ordre supérieur. On peut remarquer que si on considère un
symbole $\lambda$ d'arité de liaison $(1)$ alors nous obtenons la règle $Lambda$
du $\sigma_{\Uparrow}$-calcul.

\DEF%------------------------------------------------------------------------
Etant donnée une signature $\Gamma$, la relation de réduction $\sigma_\Gamma$
est définie par l'ensemble de règles de réécriture du $\sigma_{\Uparrow}$-calcul
moins celles relatives aux symboles d'application et d'abstraction et par
les règles $f_{\Uparrow}$ pour tout symbole $f$ de $\Gamma$.\\
Le système de réécriture défini par les termes de $\Gamma_{\Uparrow}$ et par la
relation $\sigma_\Gamma$ est appelé $\Gamma_{\Uparrow}$-calcul.
\FDEF%------------------------------------------------------------------------

Le $\Gamma_{\Uparrow}$-calcul décrivant le calcul de substitution est enrichi
par un ensemble de règles de réécriture décrivant les calculs effectifs. Ces
règles de réécriture sont définies non seulement sur l'algèbre engendrée par
$\Gamma$ mais sur les termes de $\Gamma_{\Uparrow}$. On obtient ainsi les
\textit{systèmes de réécriture avec substitutions explicites}.

\DEF%------------------------------------------------------------------------
Etant donnée une signature $\Gamma$ et un ensemble de règles de réécriture $\RR$
sur $\Gamma_{\Uparrow}$ tel que toute règle vérifie les conditions suivantes~:
\begin{itemize}
\item le membre gauche est linéaire,
\item les membres gauche et droit sont de la sorte \textBNF{Terme},
\item le membre gauche est un terme de l'algèbre engendrée par $\Gamma$.
\end{itemize}
Le système de réécriture défini par $\Gamma_{\Uparrow}$ et la relation
$\sigma_\Gamma\cup\RR$ est appelé un \textit{système de réécriture avec
substitutions explicites} et noté {\em XRS[$\Gamma,\RR$]}.
\FDEF%------------------------------------------------------------------------

Une règle de réécriture du XRS$[\Gamma,\RR]$ est dite du premier ordre si son
membre droit est un terme de l'algèbre engendrée par $\Gamma$. Les autres règles
sont dites des règles d'ordre supérieur. L'ensemble de règles du premier ordre
est noté $\RR^1$ et l'ensemble de règles d'ordre supérieur est noté $\RR^>$.

Dans~\cite{PaganoCADE15} et~\cite{PaganoThesis}, Pagano montre plusieurs
théorèmes permettant de prouver la confluence d'un XRS si différentes conditions
sont vérifiées par les règles du XRS. 
L'objectif est d'utiliser des conditions sur $\RR$ simple à vérifier,
c'est-à-dire qui se ramènent à des propriétés structurelles ou locales. Nous
présentons ici un de ces théorèmes.

\TH%------------------------------------------------------------------------
[\cite{PaganoCADE15}] \label{confXRScade15}
Etant donnée une signature $\Gamma$ et les ensembles de règles de réécriture
$\RR^1$ et $\RR^>$ sur $\Gamma_{\Uparrow}$ tel que~:
\begin{itemize}
\item $\RR^>$ est orthogonal et orthogonal avec $\RR^1$,
\item toutes les règles de $\RR^1$ et de $\RR^>$ sont linéaires à gauche,
\item $\RR^1$ est confluent sur $\Gamma_{\Uparrow}$.
\end{itemize}
Le système de réécriture avec substitutions explicites 
{\em XRS[$\Gamma,\RR^1\cup\RR^>$]} est confluent.
\FTH%------------------------------------------------------------------------

\subsection{Le \roSig\  comme XRS}
%=============================================================================== 

On peut définir immédiatement un XRS représentant le
$\lambda\sigma_{\Uparrow}$-calcul en précisant que le symbole $\lambda$ a
l'arité de liaison (1) et l'ensemble de règles $\RR$ contient une seule règle
d'ordre supérieur~: $$(\lambda t)u ~~~ \longra{Beta} ~~~ t\subs{u.id}$$ En
utilisant les théorèmes de confluence pour les XRSs on peut montrer que le
système de réécriture avec substitutions explicites XRS[$\{\lambda,(..)\},Beta$]
(i.e. le $\lambda\sigma_{\Uparrow}$-calcul) est confluent.

Nous voulons donc utiliser une approche similaire et décrire le \roSig\  comme un 
XRS. Il faut donc définir l'arité de liaison de chaque opérateur du \roSig. En
prenant les arités de liaison $(0,\ldots,0)$ pour le symbole d'ensemble
($\{\}$) et pour tous les symboles $f$ de la signature du
\roSig, $(0,0)$ pour l'opérateur d'application ($[]()$) et $(n,n)$ pour
l'opérateur d'abstraction ($\ra_n$) nous obtenons les règles $set$,
$op$, $app$ et $lam$ du \sigroCal.

Nous obtenons ainsi le XRS[$\Gamma_\rho,\RR_{\rho}$] où $\Gamma_\rho$ est la
signature contenant les symboles d'une signature du premier ordre $\FF$ et les
opérateurs $[]()$, $\{\}$ et $\ra_n$ et
$\RR_{\rho}=\RR_{\rho}^1\cup\RR_{\rho}^>$ avec l'ensemble $\RR_{\rho}^1$
contenant les règles \rname{Congruence}, \rname{Congruence\_fail},
\rname{Batch}, \rname{Switch_R}, \rname{OpOnSet} et \rname{Flat} et l'ensemble
$\RR_{\rho}^>$ contenant la règle \rname{Fire}.

Ainsi, prouver la confluence du \roSig\  revient à prouver la confluence du
système de réécriture avec substitutions explicites XRS[$\Gamma_\rho,\RR_{\rho}$]
et il suffit donc de montrer que les conditions de confluence imposées pour les
règles de $\RR_{\rho}$ sont satisfaites.

Ces conditions ne sont pas vérifiées si aucune condition n'est imposée pour la
règle \rname{Fire}. Malheureusement les théorèmes de confluence pour les XRSs
ne sont pas applicables non plus si nous utilisons la règle conditionnelle
\rname{Fire_c} présentée dans la Section~\ref{proprRoSig} et imposant une
condition de termes \ready.

Nous pouvons renforcer les conditions de cette règle afin d'utiliser les
théorèmes de confluence et en particulier le Théorème~\ref{confXRScade15} mais
la condition d'orthogonalité est très forte et nécessite une condition dans la
règle \rname{Fire} permettant d'éviter les paires critiques entre cette règle et
les autres règles du calcul. Ceci implique l'application d'une règle de
réécriture $l \ra r$ à un terme $t$ seulement si tous les termes $l,r,t$ ne
contiennent pas d'ensemble et de radical, ce qui représente une restriction
forte par rapport aux objectifs que nous nous sommes fixé au départ.

%\DontWriteThisInToc  
\subsection*{Conclusion}
%================================================================
%~

Nous avons décrit dans ce chapitre une version du \roCalE\  avec substitutions
explicites, le \roSig. 

Nous avons étendu la syntaxe du \roCal\  en introduisant les définitions des
substitutions et un opérateur décrivant leur application. Le calcul de
substitution est une extension du
\mbox{$\sigma_{\Uparrow}$-calcul} \cite{CurienHardinLevy-JACM96} et il peut être obtenu comme
une instance du $\Gamma_{\Uparrow}$-calcul~\cite{PaganoThesis}.

Une preuve de la confluence du \roSig\ utilisant une stratégie restrictive
d'évaluation est obtenue immédiatement si on le considère comme un XRS et qu'on
utilise les théorèmes de confluence de XRSs~\cite{PaganoCADE15}. Puisque les
conditions qu'on doit imposer en utilisant cette méthode sont relativement
fortes, nous avons employé une approche inspirée de celle utilisée pour la
preuve de confluence du $\lambda\sigma_{\Uparrow}$-calcul, ce qui nous a permis
d'obtenir la confluence sous les mêmes conditions que pour le \roCalE.

%% file: conclusion.tex
%%%%%%%%%%%%%%%%%%%%%%%%%%%%%%%%%%%%%%%%%%%%%%%%%%%%%%%%%%%%
% \TLtopbookmark
\chapter*{Conclusion}
\label{chap.conclusion}
%%%%%%%%%%%%%%%%%%%%%%%%%%%%%%%%%%%%%%%%%%%%%%%%%%%%%%%%%%%%

Nous nous sommes consacrés au cours de ce travail à l'étude d'un calcul
suffisamment puissant pour décrire non seulement la réécriture avec des règles
conditionnelles mais aussi leur contrôle.  Nous avons ainsi proposé le
\roCal\  qui intègre les propriétés complémentaires de la réécriture du premier
ordre et du \laCal\  ainsi que des caractéristiques permettant d'exprimer le
non-déterminisme. Ce calcul nous permet la description des langages basés sur la
réécriture et nous avons analysé plus en détail la description des règles et
stratégies du langage \elan.

Au cours de la mise en oeuvre du \roCal\  nous avons été amenés à choisir parmi
plusieurs approches possibles pour la représentation des termes, pour les règles
d'évaluation, pour le système de types, etc.. Nous avons discuté nos principaux
choix et nous avons également présenté des approches différentes à celles
choisies.

\section*{Le calcul}
%================================================================

L'application d'une règle de réécriture n'est pas une opération explicite dans
la définition de la relation la réécriture. Si l'ensemble de règles de
réécriture est confluent et terminant, l'ordre et la position d'application des
règles ne sont pas significatives par rapport au résultat final. Néanmoins, on
est souvent amenés à utiliser des systèmes de réécriture où l'ordre et la
position d'application sont essentielles soit pour la valeur du résultat obtenu,
soit simplement pour l'efficacité de l'exécution. Nous avons donc considéré
l'application comme une opération au même niveau du calcul que l'abstraction,
c'est-à-dire que les règles de réécriture.

Puisque l'application est un objet du calcul, on peut l'utiliser dans la
construction de règles de réécriture obtenant ainsi des règles où les membres
gauche et droit peuvent être autre chose que de simples termes du premier
ordre. Les applications dans le membre droit d'une règle peuvent être vues comme
des opérations locales et en effet ce type de constructions est utilisé pour
décrire les règles de réécriture avec affectations locales. Grâce à
l'expressivité du filtrage l'utilisation des applications dans le membre droit
d'une règle permet aussi la représentation des règles de réécriture
conditionnelles. Nous nous sommes concentré dans cette thèse sur l'utilisation
des règles de réécriture avec un terme du premier ordre comme membre gauche mais
l'étude des règles construites sans aucune restriction fait partie de nos
perspectives.

Nous obtenons ainsi un calcul similaire au \laCal\  avec motifs où l'abstraction
est faite non seulement par rapport à une variable mais en considérant aussi le
contexte de la variable. Ce type d'abstraction nous donne une information
purement syntaxique sur le contexte de la variable mais ne peut pas exprimer
d'autres propriétés du contexte. Nous avons donc introduit comme paramètre du
\roCalT\  la théorie $T$ décrivant le comportement des symboles de la
signature. Ceci nous permet d'exprimer implicitement des propriétés pour les
symboles comme, par exemple, l'associativité et la commutativité de l'opérateur
``+'' de l'arithmétique.

Le filtrage utilisé pour lier les variables à leurs valeurs actuelles est
effectué dans la théorie $T$ et dans certains cas peut mener à des solutions
multiples et donc à des résultats différents. Puisque nous voulons mémoriser
tous les résultats possibles d'une application nous avons choisi de représenter
l'ensemble des résultats au niveau objet du calcul. Ainsi, un ensemble vide de
résultats représente un échec de filtrage et un ensemble ayant plus d'un élément
représente les résultats non-déterministes d'une application. Les ensembles
vides ainsi que les ensembles ayant plus d'un élément sont strictement propagés
dans le sens où une sous-réduction menant à un ensemble vide ou ayant plus d'un
élément conduit à un résultat vide ou ayant plus d'un élément pour la réduction
principale.

La représentation des résultats au niveau objet du calcul permet la
construction des règles de réécriture avec un ensemble comme membre droit et
donc, avec un comportement explicitement non-déterministe. Une stratégie qui
applique une des règles d'un ensemble de règles de réécriture d'une façon
non-déterministe est représentée directement dans le \roCal\  par l'ensemble
contenant les règles respectives.

Nous avons ainsi obtenu un calcul dont les objets sont construits en partant
d'une signature et en utilisant le symbole d'abstraction ``$\ra$'', le symbole
d'application ``$[~](~)$'' et les ensembles comme moyen d'encapsulation.  Le
résultat d'une réduction dans le \roCal\  est soit un ensemble vide représentant
l'échec de l'application, soit un singleton représentant un résultat
déterministe, soit un ensemble ayant plusieurs éléments représentant un choix
non-déterministe de résultats.

L'application d'une règle de réécriture $l \ra r$ à un terme $t$, écrite $[l \ra
r](t)$, consiste à vérifier que le terme $l$ subsume (filtre) le terme $t$ et si
c'est le cas, réécrire le terme $t$ en le terme $r$ où les variables sont
remplacées par les termes obtenus en filtrant $t$ avec $l$. Le filtrage entre
les termes $l$ et $t$ ainsi que l'application de substitution résultant de ce
processus sont des mécanismes décrits au niveau méta du \roCal.

L'algorithme de \textit{filtrage} est donc utilisé pour lier les variables à
leurs valeurs actuelles. Au cours de cette thèse nous nous sommes concentré sur
les propriétés du \roCalE, c'est-à-dire le \roCal\  avec un filtrage syntaxique.
Dans le cas général, le \roCalT\  peut utiliser toute théorie et en particulier
un filtrage équationnel ou un filtrage d'ordre supérieur avec motifs, mais si
dans ces cas le pouvoir d'expression est clairement supérieur au cas syntaxique,
il est plus difficile d'étudier les propriétés des calculs obtenus.

Comme dans tous les calculs utilisant des lieurs, nous utilisons dans le \roCal\
la substitution d'ordre supérieur et non le remplacement du premier ordre et
donc, nous utilisons l'\mbox{$\alpha$-conversion} pour éviter la capture des
variables.

\subsubsection*{Substitutions explicites}
%================================================================

Nous avons aussi proposé une version du calcul où l'application de substitution
est décrite au même niveau que les autres règles d'évaluation du calcul. En
s'inspirant des $\lambda$-calculs avec substitutions explicites, et du
$\lambda\sigma_{\Uparrow}$ en particulier, nous avons développé le \roCal\  avec
substitutions explicites, appelé le \roSig.

Nous avons étendu la syntaxe du \roCal\  en introduisant les définitions des
substitutions et un opérateur d'application de substitution. En utilisant le
$\lambda\sigma_{\Uparrow}$ comme point de départ, nous avons introduit des nouvelles
règles d'évaluation pour décrire le comportement des substitutions.

\subsubsection*{Confluence}
%================================================================

Nous avons étudié la confluence du calcul en se concentrant sur le cas du
\roCalE\  et nous avons remarqué immédiatement que cette propriété n'est pas
vérifiée, principalement à cause du filtrage limité et de la manipulation des
ensembles. La théorie de filtrage syntaxique est très simple mais dans le même
temps n'est pas assez puissante pour permettre suffisamment d'applications de
règles de réécriture. Afin de résoudre ce problème nous aurions pu utiliser un
filtrage d'ordre supérieur mais nous avons préféré garder le filtrage dans
une théorie vide et imposer des stratégies d'évaluation garantissant la
confluence.

Une première stratégie confluente consiste à réduire l'application d'une règle
de réécriture seulement si le sujet de l'application est un terme clos du
premier ordre. Nous évitons ainsi tous les problèmes induits par le filtrage sur
des termes contenant des radicaux et par l'utilisation des ensembles.
Cette stratégie est naturelle et simple à implanter mais relativement restrictive.

Nous avons donc introduit des conditions plus fines sur l'application d'une règle
de réécriture et nous avons montré que ces conditions garantissent la
confluence. De plus, les stratégies que nous avons proposées deviennent triviales
(c'est-à-dire n'imposent aucune restriction) pour des restrictions du \roCalE\
général à des calculs plus simples comme le \laCal.

En ce qui concerne le \roSig, nous nous sommes limité dans cette thèse à une
version avec substitutions explicites du \roCalE\  et, en s'inspirant des
approches utilisées dans~\cite{CurienHardinLevy-JACM96} et~\cite{PaganoCADE15},
nous avons montré que le \roSig\  est confluent dans les mêmes conditions que le
\roCalE.

\subsubsection*{Typage et terminaison}
%================================================================

La seconde propriété nous permettant de conclure à l'unicité des formes normales
est la terminaison.

Le \roCal\  n'est évidement pas terminant dans le cas non-typé. Afin de récupérer
cette propriété nous avons procédé comme d'habitude et nous avons imposé une
discipline plus stricte sur la formation des $\rho$-termes en introduisant une
information de type pour chaque terme.

Nous avons présenté un système de types pour le \roCalE\  permettant de typer un
$\rho$-terme ensemble par le type de ses éléments. Nous avons montré que la
réduction de tout terme \textit{bien typé} est finie et préserve le type du
terme initial.

\section*{Extension du calcul}
%================================================================

Dans le \roCal\  nous pouvons décrire explicitement l'application d'une règle de
réécriture à une certaine position d'un terme de la même façon que l'application
d'une abstraction est explicite dans le \laCal. Dans la réécriture cette
information d'application n'est pas explicite et nous voulons représenter ce
comportement dans le \roCal.

Contrairement à la réécriture où une règle peut être appliquée seulement aux
termes subsumés par le membre gauche de la règle, dans le \roCal\  l'application
d'une règle de réécriture peut échouer à cause d'un échec de filtrage.  Le
test d'un échec d'application s'avère une tâche plus difficile dans le \roCal\
et nous avons été amenés à introduire un nouvel opérateur permettant de décrire
cette propriété. Le calcul obtenu en ajoutant la description de cet opérateur
au \roCal\  est appelé le \roCalf.

En partant des opérateurs du \roCalf\  nous avons défini des opérateurs de
parcours de terme et un opérateur permettant la description de l'application
répétée d'un $\rho$-terme à un autre $\rho$-terme. Ces derniers opérateurs
nous permettent, par exemple, de définir des $\rho$-termes représentant la
normalisation \textit{innermost} et \textit{outermost} par rapport à un ensemble
de règles de réécriture.

Nous n'avons pas étudié en détail la confluence de ce calcul mais les stratégies
d'évaluation utilisées dans le cas de base peuvent être facilement adaptées pour
tenir compte des nouveaux opérateurs introduits dans le \roCalf.

\section*{Expressivité du calcul}
%================================================================

Le \roCal\  est conceptuellement simple tout en étant très expressif. Ceci nous a
permis de représenter les termes et les réductions du \laCal\  et de la
réécriture conditionnelle. Nous avons proposé une traduction entre les
$\lambda$-termes et les $\rho$-termes et nous avons montré que les réductions
des termes correspondants sont équivalentes modulo la manipulation des
ensembles, manipulation qui est triviale dans ce cas.

Nous avons présenté une méthode de construction d'un $\rho$-terme avec une
$\rho$-réduction similaire à la réduction d'un terme par rapport à un système de
réécriture. Un tel terme est construit en utilisant les étapes intermédiaires de
la réduction ou autrement dit, les termes de preuve de la logique de
réécriture. Puisque toute l'information de réduction est explicite dans les
$\rho$-termes ainsi obtenus, ces termes sont très compliqués et particulièrement
dans les cas où des règles de réécriture conditionnelles sont utilisées.
Les termes du \roCalf\  représentant des stratégies de normalisation nous ont permis
d'obtenir un codage naturel et plus concis de la réécriture conditionnelle.

En partant de la représentation des règles de réécriture conditionnelles nous
avons montré comment le \roCal\  peut être employé pour donner une sémantique à
l'application des règles du langage \elan. Puisque les règles de réécriture du
\roCal\  peuvent contenir des applications dans le membre droit, la
représentation des règles \elan\  avec affectations locales est immédiate, mais
la combinaison des méthodes utilisées pour les règles purement conditionnelles
et pour les règles contenant que des affectations locales doit être faite
soigneusement.  La construction de factorisation d'\elan\  permettant de définir
plusieurs réductions possibles pour un même terme de départ est naturellement
représentée en \roCal\  par une règle de réécriture avec un ensemble représentant
les réductions possibles comme membre droit.

Les stratégies \elan\  sont, dans la plupart des cas, représentées directement
dans le \roCal. En effet, l'expressivité du \roCal\  nous a permis de représenter
explicitement le mécanisme d'évaluation \elan, basé sur l'utilisation d'une
stratégie de normalisation \textit{innermost} pour les règles non-nommées et des
stratégies définies pour les règles nommées. Nous pouvons donc représenter un
programme \elan\  par un $\rho$-terme approprié~; le \roCal\  fourni donc une
sémantique opérationnelle complète au langage \elan.

\section*{Perspectives}
%================================================================

Nous voyons deux directions principales pour continuer ce travail~: d'une part
l'analyse des {\em propriétés} du \roCalT\  où $T$ est une théorie de filtrage plus
élaboré que la théorie syntaxique et d'autre part, l'utilisation et
éventuellement l'amélioration de l'{\em expressivité} du calcul afin de décrire
d'autres formalismes et langages. %\\

\subsubsection*{Propriétés du \roCalT}
%================================================================

Nous avons montré la {\em confluence} du \roCalE\  ainsi que la {\em terminaison}
du \roCalE\  typé et nous voulons étendre ces propriétés pour d'autres instances
du \roCalT\  général et principalement pour les cas des théories
équationnelles. \\

La preuve de la confluence du \roCalE\  a été réalisée en séparant les propriétés
qui utilisent les caractéristiques du filtrage employé et les propriétés
indépendantes de la théorie du filtrage. Les propriétés de la relation
\rname{Set} engendrée par les règles d'évaluation décrivant le comportement des
ensembles ne sont pas influencées par la théorie de filtrage utilisée mais la
confluence de la relation \rname{FireCong} décrivant l'application d'une règle
de réécriture dépend fortement des caractéristiques du filtrage utilisé. Afin de
montrer la confluence du \roCalT\  nous devons exhiber les conditions sur la
théorie $T$ de filtrage nous permettant de prouver la stabilité du nombre de
solutions par rapport à la réduction ainsi que de montrer que les
lemmes du cas syntaxique sont valides pour chaque solution.  La théorie de
filtrage intervient aussi dans la cohérence entre les relations
\rname{FireCong} et  \rname{Set} et le lemme montrant cette propriété dans le
cas syntaxique doit être étendu afin de prendre en compte un filtrage
éventuellement non-unitaire. 

Nous envisageons une description générique des conditions qui doivent être
imposées pour le filtrage dans la théorie $T$ afin d'obtenir la confluence du
\roCalT\  et ensuite montrer que ces conditions sont satisfaites pour des
théories particulières telles que l'associativité et la commutativité.

De la même manière, des conditions génériques pour le filtrage seront utilisées
pour la preuve de terminaison du \roCalT\  typé. En plus, l'extension dans un
cadre typé de l'algorithme de filtrage utilisé dans le cas non-typé doit
garantir que les substitutions obtenues sont bien bien typées.  Une autre
extension possible du système de types existant est la description du
polymorphisme. %\\

\subsubsection*{Expressivité du \roCalT}
%================================================================

L'analyse de l'{\em expressivité} du \roCal\  doit être certainement continuée. Nous
avons montré que le \roCal\  de base est suffisamment puissant pour exprimer
explicitement des réductions dans différents formalismes et nous avons introduit
de nouveaux opérateurs dans le \roCalf\  permettant d'exprimer certains
réductions d'une manière plus concise. Nous voulons, d'une part représenter
d'autres formalismes dans le \roCal\  et d'autre part simplifier la définition des
opérateurs du \roCalf. \\

On doit ainsi analyser la possibilité d'exprimer les opérateurs définis dans le
\roCalf\  d'une manière plus simple et éventuellement directe.  Un premier
challenge sera la représentation d'un opérateur testant l'échec en utilisant
seulement les opérateurs du \roCalE\  mais ceci semble une tâche difficile dans
un calcul où l'échec est une propriété propagée strictement.  Deuxièmement, la
plupart des opérateurs du \roCalf\  ont été définis en utilisant un opérateur de
point fixe inspiré du combinateur de point fixe de Turing du \laCal. Cet
opérateur n'utilise pas la puissance du filtrage et on peut se demander
si une construction plus astucieuse ne peut pas être réalisée en exploitant
toute les caractéristiques du \roCalE.

Nous avons montré que le \roCalE\  permet la représentation du \laCal\  et de la
réécriture mais aussi des langages basées sur ces formalismes. Le langage \elan\
en est un exemple et cette approche peut être appliquée à beaucoup d'autres
cadres, y compris des langages basés sur la réécriture comme
ASF+SDF~\cite{overview-ASF-BOOK}, CafeOBJ~\cite{FutatsugiN-IEEE97},
Maude~\cite{Maude-RwLg1996}, ML~\cite{MilnerML84} ou Stratego~\cite{Vis99} mais
également à des systèmes de production et systèmes de transitions
non-déterministes.  Les démonstrateurs de théorèmes utilisant la réécriture
représentent un autre domaine d'intérêt pour notre approche et plus
particulièrement nous envisageons l'utilisation du \roCal\  pour la
représentation des ``tactics'' et ``tacticals'' des prouveurs comme par exemple
LCF~\cite{paulson83,Gordon-Milner-Wadsworth-79}, HOL~\cite{gormel93} ou Coq~\cite{CoqMan96}.

En partant de la représentation du \laCal\  et de la réécriture du premier ordre
nous envisageons d'approfondir la liaison entre le \roCalT\  et les systèmes de
réécriture d'ordre supérieur. Si la formalisation de la correspondance entre le
\roCal\  et les extensions algébriques du \laCal\  doit être relativement facile,
la représentation des réductions d'un systèmes de réécriture avec abstracteur
comme les CRSs ou les HRSs semble plus compliquée et nécessite l'utilisation
d'un \roCalT\  avec une théorie $T$ d'ordre supérieur.

Nous avons utilisé les ensembles comme moyen de représenter le non-déterminisme
et nous avons mentionné que d'autres structures peuvent être utilisées. Par
exemple, si nous voulons représenter tous les résultats d'une application et non
seulement les résultats différents, alors des multi-ensembles doivent être
utilisés et si l'ordre des résultats est significatif alors une structure de
liste est plus appropriée. Nous avons donc entamé avec Claude Kirchner et Luigi
Liquori l'étude d'une autre description du \roCal\  ayant comme paramètre non
seulement la théorie de filtrage mais aussi la construction utilisée pour
structurer les résultats. Nous avons ainsi obtenu un calcul plus flexible et
nous avons déjà montré son pouvoir d'expressivité.  Plus précisément, nous avons
analysé la correspondance entre le \roCal\  et deux calculs qui ont fortement
influencé la recherche dans le domaine des paradigmes orientés objets~: le
\textit{``Object Calculus''} de Abadi et Cardelli~\cite{AC-book} et le
\textit{``Lambda Calculus of Objects''} de Fisher, Honsell et
Mitchell~\cite{FHM-94}. Nous avons défini les opérateurs des deux calculs en
utilisant les opérateurs du \roCal\  et nous avons montré qu'à toute réduction
d'un terme $t$ en $t'$ dans ces calculs il correspond la réduction dans le
\roCal\  de la $\rho$-translation du terme $t$ en la $\rho$-translation du terme
$t'$.  L'approche que nous avons proposée permet la représentation des objets
dans le style des deux calculs mentionnés mais aussi des objets plus élaborés
dont le comportement est décrit en utilisant le filtrage.  Nous envisageons de
continuer ce travail en montrant la confluence des instances du \roCal\
utilisées pour encoder les formalismes ci-dessus et en proposant un système de
types approprié dans ce contexte.\\

Sur un plan pratique les différentes instances du \roCal\  doivent être
implantées et utilisées comme outils de réécriture. Nous avons réalisé une
implantation en \elan\  du \roCalE\  et nous avons expérimenté différentes
stratégies d'évaluation. Nous voulons continuer dans cette direction et décrire
des paradigmes orientés objets en s'appuyant sur l'implantation du \roCalE.
Dans le cadre de l'étude des paradigmes orientés objets en réécriture, Hubert
Dubois a réalisé une version objet du langage \elan\  dont la sémantique
opérationnelle est fournie par le \roCal.

%% file: these.bbl
\newcommand{\etalchar}[1]{$^{#1}$}

%% file: abstractFrancais.tex
%\begin{FrenchAbstract}  

L'objet de cette thèse est l'étude d'un calcul permettant de décrire
l'application de règles de réécriture conditionnelles et de représenter les
résultats obtenus. Nous introduisons le calcul de réécriture, appelé aussi le
rho-calcul, qui généralise la réécriture du premier ordre et le lambda-calcul
tout en permettant d'exprimer le non-déterminisme.  Dans notre approche, l'opérateur
d'abstraction ainsi que l'opérateur d'application sont des objets du calcul. Le
résultat d'une réduction dans le calcul de réécriture est soit un ensemble vide
représentant l'échec de l'application, soit un singleton représentant un
résultat déterministe, soit un ensemble ayant plusieurs éléments représentant un
choix non-déterministe de résultats.

Au cours de cette thèse nous nous concentrons sur les propriétés du calcul de
réécriture utilisant un filtrage syntaxique pour lier les variables à leurs
valeurs actuelles. Nous définissons des stratégies d'évaluation garantissant la
confluence du calcul et nous montrons que ces stratégies deviennent triviales
pour des restrictions du calcul de réécriture général à des calculs plus simples
comme le lambda-calcul.  Le calcul de réécriture n'est pas terminant dans le cas
non-typé mais la terminaison forte est obtenue pour le calcul simplement typé.

Dans le calcul de réécriture étendu par un opérateur permettant de tester 
l'échec de l'application nous définissons des termes représentant la normalisation
\textit{innermost} et \textit{outermost} par rapport à un ensemble de règles de
réécriture. En utilisant ces termes, nous obtenons un codage naturel et concis
de la réécriture conditionnelle.  Enfin, à partir de la représentation des
règles de réécriture conditionnelles, nous montrons comment le calcul de
réécriture peut être employé pour donner une sémantique au langage \elan\ basé
sur l'application de règles de réécriture contrôlées par des stratégies.

  %\end{FrenchAbstract}

%% file: abstractAnglais.tex
%\begin{EnglishAbstract}  

This thesis is devoted to the study of a calculus that describes the application
of conditional rewriting rules and the obtained results at the same level of
representation. We introduce the rewriting calculus, also called the
rho-calculus, which generalizes the first order term rewriting and
lambda-calculus, and makes possible the representation of the non-determinism.
In our approach the abstraction operator as well as the application operator are
objects of calculus. The result of a reduction in the rewriting calculus is
either an empty set representing the application failure, or a singleton
representing a deterministic result, or a set having several elements
representing a not-deterministic choice of results.

In this thesis we concentrate on the properties of the rewriting calculus where
a syntactic matching is used in order to bind the variables to their current
values.  We define evaluation strategies ensuring the confluence of the calculus
and we show that these strategies become trivial for restrictions of the general
rewriting calculus to simpler calculi like the lambda-calculus. The rewriting
calculus is not terminating in the untyped case but the strong normalization
is obtained for the simply typed calculus.

In the rewriting calculus extended with an operator allowing to test the
application failure we define terms representing \textit{innermost} and
\textit{outermost} normalizations with respect to a set of rewriting rules.
By using these terms, we obtain a natural and concise description of the
conditional rewriting. Finally, starting from the representation of the
conditional rewriting rules, we show how the rewriting calculus can be used to
give a semantics to \elan, a language based on the application of rewriting
rules controlled by strategies.


%\end{EnglishAbstract}  